\documentclass[11pt,makeidx]{phdthesis}
\usepackage{a4wide}

\pdfoutput=1
\usepackage{times}

\usepackage{makeidx}     
\makeglossary
\makeindex

\usepackage[T1]{fontenc}
\usepackage[francais,english]{babel}
\usepackage{fancyheadings}
\usepackage{amsmath}
\usepackage{amssymb}
\usepackage{pstricks}
\usepackage{aas_macros}
\usepackage{epsfig}
\usepackage{subfigure}
\usepackage{graphicx}
\usepackage{epstopdf}
\usepackage{float}
\usepackage{rotating}
\usepackage{multirow}
\usepackage{comment}
\usepackage{captionhack}
\usepackage{epigraph}
\usepackage{multicol}
\usepackage{mdwlist}
\usepackage{enumerate}
\usepackage{dcolumn}
\usepackage{url}
\usepackage{natbib}
\usepackage{pdfpages}
\usepackage[normalem]{ulem}
\usepackage{wrapfig}
\usepackage[bookmarks=false,colorlinks=true, citecolor=magenta, linkcolor=blue, urlcolor=green]{hyperref}
\usepackage{minitoc}

\def\apjl{ApJL }

\def\apjs{ApJS }

\def\aap{A\&A }

        \input alphabet.tex     
        \input abrege.tex        
\newsavebox{\fminibox}
\newlength{\fminilength}

  \def\+{^\dagger}

\def\nequiv{\not\kern-.05em\equiv}
\def\egal{\kern-.5em=\kern-.5em}        
\def\propt{\kern-.2em\propto\kern-.2em} 

\def\intdouble{\int\kern-0.3em\int}
\def\inttriple{\int\kern-0.3em\int\kern-0.3em\int}

\def\rond#1{\overset{\kern-0.33em~_\circ}{#1}}
\def\rondit[#1]#2{\overset{\kern#1~_\circ}{#2}}

\graphicspath{{./png/}} 

\newenvironment{Abstract}
{\begin{center}\textbf{Abstract}%
 \end{center} \small \it \begin{quote}}
{\end{quote}}

\dominitoc

\begin{document}

\pagestyle{fancyplain}
\pagenumbering{roman}

\thispagestyle{empty}

\begin{center}

\begin{tabular*}{\textwidth}[c]{l@{\extracolsep{\fill}}r}
\hline
\hline
\\
{}&{\huge {\bf {H$_{\bf 2}$ MAGIE}}} \\
{}&{}\\
{}&{\LARGE {\bf {H$_{\bf 2}$ as a Major Agent to Galaxy Interaction and Evolution}}} \\
\\
\hline
\hline
\end{tabular*}

\vspace{1cm}
{\huge PhD THESIS} \\
\vspace{0.5cm}
{\Large by} \\
\vspace{0.5cm}
{\huge {\bf Pierre Guillard}} \\

 \vspace{1cm}
 \begin{figure}[!h]
\centering
 \includegraphics[width=9cm]{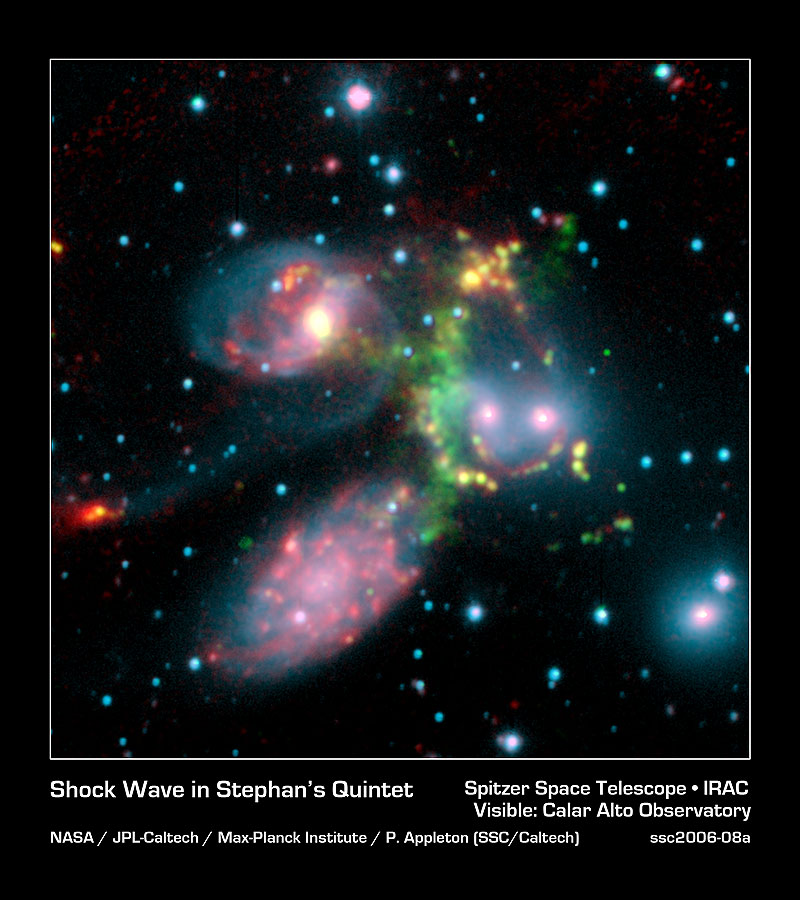}
 \label{fig_ssc2006-08a_medium}
 \end{figure}

 \vspace{1cm}

{\large
Institut d'Astrophysique Spatiale, Universit Paris-Sud 11\\ UMR 8617, Bt. 121, 91405 Orsay Cedex, France
}

 \end{center}

\newpage

 \thispagestyle{plain}
 \null
 \vspace{18cm}
 {\it Front page~:  The Universe is smiling! This spectacular picture shows the Stephan's Quintet, a compact group of interacting galaxies. A violent galaxy collision is occurring, which creates a giant shock wave highlighted in green (H$\alpha$) on the image. This is a composite image made up of optical, H$\alpha$ and infrared observations.}


\thispagestyle{empty}
\enlargethispage{1cm}

\begin{center}
\renewcommand{\arraystretch}{1.7} 
\begin{tabular}{ c l r}
 \multirow{2}*{\includegraphics[height=2cm]{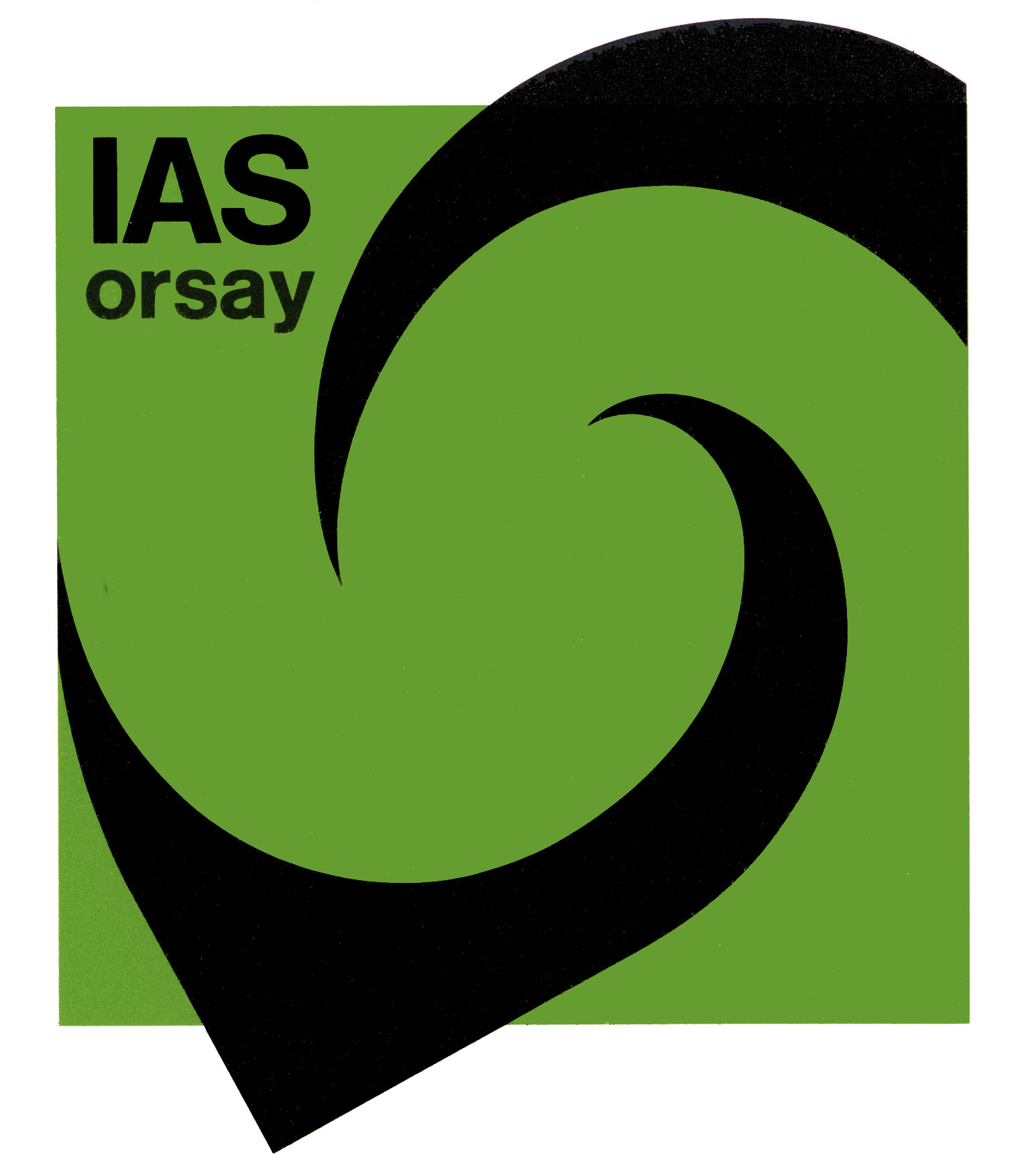}} & Institut d'Astrophysique Spatiale & Interstellar Matter and Cosmology Group \\
\cline{2-3}
& University of Paris-Sud 11 & \'Ecole doctorale d'Astrophysique d'\^Ile de France \\
\end{tabular}
\renewcommand{\arraystretch}{1} 

\vspace{0.9cm}

{\Huge {\bf ``H$_{\bf 2}$ MAGIE''}} \\
\vspace{0.3cm}
{\Huge {\bf H$_{\bf 2}$ as a Major Agent to Galaxy Interaction and Evolution
}}

\vspace{0.9cm}

{\Huge THESIS}

\vspace{0.7cm}
{\large submitted and publicly defended on November, the 12th, 2009} \\
\vspace{0.5cm}
{\large in fulfilment of the requirements for the } \\
\vspace{0.5cm}
{\LARGE {\bf Doctorat de l'Université de Paris-Sud XI}} \\
{\bf Spécialité Astrophysique et Intrumentation Associées} \\
\vspace{0.5cm}
{\large by} \\
\vspace{0.5cm}
{\LARGE {\bf Pierre Guillard}}
\vspace{1cm}

\begin{tabular}{ l l l }
  \multicolumn{3}{c}{{\large {\bf Composition of the jury}}} \\
\hline
\textit{President :} & Alain Abergel & IAS, University of Paris-Sud XI, Orsay\\

\textit{Referees:} & Cristina Popescu & University of Central Lancashire \\
								   & Amiel Sternberg & University of Tel-Aviv\\
\textit{Examiners:} & François Boulanger & IAS, Orsay, thesis supervisor \\
									  & Guillaume Pineau des Forêts & IAS, Orsay, co-supervisor \\
									  & Matthew Lehnert & GEPI, Obs. Paris-Meudon \\
									& Ronald J. Allen & STScI, Baltimore, USA\\
									& Philip N. Appleton & Herschel Science Center, IPAC, USA\\
\hline
\end{tabular}

\includegraphics[height=2cm]{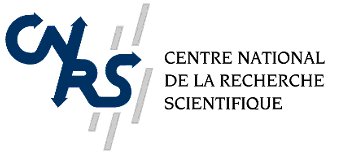}
\includegraphics[height=2cm]{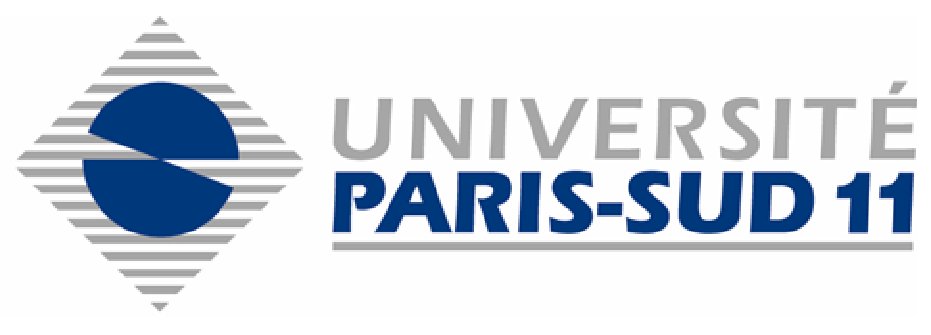}


 \end{center}
\cleardoublepage


\thispagestyle{empty}
\enlargethispage{1cm}

\begin{center}
\renewcommand{\arraystretch}{1.7} 
\begin{tabular}{ c l r}
 \multirow{2}*{\includegraphics[height=2cm]{figures/logos/logo_IAS.jpg}} & Institut d'Astrophysique Spatiale & Groupe Matière Interstellaire et Cosmologie \\
\cline{2-3}
& Université Paris-Sud 11 & \'Ecole doctorale d'Astrophysique d'\^Ile de France \\
\end{tabular}
\renewcommand{\arraystretch}{1} 

\vspace{0.9cm}

{\Huge {\bf ``H$_{\bf 2}$ MAGIE''}} \\
\vspace{0.3cm}
{\Huge {\bf L'Hydrogène moléculaire dans l'évolution des galaxies
}}

\vspace{0.9cm}

{\Huge TH\`ESE}

\vspace{0.7cm}
{\large présentée et soutenue publiquement  le 12 Novembre 2009} \\
\vspace{0.5cm}
{\large pour l'obtention du grade de} \\
\vspace{0.5cm}
{\LARGE {\bf Docteur de l'Université de Paris-Sud XI}} \\
{\bf Spécialité Astrophysique et Intrumentation Associées} \\
\vspace{0.5cm}
{\large par} \\
\vspace{0.5cm}
{\LARGE {\bf Pierre Guillard}}
\vspace{1cm}

\begin{tabular}{ l l l }
  \multicolumn{3}{c}{{\large {\bf Composition du jury}}} \\
\hline
\textit{Président :} & Alain Abergel & IAS, Université Paris-Sud XI, Orsay\\

\textit{Rapporteurs:} & Cristina Popescu & University of Central Lancashire \\
								   & Amiel Sternberg & Université de Tel-Aviv\\
\textit{Examinateurs:} & François Boulanger & IAS, Orsay, Directeur de thèse \\
									  & Guillaume Pineau des Forêts & IAS, Orsay, co-directeur \\
									  & Matthew Lehnert & GEPI, Obs. Paris-Meudon \\
									& Ronald J. Allen & STScI, Baltimore, USA\\
									& Philip N. Appleton & Herschel Science Center, IPAC, USA\\
									
\hline
\end{tabular}

\includegraphics[height=2cm]{figures/logos/LogoCNRSGrandb.jpg}
\includegraphics[height=2cm]{figures/logos/logo_ups.jpg}


 \end{center}


\include{thesis_dedication}

\cleardoublepage
\lhead[]{\fancyplain{}{\rightmark}}
\chead[\fancyplain{}{}]{\fancyplain{}{}}
\rhead[\fancyplain{}{\leftmark}]{\fancyplain{}{}}

\setcounter{tocdepth}{2}
\tableofcontents

\cleardoublepage
\addcontentsline{toc}{chapter}{Acknowledgements}
\chapter*{Acknowledgments}
\pagestyle{empty}

\vspace{-3cm} 

\epigraph{Art is I, science is we.}{Claude Bernard}

During these three years my work was extremely diversified and I had the opportunity to do widely different things such as (take a deep breath!) preparing proposals for observing time, getting married with you Carole, observing at the telescope, reducing data, going out with people I've met during schools or meetings, building empirical and analytical models, or use numerical codes to interpret the data, doubting,  working on the JWST/MIRI instrument and being part of such a huge project as MIRI, running, traveling and giving talks, teaching at the University, etc! I am aware of having been lucky and privileged, and obviously all of this would not have been possible without all the people I've been working or interacting with. So, a big big \begin{LARGE}thank you\end{LARGE}\dots

\begin{wrapfigure}{r}{110mm}
   \includegraphics[width=110mm]{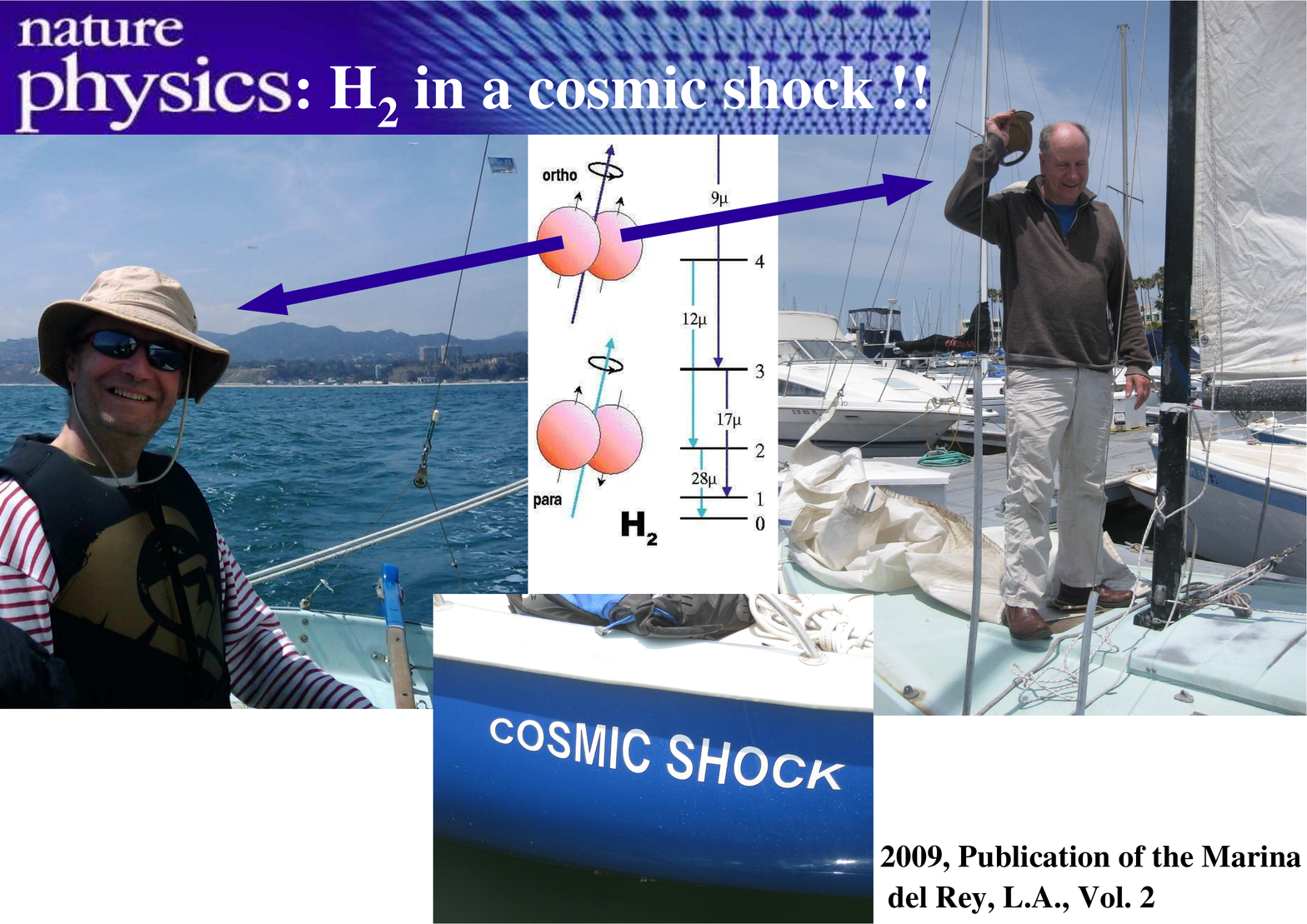}
\end{wrapfigure}
\dots to Franois Boulanger for his enthusiastic supervision of this thesis. Franois, it has been a pleasure to work with you during these three years.  I've learned so much science at your side.  I greatly benefited from your experience, your general vision of physical processes in the interstellar medium,  from your rigour and work methods. You always pushed me to the first line, ``at the shock front'', to do things by myself,  to be responsible.
After three years, I'm always impressed by your long-term vision of astrophysics and astropolitics, by your ability to raise new ideas and to question what we thought to understand. 
I greatly appreciated your humility, your optimism, and your breezy, zestful character. It is thus difficult to ask more from a supervisor! 

\dots to Guillaume Pineau des Forts, my co-supervisor. Despite your incredibly busy schedule, you were always available. I'm proud to had such an expert in shocks and chemistry on my side! I greatly appreciated the  clarity of your explanations, both during your masters courses and discussions about gas physics we had together.

\dots to the californians Phil Appleton, Patrick Ogle, and Michelle Cluver. Thanks for having invited me at Caltech and allowed me to join you for an observing run at Palomar to observe our favorite object, Stephan's Quintet. I'm grateful to Patrick and Phil for the postdoc position they offered me, and I'm particularly glad to continue our fruitful collaboration. Your team has opened a new ``H$_2$ route'', so let's explore it together\dots
Now, I'm looking forward to see your ``cosmic boat'' Phil! It looks really great! Thanks Michelle for the energy you put in reducing all these new data! I'm looking forward to meet you!

\dots to Alain Abergel for pushing me into the MIRI project. You have been always enthusiastic and giving value to the work I was doing. I do not forget the CEA team at Saclay, in particular Samuel Ronayette, Jrme Amiaux, Vincent Moreau, Pierre-Olivier Lagage and Jean-Louis Augures. I am particularly grateful to Thomas Rodet from the LSS for his skills in signal processing and deconvolution algorithms.

\dots to Vincent Guillet and Nicole Nesvadba. You were postdoc at the beginning of my PhD, and now both of you have permanent positions! Congratulations! Thanks a lot Vincent for all the nice discussions we had together about shocks and dust processing. I am also indebted to you for your help with the shock code. I also thanks Nicole for her dynamism, the efficient help she provided me to write my postdoc applications, and her careful reading of the chapters~\ref{chapter:H2_galaxies} and \ref{chapter:perspectives} of this manuscript. 

\dots to Anthony Jones, for having supervised my Masters Thesis. I greatly appreciated his kindness, and of course his hindsight view at dust evolution in the interstellar medium.
I particularly thank Anthony for having put me in contact with his colleagues Xander Tielens and Hugues Leroux, with whom we collaborated for my masters prohect.

\dots to all the people who invited me for giving talks in prestigious places, in particular Tom Abel in Standford University, Bruce Draine in Princeton, or David Neufeld at Johns Hopkins University/STScI. You listened to me with attention, and I greatly benefited from your questions and suggestions.

\dots to the IAS staff, and particularly to the members of the ``Matire Interstellaire et Cosmologie'' group. I enjoyed all the many seminars organized by the group. Relaxing coffee and lunch breaks also contributed to this friendly work atmosphere. 
Thanks Mathieu and Alexandre for the good time in Prag.
Thanks to the IAS computer scientists, for your efficiency solving problems and good humor.
I would like to thank Ghislaine Renoux for the hard-to-find articles she managed to get for me. Thanks also to the administration, and especially Alain Belvindrah for his efficient handling of all my late mission requests!

\dots to my teacher colleagues with whom I worked during these years. Thanks Alain Abergel, Laurent Verstrate, Herv Dole for the courses at the University.
I also thank Michel Dennefeld and Herv Dole for the OHP trainings and nice flying over Provence! Thanks a lot Michel for your invitation to an observing school in India.

\dots to my jury. I sincerely thank my referees, Amiel Sternberg and Cristina Popescu for examining my work and reading this manuscript with attention. They form a ideal combination, since Cristina is an expert in dust observations and modeling in extragalactic environments, and Amiel is an expert in gas physics and modeling of gas cooling processes.
I am very honoured to have such a team of scientists at my side!

\dots A glimpse to my BSc and MSc project supervisors, Karl-Ludwig Klein and Gilles Theureau. You were both models for me, and you contributed to shape my interest for astrophysics. 

\dots  to all the PhD students and postdocs at IAS, especially Benjamin B.\footnotemark[1]  (for nice relaxing moments playing tennis), Nicolas F. (thanks for your welcome in Pasadena!), Nicolas T.\footnotemark[1], Doug, Mathieu,  Eleonore\footnotemark[1] (I really enjoy your laughs!), Aurlie\footnotemark[1], Nathalie, John\footnotemark[1], Sophie\footnotemark[1] Pierre\footnotemark[1], Alexandre, Delphine, Benjamin S.\footnotemark[1] , etc. I am also indebted to Manuel Gonzalez for his help with the DUSTEM and PDR codes. 
Thanks Mathieu and Doug for being so cheerful, and for the good time spent in Les Houches!
I also thank all the students I've met during schools or conferences with whom we had good times together. Leo, Ccile, the Benjamins, Hans, Verena, thanks so much for the great time in Evora!

\footnotetext[1]{I'd like to thanks these actors for the delirium-kind-of-little-lipdub we made together. You can watch it \href{http://www.youtube.com/watch?v=Q5jaCwt8pUM}{here}.}

\dots to my family and friends. I will always be gratefull to my parents and in-laws.
Thanks to Damien, Nicolas, Benjamin G. (remember that sleepless night we spent writing our manuscripts\dots). Last but not least, a big kiss to my wife Carole, for her love, her smile, and unfailing support. I can't help thinking I was so lucky to meet you!


{\flushright Pierre Guillard\\ Orsay\\ \today\\ }

\cleardoublepage

\newcommand{\publ}{}
\pagestyle{fancyplain}
\setlength{\headrulewidth}{0.3pt}
\setlength{\footrulewidth}{0.0pt}
\setlength{\plainfootrulewidth}{0.0pt}
\setlength{\plainheadrulewidth}{0pt}
\renewcommand{\sectionmark}[1]{\markright{\it \thesection.\ #1}}
\renewcommand{\chaptermark}[1]{\markboth{
       \it \thechapter.\ #1}{}}
\lhead[\thepage]{\fancyplain{\publ}{\rightmark}}
\chead[\fancyplain{}{}]{\fancyplain{}{}}
\rhead[\fancyplain{}{\leftmark}]{\fancyplain{}{\thepage}}
\lfoot[]{}
\cfoot[]{}
\rfoot[]{}

\pagenumbering{arabic}
\mtcaddchapter

\thispagestyle{empty}

\vspace{3cm}

\begin{Abstract}

My main thesis work is to understand the origin of molecular Hydrogen (H$_2$) emission in active phases of galaxy evolution. Spitzer space telescope observations reveal a new class of H$_2$-luminous galaxies with enhanced H$_2$ line emission but where star formation is strongly suppressed. This is in sharp contrast with what is observed in standard star forming galaxies.

The Stephan's Quintet (SQ) galaxy collision is a striking example I initially focus on. We present a scenario and a detailed model to account for the presence of  H$_2$ in the SQ giant shock, to characterize its physical state, and to describe its role as a cooling agent of a violent phase of galaxy interactions. 
In this scenario, the dissipation of the mechanical energy of the collision produces a multiphase medium where molecular gas fragments coexist with a hot ($\sim 5 \times 10^6$~K), X-ray emitting plasma.
Our model quantifies the gas cooling, dust destruction, H$_2$ formation and emission in the postshock multiphase gas. The dynamical interaction between the ISM phases drives a cycle where  H$_2$ is formed out of atomic gas that cools, and is excited repeatedly before being destroyed. A cascade of energy is associated with this cycle, in which the mechanical energy powers supersonic turbulence within the molecular gas. The H$_2$ emission is associated with the dissipation of this turbulent energy.

New results of mid-infrared and radio observations in the SQ shock are presented. These observations reveal that dust and CO emission gas is associated with the warm ($\gtrsim 100$~K) H$_2$ seen by Spitzer, and that this gas is in an unusual physical state where star formation is suppressed. 
In addition, to test the scenario proposed for the formation of H$_2$ in the SQ shock, I carry on a detailed observational study and modeling of the dust emission from the H$_2$ gas. Observational perspectives with the Herschel satellite are discussed. 

These observations suggest that H$_2$ contributes significantly to the energy bugdet of galaxies which are in key phases of their evolution (galaxy interaction, gas accretion in galaxy clusters, starburst or AGN feedback).
My thesis work is a first step to understand the role that  molecular gas plays in  galaxy evolution. Our model developped for SQ is extended to the context of radio galaxies, which allow for the first time to peer at the impact of the AGN-driven jet on the multiphase ISM of the host galaxy. A natural extension of this work is the characterization of the energetics of galactic winds (in the M82 starburst galaxy for instance) and in AGN-driven winds recently discovered in high-redshift radio-galaxies. This thesis includes the tools to perform a detailed modeling of Spitzer and upcoming Herschel data. 

Besides this work, as a member of the JWST/MIRI consortium, I also report my contribution to the optical performance tests of the \textit{MIRI} instrument, which will extend the study of H$_2$-luminous galaxies to high redshifts.
The observational and theoretical work presented in this manuscript may help to develop a phenomenological ''recipe'' of the impact of H$_2$ on the energetics of galaxy evolution. This work will certainly be helpful for the preparation of future observing programs aiming at testing this phenomenology directly, thanks to spectroscopy of high-redshift galaxies with the \textit{JWST} and \textit{SPICA} missions. 

\end{Abstract}

\thispagestyle{empty}
%
%
%
%
%
%
%
%
%
%
%


\begin{Abstract}

Ce travail est dédié à la compréhension de l'émission du dihydrogène (H$_2$) dans les phases actives de l'évolution des galaxies. La découverte d'une nouvelle population de sources extragalactiques avec un spectre dans
l'infrarouge moyen dominé par les raies d'émission de H$_2$ est un résultat inattendu des
observations spectroscopiques du satellite \textit{Spitzer}. La faiblesse des bandes
d'émission des poussières et des raies du gaz ionisé par rapport à celles de H$_2$ indique la présence de grandes
quantités de gaz H$_2$ (jusqu`à $10^{10}\,$M$_{\odot}$ au centre des amas de galaxies) avec peu de formation
d'étoiles, contrairement à ce qui est observée dans les galaxies standard.

Une interprétation physique de l'émission H$_2$ associée à une collision à grande vitesse (1000 km/s) entre galaxies dans le Stephan's Quintet (SQ) est d'abord présentée. La dissipation de l'énergie cinétique de la collision crée un
milieu multi phases où des fragments de gaz moléculaire coexistent avec un
plasma de gaz chaud ($\sim 5 \times 10^6$~K) émetteur en rayons X. Notre interprétation relie la structure multi phases du gaz post-choc à la structure en densité du gaz pré-choc. L'interaction dynamique entre les
phases entretient un cycle où H$_2$ se forme à partir de gaz atomique chaud qui se refroidit
puis est excité de manière répétée avant d'être détruit. A ce cycle est associée une cascade
énergétique où l'énergie cinétique du gaz alimente une turbulence supersonique dans le gaz
moléculaire. Le rayonnement H$_2$ est associé à la dissipation de l'énergie turbulente. 

Les résultats de nouvelles observations moyen-infrarouge et radio dans le choc de SQ sont présentés. L'émission de la poussière et du gaz CO associé au gaz H$_2$ est détectée. Le gaz CO est extrêmement turbulent, ce qui pourrait expliquer pourquoi la formation stellaire est si peu efficace dans cet environnement. Pour tester notre interprétation de l'émission de H$_2$, les résultats de la modélisation de l'émission de la poussière associée au gaz H$_2$, ainsi que les perspectives observationelles apportées par le satellite \textit{Herschel}, sont discutés. 

Ces observations et ce travail théorique inscrivent l'étude du gaz moléculaire dans le cadre de la formation et de l'évolution des galaxies. Les mêmes caractéristiques d'émission H$_2$ sont observées dans les interactions entre galaxies, la rétroaction de la formation stellaire et celle des noyaux actifs de galaxies sur le milieu interstellaire, ainsi que l'accrétion de gaz dans les amas. Un dénominateur commun de ces phases violentes de l'évolution des galaxies est la libération d'énergie mécanique en
quantité suffisante pour affecter globalement le milieu interstellaire.
Cette interprétation est étendue à l'émission H$_2$ des radio-galaxies où le jet relativiste est la source d'énergie mécanique. Dans les deux cas, le gaz moléculaire apparaît comme un acteur de l'évolution dynamique des galaxies en amont de la formation stellaire.

Cette thèse présente également un travail d'analyse des tests de qualité optique réalisés au CEA sur l'instrument MIRI, une caméra moyen-infrarouge qui sera intégrée sur le futur télescope spatial JWST. Cet instrument permettra d'étendre ce travail de thèse à haut redshift, pour comprendre l'impact du gaz moléculaire sur l'évolution des galaxies lorsque l'Univers était plus jeune. Cette étude servira de base pour de futurs programmes d'observations avec le JWST.

\end{Abstract}
\addcontentsline{toc}{chapter}{Introduction}


\chapter*{Introduction}
\label{chapter:introduction_fr}


\epigraph{\`A ces belles nuits claires passées la tête en l'air \\ A ma famille et ma belle-famille pour m'avoir transmis l'amour de la nature, de la science, et la passion du vin \\ A ces joyeux gosiers qui me servent d'amis\\ A l'appellation d'origine contrôlée ``L'étoile'' \\ A Carole, pour son sourire}{}






\section*{L'Univers, c'est un peu comme un grand vin\dots}
\label{sec:intro_vin}

\PARstart{L}evez les yeux au ciel lors d'une belle nuit. Vous voyez d'abord sa belle robe scintillante. 
Quelques milliers d'étoiles, sur les cent milliards que contient notre galaxie. 
Regardez plus attentivement. Un nuage laiteux traverse le ciel,  il s'agit d'un des bras spiraux de notre ``Voie Lactée'', 
où la concentration des étoiles est telle que notre  \oe il ne peut les distinguer. 
Arrêtez-y vous un instant. Des reflets, des zones plus sombres se dévoilent. C'est le milieu interstellaire, ``entre les étoiles'', qui n'est pas vide, comme on l'a longtemps cru. Ce milieu est constitué d'un mélange de gaz (atomes, molécules) et de poussières, représentant environ $5-10$\% de la masse d'une galaxie.  

Tout apparait si tranquille, si calme. Vous commencez juste à sentir les parfums enivrants de votre nectar. Mais maintenant goutez-y. Pour cela, imaginez que vos yeux soient équipés des plus puissants instruments capables d'explorer tout le domaine du spectre électromagnétique. Vous découvrez une explosion de saveurs, un Univers bouillonnant, violent, explosif~!

Les galaxies sont loin d'être des objets ``statiques'', immuables. Elles évoluent d'un point de vue dynamique et chimique. La matière interstellaire joue précisément un rôle très important dans cette évolution, à petite échelle (celle d'une étoile), mais aussi à l'échelle de la galaxie toute entière. La gaz moléculaire, la phase la plus froide du milieu interstellaire (MIS), est le matériau de base pour la formation des étoiles. Celles-ci se forment dans les régions les plus denses des nuages moléculaires. Au cours de leur vie, les étoiles émettent du rayonnement UV, façonnent des ``bulles'' dans leur nuage parent, et lorsque ces bulles crèvent, le gaz chaud éjecté peut nourrir la phase ionisée du MIS. 
La chimie de ce vin évolue lentement avec le temps, développant des arômes de plus en plus complexes. Certaines étoiles massives explosent en supernovae, et participent à l'évolution chimique du gaz en enrichissant la phase chaude du MIS en éléments plus lourds que l'Hélium. Lorsque cette source d'ionization s'éteint, le gas chaud peut se recombiner, pour reformer du gaz neutre, et éventuellement du gaz moléculaire. Une galaxie est donc une machine ``écologique'' à recycler la matière interstellaire! Ce vin, si tranquille et si simple au premier coup d'\oe il, se révèle d'une extraordinaire richesse, d'une incroyable complexité! L'Univers, c'est comme un grand vin, plus on l'étudie, plus on l'apprécie!

L'évolution des galaxies, qui forme le cadre général de cette thèse, met en jeu des phases extrêmement énergétiques, comme par exemple la collision entre deux galaxies. Une quantité phénoménale d'énergie mécanique est libérée. Comment le MIS réagit-il à ces phénomènes énergétiques? Que devient ce cycle de la matière interstellaire, et comment la formation stellaire est-elle affectée? Ces questions sont au c\oe ur de ce travail, et sont clés pour comprendre comment les galaxies, telles que nous les voyons aujourd'hui, se sont formées.

\section*{Mais mon petit, qu'as-tu cherché?}
\label{sec:intro_contexte_problematique}

Le sujet de cette thèse a pour point de départ la découverte observationnelle d'une émission extraordinairement puissante de H$_2$, la molécule la plus simple et la plus abondante de l'Univers, dans un groupe compact de galaxies en interaction, appelé le Quintette de Stephan. Cette découverte observationnelle a été suivi par la détection d'autres objets, comme des radio-galaxies, ou des flots de gaz dans les amas de galaxies, présentant les mêmes caractéristiques spectroscopiques dans le domaine moyen-infrarouge. Le spectre de ces objets est dominé par les raies rotationelles de H$_2$, avec de faibles signatures de formation stellaire (émission de la poussière ou du gaz ionisé), contrairement à ce qui est observé dans les galaxies classiques (par exemple les spirales), formant des étoiles. De grandes quantités de gaz moléculaire sont détectés, mais avec étonnamment peu de formation stellaire. Une partie de mon travail a été de donner une interprétation physique de cette émission. 

\section*{et donc, qu'as-tu fait exactement?}

J'ai cherché à comprendre la formation et l'émission du H$_2$ dans ces phases actives de l'évolution des galaxies, en me focalisant d'abord sur le Quintette de Stephan. Ce travail a été très varié. J'ai développé et utilisé des outils théoriques pour calculer le refroidissement du gaz dans ces environnements astrophysiques, la formation et l'émission de H$_2$. Les observations existantes et ce travail de modélisation ont conduit à de nouvelles propositions d'observations. J'ai conduit plusieurs campagnes d'observations au radiotélescope de 30m de l'IRAM\footnote{Institut de Radio Astronomie Millimétrique} pour chercher le gaz CO (monoxyde de Carbone, gaz moléculaire ``froid'', i.e. $\sim 10-20$~K) associé au gaz H$_2$ ``chaud'' ($\sim 100-1000$~K) détecté dans le Quintette en infrarouge. Mon travail inclue la réduction de ces données radio et leur interprétation.

En parallèle, j'ai eu la chance de participer au projet spatial JWST, le James Webb Space Telescope, qui succèdera au télescope spatial Hubble. Une partie de mon travail est dédié aux tests de performance optique d'une caméra infrarouge, MIRI, qui sera un des 5 instruments embarqués à bord du JWST. Cette instrument sera sans nul doute le futur proche des observations du gaz moléculaire, notamment de H$_2$, dans le domaine moyen-infrarouge. Il permettra de tester si les galaxies lumineuses en H$_2$ sont présentes à haut décalage vers le rouge, c'est-à-dire lorsque l'Univers était plus jeune.

Cette thèse a aussi été une aventure humaine internationale, faite de rencontres, lors d'écoles, de conférences, ou de séminaires. J'ai eu la chance de voyager et d'interagir avec de nombreuses personnes. Vous l'aurez compris, j'ai pris beaucoup de plaisir pendant ces trois années de thèse. De manière plus générale, comme tout travail de recherche, la thèse est parsemée de moments de doute, mais aussi d'euphorie où l'on croit avoir compris ou découvert quelque chose, pour finalement le remettre en question et douter à nouveau~!

\section*{Tu as cherché, soit\dots et as-tu ``trouvé'' au moins?}

La notion de ``découverte'', de ``trouvaille'', est très subjective en astrophysique!
Les ``grandes'' découvertes sont rarement le fruit d'un programme préétabli, elles apparaissent très souvent de manière inattendue. Les percées théoriques sont aussi, bien souvent, le fruit d'un long travail et d'une grande expérience.
Guidé par mon directeur de thèse, François Boulanger, j'ai essayé de produire un travail théorique qui ne soit jamais trop éloigné des observations, c'est-à-dire qui puisse interpréter des données existantes, ou bien être vérifié par de futures observations.
Je vous invite maintenant à lire cette aventure dans les ``quelques'' pages qui suivent\dots

\part{Molecules, dust, and galaxy evolution}

\chapter{The discovery of a new population of H$_{\bf 2}$-luminous objects}
\label{chapter:H2_galaxies}

\epigraph{The most exciting phrase to hear in science, the one that heralds new discoveries, is not "Eureka!" but "That's funny..."}{Isaac Asimov}


\index{H$_2$-luminous galaxies}
\index{MOHEGs!see H$_2$-luminous galaxies}

\begin{Abstract}

One of the surprising results obtained with the Spitzer space telescope is the discovery of a significant and diverse population of low-redshift objects where the mid-infrared rotational line emission of molecular hydrogen is strongly enhanced, while star formation is suppressed. 
This is in sharp contrast with previous observations, where H$_2$ was solely associated with star formation. 
These ``H$_2$-luminous sources'' include galaxies that are in key-phases of their evolution, dominated by gas accretion, galaxy interactions, or galactic winds driven by star formation and active galactic nuclei. 
These observations open a new perspective on ISM physics, and on the role that molecular gas plays in active phases of galaxy evolution. Why is H$_2$ present in these violent environments? How is the H$_2$ emission powered? Why is the H$_2$ gas unefficient at forming stars? These questions are the core issues addressed in my thesis work.

\end{Abstract}

\minitoc



\section{Introduction}

\PARstart{T}he molecular gas plays a central role in galaxy evolution. In most galaxies, it carries the bulk of the mass of the interstellar matter and it is the fuel for star formation. Stars are indeed made from gravitationally unstable cores within molecular clouds. Therefore, the mass of molecular gas is generally associated with the star formation rate by defining a star formation efficiency\footnote{star formation rate per unit gas mass.} \citep[see][]{Schmidt1959}.
The bulk of the molecular gas in most galaxies is known from molecular line spectroscopy to be cold \citep[$\approx 10-20$~K,][]{Fixsen1999}. Most of the present studies primarily rely on tracing molecular gas through the observation of low rotational emission of carbon monoxyde (CO) in the radio domain. 

The H$_2$ molecule, being the most abundant molecule in the Universe, is the main constituent of the molecular gas. To be seen in emission, the H$_2$ gas temperature must be warmer than $\approx 100$~K. Mid-infrared (Mid-IR) rotational line H$_2$ emission traces what I will call ``warm H$_2$'' throughout my manuscript, at temperatures $T \approx 10^{2-3}$~K (see chapter~\ref{chapter:H2Molecule}). 
Warm H$_2$ was first observed within galactic star-forming regions, and associated with the illumination of the surface of molecular clouds by UV light from massive stars, creating  Photo-Dissociation Regions \citep[PDRs, see e.g.][]{Habart2003}. In the Galaxy,  H$_2$ is also observed in proto-stellar outflows \citep[e.g.][]{Neufeld2006, Maret2009} or supernovae remnants \citep[e.g.][]{Hewitt2009}, where it is shock-heated. 

\begin{figure}
\begin{minipage}{\textwidth}
    \def\thefootnote{\alph{footnote}}
      \setlength{\footnotesep}{0pt}
   \centering
   \includegraphics[width=0.7\textwidth]{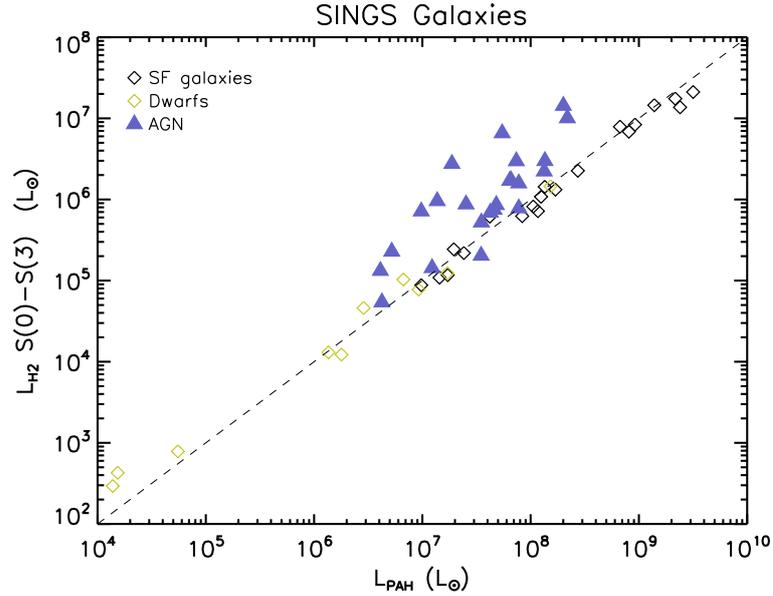}
      \caption[Relationship between H$_2$ and IR emission from PAHs]{Relationship between H$_2$ and IR emission from PAHs in nearby  (spatially resolved) star-forming (SF) galaxies, dwarf galaxies, and Active Galactic Nuclei (AGN). The data is from \citet{Roussel2007}. These galaxies belong to  the SINGS\footnotemark[1] sample of galaxies observed with the \textit{Spitzer} space telescope. The H$_2$ luminosity is summed over the S(0) to S(3) rotational lines of H$_2$.}
       \label{fig:H2_vs_PAH}
\footnotetext[1]{Spitzer Infrared Nearby Galaxies Survey, \url{http://sings.stsci.edu/}.}
\end{minipage}
   \end{figure}

The \textit{ISO}\footnote{Infrared Space Observatory, \url{http://iso.esac.esa.int/}} and \textit{Spitzer}\footnote{\url{http://www.spitzer.caltech.edu/}} IR satellites have started to peer at H$_2$ emission from external galaxies.  In star forming galaxies, rotational H$_2$ line emission is thought to come from PDRs \citep{Rigopoulou2002, Higdon2006, Roussel2007}. A general relationship between the H$_2$ and IR luminosity is inferred from these data. This correlation is shown for nearby galaxies observed with the \textit{Spitzer} space telescope on Fig.~\ref{fig:H2_vs_PAH}. Note that the Active Galactic Nuclei (AGN)  galaxies  stand above the tight correlation found for dwarf and star-forming (SF) galaxies. \citet{Roussel2007} indicate that the H$_2$ to PAH luminosity ratio in star forming galaxies is within the range of values that are expected from PDR emission. For the AGN sample, \citet{Roussel2007} suggest that the radiation from the AGN is not sufficient to power the H$_2$ emission.

Recent mid-IR spectroscopy performed with the \textit{IRS}\footnote{InfraRed Spectrograph, \url{http://ssc.spitzer.caltech.edu/irs/}} onboard the \textit{Spitzer} space telescope suggests that our census of warm molecular gas in galaxies
may be severely incomplete, revealing a new class of galaxies with strongly enhanced  
H$_2$ rotational emission lines, while classical star formation indicators (far-infrared continuum
emission, ionized gas lines, polycyclic aromatic hydrocarbons, PAHs) are strongly suppressed.
These galaxies are called in this manuscript \textit{H$_2$-luminous galaxies}.

In this chapter we paint the global observational framework of this PhD work. 
Section~\ref{sec:H2_galaxies} presents the observational discovery of H$_2$-luminous objects. We point out the key astrophysical questions that we address in this manuscript in sect.~\ref{H2Gal:issues-thesis}. The main goal of my thesis work is to interpret these surprising observations.

\section{H$_{\bf 2}$-luminous galaxies}
\label{sec:H2_galaxies}

\subsection{Global observational characteristics}
\index{H$_2$-luminous galaxies!observational characteristics}

The first extragalactic mid-IR spectrum dominated by H$_2$ rotational lines has been detected in the Stephan's Quintet (hereafter SQ) compact group of galaxies \citep{Appleton2006}. Strikingly, the H$_2$ emission is detected outside the galactic disks of the group, and is more powerful that the X-ray emission from the hot gas lying in the group halo. I started my PhD shortly after the publication of that paper. My research quickly  focused on this object, with the aim of explaining the origin of the H$_2$ gas and its power compared to that of the hot, X-ray emitting gas. To determine and model the excitation mechanisms that can power the observed H$_2$ emission was a driver to tackle the latter problem.
This observational discovery and my interpretation and modeling work are detailed in chapters~\ref{chapter:H2_SQ} and \ref{chapter:H2_SQ_mapping}.

\begin{figure}
   \centering
   \includegraphics[width=\textwidth]{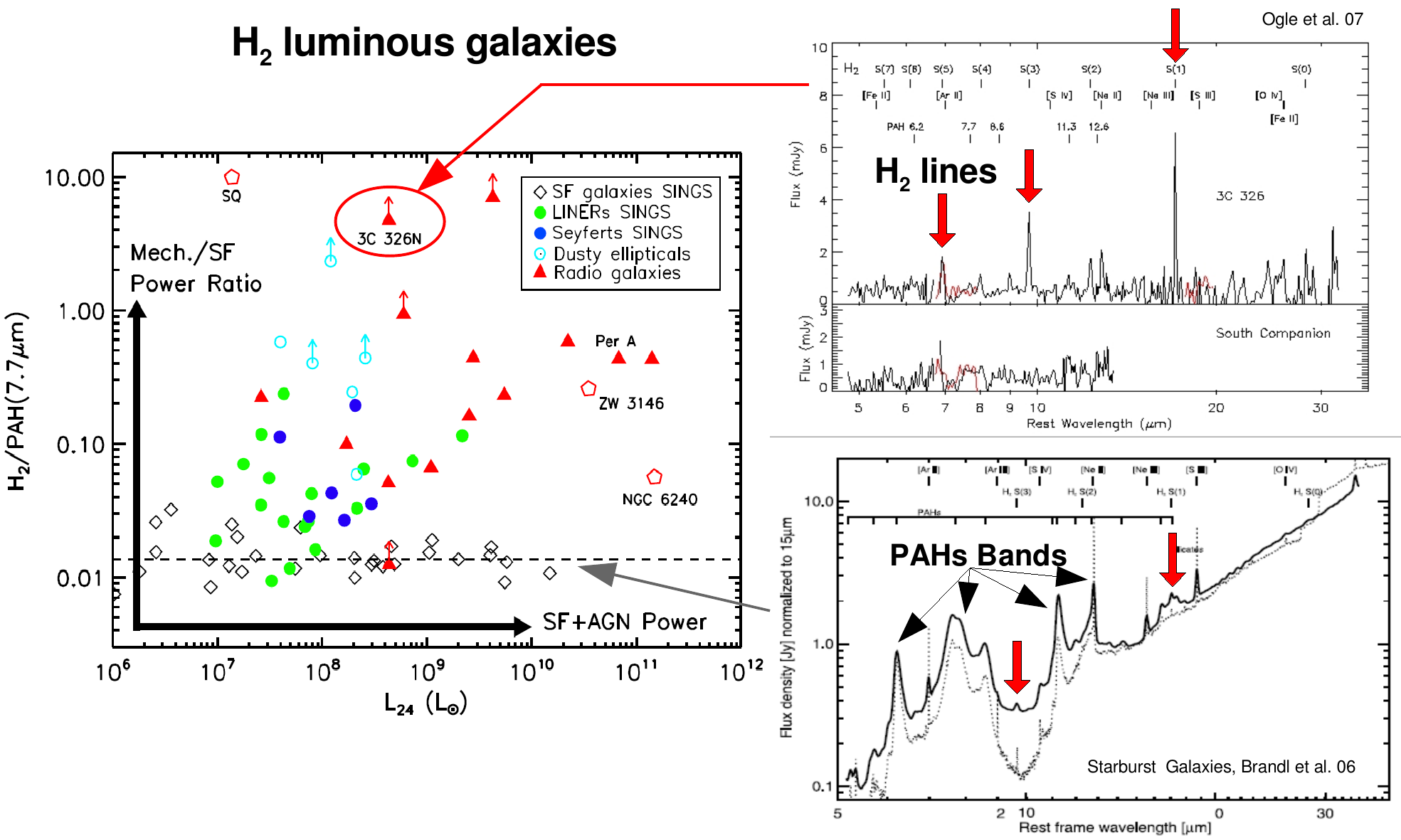}
      \caption[H$_2$-luminous galaxies: H$_2$ / PAH 7.7$\mu$m vs. $\mathcal{L}_{\rm 24\,\mu m}$]{H$_2$-luminous galaxies in one plot. \textit{Left:} Ratio of H$_{2}$ rotational emission summed over the 0-0 S(0)$-$S(3) lines to 7.7$\mu$m PAH luminosity, as a function of the 24$\mu$m continuum luminosity. 
The ratio indicates the relative importance of mechanical heating and power by star formation.
 At all IR luminosities, the observations reveal galaxies with excess H$_{2}$ emission beyond that expected from Star Formation (SF). The empty diamonds and the dashed line show the range spanned by SF-dominated galaxies. The H$_{2}$-luminous galaxies include active galactic nuclei galaxies (Seyferts, LINERs and radio galaxies), cooling flows (Perseus A, ZW3146) and interacting galaxies/mergers (Stephan's Quintet --SQ-- and NGC 6240). The H$_2$-luminous radio-galaxies are from the \citet{Ogle2009} sample. The excess H$_{2}$ emission reveals large (up to 10$^{10}$~M$_{\odot}$ ) quantities of warm ($T > 150$~K) molecular gas. On the \textit{right} side we compare the spectrum of the 3C326 H$_{2}$-luminous radio-galaxy \citep{Ogle2007} with that of starburst galaxies \citep{Brandl2006}.}
       \label{fig_H2_galaxies_H2_7_7PAH_L24uml}
   \end{figure}

 The detection of $\rm H_2 $ emission from SQ was the first of a series of discoveries of powerful H$_2$ emission in extragalactic sources that we present here.
Fig.~\ref{fig_H2_galaxies_H2_7_7PAH_L24uml} gathers a sample of this new class of extremely luminous H$_{2}$ emission galaxies, up to 10$^{10}$~L$_{\odot}$ in pure rotational molecular hydrogen emission lines and relatively weak total IR emission. 
This plot shows how H$_2$ galaxies stand out above star forming galaxies from the SINGS\footnote{Spitzer Infrared Nearby Galaxies Survey, \url{http://sings.stsci.edu/}} survey when plotting the ratio of H$_2$ rotational lines to the PAH 7.7$\,\mu$m luminosities vs. continuum luminosity at 24$\,\mu$m. 
These H$_2$-luminous galaxies are therefore out of the correlation presented in Fig.~\ref{fig:H2_vs_PAH}.
Some of the sources have undetected 7.7$\,\mu$m emission, so lower limits for $\mathcal{L}_{\rm H_2} / \mathcal{L}_{\rm PAH7.7}$ are indicated. 

The global characteristics of H$_2$-luminous galaxies are the following. \citet{Ogle2009} use empirical criteria to define H$_2$-luminous galaxies\footnote{In \citet{Ogle2009}, H$_2$-luminous galaxies are named MOHEGs, for MOlecular Hydrogen Emission Galaxies}. H$_2$-luminous galaxies have a large total H$_2$ to IR luminosity ratio:
\begin{equation}
\frac{\mathcal{L}(\rm H_2)}{\nu \mathcal{L}_{\nu}(24\,\mu\rm m)} \approx 10^{-3} - 10^{-1}
\end{equation}
and more specifically a large H$_2$ to $7.7\,\mu$m PAH luminosity:
\begin{equation}
\frac{\mathcal{L}(\rm H_2)}{\mathcal{L}_{\rm PAH7.7}} \approx 0.04 - 10
\end{equation}
The lower limit of 0.04 is an empirical value and also depends on the depth of the \textit{Spitzer} observations. 
These ratios are up to two orders of magnitude greater than the median values for the normal star-forming galaxies. This is  why they stand out of normal star-forming galaxies and luminous infrared galaxies in the plot of Fig.~\ref{fig_H2_galaxies_H2_7_7PAH_L24uml}. 

H$_2$-luminous galaxies point at a generic source of H$_2$ emission not powered by star formation. The most contrasted examples show bright H$_{2}$ emission lines with almost no spectroscopic signature (IR continuum of thermal dust emission, or ionized gas lines) of star formation. For instance, this is the case in Stephan's Quintet (see chapter~\ref{chapter:H2_SQ_mapping}) and in the 3C326 radio-galaxy \citep{Ogle2007}. The spectrum of 3C326 is shown in Fig.~\ref{fig_H2_galaxies_H2_7_7PAH_L24uml}, on top of the standard spectrum of a starburst galaxy. Note the absence of dust continuum and features in the H$_2$-luminous 3C326 radio-galaxy spectrum as compared to a star-forming galaxy. 

In many of these galaxies, molecular gas has been detected through the mid-IR H$_{2}$ rotational lines prior to any CO observation. 
This sample of H$_2$-luminous objects comprises different types of astrophysical sources. We briefly describe below some of the most important examples in each category of sources.

\subsection{Examples among different types of H$_{\bf 2}$-luminous astrophysical sources}
\index{H$_2$-luminous galaxies!examples}

Among the sample of H$_2$-luminous objects, Stephan's Quintet is certainly the object where the astrophysical context is clear enough to identify the dominant source of energy that powers the H$_2$ emission (chapter~\ref{chapter:H2_SQ_mapping}) and to associate this  H$_2$ emission  with the mechanical energy released in a galaxy collision. This is why my PhD work initially focused on this source. In other H$_2$-luminous sources, the energy source is less clear. 

\subsubsection{Interacting and infrared galaxies}

\begin{figure}
   \centering
   \includegraphics[width=\textwidth]{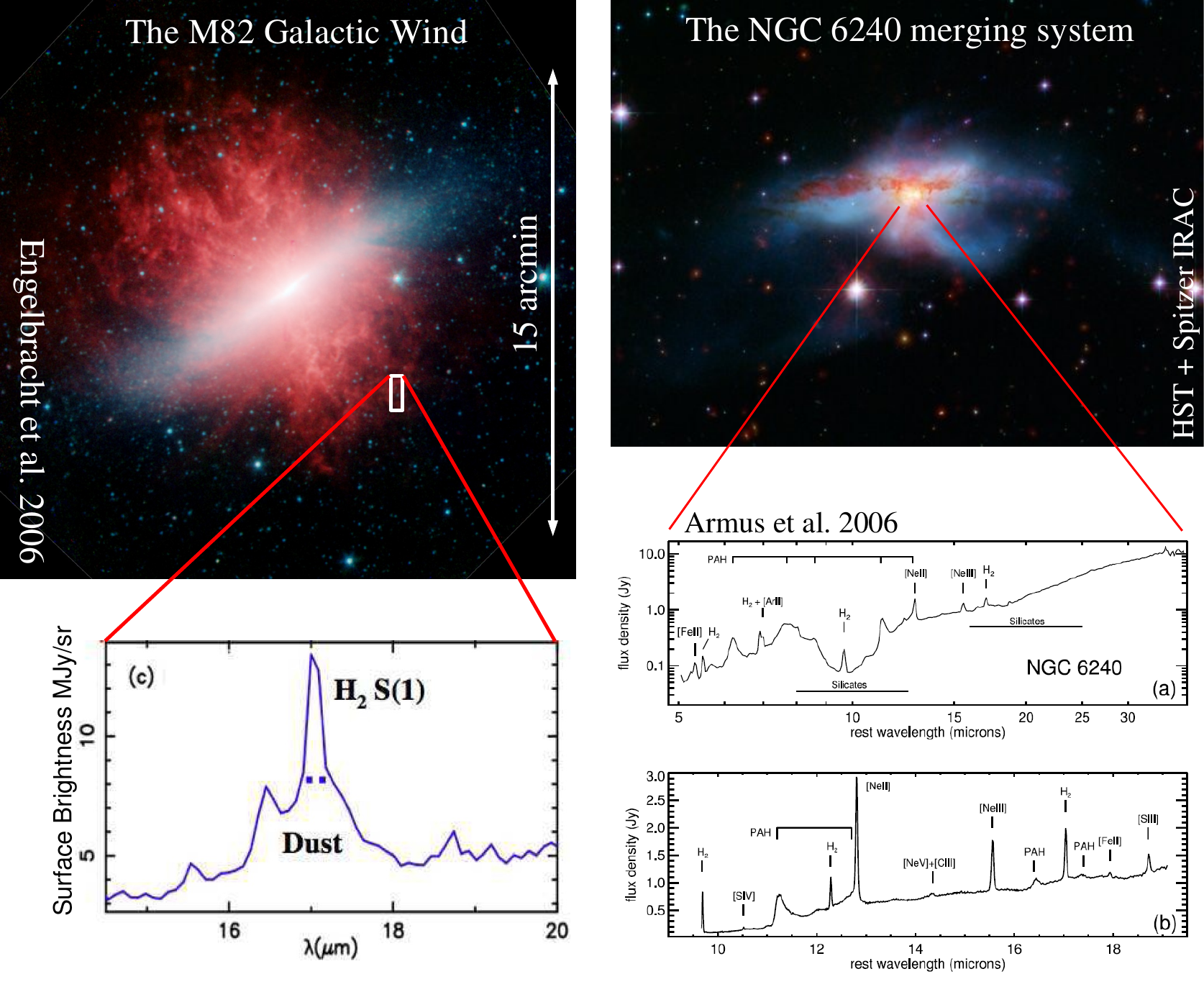}
      \caption[H$_2$ emission in M82 and NGC 6240]{H$_2$ emission in the  starburst-driven galactic wind of M82 and in the NGC~6240 merger. \textit{Left:} In
the M82 starburst galaxy, the wind is observed to be loaded with dusty molecular gas entrained and/or formed in the interaction of the hot wind with galactic halo gas. A \textit{Spitzer}  low spectral resolution ($\mathcal{R}=80$) spectrum is shown, with the H$_2$ 0-0S(1) line on top of the $17\,\mu$m PAH feature. Taken from \citet{Engelbracht2006}. \textit{Right:} Composite image (\textit{Spitzer IRAC} at 3.6 and 8$\,\mu$m --red-- and visible light from \textit{HST} --green and blue--) of the NGC~6240 pair of colliding galaxies. The bottom plot shows \textit{IRS} Short-Low and Long-Low (a), Short-High (b) spectra from \citet{Armus2006}.}
       \label{fig_H2_M82_NGC6240}
   \end{figure}

The first $\rm H_2 $ luminous galaxy, NGC~6240, was identified from ground based near-IR $\rm H_2 $ spectroscopy by \citet{Joseph1984}. Later, \citet{Hawarden2002} report the discovery of four Seyfert 2 galaxies (optically classified as LINERs\footnote{Low Ionization Emission Regions}) whose near-IR spectra are dominated by ro-vibrational emission of H$_2$.

In Fig.~\ref{fig_H2_M82_NGC6240} we show two iconic infrared galaxies, NGC~6240 and M82 that show strong H$_2$ emission. 
NGC~6240 is a ultra-luminous infrared merger ($\mathcal{L}_{\rm IR} \approx 6 \times 10^{11}$~L$_{\odot}$) with optical and mid-IR signatures of AGN activity. The outstanding property of NGC~6240 is the presence of a double-AGN \citep{Komossa2003}.
\citet{Lutz2003} report mid- and far-IR spectroscopy with the \textit{ISO} SWS\footnote{Short Wavelength Spectrometer.} and LWS\footnote{Long Wavelength Spectrometer.} instruments, and find unusually strong H$_2$ mid-IR lines. This result is confirmed by \citet{Armus2006}, who find $\approx 2 \times 10^9$~M$_{\odot}$ of warm H$_2$ within the $11.3'' \times 4.7''$ aperture of the SH\footnote{Short High module of the Infrared Spectrometer onboard the \textit{Spitzer} space telescope, see \url{http://ssc.spitzer.caltech.edu/irs/} for informations about the spectrometer.} slit of the \textit{IRS}, that include the two nuclei and part of an extended region where near-IR ro-vib H$_2$ emission is detected \citep{VanderWerf1993, Max2005}.

The M82 starburst galaxy is not \textsl{stricto sensu} an H$_2$ luminous galaxy, but we show it in Fig.~\ref{fig_H2_M82_NGC6240} because it may provide insights for the interpretation of other H$_2$-luminous objects. The well-known \textit{Spitzer} image reveals the dust emission in the starburst-driven wind. Interestingly, H$_2$ is detected in the outflow of the galaxy \citep{Engelbracht2006}. \citet{Veilleux2009} also detected emission in the H$_2$ 1-0S(1) 2.12$\,\mu$m  rovibrational line from the starburst-driven wind. 
M82 is clear example where H$_2$ gas and dust  coexist with the hot plasma outflowing gas.

Up to now, Stephan's Quintet is the only  example among the Hickson Compact Groups (HCGs) catalog \citep{Hickson1982} that show powerful emission \citep{Appleton2006, Cluver2009}. HCGs are isolated high-surface-brightness galaxy groups containing at least 4 members within a narrow optical magnitude range ($\approx 13-14$~mag), and almost all of them involve extreme galaxy tidal interactions \citep{Hickson1982, Sulentic2001}. A \textit{Spitzer} observations have been carried on to search for H$_2$ in HCGs and the data are being analysed (see chapter~\ref{chapter:perspectives}).

\subsubsection{Radio-galaxies}
\label{H2-gal-radio-galaxies}

\begin{figure}
   \centering
   \includegraphics[width=\textwidth]{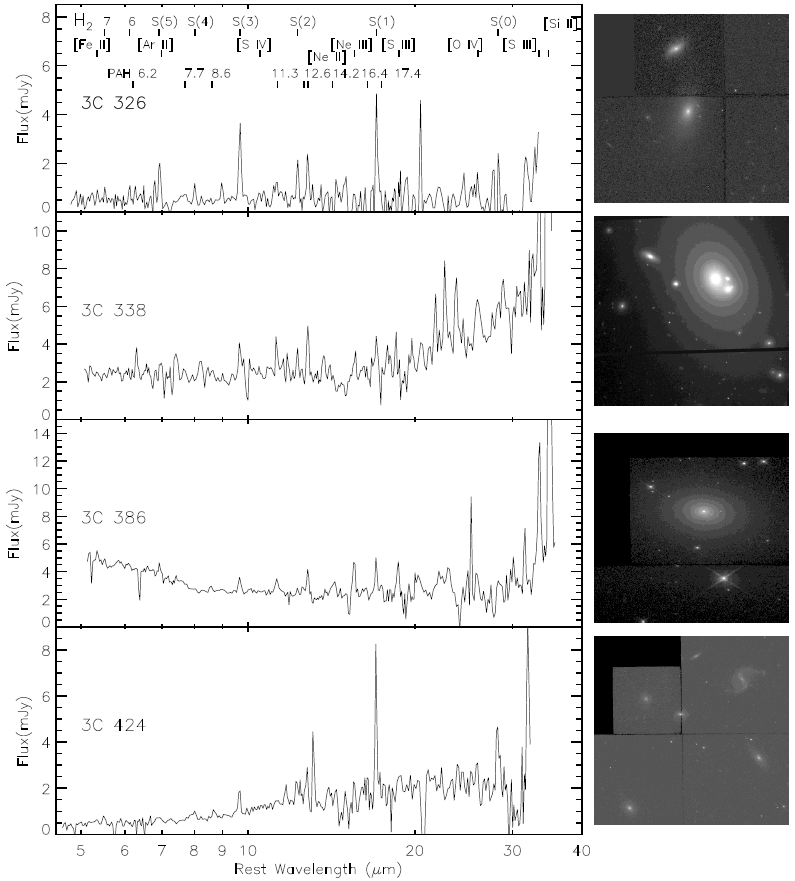}
      \caption[\textit{Spitzer IRS} low-resolution spectra of examples of MOHEGs]{Examples of \textit{Spitzer IRS} low-resolution spectra of MOHEGs (3C~326, 338, 386, 424). These sources show no silicate absorption and very weak PAH emission. On the right, we show \textit{HST WFPC2} and \textit{ACS} images. Taken from \citet{Ogle2009}.}
       \label{fig_Ogle09_spectra_RGs}
   \end{figure}

The first H$_2$-luminous radio-galaxy, 3C326, was discovered by \citet{Ogle2007}. It is one of the most extreme H$_2$-luminous sources (see Fig.~\ref{fig_H2_galaxies_H2_7_7PAH_L24uml}). We will discuss this object in more detail in chapter~\ref{chapter:perspectives}. 3C326 is one example of a sample of H$_2$-luminous radio-galaxies (selected such that $z < 0.22$) from the 3CR\footnote{Third Revised Cambridge Catalog, \url{http://3crr.extragalactic.info/}} catalog, reported in \citet{Ogle2009}. They find that, among 57 sources, 17 ($\approx 30$\%) have strong H$_2$ rotational lines. These galaxies are the red triangles in Fig.~\ref{fig_H2_galaxies_H2_7_7PAH_L24uml}. In Fig.~\ref{fig_Ogle09_spectra_RGs} we show four examples of \textit{IRS} spectra. Note the strong H$_2$ lines, as opposed to the weakness of the PAH features and the absence of silicate absorption.
The summed H$_2$  S(0)-S(3) line luminosities are $\mathcal{L}(\rm H_2) = 10^{32} - 2 \times 10^{35}$~W. \citet{Ogle2009} estimate the mass of warm molecular gas in these galaxies to be\footnote{They estimate by fitting the H$_2$  S(0) to S(7) rotational line fluxes with 3 temperatures components with thermalized H$_2$ excitation and ortho-to-para ratios. See chapters~\ref{chapter:H2Molecule} and \ref{chapter:perspectives} for details.} $M(\rm H_2) = 10^{5} - 2\times 10^{10}$~M$_{\odot}$. 

Radio-galaxies have an AGN and a source of mechanical energy that is the AGN-driven jet. It is not clear how the H$_2$ luminosity is powered. 
The large majority (16/17) of these radio-galaxies seem to belong to pairs or clusters (see the \textit{HST} images on the right of Fig.~\ref{fig_Ogle09_spectra_RGs}). This suggests that galaxy interaction  may also play a role in powering the H$_2$ emission, although radio  activity may be triggered by interactions.

\subsubsection{Cooling flows in clusters of galaxies}
\label{H2-in-cooling-flows-galaxy-clusters}
\begin{figure}
   \centering
   \includegraphics[width=\textwidth]{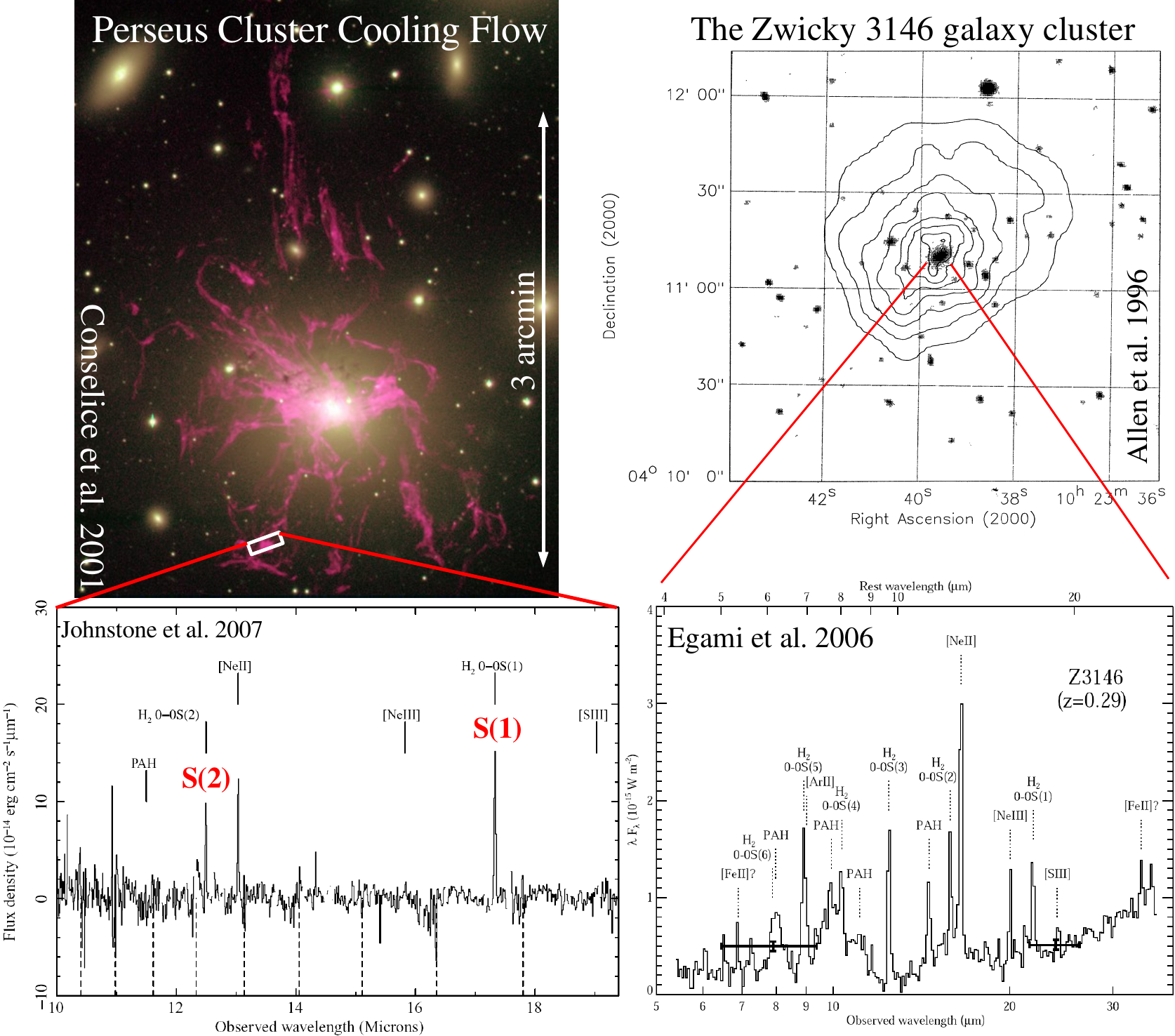}
      \caption[H$_2$ emission in the Perseus cooling flow and the Zwicky 3146 cluster]{The Perseus A (NGC 1275) and Zwicky galaxy clusters  illustrate the presence of H$_2$ in the gas that flows in/out galaxies. \textit{Left:} In the Perseus A cooling flow, luminous H$_2$ line  emission \citep[see inset spectrum,][]{Johnstone2007} is associated with the network of filaments of ionized gas in the halo  \citep[ H$\alpha$  image][]{Conselice2001}. These observations suggest that the cooling of the inter-cluster gas may occur through a phase transition from hot to molecular gas. \textit{Right:} the top panel shows an $R$-band image of the $z=0.3$ cluster Z3146 taken with the ESO 3.6-m telescope with X-ray \textit{ROSAT} contours overlaid from \citet{Allen1996}. The mid-IR spectrum shows powerful H$_2$ line emission \citep{Egami2006}.}
       \label{fig_H2_Perseus_Z3146}
   \end{figure}

\textit{Spitzer} observations reveal that warm ($\approx 10^{2-3}$~K) H$_2$ is present in the central regions of galaxy clusters. The most striking examples, shown on Fig.~\ref{fig_H2_Perseus_Z3146}, are the Zwicky 3146 \citep{Egami2006} and the Perseus A (NGC 1275) \citep{Johnstone2007} clusters. In the case of Perseus A, the H$_2$ emission seems to be associated with an extended network of optical line filaments (see the beautiful H$\alpha$ image in Fig.~\ref{fig_H2_Perseus_Z3146}), the so-called ``cooling flows''. These filaments are commonly found in X-ray luminous clusters \citep[e.g.][]{Heckman1989}, but their origin is still debated. 
\citet{DeMessieres2009} report the  detection with \textit{Spitzer IRS} of strong H$_2$ rotational line emission from nine cool-core galaxy clusters, with surprisingly weak dust features in 6 (out of 9) of them.
Again, the H$_2$ detection in these hostile environments show that warm molecular gas is coexisting with the hot (a few million K) intra-cluster gas.

\subsubsection{Elliptical galaxies}

\citet{Kaneda2008} report the detection of strong H$_2$ emission in a sample of dusty elliptical galaxies. These galaxies are reported in the left plot of Fig.~\ref{fig_H2_galaxies_H2_7_7PAH_L24uml} (light blue open circles). They also show an enhancement of the $\frac{\mathcal{L}(\rm H_2)}{\nu \mathcal{L}_{\nu}(24\,\mu\rm m)}$ luminosity ratio with respect to star-forming galaxies, in spite of prominent PAH features. Many of them have radio-loud AGN. These elliptical galaxies were previously detected by the \textit{IRAS}\footnote{InfraRed Astronomical Satellite, \url{http://irsa.ipac.caltech.edu/IRASdocs/iras.html}} satellite \citep{Goudfrooij1995}.

\section{Main questions addressed in this dissertation}
\label{H2Gal:issues-thesis}

\subsection{Astrophysical implications of observations of H$_{\bf 2}$-galaxies}

The observations of H$_2$-luminous galaxies are puzzling for several reasons, and raise these two main astrophysical  questions:

\subsubsection{Why is H$_{\bf 2}$ present in violent phases of galaxy evolution?}

First, in galaxy clusters or high-speed galaxy collisions like Stephan's Quintet, the presence of large quantities of warm H$_2$ gas in a hot gas environment is unexpected.
In starburst-driven winds like M82, or AGN-driven outflows like radio-galaxies (3C326 in particular), it is not clear whether the H$_2$ gas is molecular material that has been entrained from the galactic disk by radiation pressure and/or the outflow ram pressure, or whether it is formed \textit{in situ}, meaning formed out of  warm outflowing gas. 
Observations of H$_2$-luminous galaxies thus raise the question of the origin and the physical state of the molecular gas in these environments. In particular, is the warm H$_2$ physically associated with a more massive reservoir of cold ($T \lesssim 100$~K) molecular gas, as it is the case in ``standard'' galaxies?

\subsubsection{How is the H$_{\bf 2}$ emission powered?}

In H$_2$-luminous galaxies, H$_{2}$ lines appear to have an unexpected contribution to the gas cooling. The energy source powering the H$_2$ emission and the dominant H$_2$ excitation mechanisms remains to be identified. 
In the Stephan's Quintet galaxy collision, the astrophysical context is clearer than in distant objects, essentially because it has been extensively observed, from X-ray to radio wavelengths.
This allows to study in more details the physics of the H$_2$ formation and emission, and to infer a physical framework to interpret other observations of H$_2$-luminous sources.

In conclusion, the core of this thesis is to understand the origin of the H$_2$ emission in these H$_2$-luminous objects, to identify and to study the underlying physical processes that drive H$_2$ formation and excitation in these hostile environments. 

\subsubsection{Why is the H$_{\bf 2}$ gas inefficient at forming stars?}

In the introduction, we have stated that, up to now, H$_2$ observations have been mostly associated with stellar activity.
H$_2$-luminous galaxies are revealing molecular gas in a physical state where it is inefficient at forming stars, which does not straigthforwardly fit with our current understanding of the link between molecular gas and star formation. These H$_2$ observations raise the question of the identification of the  physical processes that inhibit star formation in H$_2$-luminous galaxies. 
I have addressed this question within the context of the Stephan Quintet galaxy collision (see chapters~\ref{chapter:H2_SQ}, \ref{chapter:SQ_CO} and \ref{chapter:SQ_dust}), and the 3C326 radio galaxy (chapter~\ref{chapter:perspectives}).

\subsection{What can we learn from H$_{\bf 2}$ galaxies?}

A common characteristic links the heterogeneous sample of H$_2$-luminous galaxies presented above: they are all in active and key phases of galaxy evolution, dominated by accretion, galaxy interactions, energy plus gas ejection due to star formation or to the action of the central black hole on its environment (AGN feedback).
The exceptional H$_2$ luminosities observed in this  variety of sources suggest that molecular gas is ``responding'' to these active phases of galaxy evolution. The discovery of H$_2$ luminous galaxies then suggests that molecular gas has an important impact on  the energetics of galaxy formation and evolution, which is, up to now,  largely unexplored.
Most of the analytical models or simulations that describe galaxy evolution within the context of cosmological structure formation ignore the physics of the molecular gas.

Although H$_2$ emission does not trace the bulk of the molecular  mass directly, these H$_2$ observations  reveal key aspects of the energetic interplay between galaxies and the intergalactic
medium, and between stars and their nascent clouds, that are missed by CO observations.
The low star-formation efficiency observed in H$_2$-luminous galaxies may also provide insights in the processes that trigger or regulate star formation. 

To sum up, H$_2$-luminous objects may represent an unexplored step in galaxy evolution, which provide insights in the role molecular gas plays in galaxy evolution. 

\section{Outline of the dissertation}

The three following chapters present the background needed to tackle the problems raised above. Chapter~\ref{chapter:dust_gas_galaxies} discusses the dynamics and thermal properties of the multiphase ISM, as well as the result of the calculation of the time-dependent cooling of a dusty plasma, taking into account dust destruction processes. In Chapter~\ref{chapter:H2Molecule} I describe the properties of the H$_2$ molecule, focusing on its excitation mechanisms. The physics and chemistry of shocks in a multiphase ISM are presented in chapter~\ref{chapter:shocks}. In particular, the codes used to interpret the observations are introduced, with emphasis on some results about molecular emission in MHD shocks. 

In the second part of this dissertation, I essentially focus on the Stephan's Quintet galaxy collision. Chapter~\ref{chapter:H2_SQ} presents the observational discovery of H$_2$ emission in the group, and the scenario I propose to explain the presence of H$_2$ in this violent environment. Then I address the question of H$_2$ excitation   (chapter~\ref{chapter:H2_SQ_mapping}), and extend the analysis to the emission from the ionized gas in the SQ shock. Chapter~\ref{chapter:SQ_CO} present the results of an observational campaign I have conducted to search for the CO counterpart of the H$_2$ emission, and chapter~\ref{chapter:SQ_dust} discusses dust emission. This part ends with chapter~\ref{chapter:perspectives}, which presents an extension of this work to the H$_2$-luminous radio galaxy 3C326, and discusses the role molecular gas plays in galaxy evolution. 

The third part of this thesis is dedicated to MIRI, the Mid-IR Instrument that will be part of the scientific payload of the James Webb Telescope. Chapter~\ref{chapter:JWST} introduces the JWST and describes the MIRI instrument, and chapter~\ref{chapter:miri_test} presents my contribution to the testing of the instrument, focusing on the analysis of the Point Spread Function (PSF) of the instrument. In the context of the scientific preparation of the guaranteed time projects for MIRI, we have proposed to use MIRI to study high-redshift H$_2$-luminous sources (chapter~\ref{chapter:science_JWST}). Finally, I discuss the observational and theoretical perspectives of this work (chapter~\ref{chapter:perspectives2}).




\chapter{Gas and dust in galaxies}
\label{chapter:dust_gas_galaxies}

\epigraph{If there's a bright center of the universe, you're on the planet that it's farthest from.}{
Luke Skywalker to C-3PO}



\begin{Abstract}
The astrophysical questions raised by the observational discovery of H$_2$-luminous galaxies involve the rich physics and chemistry of the interstellar medium (ISM). The presence and excitation of molecular gas (H$_2$) in the ISM heated by the release of mechanical energy on galactic scales (e.g. galaxy interactions, starburst and AGN feedback, cooling flows in galaxy cluster cores, etc.) is related to thermal and dynamical interactions between the phases of the ISM. 
This chapter gives an overview of the ISM phases and of the dynamical and thermal processes that drive mass and energy exchange between these phase. 
As H$_2$ forms on the surface of interstellar grains, dust evolution processes are also described in the context of this dynamical picture. This chapters ends with a description of the calculation of the cooling of a dusty, hot plasma. This result provides interesting results on dust survval in a multiphase ISM, and will be used in chapter~\ref{chapter:H2_SQ} to interpret the observations of Stephan's Quintet.

\end{Abstract}

\minitoc


\section{Introduction}

\PARstart{A}lthough William Herschel and Edward Barnard produced the  first images of dark nebulae silhouetted against the background star field of the galaxy  \citep[e.g.][]{Barnard1919}, the first direct detection of diffuse matter in interstellar space was made by \citet{Hartmann1904} through absorption spectroscopy of the $K$-line of calcium at 3934~\AA~toward the line of sight of the $\delta$-Orionis star. This discovery lauched more than one century of studies on the interstellar medium (henceforth ISM).

The ISM, the matter that exists between the stars, plays a crucial role in astrophysics precisely because of its intermediate role between stellar and galactic scales. The study of the ISM is closely related to the lifecycle of stars. Stars form within molecular clouds, and replenish the ISM with matter and energy when they die, through supernovae explosions, planetary nebulae, stellar winds, etc. This interplay between stars and the ISM helps determine the rate at which a galaxy depletes its gaseous content, and therefore its lifespan of active star formation.

Much of my PhD work is dedicated to the modeling of the  formation of molecular hydrogen and its emission in H$_2$-luminous sources (these sources have been introduced in chapter~\ref{chapter:H2_galaxies}). This involves the physics of the gas and dust in the ISM of these galaxies.
In particular, this modeling is based on the heating and cooling mechanisms of the gas, and on the exchanges of energy between the different phases of the ISM. 

This chapter is not an exhaustive review of ISM physics. We direct the reader to the books by  \citep{Dopita2003, Lequeux2005, Tielens2005} and to the review by \citet{Ferri`ere2001}. I  introduce here some of the key astrophysical processes on which we will rely in the following chapters. In section~\ref{sec:multiphase_ISM} we present the different gaseous phases of the ISM and their interactions. Section~\ref{sec_life_dust} focuses on  dust evolution in galaxies. Then I focus on the modeling of the cooling of a dusty, hot plasma. 
This calculation is illustrated within the context of H$_2$  studies in the multiphase medium of the Stephan's Quintet galaxy collision (see chapter~\ref{chapter:H2_SQ}), but \textit{(i)}  it can apply to many other astrophysical situations, and \textit{(ii)} it provides interesting results on dust survival in the multiphase ISM.

\section{The multiphase interstellar medium (ISM) in galaxies}
\label{sec:multiphase_ISM}

Interstellar matter comprises of gas, mostly composed of hydrogen, and solid particles of dust that contain half of the elements heavier than Helium, the so-called ``metals'' in astrophysics. The matter consists of about 99\% gas and 1\% dust by mass. If the ISM fills most of the volume of a galaxy,  it only represents a tiny fraction of its mass (5\% of the stellar mass, and 0.5\% of the total mass of the galaxy that is dominated by the dark matter). 

\subsection{Constituents,  phases, and structure of the ISM}
\label{phases-structure-ISM}
\index{Interstellar medium!constituents}
\index{Interstellar medium!thermal phases}

\begin{figure}
  \begin{center}
    \includegraphics[width=\textwidth]{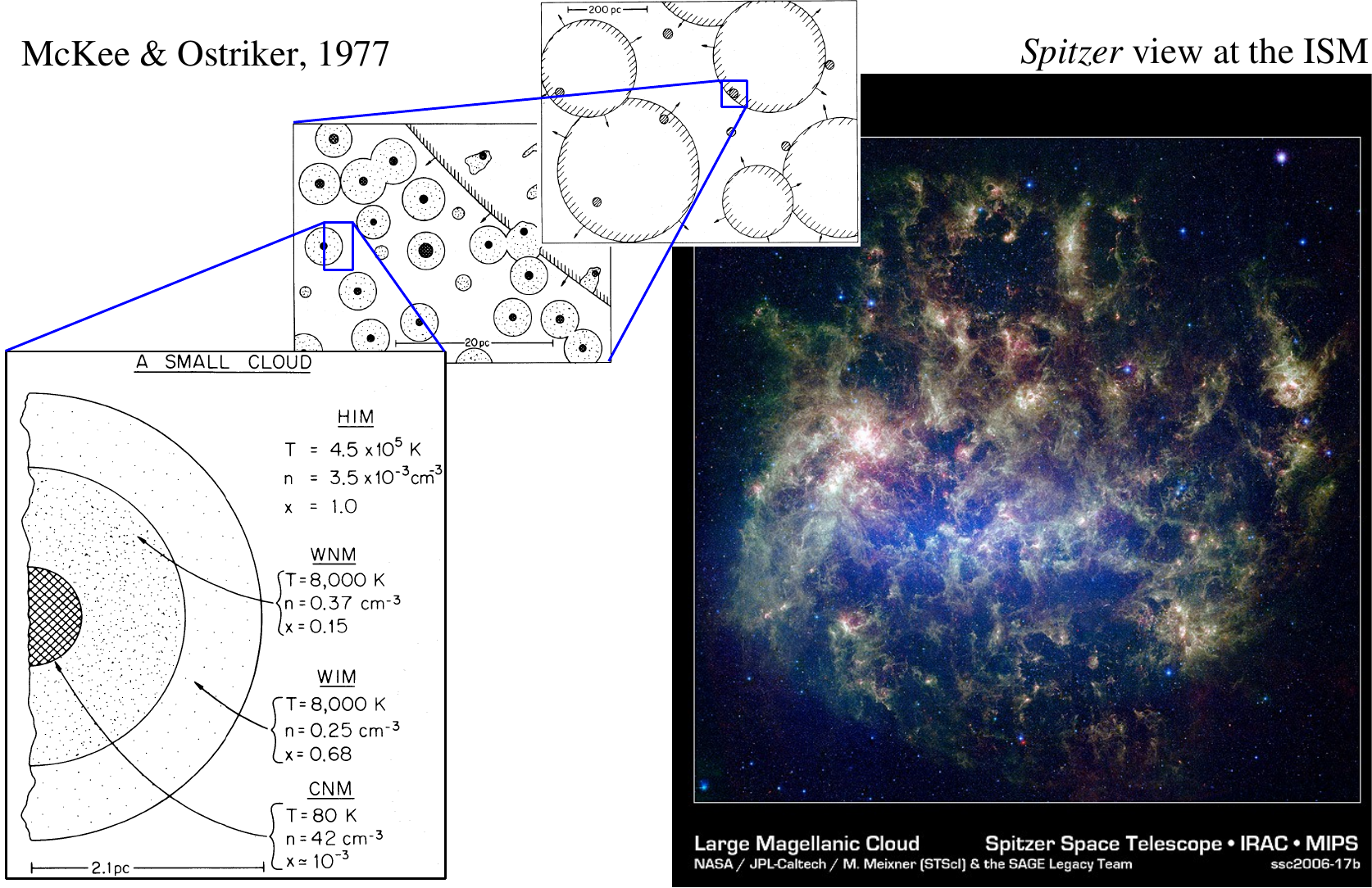}
    \caption[The structure of the ISM as seen by \citet{McKee1977} and the \textit{Spitzer} telescope]{The structure of the ISM as seen by \citet{McKee1977} and the \textit{Spitzer} telescope. In the schematic view \textit{(left)}, diffuse clouds have an \textit{onion-like} structure, in which teneous and hotter phases are surrounding denser and colder phases. The \textit{Spitzer} image of the ISM in the Large Magellanic Cloud  \textit{(right)} shows a contrasted view, where the ISM is a complex network of filaments, sheets and cavities, structured by star formation.}
    \label{fig:ISM_structure}
  \end{center}
\end{figure}

Because of the variety of processes involved in the ISM, an extremely wide range of physical conditions are present in the interstellar gas. 
The average density of the ISM is 1 particle per cm$^{3}$ in the Milky Way Galaxy\footnote{This average density of 1 particle per cm$^{-3}$ in the ISM is $10^{19}$ less than the air we are breathing!}, but this value varies by 8 orders of magnitude, from the hot ($T>10^6$~K), tenuous plasma produced by supernovae shocks, to cold ($T < 10$~K) prestellar condensations.

\renewcommand{\arraystretch}{1.1} 
\begin{table}
\begin{center}
\begin{minipage}[t]{\textwidth}
\renewcommand{\footnoterule}{}
\def\thefootnote{\alph{footnote}}
\centering
\small
\caption[Components of the interstellar medium]{Description of the phase components of the interstellar medium\footnotemark[1].}
    \begin{tabular}{c c c c c c}
	\hline
	\hline
\multirow{2}*{Phase} & Density & Temperature & Volume & Mass\footnotemark[2] & \multirow{2}*{Hydrogen\footnotemark[3]} \\
									 & [cm$^{-3}$] & [K] & fraction & [M$_{\odot}$] & \\
\hline
Molecular Cloud      		& $10^2 - 10^6$ & $20-50$ & $< 1\,$\% & $\approx 10^9$ & H$_2$ \\
Cold Neutral Medium   & $1 - 10^3$       &  $50-100$ & $1-5\,$\% & $1.5 \times  10^9$ & H$\,${\sc i} \\
Warm Neutral Medium & $0.1 - 10$ 		 & $10^3 - 10^4$  & $10-20\,$\% & $1.5 \times 10^9$  & H$\,${\sc i}  \\
Warm Ionized Medium  & $10^{-2}$ 		 &  $10^3 - 10^4$  & $20-50\,$\% & $\approx 10^9$  & H$\,${\sc ii}  \\
H$\,${\sc ii} region		 & $10^2 - 10^4$&  $10^4$  & $10\,$\% & $5\times 10^7$ & H$\,${\sc ii}  \\
Hot Ionized Medium		 & $10^{-4} - 10^{-2}$ &  $10^5 - 10^7$  & $30-70\,$\% & $\approx 10^8$  & H$\,${\sc ii}  \\
\hline
\end{tabular}
\label{tab_ISM_phases}
\footnotetext[1]{This is a very schematic classification, and the separation between phases is debated. The numbers are orders of magnitudes.}
\footnotetext[2]{These masses are  rough estimates for our Galaxy.}
\footnotetext[3]{This column lists the main state of the hydrogen atom in the different phases.}
\normalsize
\end{minipage}
\end{center}
\end{table}
\renewcommand{\arraystretch}{1.0}

In table~\ref{tab_ISM_phases} we gather the physical conditions (density, temperature, etc.) of the different components of the ISM. In the following I describe these components from an observational point of view. This description, based on observations, is very schematic, the ISM properties being a continuuous distribution. 
\begin{description}
\item[The Hot Ionized Medium (HIM):] this hot, ionized, volume-filling phase has been introduced by \citet{Spitzer1956} to explain the stability of the cold clouds that are not gravitationally bound.  
This phase results mostly from supernovae explosions, that produce hot ($10^{5-7}$~K) bubbles of expanding gas, or stellar winds. The gas at $\approx  10^5$~K is predominantly traced by soft ($< 1$~keV) X-ray emission and UV absorption lines (O{\sc vi}, N{\sc v}, etc.). Hotter gas ($\approx  10^6$~K) can be seen through X-ray emission lines (O{\sc vii}, O{\sc viii}). This diffuse, hot  plasma phase is also characterized by a thermal continuum (free-free ot free-bound) X-ray emission. 

\item[The Warm Ionized Medium (WIM):] this diffuse phase is mostly produced by the ionization of the diffuse neutral medium (by stellar radiation or shocks) or by ejection of matter from H{\sc ii} regions \citep{Yorke1989}. This gas at $10^{3-4}$~K is traced by recombination of forbidden line emission. In the  \citet{McKee1977} view, the WIM is associated with the external shell of a cloud. 

\item[H$\,${\sc ii} regions:] unlike the HIM and the WIM, this hot, ionized gas phase is  \textit{localized} around the hot stars (O- or B-type) that create ionized shells  by their UV radiation. An idealized model for these regions is the  Strömgren sphere, which results from the equilibrium between the ionizing flux ($> 13.6$~eV) and the electronic recombination rate. Some of these regions can be compact (less than 1~pc) or extremely extended (1~kpc or more) in case of very massive star clusters.
These regions are traced by free-free or free-bound emission \citep[see the book by][for details]{Osterbrock1989}, like in the HIM, and by UV absorption and IR emission due to the heating of dust by the star. Diverse optical recombination lines are also produced (like H$\alpha$) and also forbidden lines ([N{\sc ii}], [S{\sc ii}], [O{\sc iii}], etc.).

\item[The Warm Neutral Medium (WNM):] within the inner shells of the clouds, the gas is shielded from the background ionizing radiation and is mostly neutral, while photoelectric  heating maintains the gas at temperatures around 8000~K. This phase is mostly traced by H$\,${\sc i} 21~cm line  emission.

\item[The Cold Neutral Medium (CNM):] its atomic component is traced by the H$\,${\sc i} 21~cm line emission, while UV absorption spectroscopy reveals \textit{diffuse} H$_2$ gas \citep{Richter2003}. This gas is cold because it is denser than the WNM.

\item[Molecular Clouds (MCs):] these gravitationally bound structures result from the condensation of a diffuse cloud under the effect of gravity. These clouds are dense enough for hydrogen to be molecular (H$_2$). MCs occupy a tiny fraction of the total ISM volume (1\%), but account for a substantial fraction of the total mass of the ISM ($30-60$\%). Molecular clouds are structured over a wide range of spatial scales, from giant complexes (GMCs) of more than $10^6$~M$_{\odot}$ and sizes of $100$~pc, to smaller condensations of 50~M$_{\odot}$ ($<1$~pc).

\end{description}

Classically, the ISM has been modeled by a mixture of different phases. \citet{Field1969} put forward a static \textit{two-phase} equilibrium model to explain the observed properties of the ISM. Their modeled ISM consists of a cold dense phase ($T < 300$~K), made of clouds of neutral and molecular hydrogen, and a warm intercloud phase ($T \approx  10^4$~K), consisting of rarefied neutral and ionized gas. \citet{McKee1977} added a  third phase that represents the hot ($T \gtrsim10^6$~K) gas which is shock heated by supernovae and constitutes most of the volume of the ISM. 
This schematic view at the ISM is presented on the left panel of Fig.~\ref{fig:ISM_structure}.
These phases are associated with temperatures where heating and cooling can reach a stable equilibrium, except for the hot phase which is metastable and needs to be continuously replenished. Their paper formed the basis for further study over the past three decades!

Today, our view at the ISM is dynamical and based on numerical simulations. The dynamical evolution of the multiphase ISM in galaxies has been extensively investigated with
numerical simulations, within the context of the injection of
mechanical energy by star formation, in particular supernovae
explosions \citep{deAvillez2005a, Dib2006}. This injection of energy by supernovae is shown to be able to maintain a fragmented, multiphase and turbulent ISM, in which the gas phases are not in pressure equilibrium. Theoretical studies and numerical
simulations of a supersonically turbulent isothermal gas show that the pressure may vary over two orders of magnitude, the distribution function of the pressure values being lognormal, with a dispersion that increases with the Mach
number \citep{Vazquez-Semadeni1994, Padoan1997, Passot1998, MacLow2005}.

\subsection{Structuring the ISM}
\index{Interstellar medium!structure}

The CNM, WNM, WIM and HIM form the \textit{diffuse} ISM phases. In the \citet{McKee1977} equilibrium model, the diffuse clouds are structured like ``onions'', or ``Russian dolls''. Due to heat conduction, the teneous phases surround the denser ones. However, as shown in Fig.~\ref{fig:ISM_structure}, the ISM structure is much more complex. It is organized in filaments, or sheets, continuously stirred by turbulent motions. 
This structure is observed from large scales (see the \textit{Spitzer} image of the LMC\footnote{Large Magellanic Cloud} in Fig.~\ref{fig:ISM_structure}) to small scales around the solar neighbourhood. This \textit{self-similar} structure is also observed within molecular clouds \citep[e.g.][]{Falgarone2002}.

\index{Interstellar medium!self-similarity}

It has often been proposed that the condensation of matter by \textit{thermal instability}, that will be discussed in sect.~\ref{subsec:thermal-instability}, was the origin of the multiphase structure of the ISM \citep[e.g.][]{McKee1977, Wolfire1995}.
On the other hand, since almost 30 years, it is known that the ISM is a turbulent medium \citep{Larson1981, Scalo1987}. Measurements of line velocity dispersions \citep{Solomon1987, Caselli1995} show that supersonic flows permanently skim through the ISM. These observations led to models where the ISM clouds are not the product of thermal condensation, but the result of short-lived over-densities due to turbulent motions. It has even been proposed that the pressure equilibrium between the ISM phases was not relevant for the ISM \citep{Ballesteros-Paredes1999, V'azquez-Semadeni2000}. However, these statements are moderated by \citet{Hennebelle1999, Audit2005}, who show that thermal instability can be triggered by large-scale motions in the ISM.  Therefore, it seems that both thermal instability and turbulence are not exclusive with each other in the role they play in the structuring of the ISM \citep[e.g.][]{Dib2002}.

\subsection{Energy content of the ISM}
\index{Interstellar medium!energy content}


The ISM is penetrated by photons and high-energy particles. In addition, it is dynamically tied to the magnetic field of the galaxy. 
I briefly summarize here  the main energetic components of the ISM

In the solar neighbourhood, the average total \textit{electromagnetic energy} density is $\approx 1$~eV~cm$^{-3}$. This radiation is dominated by the UV and visible stellar radiation, and also the strong infrared radiation. In addition, the cosmological blackbody radiation of the Universe at 2.7~K correspond to 0.26~eV~cm$^{-3}$. 

The average strength of the \textit{magnetic field} is $B \approx 5\,\mu$G \citep{Heiles2005}, which corresponds to an average  magnetic energy density of $\approx 1$~eV~cm$^{-3}$. 
Although partly organized at the galactic scale, the magnetic field in the ISM comprises of a random component and a regular component, both of comparable strengths.

\index{Cosmic-rays}

\textit{Cosmic-rays} are high-energy (mostly relativistic) particles (electrons, protons, heavier atoms) propagating through the ISM gas with little interactions (except at energies $\lesssim 0.1$~MeV). Assuming equipartition of energy between cosmic-ray and magnetic energy densities, the average cosmic-ray energy density derived from synchrotron emission is $\approx 1$~eV~cm$^{-3}$. 

The \textit{kinetic} energy may be non-thermal (for instance in case of supersonic motions of the gas) or thermal, and the distribution between the two forms of kinetic energy depends on the gas phase. In the CNM, the non-thermal energy is observed to dominate the thermal energy. Assuming an average density in the solar neighbourhood of $n_{\rm H} \simeq 25$~cm$^{-3}$ and a temperature of $T \simeq 100$~K, the average thermal energy is $\approx 0.3$~eV~cm$^{-3}$.

All the energy densities  listed above (photons, magnetic field, cosmic-rays, random motions) are commensurate, and of the order of 1~eV~cm$^{-3}$ in the vicinity of the Sun. This is not a fortuitous coincidence, but the result of energy transfers between this different forms of energy.  
Consequently, although there are exceptions, one cannot neglect any of these energies, which makes the study of the ISM \sout{so difficult} so exciting!

\subsection{Mass and energy transfers between the ISM phases}
\label{mass-energy-transfer-ISM-phases}
\index{Interstellar medium!energy transfers between phases}

\begin{figure}
  \begin{center}
    \includegraphics[width=\textwidth]{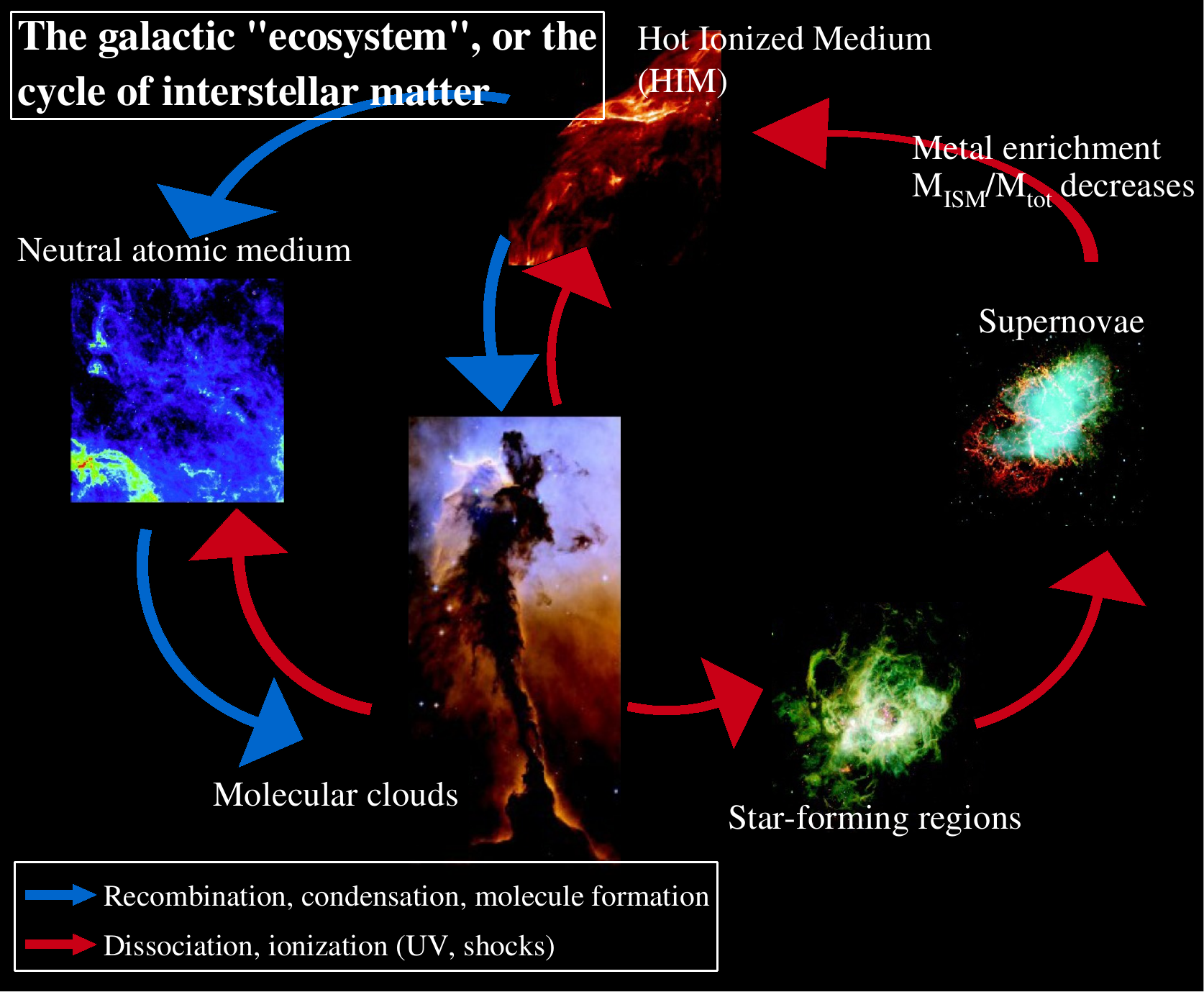}
    \caption[The cycle of the interstellar matter]{The cycle of the interstellar matter within a galaxy. The ISM is not  a passive substrate within which stars evolve. It constitutes their direct partner in the ``ecosystem of a galaxy'', continually exchanging matter and energy with them.
It is the spatial distribution ot the interstellar material together with its thermal and chemical characteristics that determines the locations where new stars form as well as their mass and luminosity spectra. Therefore, the lifecycle of interstellar matter governs the overall structure and emission properties of a galaxy.}
    \label{fig:cycle_ISM}
  \end{center}
\end{figure}

The ISM is a restless medium.  None of the ISM components introduced above is static in time or space. A permanent exchange of matter between stars and the matter between them occurs in the galactic disk, but also between the galaxy and the intergalactic medium.
The interstellar matter is thus constantly flowing from one phase to another. Depending on the physical process , and on the spatial scale we consider, the timescales associated with  these mass and energy transfers can be very different, from e.g. $10^6$~yr to form a protostar out of its parent molecular cloud, to e.g. $10^9$~yr to consume all the molecular gas content by star formation on galactic scales. 

\subsubsection{The lifecycle of interstellar matter}
\index{Interstellar medium!lifecycle of interstellar matter}

The cycle of interstellar matter, sketched in Fig.~\ref{fig:cycle_ISM}, is closely related to the lifecycle of stars. Stars are born in the densest regions of molecular clouds, when the gravity takes over the turbulent motions and condenses the gas. During their life, they emit UV radiation and hot winds that create local H{\sc ii} regions, and may eventually break through the parent cloud and fuel the WIM and HIM phases. Some of these stars, at the end of their life, explode in supernovae. The supernovae shocks heat and ionized the gas, enriching the gas with metals, and may eventually destroy the molecular cloud from which they were born. When the ionizing source has switched off, recombination occurs and, in turn, the WIM and HIM can cool. Under the effect of gravitational contraction, diffuse neutral and molecular clouds may re-form.  
Within this dynamical picture of the gas evolution, the dust is also experiencing a lifecycle, which contributes to its evolution and this to the evolution of the galaxy properties in general. This will be discussed in sect.~\ref{sec_life_dust}.

To sum up, the physical processes involved in the mass and energy transfers between ISM phases can be splitted into two categories:

\begin{description}
\item[Dynamical processes:] large- and small-scale fluid motions in a multiphase medium  transfer  momentum and kinetic energy from one phase to another. Large-scale motions are dominated by galaxy interactions, disk-halo circulation, galactic rotation, stellar-driven winds and supernovae, and gravitational collapse of giant molecular complexes. At small scales, the dynamics of the ISM is controlled by turbulence and fluid instabilities (Kelvin-Helmholtz and Rayleigh-Taylor). 
As introduced in sect.~\ref{phases-structure-ISM}, the fact that the structure of the ISM is fragmented and self-similar\footnote{This property is true over four orders of magnitude in size, from galactic to proto-stellar scales.}, suggest that turbulence plays an important role in the dynamical coupling between the different scales of the ISM. 
The kinetic energy cascade from the bulk motions on large scales to small-scale turbulence is a key aspect of ISM dynamics that is still poorly understood \citep[see][for reviews about interstellar turbulence]{Elmegreen2004, Scalo2004}. 

\item[Thermal processes:] thermal exchanges (heating and cooling mechanisms) correspond to a transfer of thermal energy to or from atoms, molecules and ions of the interstellar gas. Most of the heating mechanisms involves the release of a suprathermal particle (electron, molecule) from a gas or grain specie by an energetic particle or a photon. This suprathermal particle heats the gas by thermalization through collisions. The cooling mechanisms mainly occur through 
inelastic collisions between light gaseous colliders (e$^-$, H, H$^+$, etc.) and heavier atoms, molecules, or grains. If energy levels of these targets can be excited by the collider, the collider loses kinetic energy and the gas cools by thermalization with the collider. The target then dissipates its energy through the emission of radiation. In the neutral ISM, this line emission mostly occur in the infrared (rotational or fine-structure line emission of e.g. C$^+$, C, O, H$_2$, CO, H$_2$O, etc. in the cold phase). Except in dense molecular clouds, the opacity of the medium is small in the infrared, so that the radiation can escape and the gas can effectively cool.
  
The heat conduction allows to redistribute the thermal energy content from one thermal phase to another. Heat conduction occurs through electron conduction (collisions) or turbulent transport. We will study these processes in more detail in  sect.~\ref{evolution-shocked-molecular-cloud} when we will discuss the fate of molecular gas in a hot plasma, which is particularly relevant for H$_2$ studies in the multiphase ISM  of H$_2$ galaxies.

Within a range of physical conditions (density, temperature), the gas is  \textit{thermally unstable}. A dynamical or thermal perturbation can bring the gas from one stable phase to a thermally unstable state which is sufficiently far from its initial state that it evolves towards another stable phase. The thermal instability appears to be an important source of mass and energy exchange between phases in the ISM. This process is discussed in the next section (sect.~\ref{subsec:thermal-instability}).

\end{description}

\subsection{Thermal balance in the ISM}
\label{subsec:thermal-instability}

The ISM is shaped by the gas dynamics and thermal properties. As introduced before, the thermal instability  plays an important role in the physics and in structuring the ISM. Thermal instabilities arises from the properties of the thermal equilibrium of the gas. I will not review here all the cooling and heating mechanisms \citep[see e.g. the book by][]{Lequeux2005}, but rather focus on the properties of a thermally unstable gas.


\subsubsection{Thermal instability}
\index{Thermal instability}
\index{Instabilities!thermal}

\begin{figure}
  \begin{center}
    \includegraphics[width=0.9\textwidth]{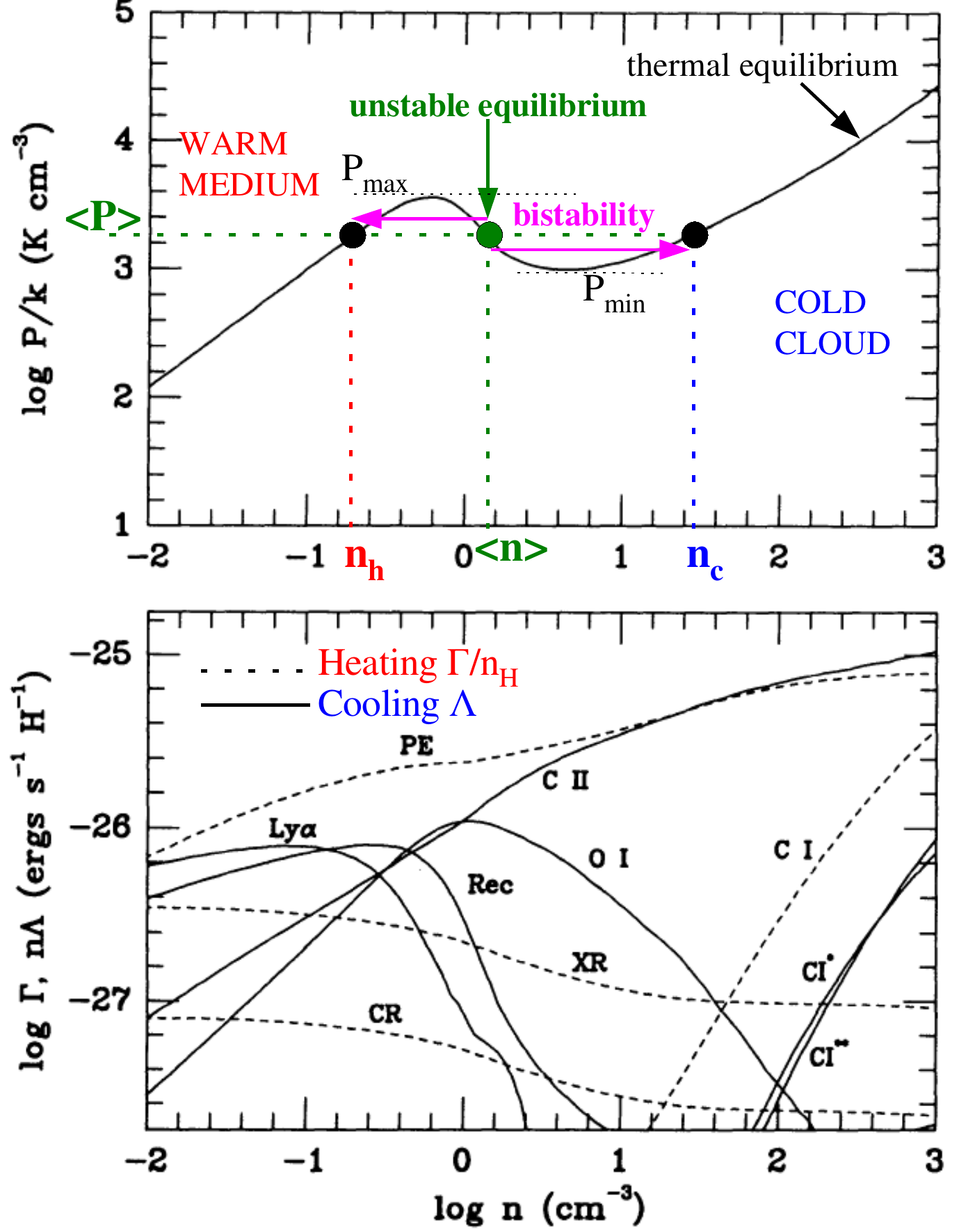}
    \caption[Thermal equilibrium and instability in the ISM]{Thermal equilibrium and  instability in the ISM. \textit{Top:} Thermal pressure ${\rm P}/k_{\rm B}$ at equilibrium as a function of hydrogen density $n_{\rm H}$ in the solar neighbourhood ISM, the X-ray heating being evaluated for $N_{\rm H} = 10^{19}$~cm$^{-2}$. If the gas at ($P, n_{\rm H}$) is above the curve, the representative point ($P, n_{\rm H}$) will move to right because its pressure is too high, and cooling dominates over heating. Inversely, a medium below the equilibrium curve will move to the left. \textit{Bottom:} Contributions of the different heating ($\Gamma / n_{\rm H}$ in our notation, dashed lines)  and cooling ($\Lambda$, solid line) mechanisms. For the \textit{heating}, PE$=$photoelectric effect on grains; XR$=$X-ray heating; C$\,${\sc i}$=$carbon photoionization. For the \textit{cooling},  C$\,$II$=$cooling by the [C$\,${\sc ii}]$\lambda 158\,\mu$m line; O$\,$I$=$cooling by the [O$\,${\sc i}]$\lambda 63\,\mu$m; CI$^{*}$ and CII$^{*}=$cooling by [C$\,${\sc i}]$\lambda 609\,\mu$m and 370$\,\mu$m; Ly$\alpha=$cooling by excitation of Lyman $\alpha$ and other transitions; Rec$=$cooling by recombination on the grains. Figures adapted from \citet{Wolfire1995}.
}
    \label{fig:thermal_instability}
  \end{center}
\end{figure}

\index{Heating rate}
\index{Cooling rate}
I will define the global cooling function [erg~s$^{-1}$~H$^{-1}$], i.e. the net heat loss function, by:
\begin{equation}
\mathcal{L} (\rho, \rm T) = \Lambda - \Gamma \ ,
\end{equation}
where $\Lambda$ and $ \Gamma$ are the total cooling and heating rates, respectively. At thermal equilibrium we have $\mathcal{L} (\rho, \rm T) = \Lambda - \Gamma =0$, which determines the temperature of the medium for a given density. 
This condition is plotted in the top panel of Fig.~\ref{fig:thermal_instability}. Above the thermal equilibrium curve, the cooling dominates the heating: $\Lambda > \Gamma$. Below, $\Lambda < \Gamma$. The bottom panel shows the main contributions to the $\Lambda $ and $\Gamma$ functions. 

The first discussion of thermal instability in low-density media like the ISM has been made in a classical paper by \citet{Field1965}. I will follow his analysis to derive the thermal instability criterion. Let us consider a volume of interstellar gas (the system) in thermal equilibrium at (P, T) and introduce a perturbation of density (or temperature) while keeping constant another thermodynamic variable, $A$, for instance the pressure or the density. Since the entropy of the medium is defined by
\index{Entropy}
\begin{equation}
\label{eq:def_entropy}
\frac{1}{{\rm T}}  = \left[ \frac{ \partial {\rm S}}{\partial {\rm E}}\right]_A \ ,
\end{equation}
$\rm E$ being the internal energy of the system. The perturbation yields to  a change $\delta \mathcal{L}$ in the cooling efficiency:
\begin{equation}
\mathcal{L}\longrightarrow \mathcal{L} + \delta \mathcal{L}
\end{equation}
Rewriting Eq.~\ref{eq:def_entropy}, the associated change $\delta {\rm S}$ in the entropy can be expressed as follows
\begin{equation}
d(\delta {\rm S}) = - \frac{1}{{\rm T}} \, \delta \mathcal{L} \, d t \ ,
\end{equation}
which can be re-written as
\begin{equation}
\frac{d(\ln \vert\delta {\rm S}\vert)}{d t} = - \frac{1}{{\rm T}} \left[ \frac{\delta \mathcal{L}}{\delta {\rm S}} \right] _A \ ,
\end{equation}
This equation gives the time-evolution of the perturbation of the entropy of the system. If this perturbation $\delta {\rm S}$ grows with time, the thermal equilibrium of the system is \textit{unstable}. This \textit{criterion of thermal instability} can thus be written as
\begin{equation}
 \left[ \frac{\delta \mathcal{L}}{\delta {\rm S}} \right] _A < 0
\end{equation}
\index{Thermal instability!criterion}
For the isobaric case, $A = {\rm P}$, and since ${\rm T} d {\rm S} = c_{\rm P} d {\rm T}$, the instability criterion can be re-written as 
\begin{equation}
\left[ \frac{\delta \mathcal{L}}{\delta {\rm S}} \right] _{\rm P} < 0 \Longleftrightarrow  \left[ \frac{\delta \mathcal{L}}{\delta {\rm T}} \right] _{\rm P} < 0 \Longleftrightarrow   \left[ \frac{\delta \rm{P}}{\delta \rho} \right] _{\mathcal{L} = 0} < 0 \ .
\end{equation}
The last form of the instability criterion\footnote{This last form is obtained by using $\displaystyle \left( \frac{d \log {\rm P}}{d \log \rho}\right) _{\mathcal{L}=0} = \left(\frac{\partial \mathcal{L}}{\partial {\rm T}} \right)_{\rm P}  / \left(\frac{\partial \mathcal{L}}{\partial {\rm T}} \right)_{\rm \rho} $.} can be interpreted by looking at the top panel of  Fig~\ref{fig:thermal_instability}. Let us consider the green point, located in the portion of the equilibrium curve where its slope is negative. If density increases, the pressure of the system  diminishes, and becomes smaller than the ambient, external pressure. Thus the system is compressed, and the density further increases. The equilibrium is \textit{unstable} in this zone. 
This \textit{positive feedback} will stop when the gas reaches a stable equilibrium (the black point to the right in this case).

\subsubsection{When thermal instability becomes bi-stability}
\index{Thermal instability!bistability}

The thermal instability is the core of the \citet{Field1969} two-phase model of the ISM, and I have sketched the principle of the bi-stability in the top panel of Fig.~\ref{fig:thermal_instability}. 
Let us consider again our green point, at ($\langle \rm P \rangle$, $\langle n_{\rm H} \rangle$). 
Any perturbation will make the system leave this unstable equilibrium and evolves toward a stable one. If the pressure of the system falls within the [P$_{\rm min}$, P$_{\rm max}$] range, 
 Fig.~\ref{fig:thermal_instability} shows that there are three equilibrium configurations: an unstable one (green point), and two stable (black points). The stable configurations are identified as the \textit{warm intercloud medium} and the \textit{cold atomic clouds} by \citet{Field1969}.

A few remarks need to be given about this simple model. First, the range of pressures observed in the ISM is not always compatible with the unstable part of the thermal equilibrium curve.  Secondly, the pressure equilibrium is not always reached. This is the case at small scales, because the timescale of the propagation of the perturbation (speed of sound times the length-scale) is shorter than the characteristic interval between two perturbation (typically a few $10^5$~yr for supernovae explosions, yielding to sizes of 0.4~pc for the CNM and 4~pc for the WNM). Third, in the warm medium, the cooling time is larger than the perturbation time, and therefore the thermal equilibrium may not be reached for the WNM. 
In addition, note that the criterion for thermal instability has to be modified for \textit{dynamical} systems, out of thermal equilibrium. The interested reader can refer to \citet{Balbus1986}. 
And last, depending on the cooling/heating process at work, the metallicity of the gas, the presence of dust or not, the shape of the cooling function may vary as function of the environment, and sometimes it may not exhibit and unstable part. We will come back to this point in sect.~\ref{sec:time-dependent-cooling-plasma}.

\section{Evolution of interstellar dust}
\label{sec_life_dust}
\index{Dust!see Interstellar dust}
\index{Interstellar dust!dust evolution}


\subsection{The importance of dust for our study of H$_ {\bf 2}$-luminous galaxies}

Dust grains are nanometer- to micrometer-sized solid particles that are mixed with the interstellar gas\footnote{At the end of the 18th century, William Herschel made the first observations of ``dark nebulae'' (dusty interstellar clouds). It is only at the beginning of the
20$^{\mbox{\scriptsize th}}$ century that astronomers began to consider the absorption and scattering of light by dusty clouds. Today, dust particles are known to be ubiquitous in galaxies.}.  Although dust represents only $\approx 1$\% of the total mass of the ISM, it plays a central role in the physics and chemistry of this medium. Dust grains are particularly important for our study for two main reasons:
\begin{description}
\item[H$_{\bf 2}$ formation:] dust grains are the sites for H$_2$ formation (this will be discussed in  chapter~\ref{chapter:H2Molecule}). More generally, dust grains act as catalysts for chemical reactions in the ISM.
\item[Thermal balance of the gas:] dust grains participate to the heating (via the photoelectric effect) and the cooling of the hot gas (via inelastic collisions with electrons). 
\end{description}

\index{Interstellar dust!lifecycle of dust}

Within the cycle of interstellar matter (sect.~\ref{mass-energy-transfer-ISM-phases} and Fig.~\ref{fig:cycle_ISM}), dust is prone to evolutionary processes, being constructive and destructive,  which lead to an exchange of metals with the gas. Dust nucleation occurs in cool atmospheres of dying stars, in supernovae, and dust grains grow in size in the ISM by condensation of metals. They are released to the diffuse medium by stellar winds or supernovae explosions, where they experience processing (in particular by shock-waves). Dust is then re-incorporated  into newly-forming stars out of molecular clouds.

In the multiphase ISM of H$_2$-luminous galaxies, we also expect dust destruction to occur. On the other hand, the presence of dust is required for H$_2$ formation. Therefore, within the framework of my modeling of the gas cooling and H$_2$ formation, I have been driven to study and model dust destruction processes. I have carried on a detailed calculation of the time-dependent cooling of a shock-heated dusty plasma, taking into account dust destruction. This calculation has been published in the appendix of \citet{Guillard2009} (hereafter \hyperref[paper_SQ_H2]{paper~{\sc i}} ) in the context of the Stephan's Quintet galaxy collision. Before presenting these results, that may apply to other astrophysical situations, we introduce the properties of interstellar dust that are relevant for dust evolution modeling (sect.~\ref{dust-composition-size-distribution}), and the physical processes of dust destruction (sect.~\ref{dust-destruction-processes}). 

\subsection{Dust composition and size distribution}
\label{dust-composition-size-distribution}
\index{Interstellar dust!composition}
\index{Interstellar dust!size distribution}

The composition of dust is only loosely constrained by observations. The most direct information on the composition of dust\footnote{Although meteorites provide us with genuine specimens of interstellar grains for examination, these are subject to severe selection effects, and cannot necessarily be considered representative of interstellar grains.} comes from spectral features observed in absorption or emission\footnote{The amplitude of the radiated electromagnetic field from small dust particles depends not only on the incident radiation, but also on the size, shape, refractive index and polarizability of the grain.}.

\begin{figure}
\begin{minipage}{\textwidth}
    \def\thefootnote{\alph{footnote}}
      \setlength{\footnotesep}{0pt}
  \centering
    \includegraphics[width=\textwidth]{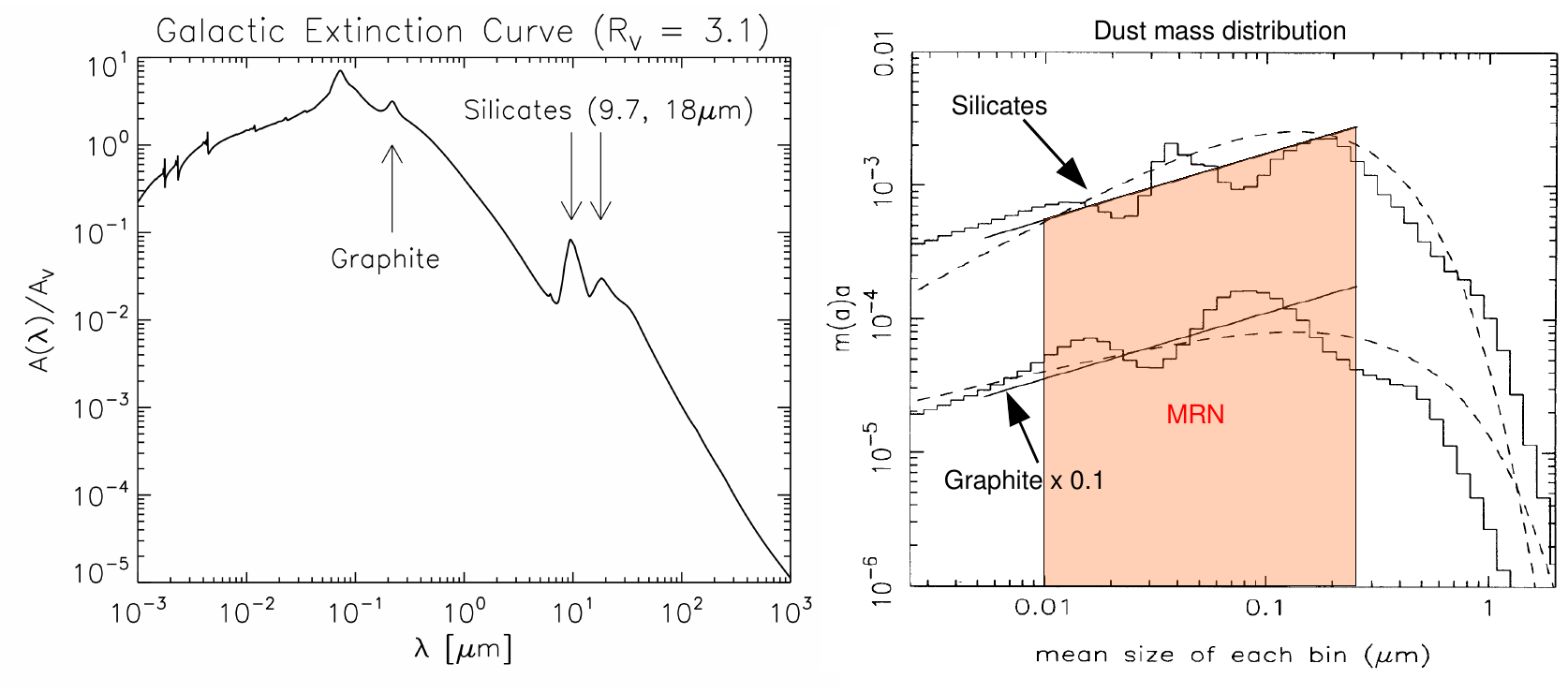}
    \caption[Interstellar dust extinction and mass distribution]{Dust extinction and mass distribution\footnotemark[1]. \textit{Left:} Extinction law computed from the \citet{Weingartner2001} model fitted to the Galactic extinction curve. \textit{Right:} Mass distributions of interstellar grains for the difuse interstellar medium ($R_{\rm V} = A_V / E(B-V) =  3.1$), expressed relative to the mass of hydrogen. The power-law ($n(a) \propto  a^{-3.5}$ or $m(a) \propto a^{-0.5}$) is the MRN distribution. The histograms show the mass distributions deduced by \citet{Kim1994} from the extinction curve. The upper histogram is for silicates, and the lower for graphite (scaled down by a factor of 10 for clarity).}
    \label{fig:dust_extinction_size_dist}
\footnotetext[1]{Note that the two plots presented in this figure do not belong to the same model. These plots are used here for an illustration purpose.}
  \end{minipage}
\end{figure}

\subsubsection{Dust extinction and size distribution}
\index{Interstellar dust!extinction}
\index{Extinction}

The left panel of Fig~\ref{fig:dust_extinction_size_dist} shows the {\it extinction law} (absorption plus scattering) from the \citet{Weingartner2001} model. 
The extinction $A_{\lambda}$, which is the ratio of the flux $I(\lambda)$ after crossing the 
dust cloud to the incident flux $I_0(\lambda)$, is, in terms of magnitude,
\begin{equation}
  I(\lambda) = I_0(\lambda)\,10^{-A_{\lambda}/2.5} = I_0(\lambda)\,e^{-\tau_{\lambda}}\; ,
\end{equation}
\index{Optical depth}
\index{Color excess}
where $\displaystyle \tau_{\lambda} = 0.921\,A_{\lambda}$ is the {\it optical depth}.
Observationally, the extinction is mostly obtained by comparing the reddened spectrum of stars to the relevant intrinsic spectrum from a library of spectral models \citep[e.g.][]{Fitzpatrick1990}. This gives a relative curve, which is usually normalized to the extinction in 
some broad spectral bands, like B and V. For instance, $A_{\lambda}$ is often divided by the {\it 
color excess} $E(B-V) = A_B - A_V$ or simply by $A_V$ like in Fig.~\ref{fig:dust_extinction_size_dist}.

The extinction curve  provides constrains on the size distribution of dust particles.
\citet{Mathis1977} showed that the average interstellar extinction could be satisfactorily reproduced by a grain model containing two components: graphite grains and silicate grains. Remarkably, the extinction curve was reproduced very well if both grain components had power-law size distribution, so-called $MRN$\footnote{Mathis J.S., Rumpl W., Nordsieck K.H.} $distribution$:
\begin{equation}
  dn_i = A_i n_H a^{-3,5} da \; ,
\end{equation}
where $dn_i$ is the number of grains of the species $i$ which radii between $a$ and $a+da$, 
$n_H$ is the reference hydrogen nuclei density, and $A_i$ is a normalization constant. This 
distribution is truncated at a minimum grain size \mbox{$a_{min} \simeq 50$~\AA} and a maximum size $a_{max} \simeq 2500$~\AA. The right panel of Fig~\ref{fig:dust_extinction_size_dist} shows the MRN mass distribution, overlaid on calculations by \citet{Kim1994}. Although big grains are less numerous than small ones, they carry the bulk of the dust mass.
\index{Extinction!and dust size distribution}

The extinction curve is often represented by plotting the optical depth per unit total hydrogen column density $N_H$. This quantity is linked to the extinction by the relation between 
the color excess and the hydrogen column density: $\displaystyle N_H / E(B-V) = 5.8\times 10^{21}$~atoms~cm$^{-2}$~mag$^{-1}$. We briefly give the principle of the theoretical calculation of the extinction law. The optical depth can be written as
\begin{equation}
  \tau _{\lambda}  = \sum _i \int _{a_{min}}^{a_{max}} Q_e (a_i,\lambda)\pi a_i^2 N_i(a_i)da_i \; ,
\end{equation}
where $N_i(a_i)$ is the grain column density for the species $i$ and $Q_e$ is the extinction 
efficiency.  For spherical particles of radius $a$, these quantities can be expressed from the absorption and scattering cross sections: $\displaystyle \sigma _a = \pi a^2 Q_a$ and 
 $\displaystyle \sigma _s = \pi a^2 Q_s$. The calculation of $Q_a$ and $Q_s$ as 
a function of the grain properties and the wavelength was first performed by \citet{Mie1908}.
The success of the \citet{Mathis1977} grain model led \citet{Draine1984}
to refine the optical constants and extend the treatment into the mid- and far-infrared. In a more recent study, \citet{Weingartner2001}  added very small carbonaceous grains which play a role in the UV extinction and IR emission (Fig.~\ref{fig:dust_extinction_size_dist}).

\subsubsection{Composition of interstellar dust}

Interstellar dust grain models have been improved for 30 years in order to fit observational constraints such as elemental abundances of the heavy elements, UV, visible and
infrared absorption and scattering properties, infrared emission, polarization properties of the
absorbed and emitted light. The models include \textit{a mixture of silicate grains
and carbonaceous grains}, each with a wide size distribution ranging from molecules containing
tens of atoms to large grains $\gtrsim 0.1\,\mu$m in diameter that can be coated with ices in dense clouds  \citep{Desert1990, Li2001}.
\index{Interstellar dust!models}


Immediately following the discovery of the 2175~\AA~bump by \citet{Stecher1965}, \citet{Stecher1965a} pointed out that small graphite particles would produce absorption very similar to the observed feature.  
However, the presence of crystalline grains is considered unlikely in interstellar space where grains are exposed to cosmic rays. Other forms  of carbonaceous grains (PAHS, amorphous carbons, and organic mantles on silicate grains) are being considered. 

The absorption feature at $\sim$ 9.7 $\mu$m is due to interstellar silicate minerals, these having
strong resonances near 10 $\mu$m due to the Si-O stretching mode. This conclusion is
strengthened by the fact that a 10 $\mu$m emission feature is observed in outflows from cool
oxygen-rich stars (which would be expected to condense silicate dust)
but not in the outflows from carbon-rich stars (where silicates do not form because
all of the oxygen is locked up in CO). There is also a broad feature at 18 $\mu$m that
is presumed to be the O-Si-O bending mode in silicates \citep{McCarthy1980}.

The extinction curve is strongly modified when one looks along different directions, which probe  the spatial variations of the dust grain properties. Some of these variations may correspond to the coagulation into bigger grains \citep{Dominik1997, Stepnik2003}, or on the contrary to some erosion or destruction of grains. 

\subsection{Dust processing in shocks} 
\label{dust-destruction-processes}
\index{Interstellar dust!destruction processes}

I focus here on the dust destruction mechanisms relevant for the calculation of the cooling of a dusty, shock-heated plasma.  I direct the reader to the review by \citet{Jones2004} and references therein for an exhaustive presentation of all the destruction processes in the ISM evoked above. For a detailed modeling of dust evolution in shocks, please see the PhD thesis of \citet{Guillet2008}.

\begin{figure}
\begin{minipage}{\textwidth}
    \def\thefootnote{\alph{footnote}}
      \setlength{\footnotesep}{0pt}
   \centering
\includegraphics[width=\textwidth]{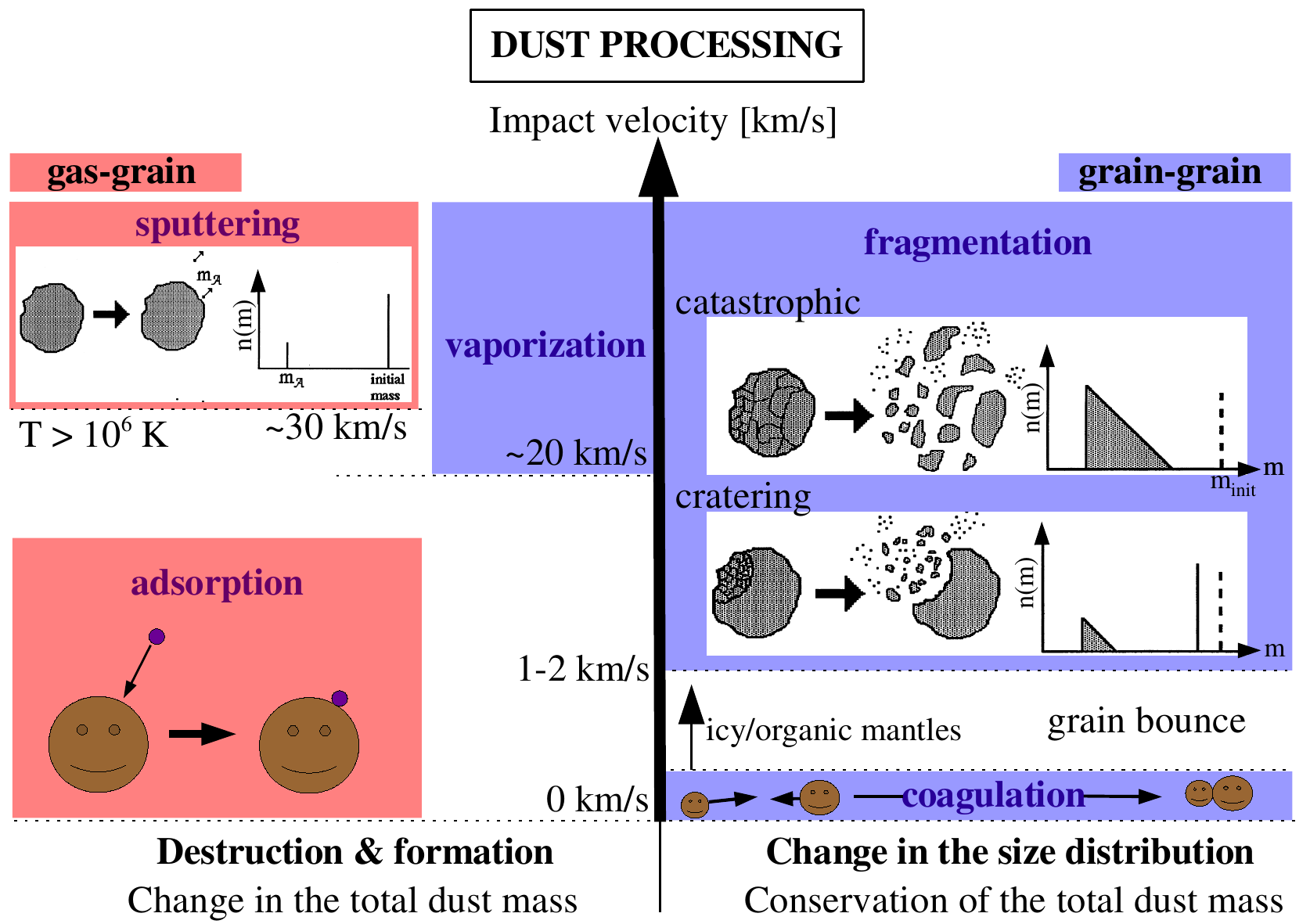}
    \caption[Overview of the dust processing in shocks]{Overview of the dust processing in shocks\protect \footnotemark[1]. The mechanisms are ordered as a function of the impact velocity and separated in two groups: those which modify the total dust mass (formation or destruction), and those which affect the size distribution while conserving the total dust mass. The small inset figures are from \citet{Borkowski1995} and show schematic description of the processing of refractory grains: sputtering, craterisation (with partial vaporization), and catastrophic fragmentation (accompanied with total vaporization). On the right of the cartoons the mass distribution of the fragments is sketched. Note that the velocity threshold for coagulation (that may lead to formation of icy and organic mantles) depends on the grain size.}
    \label{fig:dust_processing_overview}
\footnotetext[1]{This figure is inspired from a slide of Vincent Guillet.}
\end{minipage}
\end{figure}

The Fig.~\ref{fig:dust_processing_overview} gathers the mechanisms responsible for dust processing in shocked gas. Grain processing arises from interactions between the grain and ``particles'' (grain, atom, ion, atomic nucleus, electron, photon). They can be \textit{destructive}, i.e. leading to  a transfer of the grain atoms to the gas and thus to a net mass loss from the grain, or \textit{non-destructive}, the grain size distribution being modified but the total dust mass conserved. These processes are counterbalanced by formation, like the accretion and nucleation, but also coagulation between grains.

\index{Interstellar dust!sputtering}
\index{Interstellar dust!ion field emission}
\begin{description}
\item[Destructive processes:] 
destructive interactions can expulse directly atoms and ions from the grain surface (via \textit{sputtering}, ion field emission\footnote{direct expulsion of atoms and ions from the surface at extreme surface irregularities \citep{Draine1979}} or direct Coulomb explosion via extreme charging effects, coming from electron-grain interactions or photoelectric effect), or vaporize/sublimate the grain in case of $\gtrsim 20$~km~s$^{-1}$ grain-grain collisions.  Vaporisation/sublimation via absorption of energetic photons (UV, $\gamma$) or interaction with cosmic rays, may also be an important destruction mechanism for volatile material, such as ice mantles. 

\index{Interstellar dust!fragmentation}
\index{Interstellar dust!shattering}
\item[Non-destructive processes:] 
Grain-grain collisions at velocities $\gtrsim 1-2$~km~s$^{-1}$ lead to \textit{fragmentation} (or \textit{shattering}) of both grains \citep[see Fig.~\ref{fig:dust_processing_overview} and][]{Tielens1994}. The size distribution of the fragments is a power-law, suggesting that dust fragmentation in shocks is responsible for the observed power-law (MRN) in the diffuse ISM. By definition, fragmentation is not a destructive process. It re-distributes the dust mass towards smaller grains, increasing the total cross-section of the distribution of grains, and therefore the optical and UV extinction. 
\end{description}

In the following I will focus on the sputtering of grains by ion-grain interaction because it is the dominant destructive process in a hot ($T=10^6 - 10^8$~K) plasma \citep[e.g.][]{Draine1979, Dwek1996}. 

\subsubsection{Sputtering of interstellar grains}

Behind a shock, gaseous ions impacting on a grain at high velocities will sputter atoms from the 
surface layer. We distinguish 2 cases of sputtering, depending on the nature of the shock and 
physical conditions of the environment.

\index{Sputtering!thermal}
\begin{itemize}
\item For high velocity ($V_{\rm s} > 300$~km~s$^{-1}$), 
grains are dragged to rest with respect to the gas  before the postshock gas cools and compresses,  so that sputtering arises from the thermal motion of ions in the gas. It is then the most  abundant H$^+$ protons that dominate this \emph{thermal} sputtering. This case happens in a hot gas ($T \gtrsim 10^6$ K), for example in the postshock of a supernova-generated bubble.
\begin{figure}
   \centering
\includegraphics[width=0.6\textwidth, angle=90]{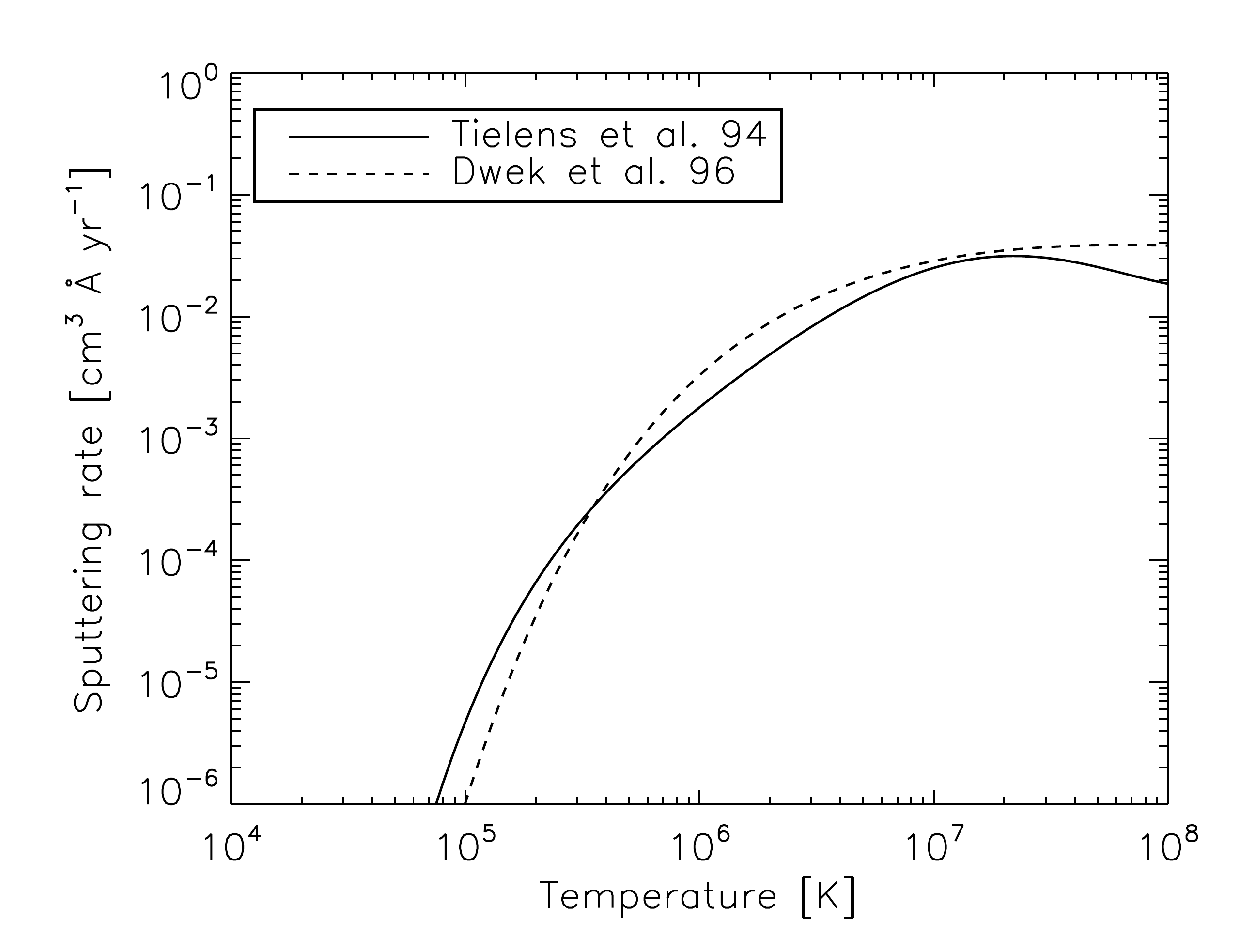}
    \caption[Thermal sputtering rates]{Thermal sputtering rates $\frac{1}{n_{\rm}} \frac{da}{dt}$ for silicate grains as a function of the gas temperature we use in our calculation of the time-dependent cooling of a dusty plasma (see sect.~\ref{sec:time-dependent-cooling-plasma}). The solid line shows the sputtering rates I adopt in my modeling \citep[from][]{Tielens1994}. For comparison, the dashed line show the exponential form (Arrhenius law) of  \citet{Dwek1996} that is practical for analytical calculations. Impacts with H, He, C, N and O are considered in the calculations. Below $10^6$~K, the sputtering of dust grains in a hot gas drops drastically.}
    \label{fig:yields}
\end{figure}
Therefore, for $V_{\rm s} > 300$~km~s$^{-1}$, thermal sputtering dominates over inertial sputtering, so we ignore inertial sputtering. In this case, the grain sputtering rate is nearly constant for  $T \gtrsim 3 \times 10^6 \,$K and strongly decreases at lower temperatures \citep[see Fig.~\ref{fig:yields} and][]{Draine1979, Tielens1994, Dwek1996}. The rate of decrease in grain size through sputtering is given by
\begin{equation}
\dot{a}_{s} = \frac{d a}{d t} = - \frac{m_{\rm sp}}{2 \, \rho _g} \, n_{\rm H} \, \sum A_i \left\langle Y_i \, v \right\rangle \ ,
\end{equation}
where $m_{\rm sp}$ and $\rho _g$ are the average mass of the sputtered atoms  and the density of the grain material, $A_i$ the abundance of impacting ion $i$ and $ Y_i $ is the sputtering yield\footnote{The yield is the number of atoms ejected from the grain surface per incident particle.} of ion $i$. The quantity $\left\langle Y_i \, v \right\rangle$ is the impacting ion velocities $v$ times the velocity-dependent yield averaged over the Maxwellian distribution. 
Fig.~\ref{fig:yields} shows the sputtering rates I use in my modeling. These yields are from \citet{Tielens1994}. Then, assuming that the grains are spherical, the rate of grain mass destruction by sputtering, $\dot{m}_{s}$, can be easily calculated from $  \dot{m}_{s} = 4 \pi a^2 \rho _g \dot{a}_{\rm s}$.
\index{Sputtering!yields}

\index{Sputtering!inertial}
\index{Sputtering!non-thermal}
\item
For low velocity ($v_s \le 300\, \mbox{km}\,\mbox{s}^{-1}$) shocks, grains 
are moving at high velocities with respect to the gas\footnote{In this case, charged grains are 
betatron-accelerated, because of their gyration around the magnetic field lines.}. The relative gas-grain velocities result in atoms and ions collisions on grain surfaces. These collisions lead to the so-called \emph{nonthermal} or \emph{inertial} sputtering. 
If a grain is moving through a stationary gas, all particles incident upon the grain have the same relative velocity.
This occurs in shock waves where the gas is rapidly swept up but the dust lags due to its 
greater inertia. In general, the  He$^+$ ions  dominate the erosion 
process \citep{Tielens1994,  Jones1994}.
In the case of inertial sputtering, the parametrization of the erosion rate of the grain is more complex. I have followed the empirical approach detailed in \citet{Borkowski1995} to compute the rate of destruction of a grain slowing down in a dusty gas. In the following, I give the principle of this calculation.

\index{Sputtering!collisional drag}
\index{Craterisation}
\index{Interstellar dust!craterisation}
\index{Interstellar dust!vaporisation}
The deceleration of a grain through the gas by collisions with the atoms or ions of the gas can be written as
\begin{equation}
\label{eq_dust_decel}
\frac{d v_g}{d t}  = - \frac{\beta \, \pi \, a ^2 \, \rho \, v_{g}^{2}}{m} \ ,
\end{equation}
where $v_g$ is the grain velocity with respect to the gas, $m$ and $a$ are its mass and radius, $\rho$ is the gas density, and $\beta$ is the enhancement of the collisional drag in a plasma relative to that of a neutral medium. The initial grain velocity is set to $3/4$ of the shock velocity \citep{Jones1994}. We take $\beta =1$, which maximizes the dust destruction\footnote{The collisional drag of the grain in an ionized plasma can be enhanced ($\beta > 1$) because of charge effects \citep[see][]{Shull1978, Draine1979b}}. Therefore, our computation of the dust survival is conservative.
As the grain is decelerating through the gas, it also experiences collisions with the other grains, which results in craterization or complete vaporization of the grain (see Fig.~\ref{fig:dust_processing_overview}). Along its trajectory, the rate of reduction of the grain mass is the sum of the inertial sputtering and craterization rates:
\begin{equation}
\label{eq:dmdt_s_cr}
\frac{d m}{d t} = \dot{m}_{is} + \dot{m}_{cr}
\end{equation}
For a given initial grain size, I have calculated the evolution of the dust-to-gas mass ratio by integrating Eq.~\ref{eq_dust_decel} and \ref{eq:dmdt_s_cr}. 
The inertial mass sputtering rate is given by \citep{Tielens1994}
\begin{equation}
 \dot{m}_{is} = 2 \pi a^2 m_{\rm sp} \, v_g \, n_{\rm H} \sum A_i  Y_i  
\end{equation}
and the mass erosion rate due to cratering collisions with  dust particles (field) of mass $m_f$ and  mass distribution $n(m_f)$  can be expressed as  \citep{Borkowski1995}
\begin{equation}
 \dot{m}_{cr} = - \mathcal{A} m_{\rm H} n(m_f) \pi a_g^2 v_g Y_{cr} = - n(m_f) \pi a^2 v_g f_c m_g \ ,
\end{equation}
where $a_g$ and $m_g$ are the radius and mass of the target grain, and $v_g$ is the relative velocity between the target grain and the impinging dust particles ($m_f$). 
$\mathcal{A}$ is the mean atomic mass in a.m.u. of a constituent atom of the fragments. $Y_{cr}$ is a dimensionless cratering yield equal to $m_f v^2 / 2 E_{cr}$, where $E_{cr}$ is the specific energy in eV~atom$^{-1}$ for the ejection of one atom in a cratering event. The quantity $ \mathcal{A} m_{\rm H} Y_{cr}$ represents the average mass excavated in each collision, which can be written as $f_c m_g$, where $f_c$ is the fraction of the target grain mass $m_g$ that is ejected in the collision. We have adopted $f_c = 0.1$ and $E_{cr} = 2$~eV for silicate grains \citep[see][for details]{Borkowski1995}.

\end{itemize}

I have described the physical processes that are taken into account for dust destruction and detailed how I have calculated the evolution of the dust-to-gas mass ratio in the postshock gas.  This calculation was based on previous studies. 
Let us now couple this calculation to that of the cooling of shock-heated gas.

\section{Time-dependent cooling of a dusty plasma}
\label{sec:time-dependent-cooling-plasma}

\index{Plasma!cooling}
\index{Plasma!dusty plasma}
In the extreme environments of H$_2$-luminous galaxies, like the Stephan's Quintet galaxy collision, starburst- or AGN-driven winds, galaxy cluster cores, the warm H$_2$ gas is coexisting with a hot ($T > 10^6$~K) plasma.  I have computed the time-dependent cooling of a gas that is shock-heated to these high temperatures. 
Initially, the gas is dusty, with a Galactic dust-to-gas mass ratio. The calculation of dust destruction detailed above is coupled to that of the gas cooling. This calculation is presented in the appendix of \citet{Guillard2009}. I briefly remind the principle of the calculation and summarize the main results, emphasizing some points that were not stressed in the paper.

\subsection{Calculation  method}

The time-dependent total cooling function is the sum of the dust cooling (weighted by the remaining fraction of dust mass at the time $t$) and the gas cooling contributions. 
From a range of postshock temperatures up to more than $10^7$~K, the isobaric gas cooling is calculated by integrating the energy balance equation which gives the rate of decrease of the gas temperature:
\begin{equation}\label{Eq_ratetemp}
\frac{5}{2}  \, k_{\rm B} \, \frac{d T}{d t} = - \mu \, n_e \left(f_{\rm dust} \, \Lambda _{\rm dust} + \Lambda _{\rm gas}\right) \ ,
\end{equation}
where $\mu$ is the mean particle weight ($\mu = 0.6~\rm a.m.u$ for a fully ionized gas), $n_e$ is the electron density, $k_{\rm B}$ the Boltzman constant, $f_{\rm dust}$ the dust-to-gas mass ratio, $\Lambda _{\rm dust}$ and $\Lambda _{\rm gas}$ are respectively the dust and gas cooling efficiencies per unit mass of dust and gas, respectively.
For the gas cooling efficiency, $ \Lambda _{\rm gas}$, I use the recent non-equilibrium, isobaric cooling curves calculated by \citet{Gnat2007}. The dust cooling efficiency (dust heating by electrons), $ \Lambda _{\rm dust} $ is calculated following the method of \citet{Dwek1987}. The calculation is stopped when the gas has cooled to $10^4$~K.

At each time step, the fraction of dust remaining in the gas $f_{\rm dust}$, is computed in parallel to the postshock gas cooling:
\begin{equation}
f_{\rm dust} = \left( \frac{a}{a_{\rm eff}} \right) ^{3} \ \rm with \ a =  a_{\rm eff} - \int _{t_0} ^{t} \dot{a} \, dt \ ,
\end{equation}
where $ \dot{a} = da / dt$ is the rate of reduction of grain size that we have parametrized above. In the following, it is assumed that dust grains have an initial radius equal to the effective  (mean)  radius of $a_{\rm eff} = 0.05 \, \mu$m. This radius is the effective mean radius computed for the MRN dust size distribution. 

\subsection{Results}

\begin{figure}
   \centering
\includegraphics[height=\textwidth, angle=90]{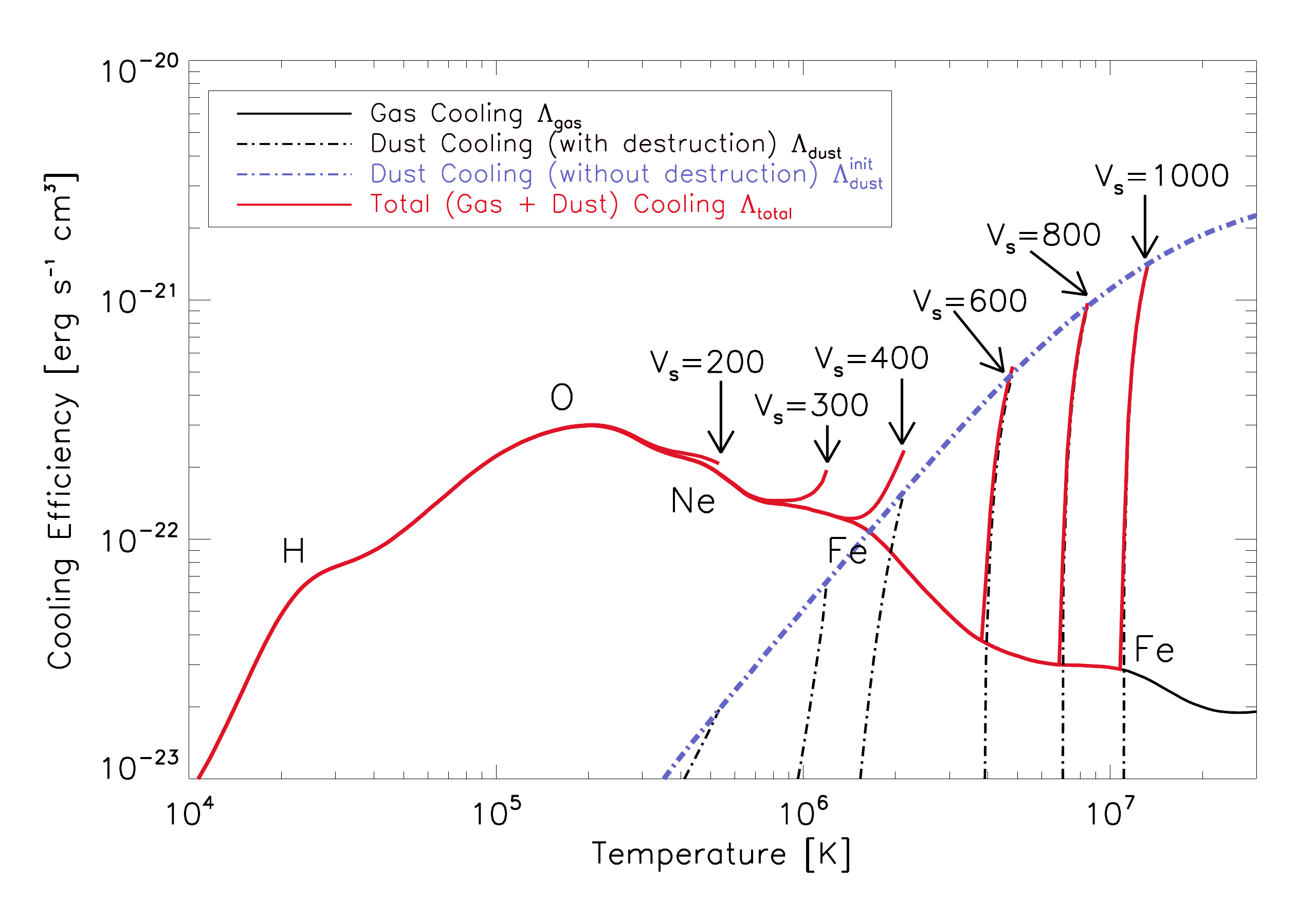}
    \caption[Cooling function of a dusty plasma]{Time-dependent cooling efficiency [erg cm$^3$ s$^{-1}$] as a function of temperature during the gas cooling, for different initial conditions. The blue dashed line, $\Lambda _{\rm dust}^{\rm init} (T)$, represents the \emph{initial} cooling function of the gas including the dust contribution, computed for a MRN interstellar dust size distribution ($0.01$ to $0.25 \; \mu$m dust particles).
The black dashed lines are the dust cooling functions for different shock velocities  (i.e. different initial temperatures) which take into account the destruction by sputtering during the cooling.
The cooling function due to atomic processes, $\Lambda _{\rm gas} (T)$, is displayed for an isobaric and non-equilibrium (time-dependent) cooling, and for solar metallicities ($Z=1$). The dominant cooling elements at various temperatures are indicated on the curves. The red lines are the  total cooling functions, $\Lambda _{\rm total} = \Lambda _{\rm dust} + \Lambda _{\rm gas}$, for different initial temperatures or shock velocities. The shock velocities are indicated in km~s$^{-1}$. Their starting points are indicated by the arrows.}
    \label{fig:cool_function_dyn}
\end{figure}

The resulting cooling functions are shown on Fig.~\ref{fig:cool_function_dyn}. 
The different initial temperatures corresponds to different shock velocities. 
The blue dashed curve is the dust cooling function assuming a constant dust-to-gas mass ratio. The red curves are the time-dependent cooling curves taking into account dust processing.  

I only briefly summarize here the main results, discussed in more detail 
 in \hyperref[paper_SQ_H2]{paper~{\sc i}}. 
 If the preshock gas is dusty, the dust is initially the most efficient coolant. Indeed, at temperature hotter than $\approx 3\times 10^6$~K, the dust cooling function (blue curve on Fig.~\ref{fig:cool_function_dyn}), dominates the gas cooling. 
This produces a ``flash'' of IR emission. The duration of this ``flash'' depends on the initial gas temperature and dust-to-gas mass ratio. This collisional heating of dust is likely to be responsible for the far-IR emission in the intracluster medium of galaxy groups  \citep[see][for a model of dust emission within the contet of the Virgo cluster]{Popescu2000}.

Significant dust emission from a hot gas does not last forever. As discussed in \citet{Smith1996} and in \citet{Guillard2009} (\hyperref[paper_SQ_H2]{paper~{\sc i}}), the dust cooling efficiency drops with time because of the reduction of dust mass (black dashed line in Fig.~\ref{fig:cool_function_dyn}). 
When the grains are significantly sputtered, the dust cooling efficiency becomes lower than that of the gas. This is why the total cooling curve (red line) ``rejoins'' the gas cooling curve (black line). As a result, the cooling times of the dusty plasma are significantly longer than those calculated with a constant dust-to-gas mass ratio (see Fig.~B.2 of \hyperref[paper_SQ_H2]{paper~{\sc i}}). 
Consequently, the cooling timescale of the hot gas is longer than the survival timescale of dust grains (see Fig.~2 of \hyperref[paper_SQ_H2]{paper~{\sc i}}).

\index{Plasma!thermal instability}
Looking at the cooling functions shown on Fig.~\ref{fig:cool_function_dyn}, we can see that the reduction of the dust-to-gas mass ratio in a cooling plasma also favors the development of thermal instability. If the dust-to-gas mass ratio was constant, the slope of the total ($\Lambda _{\rm total} = \Lambda _{\rm dust} + \Lambda _{\rm gas}$) cooling function would be almost always positive, whereas in the case of dust destruction, the total cooling curve has an unstable part over a wide range of t emperatures, between $\approx 10^5 - 10^7$~K.

Note that a significant dust mass can still remain in the hot gas, because very large grains may survive sputtering in the hot gas. 
These results are in very good agreement with that of \citet{Smith1996}, whom treat carefully the effect of the dust size distribution, but only consider thermal sputtering (which is anyway the dominant destruction mechanism for shock-heated plasma by high velocity ($V_{\rm s} \gtrsim 300$~km~s$^{-1}$) shocks.


\chapter{The H$_{\bf 2}$ molecule}
\label{chapter:H2Molecule}

\epigraph{Some scientists claim that hydrogen, because it is so plentiful, is the basic building block of the universe. I dispute that. I say that there is more stupidity than hydrogen, and that is the basic building block of the universe!}{Frank Zappa}

\index{H$_2$ molecule}


\begin{Abstract}

The hydrogen molecule, H$_2$, is the simplest and most abundant molecule in the Universe. H$_2$ has a key-role in many astrophysical processes, from galaxy formation in the primordial universe to star and planet formation. As an efficient coolant of the ISM, it is an actor of the formation of cold and dense molecular clouds, the fuel for star formation, and participates to the formation of more complex molecules.
In this introductory chapter, we review the basic properties of H$_2$, highlighting its formation mechanisms and excitation processes. We also present the tools for observing molecular gas in space, and detail how to derive key parameters of the gas physical conditions (temperature, gas masses, ortho-to-para ratio, etc.) from observations.

\end{Abstract}

\minitoc



\index{Molecular hydrogen (H$_2$)}

\section{Introduction: the role of H$_{\bf 2}$ in the Universe}
\index{Molecular hydrogen (H$_2$)!role in the Universe}

\PARstart{T}oday, molecules are known to be ubiquitous in the Universe, with more than 150 different molecules detected in the interstellar medium (henceforth ISM). H$_2$ is the simplest one, and the most abundant in the Universe.
From an astrophysical point of view,  H$_2$  is  an efficient cooling agent in a large variety of environments, which is perhaps the ``guiding thread'' of this PhD work.  
Therefore, H$_2$ occupies a central place in astrophysics, from large- to small-scale processes. 
It is impressive how such a simple molecule can touch most of the important questions in astronomy!
In the following list, I briefly summarize some of the ``hot topics'' where H$_2$ has a word to say. 

\begin{itemize}

\item
Stars are made from  the gravitational condensation  of molecular gas. Then H$_2$ can be viewed as  the fuel for star formation. Consequently, the star formation efficiency in ``standard'' galaxies correlates with the mass surface density of H$_2$ gas. This translates in a relationship between H$_2$ and IR luminosities, which has been inferred from \textit{Spitzer} data. Furthermore, these observations disclosed the existence of sources with enhanced H$_2$ emission with weak spectroscopic signatures (dust or ionized gas lines) of star formation (the H$_2$-luminous galaxies presented in chapter~\ref{chapter:H2_galaxies}). 
 In these galaxies, H$_2$ seems to be a tracer of mechanical energy dissipation associated with gas accretion, galaxy collisions, or feedback. This is the core topic of this PhD work.

\item 
H$_2$ was the first neutral molecule to be formed in the Universe, and it had influenced the collapse of the first cosmological objects, and thus had an important impact on the energetics of the formation of cosmological structures and galaxies \citep[see chapter~\ref{chapter:perspectives} and][for a short review]{Abel2000}.

\item 
Most of the baryons in the Universe are ``dark''. Indeed, the mass of observed baryonic matter (stars and gas) in galaxies makes up at most 8$\,$\% of the total mass expected from dynamics or weak lensing studies \citep{Fukugita2004}.
Although alternative scenarii are being proposed, H$_2$ is now considered as a potential significant contribution to the problem of missing baryons in galaxies \citep[e.g.][]{Pfenniger1994, Combes1997}. 
Some of the missing baryons may be in the form of H$_2$ lying beyond the optical galactic disk. This gas would have remained undetected in CO because it is metal poor, clumpy, or too cold to be seen in emission. 
First evidence for the presence of warm H$_2$ in the outer parts of galaxies comes from the
detection of H$_2$ S(0) and S(1) emission with $T\sim 80$~K in the outer disk of NGC\,891, as far out as 11~kpc \citep{Valentijn1999}.

\item
For a star to be made, the turbulent kinetic energy of the parent molecular cloud has to be dissipated so that the gravity takes over and further condense the gas. The ranges of gas temperatures ($\approx 10^{2-3}$~K) and velocities ($\approx 1-10$~km~s$^{-1}$) in the parent turbulent cloud make H$_2$ a major agent of the dissipation of this energy. 
Once stars have formed, they emit jets and winds that in turn inject some turbulent kinetic energy in   their nascent cloud. This stellar \textit{feedback} process remains largely unexplored.  Again, H$_2$, as a tracer of dissipative processes,  is certainly a key to understand what controls the star formation efficiency.
The importance of H$_2$ in star formation is confirmed by observations of H$_2$ in galactic star forming regions (e.g. Orion), in protostellar molecular outflows, or infrared luminous galaxies.

\item
H$_2$ is also a major ingredient of giant planet formation. Planets are thought 
to form in the circumstellar disks that surround proto-stars and pre-main-sequence  stars. Such disks, a 
natural outcome of the star formation process, are composed of 99$\,$\% of gas (mostly H$_2$) and 1$\,$\% of dust. 
H$_2$ is therefore the building block of giant planets. H$_2$ dominates the disk mass and dynamics while the dust is the main source of opacity and controls the thermodynamics of the disk. H$_2$ observations in protoplanetary disks may be used to constrain the scenario of planet formation, which is, up to now, not well determined.

\item
H$_2$ initiates a complex chemistry within the shielded interiors of dense clouds.
Some of these reactions take place at the surface of dust grains, and the formed molecules 
can become constituents of protoplanetary disks, where they contribute
to the formation of icy planetesimals such as the comets in our Solar System.
The richness of  molecular chemistry in dense gas is highlighted by observations 
of large organic molecules in proto-stellar cores heated by nascent stars, where
ices return to the gas, and in the circumstellar environment of evolved stars \citep[e.g.][]{Cernicharo2001}.
H$_2$, as tracer of the sites of formation of complex molecules, may be a key to elucidate the open question about the connection between molecular species present in circumstellar environments and those present in the ISM. 

\end{itemize}

All these astrophysical questions are closely related to the dynamics and energetics of dense gas, of which H$_2$ is the main consituent.
To understand why H$_2$ is a probe as such a variety of astrophysical processes, this chapter introduces the basis of the radiative and collisional properties of H$_2$.  We will rely on these properties in the next chapters. 
Part of my PhD work is dedicated to the modeling of H$_2$ formation and excitation in the extreme environments of H$_2$-luminous galaxies (chapters~\ref{chapter:H2_SQ} and \ref{chapter:perspectives}) by mean of shock and chemical network models \citep{Flower2003} described in chapter~\ref{chapter:shocks}.
While going through H$_2$ properties, I will give details about how some specific processes are treated in our shock model.
In particular, we discuss the ortho-to-para ratio of H$_2$ (sect.~\ref{subsec:H2-ortho-para-ratio}), and describe H$_2$ formation (sect.~\ref{sec:H2formationtheoryexp}) and excitation mechanisms (sect.~\ref{subsec:H2excitation-mechanisms}). This chapter ends with  a  discussion about how to observe molecular gas in space (sect.~\ref{sec:observingH2}). 

\section{Portrait of the  H$_{\bf 2}$ molecule}
\label{sec:H2-portrait}
\index{Molecular hydrogen (H$_2$)!portrait}

This section introduces some basic properties of the H$_2$ molecule that are of astrophysical interest.  We direct the reader to  \citet{Shull1982} for an historical review, and to the specialized books by \citet{Bransden1983} and \citet{Flower2007} for a fully developed theory of rotational and vibrational excitation of linear molecules. 
In this manuscript, we will mainly focus on the rovibrational transitions of H$_2$ in its electronic ground state,  denoted\footnote{The notation is $^{2S+1}\Lambda$, where $S=0$ or 1 is the total electronic spin and $\Lambda$ is the projection of electronic orbital angular momentum on the internuclear axis}  $X^1 \Sigma_g ^+$.

\subsection{H$_{\bf 2}$ rovibrational transitions}
\label{subsec:H2-transitions}
\index{Molecular hydrogen (H$_2$)!transitions}

H$_2$, the simplest molecule, consists of two hydrogen atoms linked by a covalent bond.
H$_2$ is an homonuclear molecule, i.e. the barycenter of its electrical charges concides with its center of mass. So H$_2$ has no permanent dipole moment and rovibrational transitions are forbidden electric quadrupole transitions. 
Molecular hydrogen thus exists in two different forms, depending on the value of its nuclear spin, $S$:
\begin{description}
\item[para H$_{\bf 2}$] the $S=0$ singlet state, where the two protons have opposite directions of spin (anti-parallel). This corresponds to \textit{even} values of $J$.
\item[ortho H$_{\bf 2}$] the $S=1$ triplet state, where the proton spins are parallel to each other. This corresponds to \textit{odd} values of $J$.
\end{description}
The  the parity (oddness or evenness) of the rotational quantum number $J$ is constrained by the fact that the nuclear wave function must be asymmetric under exchange of protons. 

\begin{figure}
   \centering
    \includegraphics[width=0.7\textwidth]{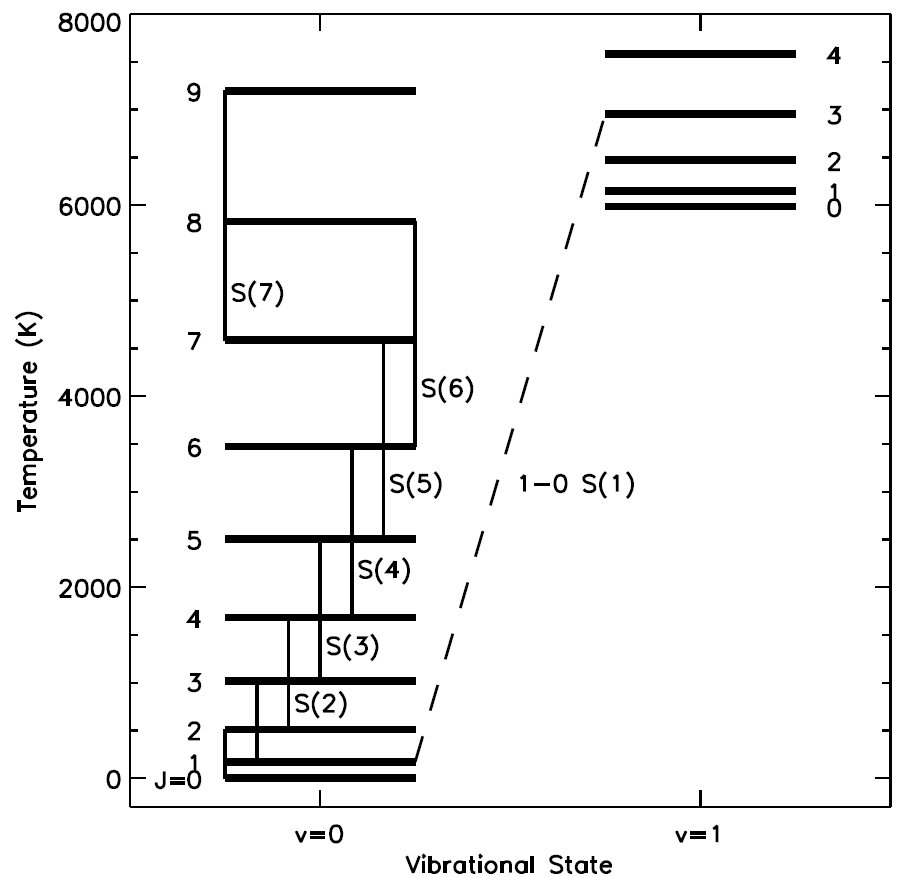}
      \caption[Rovibrational diagram ]{Rovibrational energy level diagram for molecular hydrogen with rotational $J$ states labeled. The S(0) to S(7) transitions available to \textit{Spitzer}  \textit{IRS} observations are indicated with solid lines. All ``S'' transitions are from the upper to lower J state with $\Delta J=2$. The $v=1$-0S(1) transition commonly
observed at 2.12$\,\mu$m is plotted as a das~hed line. Note that the S(7) transition traces a higher upper excitation state than the 1-0~S(0) and S(1) lines commonly observed in the near-infrared. Taken from \citet{Hewitt2009}.}
       \label{fig:H2_diagram_energies}
\end{figure}

Another consequence of the absence of permanent dipole moment is that rovibrational transitions within the $X^1 \Sigma_g ^+$ electronic ground state occur by electronic quadrupole radiation ($\Delta J = 0,\  \pm 2$). These quadrupole transitions do not allow to go from one H$_2$ form to another (from ortho to para). Because the electronic potential is anharmonic, there are no selection rules for vibrational quantum numbers, $v$. Within the harmonic approximation and if we neglect the centrifugal distorsion, the energy of the rovibrational levels are given by
\begin{equation}
E(v, J )  = B  J (J + 1) + \hbar \omega \left( v + \frac{1}{2} \right) \ ,
\end{equation}
where $B$ is the rotational constant of the molecule and $\omega$ is characteristic vibration pulsation, and $\hbar = h/2\pi$ the reduced Planck's constant.
$B$ is related to the moment of inertia of the molecule, $\mathcal{I}$, and $\omega$ can be expressed as a spring pulsation:
\begin{equation}
B = \frac{h}{8 \pi ^2 c \, \mathcal{I}} \quad \rm and \quad \omega = \sqrt{\frac{k}{\mu}} \ ,
\end{equation}
where $\mu$ is the reduced mass (here it is half the mass of an hydrogen atom, $\mu = 0.5 m_{\rm H}$) and $k$ is the spring force constant of the Hooke's law. Because H$_2$ is a light molecule, it has a small moment of inertia, and therefore a large rotational constant $B / k_{\rm B} = 85.25$~K. Its rotational levels are thus widely spaced. 
The diagram of Fig.~\ref{fig:H2_diagram_energies} shows the level energies of the $v=0$ and $v=1$ vibrational states of H$_2$, with the relevant lines for our study.
The $ J = 2$ state lies 510~K above the $ J = 0$ state, and $J = 3$ lies
845~K above $J = 1$. Consequently, the rotational excitation of H$_2$ becomes important only
for temperatures $T \gtrsim 100$~K. The $v = 1$ vibration level is $\approx 6000$~K above the
ground state, so that rovibrational excitation (such as that at the 2.2$\,\mu$m line) requires
kinetic temperatures $T > 1000$~K.

The statistical weight of a given level is
\begin{eqnarray}
g(J) g(I) &=& (2J + 1)(2I + 1) \\
     g_J        &=& 2J + 1 \quad  \quad  \quad \ \   \mbox{for the para state} \\
             &=& 3\times (2J + 1) \quad \mbox{for the ortho state} 
\end{eqnarray}
Under the assumption of Local Thermodynamic Equilibrium (LTE), the level populations are proportional to the statistical weight and the Boltzman factor, so that the rotational partition function is written
\begin{equation}
\label{eq:boltzman_rot_distrib}
Z_{\rm rot} = \sum _{J} g_J  \, e^{\dfrac{-E(v, J)}{k_{\rm B} T}}
\end{equation}
However, because of the low temperature and density conditions in the ISM, the LTE hypothesis is not valid, and the populations of the $(v,J)$ levels do not follow the Boltzman distribution. 

\begin{table}
\small
\begin{center}
\begin{minipage}{\textwidth}
 \renewcommand{\footnoterule}{}
\def\thefootnote{\alph{footnote}}
 \caption[Properties of important H$_2$ lines]{Properties of important near- and mid-infrared H$_2$ transitions \footnotemark[1]}
\centering
\begin{tabular}{c c c c c c c c c c}
\hline
\hline
line   &    Wavel.  &  Freq.  &  $g_J$ & $E_{\rm upper}$ &    $A$  &   \multicolumn{4}{c}{LTE $\quad \mathcal{I}_{\rm line} / \mathcal{I}_{\rm 1-0 S(1)}$} \\
\cline{7-10}
name   &    [$\mu$m]  &  [cm$^{-1}$] & 	&      [K]  & $10^{-7}$~s  &  1000 K & 2000 K & 3000 K & 4000 K\\
\hline
0-0 S(0)  &   28.221   &    354.35   &   5    &   510    & 0.0003  &  0.001 & &  & \\
0-0 S(1)  &   17.035  &     587.04  &   21   &   1015  &   0.0048  &  0.065  &  0.003  &  0.001  &  0.001\\
0-0 S(2)   &  12.279   &    814.43  &    9   &   1682 &    0.0276 &   0.11   &  0.008   & 0.003  &  0.002\\
0-0 S(3)  &    9.6649  &   1034.67  &   33   &   2504   &  0.0984  &  0.84  &   0.091  &  0.043  &  0.030\\
0-0 S(4)  &    8.0258  &   1245.98  &   13  &    3474   &  0.264  &   0.40  &   0.071 &   0.040 &   0.030\\
0-0 S(5)   &   6.9091  &   1447.36  &   45  &    4586   &  0.588 &    1.19  &   0.36  &   0.25  &   0.20\\
0-0 S(6)   &   6.1088   &  1636.97  &   17  &    5829   &  1.14   &   0.29  &   0.16  &   0.13 &    0.12\\
0-0 S(7)  &    5.5115  &   1814.40  &   57  &    7197  &   2.00   &   0.47  &   0.53  &   0.56 &    0.57\\
\hline
1-0 S(0)  &    2.2235  &   4497.41  &    5   &   6471  &   2.53   &   0.27  &   0.21   &  0.19  &   0.19\\
1-0 S(1)   &   2.1218  &   4712.91  &   21   &   6956   &  3.47   &   1.00   &  1.00  &   1.00   &  1.00\\
\hline
2-1 S(0)  &    2.3556  &   4245.15   &   5  &  12095   &   3.68  &    0.001  &  0.017  &  0.041  &  0.063\\
2-1 S(1)  &    2.2477  &   4448.95  &   21  &  12550    &  4.98   &   0.005  &  0.083  &  0.21  &   0.33\\
2-1 S(2)  &    2.1542   &  4642.04   &   9  &  13150   &   5.60   &   0.001 &   0.031 &   0.086  &  0.14\\
\hline
\end{tabular}
\footnotetext[1]{We list the wavelength (in vacuum) of the line, the wave number, the statistical weight of the upper level (degeneracy), the energy of the upper level $E_{\rm upper}$, the Einstein coefficient $A$. References: \citet{Turner1977, Dabrowski1984, Black1987, Wolniewicz1998}}
\label{table_H2lines}
\end{minipage}
\end{center}
\normalsize
\end{table}

In this manuscript, we will mostly focus on the first pure rotational lines of H$_2$ ($v=0$). In Table~\ref{table_H2lines} we list the properties of the main rovibrational lines we will consider. These transitions are within the near- and mid-infrared part of the spectrum. 
A transition is denoted by first writing the vibrational transition followed by the relevant branch and the lower rotational level. Thus the transition from $v=1$ to $v=0$, $J=3$ to $J=1$ is written 1-0S(1). 
More generally, the $S$-branch corresponds to $\Delta J = -2$, the $Q$-branch to $\Delta J = 0$, and the $O$-branch to $\Delta J = +2$.
Table~\ref{table_H2lines} also list the Einstein coefficients (see Eq.~\ref{eq:crit_density}) and line ratios assuming LTE (see Eq.~\ref{eq:boltzman_rot_distrib}) for 4 different gas temperatures. 

\subsection{H$_{\bf 2}$ ortho-to-para ratio}
\label{subsec:H2-ortho-para-ratio}
\index{Molecular hydrogen (H$_2$)!ortho-to-para ratio}
\index{Ortho-to-para ratio!definition}
\subsubsection{Definition and LTE values}

\begin{figure}
   \centering
   \includegraphics[width=0.7\textwidth]{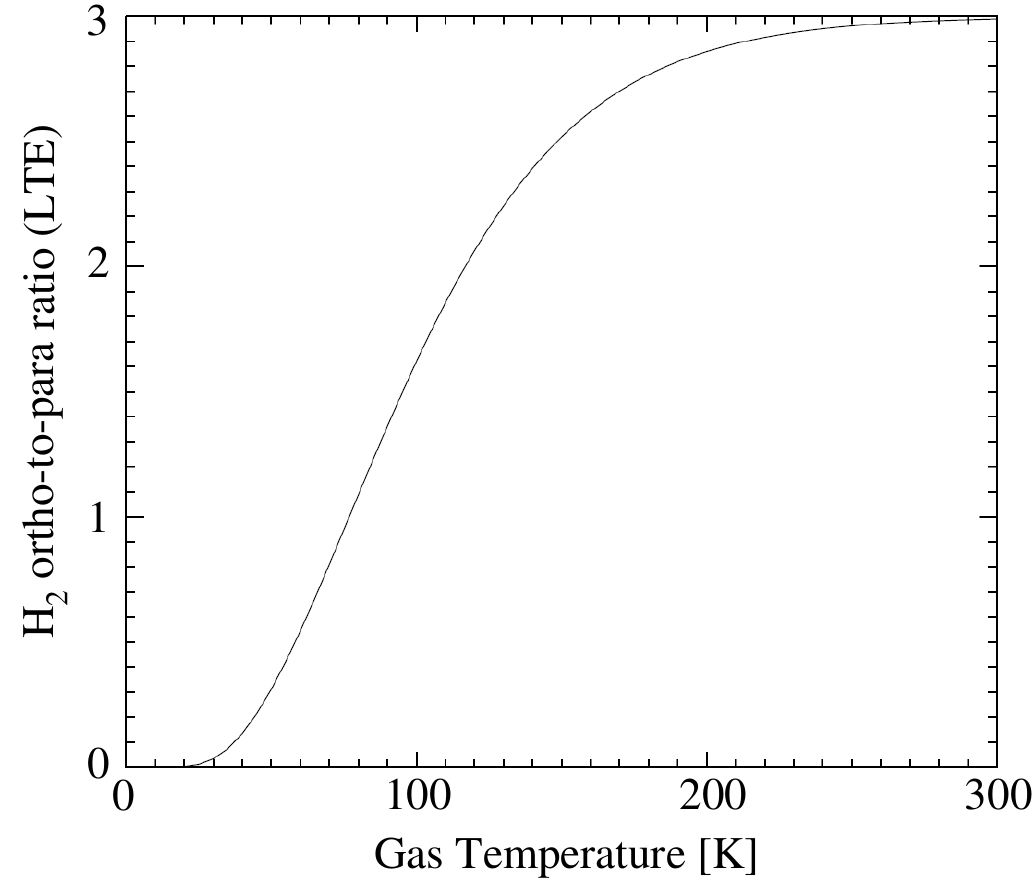}
      \caption[H$_2$ ortho-to-para ratio vs. temperature at LTE]{H$_2$ ortho-to-para ratio as a function of the gas temperature, assuming Local Thermodynamic Equilibrium (LTE) conditions. Below 50~K, the ortho-to-para ratio can be expressed as in Eq.~\ref{eq:H2_ortho-para_LTE_lowT}. }
       \label{fig_H2_ortho-para-ratio}
   \end{figure}


The \textit{local} H$_2$ ortho-to-para ratio (hereafter $\rm o/p$) is the relative abundance of the ground state ortho- and para-H$_2$.  In conditions of LTE, this ratio can be written as
\begin{equation}
\dfrac{\displaystyle Z_{\rm ortho}}{\displaystyle Z_{\rm para}}({\rm LTE}, T) = \dfrac{\displaystyle \sum_{J \ \rm odd} 3(2J+1)e^{-E_J / (k_{\rm B} T)}}{\displaystyle \sum_{J \ \rm even} \  (2J+1)e^{-E_J / (k_{\rm B} T)}}
\end{equation}
At ``high'' temperatures, $T \gtrsim  300$~K, the $\rm o/p$ ratio tends towards the value of 3, which is the ratio of the statistical weights of all the odd and even levels. This is illustrated in Fig.~\ref{fig_H2_ortho-para-ratio}, where we show the ortho-to-para ratio as a function of $T$.
At low temperatures, $T \lesssim  50$~K like in a quiet molecular cloud, only the levels ($v=0, J=0,1)$ are populated, so that, at low $T$ and in LTE conditions, the ortho-to-para ratio can be written
\begin{equation}
\label{eq:H2_ortho-para_LTE_lowT}
\frac{Z_{\rm ortho}}{ Z_{\rm para}}({\rm LTE}, T  \lesssim  50\,\rm K) = \frac{Z(J=1)}{Z(J=0)} = 9 \times e^{\dfrac{-170.5}{T}} \ ,
\end{equation}
where we get  $ \frac{2 \, B}{ k_{\rm B}} = \frac{\hbar ^2}{k_{\rm B} \mathcal{I}} = 170.5$~K.

\index{Ortho-to-para ratio!and equilibrium}
The $\rm o/p$ ratio is an important astrophysical parameter to characterize the physical state, and the history of the molecular gas. The $\rm o/p$ ratio of a given molecular cloud indeed depends on its  age and on the timescale of para-ortho conversion, $\tau _{\rm o/p}$. If the age of the molecular cloud is longer than the $\rm o/p$ conversion timescale, the $\rm o/p$ ratio equals its LTE value at the given temperature (Fig.~\ref{fig_H2_ortho-para-ratio}). Otherwise, if the cloud age is shorter than the $\rm o/p$ conversion timescale, the $\rm o/p$ ratio keeps its value during the process of H$_2$ formation. Our model assumes that H$_2$ molecules are formed with an $\rm o/p$ ratio of 3, although this value is poorly constrained.
The conversion mechanisms are briefly discussed in the following paragraph.

\subsubsection{Ortho to para conversion}
\index{Molecular hydrogen (H$_2$)!ortho-to-para conversion}
\index{Ortho-to-para ratio!conversion}

The conversion ortho-H$_2$ $\longleftrightarrow$ para-H$_2$ can occur through four main mechanisms:

\begin{enumerate}
\item 
the proton exchange with H$^+$, H$_3^+$, or other cations \citep{Dalgarno1973}:
\begin{equation}
\label{Eq:H2-Hplus-o-p-conversion}
\rm H_2 (J=1) + H^+ \rightleftarrows \rm H_2 (J=0) + H^+ (+ 170.5~\rm K)
\end{equation}
This is the dominant process at low ($T \lesssim  50$~K) temperatures \citep{Flower2006}. The reaction rates are of the order of $k_{\rm H^+, H_2} = k_{\rm H_3^+, H_2} \simeq 3 \times 10^{-10}$~cm$^{3}$~s$^{-1}$ \citep{Gerlich1990}. For instance, in a molecular cloud of hydrogen density $n_{\rm H} = 10^4$~cm$^{-3}$ and temperature $T=10$~K, H$_3^+$   being more abundant than H$^+$, H$_3^+ - \rm H_2$ reactions will set the $\rm o/p$ conversion timescale to be of the order of 
\begin{eqnarray}
\label{eq:timescale-ortho-to-para-conversion}
\tau _{\rm  o/p} &=& \frac{1}{n(\rm H_3^+) \, k_{\rm H_3^+, H_2}} \\
						  &=& 3 \times 10^6 \left( \frac{4 \times 10^{-5} \, \rm cm^{-3}}{n(\rm H_3^+)}  \right) \left( \frac{3 \times 10^{-10}\, \rm cm^{3}\, s^{-1}}{k_{\rm H_3^+, H_2}}  \right) \ [\rm yr]
\end{eqnarray}
\citet{Flower1984} carried on detailed calculations of the evolution of the $\rm o/p$ ratio as a function of the cosmic-ray ionization rate, which has an impact on the proton density, and therefore on the rate of Eq.~\ref{Eq:H2-Hplus-o-p-conversion}, and on the chemical history of the cloud. 

\item
the atom exchange between H$_2$ and H:
\begin{equation}
\label{Eq:H2-H-o-p-conversion}
\rm H_2 (\rm para) + H \rightleftarrows \rm H_2 (\rm ortho)  + H
\end{equation}
 \citet{Schofield1967} give an estimate of the collision rate of the order of 
\begin{equation}
 k_{\rm H, H_2} \simeq 8 \times 10^{-11} e ^{\dfrac{-3900}{T}} \ \ \rm [cm^{3}~s^{-1}] \ .
\end{equation}
Because of the high activation energy of $\approx 3900$~K, this process is negligible in cold molecular clouds.  In a cold dark cloud at $T = 10$~K, $n_{\rm H} = 10^4$~cm$^{-3}$, for a typical 
cosmic-ray ionization rate $\zeta = 5\times 10^{-17}$~s$^{-1}$~H$^{-1}$, and an initial degree of ionization of $\approx 10^{-8}$, it will take more than $10^7$ years to go from an $\rm o/p$ ratio
of 3 to the equilibrium value at 10 K of $\approx 3\times 10^{-7}$. The conversion timescale is only weakly dependent on density  \citep{Kristensen2007a}.

However, this mechanism can be important, and even dominant, when molecular gas is subject to heating and compression by a shock wave. Because the reaction~\ref{Eq:H2-H-o-p-conversion} is rapid, the $\rm o/p$ ratio in the hot postshock gas may differ significantly from the $\rm o/p$ ratio in the cold ambient medium. 
The evolution of the $\rm o/p$ ratio in shocks has been discussed by  \citet{Wilgenbus2000}. For application of these studies to H$_2$ observations in star forming regions, we direct the reader to the PhD thesis by \citet{Kristensen2007}.
The reactive collision rates used in the \citet{Flower2003} shock model are more complex, and are described in \citet{LeBourlot1999}. 

\item

the interaction of the H$_2$ nuclear spin with the inhomogeneous magnetic field at the surface of dust grains \citep{Tielens1987}. 
An upper limit of the ortho-to-para conversion rate by this process is given by the accretion rate of H$_2$ onto dust grains, and \citet{Timmermann1998} show that this mechanism is inefficient in low-velocity (a few 10~km~s$^{-1}$) molecular shocks.

\item

the ortho to para conversion may occur when an H$_2$ molecule is destroyed and then reformed onto the surface of a dust grain. However, given that the H$_2$ formation is slow, this process is negligible here.    

\end{enumerate}


\subsection{Critical densities: H$_2$ as a thermometer}
\label{subsec:H2-thermometer-critical-densities}
\index{Molecular hydrogen (H$_2$)!thermometer}
\index{Molecular hydrogen (H$_2$)!critical densities}

The LTE hypothesis is valid when the density of the medium reaches a threshold density, so-called \textit{critical density}, $n_{\rm crit}$, above which collisionnal deexcitations dominate over radiative deexcitations. For a $j\rightarrow i$ transition, this condition can be written
\begin{equation}
\label{eq:crit_density}
n q_{j \rightarrow i} \gg A_{j \rightarrow i} = n_{\rm crit} q_{j\rightarrow i} \ ,
\end{equation}
where $A_{j \rightarrow i}$~[s$^{-1}$] is the probability of radiative deexcitation (or so-called Einstein coefficient), and $q_{j \rightarrow i}$ the probability of collisional deexcitation [m$^{3}$~s$^{-1}$]. $A_{j \rightarrow i}$ is proportional to $(E_j - E_i)^5$, so for high values of $J$, we have $A_{j \rightarrow i} \propto J^5$, which shows that the lowest energy levels (with smallest $J$ values) are thermalized first.

\begin{figure}
   \centering
\includegraphics[angle=90, width=0.49\textwidth]{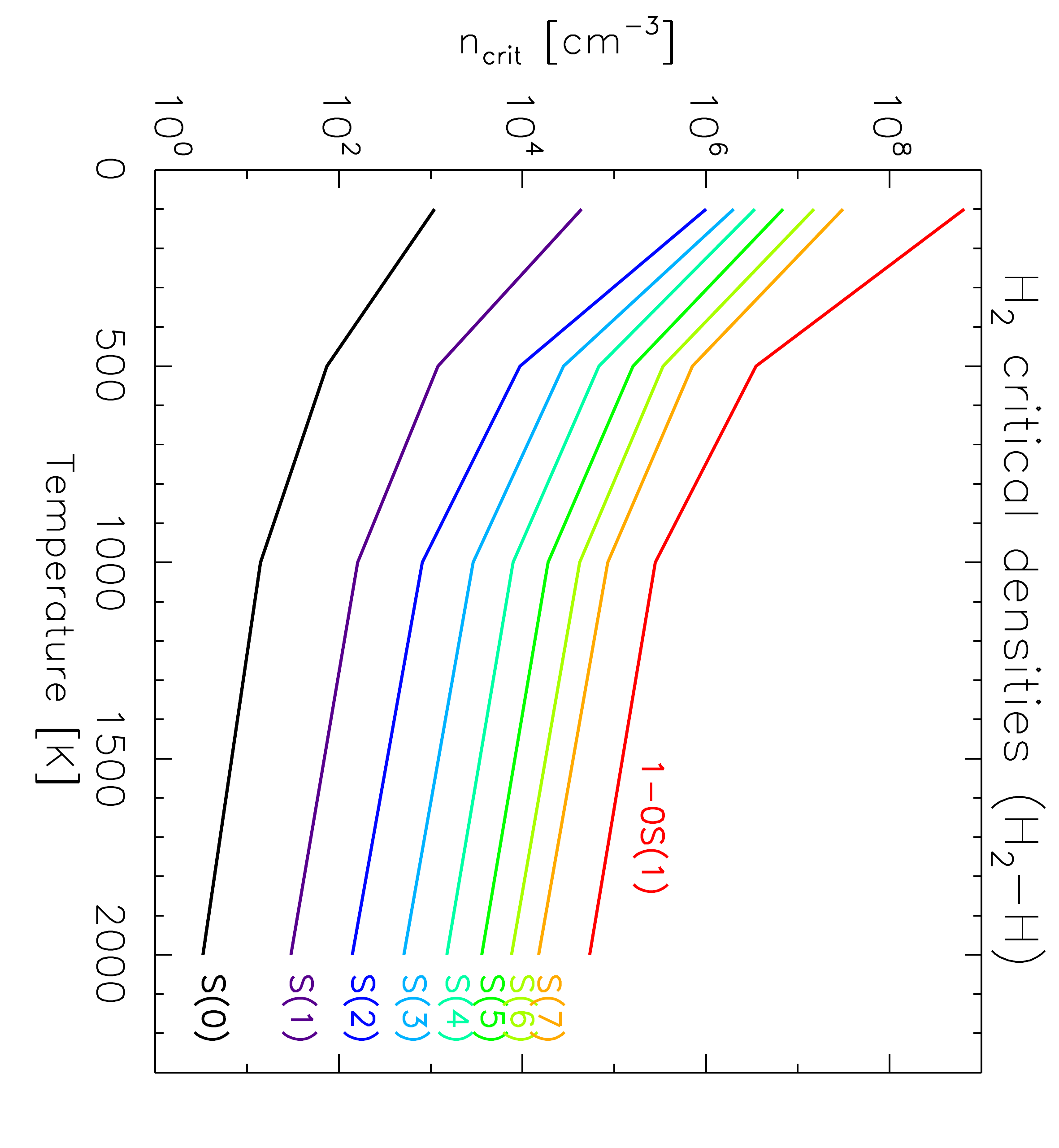}
\includegraphics[angle=90, width=0.49\textwidth]{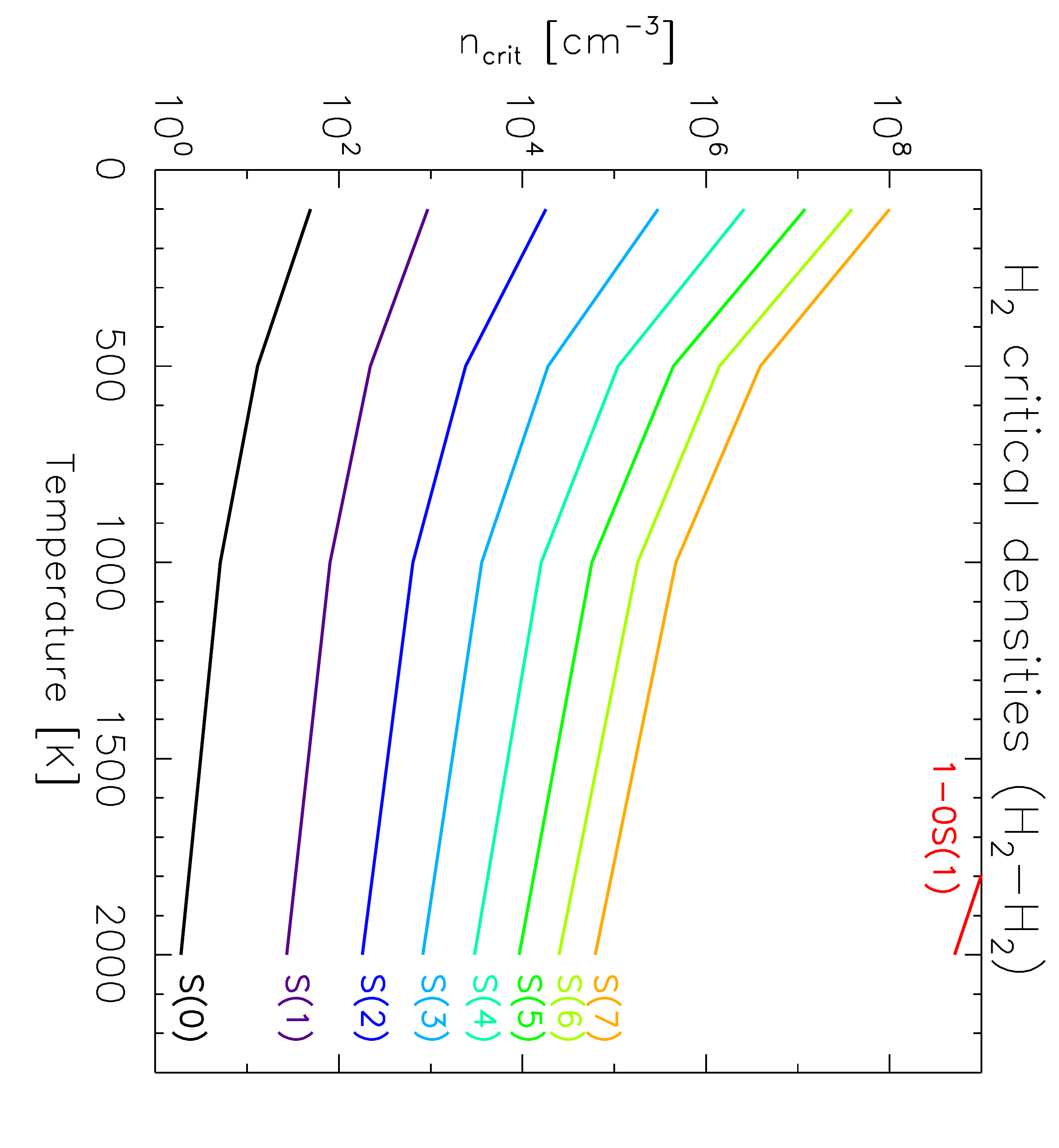}
      \caption[Critical densities of the H$_2$ rotational lines]{Critical densities, $n_{\rm crit}$,  for the S(0) to S(7) H$_2$ rotational lines, and the 1-0S(1) rovibrational line. The corresponding upper $J$ levels are thermalized if the gas density is higher than $n_{\rm crit}$. The collisional partner is H on the \textit{left} plot and H$_2$ on the \textit{right}. The critical densities are calculated at 100, 500, 1000 and 2000~K from \citet{Wrathmall2007}.}
       \label{fig:H2_crit_densities}
\end{figure}

This is illustrated in Fig.~\ref{fig:H2_crit_densities}, which shows the critical densities, $n_{\rm crit}$, at which the probabilities of collisional and radiative deexcitation are equal, for the first 7 rotational lines of H$_2$ and the 1-0S(1) rovibrational line. The collisional partner is H on the left panel and H$_2$ on the right\footnote{Note that a similar figure can be found in \citet{Roussel2007}, but the collision rates used in that paper are outdated, which lead to significant differences with the critical densities indicated here.}. At a given temperature, a given level is thermalized if the gas density is higher than $n_{\rm crit}$. On the other hand, if the gas density is much less than the critical density, collisional excitation is followed by radiative decay, and so the collisional excitation rates determine the emission line intensities. The critical densities decrease with temperature because the collision rates also decrease with $T$. Note that higher densities appear necessary to thermalize rovibrational (not shown here)  as compared with pure rotational transitions.

The excitation and deexcitation collisional rates of H$_2$ with H we use in our shock model are discussed extensively in \citet{LeBourlot1999, LeBourlot2002, Wrathmall2007}, and we also refer the reader to the PhD thesis by \citet{Gusdorf2008a} for more details and calculations of $n_{\rm crit}$ with other collisional partners (H, H$_2$, He). 

\section{H$_{\bf 2}$ formation and chemistry}
\label{sec:H2formationtheoryexp}
\index{Molecular hydrogen (H$_2$)!formation}

In spite of the simplicity of the H$_2$ molecule, its formation process in space has been a source of puzzles since more than 40 years, and H$_2$ formation is still an active research topic. I summarize here some of the basic ideas about H$_2$ formation that are crucial to have in mind for the next chapters, when we will deal with H$_2$ formation issues in the extreme environments of H$_2$-luminous galaxies.

\subsection{H$_{\bf 2}$ formation in the gas phase?}
\index{Molecular hydrogen (H$_2$)!formation in the gas phase}
\begin{figure}
   \centering
    \includegraphics[width=0.6\textwidth]{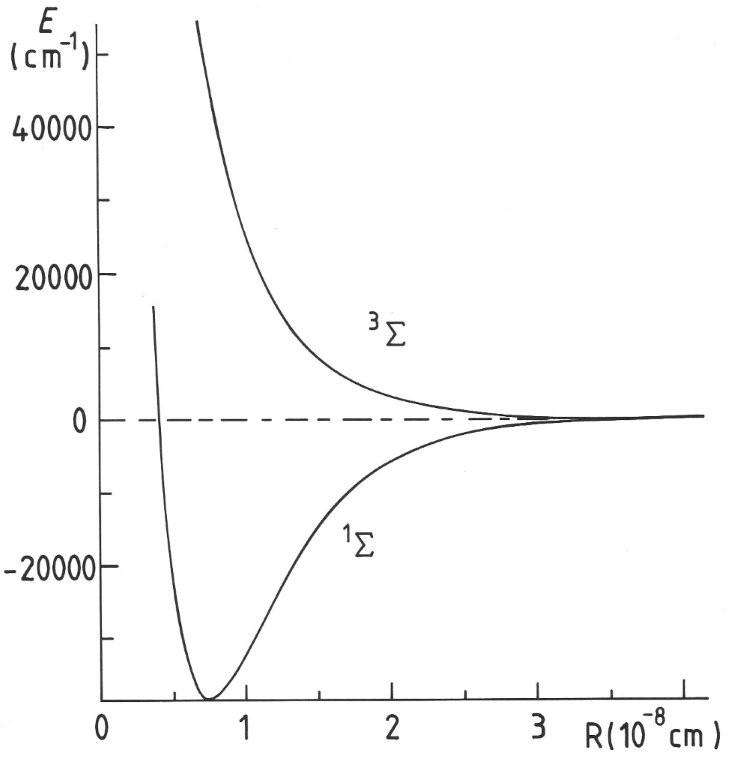}
      \caption[The lowest electronic potential energy curves of H$_2$]{The lowest electronic potential energy curves of H$_2$. The curves shows the energy of the $\rm H(1s) + H(1s)$ system at large internuclear separation $R$. The singlet and triplet molecular states are denoted $^{1}\Sigma$ and $^{3}\Sigma$ because the projection of the electronic orbital angular momentum $\Lambda = 0$. The $^{1}\Sigma$ state is attractive whilst the $^{3}\Sigma$ is repulsive. Figure taken from the book by \citet{Flower2007}, with kind permission. }
       \label{fig:H2_elec_pot}
\end{figure}

We will briefly demonstrate that the direct formation H$_2$ with gaseous elements only cannot explain the observed presence of molecular hydrogen in the ISM. 
The first process would consist in a \textit{radiative association of two hydrogen atoms}:
\begin{equation}
{\rm H + H \longrightarrow  H_2} + h \nu \ (4.478 \ \rm eV)
\end{equation}
The Fig.~\ref{fig:H2_elec_pot} shows the electronic potential energy curves of an association of two hydrogen atoms as a function of their distance. Initially, H atoms are unbound and their total energy $E$ is positive. To stabilize, the system must lose energy so that $E \leqslant 0$.
This formation process  would be possible if the molecule forms in a state that can deexcite to the ground state in order to lose its excess of energy by the emission of a photon $h \nu$. However, Fig.~\ref{fig:H2_elec_pot} shows that a molecule formed by an encounted or two H atoms may be formed in a repulsive state. Transitions between the $^{3}\Sigma$ and $^{1}\Sigma$ levels are forbidden to electronic dipole radiation because they would involve a change in the total spin quantum number. Rotational and vibrational transitions are also forbidden because H$_2$ has no permanent dipole moment. Therefore, the molecule formed by radiative association can only dissociate. H$_2$ cannot be formed by radiative association of two hydrogen atoms.

The formation H$_2$ in the gas phase (meaning only with gaseous collision partners) cannot occur through \textit{three-body collisions}, because the probability of a simultaneous encounter between 3 Hydrogen atoms is extremely small. 
When a sufficiant abundance of electrons and ions exists, ion-atom reactions can be effective. Two routes for H$_2$ formation are then possible. The ``H$_2^+$ route'', that is a radiative association followed by a charge exchange reaction:
\begin{eqnarray}
 {\rm H + H^+} & \longrightarrow & {\rm H_2^+} + h \nu \\
 {\rm H_2^+ + H} & \longrightarrow & {\rm H_2 + H^+}
\end{eqnarray}
The alternative route is through $H^-$, with a radiative attachment followed by a associative detachment:
\begin{eqnarray}
{\rm H} + e^- & \longrightarrow & {\rm H^-} + h \nu  \\
{\rm H^-} + {\rm H} & \longrightarrow & {\rm H_2} + e^- 
\end{eqnarray}
These reactions require very high densities (comparable to the density of a star) to be probable and contribute significantly to H$_2$ formation. Such processes are likely to be important for the chemistry of the primordial Universe, and we direct the reader to \citet{Stancil1998, Galli1998} for more details, in particular about the reaction rates.
However, densities in the ISM, even in the denser molecular clouds, are too small for this process to be probable. 

\subsection{The H$_{\bf 2}$ formation on interstellar dust grains}
\label{subsec:H2-formation-on-grains}
\index{Molecular hydrogen (H$_2$)!formation on dust grains}

We have shown that the H$_2$ formation in the ISM cannot occur only with gaseous contituants. To overcome this difficulty, it has been proposed by \citet{Gould1963, Hollenbach1971, Jura1975} that interstellar dust grains play the role of catalysts for H$_2$ formation:
\begin{equation}
{\rm grain + H + H \longrightarrow grain + H_2} + E_{\rm f} \ (4.478 \ {\rm eV})
\end{equation}
The grain plays the role of the third-body in a three-body reaction. The mechanism for H$_2$ through grain surface reactions is poorly known, as well as the distribution of the energy released by the formation, $E_{\rm f} \approx 4.5$~eV, among the internal, kinetic and grain energy components. The part that goes into H$_2$ internal  energy contribute to H$_2$ excitation, and we will come back to this point in sect.~\ref{subsec:H2-excitation-during-formation}.

\index{Molecular hydrogen (H$_2$)!mechanisms of formation on grains}
Two main mechanisms exist for the heterogeneous catalysis on a solid:
\begin{description}
\item[The Langmuir-Hinshelwood reaction:] H$_2$ is formed by physisorption, i.e.  through a reaction between two atoms (or radicals) that are already adsorbed on the grain surface.
\item[The Eley-Rideal reaction:] H$_2$ is formed by chemisorption, i.e. by the direct interaction of an adsorbed radical with an atom hitting it upon arrival from the gas phase.
\end{description}
\index{Molecular hydrogen (H$_2$)!formation rate}
In both cases, the H$_2$ formation rate per unit volume can be written as
\begin{equation}
R({\rm H_2}) = R_f ({\rm H_2}) \times n_{\rm H} \times n_{\rm gr} \quad \rm [cm^{-3}~s^{-1}]\ \ ,
\end{equation} 
where $R_f ({\rm H_2})$~[cm$^{3}$~s$^{-1}$] is the H$_2$ formation rate, $n_{\rm H}$ is the gas hydrogen density, and $n_{\rm gr}$ is the number density of grains. Empirically the H$_2$ formation rate  [cm$^{3}$~s$^{-1}$] is often expressed as:
\begin{eqnarray}
R_f ({\rm H_2})    &=& \frac{1}{2} \sigma v_{\rm H} \times S(T) \times f \times \eta \\
							  &=& \pi a_{\rm gr}^{2} \left(\frac{8 k_{\rm B} T}{\pi m_{\rm H}} \right)^{1/2}  \times S(T) \times f \times \eta \quad ,
\end{eqnarray}
where $\sigma$ is the average cross-section [cm$^{2}$] of the interaction between H atoms and grains, $ v_{\rm H}$ is the velocity of H, $S(T)$ is the temperature-dependent probability that an H atom sticks onto the grain surface, $f$ is the probability that the incident H atom stays on the grain and migrates towards the other adsorbed H atom, and $\eta$ is the probability of recombination to form H$_2$. $\sigma$ is calculated by averaging the quantity $\displaystyle \left\langle  \frac{n_{\rm gr} (a_{\rm gr})}{n_{\rm H}} \pi a_{\rm gr}^{2} \right\rangle $ over the grain size distribution ($a_{\rm gr}$ is the grain radius).
It is generally assumed that the newly-formed H$_2$ molecule has a probability of 1 to be ejected from the grain because the potential wells of the physisorption ($\approx 50$~meV) and chemisorption ($\approx 1$~eV) are much smaller than the energy released by the formation (4.5~eV).

\index{Molecular hydrogen (H$_2$)!sticking coefficient}
Several expressions have been proposed for the dependence of the sticking coefficient on the temperature. In our model we use the following expression, given in  \citep{Hollenbach1979}:
\begin{equation}
S(T) = \frac{1}{1 + 0.04 \sqrt{T + T_{\rm gr}} + 2\times 10^{-3} \, T + 8 \times 10^{-6} \, T^{2}}  \ ,
\end{equation}
where $T$ is the kinetic gas temperature and $T_{\rm gr}$ is the grain temperature. In addition, we assume $f=\eta=1$.

From UV observations of diffuse interstellar clouds with the Copernicus satellite\footnote{\url{http://heasarc.gsfc.nasa.gov/docs/copernicus/copernicus_about.html}},  \citet{Jura1975} estimates an H$_2$ formation rate of $R_f ({\rm H_2})   \approx 3 \times 10^{-17}$~cm$^{3}$~s$^{-1}$ at $T = 70$~K.
For more details about H$_2$ formation and recent estimates of $R_f ({\rm H_2})$ from photo-dissociation regions (PDRs), we direct the reader to the the PhD thesis by  \citet{Habart2001a} and to \citet{Habart2004}.

Note that, in cold clouds, the dust grains may be covered by ices, primarily composed of H$_2$O and CO, whereas in hotter regions, such as circumstellar environments or shock-heated gas, the  icy mantles evaporate. The nature of the grain surface affects the H$_2$ formation rate, and an active experimental work is being done on this topic \citep[e.g.][]{Pirronello1997, Manico2001, Hornekaer2003, Amiaud2007}.
%
%

\subsection{H$_{\bf 2}$, initiator of an impressive molecular complexity}
\label{subsec:H2-and-complex-chemistry}
\index{Molecular hydrogen (H$_2$)!and molecular complexity}

We have stated in the introduction that H$_2$ initiates the formation of more complex molecules in cold clouds. Obviously, I will not review here the full theory of the formation of heavy molecules in the ISM \citep[see the book by][for details]{Duley1984}. I will only briefly give the tracks  towards this molecular complexity that we observe in the ISM.

\begin{figure}
   \centering
    \includegraphics[width=0.6\textwidth]{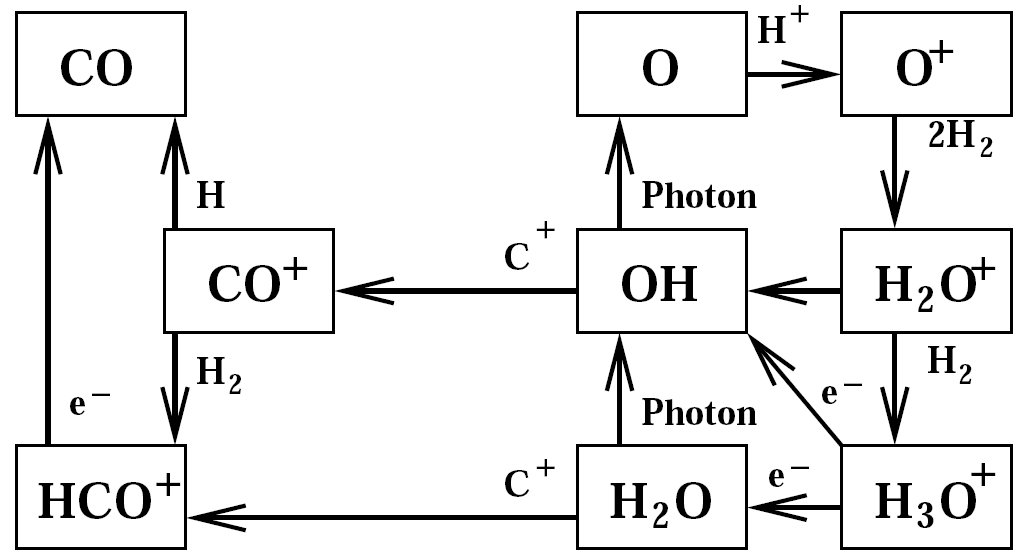}
      \caption[Sketch of the CO formation routes]{H$_2$ as an initiator of the formation of CO. This figure sketches the CO formation routes in a low UV radiation field. Destruction is still dominated by photo-dissociation. From \citet{Nehm'e2008a}.}
       \label{fig:CO_formation_routes}
\end{figure}

In fact, the formation of molecules more complex than H$_2$ requires the H$_3^+$ ion, which is primarily formed by cosmic-ray (CR) ionization of H$_2$ \citep{Herbst1973}:
\begin{equation}
\rm H_2 + CR \longrightarrow H_2^+ + e^- + CR
\end{equation}
This reaction is rapidly followed by 
\begin{equation}
\rm H_2^+ + H_2 \longrightarrow H_3^+ + H
\end{equation}
Then, the way for complex chemistry is open. Heavier molecules can form via proton transfer reactions. Let us give one example of a chain of reactions  which leads to the formation of CO. The CO formation routes are schematically illustrated in Fig.~\ref{fig:CO_formation_routes}. One of the possibilities, the ``H$_2$O route'', can be summarized as:
\begin{eqnarray}
\rm 2 O^+ + 3 H_2 & \longrightarrow & \rm  2 H_3O^+  \nonumber \\ 
\rm H_3O^+ + e^-  & \longrightarrow &\rm   H_2O + H  \nonumber \\ 
\rm C^+ + H_2O   & \longrightarrow &\rm HCO^+ + H   \nonumber \\ 
\rm HCO^+ + e^-  & \longrightarrow &\rm  CO + H
\end{eqnarray}
CO is then formed! The importance of the CO molecule is stressed in sect.~\ref{sec:observingH2}, and we will come back to CO emission in Stephan's Quintet in chapter~\ref{chapter:SQ_CO}.

\section{ H$_{\bf 2}$ excitation mechanisms and diagnostics}
\label{subsec:H2excitation-mechanisms}
\index{Molecular hydrogen (H$_2$)!excitation mechanisms}

Excitation of rotation-vibration levels of H$_2$ can occur through different mecanisms, which can be separated in three categories: collisional excitation, radiative (UV pumping) excitation, or excitation directly in the process of forming the molecule. 
Obviously, it would be impossible to give an exhaustive review about each processes here. We summarize  the main ideas and classify these processes by order of relevance for our astrophysical context. I also give a few details about the chemical reaction rates incorporated in the shock model  we use in the following chapters (in particular chap.~\ref{chapter:shocks} and \ref{chapter:H2_SQ}).

\subsection{Collisional excitation and dissociation}
\index{Molecular hydrogen (H$_2$)!collisional excitation}
\label{subsec:H2-collisional-excitation}

H$_2$ levels may be excited by collisions in warm gas, and the resulting spectrum is often invoked as an indicator that shocks are occurring. The calculation of the collisional rates is complex because it involves detailed quantum mecanics calculations of the interaction potential between the collision partner and the target. For a comprehensice description of the collisional processes in the ISM, we direct the reader to the book by \citet{Flower2007}.

\subsubsection{ H$_{\bf 2}$ collisions with atoms or molecules}

The most important collisional partners for H$_2$ excitation with atoms or molecules are the light and most abundant partners H, He, ortho- and para-H$_2$, that lead to rotational and vibrational excitation of H$_2$. 
The cross sections for rovibrational transitions of H$_2$, induced by collisions with H and He atoms have been discussed in \citet{LeBourlot1999, LeBourlot2002}. An update of H$_2$-H collision rates is given in \citet{Wrathmall2007a}. H$_2$-H$_2$ collisions are treated in \citep{Flower1998}. 
For the particular interest of this PhD work, the reader will find the cross-sections and rate coefficients that are used in our models for collisional excitation of H$_2$ rotational transitions in the appendices of \citet{Flower2007}.

Collision with H can also lead to the dissociation of H$_2$. The rate coefficient calculated by \citet{Dove1986} can be fitted as a function of the temperature of the neutral fluid, by the following form \citep{Flower1996}:
\begin{equation}
k_{\rm diss.}^{\mbox{{\tiny H-H}}_2} (T_n) = 10^{-10} \times e^{\dfrac{-5.2\times 10^4}{ T_n}} \quad \rm [cm^3~s^{-1}] \ ,
\end{equation}
where the temperature of $5.2\times 10^4$~K in the exponential corresponds to the dissociation energy of H$_2$ (4.48~eV). In our model, the above expression has been adapted to allow for the excitation energy $E(v, J)$ of the initial rovibrational level $(v, J)$. The following rate is adopted in the code \citep{LeBourlot2002}:
\begin{equation}
k_{\rm diss.}^{\mbox{{\tiny H-H}}_2} (T_n) = \frac{10^{-10}}{n(\rm H_2)} \sum_{v, J} n(v, J) e^{\dfrac{-(56\,644 - E(v, J))}{T_n}} \quad \rm [cm^3~s^{-1}] \ ,
\end{equation}
where $E(v=4, J=29) = 56\,644$~K is the energy of the highest bound level of H$_2$, relative to the $v=0$, $J=0$ ground state \citep{Dabrowski1984a}. $n(v, J)$ is the level population density. By definition, 
\begin{equation}
n({\rm H_2}) = \sum_{v, J} n(v, J) \ \ .
\end{equation}
The rate coefficient for the collisional dissociation of H$_2$ by H$_2$ is taken to be \citep{Jacobs1967}:
\begin{equation}
 k_{\rm diss.}^{\mbox{{\tiny H}}_2\mbox{{\tiny -H}}_2} = \dfrac{k_{\rm diss.}^{\mbox{{\tiny H-H}}_2}}{8} 
\end{equation}
The rate coefficient for the collisional dissociation of H$_2$ by He is taken to be \citep{LeBourlot2002}:
\begin{equation}
 k_{\rm diss.}^{\mbox{{\tiny H}}_2\mbox{{\tiny -He}}} \approx \dfrac{k_{\rm diss.}^{\mbox{{\tiny H-H}}_2}}{10} 
\end{equation}

\subsubsection{Cosmic-ray ionization of H$_{\bf 2}$}

\begin{figure}
   \centering
    \includegraphics[width=0.7\textwidth]{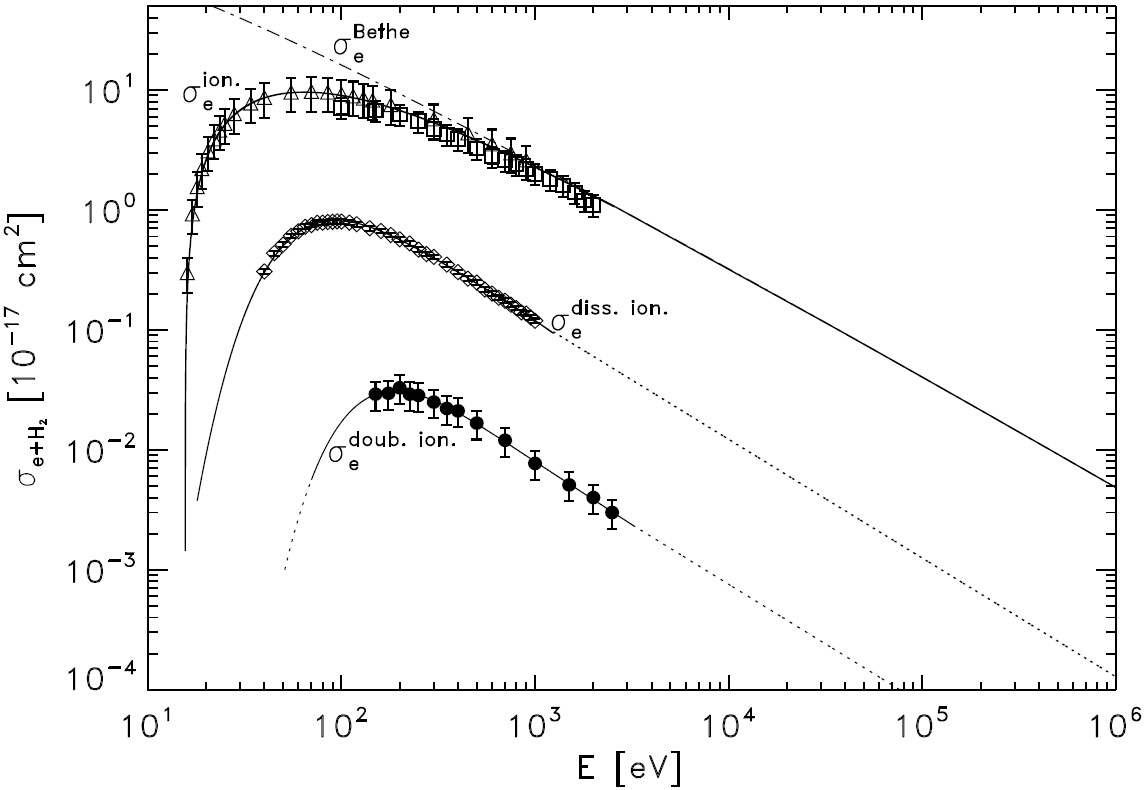}
      \caption[Critical densities of the H$_2$ rotational lines]{Cross sections for electron impact on H$_2$ from \citet{Padovani2009}: ionization cross section $\sigma _{P}^{\rm ion.}$ (Rudd 1991), dissociative ionization $\sigma _{P}^{\rm diss.ion.}$, and double ionization cross section $\sigma _{P}^{\rm doub.ion}$. The solid curves are polynomial fits. For comparison, the dot-dashed line shows the Bethe ionization cross section multiplied by a factor of 2. The points are experimental data (see \citet{Padovani2009} for references). }
       \label{fig:H2_e_cross_section}
\end{figure}

H$_2$ excitation by electrons is a process that occurs in the astrophysical context of the interaction of  cosmic rays or X-ray photons or cosmic rays with molecular gas. 
When X-ray photons or cosmic ray particles (electrons, protons, and heavy nuclei)  penetrate into a molecular cloud, photoelectrons or secondary electrons are produced \citep[see][for a review]{Padovani2009}. 
\index{Cosmic rays!H$_2$ ionization}
Let us consider the direct ionization of H$_2$ by a cosmic-ray particle (electrons, protons, and heavy nuclei) impact:
\begin{equation}
{\rm H}_2 + {\rm CR} \longrightarrow {\rm H}_2 ^+  + {\rm e^-} + {\rm CR}
\end{equation}
The Fig.~\ref{fig:H2_e_cross_section} shows the theoretical and experimental cross-sections for the impact of electrons onto H$_2$ from \citet{Padovani2009}. The experimental data for electron-impact ionization of H$_2$ have been reviewed by \citet{Liu2004}. The curves show a semi-empirical model developped by Rudd (1991). Note that the  Bethe (1933) cross section for primary ionization of atomic hydrogen, multiplied by a factor of 2, reproduces very well the behavior of the ionization cross section at energies above a few tens of keV.

\subsubsection{Collisional excitation of H$_{\bf 2}$ by electrons}

The scattering of these photoelectrons or secondary electrons on H$_2$ molecules can lead to thermal \citep{Lepp1983} or non-thermal excitation to singlet and triplet electronic states of H$_2$. The triplet states lead to dissociation but in some fraction of the excitations the singlet states cascade down by allowed radiation transitions
into the array of rotation-vibration levels of the ground electronic state.

We will show below that the H$_2$ excitation by thermal electrons can be very important in shocked molecular gas because of the shock-induced rise of the electron temperature.
We give a brief review of the theoretical studies about H$_2$ excitation by collision with thermal and non-thermal electrons, and we give some key-examples of excitation rates incorporated in the shock code we use later. The global properties of the efficiencies of H$_2$ excitation by cosmic or X-rays will be addressed in section~\ref{subsec:H2-cosmic-X-ray-heating}.

\index{Cosmic rays!secondary electrons}
The rate of production of secondary electrons by cosmic ray ionization of hydrogen is approximately $\zeta n(\rm H_2)$, where $\zeta$ is the cosmic ray ionization rate (a few $10^{-17}$~s$^{-1}$ is a classical value in molecular clouds) and $n(\rm H_2)$ the  number density of H$_2$ molecules. Typically, $\sim 15\,$\% of the secondary electrons subsequently excite H$_2$ molecules by collision \citep{Cravens1975, Sternberg1987}. Therefore the rate of H$_2$ excitation by secondary electrons is $0.15 \zeta n(\rm H_2)$.

We consider now the H$_2$ excitation by thermal electrons. 
If we denote $k_1$ the rate coefficient for the excitation of the singlet states, then the rate of excitation can be written as $n_e n({\rm H}_2) k_1 (T_e)$.
The cross-sections for the excitation of singlet states of H$_2$ by thermal electrons are linear as a function of energy for the dominant B and C states \citep{Ajello1984}. \citet{Flower1996} provide the following form for the rate coefficient for electron excitation of the B and C states:
\begin{equation}
k_1(T_e) = 10^{-16} \left( \frac{8 k_{\rm B} T_e}{\pi m_e} \right) ^{1/2} \left( 0.429 + 6.14 \times 10^{-6} \, T_e \right) \times e^{\dfrac{-1.4 \times 10^5}{T_e}} \ \rm cm^3 \ s^{-1}
\end{equation}
The factor $ \left( \frac{8 k_{\rm B} T_e}{\pi m_e} \right) ^{1/2}$ represents the mean velocity of electrons  at temperature $T_e$. The $1.4 \times 10^5$~K temperature in the exponential factor is the mean excitation temperature of the B and C states from the $X^1 \Sigma _{g} ^{+}$ ground state.

Therefore, the total rate of H$_2$ excitation by (thermal + non-thermal) electrons, $R_1 (T_e)$~[cm$^{-3}$~s$^{-1}$],  is
\begin{eqnarray}
R_1 (T_e)  \!  &\! =\! & \!   0.15 \, \zeta \, n({\rm H}_2) +  n_e\,   n({\rm H}_2)\,   k_1 (T_e) \\
                \!   & \!   \approx \! & 7.5 \, 10^{-15}  \frac{ n_{\rm H}}{10^4}\! \left[\! \frac{\zeta}{10^{-17} } + 2 \,  10^6 \frac{n_e\! /\! n_{\rm H}}{10^{-7}} \frac{ n_{\rm H}}{10^4} \frac{T_e}{10^5} \left( 0.43 \!  + \!  0.61 \frac{T_e}{10^5} \right) e^{\dfrac{-1.4}{T_e / 10^5}} \!\right] 
\label{eq:rate_H2_electron_1}
\end{eqnarray}
This is the rate we use in our model for the H$_2$ excitation by electrons. In eq.~\ref{eq:rate_H2_electron_1} we have assumed that the gas is molecular, so $n({\rm H}_2) = 0.5 n_{\rm H}$. For a gas of density $n_{\rm H}=10^4$~cm$^{-3}$ and fractional ionization $n_e / n_{\rm H} = 10^{-7}$, the  H$_2$ excitation rate by thermal electrons exceeds by a factor $10^6$ the 
cosmic ray electrons rate if we assume $T_e = 10^5$~K. If $T_e = 10^4$~K, the two contributions are comparable. More generally, because magneto-hydrodynamical shocks can easily increase the electron temperature above $10^4$~K,  the photodissociation and photoionization of molecular species induced by a shock wave can greatly exceed the rate of cosmic ray-induced processes \citep{Flower1996}.

\subsection{X-ray and cosmic-ray heating efficiencies}
\label{subsec:H2-cosmic-X-ray-heating}
\index{X-ray!heating of H$_2$}
\index{Molecular hydrogen (H$_2$)!X- and cosmic-ray heating}

\subsubsection{X-rays}

\begin{figure}
   \centering
    \includegraphics[width=0.7\textwidth]{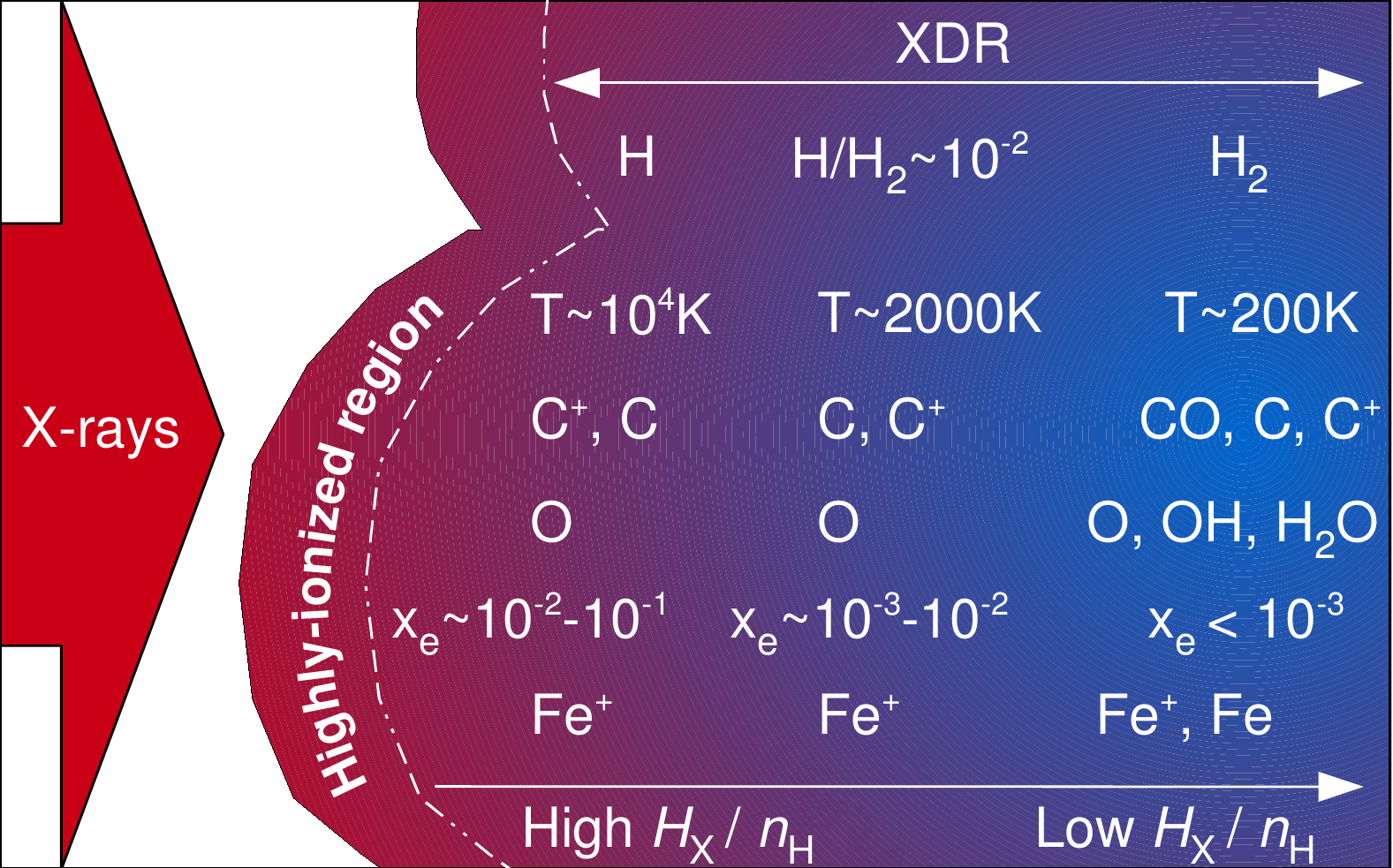}
      \caption[Structure of an XDR]{Schematic structure of an X-ray dissociation region (XDR), where an X-ray flux \textit{(arrow on the left)} illuminates a molecular cloud \textit{(on the right)}. As we penetrate into the XDR, we indicate the approximate temperature, chemical composition and ionization fraction $x_e$ as a function of the ratio of the local X-ray energy deposition rate per particule, $H_{\rm X}$, to the total hydrogen density $n_{\rm H}$.}
       \label{fig:XDR_sketch}
\end{figure}

The excitation by X-ray photons was quantified by a number of authors  \citep[e.g.][]{Voit1991, Tine1997, Dalgarno1999}. 
Let us estimate the maximum X-ray to H$_2$ luminosity conversion factor. We consider that an X-ray source heats the H$_2$ gas (see Fig.~\ref{fig:XDR_sketch} for a sketch of an XDR), by assuming that 100$\,$\% of the X-ray flux penetrates into the molecular gas. 
 Models of X-ray dissociation regions (XDRs)  show that $30 - 40\,$\% of the absorbed X-ray flux goes into photoelectron heating \citep{Maloney1996}. The atomic photoelectric cross section is a steep decreasing function of the energy ($E^{-8/3}$). Photoionization in the energy range $1-30$~keV are the most important. At a temperature of 200~K for the H$_2$ gas, the cooling by H$_2$ rotational lines in XDRs models is $\sim 2\,$\% of the total cooling \citep[see Fig~3a and 5 of][]{Maloney1996}.
The maximum X-ray to H$_2$ luminosity conversion factor is therefore
\begin{equation}
\frac{\mathcal{L}(\rm H_2 \ \mbox{0-0S(0)-S(3)})}{\mathcal{L}_{\rm X}(2-10 \rm \ \mbox{keV})} < 8 \times 10^{-3}
\end{equation}
This estimate will be useful in the next chapter to see whether X-rays can power the observed H$_2$ emission in the extreme environments of H$_2$-luminous galaxies. 

\subsubsection{H$_{\bf 2}$ heating by cosmic-ray ionization}
\index{Cosmic rays!H$_2$ heating}

We estimate the molecular heating associated with cosmic-ray ionizations (see sect.~\ref{subsec:H2-collisional-excitation} for a description of the process). Each ionization of a hydrogen molecule is associated on average with 40.1~eV energy loss by electrons, of which 11$\,$\% ends up as heat \citep{Dalgarno1999}. This gives a heating energy of 4~eV per ionization. This is the value adopted in the Meudon PDR code \citep{LePetit2006}. 
If we add the energy associated with the recombination of  H$_{3}^{+}$ that may follow the H$_2$ dissociation, another 8~eV is released per recombination  \citep{Maloney1996}. Thus, each ionization of a hydrogen molecule is associated with the deposition of 12~eV into the gas. 
In the \textit{CLOUDY}\footnote{\url{http://www.nublado.org/}} gas code, last described by   \citet{Ferland1998}, the heating efficiency increases with the ionization fraction of the gas. For an ionization fraction of $10^{-4}$, a classical value for the diffuse ISM \citep{Shaw2005}, the heating energy per H$_2$ ionization is $\approx 7~eV$.
As the ionization rate per hydrogen nucleus is half the H$_2$ ionization rate $\zeta_{\rm H}$, the heating rate per hydrogen nucleus $\Gamma/n_{\rm H}$ is $\approx 14$~eV$\times \zeta_{\rm H}$, or
\begin{equation}
\Gamma/n_{\rm H} = 2.2 \times 10^{-33} \, \frac{\zeta_{\rm H}}{10^{-15} \, \rm s^{-1} \, } \rm \quad  W \   .
\end{equation}
Using the cooling rates calculated by \citet{Neufeld1995} for $n(\rm H) =10^4$~cm$^{-3}$ and $N(\rm H) / \Delta v = 10^{22}$~cm$^{-2}$~km$^{-1}$~s, the equilibrium temperatures are approximately 50, 150, and 300 K for $\zeta_{\rm H} = 10^{-15}$, $10^{-14}$, and $10^{-13}$~s$^{-1}$, respectively.

\subsection{Excitation during the H$_{\bf 2}$ formation process}
\index{Molecular hydrogen (H$_2$)!excitation during formation}
\label{subsec:H2-excitation-during-formation}

During the H$_2$ formation process on grain surfaces (see sect.~\ref{subsec:H2-formation-on-grains}), the binding energy of H$_2$ ($\approx 4.5$~eV, or $51\,000$~K) is released and partitioned  
among the grain (internal heating), kinetic energy of the ejected H$_2$ (gas heating), and internal
energy in H$_2$ (i.e. the molecule is formed in a rovibrationally excited state). 
Therefore, H$_2$ excitation can also occur directly during the process of its formation.

How is the H$_2$ binding energy divided between the constituents? What is the internal
energy distribution? What the ortho-to-para ratio of the newly-formed molecule? In spite of many theoretical work and experiments \citep[e.g.][]{Hornekaer2003, Amiaud2007}, all these questions remain open. 

In our model, we choose to distribute the H$_2$ binding energy with \textit{equipartition}: one third of the formation energy (1.4927~eV, or 17249~K) goes into internal energy of H$_2$ and levels are populated in proportion to the Boltzmann factor at a temperature of 17249~K. Another third goes into kinetic energy and the last third goes into grain heating. In our model we assume that the H$_2$ molecule is formed with an ortho-to-para ratio of 3.

Several surveys of cold dark clouds have been performed to observe directly the H$_2$ excitation by formation, but so far without results \citep[see][and references therein]{Tin'e2003}.

%
%

\subsection{Radiative excitation of H$_{\bf 2}$ (UV pumping)}
\index{Molecular hydrogen (H$_2$)!radiative excitation}
\index{Molecular hydrogen (H$_2$)!UV pumping}

When a UV source is present, for instance in Photo-Dissociation Regions (PDRs), H$_2$ may be excited rovibrationally by first being UV-pumped to an electronically excited state \citep{Black1973, Black1987} via the Lyman and Werner transitions within the range $912-1110$~\AA \citep[see also][for a review about dense PDRs]{Hollenbach1997}. The excited molecule will then deexcite (fluorescence) back into  vibrationally excited levels of the ground electronic state. If the densities are not too large ($n_{\rm H} < n_{\rm crit} \approx 10^{4-5}$~cm$^{-3}$) these levels will cascade down through rovibrational transitions. The corresponding UV spectrum has been calculated by   \citet{Sternberg1989}. 
This fluorescence occurs with a probability of $\approx 90\,$\%. 
If $n_{\rm H} > n_{\rm crit}$, the collisional deexcitation of H$_2$ contributes to heat the gas.
In $\approx 10\,$\% of the
electronic excitations,  the molecule will dissociate because the absorption of a UV photon is followed by a cascade towards rovibrational levels of the ground electronic state.


\subsection{ H$_{\bf 2}$ excitation diagrams}
\label{subsec:H2excdiagrams}
\index{Molecular hydrogen (H$_2$)!excitation diagrams}
\index{Excitation diagrams}

Excitation diagrams are a way to represent the SED of the H$_2$ emission  that provides a practical  visualization of the physical conditions of the emitting medium. We detail how one builds an excitation diagram and we give canonical examples within the context of H$_2$ emission in shocks.

\subsubsection{Building excitation diagrams}

Excitation diagrams plot the logarithm of the column densities of the upper H$_2$ rovibrational levels ($v$, $J$) divided by
their statistical weights, $\ln( N_{v J} /g_J )$, against their excitation energies, $E_{vJ} / k_{\rm B}$, expressed in K. The column density is expressed in cm$^{-2}$. 
The statistical weight $g_J  = g_s \, (2J+1)$ is the product of the nuclear spin statistical weight, $g_s$, which
has value of 1 and 3 for even and odd rotational levels $J$, respectively, and the rotational
statistical weight, which is $2J+1$.

Such diagrams are very useful to get a first idea of the temperature conditions that exist in the emitting gas.
If one assumes that a given  line is optically thin to  H$_2$ emission, one 
can calculate the gas column density, $N$ from the observed  brightness of the line, $\mathcal{I}$. The
probability for spontaneous emission is given by the Einstein $A$-coefficient.
The measured H$_2$ line intensities $\mathcal{I}_{v J}$ are converted into column densities
$N_{v J}$  as follows:
\begin{equation}
\label{eq:column-density-excit-diagram}
N_{v J} =\frac{4 \, \pi}{h \, c} \frac{\lambda}{A _{v J}}    \mathcal{I}_{v J}  
\end{equation}
where $\lambda$ is the central wavelength of the line transition, $A_{vJ}$ the transition probability (Einstein coefficient) for the rovibrational level $(v ; J)$, $h$ is Planck's constant, and $c$ is the velocity of light in the vacuum. 

To estimate whether the assumption that the line is optically thin, we may calculate
the optical depth, $\tau$, for a transition $ j \rightarrow i$ between an upper $j$ and lower $i$ level:
\begin{equation}
\tau = \frac{A_{v J}}{8 \pi} \frac{1}{\Delta v} \frac{g_j}{g_i} \lambda ^3 \, N_{v J}
\end{equation}
where $\Delta v$ is the linewidth (dispersion velocity) and $g_j$, $g_i$ are the degeneracies of the upper and lower levels, respectively.
For the $v=0$-0~S(1) transition we find
\begin{equation}
\tau =3.94 \times 10^{-5} \left( \frac{10\,\rm km\,s^{-1}}{\Delta v} \right) \left(\frac{N_{\rm H}}{10^{20}\,\rm cm^{-2}} \right) \ .
\end{equation}
Therefore the assumption that the line is optically thin to H$_2$ emission is fulfilled for most of the situations in the ISM. 

Column densities can then be compared with those expected from thermal excitation.
The method is the following one. In a thermal distribution, $N_{vJ}$ is proportional to
the statistical weight $g$ of the level and to the Boltzmann factor $e^{-\frac{E_{vJ}}{k_{\rm B}T_{\rm exc}}}$, where $E_{v J}$ is the excitation energy of the respective level. The Boltzman law allows to define the excitation temperature associated to a transition $ j \rightarrow i$, $T_{\rm exc}$ from the ratio of the level populations:

\begin{equation}
e^{-\dfrac{h \nu}{k_{\rm B}T_{\rm exc}}} = \dfrac{g_j / n_j}{g_i / n_i}
\end{equation}

where $\nu$ is the frequency associated to the $ j \rightarrow i$ transition, and  $n$ the level populations
associated to the connected levels. In a situation
of local thermal equilibrium, the excitation temperature equals that of the gas. For a
uniform excitation temperature, the values $\ln N_{v J} =g_J$ should thus all fall on a straight
line plotted versus $E_{v J}$, with a slope proportional to $T_{\rm exc}^{-1}$. This diagram thus allows to roughly estimate the temperature conditions in the emitting gas.

From an observational point of view, deriving the excitation temperature is a rough estimate of the temperature conditions in the considered region. With this diagram, deviations from LTE can be easily be probed. Generally, multiple temperature fits are done, corresponding to several excitation temperatures. 
In a shocked medium, this reflects the distribution of temperatures involved in the postshock medium.

\subsubsection{Examples of observed  H$_{\bf 2}$ excitation diagrams}

\begin{figure}
   \centering
    \includegraphics[width=\textwidth]{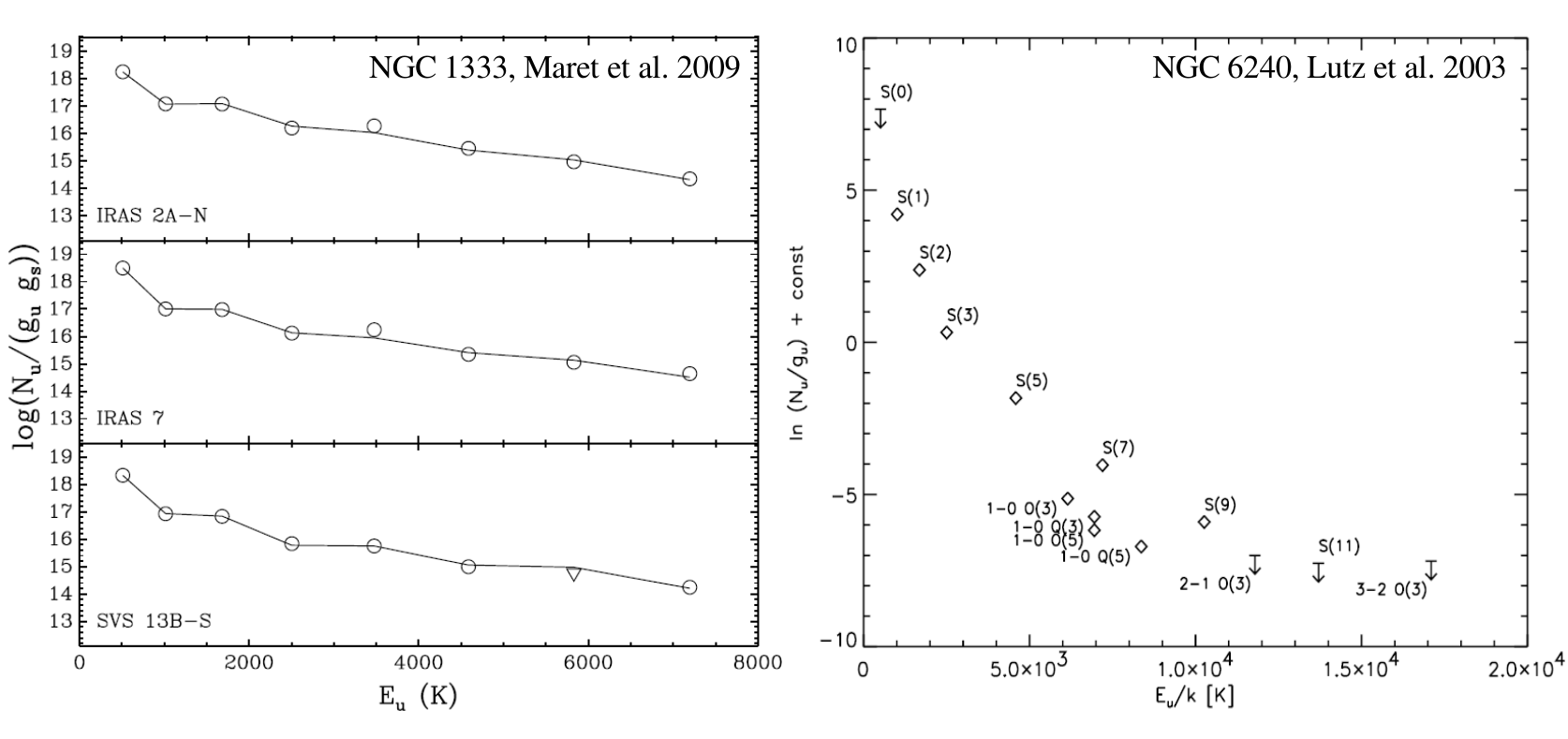}
      \caption[Examples of observed H$_2$ excitation diagrams in NGC 1333 and 6240]{Examples of observed H$_2$ excitation diagrams in NGC 1333 and 6240. \textit{(Left)} Excitation diagrams of pure rotational H$_2$ lines observed with \textit{Spitzer} of three sub-regions in NGC 1333, a star-forming region with molecular outflows \citep[from][]{Maret2009}. \textit{(Right)} H$_2$ excitation diagram of NGC~6240, a luminous infrared galaxy observed with \textit{ISO}, by \citet{Lutz2003}.}
       \label{fig_H2_excit_diag_obsNGC1333_6240}
   \end{figure}

For illustration, we show in Fig.~\ref{fig_H2_excit_diag_obsNGC1333_6240} two examples of observed excitation diagrams of pure rotational H$_2$ lines in very different environments. NGC~1333 (on the \textit{left}) is a Galactic star forming region comprising mutiple outflows \citep{Neufeld2006, Maret2009}. The diagrams show a ``zigzag'' pattern, indicating that the ortho-to-para ratio is lower than 3, the high-temperature limit (see sect.~\ref{subsec:H2-ortho-para-ratio}). In addition, the curvature of the diagram show that the observed values cannot be fitted with a single LTE excitation temperature. This suggest that the gas, shocked by the proto-stellar outflows, is a mixture of multiple temperature gas components.

NGC~6240, an IR-luminous merger, shows the same  non-linear decline of the H$_2$ excitation diagram than what is commonly seen in shocks within the Galaxy,  as well as in external galaxies \citep[e.g.][]{Rigopoulou2002, Lutz2003}.

\section{Observing H$_{\bf 2}$ in space}
\label{sec:observingH2}
\index{Molecular hydrogen (H$_2$)!observations}

\subsection{Direct H$_2$ observations}

The first detection of H$_2$ beyond the Solar System was made by \citet{Carruthers1970} through absorption spectroscopy of UV light with a rocket-borne UV spectrometer. 
This discovery was followed by UV observations with the Copernicus space mission that demonstrated the presence of the hydrogen  molecule in diffuse interstellar clouds \citep[for a first review on this subject, see][and references therein]{Spitzer1975}. These observations revolutionized our understanding of the origin of
 astronomical objects.

A decade later, \citet{Gautier1976} detected H$_2$ in emission for the first time through near-infrared rovibrational lines from a massive young stellar object in the Orion Nebula. This demonstrated the importance of H$_2$ in dense environments as well. However, because of lack of sensitivity, such observations remained of limited scope over the next three decades.  Today, the near-infrared rovibrational transitions are routinely imaged from ground-based telescopes.

\begin{figure}
   \centering
    \includegraphics[width=\textwidth]{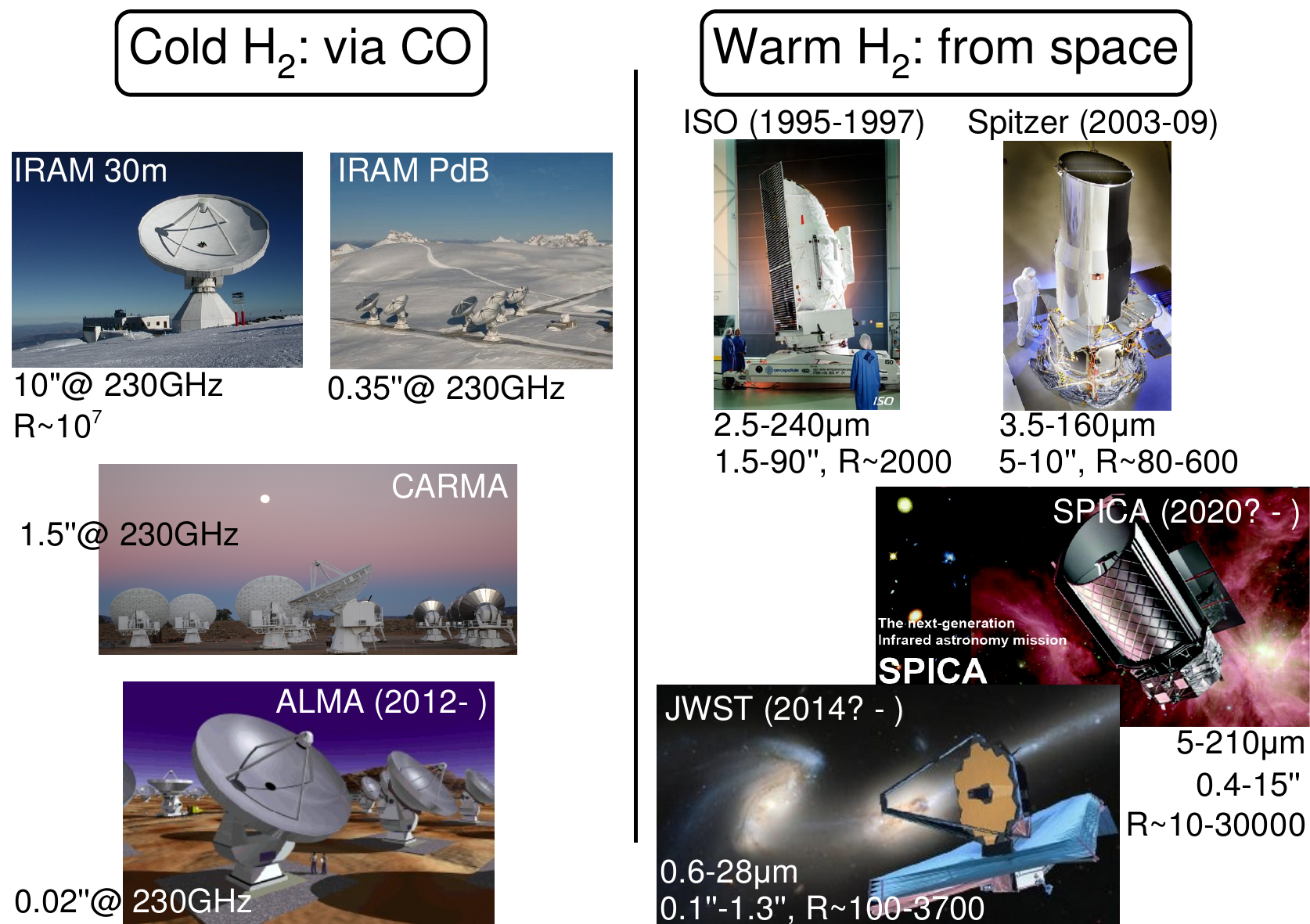}
      \caption[Observing cold and warm molecular gas from the ground and from space]{Cold and warm molecular gas are observed with two different kinds of observatories. Cold molecular gas is essentially derived from CO ground-based observations with single-dish radio-telescopes or interferometers. On the other hand, the IR emission from warm H$_2$ must be observed from space. This figure shows a few examples (past, present and future) of these observatories, with their basic characteristics (wavelength coverage, spatial and spectral resolution).}
       \label{fig_telescopes_H2}
   \end{figure}

The H$_2$ mid-infrared rotational transitions, which trace warm ($\approx 10^{2-3}$~K) molecular gas, must be observed from space (see Fig.~\ref{fig_telescopes_H2}, right panel). Indeed, the Earth's atmosphere is mostly opaque to the mid-IR lines of
ortho and para H$_2$. For the transitions occurring at wavelengths where the
atmosphere is partially transparent, the thermal background from the warm
telescope, its spectrometer, and the sky, limit the sensitivity.  The H$_2$ infrared
lines were observed by \textit{ISO} and \textit{Spitzer} with limited imaging capabilities
and low spectral resolution.  
The next future ($\approx 2015$) of mid-IR H$_2$ observations will the James Webb Space Telescope (JWST, see Fig.~\ref{fig_telescopes_H2}) in which I am involved (see chapters~\ref{chapter:JWST} to \ref{chapter:science_JWST}). On a longer timecale, the \textit{SPICA} mission\footnote{Space Infrared Telescope for Cosmology and Astrophysics} \citep{Swinyard2009}, an actively-cooled \textit{Herschel}-like telescope, would be capable of doing MIR and FIR spectroscopy at high spectral resolution with an unprecedent sensitivity.

\citet{Boulanger2009} have proposed a mission to the Cosmic Vision 2015-2025 call, so-called \textit{H$_2$Ex,} dedicated to wide-field ($20'\times 20'$) observations of the lowest H$_2$ rotational transition lines, S(0) to S(3), at high spectral resolution (up to $\lambda/\Delta\lambda \approx 30\,000$, in our Galaxy and up to redshift $2-3$. Actually, the start of my PhD work about modeling H$_2$ formation and emission in active phases of galaxy evolution was motivated by the preparation of this Cosmic Vision proposal, which was not selected, unfortunately.

\subsection{CO, a proxy for H$_2$}
\label{subsec:CO-proxy-for-H2}
\index{Molecular hydrogen (H$_2$)!H$_2$ to CO conversion factor}

The bulk of the molecular hydrogen in a galaxy is cold ($10-20$~K), too cold to be seen in emission. This is why the presence of cold H$_2$ is inferred essentially from carbon monoxyde (CO) observations.
CO is the most abundant molecule after H$_2$. Its dipole moment is small (0.1 Debye), so CO is easily excited. The emission of the CO(1-0) line at 2.6~mm (corresponding to its first level at 5.52~K) is ubiquitous in galaxies. 

The CO lines (mainly 1-0 at 115~GHz and 2-1 at 230~GHz) are observed with radiotelescopes (see left panel of Fig.~\ref{fig_telescopes_H2}), being single-dish or interferometers. During my PhD, I have made use of the IRAM 30m single-dish telescope. The high spectral resolution of radiotelescopes provides crucial information about the molecular gas kinematics which, in general, cannot be inferred from mid-IR H$_2$ line observations alone because of limited spectral   resolution.

In order to derive the cold H$_2$ masses, we generally assume  a fixed CO luminosity to H$_2$ mass conversion factor, the so-called $X = N(\rm H_2) / \mathcal{I}_{\rm CO}$ factor, where $N(\rm H_2) $ is the column density of H$_2$ and $ \mathcal{I}_{\rm CO}$ the integrated CO line intensity.
Several methods are employed to derive the $X$ factor \citep[see][]{Solomon1991}, leading to similar results.
In our Galaxy, the most common value is \cite{Dame2001}:
\begin{equation}
\label{eq:X-factor-CO-H2}
X = \frac{N(\rm H_2) \ [cm^{-2}]}{ \mathcal{I}_{\rm CO} \ [\rm K \ km \ s^{-1}]} = 2.2 \times 10^{20} \ \rm cm^{-2} \ [K \ km \ s^{-1}]^{-1}
\end{equation}
In extragalactic environments, the CO(1-0) line luminosity seems also to be a reliable mass
tracer \citep[e.g.][]{Dickman1986, Solomon1991}. This allows us to convert the integrated line flux of
 CO(1-0) line into molecular gas mass. Including a factor of 1.36 to account for helium,
the gas mass bound into molecular clouds can be calculated as:
\begin{equation}
\label{eq:CO10-to-H2mass}
M_{\rm gas} = 2.94 \times 10^7 \left( \frac{\mathcal{F}_{\rm CO(1-0)}}{1 \, \rm Jy\, km\, s^{-1}} \right) \left( \frac{X}{2.2 \times 10^{20}} \right) \left( \frac{d}{50 \, \rm Mpc} \right) ^2  \ \rm M_{\odot} \ \ .
\end{equation}
We will make use of this useful expression later in this manuscript.


\chapter{Shocks and chemistry in the multiphase ISM}
\label{chapter:shocks}

\epigraph{Nothing shocks me. I'm a scientist.}{Indiana Jones, \textit{in Indiana Jones and the Temple of Doom}}

\index{Shocks}


\begin{Abstract}

H$_2$-luminous galaxies appear to be in active phases of their evolution (galaxy collisions, galaxy cluster cooling flows, starburst- or AGN-driven winds, etc., see chapter~\ref{chapter:H2_galaxies}) that release a large amount of kinetic energy in the interstellar medium . Shocks are thus expected to be ubiquitous in these media, and may account for a significant, if not dominant, part of the powerful H$_2$ emission observed in these sources. Consequently, I have been driven to study H$_2$ emission in shocks, and use MHD shock models to interpret these observations. This has been done first within the context of the Stephan's Quintet galaxy collision (chapter~\ref{chapter:H2_SQ} and \ref{chapter:H2_SQ_mapping}), and is being extented to other astrophysical objects (chapter~\ref{chapter:perspectives}).
In this chapter, I present the necessary background on the physics and chemistry of MHD shocks, as well as a few results extracted from the grid of shock models I have built to interpret the observations. Then the discussion is extented to shocks in inhomogeneous media. I focus on two situations: the evolution of a shocked dense cloud, and the collision between two gas streams. The physical processes that control the evolution of the shocked gas will be discussed, as well as recent results for numerical simulations.

\end{Abstract}

\minitoc



\section{Introduction}

\PARstart{S}hock waves are ubiquitous in the interstellar matter settled in galactic disks, but also in the intergalactic medium. They are produced by violent pressure disturbances such as star- or AGN-driven jets and winds, supernovae explosions, collisions between molecular clouds or, at larger scales, between galaxies.


\begin{wrapfigure}{r}{62mm}
  \begin{center}
    \includegraphics[width=62mm]{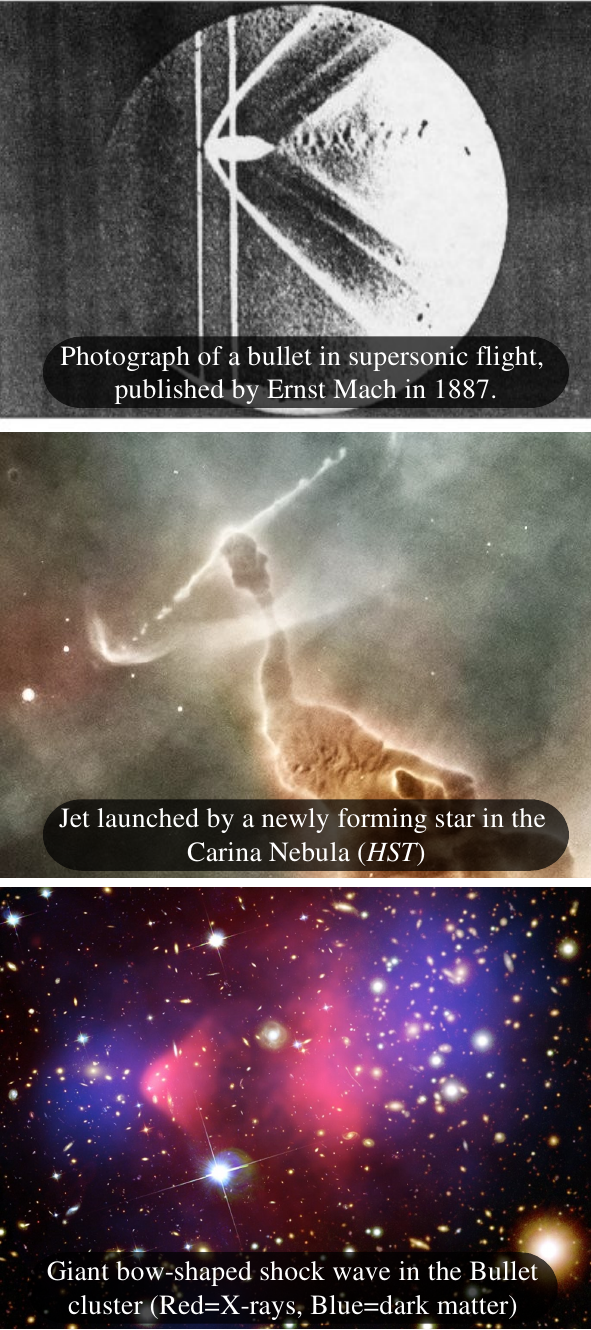}
  \end{center}
\end{wrapfigure}

A shock wave is \textit{an overpressure region that propagates in the fluid at a supersonic velocity}. These shock waves accelerate, heat and compress the gas. This condensation may lead to star formation, as it is the case for giant shock waves propagating in the spiral arms of a galaxy. On the other hand, shocks may have destructive effects on the molecular clouds, thus quenching star formation, as it is perhaps the case in AGN jets driven in the ISM of their host galaxy. 

Shocks have an important impact on the evolution of the interstellar medium (ISM), from a dynamical but also from a chemical point of view. By heating the gas, shocks driven into molecular clouds initiate chemical reactions that would not occur in peaceful, cold molecular gas. These chemical processes play a role in the thermal balance of the shock, which in turn has an impact on the gas dynamics. 

H$_2$ appears to be a dominant coolant of MHD shocks. H$_2$-luminous galaxies (chapter~\ref{chapter:H2_galaxies}) are be precisely in some of the energetic phases of strong dissipation of kinetic energy quoted above. Therefore, a large part of my modeling work of the dissipation of mechanical energy involves the physics and chemistry of shocks. 

In the multiphase environment of these H$_2$-luminous galaxies, different types of shocks are expected. Fast shocks, heating the gas to high ($T \approx 10^6 - 10^8$~K),  produce EUV and X-ray photons that photoionize the tenuous medium. These shocks are one source of energy that produce the HIM\footnote{Hot Ionized Medium, see chapter~\ref{chapter:dust_gas_galaxies}, sect.~\ref{phases-structure-ISM}.} in galaxies. On the other hand, low-velocity shocks (typically $\lesssim 50$~km~s$^{-1}$) are known to \textit{(i)} initiate the formation of molecules in the gas that cools behind the shock \citep[e.g][]{Hollenbach1979}, and \textit{(ii)} be a very efficient process to excite molecules, especially H$_2$, via collisions in the dense gas compressed and heated by the shock. 

To interpret the H$_2$ observations presented in chapter~\ref{chapter:H2_galaxies}, I have been driven to study the formation of H$_2$ during the cooling of a shock-heated gas, and H$_2$ excitation in magnetohydrodynamical (MHD) shocks driven into dense molecular gas. I have used the \citet{Flower2003} shock model, a 1-D\footnote{One dimension} MHD code coupled to a extensive network of chemical reactions, to build a large grid of shock models for different values of key parameters such	as the shock velocity, the pre-shock density, the magnetic field, etc. In particular, the results of the modeled emission of molecular hydrogen (whose treatment is included inside the shock code) to the observations. 

For a review about the physics and chemistry of interstellar shocks, I direct the interested reader to \citet{Draine1980, McKee1980, Chernoff1987, Hollenbach1989, Draine1993, Hartigan2003}. For a comprehensive and didactic presentation of hydrodynamical and MHD shocks, the reader will consult the PhD thesis (in french) by \citet{Guillet2008}. I assume that the reader is familiar with the basic properties of hydrodynamical shocks (in which the magnetic field is absent $B=0$). This chapter first focuses on the basic properties of MHD (\textit{single-fluid} or \textit{multi-fluids}) shocks (sect.~\ref{sec:MHD_shocks}). I present the shock code and the grid of shock models I use in the following chapters (sect.~\ref{sec:MHD-shock-models-molecular-gas}), and mainly discuss H$_2$ cooling in  MHD shocks (sect.~\ref{subsec:shocks-molecular-cooling}). Then I extend the discussion to shocks into multiphase media. I describe the evolution of a cloud (embedded in a tenuous gas) that is overrun by a shock wave, from a theoretical and numerical simulation point of view.  Finally, I consider the collision between two gas stream and point out the main differences between  1-D  and 3-D models.

\section{Magnetohydrodynamic (MHD) shocks in an homogeneous medium}
\label{sec:MHD_shocks}
\index{Shocks!MHD shocks}
In the presence of a magnetic field, ubiquitous in the ISM, the shock wave satisfies not only the equations of the fluid dynamics, but also the Maxwell's equations. 
The magnetic field does not interact the same way with charged and neutral particles, which can result in a \textit{decoupling} of the charged (ions and electrons) and neutral fluids. Indeed, the magnetic field modifies directly the dynamics of the charged particles\footnote{For instance, if the magnetic field is oriented parallel to the shock front, the Lorentz's force will make the particles girate around the magnetic field lines, thus leading to a coupling of the charged fluid to the magnetic field.} (ions, electrons or grains), whereas it affects only indirectly the neutrals through collisions with the positive ions (and electrons). 
Therefore, the magnetic field and the ionization fraction of the gas have thus an important impact on the structure of shock waves.

Before discussing the structure of MHD shocks, I shall recall the basic properties of the types of MHD waves that can propagate in a plasma.

\subsection{Waves propagating into a plasma}
\index{MHD shocks!waves}

In a magnetized plasma, different types of waves can propagate at distinct velocities:

\index{Sound wave}
\index{MHD waves!sound wave}
\begin{description}
\item[Sound waves] are (\textit{longitudinal}) compression waves that propagate pressure disturbances at a speed  
\begin{equation}
\displaystyle c_{\rm s} = \sqrt{\frac{\gamma k_{\rm B} {\rm T}}{\bar{\mu}}} \ ,
\end{equation}
where $\gamma$ is the \textit{adiabatic index} (the heat capacity ratio, with $\gamma = 5/3$ for a monoatomic gas and $\gamma = 7/5$ for a diatomic gas). T is the gas temperature and $\bar{\mu}$ is the mean molecular mass\footnote{$\bar{\mu} = 0.6$~amu for a fully ionized gas, or 2.33 for a molecular gas. $1~\rm amu = 1.66 \times 10^{-24}$~g} in atomic mass unit (amu).

\index{MHD waves!Alfvn  waves}
\item[Alfvn waves] are \textit{transverse} waves that propagate the distortion of the magnetic field lines \textit{along} these lines \citep{Alfven1950}. Alfvn waves are thus incompressible, dispersionless waves. The neutral species can propagate Alfvn waves because of their coupling to the charged fluid. The \textit{Alfvn speed} of these waves are thus
\begin{equation}
V_{\rm A}^n = \sqrt{\frac{B^2}{\mu_0 \rho _n}} \  \ \mbox{for the neutrals} \quad {\rm and} \quad V_{\rm A}^c = \sqrt{\frac{B^2}{\mu_0 \rho _c}}\ \  \mbox{for the charged fluid} \ \ ,
\end{equation}
where $\rho _n$ and $\rho _c$ are the mass density of the neutral and charged (ions, electrons, grains) fluids, respectively. 

\index{MHD waves!magnetosonic  waves}
\item[Magnetosonic waves] are \textit{longitudinal}, \textit{compressible}, dispersionless waves that propagate pressure variations in the direction perpendicular to the magnetic field lines. The propagation of these waves involves variations of the strength of the magnetic field.
In the neutral fluid (most of the gas mass), the magnetosonic speed is
\begin{equation}
V_{\rm ms}^n = \sqrt{c_{\rm s}^2 + \frac{B^2}{4  \pi \rho _n}} \simeq \frac{B}{\sqrt{4 \pi \rho _n}}
\end{equation}
In the charged fluid, magnetosonic waves can propagate at a speed
\begin{equation}
\label{eq:magnetosonic-vel-charged}
V_{\rm ms}^c = \sqrt{c_{\rm s}^{2} + \frac{B^2}{4  \pi \rho _c}} \simeq \frac{B}{\sqrt{4 \pi \rho _c}} \gg  V_{\rm ms}^n \ \ ,
\end{equation}
In a typical molecular cloud, $\rho _c \approx 10^{-2} \rho _n$ because the dust grains are charged. So $V_{\rm ms}^c \approx 10 V_{\rm ms}^n$.
\end{description}

\index{MHD shocks!stationary}
\index{MHD shocks!transverse}
In the following I will only consider \textit{stationary} shocks, where the structure of the shock (density, temperature profiles) in the reference shock frame do not evolve with time. In addition, \textit{the direction of the magnetic field is perpendicular to the direction of propagation of the wave}, or parallel to the shock front. In this case, the magnetic field limits the compression of the gas at a maximum efficiency, since the fluid cannot flow along the magnetic field lines. 

\index{MHD shocks!C-shocks}
\index{MHD shocks!J-shocks}
\subsection{C- and J- type shocks}
\label{C-J-type-shocks}
\index{Shocks!$C$- and $J$- type shocks}
Depending on the intensity of $B$ and the ionization fraction of the gas, we distinguish between two types of MHD shocks, $C$- and $J$-types:
\begin{description}
\item[$J$-shocks:] if the $B$ is weak or absent, or if the ionization fraction of the gas is high, the collisional coupling between charged and neutral particles is strong. These particles behave like a \textit{single fluid}, coupled to the magnetic field. The properties of these shocks are similar to that of the hydrodynamical shocks. Across the shock front, the variables (pressure, density, velocity, etc.) change abruptly (see the top plot of Fig.~\ref{fig:types_shocks_Draine1980}). The transition region, which scale is of the order of the mean free path, is much narrower than the other dimensions in the gas. This transition is  treated as a discontinuity, where the change of the variables of the gas is discrete ($J$ being for \textit{Jump}). The preshock and postshock  values of the gas density, pressure and temperature are related by the Rankine-Hugoniot jump conditions (see sect.~\ref{Rankine-Hugoniot-MHD}).

\item[$C$-shocks:] if the magnetic field is present and the ionization fraction low (typically $x_e = 10^{-7} - 10^{-8}$ in dense $n_{\rm H} = 10^{3} - 10^{4}$~cm$^{-3}$ molecular clouds), the neutral and charged fluids are decoupled from each other. These shocks are named \textit{multi-fluids} or $C-$shocks \citep{Mullan1969, Mullan1971}, $C$ being for \textit{Continuous}, because in this case the discontinuity is smoothed and the gas parameters vary continuously across the shock front (see details below and Fig.~\ref{fig:types_shocks_Draine1980}).

\end{description}

\subsubsection{Structures of $C$- and $J$-shocks} 
\label{structure-C-J-shocks}

\begin{figure}
   \centering
    \includegraphics[width=0.6\textwidth]{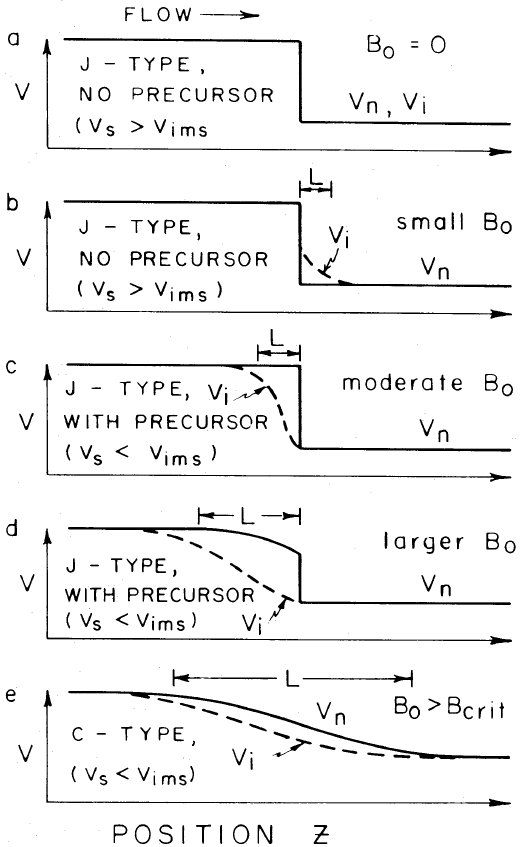}
      \caption[Schematic structure of different types of shock waves]{Schematic structure of the velocity of a stationary shock wave
in poorly ionized medium, as a function of the local magnetic field strength $B_0$. $B_0$ increases from top to bottom. The
velocities are given in the frame where the shock front is idle: $V_n$ is the neutral velocity
in the propagation direction of the shock, $V_i$ that of the ions and electrons. The preshock
medium is on the left hand side of the shock front, the postshock on its right hand side. $L$
is the shock length, and $V_{\rm ims}$ is the ions magnetosonic speed given by Eq.~\ref{eq:magnetosonic-vel-charged} (the dust grains are ignored in this model). Taken from \citet{Draine1980}.}
       \label{fig:types_shocks_Draine1980}
\end{figure}

The Rankine-Hugoniot relations are conservations laws between any points behind and ahead of the shock front. They do not give any information about the internal structure of the shock. 
This structure depends on the strength of the magnetic field and the ionization fraction of the gas, and has a direct impact on the characteristics of the excitation of species in the shock, in particular H$_2$. This is why I briefly discuss here the Fig.~\ref{fig:types_shocks_Draine1980}, which shows schematically the structure of a stationary shock as a function of the preshock strength of the (transverse) magnetic field.

As shown in Eq.~\ref{eq:magnetosonic-vel-charged}, in a molecular cloud of density $n_{\rm H} = 10^4$~cm$^{-3}$ and ionizing fraction $\chi \approx 10^{-5} - 10^{-9}$,  magnetosonic waves propagate faster in the charged fluid than the shock front. Typically for this medium we have:
\begin{equation}
c_{\rm s} \lesssim V_{\rm A}^{n} < V_{\rm s} \ll V_{\rm A}^{c} \ \ ,
\end{equation}
where $c_{\rm s}$ is the sound speed, $ V_{\rm s}$ is the shock velocity, $V_{\rm A}^{n}$ and $V_{\rm A}^{c}$ being the Alfvn speed for the neutral and charged fluids, respectively. 
So, if the strength of the magnetic field is high enough, the information can be transported ahead of the shock by the charged fluid, and a so-called  \textit{magnetic precursor} appears. This precursor slows and compresses the charged fluid of the preshock gas before the arrival of the shock front, on spatial scales much larger than the mean free path. This magnetic precursor decouples the neutral and charged fluid, generating \textit{ambipolar diffusion}. The resulting friction between the fluids heats mostly the charged fluid. The variables of the charged fluid are continuous across the shock front, whereas those of the neutral fluid are discontinuous. Such a shock is called\textit{ J-shock with magnetic precursor}. This is the case \textit{(c)} of Fig.~\ref{fig:types_shocks_Draine1980}.

If the magnetic field is more intense, the magnetosonic speed further increases and the length of the magnetic precursor, $L$, too. The exchanges of momentum and energy between the neutral and the charged fluids in the precursor are more important, and the neutral fluid itself is likely to be heated, compressed and slowed before the arrival of the shock front, through elastic collisions with the charged fluid.   
Because of this slowing down, the neutral fluid also undergoes a velocity jump of a
smaller amplitude than in the situation where there is no magnetic field. This is the case \textit{(d)}, in Fig.~\ref{fig:types_shocks_Draine1980}.

When the magnetic field strength is higher than a critical value, $B_{\rm crit}$, the flow of the neutral fluid becomes continuous across the shock front.  The discontinuity vanishes, and
the shock turns into a $C$ shock (case \textit{(e)} in  Fig.~\ref{fig:types_shocks_Draine1980}).

%
%
%

\subsubsection{Rankine-Hugoniot jump conditions for an adiabatic MHD shock}
\label{Rankine-Hugoniot-MHD}
\index{Shocks!Rankine-Hugoniot jump conditions}
\index{Rankine-Hugoniot relations!adiabatic MHD shock}
\index{Adiabatic index}

The jump conditions across an MHD shock or discontinuity are referred as the Rankine-Hugoniot equations for MHD. Before giving these equations, let us  recall
some thermodynamical relations and definitions. We denote by $c_{\rm P}$ and $c_{\rm V}$ the
specific heats of the gas at constant pressure and volume respectively. The \textit{adiabatic index} is defined by $\gamma =c _{\rm P} / c_{\rm V}$. In the case of a perfect gas, the pressure $\rm P$, density $\rho$ and temperature $\rm T$ are connected by the classical equation
\index{Perfect gas!perfect gas law}
\begin{equation}
{\rm P} = \frac{\rho R {\rm T}}{\bar{\mu} \, m_{\rm H}}
\end{equation}
\index{Perfect gas!universal gas constant}
where $R=8.31$~J~K$^{-1}$~mol$^{-1}$ is the universal gas constant, $\bar{\mu}$ is the mean molecular mass in atomic mass unit (amu) and $m_{\rm H}$ is the mass of an hydrogen atom. 
For an adiabatic evolution, the additional relation holds:
\index{Perfect gas!adiabatic relation}
\begin{equation}
{\rm P} = K \rho ^{\gamma}
\end{equation}
\index{Perfect gas!entropy}
where $K$  is a constant. Apart from a constant, the entropy of a perfect gas is defined as
\begin{equation}
\label{eq:entropy-perfect-gas}
{\rm S} = c_{\rm V}  \log \left( \frac{{\rm P}}{\rho ^{\gamma}} \right)
\end{equation}

We place ourselves then in the reference frame of the shock (the surface discontinuity is at rest), and we take the suffixes $_1$ and $_2$ to refer to the preshock and postshock regions, respectively. The jump conditions are obtained by developing the conservation laws for mass, momentum, and enthalpy\footnote{Because of the pressure forces on the fluid element, the enthalpy fux is conserved, not the energy flux},  with the few relations above \citep[see e.g. the book by][]{Shu1991}:

\begin{eqnarray}
\rho _1 v_1 & = & \rho _2 v_2 \\
 \label{eq:RH_cons_momentum}
\rho _1 v_1^2 + {\rm P}_1 + \frac{B_1^2}{8 \pi} & = & \rho _2 v_2^2 + {\rm P}_2 + \frac{B_2^2}{8 \pi} \\ 
\left(  \rho _1 \frac{v_1^2}{2} + \frac{\gamma}{\gamma - 1} {\rm P}_1 +  \frac{B_1^2}{4 \pi} \right)   v_1 & = & \left(  \rho _2 \frac{v_2^2}{2} + \frac{\gamma}{\gamma - 1} {\rm P}_2 +  \frac{B_2^2}{4 \pi} \right)   v_2
\end{eqnarray}
where $v$ the velocity of the flow in the reference frame of the shock, $\rho$ the mass density, and P the thermal pressure of the gas. The term $B^2 / 8 \pi$ is the magnetic pressure that is added to the thermal pressure in the direction perpendicular to the magnetic field lines. The terms $\gamma {\rm P} / (\gamma - 1)$ and $B^2 / 4 \pi$ are the specific enthalpy\footnote{entropy per unit volume.} and the magnetic energy, respectively. 
The conservation of momentum (Eq.~\ref{eq:RH_cons_momentum}) can also be viewed as the conservation of the sum of the ram pressure ($\rho v^2$) and the thermal pressure P. 

We remind that the preshock (respectively postshock) flow is \textit{supersonic} (resp. \textit{subsonic}) in the reference frame of the shock front. In this frame, the gas appears to be decelerated. Knowing the Rankine-Hugoniot relations, and considering the relation defining the entropy for a perfect gas (see above Eq.~\ref{eq:entropy-perfect-gas}), one can show that the entropy increases during the shock if and only if ${\rm P}_2 > {\rm P}_1$, i.e. whenever a shock can form \citep[e.g.][]{Zeldovich1966}. It follows that all transverse shock waves lead to compression of the gas.
In the reference frame of the preshock medium, a fluid parcel that crosses the shock front, is compressed (${\rm P}_2 > {\rm P}_1$),  accelerated and heated (${\rm T}_2 > {\rm T}_1$) by the work of the pressure forces applied by the postshock gas, just behind the shock front. 

\index{Magnetic field!frozen}
\subsubsection{Frozen magnetic field}
\label{frozen-magnetic-field}
In the interstellar medium, the characteristic time of diffusion of the magnetic field is
generally much longer than  the characteristic time of advection. One can therefore
assume that the field is \textit{frozen} in the charged fluid. The charged grains are also coupled to the magnetic field because their slowing-down time in the shock is longer than their timescale of gyration around the magnetic field lines \citep{Pineau1997}.
For a one-dimension (1D) transverse shock, the jump relations for the magnetic field are the following:
\begin{equation}
B_2 = B_1 \frac{v_1}{v_2} = B_1 \frac{\rho _2}{\rho _1}
\end{equation}
The compression of the magnetic field follows that of the charged fluid.

\subsubsection{Compression rate and postshock temperature}
\index{MHD shocks!compression}
\index{MHD shocks!postshock temperature}

\citet{Roberge1990} gives the compression factor of the gas:
\begin{eqnarray}
\chi = \frac{\rho _2}{\rho _1} &=& \frac{2(\gamma + 1)}{D + \sqrt{D^2 + 4(\gamma +1)(2-\gamma)M_{\rm A}^{-2}}} = \frac{v_1}{v_2} \\
D &=& (\gamma -1) + 2 M_{\rm s} ^{-2} + \gamma M_{\rm A}^{-2} \ \ ,
\end{eqnarray}
where $M_{\rm s} = V_{\rm s} / c_1$ is the sonic Mach number, $V_{\rm s}$ being the shock velocity (velocity of the shock front in the reference frame of the preshock medium) and $c_1$ the sound speed in the preshock gas. $M_{\rm A}$ is the Alfvn Mach number $M_{\rm A} = V_{\rm s} / V_{\rm A}$.

For a \textit{single-fluid} MHD shock, the postshock temperature (and thus the postshock pressure) 
can be expressed as a function of the compression factor $\chi = \rho _2 / \rho _1$ \citep{Roberge1990}:
\begin{equation}
{\rm T}_2 = \frac{\bar{\mu}_2 V_{\rm s}^{2}}{k_{\rm B}} \frac{1}{\chi} \left( 1 + \frac{1}{\gamma M_{\rm s}^2} -  \frac{1}{\chi} +  \frac{1}{2 M_{\rm A}^2} (1 - \chi ^2) \right) 
\end{equation}

More generally, in case of hydrodynamical shocks ($B=0$), or single-fluid MHD shocks, the physical preshock and postshock quantities can be expressed as a function of the pressures ${\rm P}_1$ and ${\rm P}_2$, or as a function of the Mach number of the shock $M_{\rm s}$ \citep[see][]{Landau1959, Roberge1990}.

\section{Modeling MHD shocks in molecular gas}
\label{sec:MHD-shock-models-molecular-gas}

My interpretation work of H$_2$ emission in H$_2$-luminous galaxies has involved the use of a sophisticated shock model developed by David Flower,  Guillaume Pineau des Forts and Jacques Le Bourlot. The first paper describing the code is \citet{Flower1985}, and the most recent developments are published in \citet{LeBourlot2002, Flower2003, Flower2003a}. 

I run a grid of shock models that I use to fit the observations (see chapters~\ref{chapter:H2_SQ}, \ref{chapter:H2_SQ_mapping} and \ref{chapter:perspectives}). 
In this section, I present illustrative results for $C$ and $J$ shock models, and focus on molecular emission (especially H$_2$) in $C$ shocks, which are relevant for our study.

\subsection{The Flower et al. code and the grid of shock model}
\label{subsec:grid-models-Flower}
\index{Shocks! the Flower et al. shock code}
\index{MHD shocks!description of the Flower et al. code}
In a few words, this 1-D\footnote{The Flower et al. code is a one dimensional code, i.e. it considers a plane-parallel geometry.} code integrates\footnote{The algorithm is base on the VODE integrator \citep{Brown1989}.} the dynamical and chemical rate equations, together with those for the level populations of the rovibrational levels of the electronic ground state of H$_2$. In my grid of models, I use 137 chemical species, connected by a set of 1040 chemical reactions. 
The code can be used to model hydrodynamic shocks or transverse MHD shocks ($C$- or $J$-type). 

The code has been described extensively in many papers and PhD thesis, so I will not repeat this description in this manuscript.  The interested reader will consult  \citet{Flower1985} for a first description of the equations solved by the code, and to the PhD thesis  by \citet{Guillet2008} for a detailed formulation of these equations closer to the current version  of the code. For a description of the treatment of H$_2$ in the code, please see \citet{Flower1986}, \citet{LeBourlot2002} and the PhD thesis by \cite{Gusdorf2008} for comparison of the model predictions with observations of proto-stellar outflows and a description of other chemical aspects. For the treatment of dust grains, see \citet{Flower2003a} and \citet{Guillet2008, Guillet2009} for the implementation of a sophisticated module that treats in details the dynamics and evolution of the dust.

\subsubsection{Hypothesis}
\index{MHD shocks!stationarity}

In this section I  consider MHD shocks  \textit{without radiative precursor}, i.e. the  shock velocities are low enough \citep[typically $V_{\rm s} \lesssim 80$~km~s$^{-1}$ in molecular gas][]{Hollenbach1989}. Shocks with radiative precursor will be considered later in sect.~\ref{sec:fast-shock-models}. 
In the following, I list a few important features and assumptions used in the MHD shock model I have been using. 
\begin{description}
\index{MHD shocks!flow time}
\item[Stationary shock:] the model assumes that a \textit{stationary state} is reached in the 1D flow at velocity $v_n(z)$ (neutrals velocity) in the $z$-direction, so that $d/dt = v_n d/ dz$.
The flow time, $t$, of the neutrals is thus related to the distance, $z$, by the relation:
\begin{equation}
t = \int \frac{1}{v_n} \, d z
\end{equation}
The following results often present the physical quantities as a function of the flow time. 

\index{MHD shocks!preshock magnetic field}
\item[Preshock magnetic field:] we adopt the classical scaling relation for the preshock magnetic field strength:
\begin{equation}
\label{eq:magnetic-field-preshock}
\frac{B_0}{1\,\mu\rm G} = b \sqrt{\frac{n_{\rm H}}{1~\rm cm^{-3}}} \ \ ,
\end{equation} 
where $b$ is the magnetic scaling factor and $n_{\rm H}$ the preshock hydrogen density ($n_{\rm H} = n({\rm H}) + 2 n({\rm H_2}$). 
This expression is in agreement with the Zeeman effect observations of galactic molecular clouds that lead to $B \propto n_{\rm H}^{0.47 \pm 0.08}$ \citep{Crutcher1999}. This observational result seems to be in agreement with numerical simulations of the condensation of molecular clouds under the effect of ambipolar diffusion \citep{Fiedler1993, Hennebelle2008}, although the densities considered in these studies (prestellar cores) are much higher than those we consider here.
For the following illustrative results, we adopt $b=1$ for $C$-shocks, and $b=0$ for $J$-shocks (no magnetic field). 

For $C$-type shocks, the gas is treated as 3 fluids consisting
of neutral species, positively and negatively charged species. We assume
that the  transverse magnetic field remains frozen into the charged fluid of
the preshock gas all through the shock (see sect.~\ref{frozen-magnetic-field}).

\item[H$_{\bf 2}$] is treated in a detailed manner in the code. The excitation mechanisms taken into account in the model, as well as associated hypothesis, have been presented in chapter~\ref{chapter:H2Molecule}. We remind that collisional excitation with H, H$_2$ and He is included. When an H$_2$ molecule forms on the surface of a grain, we assume that the binding energy of H$_2$ is distributed with equipartition between heating of the grain, kinetic energy of the newly-formed H$_2$ molecule, and internal energy of H$_2$.

\index{MHD shocks!and dust grains}
\item[Dust grains]  are assumed to be composed of olivine-like material, (same stoechiometry as MgFeSiO$_4$). The size distribution
is assumed to be the MRN \citep{Mathis1977} distribution $dn_g(a)/da \propto a^{-3.5}$, where $dn_g (a)$ is the  density of grains having a radius between $a$ 
and $a+da$. The radius is taken to be in the range of $10 - 300$~nm.
The total mass density (including mantles) of the grains is taken to be
$8.5 \times 10^{-3} \times 1.4 \, n_{\rm H} m_{\rm H}$, where $8.5 \times 10^{-3}$ is the dust-to-gas mass ratio. The grain temperature is not calculated in the models but remains constant at a user specified value. Here we use 15 K. The rate coefficients for
charge transfer with grains is also taken into account into the model, allowing for
the grain charge distribution to be calculated for each step of the model. 

In the version I use, gas-grain collisions are also taken into account \citep{Flower2003a}, which may lead to sputtering of icy mantles and grain cores. It is possible to release the grain core elements (Mg, Fe, Si and O) into the
gas phase through sputtering. Sputtering yields are given in \citet{May2000}.
It is an important process both in $J$- and $C$-type shocks. In $J$-type shocks the
high kinetic temperature ensures that sputtering is an efficient process, while in
$C$-type shocks it is due to the velocity difference between neutral and charged
species.

\index{MHD shocks!artificial viscosity}
\index{Viscosity!artificial}
\item[An artificial viscosity] is added to the terms of momentum and energy in the case of $J$-shocks. This allows, rather than applying the Rankine-Hugoniot jump conditions at the shock discontinuity, to integrate and resolve the discontinuity

\end{description}

\subsubsection{Description of the grid of models}

I have run a grid of $C$- and $J$- shock models by varying the shock velocity,
preshock density, magnetic scaling factor, initial H$_2$ ortho-to-para ratio, and dust-to-gas mass ratio. 
This grid was  primarily designed to interpret the H$_2$ observations in the Stephan's Quintet galaxy collision, where we expect significant dust destruction  (chapters~\ref{chapter:H2_SQ} and \ref{chapter:H2_SQ_mapping}). Note that the output also includes atomic lines such as the intensity of the far-IR [O$\,${\sc i}]$\lambda \,63.2\,\mu$m line to be observed with the \textit{Herschel Space Telescope}.
The Table~\ref{table:grid-shock-parameters} lists the range of physical parameters of the grid of shock models.

\begin{table}
\begin{center}
\begin{minipage}{\textwidth}
 \renewcommand{\footnoterule}{}
\def\thefootnote{\alph{footnote}}
 \caption[Description of the grid of MHD shock models]{Description of the parameter space of the grid of MHD shock models}
\centering
\begin{tabular}{l l }
\hline
\hline
\multicolumn{1}{c}{parameter} & \multicolumn{1}{c}{values} \\
\hline
shock velocity & $V_{\rm s} = 3 - 50$~km~s$^{-1}$ (step: 1~km~s$^{-1}$) \\
preshock density & $n_{\rm H} = 10^2$, $10^3$, $10^4$, $10^5$~cm$^{-3}$ \\
\multirow{2}{*}{Magnetic scaling factor\footnotemark[1]} & $J$-shocks: $b=0$, 0.1 \\
&$C$-shocks: $b=1$  \\
Initial H$_2$ ortho/para ratio & $\rm o/p = \rm LTE \ value$\footnotemark[2] at 10~K, 0.01, 0.1, 3 \\ 
dust to gas mass ratio & $Z = 7.5 \times 10^{-3}$ (Galactic), $7.5 \times 10^{-4}$ \\
cosmic-ray ionization rate & $\zeta = 5 \times 10^{-17}$, $5 \times 10^{-16}$, $5 \times 10^{-15}$, $5 \times 10^{-14}$~s$^{-1}$ \\
\hline
\end{tabular}
\footnotetext[1]{The strength of the preshock magnetic field is given by Eq.~\ref{eq:magnetic-field-preshock}.}
\footnotetext[2]{The LTE value at low temperatures is given in chapter~\ref{chapter:H2Molecule}, Eq.~\ref{eq:H2_ortho-para_LTE_lowT}. In a molecular cloud at 10~K, $\rm o/p = 3.5 \times 10^{-7}$. }
\label{table:grid-shock-parameters}
\end{minipage}
\end{center}
\end{table}

The initial abundances of the 137 species are determined before any
shock model calculation, by running a chemical steady-state
model where we do not include adsorption on grains to avoid complete freezeout of molecules 
onto grains. The output abundances of this chemical steady-state model are
then used as input abundances in the shock models.

I do not give here a complete review of all the model results of the grid, but I  rather comment a few examples. The comparison with observations is presented in the following chapters.
For a description of an other extensive grid of models, the interested reader may consult the PhD thesis by \citet{Kristensen2007}.

\begin{table}
\begin{center}
\begin{minipage}{\textwidth}
 \renewcommand{\footnoterule}{}
\def\thefootnote{\alph{footnote}}
 \caption[Abundances of some of the refractory elements in the gas phase used in the shock code]{Initial repartition of the refractory elements in the gas phase and in the
grain mantles and cores used in the shock code\footnotemark[1]. }
\centering
\begin{tabular}{c c c c c}
\hline
\hline
element & fractional abundance\footnotemark[2] & gas phase & grain mantles & grain cores \\
\hline
C & $3.55 \times 10^{-4}$ & $8.27 \times 10^{-5}$ & $5.53 \times 10^{-5}$ & $ 2.17\times 10^{-4}$ \\
O & $ 4.42 \times 10^{-4}$ & $1.24 \times 10^{-4}$ & $1.78 \times 10^{-4}$ & $ 1.40\times 10^{-4}$ \\
Mg & $ 3.70\times 10^{-5}$ & 								& 									& $ 3.70\times 10^{-5}$ \\
Si & $ 3.37\times 10^{-5}$ & 								&									 & $3.37 \times 10^{-5}$ \\
Fe &$ 3.23\times 10^{-5}$ & $ 1.50 \times 10^{-8}$ & 							& $3.23 \times 10^{-5}$ \\
\hline
\end{tabular}
\footnotetext[1]{A more complete table can be found in Table~1 of  \citet{Flower2003a}.}
\footnotetext[2]{relative to Hydrogen: $n_{\rm X} / n_{\rm H}$ with $n_{\rm H} = n(\rm H) + 2 n(\rm H_2)$.}
\label{table:abundances-shock-models}
\end{minipage}
\end{center}
\end{table}

In the following illustrative results, I have used the initial elemental abundances of \citet{Flower2003a}, listed in Table~\ref{table:abundances-shock-models}. These abundances are those given by the output of the steady-state calculation. The initial repartition of the chemical species in the grain mantles can be found in Table~2 of \citet{Flower2003a}. The preshock medium is molecular gas at $n_{\rm H} = 10^4$~cm$^{-3}$. The cosmic-ray ionization rate is the standard value $\zeta = 5 \times 10^{-17}$~s$^{-1}$.

\subsection{Impact of the chemistry on the shock structure}
\index{Shocks!timescales and chemistry}

\begin{figure}
   \centering
    \includegraphics[width=0.495\textwidth]{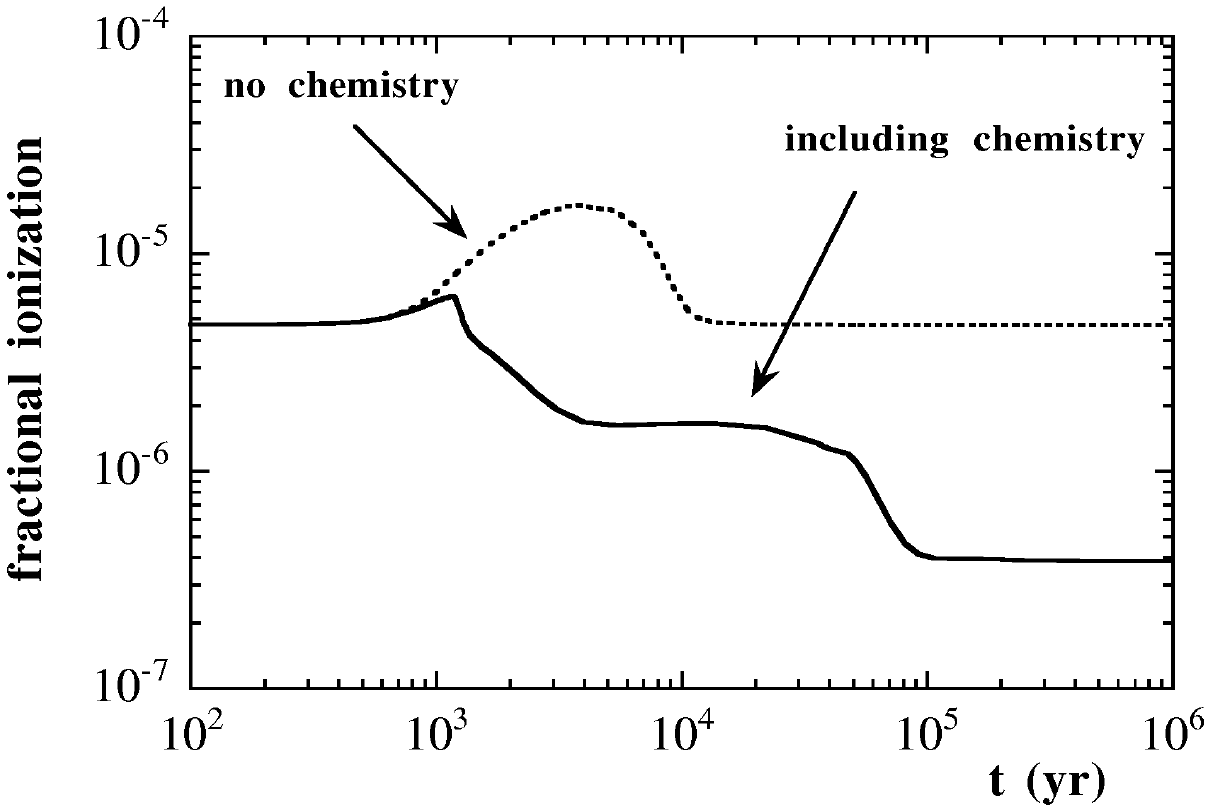}
 \includegraphics[width=0.495\textwidth]{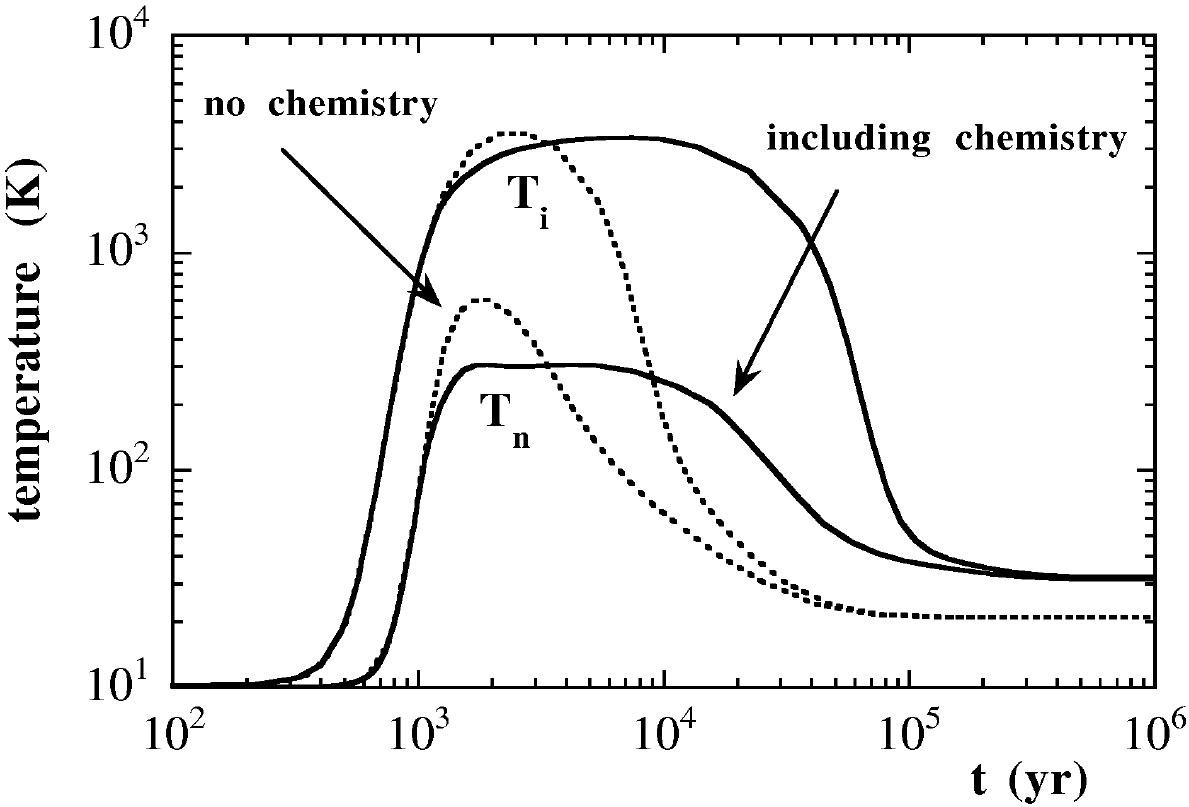}
      \caption[Impact of the chemistry on the ionization and temperature of a C-shock]{Impact of the chemistry on the  fractional ionization (\textit{left}) and temperature  (\textit{right}) of a $C$-shock. The shock velocity is 10~km~s$^{-1}$. The preshock medium is molecular, at a density $n_{\rm H} = 10^3$~cm$^{-3}$. The preshock magnetic strengh is $B_0=25\,\mu$G, perpendicular to the direction of propagation of the shock. $T_n$ and $T_i$ are the temperatures of the neutral and ionized fluids, respectively. The plots show model results with (\textit{solid line}) and without (\textit{dotted line}) chemical reactions. Taken from \citet{Pineau1997}. }
       \label{fig:impact-chemistry}
\end{figure}

\index{MHD shocks!and chemistry}
Fig.~\ref{fig:impact-chemistry} illustrates the impact of the chemical processes on the shock structure. Chemistry plays a role in the thermal balance of the shock, which in turn impacts the number of reactions the shock can initiate and the reaction rates. Fig.~\ref{fig:impact-chemistry} shows the difference of the
results obtained with or without taking the chemistry into account:
\begin{description}
\item[without chemistry,] the ionization fraction varies only through the differential compression
of the ionized and neutral fluids by the shock wave, and has the same value
in the preshock and postshock regions.

\item[with chemistry,] the ionization degree of the postshock in one order of magnitude lower. This is essentially due to the neutralization of C$^+$ into C, initiated by H$_2$:
\begin{equation}
\rm C^+ + H_2  \longrightarrow  \rm CH^+ + H 
\end{equation}
followed by a chain of reactions, which balance is
 \begin{equation}
\rm CH^+ + e^-  \longrightarrow  \rm C + H
\end{equation}
Consequently, the ion-neutral coupling is weaker, hence broadening the size of the precursor (the width of the shock is $\approx 5$ times larger). Since
the energy is dissipated on a larger region as compared to the first case, the maximum
temperature is lower (here approximately twice lower).
\end{description}
This shows that chemistry and hydrodynamics are coupled, and have to be treated in parallel.

\index{MHD shocks!and dust grains}
Dust grains, through gas-grain interactions, have also an important impact on the physical and chemical properties of the shock. They participate to the formation or removal of molecules of the gas, and their inertia can modify the propagation of the shock\footnote{The mass of the dust grains makes the bulk of the mass of the ions.}, especially in the
case of $C$-shocks. Some details about the nature and treatment of the grains in the model have been given in sect.~\ref{subsec:grid-models-Flower}. 
A detailed review of the processes involving grains taken into account in the code, as well as
their impact on the structure of $C$-shocks can be found in \citet{Flower2003a}. For detailed calculations of  the evolution of the dust size distribution  in MHD shocks, please consult the PhD thesis by \citet{Guillet2008}.  

\subsection{Energetics and molecular emission in shocks}
\label{subsec:shocks-molecular-cooling}
\index{Shocks!molecular cooling}

\index{MHD shocks!energetics}
In this section, a few results from my grid of models are presented. 

\subsubsection{Energy budgets}

\begin{figure}
   \centering
    \includegraphics[height=0.495\textwidth, angle=90]{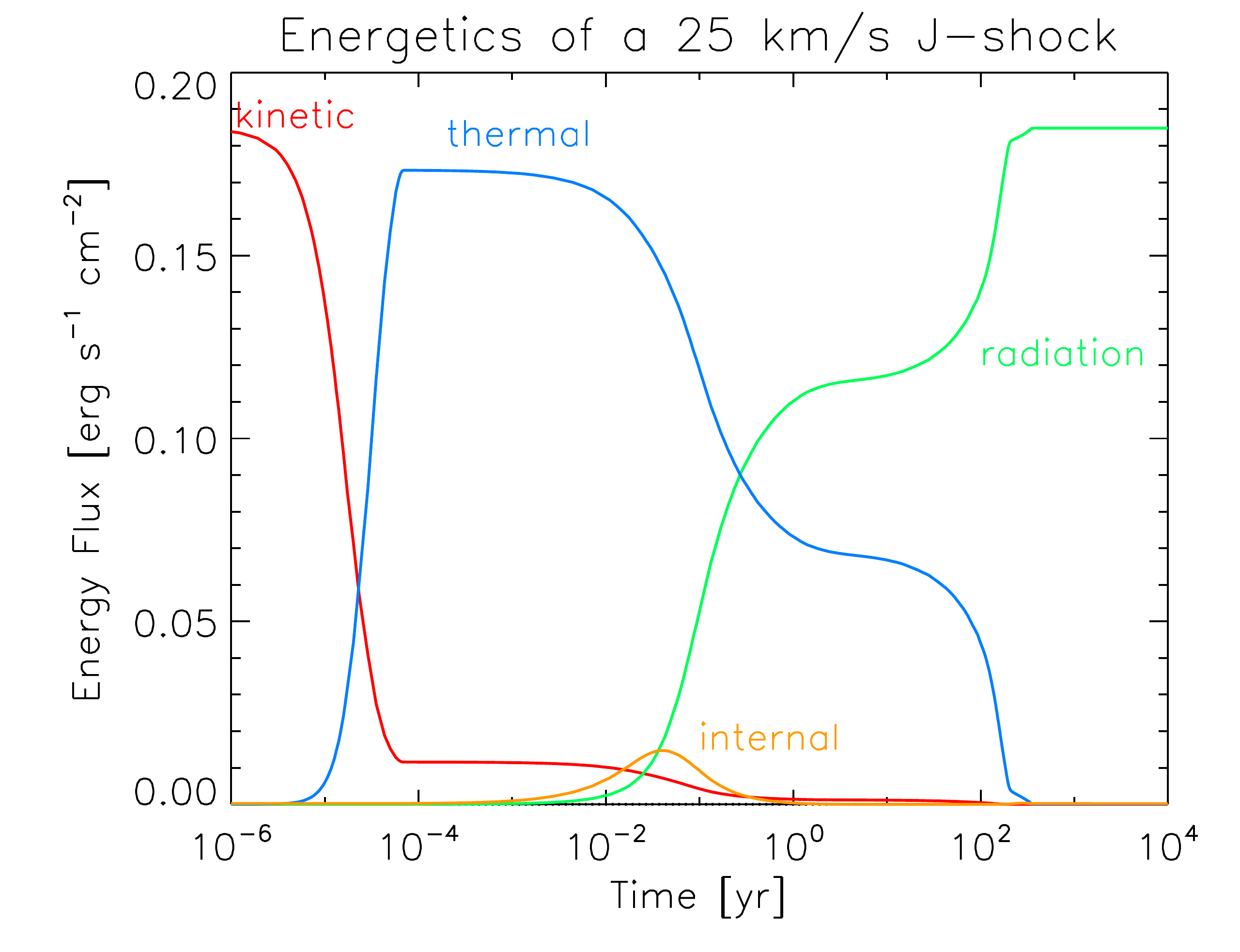}
    \includegraphics[height=0.495\textwidth, angle=90]{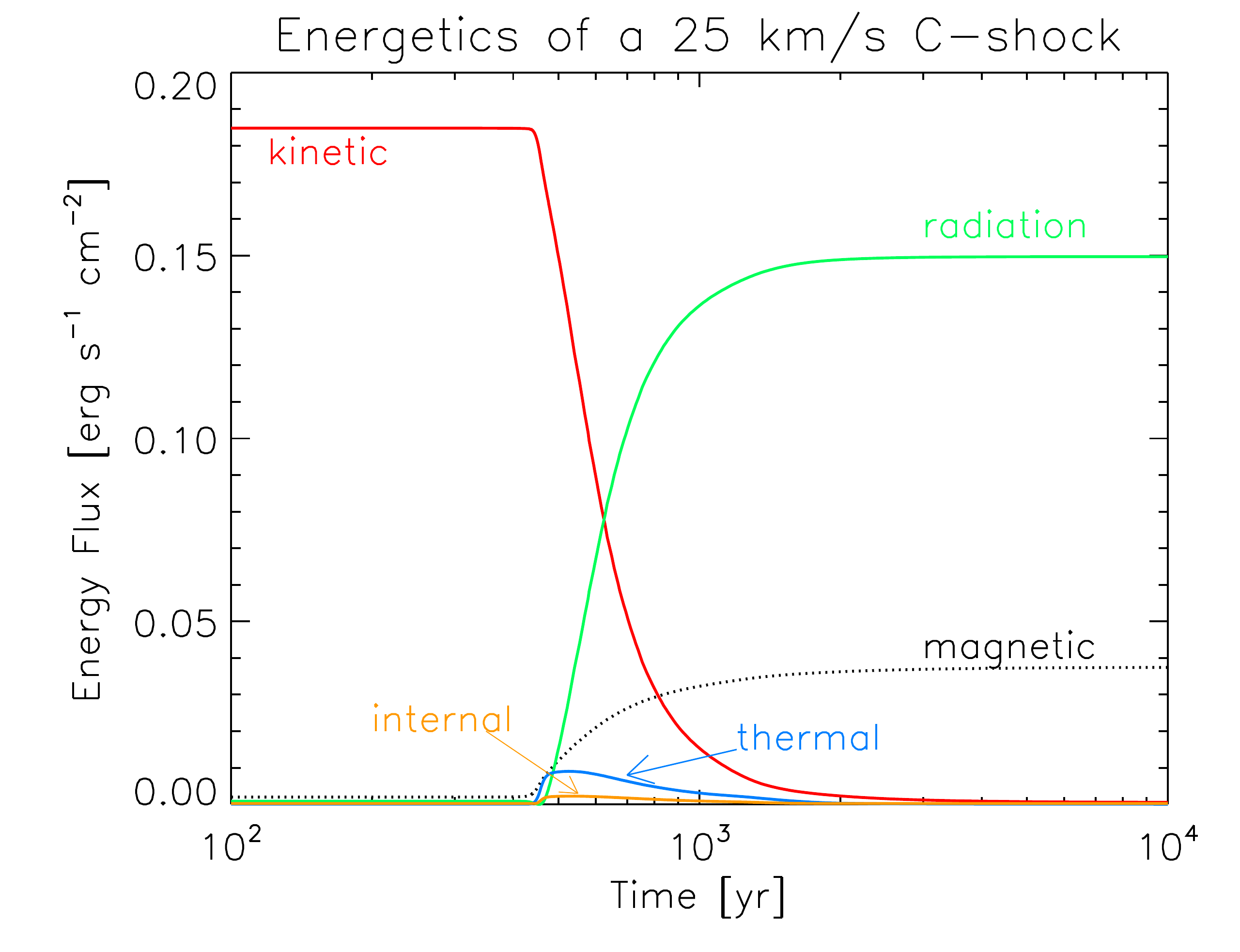}
      \caption[Energetics of 25 km~s$^{-1}$ $C$- and $J$-shocks]{Contributions of the kinetic, thermal, internal and magnetic energies as a function of the flow time for $J$- (\textit{left}) and $C$-shock (\textit{right}), both at $V_{\rm s}=25$~km~s$^{-1}$. 
The two shocks have the same total energy.
The preshock medium is molecular, at a density $n_{\rm H} = 10^4$~cm$^{-3}$. The preshock magnetic strengh is $B_0=100\,\mu$G for the $C$-shock and $B_0=0$ for the $J$-shock. The magnetic field is perpendicular to the direction of propagation of the shock.}
       \label{fig:energetics_C_25kms}
\end{figure}

Fig.~\ref{fig:energetics_C_25kms} compares the energy budgets for two types of shocks ($C$ and $J$) of the same energy. In the reference frame of the shock front, the energy of the preshock gas is pure bulk kinetic energy. 
In the hydrodynamic case ($J$-shock with $B_0=0$), when the gas crosses the shock front, most ($15/16$) of the bulk kinetic energy is converted into thermal energy. Then, at $t\approx 10^{-4}$~yr on the left panel of Fig.~\ref{fig:energetics_C_25kms}, the shock dissipates its energy in radiation form (collisional excitation of species in the shock, followed by radiative deexcitation).

In $C$-shocks, the thermal energy stays low as compared to other forms of energy. The reason is that the dissipation of energy is not instantaneous and occurs over a timescale comparable to the cooling time through radiation. 
The heating of the neutral gas is due to inelastic collisions with the charged fluid. 
Unlike hydrodynamic shocks, part of the postshock gas energy is stored as \textit{magnetic energy}. This magnetic energy can even dominate the postshock energy budget for very low-velocity shocks ($< 7$~km~s$^{-1}$ for $n_{\rm H} = 10^4$~cm$^{-3}$). The internal energy corresponds to the excitation of molecules, essentially H$_2$.

\begin{figure}
    \includegraphics[width=0.45\textwidth, angle=90]{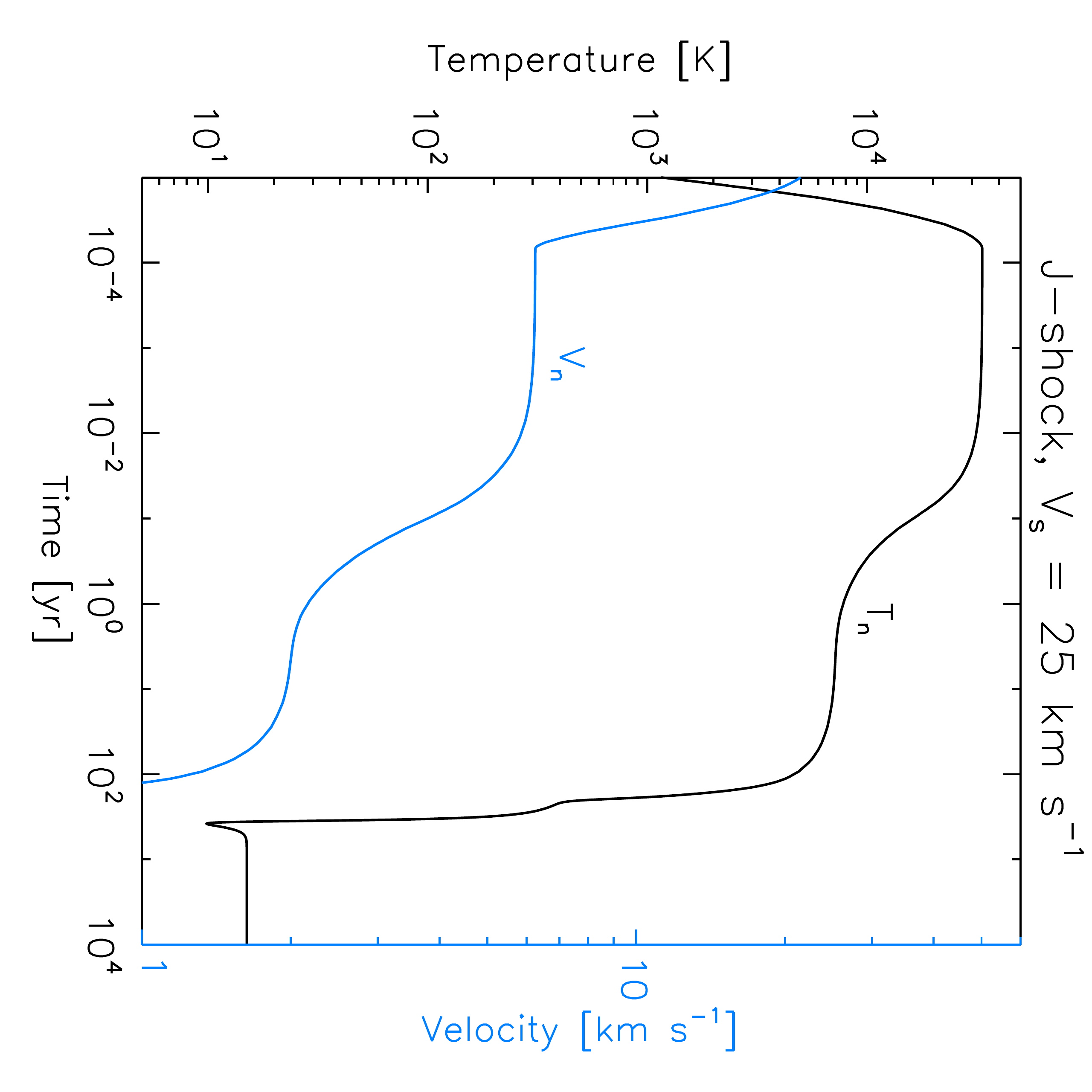}
    \includegraphics[width=0.45\textwidth, angle=90]{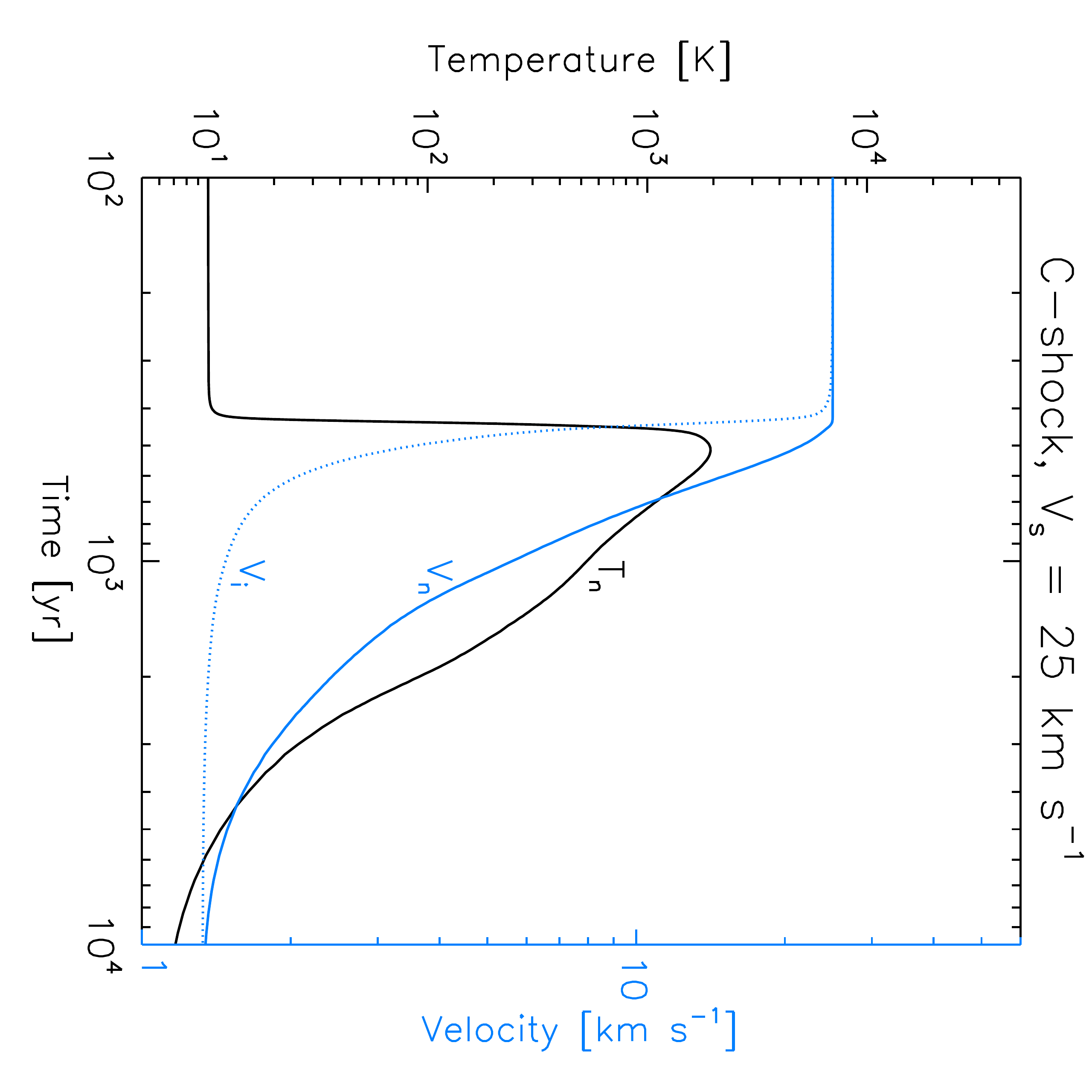}
    \includegraphics[width=0.45\textwidth, angle=90]{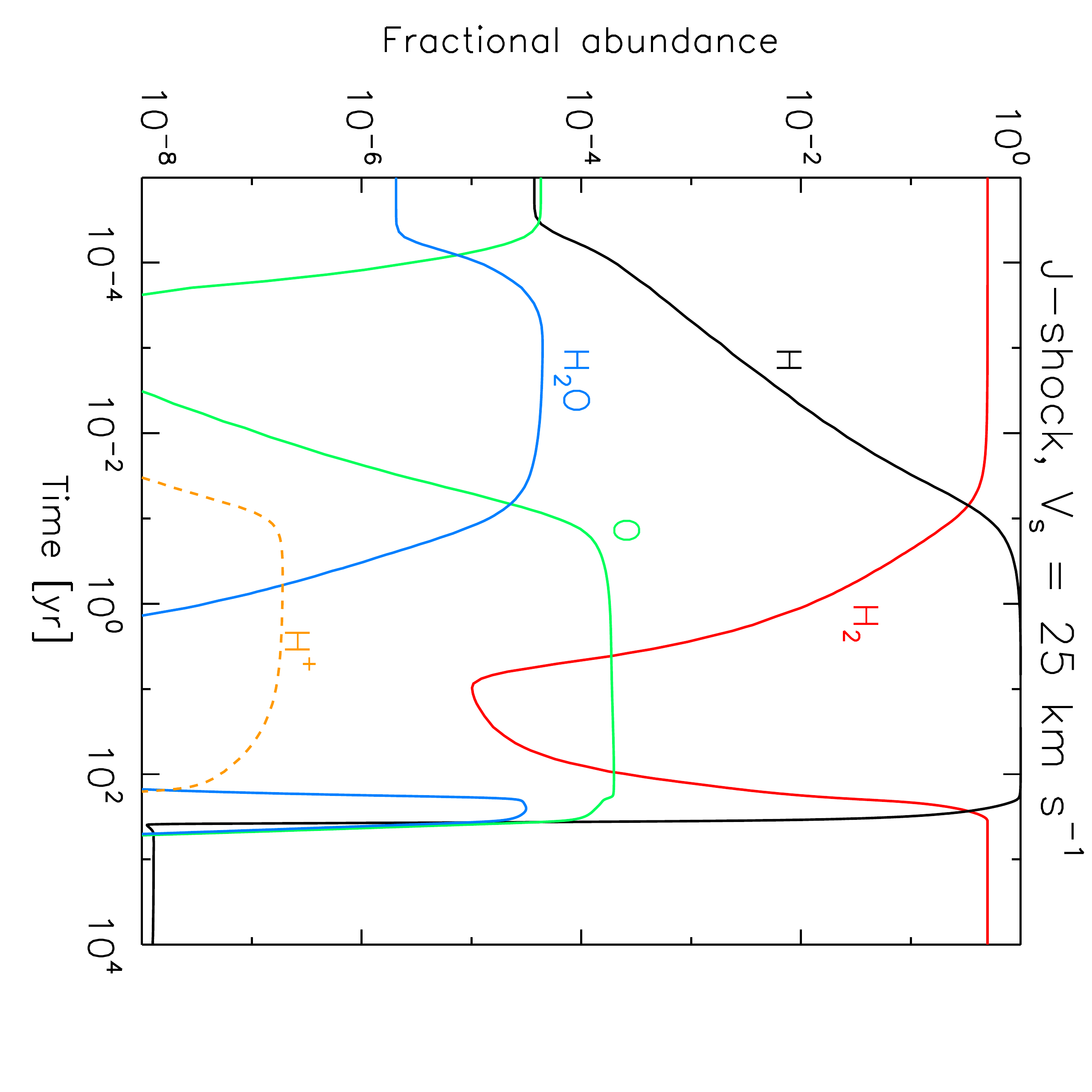}
    \includegraphics[width=0.45\textwidth, angle=90]{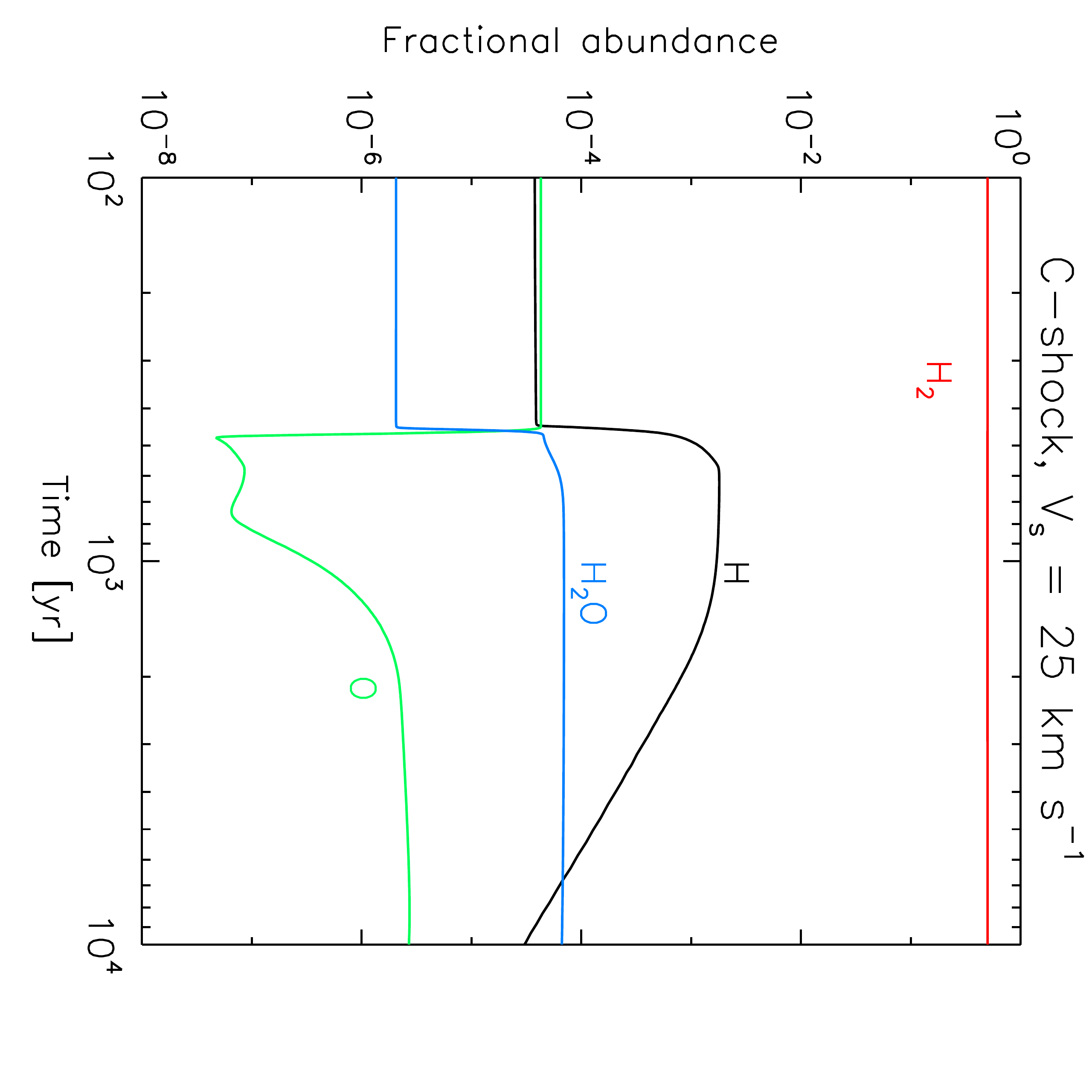}
    \includegraphics[width=0.45\textwidth, angle=90]{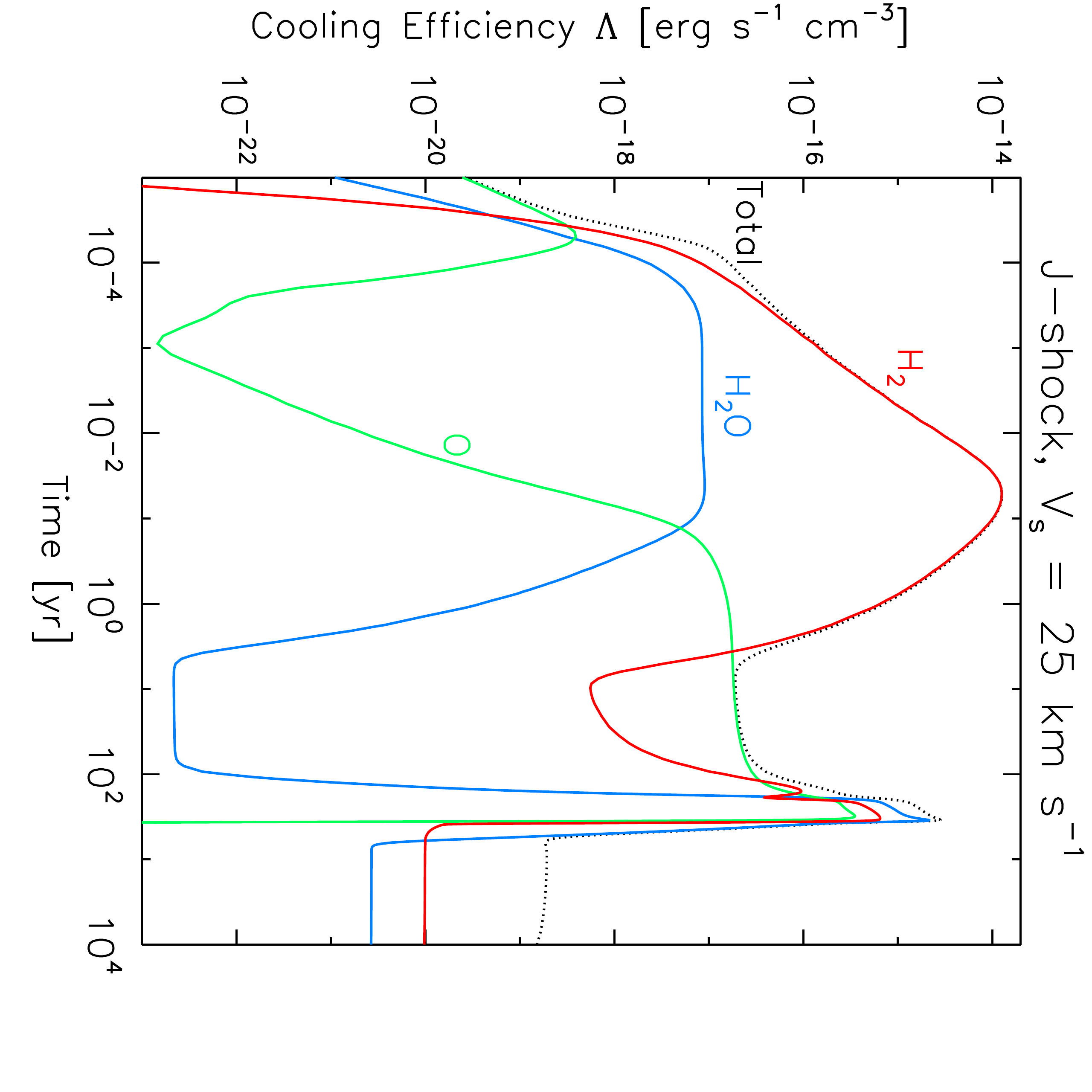} \quad \quad  \quad  \quad
    \includegraphics[width=0.45\textwidth, angle=90]{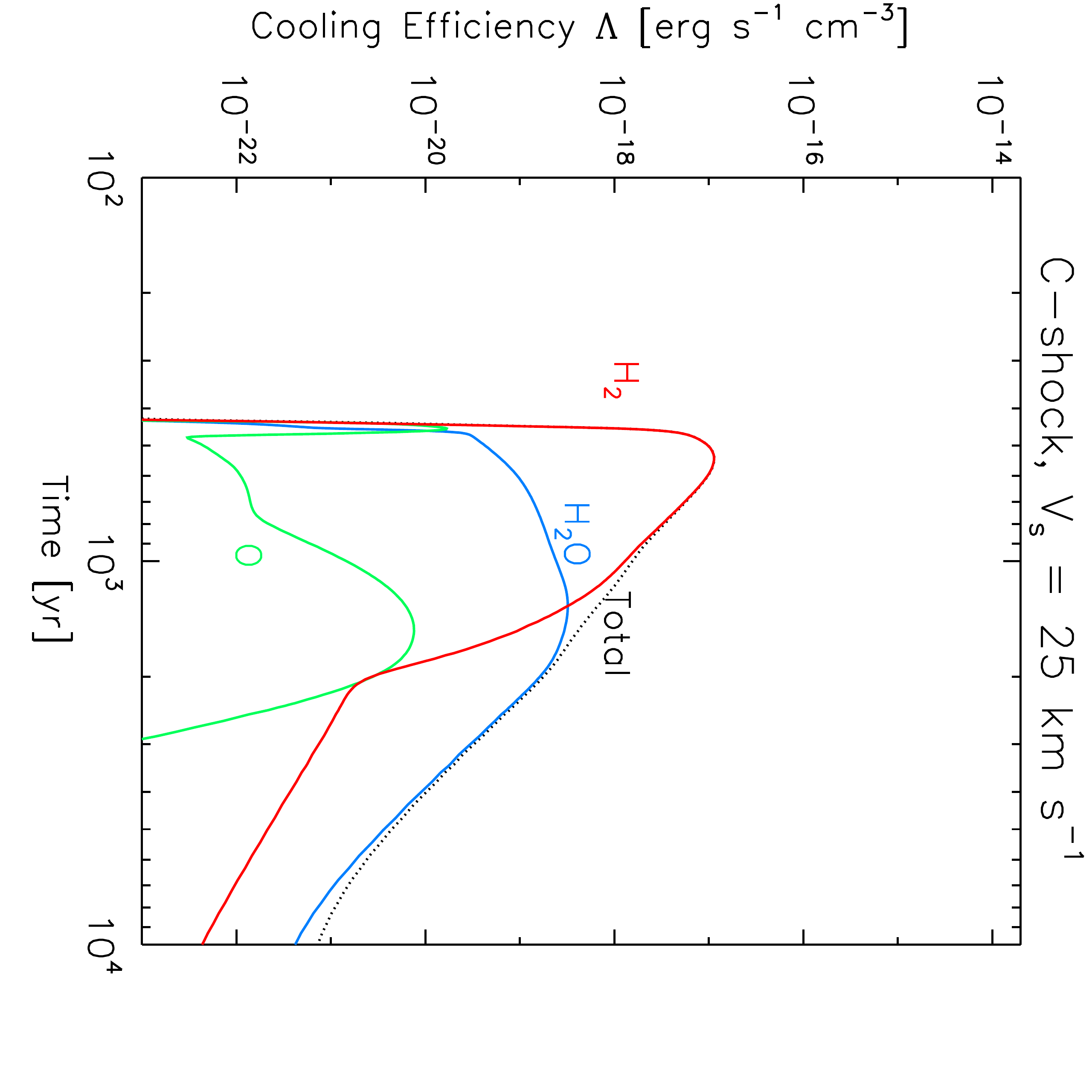}
      \caption[Overview of $C$ and $J$ shock profiles]{Comparison between $J$ (\textit{left}) and $C$ (\textit{right}) shock profiles. The shock velocity is 25~km~s$^{-1}$ and the preshock gas density $n_{\rm H} = 10^4$~cm$^{-3}$ for both shocks.  The preshock magnetic field is $B_0=0\,\mu$G for the $J$ shock and $B_0 = 100\,\mu$G for the C shock. Abundances are relative to H ($n(X) / n_{\rm H}$).  See text for details.}
       \label{fig:shocks_C-J-comparison}
\end{figure}

\subsubsection{Temperature, velocity and abundance profiles}
\index{MHD shocks!profiles}

Fig.~\ref{fig:shocks_C-J-comparison} compares two shocks of same energy, one $J$- and one $C$-shock. The profiles show some important physical quantities across the shock as a function of the flow time. The reference frame is the shock front. On the left colum we show $J$-shock profile with no magnetic field. On the right column, we show $C$-shock profiles with a preshock $B_0=100\,\mu$G. 
The models are the same than those used to make Fig.~\ref{fig:energetics_C_25kms}. In both cases, the shock velocity is $V_{\rm s} = 25$~km~s$^{-1}$ and the preshock gas density $n_{\rm H} = 10^4$~cm$^{-3}$. To facilitate the comparison, the $y$-axis are the same for both shocks. The $x$-axis has been extended to much shorter flow times for the $J$-shock, as the physical quantities start to vary much faster   than for $C$-shocks.

The top panel of Fig.~\ref{fig:shocks_C-J-comparison} shows the temperature of the neutrals and velocities (of the neutrals, $V_n$, and the charged species, $V_i$ in the case of the multi-fluids shock). As discussed in sect.~\ref{structure-C-J-shocks} (see also Fig.~\ref{fig:types_shocks_Draine1980}), the gas is slowed down abruptly when  crossing the shock front, whereas in $C$-shocks the charged fluid, and then the neutrals by friction, slow down progressively.

In $C$-shocks, the energy is dissipated over a longer timescale than in $J$-shocks, which implies that the maximum temperature attained in $C$-shocks ($\approx 2000$~K here) is much lower than in $J$-shocks ($\approx 30\,000$~K). 
For strong ($M_{\rm s} \gg 1$) shocks with $B = 0$, the postshock temperature is given by \citep[see e.g.][]{Draine1993}:
\begin{equation}
\label{eq:postshock-temperature-Vs}
T_{\rm ps} = \frac{2(\gamma - 1)}{(\gamma + 1)^{2}} \, \frac{\bar{\mu}}{k_{\rm B}} \, V_{\rm s}^{2} \simeq 2.6 \times 10^4 \, \left( \frac{V_{\rm s}}{25 \ \rm km \, s^{-1}}\right) ^{2} \ \ \rm K \; ,
\end{equation}
where $V_{\rm s}$ is the velocity of the shock wave, $\mu$ the mean molecular weight,  $k_B$ the Boltzmann constant and $\gamma$ the adiabatic index.
$\gamma = 7/5$ in a fully molecular medium in which the internal (rotational) degrees of freedom are thermalized. 
 In the second member of Eq.~\ref{eq:postshock-temperature-Vs}, we indicate typical values for shocks driven into molecular gas. Thus we assume $\bar{\mu} = 2.33$~a.m.u. and $\gamma = 1.44$, since we take into account the presence of atomic Helium\footnote{instead of $\gamma = 1.4$ for a pure H$_2$ gas.}, with a fractional abundance $n({\rm He}) / n_{\rm H} = 0.1$.  

The \textit{middle panel} of Fig.~\ref{fig:shocks_C-J-comparison} shows the evolution of the fractional abundances of a few important species (H$_2$, O, H$_2$O, H, H$^+$) in the shock. The $J$-shock is fast enough to dissociate H$_2$ (4.5 eV), but not enough to ionize significantly H (13.6 eV).
H$_2$ molecules reform over a short timescale in the postshock cooling gas, which releases energy in the gas, and induces a small ``knee'' at $\approx 300$~K on the temperature profile. 

Unlike $J$-shocks, $C$-shocks do not generate significant dissociation of molecular hydrogen, as shown on the profiles of the fractional abundances\footnote{The fractional abundances shown here are defined relative to hydrogen: $n(X) / n_{\rm H}$.}.  
The most active molecular chemistry  takes place within these shocks,
as the temperature jump is high enough to trigger endothermic chemical reactions without disrupting the involved molecules. Therefore I shall concentrate on $C$-shocks. 

\index{MHD shocks!and depletion}
Note that depletion of gaseous species onto dust grain surfaces in the postshock region reduces the abundances of some of the newly formed molecules. This affects the thermal balance of the shock. This is clearly seen for the cooling rates profiles of the $J$-shock, where cooling efficiency of all the molecules drops at about $t = 3 \times 10^3$~yr because of adsorption onto dust grains. 

Both kinds of shocks generate an intensive grain processing, that is still the subject of
studies \citep{Flower2003, Guillet2007}, and depletion onto dust grain surfaces in the postshock region. This depletion reduces the abundances of some of the newly formed molecules.

\subsubsection{Cooling functions and main coolants}

\begin{figure}
   \centering
 \includegraphics[height=0.495\textwidth, angle=90]{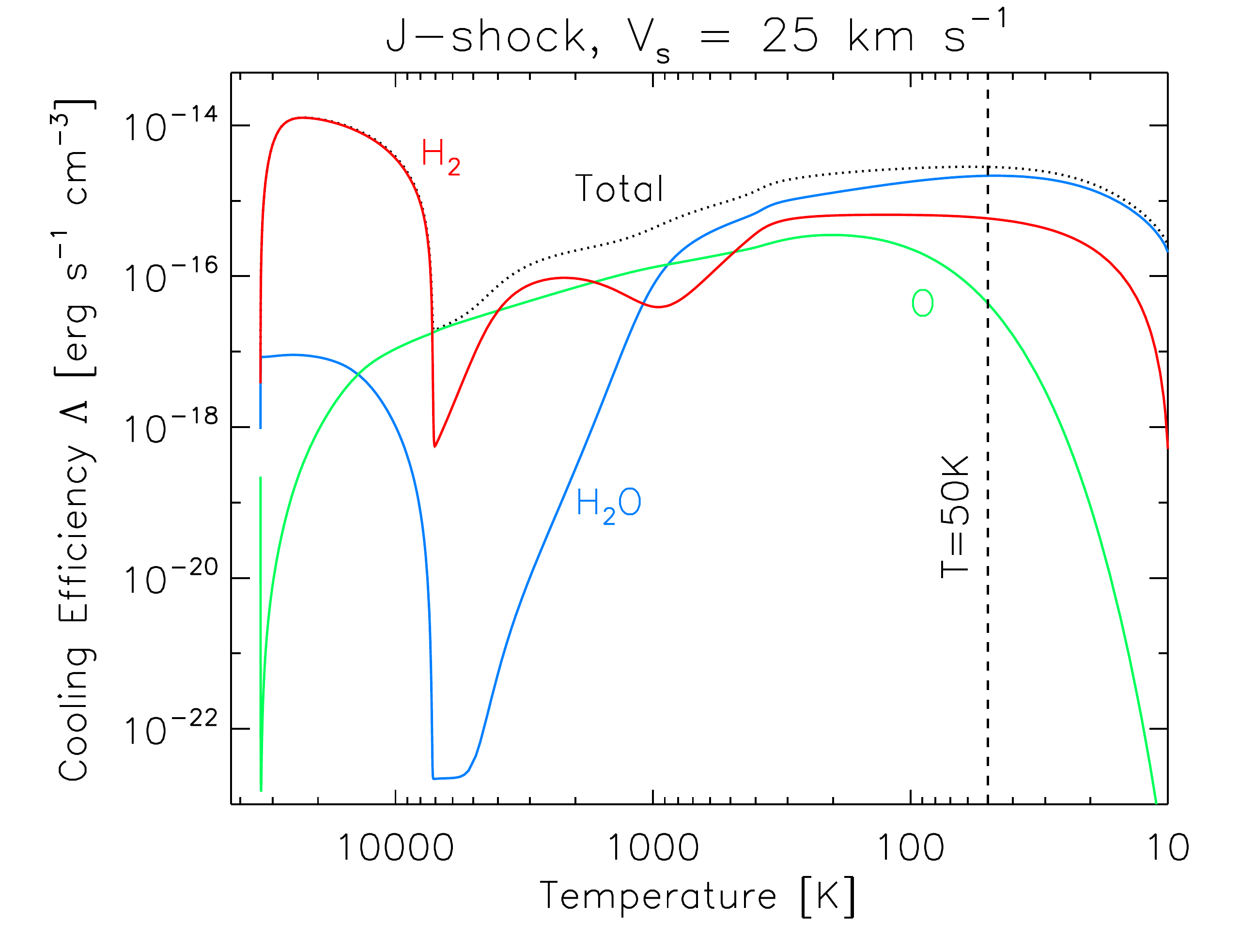}
    \includegraphics[height=0.495\textwidth, angle=90]{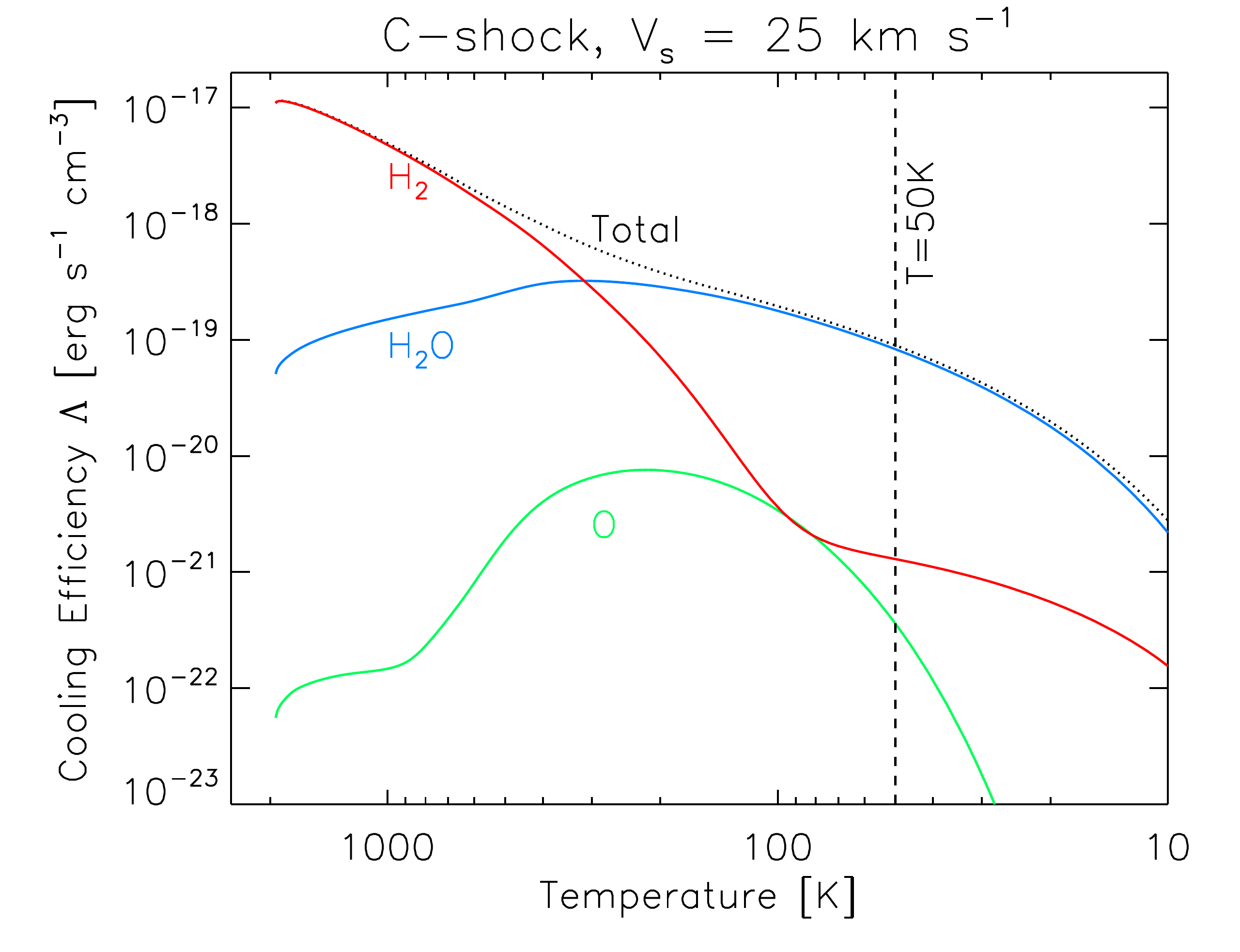}
      \caption[Cooling efficiencies for 25 km~s$^{-1}$ $J$- and $C$-shocks vs. T]{Contribution of the main coolants to the total cooling function as a function of the gas temperature for 25 km~s$^{-1}$ $J$- (\textit{left}) and $C$- (\textit{right}) shocks. The models are the same than in Fig.~\ref{fig:energetics_C_25kms} and \ref{fig:shocks_C-J-comparison}. The gas is cooling from the left to the right of the plots, from the maximum postshock temperature to 10~K. }
       \label{fig:cooling_Eff_vs_T_C_25kms}
\end{figure}

We now discuss the radiative properties of the shock. The \textit{bottom panel} of Fig.~\ref{fig:shocks_C-J-comparison} shows the local cooling rates of the main coolants relevant for our study, as a function of the flow time.
 Fig.~\ref{fig:cooling_Eff_vs_T_C_25kms} shows these cooling efficiencies as a function of the temperature of the cooling postshock gas.
More generally, the main coolants are O, H$_2$, H$_2$O, C$^+$, OH, CO. 
Their relative contributions to the cooling depend on the shock profile, and determine the radiative properties of the shock, which are very different in $J$ and $C$ shocks. 

\index{MHD shocks!main coolants}
For $J$-shocks, the cooling is first dominated by H$_2$ over very short period of time, at high temperatures where it experiences dissociation. Then, at $\approx 8000$~K, the [O$\,${\sc i}]$\lambda 6300\,$\AA~line takes over. When the gas has cooled down sufficiently, H$_2$, H$_2$O, O and C$^+$ dominate the cooling. More precisely, at $T \approx 1000$~K, the fine-structure line [O$\,${\sc i}]$\lambda 63.2\,\mu$m dominates. Other important lines in this temperature
 range are [C$\,${\sc i}], [N$\,${\sc i}]$\lambda 5200\,$\AA, [C$\,${\sc ii}]$\lambda 158\,\mu$m, [N$\,${\sc ii}]$\lambda 121.8\,\mu$m, [S$\,${\sc ii}]$\lambda 6731\,$\AA~and [Fe$\,${\sc ii}]. At lower temperatures, the oxygen is converted into CO, H$_2$O, and OH, which become the dominant coolants. 

For $C$-shock, H$_2$ is the main coolant for $T > 300$~K, which is the excitation range of the mid-IR rotational lines in which we are interested for our study. Remarkably, the cooling is dominated by H$_2$O emission at $T < 200$~K.

\begin{figure}
   \centering
 \includegraphics[height=0.495\textwidth, angle=90]{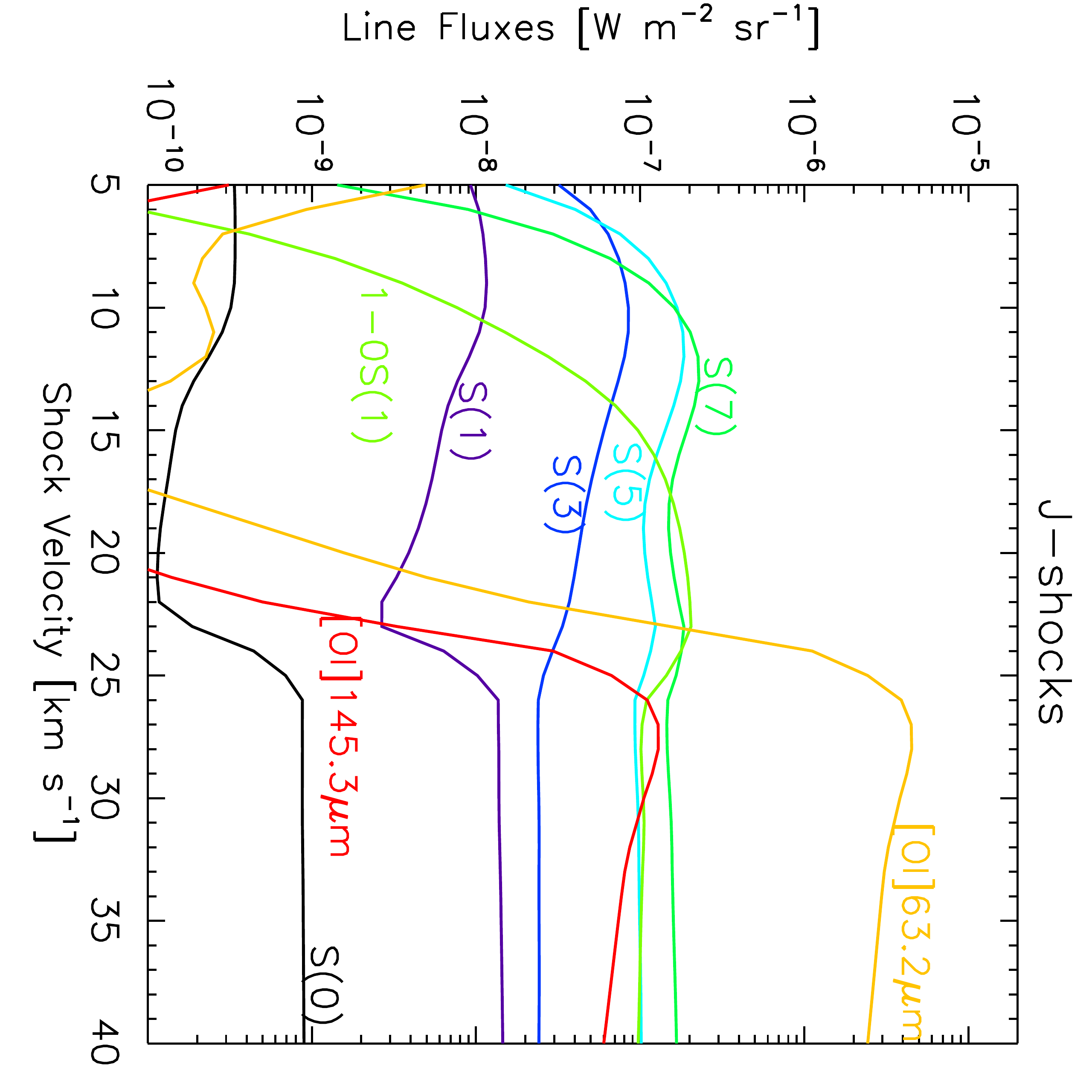}
    \includegraphics[height=0.495\textwidth, angle=90]{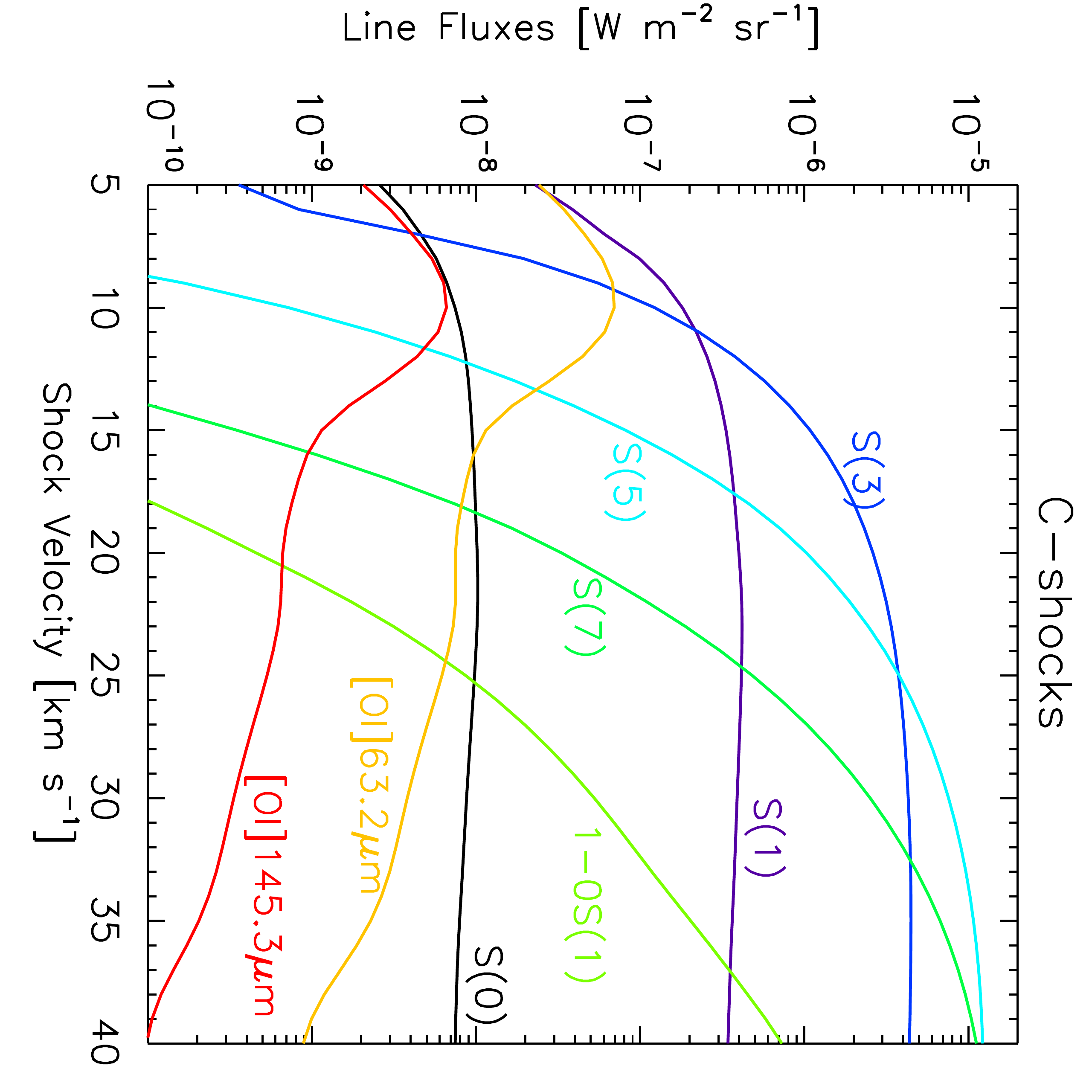}
      \caption[Line fluxes as a function of shock velocity]{Integrated line fluxes as a function of shock velocity for $J$- (\textit{left}) and $C$- (\textit{right}) transverse shocks. The models are the same than in Fig.~\ref{fig:energetics_C_25kms} and \ref{fig:shocks_C-J-comparison} (preshock density $n_{\rm H} = 10^4$~cm$^{-3}$). The integrated line fluxes are calculated after a cooling down to 50~K.}
       \label{fig:Line_Fluxes_shocks}
\end{figure}

\subsubsection{Line fluxes}

\index{MHD shocks!line fluxes}
The Fig.~\ref{fig:Line_Fluxes_shocks} shows the integrated (down to 50~K) line intensities as a function of shock velocity, for the two types of shock ($C$ and $J$). Note that in $J$-shocks, the higher H$_2$ rotational excitation levels, like the corresponding upper levels of the  S(5) and S(7) lines, are populated efficiently for low-velocity shocks. For $J$-shocks of velocities such that $25 < V_{\rm s} < 40$~km~s$^{-1}$, the integrated intensities are roughly constant with shock velocity, whereas for $C$-shocks, line fluxes increases smoothly with shock velocity.

Note that the [O$\,${\sc i}]$\lambda \, 63\,\mu$m and $145\,\mu$m lines are good tracers to distinguish between the two types of shock (see Fig.~\ref{fig:Line_Fluxes_shocks}). These lines are particularly sensitive to low velocity ($V_{\rm s}\lesssim 20$) shocks, where their intensities are much higher for MHD shocks than for pure hydrodynamical shocks ($B_0 = 0$).

\subsubsection{Examples of H$_{\bf 2}$ excitation diagrams for C- and J-type shock models}

\begin{figure}
   \centering
    \includegraphics[width=0.5\textwidth, angle=90]{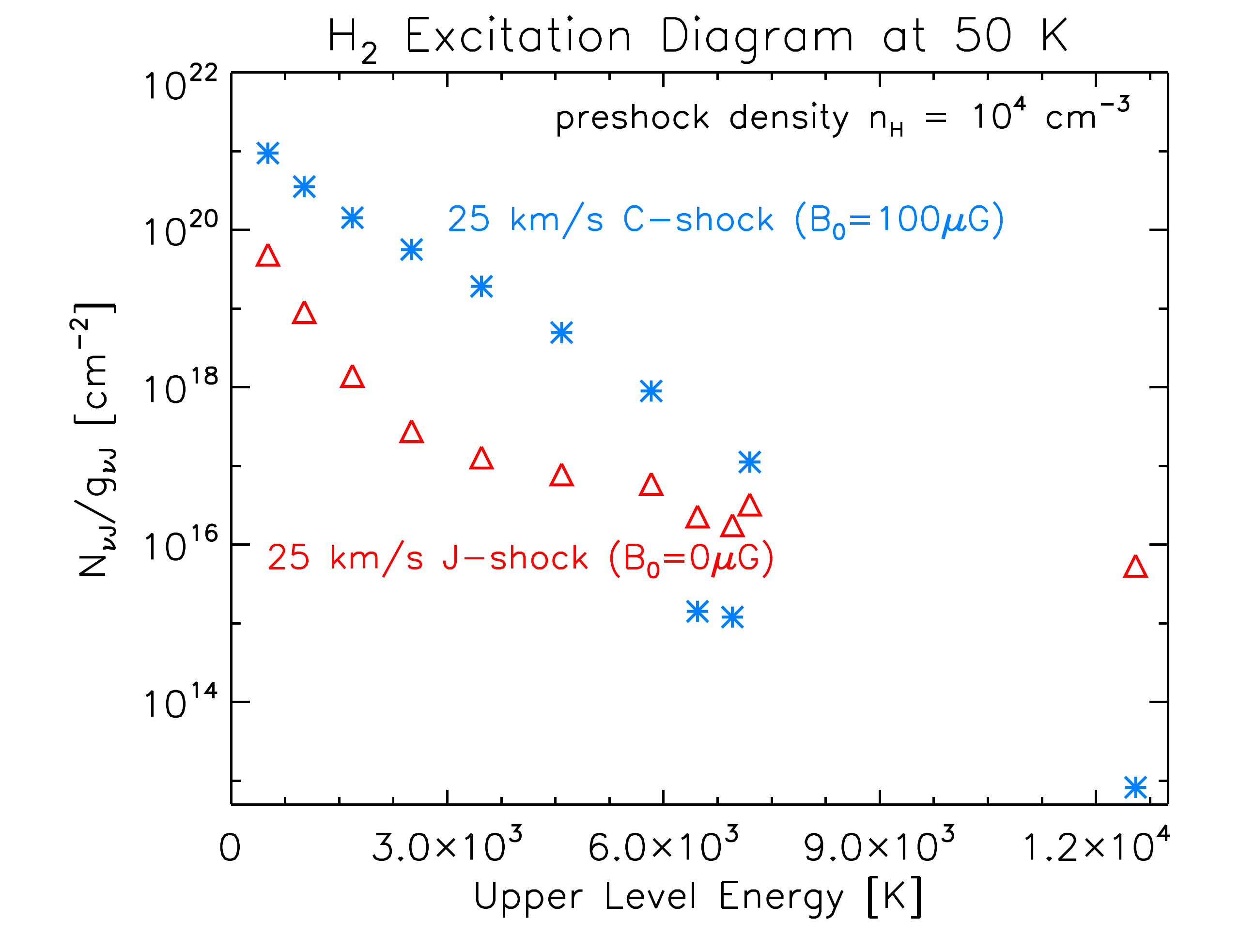}
      \caption[H$_2$ excitation diagram for 25 km~s$^{-1}$ $C$- and $J$-shocks]{H$_2$ excitation diagram for 25 km~s$^{-1}$ $C$- and $J$-shocks. The preshock medium is molecular, at a density $n_{\rm H} = 10^4$~cm$^{-3}$. The preshock magnetic strengh is $B_0=100\,\mu$G for the $C$-shock and $B_0=0\,\mu$G for the $J$-shock. The first 8 rotational levels of H$_2$, corresponding to the upper levels of the S(0) to S(7) transitions, plus three rovibrational levels, corresponding to the upper levels of the 1-0S(0), 1-0S(1),  and 2-1S(1) lines (these are the 3 blue points at the bottom of the plot for the $C$-shock, showing that pure rotational levels are not thermalized with rovibrational levels). }
       \label{fig:H2_ortho-para_C_25kms}
\end{figure}

\index{MHD shocks!excitation diagram}
\index{Excitation diagrams!and MHD shocks models}
The Fig.~\ref{fig:H2_ortho-para_C_25kms} shows theoretical H$_2$ diagrams for our two shock models (see chapter~\ref{chapter:H2Molecule}, sect.~\ref{subsec:H2excdiagrams} for the definition and calculation of the diagram). 
This diagram is computed at a temperature of 50~K in the postshock gas. If the gas temperature is below 50~K, almost no H$_2$ emission is produced. 
The integrated H$_2$ line intensities at 50~K are converted into column densities using Eq.~\ref{eq:column-density-excit-diagram}.
I only show on the plot the first 8 rotational levels relevant for our study, corresponding to the upper levels of the S(0) to S(7) transitions, plus three rovibrational levels, corresponding to the upper levels of the 1-0S(0), 1-0S(1),  and 2-1S(1) lines. The characteristics of these lines have been given in chapter~\ref{chapter:H2Molecule}, Table~\ref{table_H2lines}. 

This figure illustrates the impact of the type of shock on the excitation of H$_2$.
For the $C$-shock, the rotational lines are much brighter than for the $J$-shock, because the temperature is lower and the shock is broader, thus increasing the column density of emitting gas (see the bottom panel of Fig.~\ref{fig:shocks_C-J-comparison}). The high temperatures reached in the $J$-shock favor the ro-vibrational lines. Note that if the $J$-shock velocity further increases, the dissociation of H$_2$ will be so important that the ro-vibrational excitation will be much weaker. 

\textit{To conclude, MHD low-velocity shocks, driven into molecular gas,  are very efficient in channelling the kinetic energy of the shock into the rotational H$_2$ lines.} This characteristic is essential within the framework of my interpretation of H$_2$ emission in H$_2$-luminous sources.

I have used these diagrams to characterize the excitation characteristics in H$_2$-luminous objects, such as the Stephan's Quintet and the 3C326 radio galaxy. 
This will be discussed in details in chapters~\ref{chapter:H2_SQ}, \ref{chapter:H2_SQ_mapping} and  \ref{chapter:perspectives}.
\citet{Flower2003, Giannini2004, Giannini2006} provide examples of
celestial sources (proto-stellar outflows, Herbig-Haro objects) for which use was made of such excitation diagrams to describe both the  type of shock that is occuring and the physical parameters in these regions.

\subsubsection{Evolution of the H$_{\bf 2}$ ortho-to-para ratio in $C$-shocks}
\index{Ortho-to-para ratio!evolution in C-shocks}
\begin{figure}
   \centering
    \includegraphics[width=0.5\textwidth, angle=90]{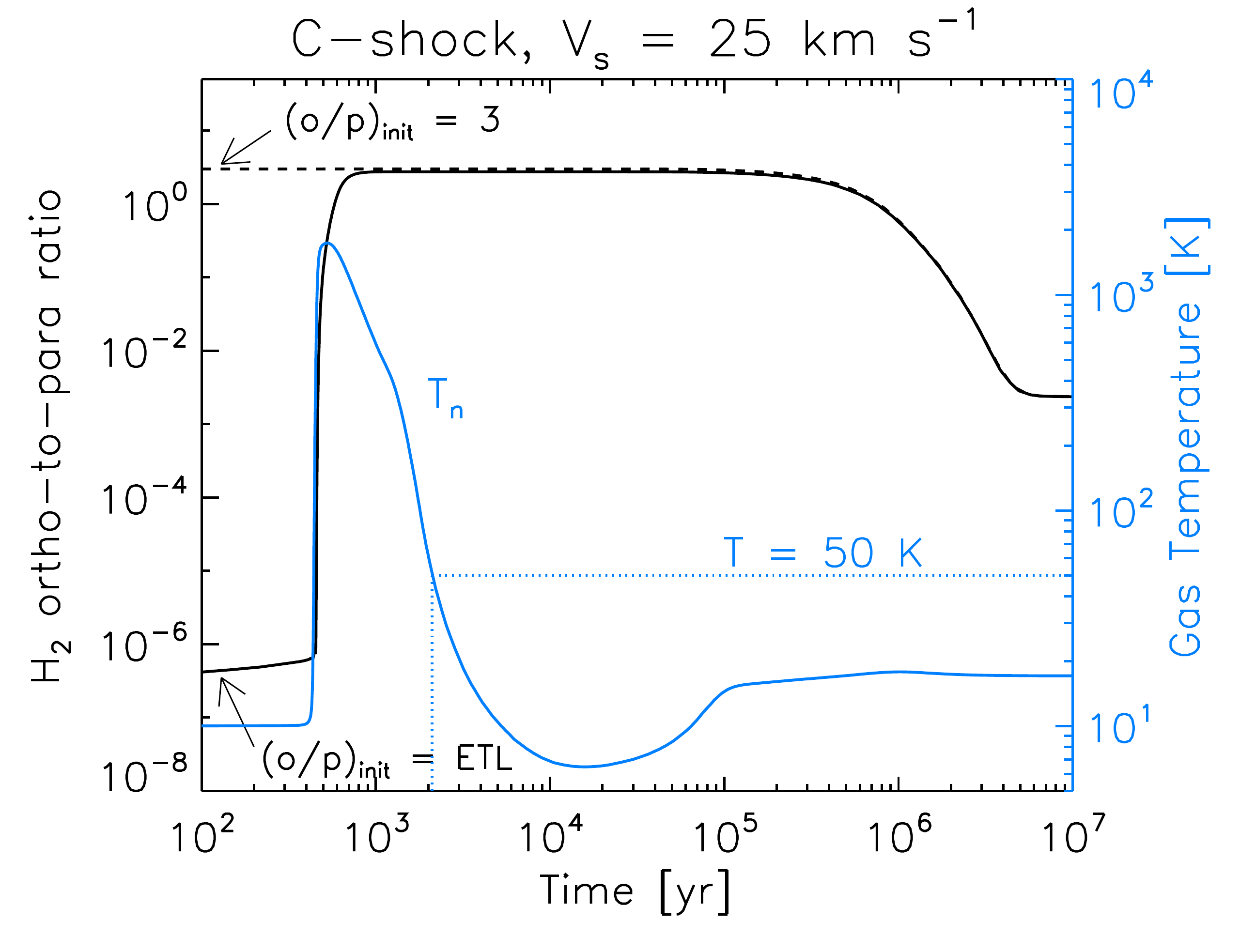}
      \caption[Evolution of the H$_2$ ortho-to-para for a 25 km~s$^{-1}$ $C$-shock]{Evolution of the H$_2$ ortho-to-para ratio and temperature (in blue, labeled on the right)  for a 25 km~s$^{-1}$  $C$-shock. Two initial values of the $\rm o/p$ ratio are shown: the Local Thermodynamical Equilibrium (LTE) value at 10~K ($3.5 \times 10^{-7}$), and the high-temperature limit (3). 
The preshock medium is molecular, at a density $n_{\rm H} = 10^4$~cm$^{-3}$. The preshock magnetic strengh is $B=100\,\mu$G, perpendicular to the direction of propagation of the shock.}
       \label{fig:H2_ortho-para_C_25kms}
\end{figure}

So far, the initial H$_2$ ortho-to-para  ratio is assumed equal to the statistical equilibrium value of 3 in our shock models. I also adopt this value in the following chapters because the  H$_2$ excitation diagrams observed in Stephan's Quintet or in the 3C326 radio-galaxy do not show any significant departure from this value, which would result in a systematic
displacement of the ortho with respect to the para levels (``zig-zag pattern''). 

However, we have shown in chapter~\ref{chapter:H2Molecule}, sect.~\ref{subsec:H2-ortho-para-ratio} that the LTE value of the $\rm o/p$ ratio is $3.5\times 10^{-7}$ at $T=10$~K, and 0.03 at $T=30$~K. To justify the high value of the initial $\rm o/p$ ratio, I briefly discuss the timescale of H$_2$ $\rm o/p$ conversion in shocks. This point has been discussed in more details by \citet{Wilgenbus2000} and \citet{Kristensen2007a}.

Fig.~\ref{fig:H2_ortho-para_C_25kms} illustrates the evolution of the  H$_2$ $\rm o/p$ ratio in a 25~km~s$^{-1}$  $C$-shock, for two initial values of the $\rm o/p$ ratio: $3.5 \times 10^{-7}$ (the LTE value at 10~K) and 3 (high-temperature limit). 
In the case of the low initial value, the interconversion between para- and ortho-H$_2$ starts when the temperature is rising, and, excepted within the first 10$^3$~years, the evolution of the $\rm o/p$ ratio is similar for both cases. The timescale of thermalization of the H$_2$ $\rm o/p$ ratio is long, $\approx 4 \times 10^6$~yr in this case, which is in good agreement with Eq.~\ref{eq:timescale-ortho-to-para-conversion}. 

As a result, if the molecular gas experiences multiple shocks, and if the time interval between two shocks is smaller than the  H$_2$ $\rm o/p$ conversion timescale, the $\rm o/p$ will not have time to thermalize to the low equilibrium value in the cold H$_2$. This may be the case in the   H$_2$-luminous objects I have studied. Note that there are objects that favor a low initial $\rm o/p$ value, like proto-stellar bipolar outflows for instance \citep{Wilgenbus2000}.

The initial value of the H$_2$ ortho-to-para ratio has an impact on the H$_2$ line brightnesses. Generally speaking, the lower the initial $\rm o/p$ ratio, the lower the brightness is. This impact is a timescale effect, and is more important in $C$-shocks than in $J$-shocks. 
The fact that the ortho-H$_2$ levels cannot begin to be populated until the temperature reaches $\approx 800$~K causes a delay. This delay is responsible for the lower brightness in transitions between ortho-levels at low initial $\rm o/p$ ratio. Inversely, the brightness from transitions between para-levels is higher for low initial $\rm o/p$ ratio.

\section{Fast shock models}
\label{sec:fast-shock-models}

So far, we have mostly discussed low-velocity shocks driven into dense molecular gas, that power molecular emission. These shocks are well traced through rotational line emission of H$_2$. 
Obviously, much faster shocks are likely to occur in the violent environments of H$_2$-luminous galaxies. Within the framework of understanding their multi-wavelength observations, I have been driven to use an other shock model that describes \textit{fast shocks with radiative precursor}. This section briefly discuss the properties of these shocks, very different from the shocks discussed above. 

\subsection{Structure of a fast shock with radiative precursor}

\begin{figure}
   \centering
    \includegraphics[width=0.6\textwidth]{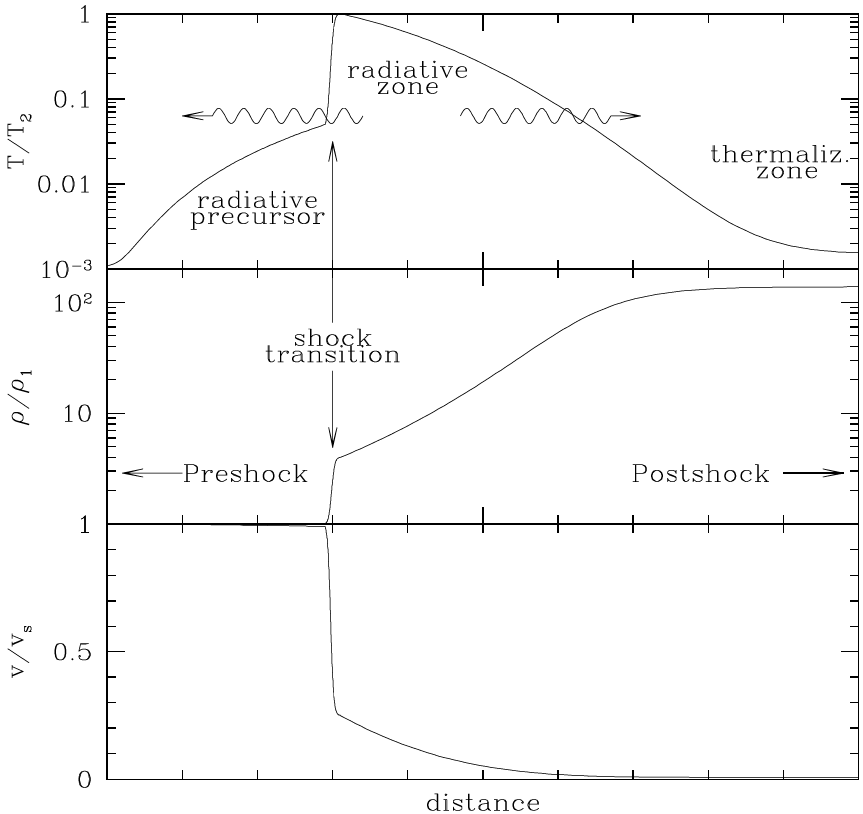}
      \caption[Stationary profile of a shock with radiative precursor.]{Stationary profile of a shock with radiative precursor, in the reference frame of the shock front. In this reference frame, the preshock gas comes from the left at a high speed, and is slowed down in the shock front (\textit{bottom}), compressed (\textit{middle}) and heated (\textit{top}). The radiative emission of the postshock gas  produces a \textit{radiative precursor} that heats and ionize the gas ahead of the shock. Taken from \citet{Draine1993}.}
       \label{fig:shock_structure_radiative_precursor}
\end{figure}

Fig.~\ref{fig:shock_structure_radiative_precursor} schematically illustrates the structure of the fast shocks we now consider. In fast shocks, the shocked medium emits UV of soft X-ray radiation that propagates in all directions, including that of the preshock gas. This UV emission produces a \textit{radiative precursor} that heats, dissociates (for $V_{\rm s} \gtrsim 90$~km~s$^{-1}$) and ionizes (for $V_{\rm s} \gtrsim 120$~km~s$^{-1}$) the preshock gas before it is shocked \citep[see][]{Hollenbach1989}.

\subsection{The MAPPINGS shock model library}

In my PhD work, I have made use of the shock model library named \textit{MAPPINGS III}, described in \citet{Allen2008}, which is an updated version of the \textit{MAPPINGS II} code described in \citet{Sutherland1993}. I do not give a complete review of this model. Here I only present a short description of the basic properties and underlying physics included or not included in the model. The theory of fast, photoionizing shocks is described in \citet{Sutherland1993, Dopita1995, Dopita1996}. I also direct the reader to the complementary book by \citet{Dopita2003}. 

\subsubsection{Modeling technique}

Shocks with radiative precursor are difficult to solve numerically because 
\begin{enumerate}
\item the UV radiation inside the shock imposes to carry on radiative transfer calculations to compute the ionization state of the gas, as well as the excitation state of the species in the shock
\item the shock has a \textit{feedback} effect on the preshock gas in which it propagates. Therefore, one cannot integrate the shock structure (meaning integrate the MHD and chemical rate reactions) starting from any point in the preshock. 
\end{enumerate}
 
The MAPPINGS 1-D code couples an MHD shock code to a radiative transfer photoionizing code. At each time step of the integration of the dynamical conservation equations in the flow, the rate equations for non-equilibrium ionization, recombination, excitation, and radiative transfer and cooling are solved at each time step of the flow.
A total of four iterations (shock integration followed by calculation of the precursor) are performed to allow the temperature and ionization state of the precursor gas to stabilize at a constant value.

\subsubsection{The physics included in the code}

The cooling and radiative emission is calculated using a very
large atomic data set (16 atoms) which allows treatment of all ionization stages (225 ions), up to fully ionized nickel.

\begin{itemize}
\item \textit{The atomic processes affecting the ionization balance} of the plasma taken into account in the code are photoionization, ionization by electron collision, charge transfer reactions, and various forms of recombination channels, including radiative and dielectronic recombination,    

\item \textit{Radiative processes} includes fine-structure, inter-system, and forbidden emission, as well as resonance transitions and continuum calculations. 
Fine-structure, inter-system and resonance lines are approximated by a two-level system. The model assumes that the resonance lines are primarily excited by collisions. 
The continuum calculations includes free-free, free-bound and two-photon processes. 

\item \textit{Additional heating and cooling processes}, including recombination and Compton heating, are integrated over the electron thermal distribution or the photon spectral distribution. 

\end{itemize}

Physically, the ionizing radiation produced in the cooling zone behind the shock front shock is mostly composed of thermal bremsstrahlung (free-free) continuum, resonance
lines, bound-free continuum of hydrogen, and the strong
hydrogen two-photon continuum produced mostly by the
down-conversion of Ly$\,\alpha$ photons trapped in the recombination region
of the shock structure. Recombination lines and bound-free continuum from Helium and heavier elements are also contrinution to the ionizing radiation of the precursor.

The hydrogen ionizing radiation flux, $\mathcal{L}_{\rm UV}$, integrated
for all energies $h\nu > 13.6$~eV, and over $2\pi$~sr follows the following scaling relation \citep{Allen2008}:
\begin{equation}
\mathcal{L}_{\rm UV} = 2.44 \times 10^{-7} \left(\frac{V_{\rm s}}{100\,\rm km\,s^{-1}} \right)^{3.02} \left(\frac{n_{\rm H}}{1\,\rm cm^{-3}} \right)  \ \rm [W~m^{-2}] \ \,
\end{equation}
which is almost scaled as $n_{\rm H} V_{\rm s}^3$.

\subsubsection{What is not included: shortcomings}

\begin{itemize}

\item \textit{Dust grains} are ignored in these models. The shocks propagate in a dust-free gas. However, we have seen in sect.~\ref{sec:MHD-shock-models-molecular-gas} that grains have an important impact on the structure, and thereby on the emission, of MHD shocks. The main difference is that in fast ($> 200$~km~s$^{-1}$ shocks, the dust destruction is expected to be very efficient. As shown in chapter~\ref{chapter:dust_gas_galaxies}, sect.~\ref{sec:time-dependent-cooling-plasma}, if the postshock temperature is hotter than $\approx 10^6$~K, dust grains may affect the \textit{initial} stages of the cooling of the postshock gas, and thus the structure of the shock. However, their cooling efficiency will drop rapidly because of thermal sputtering.

\item \textit{Molecules} are not included although the computation of the shock structure is allowed to proceed until the gas has cooled to 1000~K. Therefore, the cooling functions are valid for $T \gtrsim 10^4$~K. 

\end{itemize}

The shock model library is available online\footnote{\url{http://cdsweb.u-strasbg.fr/~allen/mappings_page1.html}}. In my PhD work, I have used this library to produce diagnostics of mid-IR fine-structure line ratios in order to contrain the physical properties of the ionized gas in the Stephan's Quintet. Examples of line ratio diagnostics and comparison to observations are discussed in chapter~\ref{chapter:H2_SQ_mapping}.

I have also used these shock models to obtain a synthetic ionizing spectrum  to model the emission from dust that would be heated by this radiation. This modeling has been performed within the context of the Stephan's Quintet, and this is discussed in chapter~\ref{chapter:SQ_dust}.

\section{Shocks propagating into an inhomogeneous medium}
\label{sec:shocksinmultiphasemedia}
\index{Shocks!inhomogeneous medium}

So far, we have considered the ideal case of a shock that propagates in an homogeneous medium. However, as introduced in chapter~\ref{chapter:dust_gas_galaxies}, the interstellar or intergalactic media have a  multiphase and a self-similar structure. 
The interaction of shock waves with density inhomogeneities (I will call them ``clouds'') in the ISM is thought to be an important dynamical process in a multiphase medium \citep[e.g.][]{McKee1977}. As discussed in chapter~\ref{chapter:dust_gas_galaxies}, it contributes to mass exchange between the dense and diffuse phases. On one hand, it may trigger gravitational
collapse and star formation \citep[e.g.][]{Elmegreen2004}. On the other hand, shocks may have a net destructive effects on molecular clouds, suppressing star formation.


Chapter~\ref{chapter:H2_SQ} and \citet{Guillard2009} (\hyperref[paper_SQ_H2]{paper~{\sc i}}) discuss the  formation and excitation of H$_2$ in the context of the Stephan's Quintet galaxy collision, where observations indicate that these processes may be the response of a multiphase medium to a galactic-scale shock. 
It would be impractical to give an exhaustive review of the abundant litterature on such a vast topic. I will rather focus on  two situations that are relevant for the context of H$_2$-galaxies, but also in many other astrophysical environments: the evolution of a cloud hit by a strong shock wave (sect.~\ref{evolution-shocked-molecular-cloud}), and the supersonic collision between two gas streams of different densities and velocities (sect.~\ref{subsec:collision-two-gas-flows}). 
Before describing these two situations and the associated physical processes, I shall recall a few properties of a shock wave that encounters a discontinuity of density.

\subsection{Reflexion and transmission of a shock through a discontinuity}
\label{reflexion-transmission-shock-discontinuity}

\begin{figure}
   \centering
    \includegraphics[width=0.8\textwidth]{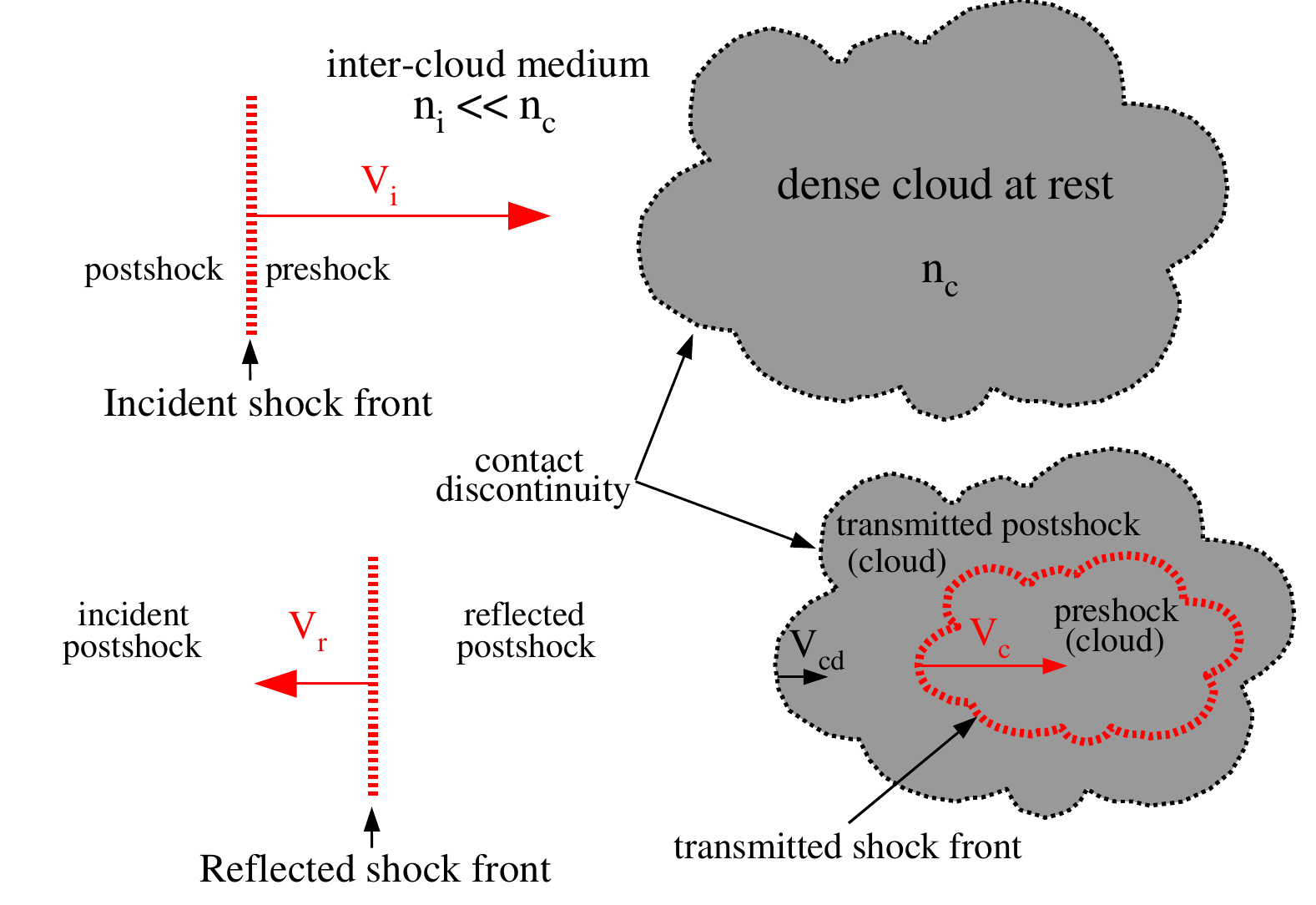}
      \caption[Reflexion and transmission of a shock through a cloud]{Reflexion and transmission of a shock through a discontinuity of density (cloud of density $n_{\rm H} = n_{\rm c}$). The incident shock wave propagates in the tenuous inter-cloud medium ($n_{\rm i} \ll n_{\rm c}$). The incident shock is transmitted into the cloud and reflected  into the inter-cloud medium because of the abrupt density contrast. The initial contact discontinuity remains and is being accelerated at a velocity $V_{\rm cd}$ in the direction of the propagation of the incident shock wave. The postshock media of the reflected and transmitted shocks are in pressure equilibrium and propagate at the same velocity $V_{\rm cd}$.}
       \label{fig:shock_on_cloud}
\end{figure}

When the propagation medium in  which the shock wave is going through changes, the shock wave may enter
\begin{enumerate}
\item \textit{a denser medium:} a reflected shock wave is driven back into the postshock flow of the incident wave. A shock wave is transmitted into the dense medium, and the initial discontinuity is accelerated. This case is schematically presented in Fig.~\ref{fig:shock_on_cloud}. The reflected and transmitted  postshock gas, separated by the contact discontinuity, are in pressure equilibrium.
\item   \textit{a less dense medium, or a medium that moves in the same direction of the shock:} a rarefaction wave is reflected into the postshock gas of the incident shock, and a transmitted shock is produced. The contact discontinuity is also accelerated. 
\end{enumerate}

In the following I shall consider the case (1), illustrated in Fig.~\ref{fig:shock_on_cloud}. I take the suffixes $_{\rm i}$ and $_{\rm c}$ to refer to the inter-cloud and cloud media. If the age of the incident shock is much shorter than its cooling timescale (\textit{adiabatic} shock), \citet{McKee1975} estimate that the velocity of the transmitted shock into the cloud, $V_{\rm c}$,  is of the order of \footnote{This approximate relation is obtained from the conservation of the kinetic pressure $\rho V^2$. More detailed expressions are given in \cite{Klein1994}} 
\begin{equation}
\label{eq:velocity-transmitted-shock-adiabatic}
V_{\rm c} \simeq \sqrt{\frac{n_{\rm i}}{n _{\rm c}}} \,  V_{\rm i} = \frac{V_{\rm i}}{\sqrt{\chi}}
\end{equation}
where $V_{\rm i}$ is the shock velocity in the inter-cloud medium, and $n_{\rm c}$ and $n _{\rm i}$ are the densities of the cloud and intercloud medium, respectively. The ratio between the densities in the two media is $\displaystyle \chi = \frac{n _{\rm c}}{ n_{\rm i}} > 1$.

Note that if the shock is \textit{isothermal}, i.e. if the shock has cooled down through radiation significantly to reach the preshock temperature, and if the incident shock is fast enough (such that $M_{\rm i} \gg \sqrt{\chi}$), the shock is not slowed down when penetrating into the cloud  \citep{Miesch1994}:
\begin{equation}
V_{\rm c}  \simeq V_{\rm i} \quad \mbox{if} \quad M_{\rm i} = \frac{V_{\rm i}}{c_{\rm i}} \gg \sqrt{\chi} \ ,
\end{equation}
where $M_{\rm i}$ is the Mach number of the inter-cloud shock and $c_{\rm i}$ the sound speed of the inter-cloud medium. This puzzling result can be understood as follows. If the incident shock is isothermal, the compression factor is much higher ($M_{\rm i}^2$) than for the adiabatic case (equal to 4). 
If $M_{\rm i}^2 \gg \chi$, the  incident postshock gas (the piston) is denser than the cloud\footnote{The density of the piston is $M_{\rm i} \rho _i \gg \chi \rho _i = \rho _c$. This is not the case for the adiabatic case if $\chi > 4$, where the density of the piston is lower than the shocked cloud ($4 \rho _i < \chi \rho _i$). }. 
Therefore, the ram pressure $\rho V^2$ of the incident postshock gas, which is much higher for the isothermal  than for the adiabatic case, drives the transmitted shock into the cloud at a higher velocity than the adiabatic shock. 


\textit{To conclude}, a shock wave that hits a medium that is denser than its propagation medium, is slowed down at the boundary of the dense medium (adiabatic phase) and then travels into the denser inner regions of the cloud at a quasi-constant velocity (isothermal phase).

\subsection{Evolution of a shocked molecular cloud}
\label{evolution-shocked-molecular-cloud}

I will now introduce the relevant timescales for the description of the evolution of a cloud that is run over by a fast shock wave, illustrated in Fig.~\ref{fig:shock_on_cloud}. 
How does the shock wave propagate inside and outside the cloud? What happens to the cloud? This section first adresses these questions, both from an analytical and numerical point of view.

This problem  has been extensively studied by means of numerical simulations, which is the only quantitative way to study the development of instabilities that are produced when the flow of background gas establishes around the cloud and interacts with it.
The 2D dynamical evolution of such a cloud has been described in many papers, and I direct the interested reader to \citet[e.g.][]{Klein1994, MacLow1994, Poludnenko2002}. 

It is important to realize that the evolution of a shocked cloud does not only depend only on the dynamics, but also on its thermal and radiative properties. 
Thermal and radiative properties are just being explored in multi-dimensional numerical simulations, and have been largely neglected in the past. I will briefly discuss these recent numerical results.

\begin{figure}
\begin{minipage}{\textwidth}
    \def\thefootnote{\alph{footnote}}
      \setlength{\footnotesep}{0pt}
   \centering
    \includegraphics[width=\textwidth]{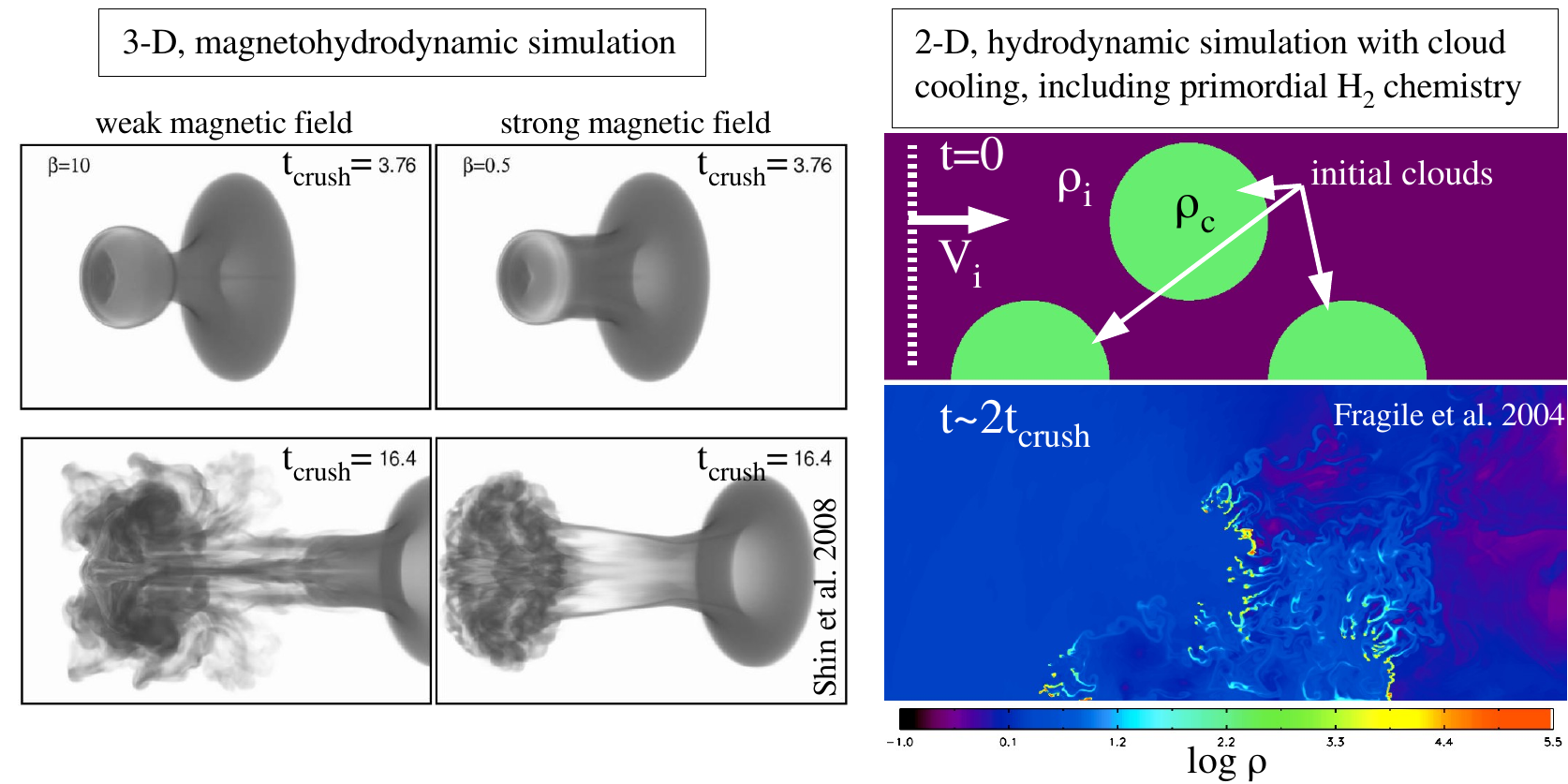}
      \caption[Numerical simulations of shocked clouds]{Numerical simulations of shocked clouds. \textit{Left:} 3-D MHD simulation of a non-cooling cloud \citep{Shin2008}. The plots show volumetric renderings of the cloud density for a weak magnetic field value\footnotemark[1] ($\beta=10$, left column) and strong-field ($\beta=0.5$, right column) parallel shock simulations. The Mach number of the shock is $M_{\rm s}=10$. The cloud boundary is not a discontinuity but a smooth density gradient. The density contrast between the cloud and the inter-cloud medium is $\chi = \rho _c / \rho _i = 10 $. \textit{Right:} 2-D hydrodynamical simulation ($B=0$) of three cooling clouds hitted by a shock wave \citep{Fragile2004}. The above plot shows the initial conditions. The dimensions of the computational grid are $900 \times 400$~pc. The initial cloud radii are 100~pc and their temperature $10^4$~K. The initial cloud density is $\rho _c = 1.7 \times 10^{-24}$~g~cm$^{-3}$ ($n_{\rm c} = 1$~cm$^{-3}$), with a density contrast of $\chi = 10^3$. The bottom plot shows contours of the logarithm of gas density, in units of $6.7 \times  10^{-27}$~g~cm$^{-3}$ on the scale. The shock velocity in the inter-cloud medium is $V_{\rm s} = 7.4 \times 10^3$~km~s$^{-1}$, for a Mach number of $M_{\rm s}=20$.}
       \label{fig:shock-cloud-simu-Shin-Fragile}
\footnotetext[1]{$\beta$ is the ratio of the gas to magnetic pressures.}
\end{minipage}
\end{figure}

I shall  consider here the case of a spherical, cooling cloud, of radius $R_{\rm c}$ and mass density $\rho_{\rm c} = m_{\rm H} n_{\rm c}$. The cloud is initially in pressure equilibrium with the inter-cloud medium (background gas). The density contrast between the two media is $\chi = \rho _c / \rho _i > 1$. 
The evolution of the shocked cloud consists of three phases:
\begin{enumerate}
\item Initially, the shock runs over the cloud. The time scale for this is the
\textit{shock passing time}, 
\begin{eqnarray}
t_{\rm sp} &=& \frac{2 R_{\rm c}}{V_{\rm i}} \\
& \simeq & 2 \times 10^5  \left( \frac{R_{\rm c}}{10\,\rm pc}\right) \left( \frac{V_{\rm i}}{100\,\rm km\,s^{-1}}\right) ^{-1}  \ \ \rm [yr]
\end{eqnarray}
where $V_{\rm i}$ is the velocity of the passing shock in the inter-cloud medium.

\item

The second phase is the compression phase, in which the
cloud finds itself inside the high pressure cocoon. It is now underpressured
compared to its environment. Shock waves are transmitted into the cloud, and 
start to travel into it from all sides. The timescale for the cloud compression is named the \textit{cloud crushing time} and is simply given by
\begin{eqnarray}
t_{\rm crush} &=& R_{\rm c} / V_{\rm c} \simeq \sqrt{\chi} \frac{R_{\rm c}}{V_{\rm i}} \\
& = & 3.1 \times 10^6 \left( \frac{\chi}{10^3} \right)^{1/2} \left( \frac{R_{\rm c}}{10\,\rm pc}\right) \left( \frac{V_{\rm i}}{100\,\rm km\,s^{-1}}\right) ^{-1} \ \ \rm [yr]
\end{eqnarray}
 where $V_{\rm c}$ is the velocity of the shock travelling into the cloud, which can be calculated from Eq.~\ref{eq:velocity-transmitted-shock-adiabatic} (see sect.~\ref{reflexion-transmission-shock-discontinuity}). 

\item The third phase starts when the transmitted shocks have passed over the cloud. During this phase the cloud is subject to both \textit{Rayleigh-Taylor} and \textit{Kelvin-Helmholtz} instabilities. Meanwhile, during phases 2 and 3, the radiative cooling of the cloud impacts the evolution of the gas, which may also become \textit{thermally unstable}. These processes lead to a fragmentation of the shocked cloud, and the fragments may end up mixed with the inter-cloud gas. Taking a first, qualitative look at Fig.~\ref{fig:shock-cloud-simu-Shin-Fragile}, one can clearly see the effect of fragmentation on the shocked cloud.

\end{enumerate}

In the following I discuss the key physical processes introduced above that control the evolution of the shocked cloud and its fragments.

\index{Shocks!hydrodynamical instabilities}
\index{Instabilities!Rayleigh-Taylor}
\paragraph*{Cloud acceleration and growth of Rayleigh-Taylor instabilities}
The growth of the Rayleigh-Taylor instability\footnote{The Rayleigh-Taylor instability (after Lord Rayleigh and G. I. Taylor), is an instability of an interface between two fluids of different densities, which occurs when the lighter fluid is pushing the heavier fluid. As the instability develops, downward-moving irregularities (``dimples'') are quickly magnified into sets of inter-penetrating Rayleigh-Taylor ``fingers''. The upward-moving, lighter material is shaped like mushroom caps. This phenomenon is clearly visible in the Crab and Helix nebulae.} is related to
the acceleration of the cloud by the postshock background gas.
The acceleration timescale is the amount of time it takes to
accelerate the cloud gas to the velocity of the postshock background.
Let us write the momentum transfer from an inter-cloud gas flowing by a spherical cloud (of cross-sectional
area $\pi R_{\rm c}^2$ ) at a relative  velocity $V_{\rm ps, i}$.
The postshock force acting on the cloud can be written 
\begin{eqnarray}
F & = & \pi R_{\rm c}^2 \rho _i \, V_{\rm ps, i}^2 \\
  & =&  \pi R_{\rm c}^2 \rho _i  \frac{3}{4} V_{\rm i}^2  \ ,
\end{eqnarray}
where I have assumed the jump condition $\rho _{ps, i} = 4 \rho _i$ with $\gamma = 5/3$. 
Then the \textit{acceleration timescale} is given roughly by $t_{\rm acc}  = V_{\rm i} / (F/M_{\rm c})$, $M_{\rm c}$ being the mass of the cloud. We find that 
\begin{eqnarray}
\label{eq:cloud-acceleration-timescale}
t_{\rm acc} & =& \frac{4}{3} \,  \frac{\rho _{\rm c}}{\rho _i} \,   \frac{R_{\rm c}}{ V_{\rm i}} =  \frac{4}{3}  \,   \sqrt{\chi} \,   t_{\rm crush} \\
& \simeq & 1.3 \times 10^8 \left( \frac{\chi}{10^3} \right) \left( \frac{R_{\rm c}}{10\,\rm pc}\right) \left( \frac{V_{\rm i}}{100\,\rm km\,s^{-1}}\right) ^{-1} \ \ \rm [yr]
\end{eqnarray}
The above equation show that it is very difficult to couple dynamically dense gas with a background flow of tenuous gas. The crushing time to acceleration time ratio is $t_{\rm crush} / t_{\rm acc}  = 3 / 4 \sqrt{\chi}$, which shows that as soon as the density contrast is higher than $\approx 10^2$, the cloud will be compressed on a timescale much shorter than the acceleration time. 
The \textit{growth timescale of Rayleigh-Taylor instabilities} is given by $t_{\rm RT} = (a \, k)^{-1/2}$, where $a$ is the acceleration of the cloud and $k = 1 / \lambda$ is the wavenumber of the perturbation \citep{Chandrasekhar1961}. In our case, $a \sim V_{\rm i} / t_{\rm acc}$, so we obtain
\begin{equation}
t_{\rm RT} =\frac{ t_{\rm crush}}{\sqrt{k \, R_{\rm c}}} \approx  t_{\rm crush} \ .
\end{equation}
Perturbation wavelengths such that $k \, R_{\rm c} \sim 1$ are the most disruptive\footnote{The growth rate of the RT instabilities is highest for the shortest wavelengths $(\lambda \ll  R_{\rm c})$, but the amplitude of these perturbations ($\sim \lambda$) saturates rapidly.}, so we conclude that Rayleigh-Taylor instabilities develop on a timescale comparable to the crushing timescale, and more than one order of magnitude  shorter than the acceleration timescale for $\chi \gtrsim 100$. 

\paragraph{Growth timescale of Kelvin-Helmholtz instabilities}
\index{Instabilities!Kelvin-Helmholtz}
A similar analysis can be applied to derive the growth timescale of the Kelvin-Helmholtz instabilities\footnote{The Kelvin-Helmholtz instability occurs when velocity shear is present within a continuous fluid or, when there is sufficient velocity difference across the interface between two fluids.}, $t_{\rm KH}$. For a high density contrast ($\chi \gg 1$), $t_{\rm KH} = \sqrt{\chi} / k v_{\rm rel}$, where $v_{\rm rel}$ is the relative velocity between the postshock background and the cloud  \citep{Chandrasekhar1961}. Assuming that $\displaystyle v_{\rm rel} \approx  V_{\rm ps, i} = \frac{3}{4} V_{\rm i}$, we obtain:
\begin{equation}
t_{\rm KH} \simeq \frac{ t_{\rm crush}}{\sqrt{k \, R_{\rm c}}} \approx  t_{\rm crush} \ .
\end{equation}
As Rayleigh-Taylor instabilities, Kelvin-Helmholtz instabilities growth on a timescale comparable to the cloud crushing time. 

\textit{To sum up}, if we assume an adiabatic evolution of the gas, the timescales calculated above suggest that the cloud is destroyed over a timescale comparable to the compression timescale, or crushing time. Destruction is enhanced by the fact that  the 
shocks travelling into the cloud meet and interact, or reach the opposite boundary of the cloud. This produces a rarefaction wave travelling through the shocked cloud material. 
 The cloud, which was compressed by the hot cocoon, now starts expanding again. 
This  increases the pressure contrast between the compressed cloud and the
postshock external gas, effectively accelerating the growth of
destructive instabilities.

\paragraph{Impact of the magnetic field on the evolution of the shocked cloud}

Because of the coupling between the magnetic field and the cloud material, the magnetic field may have strong effects on the cloud evolution and on the development of instabilities. This effect is subtle, because it depends on the local orientation of the magnetic field in the inter-cloud medium.

If the background medium is magnetized, then magnetic
field lines can become trapped in deformations on the surface
of the cloud. As these field lines are stretched, the magnetic
pressure along the leading edge of the cloud can increase
enough to accelerate the disruption of the cloud through the
Rayleigh-Taylor instability \citep{Gregori1999}.
On the other hand, tangled magnetic fields within the clouds would  act
to resist compression, potentially reducing cooling and enhancing
cloud destruction by shocks.

The left panel of Fig.~\ref{fig:shock-cloud-simu-Shin-Fragile} shows the results of two 3D MHD simulations, one being for a weak magnetic field, the other for a strong one \citep{Shin2008}. The effect of the magnetic field seems important only in the late stages of the evolution of the cloud. This figure illustrates that at late times, strong fields substantially alter the dynamics of the cloud, suppressing fragmentation and mixing by stabilizing the interface at the cloud surface. Even weak magnetic fields can drastically alter the evolution of the cloud compared to the hydrodynamic case.

In addition, the topology of the magnetic field has a strong impact on the thermal conduction for clouds in hot ($ > 10^6$~K) gas, because in this case the conduction is carried on by the electrons (see the discussion about the thermal conduction below). Indeed, the magnetic field channels the electron conductivity along the field lines.

\paragraph{Impact of the cooling of the cloud and thermal instability}

The above conclusion that clouds are destroyed in a time comparable to the cloud crushing time is drastically changed if the cloud is able to cool through radiative processes over a timescale $ t_{\rm cool}$ shorter than the dynamical timescale of destruction, $ t_{\rm crush}$ \citep{Mellema2002, Fragile2004}. 
\citet{Mellema2002} define a cooling-dominated regime by $t_{\rm cool} < \frac{t_{\rm crush}}{10}$, in which the evolution of the cloud is dominated by radiative processes. A simple estimate of the cooling time \citep{Fragile2004} leads to the following condition:
\begin{equation} 
R_{\rm c}  > 1.2 \times 10^{-2} \left(\frac{\chi}{10^{3}} \right)^{-2}   \left( \frac{V_{\rm i}}{ 10^3 \,\rm km\, s^{-1}} \right) ^4 \left( \frac{n_{\rm c}}{1\, \rm cm^{-3}} \right) ^{-1} \ \ ,
\end{equation}
which shows that cooling will generally govern evolution for moderate
cloud densities and shock velocities. However, sufficiently
high shock velocities ($V_{\rm i} \gtrsim 10^4$~km~s$^{-1}$) can suppress the effect of cooling over the cloud destruction time.  Cooling is also negligible at low densities ($n_{\rm c} \lesssim 10^{-4}$~cm$^{-3}$ for $V_{\rm i} \gtrsim 10^3$~km~s$^{-1}$).

Heating and cooling processes may generate thermally unstable gas in the cloud (see Fig.~\ref{fig:thermal_instability} or later in Fig.~\ref{fig:thermal_eq_simu_P_vs_n}). The thermal instability adds to the Rayleigh-Taylor and Kelvin-Helmholtz instabilities, and is an important fragmentation mechanism. 
The thermal instability has been discussed in chapter~\ref{chapter:dust_gas_galaxies}, sect.~\ref{subsec:thermal-instability}).  
Below $\sim 10^6$~K, the gas is thermally unstable. Numerical simulations show that the  thermal instability generates complex inhomogeneous structures, dense regions that cool and low density voids \citep{Sutherland2003, Audit2005}. This fragmentation occurs on the gas cooling timescale. In most numerical simulations, the thermal instability is ignored because simplifying assumptions are made on the equation of
state of the gas (adiabatic, isothermal or polytropic evolution).

The lifetime of the fragments in the hot gas depends on their ability to cool. An efficient gas cooling stabilizes the clouds. As H$_2$ is a major coolant in the warm molecular gas, it may have an important impact on the survival of molecular clouds hitted by strong shocks. This is illustrated in the right panel of Fig.~\ref{fig:shock-cloud-simu-Shin-Fragile}, that shows numerical simulations of multiple clouds overtaken by a shock wave that takes into account the radiative properties of the clouds, as well as primordial gas-phase formation of H$_2$. 
These simulations show that the cloud first breaks into fragments. Because of the cloud cooling, these fragments become dense enough to be able to survive over a timescale of the order of a few  million years \citet{Mellema2002, Fragile2004}. 

These numerical simulations ignore in particular the H$_2$ formation on grains, which may greatly enhance the cooling efficiency of the cloud, provided that dust can survive the transmitted shock. If it is the case, the cloud fragments may be even further stabilized.

\paragraph{Thermal conduction and evaporation timescale}

The heat conduction has an important and subtle impact on the evolution of the shock cloud. Depending on how the heat flux is treated and whether cloud cooling is included, heat conduction may or may not lead to the evaporation of the cloud fragments.

The heat flux is carried by electrons in hot or ionized gas, and by H in cold or warm neutral gas. This heat flux is proportional to the temperature gradient (Fourier's law):
\begin{equation}
\dot{\textbf{q}} = - \kappa \, \nabla \textbf{T} \ ,
\end{equation}
which defines the \textit{thermal conductivity}\footnote{The mean free path of the specie that carries the heat flux determines the temperature dependence of the thermal conductivity: $\kappa \propto T^{5/2}$ for electron conductivity and $\kappa \propto T^{0.8}$ for H. Note that the charge transfer reactions strongly limits the H mean free path. Thereby, ionization reduces conductivity in partially ionized warm gas.} $\kappa$. 
In case of a steep temperature gradient between the cloud and intercloud medium, the heat flux is limited by the ability of the electrons or H atoms  to diffuse across the boundary. If the spatial scale of the temperature variation is smaller than the mean free path of the electrons, the heat flux \textit{saturates}, and its value can be reduced up to a factor $\sim 10$ as compared with the classical heat flux given by \citet{Spitzer1962}. The saturation of the conduction thus  reduces  the evaporation of the cloud \citep{Cowie1977}. 

In the case of saturated heat flux, the rate at which
the cloud evaporates can be written analytically \citep{Cowie1977, Klein1994}:
\begin{equation}
\dot{M}_{\rm c} \approx 4 \pi R_{\rm c} ^2 \, \rho _{\rm i} \, c_{\rm i} \,
\end{equation}
where $\rho _{\rm i}$ and $c_{\rm i}$ are the mass density and sound speed in the inter-cloud medium.
One can define an \textit{evaporation timescale} 
\begin{equation}
\label{eq:evaporation-cloud-timescale}
t_{\rm ev} = \frac{M_{\rm c}}{\dot{M}_{\rm c}} \approx \frac{\sqrt{\chi} \, t_{\rm crush}}{7} \ \ ,
\end{equation}
which shows that for $\chi \sim 10^2$, the clouds will evaporate in a time comparable
with  the compression timescale (cloud crushing time). 

\citet{Vieser2007a} performed numerical simulations of static clouds embedded within a hot gas, taking the saturated heat flux. The main result is that the evaporation timescale is at least one order of magnitude longer than the analytical value given by Eq.~\ref{eq:evaporation-cloud-timescale}. The reason is that a transition zone forms at the cloud edge in which the steep temperature and density gradients are reduced. This results in a lower evaporation rate than predicted. 

Simulations that include both thermal conduction and radiative cooling (and heating) processes show an even more dramatic effect.
The clouds can even gain material if radiative cooling exceeds
the energy input by heat conduction. \citet{Vieser2007a} show that matter from the hot inter-cloud medium can condensate onto the clouds in a regime of cloud parameters
where evaporation is requested from the analytical approach. The reason is that the additional energy driven by heat
conduction from the hot background gas into the cloud can be transported away from the interface and radiated off efficiently from the  inner parts of the cloud.

\begin{figure}
   \centering
 \includegraphics[width=0.39\textwidth]{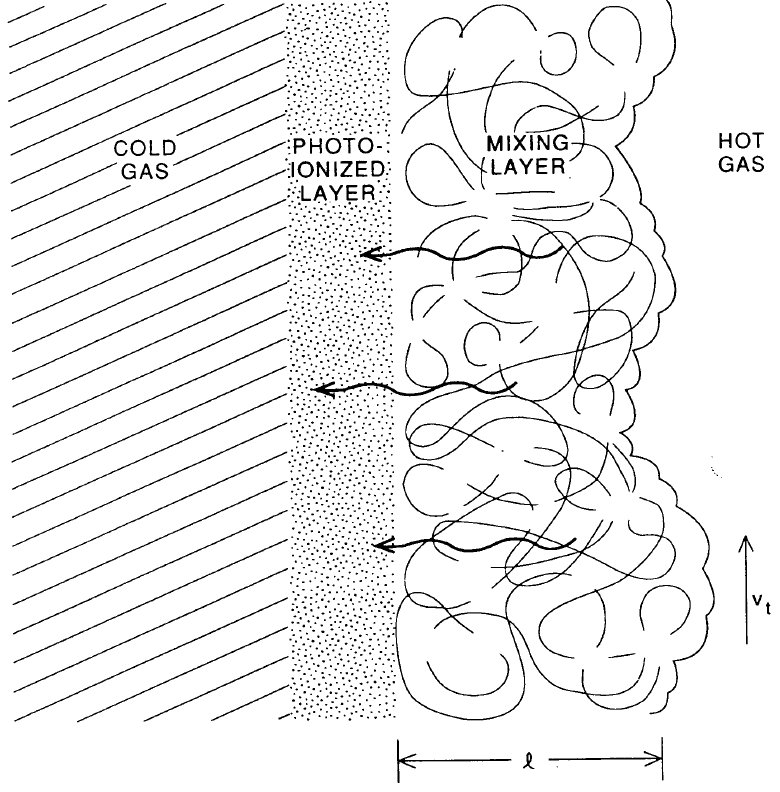}
    \includegraphics[width=0.60\textwidth]{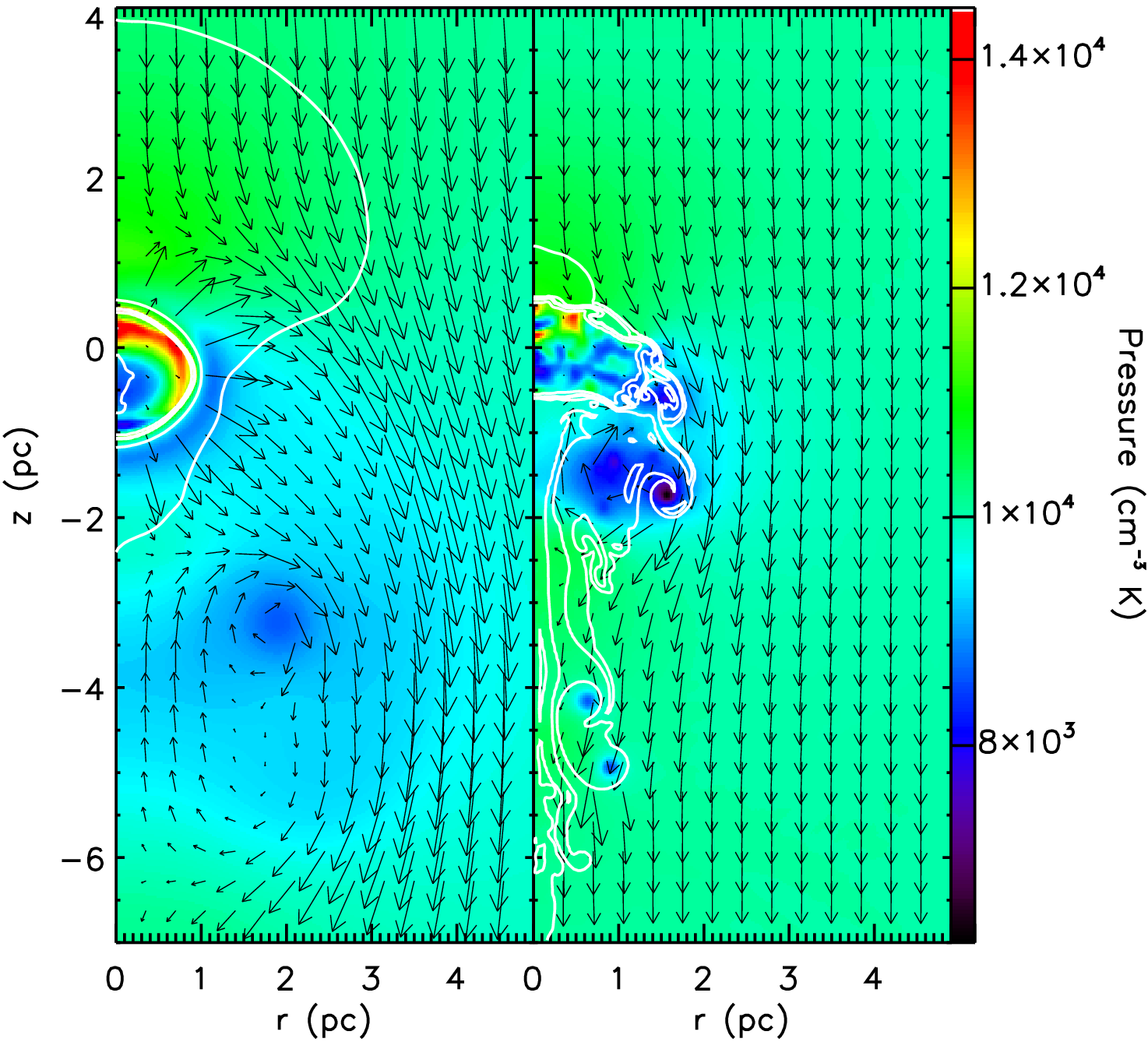}
      \caption[Numerical simulations of a cloud in a hot plasma stream]{\textit{Left:} Schematic description of a turbulent mixing layer from \citet{Slavin1993}. \textit{Right:} Numerical simulations  of a CNM cloud embedded in a hot plasma stream (J. Slavin, private communication). The colors indicate the gas pressure, and the arrows the velocity field. In the \textit{left} panel, heat conduction is taken into account, whereas in the \textit{right} panel not. The cloud with thermal conduction (\textit{left}) is much less disrupted  than the cloud with conduction turned off (\textit{right}). The hot gas has a speed of $50$~km~s$^{-1}$ relative to the cloud and a temperature $10^6$~K. The initial cloud density is $n_{\rm H} = 0.6$~cm$^{-3}$ and its radius 1~pc. The cloud to hot gas density contrast is $\chi = 125$. The ambient pressure is $10^4$~K~cm$^{-3}$. No cooling processes are included in these simulations. 
}
       \label{fig:cloud_hot_simu_thermal_conduct}
\end{figure}

\begin{figure}
   \centering
    \includegraphics[width=0.8\textwidth]{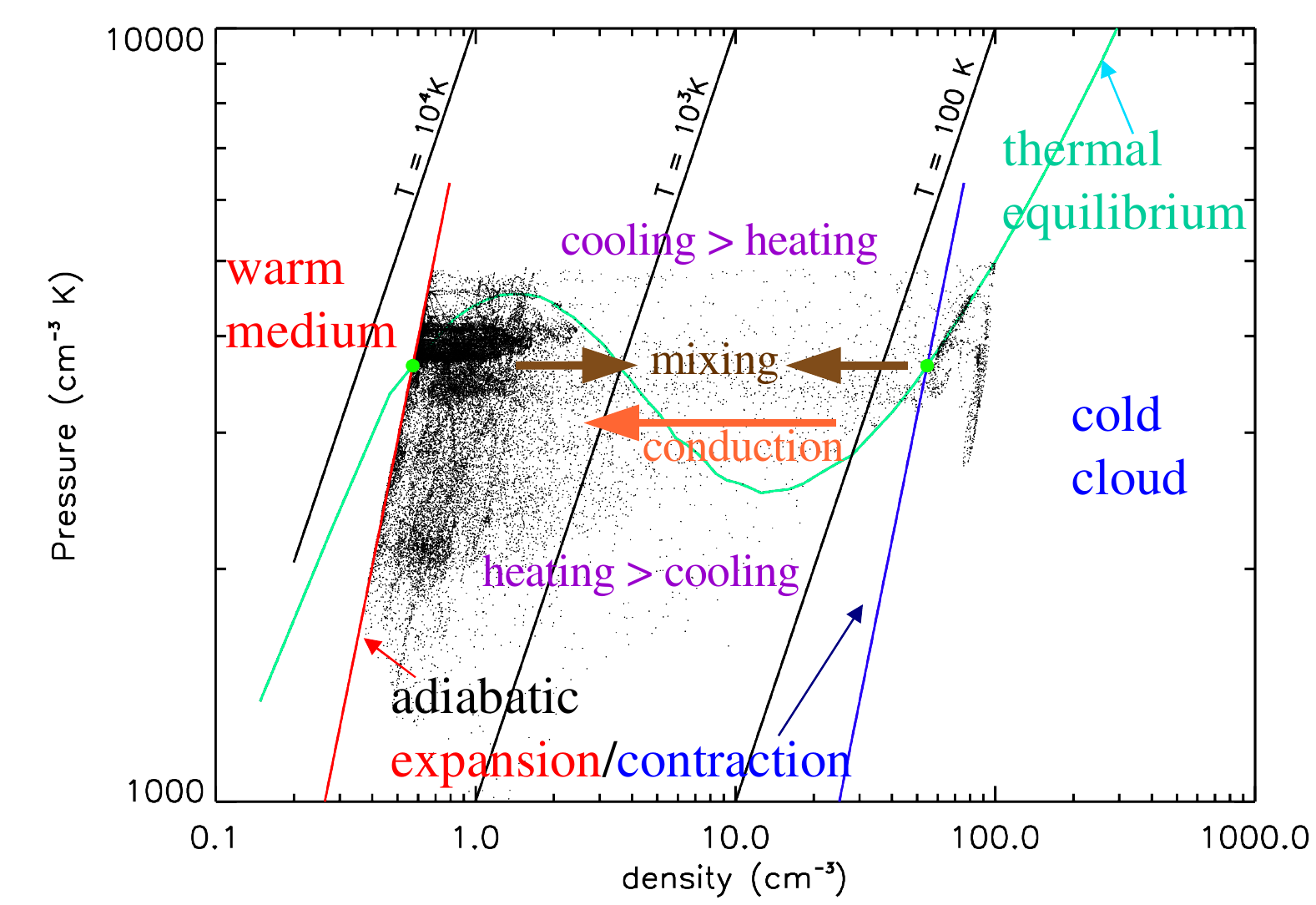}
      \caption[Clouds in a hot plasma stream: thermal balance]{CNM Clouds in a WNM stream: thermal balance. The diagram shows the phase properties of a CNM cloud being destroyed by turbulent mixing in a WNM flow. Points are gas parcels from the hydro-dynamical simulation shown on Fig.~\ref{fig:cloud_hot_simu_thermal_conduct}, at $t = 3.5\times 10^5$~yr (J. Slavin, private communication).  Part of the gas is located on the thermally unstable branch of the equilibrium cooling curve. Dynamical processes (turbulent mixing) lead to departures from equilibrium (stable or unstable). The WNM gas has a speed of $5$~km~s$^{-1}$ relative to the cloud and a temperature $6\,300$~K. The initial cloud density is $n_{\rm H} = 55$~cm$^{-3}$ and its radius 0.1~pc. The cloud to hot gas density contrast is $\chi \sim 100$. The ambient pressure is $3660$~K~cm$^{-3}$. No thermal conduction is included in these simulations. }
       \label{fig:thermal_eq_simu_P_vs_n}
\end{figure}

\paragraph{Turbulent mixing}
\label{turbulent-mixing}
\index{Turbulent mixing}
\index{Mixing layers}
Another important mean to transport the energy from the hot background gas to the cool medium is through \textit{mixing of gas phases at boundary layers}. This mixing is expected to occur wherever there is a velocity shear is present at the interface of cloud phases. A schematic description of such a boundary layer is given in Fig.~\ref{fig:cloud_hot_simu_thermal_conduct}. Such velocity shears occur in many astrophysical situations, including outflows of hot gas driven into the ISM by supernovae explosions, starburst or AGN, but also galaxy mergers, etc.

A turbulent mixing layer at the boundary of the cloud develops through the growth of Kelvin-Helmholtz instabilities, that drive transverse waves along the boundary between the two fluids. The energy of these waves is dissipated at small scales, and a layer of gas at intermediate temperatures is produced at the interface \citep[see][]{Begelman1990, Slavin1993}. 
The intermediate temperature, $\bar{\rm T}$, in the turbulent mixing layer is determined by the temperatures of the hot (${\rm T}_h$) and cloud (${\rm T}_c$) phases,   the evaporation rate of the cold gas into the intercloud medium, $\dot{m}_c$, and the entrainment rate of hot gas into the mixing layer, $\dot{m}_h$:
\begin{equation}
\bar{\rm T} = \frac{\dot{m}_h {\rm T}_h + \dot{m}_c {\rm T}_c}{\dot{m}_h + \dot{m}_c} = \zeta \sqrt{{\rm T}_c {\rm T}_c} \ ,
\end{equation}
where $\zeta$ is a dimensionless factor of the order of unity \citep[see][]{Begelman1990}. 
Typically, for warm ($T \approx 10^4$~K) clouds embedded within a hot ($ T \approx 5 \times 10^6$~K), the temperature of the mixing layer is $T_{\rm mix} \approx  10^{5-5.5}$~K. 

The  intermediate-temperature gas may be thermally unstable and any dynamical disturbance will cause the gas to rejoin the hot or warm phase (see Fig.~\ref{fig:thermal_eq_simu_P_vs_n}). 
This length of the boundary layer is obtained by assuming a stationary mixing layer, in which the energy flux input by turbulent motions is balanced by the cooling rate per unit area.

\index{Mixing layers!simulations}
\index{Mixing layers!and thermal conduction}
The right panel of Fig.~\ref{fig:cloud_hot_simu_thermal_conduct} illustrates the situation of a cold neutram medium (CNM) cloud embedded within a stream of hot gas. 
The two sub-panels compare the results with our without thermal conduction.  
These simulations have been performed by J. Slavin (private communication). When thermal conduction is ignored \textit{(right)}, the Kelvin-Helmholtz instabilities are developped.  The comparison between the two figures  clearly shows that the conduction  reduces the growth rate of Kelvin-Helmholtz instabilities. When conduction is taken into account, the evaporative outflow prevents instabilities from developing at the interface with the flow. With conduction, the cloud is less disrupted but  loses its mass faster. These illustrative results are in agreement with other simulations performed by  \citet{Vieser2007, Vieser2007a}, and we direct the reader to those papers for a detailed discussion of the impact of thermal conduction on the evolution of a cloud embedded in hot gas.


\index{Mixing layers!thermal instability}


\subsection{Supersonic collision between two gas streams}
\label{subsec:collision-two-gas-flows}
\index{Shocks!collision between 2 gas streams}

Let us consider now the collision between two gas streams at different densities and velocities. This situation may be relevant in the case of galaxy collisions, where the ISM of the galaxies are colliding with each other. I briefly discuss the plane-parallel situation, which can be solved analytically, and move to a brief discussion of numerical simulations that provide some useful feeling about how the gas and shocks behave in such collisions.  

\subsubsection{Plane-parallel collision between two gas streams}

\begin{figure}
   \centering
    \includegraphics[width=0.8\textwidth]{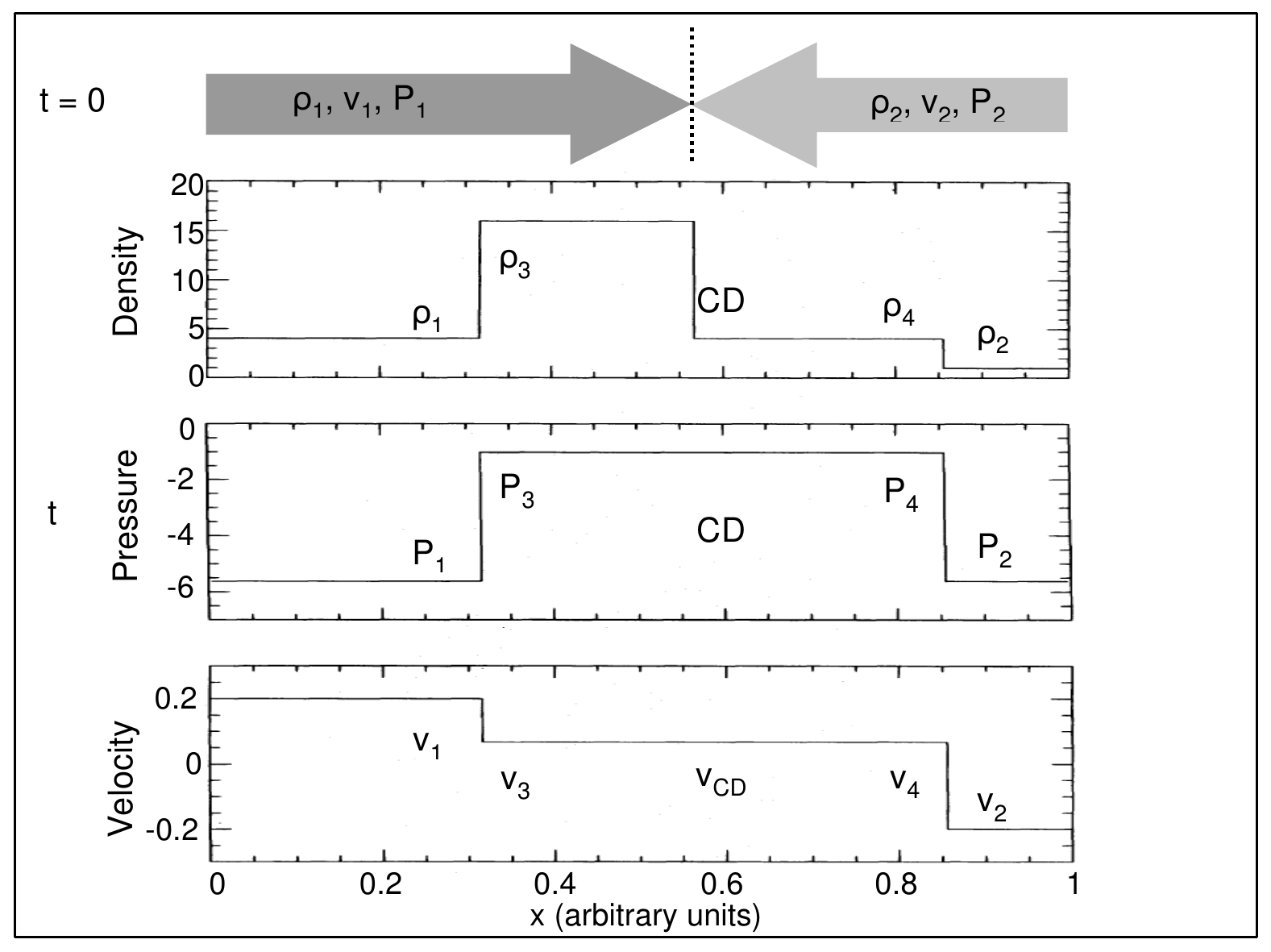}
      \caption[Flow structure of a 1D collision between two gas streams]{Flow structure of a one-dimensional (plane-parallel) collision between two gas streams. The parameters $\rho _1$, $v_1$, $P_1$ and $\rho _2$, $v_2$, $P_2$ of the two flows indicate the pre-collision parameter to the left and right of the contact discontinuity (CD). Two shocks propagate away from the discontinuity. Adapted from \citet{Lee1996}.}
       \label{fig:two_colliding_gas_streams}
\end{figure}

We consider the situation of Fig.~\ref{fig:two_colliding_gas_streams}. Two plane-parallel flows of preshock densities $\rho _1$ and $\rho _2$ are colliding at $t=0$, which produces the contact discontinuity (CD). Fig.~\ref{fig:two_colliding_gas_streams} shows the situation where $\rho _1 = 4 \rho _2$, $P_1 = P_2$ and $v_1 = - v_2$. 
To simplify the expressions of the post-collision quantities, we place ourselves in the case where the two flows are identical, i.e. $\rho _1 = \rho _2$  ($\chi =1$). The case presented in Fig.~\ref{fig:two_colliding_gas_streams} is more general. The corresponding equations for the general case can be found in \citet{Lee1996}.
Since the pressure and velocity are continuous across the surface discontinuity, we have: 
\begin{equation}
P_3 = P_4 = \frac{4}{3} \rho _1 v_1 ^2 \quad \rm  and \quad v_3 = v_4 = v_{\rm CD} \ ,
\end{equation}
where $v_{\rm CD}$ is the velocity of the contact discontinuity. For this simplified case, the bulk kinetic energy to thermal energy conversion rate can be written as
\begin{equation}
R = \frac{ \frac{4}{3} \rho _1 v_1 ^3 }{\frac{1}{2}\rho _1 v_1 ^3 + \frac{1}{2}\rho _2 v_2 ^3} = \frac{4}{3} \ ,
\end{equation}
which is the largest value attained for $\gamma = 5/3$ when the two colliding flows are identical. The fact that $R$ is greater than 1 is not surprising. Since the postshock gas is at rest in the lab frame, all the bulk kinetic energy of the gas that passes through the shock is converted into thermal energy. But the shocks move outward from the discontinuity, so $R$ is actually the ratio between the shock velocity relative to the upstream flow in region 1 to the shock velocity $v_1$, which is $4/3$ for $\gamma = 5/3$.

We will now consider the similar situation of the collision between two gas streams, but from the point of view of 3D simulations. This allows to point out the main differences between the 1D and the 3D cases, in particular in terms of conversion of energy from the bulk pre-collision energy to the postshock kinetic energy.

\subsubsection{3D simulations of an oblique collision between two gas streams}
\label{3D-simulations-collision-two-gas-streams}
\begin{figure}
   \centering
    \includegraphics[width=0.8\textwidth]{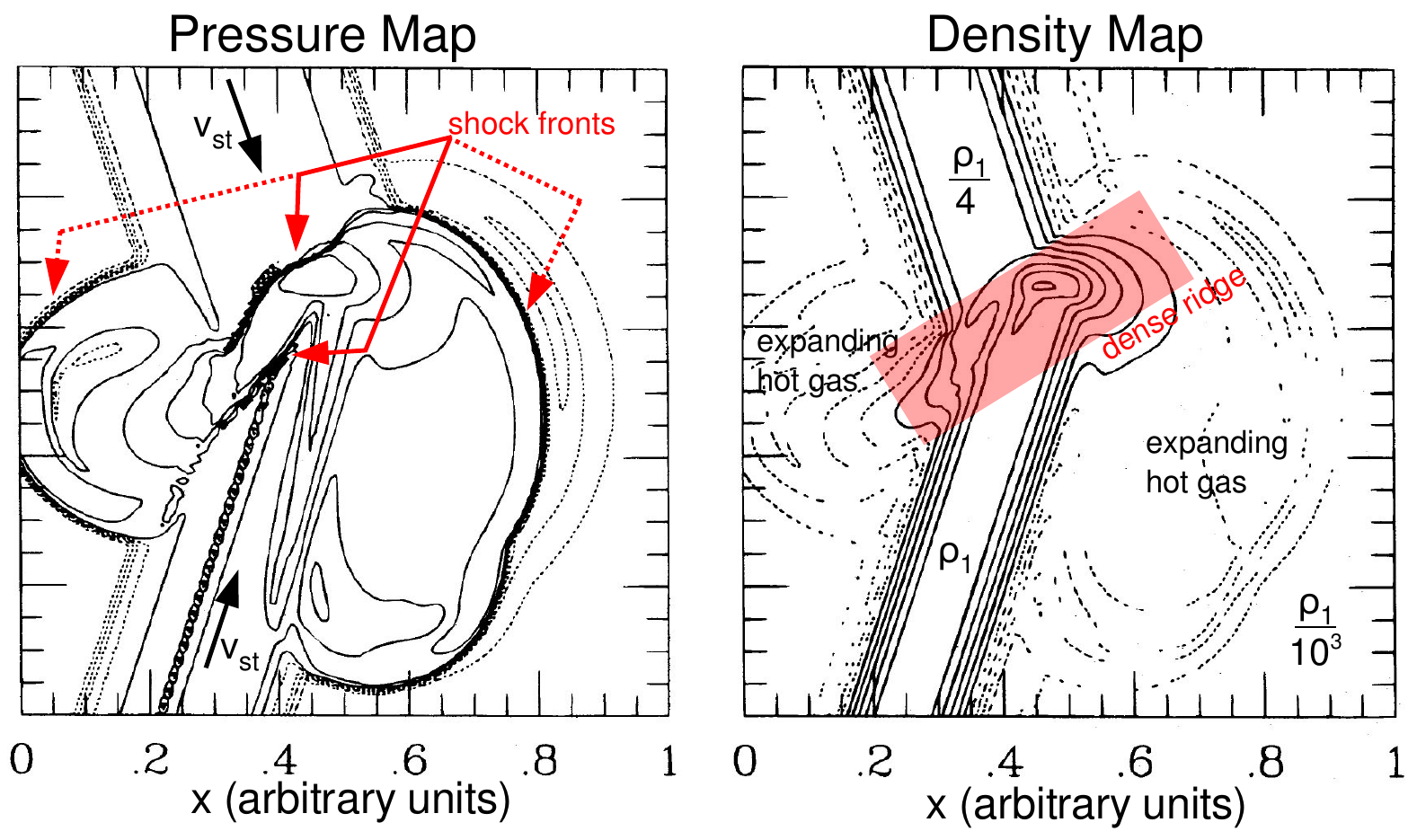}
      \caption[Simulations of the collision between two gas streams]{3D numerical simulations of an oblique,  supersonic collision between two homogeneous gas streams. The plots show pressure \textit{(left)} and density \textit{(right)} maps of a slice cut in the simulation cube. The angle between the flows is 143\degre and their velocities and pressure are identical. The bottom flow is four times denser than the top stream. The streams propagate in a background medium that is $10^3$ times less dense and pressurized. The figures are adapted from \cite{Lee1996}.}
       \label{fig:colliding_2_gas_streams}
\end{figure}

Fig.~\ref{fig:colliding_2_gas_streams} shows the result of 3D hydrodynamic simulations of the oblique collision between two gas streams. These simulations are performed by \citet{Lee1996} within the context of the collision of gas streams produced by the disruption of a massive star around a black hole. Obviously, these simulations do not directly apply to the SQ collision, in particular because in the simulations the gas flows are homogeneous and radiative processes are ignored. However, these simulations provide interesting insights about the dynamical evolution of the flows and the geometry of the shock fronts produced by the collision. 

The pressure map (left panel of Fig.~\ref{fig:colliding_2_gas_streams}) shows that oblique shocks are produced around the initial contact surface. The position of these shock fronts are indicated on Fig.~\ref{fig:colliding_2_gas_streams} by the solid red arrows. In the case of a one dimensional (1D) collision of two stream at the same velocity $v_{\rm st}$, the two shocks produced at the discontinuity surface propagate back to the upstream flows at a velocity $V_{\rm sh} = \frac{4}{3} v_{\rm st}$ (see above for an analytical description of the 1D case). In the 3D case, the shock speed is $\sim v_{\rm st}$ and the shock fronts  stand off near the initial contact surface because the shocked gas can escape in the direction perpendicular to the stream flow. 
For instance, if we assume two identical gas streams with a relative velocity of 1000~km~s$^{-1}$, the shock velocity is $\sim 500$~km~s$^{-1}$.
The density map (right panel) show that the shocked gas is accumulated in a dense ridge (red rectangle) between the shocks and accelerated into the ambient medium. This hot gas is expanding adiabatically and drives other shocks in the background medium (dashed red arrows). 

We have discussed above the conversion of the mechanical energy of the colliding gas flows into thermal energy of the gas for a plane-parallel (1D) collision. 
We remind that the ratio of the thermal energy flux produced by the shock to the flux of kinetic energy, $R$,  is maximum ($R = 4/3$) for a 1D collision between two identical stream flows. \citet{Lee1996} find drastic differences between one- and three-dimensional cases. In 3D, most of the bulk kinetic energy of the flows is converted into the kinetic energy of the accelerating hot gas instead of thermal energy. 
For the 3D simulations, \citet{Lee1996} compute the time-dependent energy conversion rate, $R(t)$, as  the ratio of the increase in the total thermal energy of the gas in the computational box to the kinetic energy added in the box through the discontinuity:
\begin{equation}
R(t) = \frac{E_{\rm th}(t) - E_{\rm th}(0) }{\frac{1}{2} (\rho _1 \sigma _1 v_1 t) v_{1}^{2} + \frac{1}{2} (\rho _2 \sigma _2 v_2 t) v_{2}^{2}} \ ,
\end{equation}
where $\rho _1$, $v _1$ and  $\rho _2$, $v _2$ are the pre-collision mass densities and the velocities of the gas flows. $\sigma _1$ and $\sigma _2$ are the  cross-sections of the two colliding streams, so the quantity $\rho \, \sigma \, v \, t$ represents the mass of gas that passes through the boundary within the simulation time $t$.
For the model shown on Fig.~\ref{fig:colliding_2_gas_streams}, \citet{Lee1996} find $R \sim 0.1$ at the end of the simulation. This value is one order of magnitude smaller than the one-dimensional model.
This is due to the possibility for the gas to expand in directions perpendicular to the gas streams. This illustrative result suggests that the dynamical and thermal properties of the shocked gas are very different between 1D and 3D models. 


\subsection{Concluding remarks and limitations of existing codes}

The modeling of shocks in inhomogeneous media is very complex, and this is due to the richness of the physical processes at work (listed above), and the fact that they almost all operate on a  comparable timescale. Here I would like to stress some key processes that one cannot ignore when describing  the survival on shocked molecular gas, as it is one key aspect of the study presented here.

So far, most of the 3D numerical simulations focus on the development of hydrodynamical instabilities and the effect of the magnetic field. They generally predict the complete destruction of clouds by fast shocks in a timescale comparable with the compression (\textit{crushing}) timescale of the cloud. This is essentially because these simulations do not include a realistic treatment of the thermal properties of the shocked gas, including H$_2$ formation on dust and H$_2$ cooling. 
On the other hand, models that are used for the comparison with observations, in which chemical processes are included, are limited to 1D geometry. Here I summarize a few important points among which some of them remain largely unexplored in numerical simulations:
\begin{enumerate}
\item thermal processes have a very important impact on the fate of shocked gas, perhaps more important than the effect of the magnetic field. 
The thermal instability, so far ignored in numerical simulations, triggers gas cooling and thereby stabilizes the cloud phase.  
Both cooling and thermal conduction seem to limit evaporation of cold and dense gas embedded in the hot and tenuous postshock medium. In the case of molecular gas, H$_2$, as a main coolant of the gas, may be a key-actor. In addition, the production of thermally unstable gas in turbulent mixing layers (at the cloud-intercloud interface) may also help in balancing the evaporation with condensation of material from the hot to the warm/cold phase.
\item the 3D geometry also influences the dynamical and thermal properties of the shocked gas. One should keep in mind that 1D models assume a maximum compression of the gas by the shock, leading to a maximum conversion of the bulk kinetic energy of the shock to thermal energy.  In 3D, this conversion may be less efficient. This is perhaps the main limitation of the models that I have been using (the Flower et al.  and the MAPPINGS shock codes) to interpret observations.
\end{enumerate}
All the effects listed above would tend to increase the lifetime of shocked clouds compared to pure hydrodynamical simulations.


\part{Detailed studies of H$_{\bf 2}$-luminous sources in space}

\chapter{Powerful H$_{\bf 2}$ emission from the Stephan's Quintet galaxy-wide shock}
\label{chapter:H2_SQ}

\epigraph{Life is a series of collisions with the future; it is not the sum of what we have been, but what we yearn to be.}{Jose Ortega y Gasset}


\begin{Abstract}
The Stephan's Quintet (SQ) is a compact group of interacting galaxies and a spectacular example of H$_2$ luminous source. The \textit{Spitzer Space Telescope} indeed detected a powerful mid-infrared H$_{2}$ line emission from the X-ray bright galaxy-wide shock created by the high-velocity ($\sim 1000$~km~s$^{-1}$) collision between a galaxy and the tidal tail of an other member of the group. Observations show that the H$_{2}$ gas is extremely turbulent and that the H$_2$ luminosity exceeds that of the X-rays. The core of my PhD work is dedicated to the interpretation and modeling of these observations. 
In this chapter I present the observational discovery of H$_2$ in the shock, and address the following question: Why is H$_2$ present in such a violent environment? I propose a scenario where H$_2$ forms out of shocked gas, and present my  results on the modeling of the cooling of the multiphase postshock gas and associated H$_2$ formation.
\end{Abstract}

\minitoc

\index{Stephan's Quintet}

\section{Introduction}
\label{sec:SQ_intro}

\PARstart{S}tephan's Quintet (hereafter SQ) is a fascinating example of an interacting galaxy system (Fig.~\ref{fig_SQ_HST_legend}). Perhaps not as famous as Arp 220 or M82, SQ is a well-known astrophysical target for amateur or professional astronomers. It is a nearby (94~Mpc) compact group of galaxies situated in the Pegasus constellation, at about half a degree to the South-West of NGC~7331\footnote{\url{http://antwrp.gsfc.nasa.gov/apod/ap071124.html}}. SQ has been observed in almost all wavebands, and it keeps revealing surprises when being looked at by new instruments.
SQ belongs to a peculiar class of interacting systems of galaxies: the compact groups \citep{Hickson1982} that are characterized by aggregates of more than $4$ galaxies in implied space densities as high as those in galaxy cluster cores. 

SQ is a complex web of galaxy-galaxy and galaxy-intragroup medium (henceforth IGM) interactions that have triggered different types of phenomena related to the interactions, including an IGM starburst, long tidal tails ($>100$~kpc) with tidal dwarf candidates, and a  large scale shock ($\sim 40$~kpc long). This shock is materialized by  a giant X-ray and radio-bright \textit{ridge}, attributed to the collision of an intruding galaxy with the tidal tail of another member of the group. 

In 2006, \textit{Spitzer Space Telescope} observations revealed a powerful mid-IR H$_2$ rotational line emission in the SQ shock structure. This discovery makes SQ the first known example of an \textit{``$\rm H_2$-luminous source''}, amongst the emerging class of H$_2$-galaxies presented in chapter~\ref{chapter:H2_galaxies}. This was perhaps one of the most exciting results of extragalactic spectroscopic observations carried out with the \textit{Spitzer} mission. 

This surprising observational result was published at the start of my PhD thesis. Much of my PhD work was dedicated to its interpretation. I have devoted a significant part of my time in building a theoretical model that attempts to explain the formation and excitation of H$_2$ in the SQ galaxy collision within the broader framework of the numerous multi-wavelength observations of the system. It was not an easy task (!), and I will try to show that our current understanding is the result of a long path, along which several working hypothoses have been explored before coming to what we think to be a ``coherent'' scenario. 

This work has been published in \citet{Guillard2009} (hereafter \hyperref[paper_SQ_H2]{paper~{\sc i}}). In this manuscript, I will introduce the two questions of H$_2$ formation and excitation almost independently. In this chapter, I only discuss H$_2$ formation. The interpretation work carried out in \hyperref[paper_SQ_H2]{paper~{\sc i}} is based on the single-pointing observations of \citet{Appleton2006}. After the publication of that paper, we have obtained new \textit{Spitzer} observations that now cover the full area of the SQ ridge. I have re-visited my previous analysis of the H$_2$ excitation in this extended region, and  complemented it by an analysis of mid-IR fine-structure lines diagnostics. These new observations and the modeling results are presented in the next chapter (chap.~\ref{chapter:H2_SQ_mapping}).

After reading chapters~\ref{chapter:H2_SQ} and \ref{chapter:H2_SQ_mapping}, I hope that the reader will be conviced that SQ is an outstanding source to study in detail phenomena such as large scale shocks and high velocity galaxy collisions. SQ is a beautiful example of astrophysical source where detailed and rich micro-physics of the ISM are needed to understand the global observational picture. These studies have implications for our understanding of the dynamics and energetics of the molecular gas in galaxy evolution.


In this chapter, I  present the observational discovery of H$_2$ emission in SQ (sect.~\ref{sec:H2obsSQ}), and point out the set of astrophysical questions raised by these observations (sect.~\ref{H2_SQ_questions}).
Then sect.~\ref{sec:H2formation_SQ}  adresses the question of the H$_2$ formation in the SQ shock. I present the long route that drove us to our current interpretation of these observations and introduce our \hyperref[paper_SQ_H2]{paper~{\sc i}}.  A quick summary of the results and concluding remarks end this chapter (sect.~\ref{sec:SQ_H2-conclusions})


\section{Observations of warm H$_{\bf 2}$ gas in Stephan's Quintet}
\label{sec:H2obsSQ}
\index{Stephan's Quintet!H$_2$ emission}

The first mid-IR spectroscopic observations of the SQ shock were performed by \citet{Appleton2006}. 
We present the observational discovery based on single-pointing observations (sect.~\ref{subsec:SQdiscoveryH2}), and their extension to new spectral maps are  presented in chapter~\ref{chapter:H2_SQ_mapping}.

\subsection{The discovery: an H$_{\bf 2}$-bright giant shock!}
\label{subsec:SQdiscoveryH2}

\begin{figure}
   \centering
    \includegraphics[width=\textwidth]{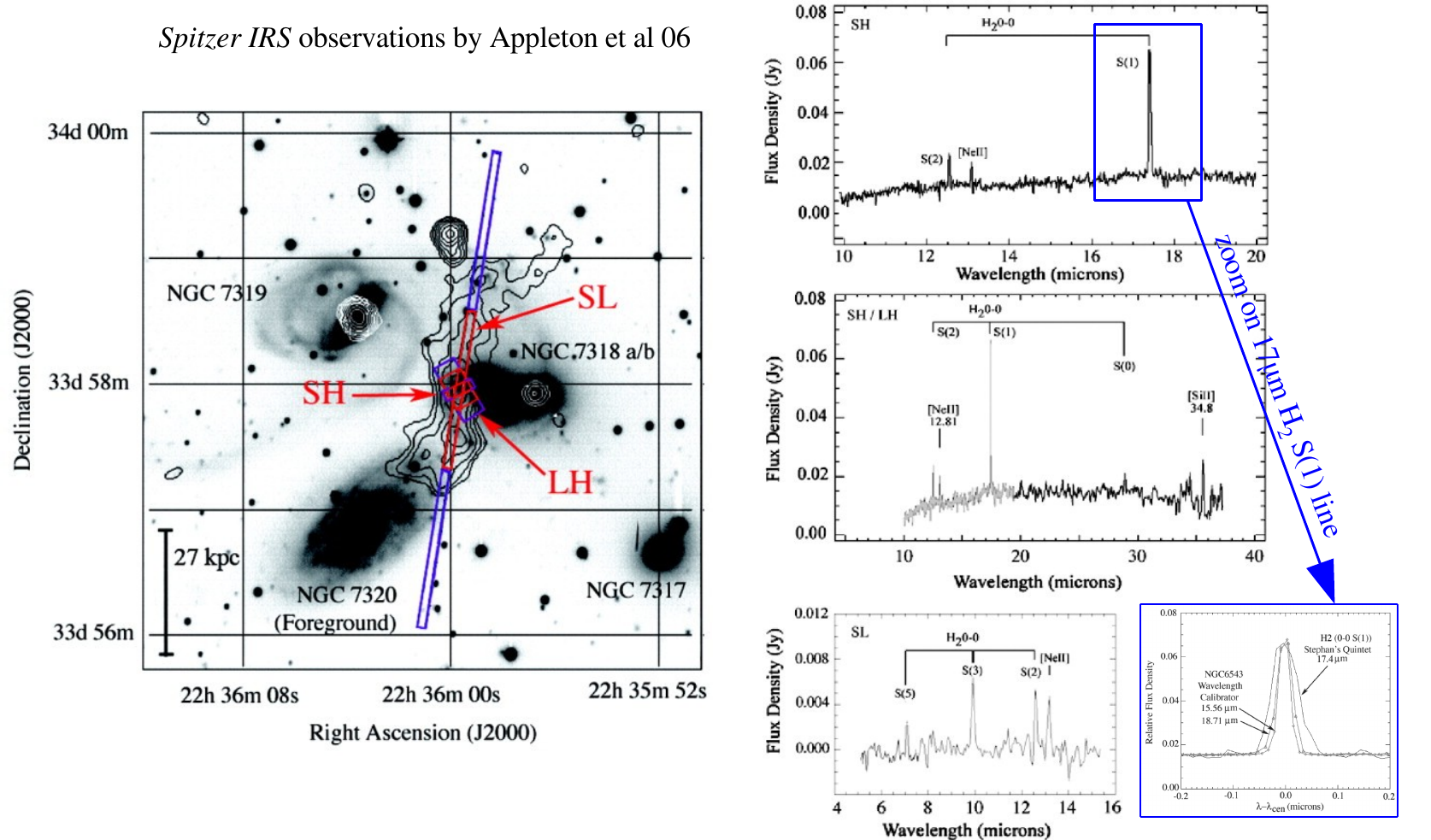}
      \caption[\textit{Spitzer IRS} H$_2$ observations by \citet{Appleton2006}]{\textit{Spitzer IRS} H$_2$ observations by \citet{Appleton2006}. \textit{Left:} Positioning of the \textit{IRS} slits overlaid on an R-band image and \textit{VLA} radio contours at 1.4$\,$GHz by \citet{Xu2003}. The slits are centered on $\alpha_{J2000} = 22$h35m59s.57, $\delta_{J2000} = 33^{\rm o}$58'1.8''. Only the central portion of the slits
(red boxes) were common to the two separate observing (nod) positions made in
each IRS module slit (the purple shows the full coverage). $\rm SL = 57 \times 3.6$~arcsec$^{2}$,
$\rm SH = 11.3 \times 4.7$~arcsec$^{2}$, and $\rm LH = 22.3 \times 11.1$~arcsec$^{2}$. \textit{Right: } top:  IRS SH spectrum of the brightest radio/IR point in the shock front. Middle: Combined SH (gray line) and LH (black line) spectrum of the
same target region. Bottom left: SL spectrum at the same position. SL2 is the gray line, and SL1 is the black line. Bottom right: zoom on the 17$\,\mu$m H$_2$ 0-0 S(1) line (detected with a signal-to-noise ratio of 33) showing that the line is resolved (intrinsic FWHM of $870 \pm 60$~km~s$^{-1}$) as compared with two unresolved spectral lines
($\mathcal{R} = 600$) in the planetary nebula NGC 6543 at 15.6 and 18.7$\,\mu$m that bracket the H$_2$ line (gray line).}
       \label{SQ_IRS_obs_Appleton06}
\end{figure}

The \emph{Spitzer} Infra-Red Spectrograph (\textit{IRS}) observations of the SQ galaxy-wide collision led to the unexpected detection of extremely bright mid-IR $\rm H_2$ rotational line emission from warm ($\sim 50 - 10^{3}\,$K) molecular gas in the center of the X-ray emitting ridge \citep{Appleton2006}.
The observational results about the first detection of H$_2$ in the SQ shock are summarized in Fig.~\ref{SQ_IRS_obs_Appleton06}. The left panel shows the position of the \textit{IRS} slits (SH, LH and SL modules; see caption for the slit sizes) overlaid on a $R$-band image with radio \textit{VLA} contours. The high-resolution modules are placed close to the ``center'' of the radio-bright ridge. The SL slit is oriented along the radio filament that traces the shock structure.

The right panel of Fig.~\ref{SQ_IRS_obs_Appleton06} gathers the low- and high-resolution spectra extracted in the SQ shock. These spectacular spectra are dominated by the mid-IR rotational emission of H$_2$. The 0-0 S(0) to S(5) lines are detected, plus atomic lines of [Ne$\,${\sc ii}] at $12.8\,\mu$m and [Si$\,${\sc ii}] at $34.8\,\mu$m. No PAH emission is seen\footnote{this conclusion will be revised in chapter~\ref{chapter:H2_SQ_mapping}. New \textit{Spitzer IRS} observations indeed show PAH emission in the ridge\protect  \citep{Cluver2009}} and the thermal dust continuum is very weak. 
This is the first time an almost pure $\rm H_2$ line spectrum has been seen in an extragalactic object. 

A two-temperatures fit of the observed H$_2$ excitation diagram reproduces the data satisfactorily and gives a rough estimation of the excitation temperatures of the gas. The warm molecular gas is constrained by the S(0)/S(1) line ratio and is consistent with $T_{\rm exc} = 185 \pm 30$~K and an H$_2$ column density of $2.06 \times 10^{20}$~cm$^{-2}$. The higher order S(3)/S(5) [ortho] transitions give $T_{\rm exc} = 675 \pm 80$~K and an H$_2$ column density of $1.54 \times 10^{18}$~cm$^{-2}$. \citet{Appleton2006} assume an ortho-to-para ratio of 3 and note no obvious deviations from this value in the data.

The SL spectroscopy shows that the H$_2$ emission is extended on the scale of 52'' (24 kpc). Assuming that the extra, undetected H$_2$ lines represent 40\% of the flux detected by \textit{Spitzer},  \citet{Appleton2006} derive an total H$_2$ luminosity of $8.4 \times 10^{33}$~W from the SH slit area ($11.3 \times 4.7$~arcsec$^{2}$).

The bottom right inset of Fig.~\ref{SQ_IRS_obs_Appleton06} shows one of the remarkable results of these observations: the H$_2$ $17\,\mu$m 0-0 S(1) line is resolved, though the relatively low resolution ($\mathcal{R} = 600$) of the \textit{IRS}. The S(1) line is extremely broad, with a FWHM linewidth of $\Delta v = 870 \pm 60$~km~s$^{-1}$. This is the broadest H$_2$ line ever observed!
The high velocity dispersion  of the warm molecular gas suggest that the postshock medium in the ridge is extremely turbulent. This is a puzzling result since H$_2$ is a fragile molecule and easily dissociated in $V_{\rm s} > 50$~km~s$^{-1}$ shocks. 

We summarize the main unexpected results from the \citet{Appleton2006} observational paper:
\begin{enumerate}
\item 
the H$_2$ gas is coexisting with a hot ($ \sim 5 \times 10^6$~K), X-ray emitting plasma.
\item 
the velocity dispersion of the H$_2$ gas seems unusually large and of the order of the velocity of the galaxy collision $\sim 900$~km~s$^{-1}$, which is of the order of the velocity of the galaxy collision.
\item
 the total H$_2$ line luminosity within the SH slit aperture exceeds that of the X-rays in this area by a factor of 3. 
\item
the spectroscopic signatures (dust or ionized gas lines) of star formation are much fainter than the $\rm H_2$ emission. This is unlike what is observed in star forming galaxies where the $\rm H_2$ lines are much weaker than the mid-IR dust features \citep{Rigopoulou2002, Higdon2006, Roussel2007}.
\end{enumerate}

\subsection{Astrophysical questions}
\label{H2_SQ_questions}

These surprising observations raise the following questions:
\begin{itemize}
\item Why is there H$_2$ gas is the extreme environment of the SQ postshock gas?
\item How can we account for the H$_2$ excitation characteristics?
\item Why is H$_2$ a dominant coolant?
\end{itemize}

Much of my thesis work has been driven by these astrophysical questions, which are addressed in \citet{Guillard2009} (\hyperref[paper_SQ_H2]{paper~{\sc i}}). In the following, I only discuss the scenario proposed to explain the formation of H$_2$ in the SQ shock.
The two last issues will be discussed in the next chapter (chap.~\ref{chapter:H2_SQ_mapping}), where I extend the analysis of \hyperref[paper_SQ_H2]{paper~{\sc i}}, which was based on the single-pointing observations presented above, to the full area of the SQ ridge.

\section{How does H$_{\bf 2}$ form in the Stephan's Quintet giant shock?}
\label{sec:H2formation_SQ}

In this section we focus on the first out of our three astrophysical questions: the formation of H$_2$ in the SQ ridge. We introduce the scenario presented in \citet{Guillard2009}  (\hyperref[paper_SQ_H2]{paper~{\sc i}}).  I will first show that we have explored several working hypothesis before coming to this scenario. Obviously, our scenario may not be the right solution, but, to our knowledge, this is the first quantitative model that attempt to explain the H$_2$ formation and emission within the broader context of multi-wavelength observations of SQ. Up to now, it has not been invalidated by observations.

\subsection{A long route to a coherent scenario\dots}
\label{subsec:route2H2formationscenario}

The main difficulty in understanding the presence of H$_2$ in the SQ shock is to explain how the gas can cool and become molecular in a hot plasma, and how it can survive in such a violent environment. Inspired by previous studies, \citet{Appleton2006} evoke two possibilities: \textit{(i)} an \textit{oblique} shock geometry, and \textit{(ii)}, the presence of \textit{dust} in the hot plasma. Both ingredients will help diminishing the gas temperature and make its cooling easier to form H$_2$ within the age of the collision. These hypothesis may also explain why the gas temperature in the ridge is not much hotter than that of the underlying halo \citep[see sect.~\ref{subsubsec:Xray_obs_SQ} and][]{Trinchieri2003, Xu2003, O'Sullivan2009}.
However, these ingredients alone are not sufficient enough to explain the coexistence of warm molecular gas with a hot plasma. I will show that the multiphased nature of the shocked gas is a key to interpret  multi-wavelength observations of SQ. 

\subsubsection{An oblique shock geometry?}
\index{Stephan's Quintet!oblique shock}

The collision between the new intruder, NGC~7318b, and the intra-group tidal tail is not necessarily head-on.
The shock associated with this collision may be oblique, i.e. the upstream fluid crosses the shock front with an incident angle $\phi \neq 90$\degre (see sect.~\ref{subsec:collision-two-gas-flows} for a  description of oblique supersonic collisions between two gas streams). 
This hypothesis is considered by \citet{Trinchieri2003, O'Sullivan2009} to explain with there is so little difference between the gas temperature in the radio ridge ($6.9 \times 10^6$~K) and outside, in the halo ($5.7 \times 10^6$~K). \citet{Trinchieri2003, O'Sullivan2009} assume a shock velocity equal to the relative line-of-sight velocity between the new intruder and NGC~7319's tidal tail ($900-1000$~km~s$^{-1}$). In the case of a perpendicular shock, a $1000$~km~s$^{-1}$ shock would lead to a post shock temperature of $\sim 1.4\times 10^7$~K (see Eq.~\ref{eq_postshock_temp_norm}), which is twice the temperature derived from X-ray observations in the ridge. If the incoming flow forms an angle $\phi$ with the shock surface, then the shock velocity is reduced by a factor $\sin \phi$. In this case, the postshock temperature can be written as
\begin{equation}
\label{eq_postshock_temp_oblique}
T_{\rm ps} = \frac{2(\gamma - 1)}{(\gamma + 1)^{2}} \, \frac{\mu}{k_{\rm B}} \, V_{\rm s}^{2} \sin ^2 \phi  \simeq 6.9 \times 10^6 \, \left( \frac{V_{\rm s}}{1000 \ \rm km \, s^{-1}}\right) ^{2} \sin ^2 \left( \frac{\phi }{30^{\circ}} \right) \ \ \rm K \; ,
\end{equation}
where $V_{\rm s}$ is the velocity of the shock wave, $\mu$ the mean particle mass (equals to $10^{-24}$~g for a fully ionized gas), $k_B$ the Boltzmann constant and $\gamma = 5/3$. 
Therefore, an incidence of $\phi \sim 30$\degre would be required to match the observed temperature of the hot gas in the ridge. 

The possibility that the large-scale shock is oblique may be an argument to reduce the X-ray power with respect to the more luminous H$_2$ emission in the ridge. However, it does not explain why part of the gas has cooled down  efficiently to form H$_2$, and part of the gas is still hot and emits X-rays. 

\subsubsection{A significant cooling of the hot gas by the dust?}
\index{Stephan's Quintet!dust cooling}

Based on \textit{ISO} IR observations, \citet{Xu2003} suggest that dust grains are the main coolant of the hot gas, which produce FIR emission. After subtraction of the IR emission from the individual sources in the group, they  detect a residual IR emission from the shock. The IR emission is extended, its morphology is similar to the radio emission in the ridge, and is one order of magnitude more luminous than the X-ray emission. They interpret this IR emission by collisional heating of dust grains by the hot electrons and ions of the plasma. Assuming a constant dust-to-gas mass ratio of $6 \times 10^{-3}$, they estimate a cooling time scale of $t_{\rm cool} \sim 2 \times 10^6$~yr, which is two orders of magnitude shorter than if the plasma is dust-free (see sect.~\ref{sec:time-dependent-cooling-plasma}). They use the collisional heating model of \citet{Popescu2000} and find that the expected FIR luminosity is in agreement with observations.

The problem is that the cooling timescale is so short that we should observe cool gas around the shock, which is not seen on the X-ray temperature maps \citep{O'Sullivan2009}. This scenario has difficulties to explain why we observed both a diffuse hot ($6-7 \times 10^6$~K) component in the ridge and in its outskirts, and warm ($100-1000$~K) molecular gas.
In fact, at the beginning of our study, we started to work in the direction of  \citet{Xu2003} to explain the cooling of the gas by the dust grains in the ridge. This is why I decided to carry on a detailed  calculation of the time-dependant cooling of the hot gas, taking into account dust destruction. This modeling has been presented in sect.~\ref{sec:time-dependent-cooling-plasma} and in \hyperref[paper_SQ_H2]{paper~{\sc i}}. The results lead us to revise the conclusions of \citet{Xu2003}, and we discuss the application of this modeling to the case of SQ in  sect.~\ref{subsubsec:gas-cooling-dust-survival-H2-formation}.

We find that the cooling timescale of the hot gas in the ridge is more than one order of magnitude longer than the shock age. We conclude that it is very unlikely that the H$_2$ gas results from the direct cooling of the hot gas. 
This result has driven us to the idea that, if the H$_2$ formation was induced by the galaxy collision, it must have cooled down from much lower temperatures. Then the gas in the ridge must have been shocked by a wide distribution of shock velocities, a thus of postshock temperatures. This is how we came to consider that the preshock medium was inhomogeneous. 
In the following, we present our scenario for H$_2$ formation. New IR observations and modeling of the dust emission in the SQ shock within the framework of this scenario will be discussed in chapter~\ref{chapter:SQ_dust}. 

\subsection{H$_2$ formation out of multiphase postshock gas}
\label{subsec:SQ-H2formation-multiphase-gas}

The difficulties evoked above suggest that it is very difficult to explain H$_2$ formation from homogeneous shocked gas at high-speed. In our view, the presence of H$_2$ in the SQ ridge is closely related to the multiphase structure of the preshock gas and may result from two possibilities:
\begin{enumerate}
\item molecular clouds are present in the preshock gas and embedded within the pre-existing halo of hot gas. Because of the density contrast between the tenous gas and the clouds, slower shocks are driven into 
the clouds (see sect.~\ref{sec:shocksinmultiphasemedia}), and H$_2$ molecules may or may not be dissociated, depending on the shock velocity. If the shock transmitted into the cloud is dissociative,
H$_2$ molecules may reform in the postshock medium \citep[see sect.~\ref{sec:MHD_shocks} and][]{Hollenbach1979}.
If gravitationally bound molecular clouds are present in the preshock medium, we expect the shock to trigger their collapse, which would lead to star formation \citep{Jog1992}. 
This process is perhaps what is happenning in the northern region of the shock, SQ-A, where optical and IR imaging show that a starburst is occuring \citep{Xu1999}.

The weakness of tracers of star formation in the region of the main shock \citep{Xu2003} suggests that preshock GMCs are not a major source of the observed postshock H$_2$ gas in the ridge. 
Although we do not exclude pre-existing molecular gas in the ridge, we explore the following second possibility.

\item

H$_2$ gas forms out of preshock H$\,${\sc i} gas. \citet{Appleton2006} also evoke this scenario by proposing that a large-scale shock overruns a clumpy preshock medium. H$_2$ molecules would form in the denser regions that experience slower shocks compared with the high-speed shock in the tenuous gas. 
In the following we summarize and complete the results presented in our  \hyperref[paper_SQ_H2]{paper~{\sc i}}.
\end{enumerate}
 
\subsubsection{Supersonic collision between two inhomogeneous gas streams} 

Multiwavelength observations of SQ suggest that two multiphase gas flows are colliding (see  \hyperref[paper_SQ_H2]{paper~{\sc i}} for an introduction to the past observations and to the astrophysical context that are  relevant for our study\footnote{A more complete review of SQ observations is differed to the sect.~\ref{subsec:obs_galaxy-wide_shock} of  chapter~\ref{chapter:H2_SQ_mapping} because this observational context is needed to interpret our new extended observations of SQ.}). One stream is associated with the intruding galaxy (NGC~7318b), and the other is associated with the tidal tail of NGC~7319. Both gas streams are likely to contain hot (a few million K), H$\,${\sc i} and perhaps molecular clouds. Therefore we consider the collision of two flows of multiphase dusty gas. 
The relative velocity between the flow is $900-1000$~km~s$^{-1}$, constrained by H$\,${\sc i} observations and optical line spectroscopy. 
Our model quantifies the gas cooling, dust destruction, and H$_{2}$ formation in the postshock medium within this framework.

If the colliding gas flows are inhomogeneous, the rise of pressure in the hot, tenuous gas drives slower shocks into the denser clouds (see sect.~\ref{sec:shocksinmultiphasemedia}).
Each cloud density corresponds to a shock velocity. 
The galaxy collision generates a range of shock velocities and postshock gas temperatures. The state of the postshock gas is related to the preshock gas density. Schematically, low density (preshock at $n_{\rm H} \lesssim 3 \times 10^{-3}\,$cm$^{-3}$) gas is shocked at high speed ($\gtrsim~700\,$km~s$^{-1}$) and accounts for the X-ray emission. Denser H$\,${\sc i} gas ($n_{\rm H}  \gtrsim \times 10^{-2}$) is heated to lower ($ \lesssim10^6$~K) temperatures. 
The fig.~1 of \hyperref[paper_SQ_H2]{paper~{\sc i}} sketches this scenario. The rise of pressure in the hot gas triggers the compression of the H$\,${\sc i} clouds that are the sites for H$_2$ formation.

In the following we compute the isobaric cooling, the evolution of the dust-to-gas mass ratio, and the formation of molecules in this multi-temperature postshock gas (see chapter~\ref{chapter:dust_gas_galaxies} and  chapter~\ref{chapter:shocks} for a description of the microphysics included in  the calculation). We show that the hot plasma did not have time to cool down significantly  since the collision was initiated, whereas the denser gas has time to cool and form $\rm H_2$.

\subsubsection{Gas cooling, dust survival and H$_2$ formation in the SQ ridge}
\label{subsubsec:gas-cooling-dust-survival-H2-formation}

We summarize here the context and main results of our calculation of the time-dependence of the gas temperature and dust-to-gas mass ratio. 
We start  the computation with a galactic dust-to-gas mass ratio and equilibrium ionization. We assume that the metallicity of the gas is solar and that the cooling gas is in pressure equilibrium with the  hot, volume-filling gas (the average thermal pressure of the hot plasma is $P / k_{\rm B} = 2 \times 10^5$~K~cm$^{-3}$).  As the gas cools, it condenses. In \hyperref[paper_SQ_H2]{paper~{\sc i}} we explore the possibility that the gas pressure vary around this average value (see in particular Fig.~3).

The H$_2$ formation, gas cooling, and dust destruction timescales are plotted as a function of the postshock temperature in the key-figure 2 of \hyperref[paper_SQ_H2]{paper~{\sc i}}.
The gas cooling timescale is defined as the time at which the temperatures coolds down to $10^4$~K. The H$_2$ formation timescale is the time when the H$_2$ fractional abundance reaches  90\% of its final (end of cooling) value, including the gas cooling time from the postshock temperature to $10^4$~K. The dust destruction timescale is the time when 90\% of the dust mass is destroyed and returned to the gas. We remind that a given initial postshock temperature is associated to a given preshock density. 
The relative values of the timescales with respect to the collision age define three gas phases:

\begin{description}
\item[The hot plasma:] the preshock gas at $n_{\rm H} \lesssim 6 \times 10^{-3}\ \rm cm^{-3}$  is shocked at $V_{\rm s} > 500$~km~s$^{-1}$ and therefore heated to temperatures $T_{\rm ps} > 5 \times 10^6$~K. Our estimate of the cooling timescale of the hot plasma takes into account the time-dependence of the dust-to-gas mass ratio. 
Fig.~2 of \hyperref[paper_SQ_H2]{paper~{\sc i}} shows that,  because the dust grains are efficiently sputtered in the hot gas, the cooling  timescale is much longer ($t_{\rm cool} = 2 \times 10^7$~yr for $T_{\rm ps} = 7 \times 10^6$~K) than if one estimates it for a constant dust-to-gas mass ratio \citep[$t_{\rm cool} = 2 \times 10^6$~yr as estimated by][]{Xu2003}. The cooling timescale of the hot gas is longer than the age of the collision, so the tenuous, shocked gas has not yet cooled down significantly.
This is the X-ray emitting plasma indicated by the red thick line.
Note that this hot gas may still contain some dust because the largest  grains (representing most of the mass) may not have been completely sputtered within the collision age. However, these grains are the less numerous, so their contribution to the gas cooling is weak (see chapter~\ref{chapter:dust_gas_galaxies}, sect.~\ref{sec_life_dust} and \ref{sec:time-dependent-cooling-plasma} for a quantitative discussion of these processes).

\item[The molecular gas:] The preshock gas at $n_{\rm H} \gtrsim 3 \times 10^{-2}\ \rm cm^{-3}$ is shocked at 
$V_{\rm s} < 250$~km~s$^{-1}$ and therefore heated to temperatures $T_{\rm ps} \lesssim 8 \times  10^{5}$~K. In this case, the gas cooling timescale to $10^4$~K and the $\rm H_2$ formation timescale are shorter than the collision age, whereas the dust destruction timescale is longer. Therefore, this gas keeps a significant fraction of its initial dust content and becomes $\rm H_2$ gas (blue thick line) within the age of the collision.

\item[Atomic and ionized  gas:] for intermediate densities ($6 \times 10^{-3} \lesssim n_{\rm H} \lesssim 3 \times 10^{-2}$) and therefore intermediate postshock temperatures ($8 \times 10^{5} < T_{\rm ps} < 5 \times 10^{6}$~K), the gas has time to cool down but it loses its dust content and does not have time to form $\rm H_2$ within the collision age. This phase is indicated with the grey thick line and may represents H$\,${\sc i} and H$\,${\sc ii} gas phases.
\end{description}

This description of the multiphase postshock ignores the fragmentation of the clouds by thermal and hydrodynamical instabilities (Rayleigh-Taylor and Kelvin-Helmholtz) that develop when a cloud is shocked. As discussed in sect.~\ref{evolution-shocked-molecular-cloud}, this fragmentation can lead to the mixing (evaporation), and therefore the destruction, of the clouds with the background hot plasma. 
The powerful H$_2$ emission in the SQ shock suggest that warm molecular gas is formed before the clouds mixing timescale. In sect.~3.3 of \hyperref[paper_SQ_H2]{paper~{\sc i}} we compare the H$_2$ formation timescale to the fragmentation and mixing timescales. The conclusions are the following:

\begin{description}
\item[Cloud fragmentation:]
we show that for a wide range and preshock cloud densities and sizes (see Fig.~4,  \hyperref[paper_SQ_H2]{paper~{\sc i}}), the crushing time is longer than the cooling, so the gas fragments by thermal instability as the shock penetrates into the cloud. 
H$_2$ forms before the shock has crossed the cloud, and therefore before hydrodynamical instabilities start to develop. 

\item[Cloud evaporation:] analytical and numerical studies suggest that the typical evaporation timescale of warm ($\sim 10^4$~K) H$\,${\sc i} clouds in hot ($> 10^6$~K) gas is a few million years (see sect.~\ref{turbulent-mixing} for a description of turbulent mixing and references).  This is long enough for H$_2$ to form within the lifetime of the fragments.
The cloud lifetime may be even longer because they are stabilized by the efficient H$_2$ cooling. Up to now, none of the simulations include both the detailed micro-physics (realistic cooling functions and H$_2$ formation on dust grains), and the full dynamic range of spatial scales and densities that are involved in the evolution of clouds embedded into a flow of hot gas (see sect.~\ref{turbulent-mixing} for the description of the state-of-art simulations).
\end{description}

\section{Summary and conclusions}
\label{sec:SQ_H2-conclusions}

One of the most interesting features of the Stephan's Quintet compact group is the presence of a galaxy-wide shock in the halo of the group, created by an intruding galaxy colliding with a tidal tail at a relative velocity of $\sim 1\,000$~km~s$^{-1}$.  Evidence for a group-wide shock comes from observations of X-rays from the hot postshock gas in the ridge strong radio synchrotron emission from the radio emitting plasma and shocked-gas excitation diagnostics from optical emission lines. \textit{Spitzer} observations show that this gas also contains molecular hydrogen with an extreme velocity dispersion. The $\rm H_2$ luminosity  is larger than the X-ray emission from the same region, thus the $\rm H_2$ line emission dominates over X-ray cooling in the shock. As such, it plays a major role in the energy dissipation and evolution of the postshock gas. 

The interpretation of these observations is the core of my PhD work. I present a scenario where H$_2$ gas forms out of the postshock gas that results from the supersonic collision between two multiphase gas streams. 
This collision leads to a multiphase medium with a distribution of shock velocities, related to the variations of the density in the preshock medium. The H$_2$ formation in the SQ ridge environment is quantified by calculating the time-dependent isobaric cooling of the gas, including dust destruction mechanisms. I show that H$_2$ forms out of the dense regions where dust survives the shocks. 
The bulk kinetic energy of the molecular gas is the main energy reservoir of the postshock gas, and it has to be dissipated within the H$_2$ gas to explain the poserful H$_2$ excitation. I propose that supersonic turbulence is sustained within the molecular gas by a transfer of momentum induced by cloud-cloud collisions and the relative shear motions between the molecular gas and the hot plasma. A phenomenological model of low-velocity MHD shocks driven into the dense H$_2$ gas is capable of explaining the observed excitation characteristics.

Within this framework, dust is a key ingredient. It is required for H$_2$ formation. Our model has implications on the expected dust emission in the SQ halo and on dust survival in a multiphase environment. These points are adressed in chapter~\ref{chapter:SQ_dust}.

\section{Publication: paper I}
\label{paper_SQ_H2}

In the following, we reproduce the  \citet{Guillard2009} paper, \textit{H$_2$ formation and excitation in the Stephan's Quintet galaxy-wide collision}, 2009, A\&A, 502, 515. This paper contain key-figures that I have discussed in this chapter, and a lot of details that I have intentionally not repeated in this manuscript.

\includepdf[noautoscale=true, scale=0.9,link=true,frame=false,pages=-]{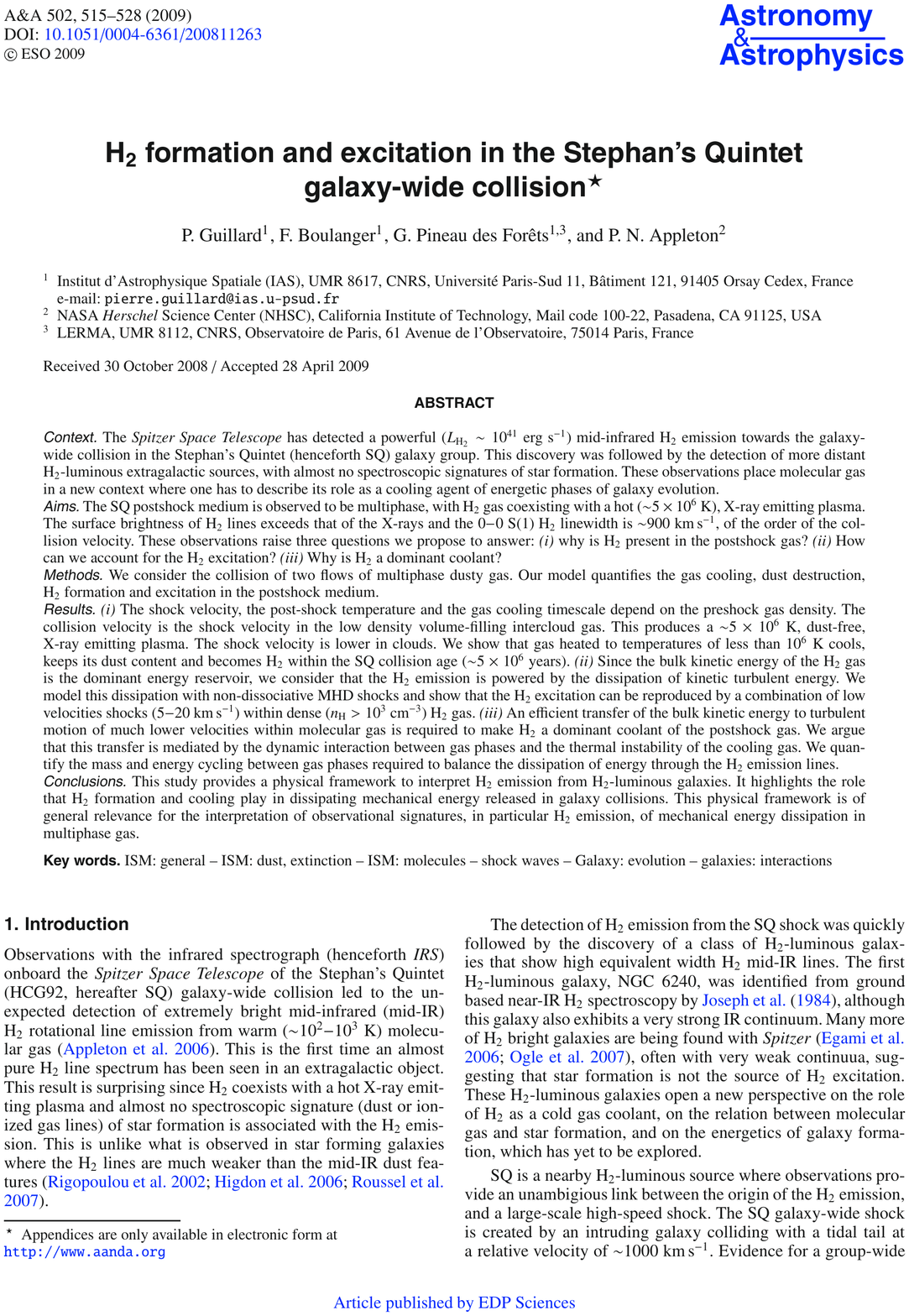}


\chapter{Mapping the mid-IR line cooling in Stephan's Quintet}
\label{chapter:H2_SQ_mapping}

\epigraph{It's just like when you've got some coffee that's too black, which means it's too strong. What do you do? You integrate it with cream, you make it weak. But if you pour too much cream in it, you won't even know you ever had coffee. It used to be hot, it becomes cool. It used to be strong, it becomes weak. It used to wake you up, now it puts you to sleep.}{Malcolm X}


\begin{Abstract}

The observations of H$_2$ rotational line emission from the Stephan's Quintet (SQ) galaxy collision presented in chapter~\ref{chapter:H2_SQ} are extended to a complete, and high signal-to-noise spectral mapping of the SQ giant shock. We find that the H$_2$ line emission is extended over the full ridge ($\approx 15 \times 35$~kpc$^{2}$).  I present these recent observations and the new astrophysical questions they raise, in particular on the dynamical history of the group. I thus introduce some observational background needed to elucidate these questions. Then, I present an update of the mass and energy budgets of the galaxy collision based on these new data. I also performed a complete re-analysis of the new data with shock models, and discuss the issue of H$_2$ excitation in the shock. This analysis is complemented by the use of fast shocks models to interpret the fine-structure line emission from the ionized gas. 

\end{Abstract}

\minitoc

\index{Stephan's Quintet}

\section{Introduction}
\label{sec:SQ_intro}

\PARstart{T}he  interpretation of H$_2$ observations presented in the previous chapter (chap.~\ref{chapter:H2_SQ}) and in  \citet{Guillard2009} (\hyperref[paper_SQ_H2]{paper~{\sc i}}) is based on the single-pointing observations by \citet{Appleton2006}, who first detected H$_2$ emission from the Stephan's Quintet (henceforth SQ) shock. 
The spectacular result of \citet{Appleton2006} led us to propose a new set of \textit{Spitzer} observations to obtain a full mid-IR spectral map of the  X-ray bright shock region. I have participated in the preparation of this proposal (P.I.: P.N. Appleton) by performing the model calculations that support it. It was accepted and observed in December 2007 and January 2008. Both mid-infrared (mid-IR) imaging and spectroscopy were performed. 

These new observations were reduced by M. Cluver and P. Appleton at the SSC and the results are presented in \citet{Cluver2009} (hereafter  \hyperref[subsec:paper_Cluver]{paper~{\sc ii}}). I have participated to the interpretation of these data and to the writing of this paper. Thanks to this new  \textit{IRS} spectral mapping of the SQ ridge, we have now an almost complete census of the mid-IR warm H$_2$ emission, fine-structure line and dust emission in the SQ shock. 
These observations allow to complement the mass and energy budget of the collision presented in \hyperref[paper_SQ_H2]{paper~{\sc i}}, which was resticted to a small area of the giant shock. This energy budget is the first step to understand the mechanisms that produce the high H$_2$ luminosity over such a large area ($\approx 40 \times 15$~kpc). 
This leads me to do a complete re-analysis of the H$_2$ data, based on my previous interpretation work. I have also complemented this work by an interpretation of the emission from the ionized gas in the group (optical and fine-structure mid-IR lines).

This chapter extends and re-visits the analysis presented in chapter~\ref{chapter:H2_SQ} and in  \hyperref[paper_SQ_H2]{paper~{\sc i}} for the full area of the SQ shock. 
In sect.~\ref{spectral_IRSmap_H2} I briefly present the results of the spectral maps, focussing on H$_2$. Obviously, these new observations raise new astrophysical questions, formulated in sect.~\ref{subsec:questions-IRSmappingSQ}. In sect.~\ref{subsec:mass-energ-budget}, I present an updated mass and energy budget in the SQ shock, similar to that presented in \hyperref[paper_SQ_H2]{paper~{\sc i}}, but for the full extension of the ridge. 
Section~\ref{sec:SQ_context} describes past observations of SQ to set the astrophysical observational context, focusing on the complex dynamics of the galaxy group. This background seems crucial for the understanding of our new set of \textit{Spitzer} observations. Then I extend my modeling of  the H$_2$ excitation for the whole ridge (sect.~\ref{sec:H2-excitation-ridge}). 
Pursuing the work of  \hyperref[paper_SQ_H2]{paper~{\sc i}}, I discuss the dynamical picture of the energy transfer between the ISM phases that emerges from this interpretation, and the physical mechanisms that may explain how the bulk kinetic energy of the galaxy collision is transfered into the warm H$_2$ (sect.~\ref{sec:SQ-Why-H2-important-coolant}). 
Then I complement the H$_2$ analysis by a study of the excitation of the mid-IR fine structure lines in sect.~\ref{sec:opt-midIR-line-excit}.
Section~\ref{SQ-NGC7319-comments} comments about the AGN member of the group, NGC~7319, which shows an interesting H$_2$ feature revealed by our new \textit{IRS} spectral maps. A quick summary and concluding remarks end this chapter (sect.~\ref{sec:SQ-conclusions}).

The \hyperref[subsec:paper_Cluver]{paper~{\sc ii}} is reproduced at the end of this chapter because it contains the figures I will rely on in this chapter, and a lot of information that I have intentionally not repeated in this manuscript.


\section{Spectral mapping of the H$_{\bf 2}$ line emission in the SQ shock}
\label{spectral_IRSmap_H2}

\begin{figure}
   \centering
    \includegraphics[width=\textwidth]{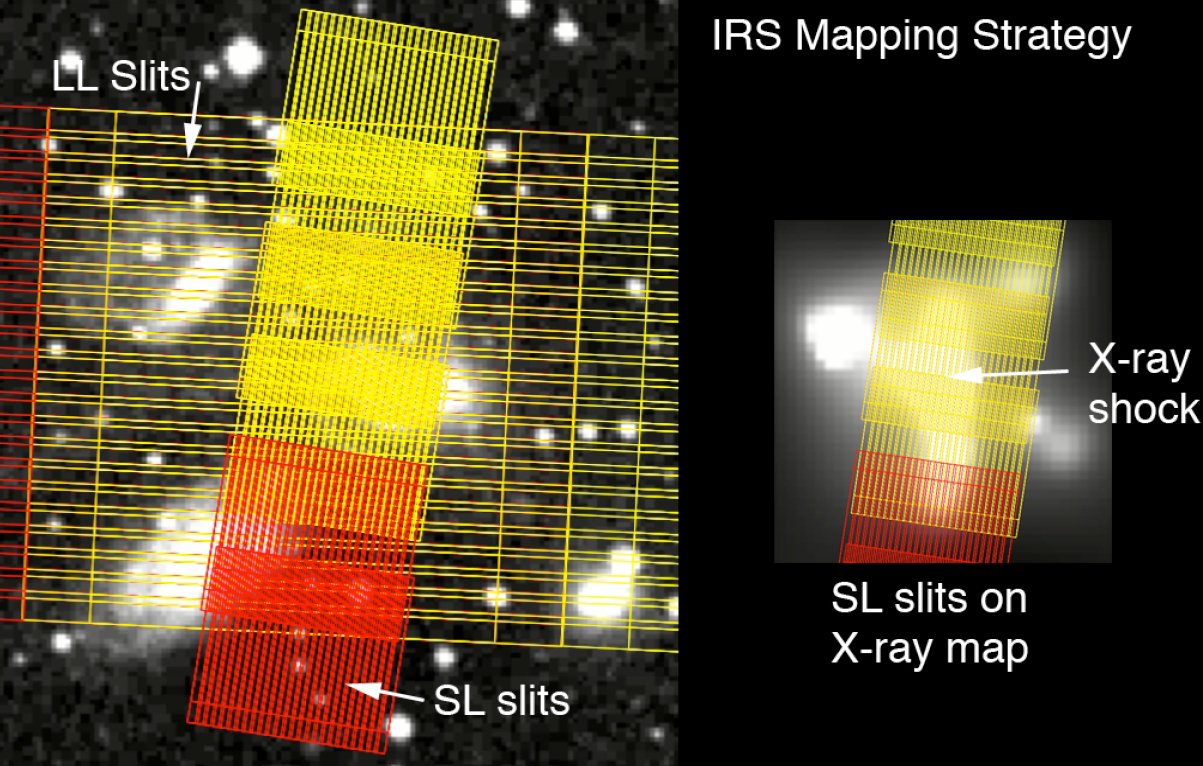}
      \caption[\textit{Spitzer IRS} mapping strategy]{The IRS mapping strategy. \textit{Left Panel:} The SL and LL slits superimposed over the DSS image of Stephan's Quintet. The SL map covers most of the intruder galaxy NGC~7318b. \textit{Right Panel: }The same SL slit-mapping superimposed on the Chandra image of the shock. The shock is the vertical X-ray structure. NGC~7318b does not show up in the X-ray image, although the brighter of the two X-ray sources just on the right-edge of the SL region is associated with NGC~7318a.}
       \label{SQ_IRS_MappingStrategy}
\end{figure}

The \textit{Spitzer IRS} mapping strategy is summarized in Fig.~\ref{SQ_IRS_MappingStrategy}. The SL and LL modules were used to map the entire shock region, the SL slits being oriented along the shock structure and the LL slits perpendicular to it.  The spectral mapping aims at spatially resolving the H$_2$ and dust across the shock.

\subsection{H$_{\bf 2}$ spatial distribution}

The new spectral maps are reported in \citet{Cluver2009} (hereafter  \hyperref[subsec:paper_Cluver]{paper~{\sc ii}}).  We summarize here the main observational results, focusing on  the distribution and characteristics of the  H$_2$ emission in the SQ intra-group. 
The main results regarded the mapping of other mid-IR fine-structure lines are discussed later in sect.~\ref{subsec:IRSmap_fine_struct_lines}.

\subsubsection{A remarkably extended H$_{\bf 2}$ emission!}

Fig.~2 of \hyperref[subsec:paper_Cluver]{paper~{\sc ii}} shows contours of the H$_2$ S(0)-S(5) line emission overlaid on an $R$-band image. The S(0) and S(1) lines were mapped by the LL modules, while S(2) - S(5) transitions were mapped by the SL modules of the \textit{IRS}. These spectral maps show an  extended emission from a giant filament  ($\sim 40$~kpc long) oriented north-south along the direction of the \textit{``main shock''}. This ``H$_2$ ridge'' spatially correlates very well with the radio or the X-ray emission. 
This provides an unambigious link between the origin of the $\rm H_2$ emission, and a large-scale high-speed shock generated by the galaxy collision between NGC~7318b and the tidal tail of NGC ~7319. We define the main shock aperture as a $\mathcal{A}_{\rm MS} = 77'' \times 30''$ rectangular aperture, corresponding to $\sim 13.7 \times  35.1$~kpc$^{2}$, centered on the coordinates $\alpha =  \rm 22h35m59.6s$, $\delta = +33$\degre$58'05.7''$. 

H$_2$ emission from the star forming region, SQ-A, as well as associated with NGC~7319 is also detected. We note that the $28.2 \, \mu$m S(0) line emission appears to be concentrated towards the SQ-A northern region. The maps also show an H$_2$ structure running eastward, from the center of the main shock to NGC~7319. We call this the \textit{``H$_2$ bridge''}, because it connects the ridge to the core of NGC~7319.  
This feature is also visible in radio,  H$\,\alpha$, and is and faintly detected in X-rays. The origin of this bridge feature is not clear, and we will come back to this point later (sect.~\ref{SQ-NGC7319-comments}). 

\subsubsection{Enormous H$_{\bf 2}$ fluxes!}

The spectra (see Fig.~8 of  \hyperref[subsec:paper_Cluver]{paper~{\sc ii}}) and the  H$_2$ line fluxes (see sect.~4 and table~2) have been calculated for different apertures in the ridge (Fig.~7). 
The spectra show the striking dominance of the H$_2$ lines in the mid-IR range of the \textit{IRS}. 
The thermal dust continuum is very weak in the center of the shock and in the bridge structure, whereas it is stronger for the northern region SQ-A where star formation is likely occurring.
For the main shock structure, the 0-0S(1) line luminosity alone, the sum of the S(0) to S(5), and the total H$_2$ (accounting for $\approx 30\,$\% of the flux in undetected lines) luminosities are respectively
\begin{eqnarray}
\mathcal{L}_{\rm S(1)}({\rm H_2}) &\approx&  2.3 \times 10^{34} \ \rm W \\
\mathcal{L}_{\rm S(0)-S(5)}({\rm H_2}) &\approx&  7 \times 10^{34} \ \rm W \\
\mathcal{L}_{\rm tot.}({\rm H_2}) &\approx&  9 \times 10^{34} \ \rm W
\label{eq:H2-line-lumin-total}
\end{eqnarray}
This exceptional power in the molecular hydrogen lines is one order of magnitude brighter than the next brightest mid-IR line, [Si$\,${\sc ii}]$\lambda$34.82$\mu$m, with a line luminosity of $\mathcal{L}(\rm Si  II)  = 8.5 \times 10^{33}$~W.

The mid-IR spectral mapping  will allow us to map the excitation of the H$_2$ along and across the X-ray shock. We have started to work on a two-dimensional modelling of the H$_2$ excitation.  We will create  maps the H$_2$ excitation temperature across the shock, and correlate this with dust emission SEDs, in order to  test models we are developing to explain the extreme H$_2$ power observed. This work will be pursued during my postdoc.

\subsection{Astrophysical questions raised by the new observations}
\label{subsec:questions-IRSmappingSQ}

The new spectral maps described above reveal the wide extension of the H$_2$ emission in the halo of the SQ group, as well as the spatial distribution of the mid-IR emission from dust and fine-structure lines. They essentially raise the main following questions:

\begin{itemize}
\item Why is the H$_2$ emission so extended? In particular, what is the origin of the H$_2$ emission in the bridge feature between the main shock and the AGN (NGC~7319) of the group?
\item Which mechanisms power the mid-IR fine-structure lines detected in the ridge? Is H$_2$ really the  dominant coolant in the shock?
\item Where does the emission come from? From the hot gas in the halo of the group, or from the molecular gas? 
\end{itemize}

The first issue is related to the geometry and dynamics of the galaxy interactions in the group. This is why we discuss this point in sect.~\ref{sec:SQ_context}. The second point is discussed in  \hyperref[subsec:paper_Cluver]{paper~{\sc ii}} and I have complemented this discussion in sect.~\ref{sec:opt-midIR-line-excit}. The dust emission in SQ is tackled in chapter~\ref{chapter:SQ_dust}.
Before addressing the two first questions, it is important to assess the energy content of the extended SQ shock. 

\subsection{Updated mass and energy budgets for the whole ridge}
\label{subsec:mass-energ-budget}
\renewcommand{\arraystretch}{1.1} 
\begin{sidewaystable}
\begin{center}
\begin{minipage}[t]{18cm}
\renewcommand{\footnoterule}{}
\def\thefootnote{\alph{footnote}}
\small
\centering
      \caption[Mass, energy and luminosity budgets of the SQ ridge]{Mass, energy and luminosity budgets of the preshock and postshock gas in the SQ ridge\protect\footnotemark[1]. }
    \begin{tabular}{c || c c c || c c c c c}
	\hline
	\hline
		& \multicolumn{3}{c ||}{{\sc Preshock Gas}} &  \multicolumn{5}{c}{{\sc Postshock Gas}} \\
  \hline
	\hline
   & Hot Plasma\footnotemark[2] & H{\sc i}\footnotemark[4] & $\rm H_2$ &  Hot Plasma\footnotemark[2] & H{\sc ii}\footnotemark[3]  & H{\sc i}\footnotemark[4] & Warm $\rm H_2$\footnotemark[5]& Cold $\rm H_2$ \\
  $n_{\rm H}$ [cm$^{-3}$] & $3.2 \times 10^{-3}$ & & & $1.17 \times 10^{-2}$ &  & & & \\
  $T$ [K] & $5.7 \times 10^{6}$ &  & & $6.9 \times 10^{6}$ &  & & $>10^2$ & $< 10^{2}$ \\
  $P / k_{\rm B}$ [ K cm$^{-3}$] &  &  & & $1.9 \times 10^{5}$ &  &  & &   \\
   $N_{\rm H}$ [cm$^{-2}$] & & $3 \times 10^{20}$ & & & $1.4 \times 10^{19}$  & $< 5.8 \times 10^{19}$ & $2 \times 10^{20}$ & \\
  \hline
	\hline
	\multirow{2}*{Masses [$\rm M_{\odot}$]} & Hot Plasma\footnotemark[2] & H{\sc i}\footnotemark[4] & $\rm H_2$ &  Hot Plasma\footnotemark[2] & H{\sc ii}\footnotemark[3]  & H{\sc i}\footnotemark[4] & Warm $\rm H_2$\footnotemark[5]& Cold $\rm H_2$ \\
     & $ 2.4 \times 10^8 $ &  $0.8 - 2.5 \times 10^{9}$ &  &  $ 5.3 \times 10^{8}$ & $4.9 \times 10^7$ & $< 2 \times 10^8$  & $ 1.2 \times 10^9$ & \\
 \hline
	\hline
	\multirow{3}*{Energy [erg]} & \multicolumn{1}{c}{Thermal} & \multicolumn{2}{c ||}{Bulk Kinetic} & \multicolumn{2}{c}{Thermal} & \multicolumn{3}{c}{Bulk Kinetic\footnotemark[6]} \\
& \multicolumn{1}{c}{(Plasma in halo)} & \multicolumn{2}{c ||}{(Shocked H$\,${\sc i} gas)} & \multicolumn{2}{c}{(X-ray emitting plasma)} & \multicolumn{3}{c}{(Turbulent $\rm H_2$)} \\
& $3.4  \times 10^{56}$ & \multicolumn{2}{c ||}{$0.4  - 1.2 \times 10^{58}$} &  \multicolumn{2}{c}{$ 9 \times 10^{56}$} & \multicolumn{3}{c}{$\gtrsim 1.8 \times 10^{57}$} \\
	\hline
	\hline
  \multirow{2}*{Flux [W m$^{-2}$]} & \multicolumn{1}{c}{X-rays} & & & \multicolumn{1}{c}{X-rays} & \multicolumn{1}{c}{H$\, \alpha$} &   \multicolumn{1}{c}{O{\sc i}} & \multicolumn{1}{c}{H$_{2}$ \footnotemark[5]}\\
& $6.6 \times 10^{-17}$ & & & $31.4 \times 10^{-17}$ & $7.8 \times 10^{-17}$ & $5.5 \times 10^{-17}$ & $74.6 \times 10^{-17}$ \\
\hline
 \multirow{2}*{$\mathcal{L}$\footnotemark[7]  [erg s$^{-1}$]} &  \multirow{2}*{$6.2 \times 10^{40}$} &  \multicolumn{2}{c||}{} &  \multirow{2}*{$2.95 \times 10^{41}$} &  \multirow{2}*{$6.5 \times 10^{40}$} &  \multirow{2}*{$4.6 \times 10^{40}$} &  \multirow{2}*{$8 \times 10^{41}$} \\
& & & & & & & &  \\
  \hline
	\hline
    \end{tabular}
    \label{table_mass_NRJ_budgets_SQ}
\footnotetext[1]{All the numbers are scaled to our main shock aperture ($A_{\rm MS} = 70'' \times 30'' = 33.8 \times 13.2$~kpc$^{2}$) where \textit{Spitzer} mapping was performed. The preshock gas is mainly the H{\sc i} gas contained in the tidal tail of NGC~7319. After the shock, the mass is distributed between the hot X-ray emitting plasma and the  $\rm H_2$ gas (see text for details). Before the shock, most of the energy available is the kinetic energy of the H{\sc i} gas that will be shocked. The shock splits the energy budget in two parts: thermal and kinetic energy. The former is stored in the hot plasma whereas the latter goes into turbulent motions that heat the $\rm H_2$ gas. Observations show that mechanical energy is the dominant energy reservoir.}
\footnotetext[2]{\textit{Chandra} observations of the extended X-ray emission in the shock and the tidal tail \citep{O'Sullivan2009}.}
\footnotetext[3]{H$\, \alpha$ and O{\sc i} optical line observations by  \citet{Xu2003}.}
\footnotetext[4]{Based on extrapolation of H{\sc i} observations in the tidal tail by \citet{Sulentic2001, Williams2002}.}
\footnotetext[5]{From \textit{Spitzer IRS} observations. The $\rm H_2$ line flux is summed over the S(0) to S(5) lines \citep{Cluver2009}.}
\footnotetext[6]{This bulk kinetic energy is a lower limit because an unknown mass of turbulent cold molecular gas can contribute to this energy.}
\footnotetext[7]{Luminosities are indicated assuming a distance to SQ of 94~Mpc.}
\normalsize
\end{minipage}
\end{center}
\end{sidewaystable}
\renewcommand{\arraystretch}{1.0} 

In \citet{Guillard2009}, we use the multiwavelength observations (see sect.~\ref{sec:SQ_context}) to estimate the masses, thermal and mechanical energy, and luminosities in the preshock and postshock medium (see table~1 of \hyperref[paper_SQ_H2]{paper~{\sc i}}). At that time, only the single-pointing H$_2$ observations by \citet{Appleton2006} were available, so the mass and energy budgets were calculated within the \textit{IRS}  SH slit aperture. Now that we have a complete spectral mapping of the ridge, we are able to estimate the physical quantities within the full shock region. The details on how we determine the quantities from the observations are given in sect.~2.2 and 2.4 of \hyperref[paper_SQ_H2]{paper~{\sc i}}. The new results for the whole ridge are gathered in table~\ref{table_mass_NRJ_budgets_SQ}. All the quantities are estimated  within the ``main shock'' $77'' \times 30''$ aperture defined above. 

The conclusions are the same as those of \citet{Guillard2009}. In the preshock medium, the thermal energy of the hot plasma ($\frac{3}{2} \frac{M_X}{m_{\rm H}} T_X$) is only a few percent of the bulk kinetic energy of the H$\,${\sc i} gas. In the postshock medium, the main energy reservoir is the mechanical energy associated with the high-velocity dispersion of the warm H$_2$ gas. All the mechanical energy of the collision is not dissipated in the hot plasma, but a substantial amount is carried by the warm H$_2$. We show that the bulk kinetic energy of the molecular gas is $> 2$ times the thermal energy of the postshock plasma.


\section{A complex astrophysical context: SQ observational constrains}
\label{sec:SQ_context}

I describe here the SQ group of galaxies and give an overview of its multi-wavelength data (sect.~\ref{sec:SQ_context}). This review is focused on the dynamics of the group and on the evidence for a high-speed, galaxy-wide shock. This is crucial to understand the complexity of this interacting system and to set the context for the interpretation of the extended H$_2$ emission. 

\subsection{Are there five musicians playing in the Stephan's Quintet?}
\index{Stephan's Quintet!early litterature}
\index{Stephan's Quintet!population}
\index{Stephan's Quintet!description}

\begin{figure}
\begin{minipage}{\textwidth}
    \def\thefootnote{\alph{footnote}}
      \setlength{\footnotesep}{0pt}
   \centering
   \includegraphics[width=0.6\textwidth]{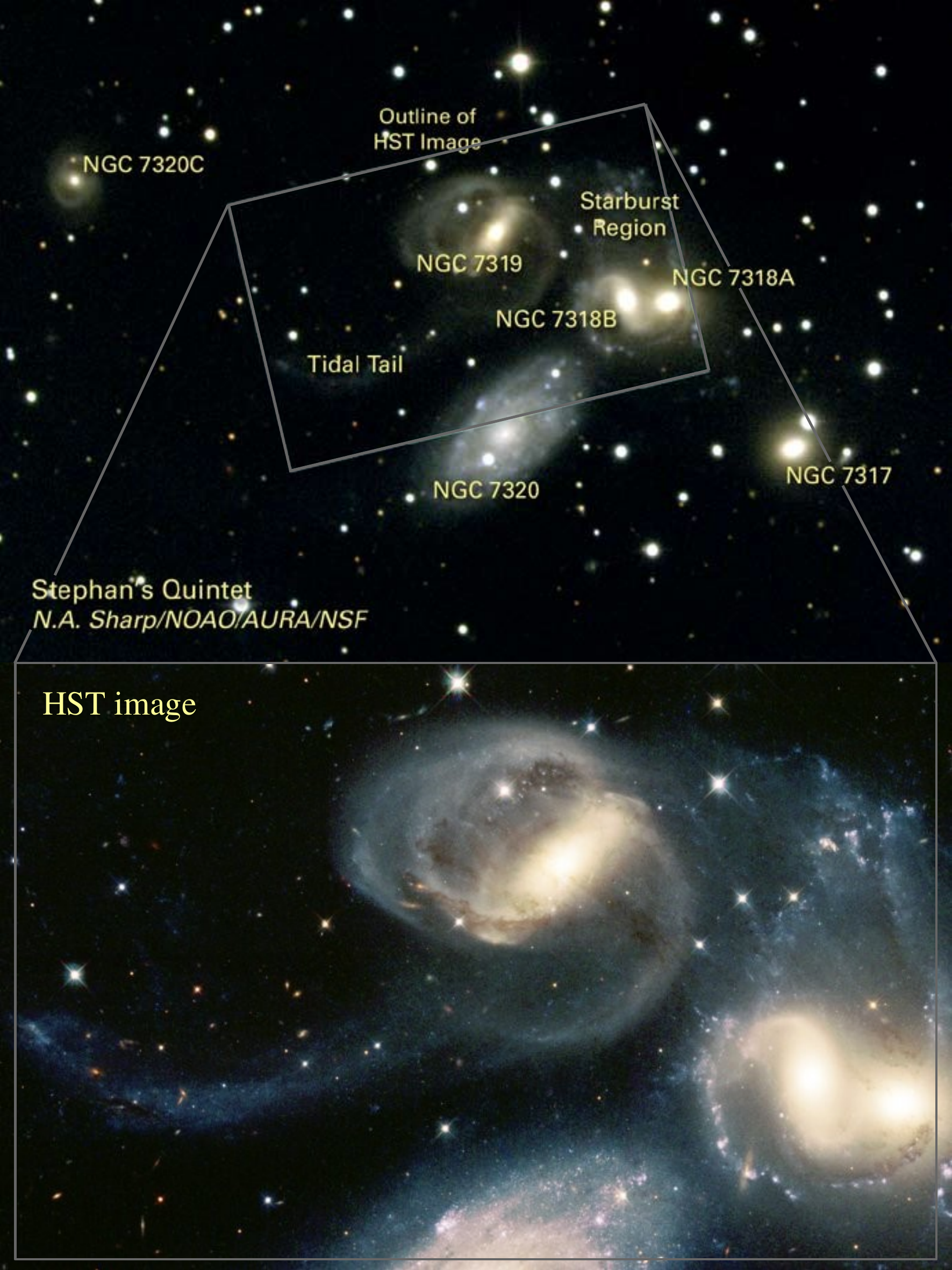}
      \caption[The main sources in Stephan's Quintet]{The main elements of Stephan's Quintet (North is up, East is left). The top panel shows a CCD image (NOAO\protect \footnotemark[1], Kitt Peak) of SQ. The five members of the quintet are: NGC~7317, NGC~7318a, NGC~7318b, NGC~7319, ans NGC~7320c. The NGC~7320 galaxy is a foreground. The bottom panel is the HST image\protect \footnotemark[2] of the center of SQ \citep{Gallagher2001}, which shows that stars are forming outside of the galactic disks of the group members, in the SQ halo. }
       \label{fig_SQ_HST_legend}
\footnotetext[1]{National Optical Astronomy Observatory, \url{http://www.noao.edu/}}
\footnotetext[2]{\url{http://hubblesite.org/newscenter/archive/releases/2001/22}}
\end{minipage}
   \end{figure}

SQ was discovered in 1877 by the french astronomer Edouard Jean-Marie Stephan ($1837-1923$) at the Observatory of Marseille, France \citep{Stephan1877}. He discovered a group of four ``elliptical nebulae'', today known as SQ, much earlier than anyone knew the existence of galaxies outside the Milky Way!
When one looks at a SQ image, one actually sees 6 galaxies: NGC~7317, NGC~7318a, NGC~7318b, NGC~7319, NGC~7320 and NGC~7320c (see Fig.~\ref{fig_SQ_HST_legend}).  
So how many galaxies belong to the group, 4, 5 or 6?

The first estimates of the redshifts of the SQ members were done through optical spectroscopy, and observations show a  wide range of values among the galaxies within the group \citep{Humason1956, Ambartsumian1958, Limber1960, Burbidge1961}. 
In particular, NGC~7320 has a recession velocity of $\sim 6000$~km~s$^{-1}$ less than NGC~7317, NGC~7318a and NGC~7319 that are at $\sim 6600$~km~s$^{-1}$ (see table~\ref{table_SQmembers}). In addition, NGC~7318b and NGC~7320c have respectively ``redshift'' $\sim 900$ and $\sim 700$~km~s$^{-1}$ less than the three members at $\sim 6600$~km~s$^{-1}$.
This puzzling result was the start of a long debate about the population of Stephan's Quintet.  Do all the galaxies belong to the group? How many galaxies are in gravitational interaction?

\begin{sidewaystable}
\small
\begin{center}
\begin{minipage}{\textwidth}
 \renewcommand{\footnoterule}{}
\def\thefootnote{\alph{footnote}}
 \caption[Characteristics of the galaxies lying in the Stephan's Quintet field of view]{Characteristics of the galaxies lying in the Stephan's Quintet field of view \footnotemark[1]}
\centering
\begin{tabular}{c c c c c c c }
\hline
\hline
			& NGC 7317 &	NGC 7318a &	NGC 7318b & 	NGC 7319 & NGC 7320c & NGC 7320 \\
\hline
Type \footnotemark[2] & 	 E4 &	 E2 pec &	 SB(s)bc pec &  SB(s)bc pec (Sy2  \footnotemark[6]) &	(R)SAB(s)0/a &  SA(s)d  \\
 	RA, $\alpha$ &	22 35' 51.9'' 	& 22 35' 56.8'' &	22 35' 58.5'' & 22 36' 03.7'' &	22 36' 20.4''& 22 36' 03.6s \\
 DEC, $\delta$ &	+33 56' 40.5" &	+33 57' 54.3" &	+33 57' 55.4" &	+33 58' 31.1" & +33 59' 06'' &	+33 56' 52.7" \\
Distance \footnotemark[3] & $93.3 \pm 6.5$  & $93.7 \pm 6.6 $ & $82.0 \pm 5.8 $  &	$95.4 \pm 6.7 $ & $ 84.9 \pm 5.9 $ & $ 13.7 \pm1.0$  \\
B-mag  \footnotemark[4] & 	15.3 & 14.4 & 14.06 & 14.8 & 17 & 13.8 \\
Ang. Size \footnotemark[5] & 	$0.512 \times 0.512$  &	$0.512 \times 0.512$  &	 $1.862 \times 1.348 $   &  $1.737 \times 1.348 $  & $ 0.575 \pm 0.489 $  & $2.041 \times 1.148$ \\
Velocity &  $6564 \pm 47 $ & $6597 \pm 32$ & $5727 \pm 24$ & $ 6578\pm49 $ & $ 5932 \pm 11$ & $738 \pm 48$ \\
Redshift & 	$0.02214 \pm 0.00016$ &	$0.02225 \pm 0.00011 $ &	$0.019289 \pm 0.00009 $ & $ 0.02219 \pm 0.00016$ &$ 0.019987 \pm 0.000037 $ & $0.00247 \pm 0.00016$ \\
\hline
\end{tabular}
\footnotetext[1]{Data extracted from the Simbad and NED databases. See Fig.~\ref{fig_SQ_HST_legend} for an image of SQ.}
\footnotetext[2]{Morphological type. E stands for elliptical galaxy, S for spiral.}
\footnotetext[3]{Galactocentric distance in Mpc}
\footnotetext[4]{Magnitude in the B-band ($0.83\,\mu$m)}
\footnotetext[5]{Angular size: major axis [arcmin] $\times$ minor axis [arcmin].}
\footnotetext[6]{ NGC~7319 is a Seyfert 2 galaxy showing a weak Active Galactic Nucleus (AGN) activity \citep{Durret1994a}}
\label{table_SQmembers}
\end{minipage}
\end{center}
\normalsize
\end{sidewaystable}

 \citet{Burbidge1961a} reported that the difference in redshift of $\sim 6000$~km~s$^{-1}$  between the high- and low-redshift members of SQ could only means one of two things: \textit{(i)} NGC~7320 is a foreground galaxy, or \textit{(ii)} NGC~7320 is literally exploding away from the other members. They estimated the probability of \textit{(i)} being true is $\sim 1/1500$. On the other hand, 
they stated that conclusion \textit{(ii)} ``is outside conventional ideas about the [dynamics] of galaxy groups''.
\citet{Arp1973} reports the first H$\alpha$ interference filter images of SQ. These images clearly separate the low- and high-redshifts systems, associated with the foreground galaxy NGC~7320 ($\sim 800$~km~s$^{-1}$) and the SQ group ($\sim 6000$~km~s$^{-1}$).
Based on the analysis of the distribution and diameters of the H$\,${\sc ii} regions, \citet{Arp1973, Arp1976} argued that NGC~7320 is physically associated with other SQ galaxies, and this was a prime case for the existence of non-Dopler redshift.

 \citet{Tammann1970, Allen1980, Moles1997} provided counter arguments against this claim. First, H$\,${\sc i} and H$\alpha$ observations show no evidence for NGC~7320 interaction with other SQ galaxies. Second, redshift independent estimates yield
to distances consistent with the redshifts. Third, the redshift of NGC~7320  is close to that of NGC~7331 ($818 \pm 5$~km~s$^{-1}$), so NGC~7320 appear to be a member of a foreground loose galaxy group which includes the large spiral galaxy NGC 7331. 
In the following, we thus assume that NGC~7320 is a foreground galaxy and it will not be further discussed hereafter.
 \citet{Arp1973a} suggested that NGC~7320c, the galaxy lying 4~arcmin to the
east of NGC 7319 (see Fig.~\ref{fig_SQ_HST_legend}, top panel) is linked to the group by optical tidal tails. Therefore the ``quintet'' status is restored!

In the bottom panel of Fig.~\ref{fig_SQ_HST_legend} we show a recent mosaic picture  taken by  \citet{Gallagher2001} with the Wide Field and Planetary Camera 2 (WFPC2) onboard the \textit{Hubble Space Telescope} (\textit{HST}), on Dec. 30, 1998 and June 17, 1999. One clearly sees on the image that galaxy interactions have distorted the galaxies' shapes, creating elongated spiral arms and long, gaseous streamers. Interestingly, this spectacular image reveals that star clusters are forming outside the galactic disks, in the SQ halo. Some of these star cluster contain a few millions of stars.

In the following, we assume the distance to the SQ group to be 94~Mpc (with a Hubble constant of 70 km~s$^{-1}$~Mpc$^{-2}$) and a systemic velocity for the group as a whole of $6\,600$~km~s$^{-1}$. At this distance, 10~arcsec$=4.56\,$kpc.

\subsection{The complex dynamical history of Stephan's Quintet}
\index{Stephan's Quintet!dynamical history}

\begin{figure}
   \centering
    \includegraphics[width=\textwidth]{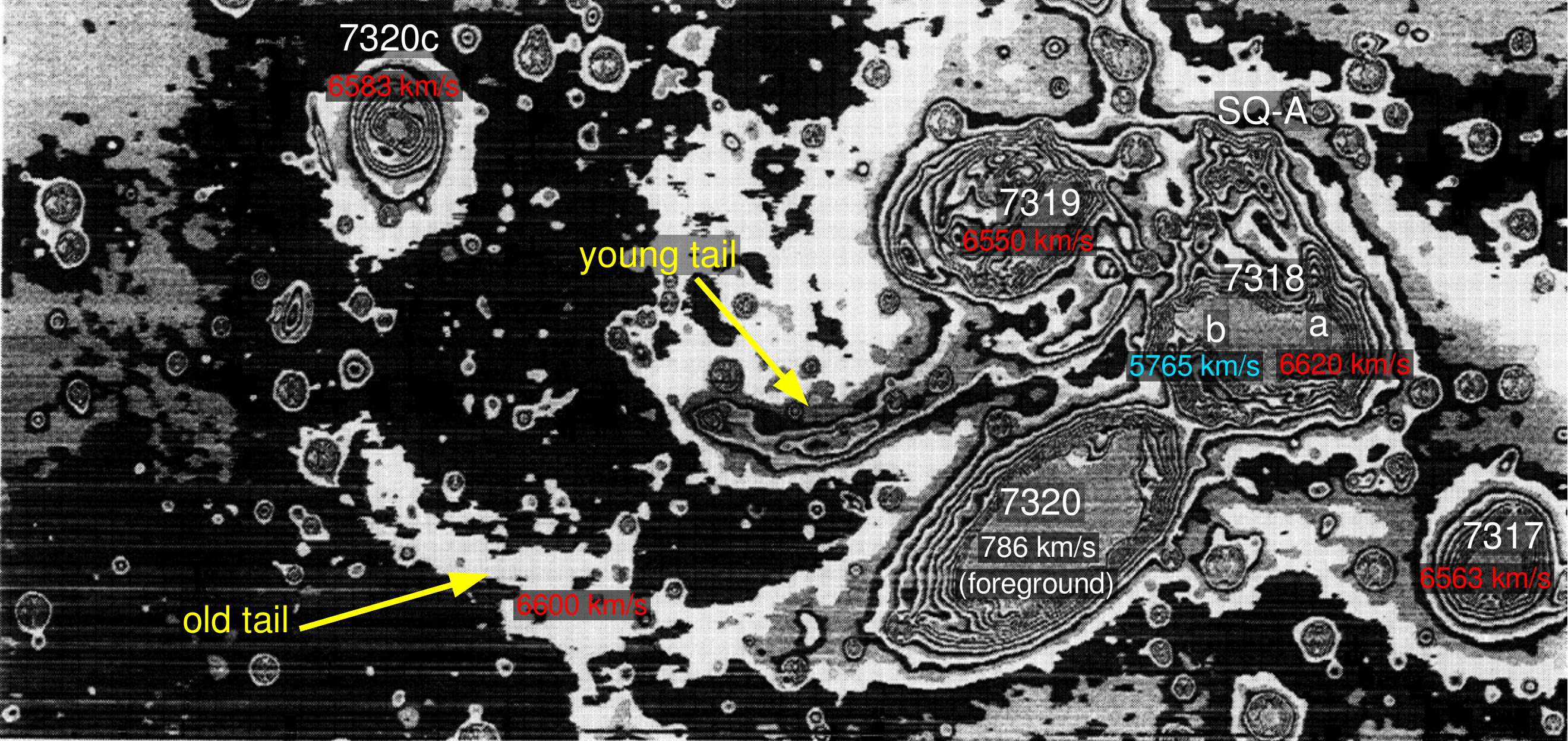}
      \caption[Tidal tails in Stephan's Quintet]{Iso-density image of an optical photographic plate (3h of observations) taken with the 200-inch telescope at Palomar \citep{Arp1973a}. We identify the members of the Stephan's Quintet group and the two tidal tails. Recent measurements of redshifts are indicated.}
       \label{fig_SQ_Palomar_Arp1973_tails}
   \end{figure}

The first deep optical images taken by  \citet{Arp1972, Arp1973a, Arp1976} show
 two concentrating parallel tidal tails, both starting from NGC~7319 and extending to
NGC~7320c, with the more diffuse one passing through the low redshift galaxy NGC 7320. Fig.~\ref{fig_SQ_Palomar_Arp1973_tails} presents an example of a photographic plate taken at the 200-inch Palomar telescope, where these two tidal tails are clearly visible. This is the first indication of a long list of observations that show that the group has a complex dynamical history.

As its population, the dynamical history of SQ has been a long debate that is not entirely solved yet. In the following, we first summarize the scenario of past and present dynamical interactions  that emerges from  multi-wavelengths observations. Then we detail the different arguments that support it, and we point out the remaining uncertainties.

\subsubsection{Summary of the scenario: multiple galaxy collisions}

In the past, an ``old intruder'', NGC~7320c, stripped most of the gas from group members (essentially NGC~7319 and NGC~7318a), and a new intruder, NGC~7318b is currently colliding with this gas and triggering a large scale shock. The old
intruder passed the core of SQ twice, first $\sim 5 - 7 \times 10^8$~yrs ago and creating the ``old tail'' (the southern tail in  Fig.~\ref{fig_SQ_Palomar_Arp1973_tails}), and second  about $2 \times 10^8$~yrs ago, triggering the ``young tail'' (the narrower one in Fig.~\ref{fig_SQ_Palomar_Arp1973_tails}).

\subsubsection{Observational constrains on past interactions with NGC~7320c}

The ``two-intruders'' scenario summarized above was clearly formulated by  \citet{Moles1997} and further confirmed by \citet{Sulentic2001}, who discussed in detail the link between the two tidal tails and the two passages of NGC~7320c through the kernel of the group. 

The \textit{HST} image (Fig.~\ref{fig_SQ_HST_legend}) highlights that star birth occurs mainly in three dynamically-perturbed sub-regions of the group: the long, sweeping tail and spiral arms of NGC 7319 [left]; the gaseous debris of the galaxy pair NGC~7318a and b [bottom right]; and an area north of those galaxies, the so-called SQ-A starburst region [top right]. Since galaxy interactions are known to trigger star formation, the SQ star formation history is closely linked to its dynamical history. Photometric analysis of the \textit{HST} images shows that the ages of these star clusters span over a wide range, spanning from $\sim 2 \times 10^{6}$~yrs to $\sim 3 \times 10^{9}$~yrs, which shows past and recent star formation \citep{Gallagher2001}. 
The southern, diffuse tail mostly contains an old population of star clusters, whereas the northern one is comprised of younger stars.

X-ray observations also support multiple episods of galaxy interaction in the SQ group. A very extended halo of hot ($\sim 6 \times 10^6$~K) gas was first detected by \citet{Bahcall1984}. They state that the extent of the halo is too big and its temperature too low to have been produced by the collision of the new intruder. 
They also show that the X-ray luminosity of the halo ($\mathcal{L}_X(0.1-3\rm \ keV) = 5 \times 10^{42}$~erg~s$^{-1}$) is 10 times the expected luminosity from the combined five SQ galaxies. 
Later, \citet{Moles1997, Sulentic2001, O'Sullivan2009} suggest that this X-ray emitting hot intracluster gas  was produced by previous galaxy encounters (with NGC~7320c for instance). We will further discuss X-ray observations in sect.~\ref{subsec:obs_galaxy-wide_shock}.

\subsubsection{The discovery of a galaxy-wide shock in Stephan's Quintet}

The first evidence for the recent collision between the intruder NGC~7318b and the IGM comes from two discoveries in the radio band, both made with the Westerbork Synthesis Telescope, Netherlands.

The first is a giant ($\sim 40$~kpc long) radio continuum ridge found in the IGM between NGC 7319 and NGC 7318b \citep{Allen1972}. Among other possible interpretations, \citet{Allen1972} suggested this may be a large scale shock triggered by an on-going collision between NGC 7318b (the recent intruder) and the rest of the group. The collision velocity is about one million mph (1000~km~s$^{-1}$)! A one hundred thousand light-year long giant shock is produced!  Stephan's Quintet is thus a multi-galaxy collision caught in action!

The second \citep{Allen1980, Shostak1984} is the revelation that nearly all the H$\,${\sc i} gas associated with SQ is lying outside galaxy disks, in the group halo. This represents a mass of neutral gas of $\sim 10^{10}$~M$_{\odot}$. These observations suggest that this gas has been stripped from late type galaxy disks due to an earlier interaction (a few $10^8$ yr ago) either between NGC~7319 and NGC~7318a \citep{Allen1980}, or between NGC 7319 and NGC 7320c \citep{Shostak1984}. In sect.~\ref{subsec:obs_galaxy-wide_shock} we discuss more recent observations of the radio-emitting ridge.

\subsubsection{Ongoing debate about the origin of the ``young'' tidal tail}

If most of the litterature now agrees on the origins of the old tail and of the main shock, the origin of the ``young tail'' is more debated. Instead of the ``two step'' scenario proposed by \citet{Moles1997, Sulentic2001}, \citet{Xu2005}
proposed a ``three-intruders'' scenario based on new UV observations.

They argue that it is unlikely that the young tail is also triggered by NGC 7320c. This is
because the recently measured redshift of NGC~7320c \citep[6583~km~s$^{-1}$][]{Sulentic2001} is almost identical
to that of NGC 7319, indicating a slow passage \citep{Sulentic2001} rather than a fast passage \citep[$\sim 700$~km~s$^{-1}$][]{Moles1997}. In order for NGC~7320c to move to its current position, the NGC 7319/7320c
encounter must have occurred $\gtrsim 5 \times 10^8$~yr ago. This is close to the age of the old tail, but older than that of the young tail. They suggested that the young tail is triggered by a close
encounter between the elliptical galaxy NGC~7318a and NGC~7319. The projected distance
between NGC~7318a/7319 is only $\sim 1/3$ of that between NGC~7320c/7319. Therefore the time
argument is in favor of the new scenario. 

Numerical N-body simulations are certainly needed to explore and test the dynamical scenarios inferred from observations. Recent work within the specific context of SQ has been made by \citet{Renaud2007}, and has to be pursued. We will not further discussed this debate and shall concentrate on the main shock. 

\subsection{Further evidence for a galaxy-wide shock}
\label{subsec:obs_galaxy-wide_shock}
\index{Stephan's Quintet!galaxy-wide shock}

In the following we further discuss multi-wavelength observations of Stephan's Quintet. These data sets allow to establish the energy and mass budget of the SQ galaxy collision, which is gathered in the Table~1 of \citet{Guillard2009} (\hyperref[paper_SQ_H2]{paper I}).

\subsubsection{Radio observations}
\label{subsubsec:SQ-radio-observations}
\index{Stephan's Quintet!radio observations}

\begin{figure}
   \centering
   \includegraphics[width=0.49\textwidth]{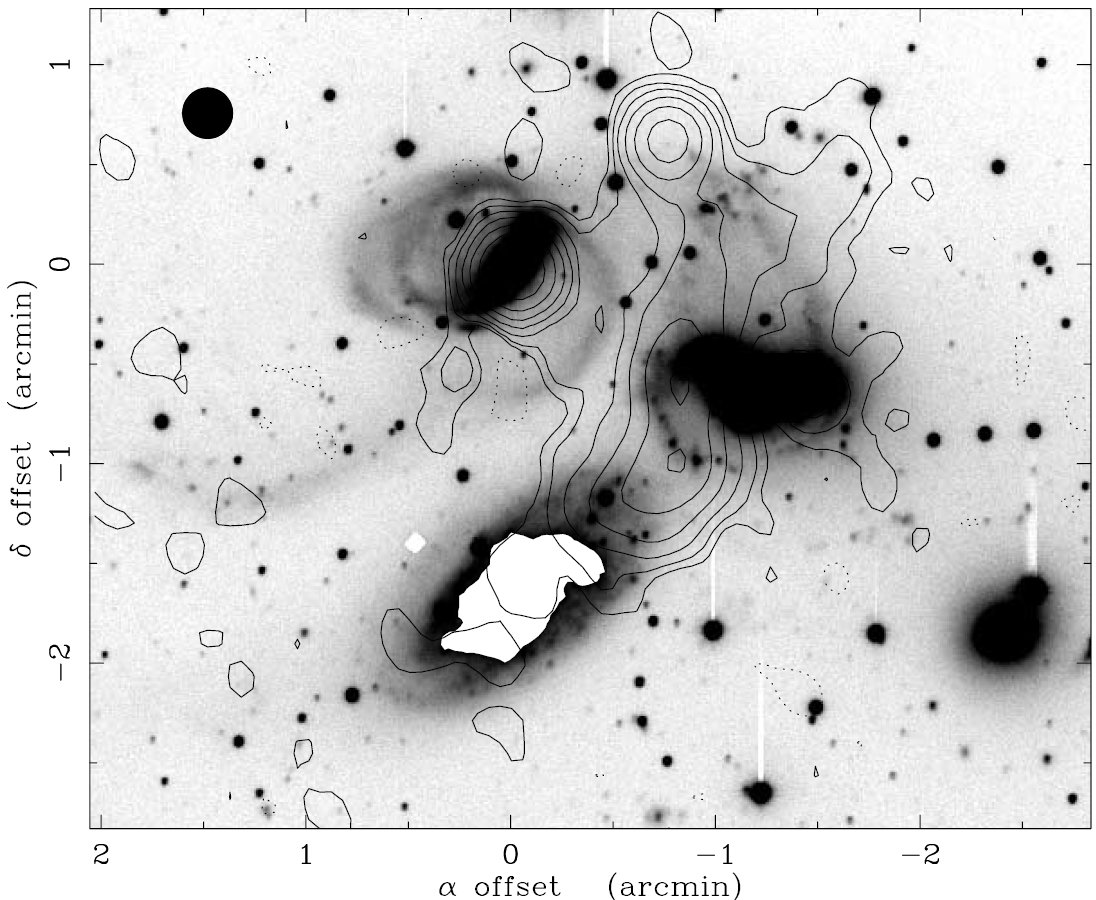}
  \includegraphics[width=0.49\textwidth]{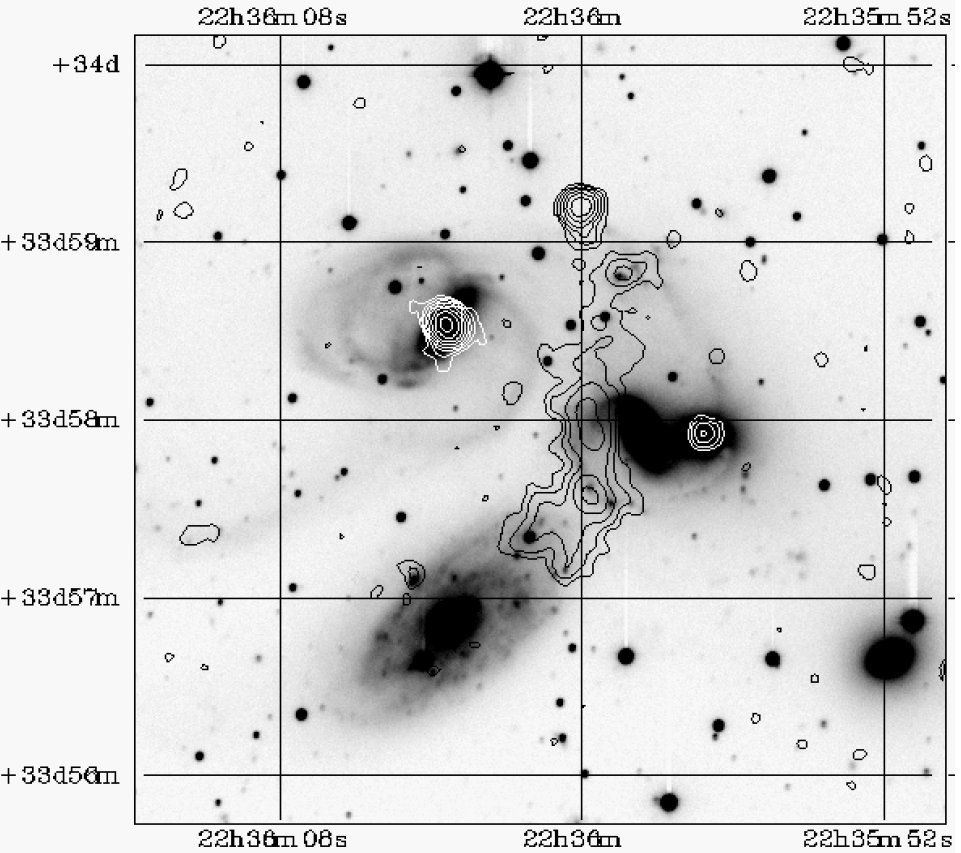}
      \caption[VLA radio continuum observations of Stephan's Quintet]{Radio continuum observations of  SQ. \textit{(Left):} Contour map of the 21 cm continuum emission at 1.4~GHz overlaid on an R-band image obtained by J. Sulentic at the 3.5 m Calar Alto telescope. The contours are $-0.2$, 0.2, 0.4, 0.8, 1.6, 3.2, 6.4, 13, and 26 mJy beam$^{-1}$, for an rms noise level of 0.1~mJy beam$^{-1}$ (0.3~K). The synthesized beam, $15.4'' \times 14.8''$, is shown on the upper left corner. Coordinate offsets are with respect to the optical nucleus of NGC 7319 at $\alpha_{\rm J 2000}$ 22 36' 03.7'', $\delta_{\rm J 2000}$ +33 58' 31.1". From \citet{Williams2002}. \textit{(Right):}   Contours of the radio continuum at 4.86 GHz (VLA B configuration) overlaid on an R-band CCD image \citep[from][]{Xu2003}. The lowest contour is 50 $\mu$Jy~beam$^{-1}$, and the spacing is equal to 2 in ratio. The FWHM of the synthesized beam is 6''.}
       \label{fig_SQ_Xrays_radio_contours}
   \end{figure}
\begin{figure}
   \centering
   \includegraphics[width=\textwidth]{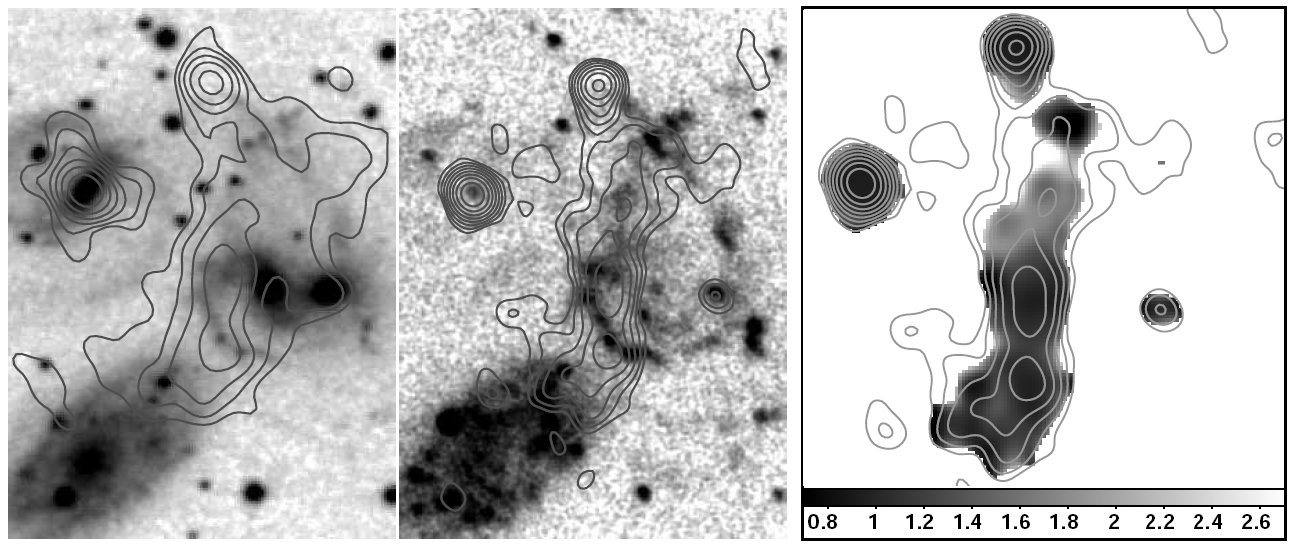}
      \caption[GMRT radio observations of Stephan's Quintet]{GMRT low-frequency observations of SQ by \citet{O'Sullivan2009a}.\textit{ Left:} 327 MHz GMRT contours overlaid on POSS blue image
of the core of Stephan's Quintet. Contours start from $3 \sigma = 0.9$~mJy~beam$^{-1}$,
HPBW (Half Power Beam Width) is $12'' \times 10''$. \textit{Center: } 610 MHz GMRT contours overlaid on a Swift UVOT UVW2-band image ($\sim \! 200$~nm). \textit{Right:} $1400-610$ MHz spectral index map (HPBW of $6''\times 5''$) with 610 MHz contours overlaid.}
       \label{fig_SQ_radio_GMRT_OSullivan2009}
   \end{figure}

The galaxy-wide radio ridge discovered by \citet{Allen1980}  has been confirmed
by  VLA\footnote{Very Large Array \url{http://www.vla.nrao.edu/}} observations by \citet{VanDerHulst1981, Williams2002, Xu2003}. Fig.~\ref{fig_SQ_Xrays_radio_contours} presents radio observations of SQ at 1.416 and 4.86~GHz. 
\citet{O'Sullivan2009a} extend these observations to lower frequencies (327 and 610~MHz) with the GMRT telescope. Fig.~\ref{fig_SQ_radio_GMRT_OSullivan2009} shows \textit{GMRT} radio contours overlaid on optical and UV images of the group core, and on a $1400-610$~MHz spectral index map. 
We summarize the main results:
\begin{itemize}
\item there is a bright north-south linear ridge of radio emission running in between NGC~7319 and NGC~7318b.  The diffuse radio emission is more extended at lower frequencies, including areas
west of the ridge. The 1.4~GHz image shows that faint emission connects this extended ridge to NGC~7319.
A faint extension to the northwest, coinciding with H$\alpha$ emission \citep{Plana1999, Sulentic2001}, is also detected.
The  total flux density associated with the extended radio ridge is $48 \pm 7$~mJy \citep{Williams2002}. 
\item  The spectral index map shows that the southern part of the
ridge has an index of $0.7-0.8$, while the northern part has a steeper spectrum \citep{O'Sullivan2009a}.
Some of the brightest spots of radio emission corresponds with knots of UV emission in the
south-eastern spiral arm of NGC 7318b. In this region the spectral index is consistent with star formation, so the underlying extended radio emission arising from shocks may be  enhanced by star formation in some areas.
\item the core of the Seyfert 2 NGC~7319 shows a radio source with a flux $28.5 \pm 0.5$~mJy. Both images also show that there is a radio source associated with the nucleus of NGC~7318a \citep[$0.95 \pm 0.05$~mJy][]{Xu2003}.
\item note that the bright source lying to the north of SQ-A is a backgroung distant binary source seen in projection behind SQ \citep{VanDerHulst1981, Williams2002}. 
\end{itemize}

The total luminosity emitted by the radio ridge at 1.4~GHz is $\mathcal{L}_{\rm1.4GHz} \sim 4 \times 10^{31}$~W from a volume extending about $90'' \times 15''$. 
This luminosity is 10 times the radio-luminosity of all star-forming regions in SQ (NGC~7318a, SQ-A, and a possible non-AGN component from 7319). \citet{VanDerHulst1981} proposed that the radio emission in the ridge is produced by an intra-cluster starburst. This would imply a star-formation rate of $\sim 30$~M$_{\odot}$~yr$^{-1}$, which is uncompatible with optical spectroscopy, H$\alpha$ imaging, or mid-IR observations (see chapter~\ref{chapter:SQ_dust}).

The radio ridge must be associated with synchrotron emission of relativistic electron accelerated in the giant shock created by the NGC~7318b galaxy that is entering into NGC~7319's tidal tail.
The ridge is the interface between the interloper NGC~7318b and the group. Indeed, the shocked, ionized gas is predicted to produce significant amounts of synchrotron emission \citep{VanDerHulst1981, Allen1972}. The synchrotron life time for the radiating electrons is estimated to be rather short, a few times $10^7$~yr, which is comparable to the crossing time for the collision.

Assuming the equipartition of energy between gas internal energy, cosmic rays and magnetic energy, \citet{Xu2003} estimate the magnetic field strength to be $B_{\rm eq} \simeq 10\,\mu$G.
The corresponding energy density of relativistic particles plus the magnetic fields is $\mathcal{U}_{\rm min} \sim 10^{11}$~erg~cm$^{-3}$. This is significantly lower than the thermal energy density implied by X-ray observations (see below). 

\subsubsection{Kinematics of the H$\,${\sc i} gas}
\index{Stephan's Quintet!H$\,${\sc i} observations}
\label{subsubsec:SQ_HI_kinematics}

\begin{figure}
   \centering
    \includegraphics[width=\textwidth]{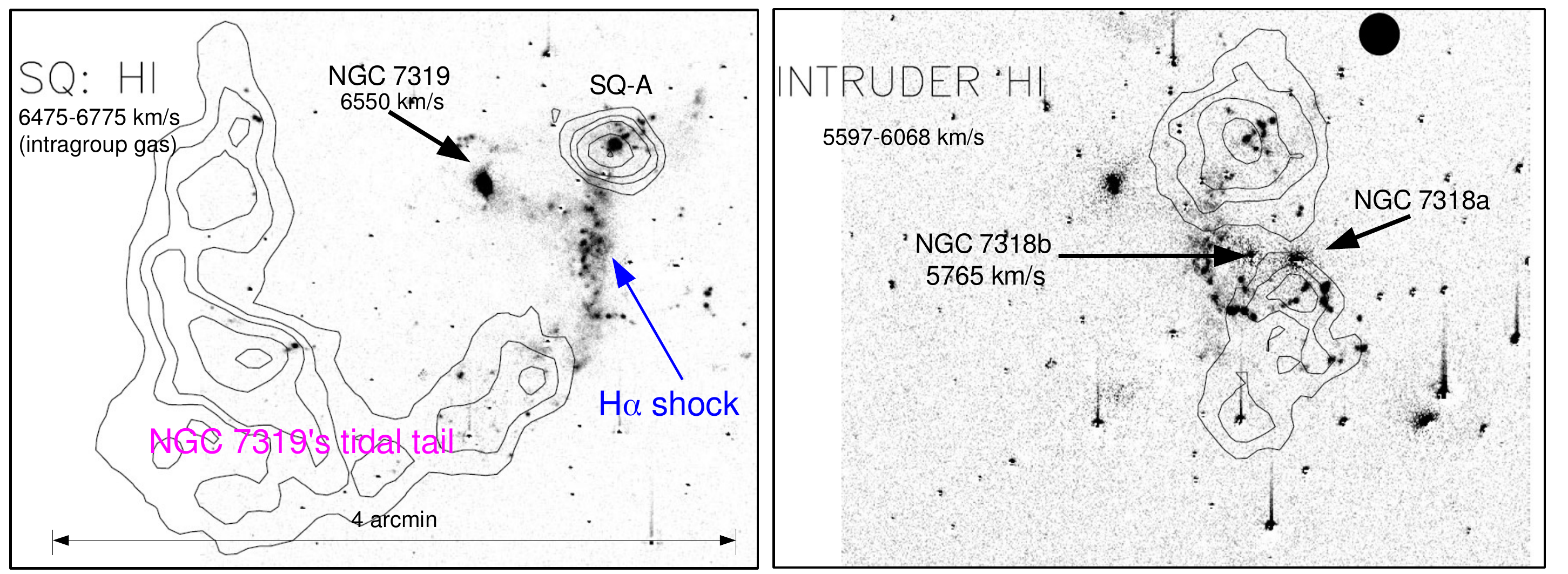}
      \caption[H$\,${\sc i} gas in Stephan's Quintet ]{H$\,${\sc i} 21 cm radio contours superposed on H$\alpha$ interference filter images. \textit{Left:} H{\sc i} contours for velocities in the SQ range $6475-6755$ km~s$^{-1}$. \textit{Right:} H{\sc i} contours for velocities in the new intruder range ($5597-6068$ km s$^{-1}$). H{\sc i} contours levels correspond to [0.526, 1.58, 2.63, 5.26, 7.89]$\times 10^{20}$~cm$^{-2}$ with last contour shown only in the left panel. Adapted from \citet{Sulentic2001}.}
       \label{fig_SQ_HI_contours_Sulentic}
   \end{figure}
\begin{figure}
   \centering
    \includegraphics[width=\textwidth]{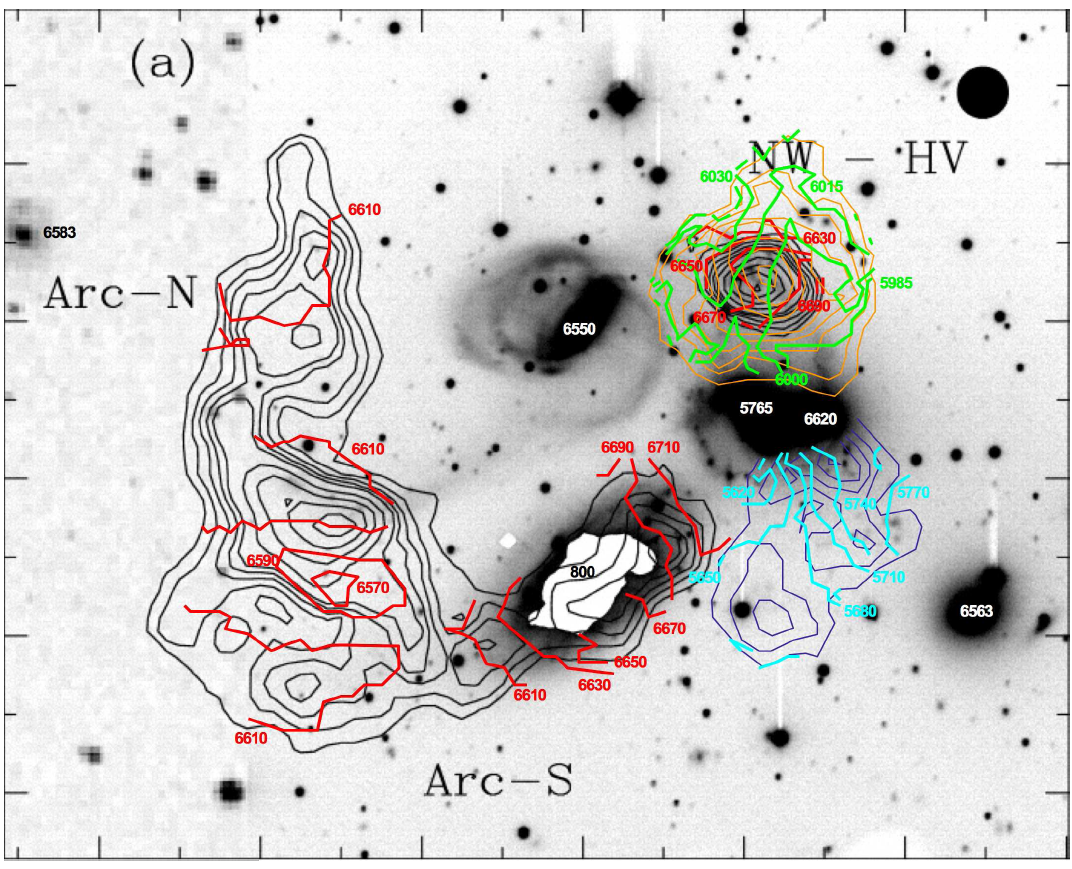}
      \caption[H$\,${\sc i} velocity maps in Stephan's Quintet]{The of H$\,${\sc i} column density contours and first-order moment of the radial velocity (lines and numbers in km~s$^{-1}$) for low (cyan and blue), intermediate (green
and orange) and high (red and black) velocity ranges. The galaxies velocities are from \citet{Sulentic2001}. This image is taken from \citet{Renaud2007}.}
       \label{fig_SQ_velocity_maps}
   \end{figure}

Fig.~\ref{fig_SQ_HI_contours_Sulentic} shows 21~cm observations of the H$\,${\sc i} line at two velocities, that of the intruding galaxy, $\sim 5700$~km~s$^{-1}$, and that of NGC~7319, $\sim 6700$~km~s$^{-1}$. 
The H$\,${\sc i} contours are overlaid on H$\alpha$ interference filter images centered at the intruder and intra-group velocities. 
These images will be discussed later, in sect.~\ref{subsec:opticallineemission_shocks}.
On Fig.~\ref{fig_SQ_velocity_maps}, the H$\,${\sc i} contours are overlaid  on an optical R-image. In colors we show the radial velocities of the gas. 
H{\sc i} observations exhibit a large stream created by tidal interactions between NGC~7320c, and NGC~7319 \citep{Moles1997, Sulentic2001}.  This stream corresponds to the ``old tidal tail'' (see Fig.~\ref{fig_SQ_Palomar_Arp1973_tails}) and the velocity of the gas in the tail is $6600 - 6700$~km~s$^{-1}$. No H$\,${\sc i} gas is detected in the younger tail. 
The SQ-A starburst region both contains H$\,${\sc i} gas at the intragroup (tidal tail) velocity ($\sim 6600$~km~s$^{-1}$) and at a velocity slightly higher that the intruder velocity ($\sim 6000$~km~s$^{-1}$), which suggests that this gas was accelerated by the shock.
Some H$\,${\sc i} gas associated with the intruder velocity ($\sim 5700$~km~s$^{-1}$) is detected to the south of NGC~7318a/b. 

The observed difference in radial velocities between the H$\,${\sc i} tidal tail and the intruder is about $1\,000$~km~s$^{-1}$. 
On Fig.~\ref{fig_SQ_HI_contours_Sulentic}, the 21~cm contours are overlaid on an H$\,\alpha$ Fabry-Pérot image. At the velocity of the intra-group medium, the H$\,\alpha$ image (left panel) clearly shows the linear feature associated with the shock.  Over the shock region, no H$\,${\sc i} gas is detected, but optical line emission from H$\,${\sc ii} gas is observed at
the H$\,${\sc i} tidal tail velocity. This suggests that \textit{the H$\,${\sc i} gas  present in the ridge before the recent galaxy collision was ionized  by the shock induced by the collision between the two gas flows}. It is likely that the H$\,${\sc i} gas at $\sim 6600$~km~s$^{-1}$ in SQ-A was physically connected to the   H$\,${\sc i}  tail.
We infer the preshock H$\,${\sc i} gas masses based on the $\,${\sc i} column densities observed on both sides of the ridge (see \hyperref[paper_SQ_H2]{paper I}). 

\subsubsection{X-ray observations}
\label{subsubsec:Xray_obs_SQ}
\index{Stephan's Quintet!X-ray observations}

\begin{figure}
   \centering
   \includegraphics[width=0.51\textwidth]{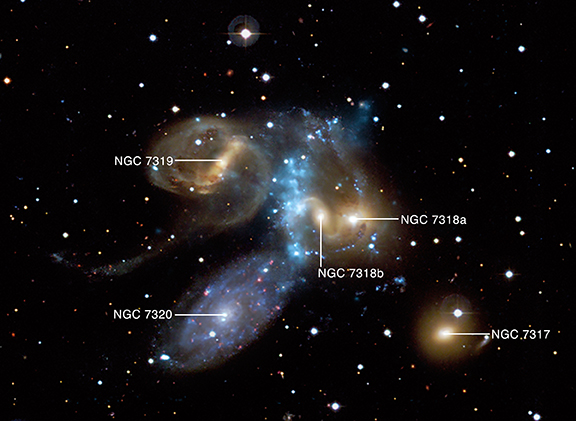}
     \includegraphics[width=0.48\textwidth]{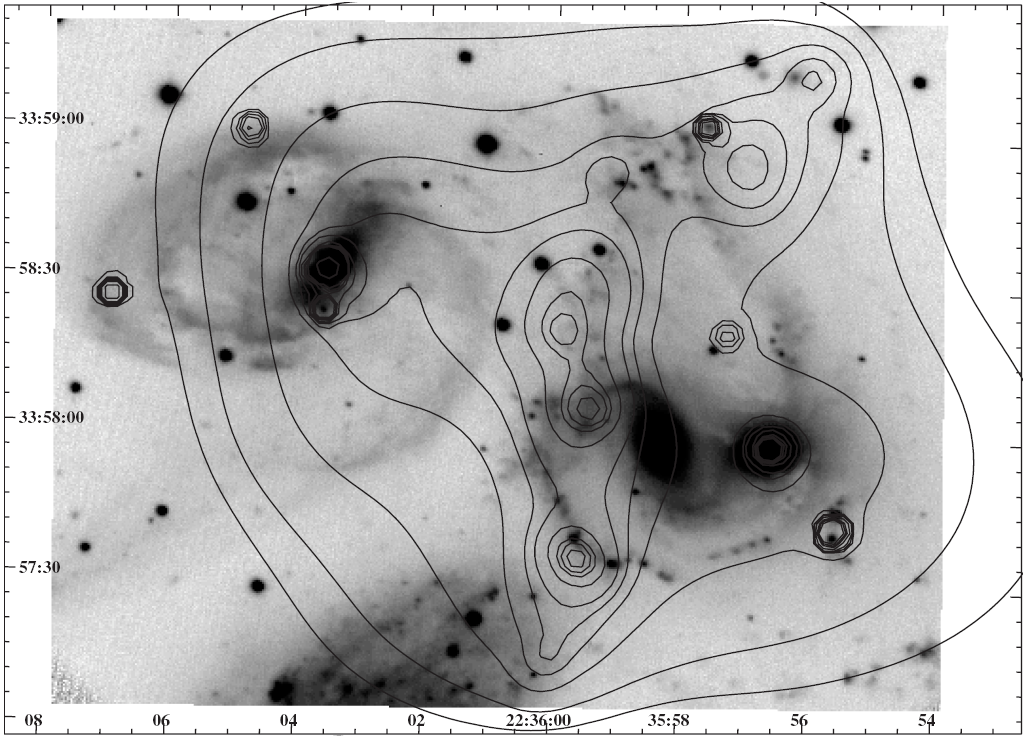}
      \caption[Chandra X-ray observations of Stephan's Quintet]{Chandra X-ray observations of SQ \textit{(Left)}: $0.3-2.0$~keV image (curved, light blue ridge running down the center of the image) overlaid on a deep CFHT image. \textit{(Right):} A zoomed X-ray contour map ($0.5-3$ keV) of the main concentration of X-ray photons \citep[from][]{Trinchieri2003}. Contours are superimposed on a CFHT B-band image of the field. The X-ray data are smoothed with an adaptive filter (FFT method) and 2.5$\, \sigma$ as the lowest significance of the signal within the kernel. 
}
       \label{fig_SQ_CFHT_Xray}
   \end{figure}

The first X-ray observations of SQ were performed with the \textit{Einstein Observatory} and were reported by \citet{Bahcall1984}. They found an extended halo but the spatial resolution was not high enough to resolve the structure of the ridge. 
The presence of shock-heated gas in the SQ halo was confirmed by X-ray \textit{ROSAT}, \textit{Chandra} and \textit{XMM} observations. 
X-ray maps \citep{Pietsch1997, Sulentic2001} revealed an extended and diffuse hot halo, plus features  associated with the radio ridge and with galaxy members of the group. 
The extent of the underlying X-ray emitting halo is too big to have been produced by the new interaction of NGC~7318b with the tidal tail, strongly suggesting that it must have been heated during previous dynamical encounters \citep{Moles1997, Sulentic2001, O'Sullivan2009}.
One third of the total X-ray luminosity, $\mathcal{L}_{X} = 5 \times 10^{41}$~erg~s$^{-1}$, is associated with the intergalactic ridge. The comparison of the relative structures in radio and X-ray bands (Fig.~\ref{fig_SQ_Xrays_radio_contours} and \ref{fig_SQ_CFHT_Xray}) show that the X-ray emission is brightest in the northern part of the ridge, while the radio emission is brightest in the south.

High spatial resolution maps \citep{Trinchieri2003, Trinchieri2005, O'Sullivan2009} and optical narrow-band H$\alpha$/[N$\,$II] maps \citep{Vilchez1998, Ohyama1998, Xu1999, Plana1999, Sulentic2001} show a similar structure, in both the hot gas and the ionized gas. The left panel of Fig.~\ref{fig_SQ_CFHT_Xray} shows a spectacular \textit{Chandra} image (blue) overlaid on an optical CFHT image.The right panel shows a zoomed gray-scale B-band image overlaid with \textit{Chandra} $0.5 - 3$~keV iso-contours. 

The width of the radio or X-ray emission in the shock is $\sim 5$~kpc, which is commensurate with the width of the H$\, \alpha$ ridge. Dividing the width of the ridge by the collision velocity (assumed to be $1\,000$~km~s$^{-1}$), we get a collision age of the order of $t_{\rm coll} \sim 5 \times 10^{6} \,$yr.

Spectral energy distribution fitting of \emph{Chandra} and \emph{XMM} observations of SQ show that the temperature of the hot plasma in the ridge is $T_{\rm X} = 6.9 \times 10^{6} \,$K \citep{Trinchieri2003, Trinchieri2005, O'Sullivan2009}. 
The postshock temperature for a perpendicular shock is given by \citep[see chapter~\ref{chapter:shocks}, Eq.~\ref{eq:postshock-temperature-Vs} or e.g.][]{Draine1993}:
\begin{equation}
\label{eq_postshock_temp_norm}
T_{\rm ps} = \frac{2(\gamma - 1)}{(\gamma + 1)^{2}} \, \frac{\mu}{k_{\rm B}} \, V_{\rm s}^{2} \simeq 6.94 \times 10^6 \, \left( \frac{V_{\rm s}}{715 \ \rm km \, s^{-1}}\right) ^{2} \ \ \rm K \; ,
\end{equation}
where $V_{\rm s}$ is the velocity of the shock wave, $\mu$ the mean particle mass (equals to $10^{-24}$~g for a fully ionized gas), $k_B$ the Boltzmann constant and $\gamma = 5/3$. Therefore, the postshock temperature of the hot plasma allows us to estimate a shock velocity of $\sim 700$~km~s$^{-1}$. 
We expect  the shock velocity to be equal to the gas velocity in the center of mass frame. 
If the volume-averaged gas densities are the same on the intruder side and on the side of the tidal tail  (see sect.~\ref{subsubsec:SQ_HI_kinematics} for details), the shock velocity would be approximately half of the observed relative velocity, so $1000 / 2 = 500$~km~s$^{-1}$. This is significantly lower than the shock velocity we deduce from X-ray observations, suggesting that the mass of the gas in the intruding galaxy is higher than in the tidal tail.

The fact that the shock velocity agrees with the line-of-sight relative velocity of the intruder suggests that its transverse velocity component is small. \citet{Sulentic2001} also argue that the distribution of the H$\,${\sc i} and H$\,${\sc ii} emissions support the idea of an ongoing collision of NGC~7318b with the tidail tail almost parallel to the line of sight. This conclusion has been revised by \citet{Trinchieri2003, O'Sullivan2009} who argued for an oblique shock geometry. They find that the temperature in the ridge ($6.9 \times 10^{6} \,$K) is close to that of the underlying hot halo ($5.7 \times 10^{6} \,$K). As they are assuming a shock velocity of $900-1000$~km~s$^{-1}$, they conclude that the temperature in the ridge is too low. The hypothesis of an oblique shock may help to decrease the main shock velocity and therefore the postshock temperature.  This will be further discussed in sect.~\ref{subsec:route2H2formationscenario}.

The only sign for any interaction between NGC 7317 with other members of SQ is a common
halo linking it to the binary NGC 7318, found in the deep X-ray and optical R-band images
\citep{Trinchieri2005}.

\section{How can we account for the H$_2$ excitation?}
\label{sec:H2-excitation-ridge}

In this section we address the question of the H$_2$ excitation in the ridge. Thanks to the  \textit{IRS} spectral mapping of the ridge presented in sect.~\ref{spectral_IRSmap_H2}, we have a complete census of the warm H$_2$ excitation over the extended region of the SQ ridge. I have extended and re-visited the analysis presented in chapter~\ref{chapter:H2_SQ} and in  \hyperref[paper_SQ_H2]{paper~{\sc i}} for the full area of the SQ shock. 

\subsection{Possible H$_2$ excitation mechanisms in the SQ shock}
\label{subsec:SQ-possible_H2-excitation-mec}

H$_2$ can be excited by several mechanisms: X-ray heating, cosmic-ray heating and shock excitation. We have discussed the micro-physics of these processes in sect.~\ref{subsec:H2excitation-mechanisms} and we discuss here which are the dominant processes for H$_2$ excitation in the environment of the SQ ridge.

\subsubsection{X-ray heating}

The physics of the H$_2$ excitation by X-rays is discussed in sect.~\ref{subsec:H2-collisional-excitation} and  \ref{subsec:H2-cosmic-X-ray-heating}. We show that the energy conversion from X-ray flux to the bolometric $\rm H_2$ emission is at most 10$\,$\% for a molecular cloud that is optically thick to X-ray photons. This excitation efficiency may be even smaller because the absorbed fraction of X-ray photons may be smaller than 1. We indeed expect that the surface filling factor of  the postshock $\rm H_2$ gas  is smaller than 1.

The \textit{Spitzer} observations show that the H$_2$ line luminosity domines the X-ray emission from the SQ shock region. This result is true for the whole SQ ridge, but also for sub-regions extracted in the center of the ridge or in the ``H$_2$ bridge structure'' towards NGC~7319 (see sect.~\ref{spectral_IRSmap_H2} and \hyperref[subsec:paper_Cluver]{paper~{\sc ii}}).
For the main shock, the total X-ray  flux is $\mathcal{F}_X (0.001-10\, \rm kev) \approx 
2.8 \times 10^{-16}$~W~m$^{-2}$, which corresponds to a luminosity of $\mathcal{L}_X (0.001-10\, \rm kev) \approx 2.95 \times 10^{34}$~W.  Therefore, we find an H$_2$ to X-ray luminosity ratio of
\begin{equation}
\frac{\mathcal{L}_{\rm tot.}(\rm H_2)}{\mathcal{L}_X (0.001-10\, \rm keV) } \approx 3
\end{equation}
Given the X-ray to H$_2$ flux conversion estimated previously, this ratio shows that the H$_2$ gas cannot be predominantly excited by the X-ray photons of the hot plasma. 

\subsubsection{Cosmic ray heating}

Warm $\rm H_2$ gas may also be heated by relativistic particles in the radio ridge. 
The physical processes related to H$_2$ excitation by cosmic-ray ionization have been discussed in sect.~\ref{subsec:H2-collisional-excitation} and \ref{subsec:H2-cosmic-X-ray-heating}. 
An enhanced cosmic ray density has been proposed to account for the high temperatures of molecular clouds in the Milky Way nuclear disk  \citep{Yusef-Zadeh2007} and for the mid-IR H$_2$ emission in the filaments of the Perseus A (3C~84) cool core cluster \citep{Ferland2008}. 
Is the cosmic-ray heating can explain the powerful H$_2$ emission in the SQ ridge?

Let us estimate the cosmic-ray ionization rate $\zeta$ that is needed to balance the cooling of the gas by H$_2$ line emission. 
In the SQ ridge, the total H$_2$ line luminosity is $9 \times 10^{34}$~W (see eq.~\ref{eq:H2-line-lumin-total}) and the estimated total mass of H$_2$ is $1.2 \times 10^9$~M$_{\odot}$ (see sect~\ref{subsec:mass-energ-budget}, table~\ref{table_mass_NRJ_budgets_SQ}). Therefore, the average total line emission per H$_2$ molecule is 
\begin{equation}
\mathcal{L}_{\rm H_2}(\rm SQ) \approx 1.3 \times 10^{-31} \quad \rm W \ H_2^{-1}
\end{equation}
We assume that each H$_2$ ionization releases $\mathcal{E}_{\rm ion.}(\rm H_2) \approx 7$~eV of energy to heat the molecular gas (see sect.~\ref{subsec:H2-cosmic-X-ray-heating} for the determination of this value). Therefore, the H$_2$ line cooling is balanced by cosmic ray heating for an ionization rate per hydrogen equal to
\begin{equation}
\zeta = 1.2 \times 10^{-13} \left( \frac{\mathcal{L}_{\rm H_2}(\rm SQ) }{ 1.3 \times 10^{-31} \, \rm W} \right) \left(\frac{7 \, \rm eV}{\mathcal{E}_{\rm ion.}(\rm H_2)} \right)  \quad \rm {-1} \ H^{-1}
\end{equation}
Therefore, the cosmic-ray ionization rate needed to balance the H$_2$ line cooling in the SQ ridge is $\approx 5 \times 10^3$ times higher than the standard  Galactic rate \citep{Shaw2005}. For a cosmic ray energy density $\approx 5 \times 10^3$ times higher the the Milky Way mean value, the cosmic ray pressure would be a factor 400 higher than the mean cosmic ray pressure in the Milky Way disk \citep[$P_{\rm cos}(\rm MW) \approx 10^4$~K~cm$^{-3}\,$,][]{Boulares1990} and $\approx 20$ times larger than the thermal gas pressure. 
Such a high difference in the pressures is unlikely on large scales. It would imply large departures from the g	as equilibrium and from the  equipartition of energy on a very large volume,  which does not seem to be confirmed by radio observations \citep[see][and paper~{\sc i}]{Xu2003, O'Sullivan2009a}.

However, enhanced cosmic-ray ionization rates may be produced at small scales, for instance in turbulent mixing layers. This turbulent mixing occurs when the flow of the background hot plasma interacts with the molecular clouds. At the interface, cosmic-ray particles can be accelerated efficiently \citep[see][for a review of these processes]{Scalo2004}. For such a high value, cosmic rays are the main destruction mechanism of H$_2$ molecules, and the molecular gas fraction depend on the ionization rate to gas density ratio $ \zeta / n_{\rm H}$. The models by \citet{Ferland2008} show that for $\zeta \approx 10^{-13}$~s$^{-1}$, the gas is molecular for $n_{\rm H} > 10^4$~cm$^{-3}$ (see their figure~2). Given that the warm H$_2$ temperature inferred from the S(1) to S(0) line ratio is $\approx 160$~K (see fig.~9 of  \hyperref[subsec:paper_Cluver]{paper~{\sc ii}}) in the main shock region, the pressure of the warm molecular gas would be $> 2 \times 10^6$~K~cm$^{-3}$, which is physically possible  in the SQ environment.


\subsubsection{H$_{\bf 2}$ emission associated with its formation}

The formation of H$_2$ during the cooling of the dense postshock gas (discussed in sect.~\ref{subsec:SQ-H2formation-multiphase-gas}) must be accompanied by H$_2$ emission. The excitation mechanism associated with H$_2$ formation is discussed in sect.~\ref{subsec:H2-excitation-during-formation}.
Can this excitation explain the powerful H$_2$ excitation in the SQ ridge? To quantify this, we calculate the H$_2$ emission associated with its formation during the isobaric cooling of the gas from $10^4$~K to $\approx 10$~K. This calculation has been discussed in sect.~5 of \hyperref[paper_SQ_H2]{paper~{\sc i}} and we summarize here our results.

Within the framework of a multiphase postshock medium, we assume that the gas that is forming H$_2$ comes from hotter, ionized gas that has recombined. 
We find that the observed H$_2$ rotational line ratios are well reproduced by the model. Fig.~5 of \hyperref[paper_SQ_H2]{paper~{\sc i}} shows that the shape of the H$_2$ excitation diagram associated with H$_2$ formation during the gas cooling matches the observations very well (blue dashed curve).
However, we find that the mass flow of cooling gas (the mass of gas that recombines per unit time, $\dot{M}_{\rm rec} = d M_{\rm rec} / d t$) needed to reproduce the absolute level of H$_2$ emission in the SQ shock is 25 times larger than the mass flow derived from H$\alpha$ observations. We use the following expression to derive $\dot{M_{\rm rec.}}$ from the H$\alpha$ luminosity:
\begin{eqnarray}
\dot{M}_{\rm rec} & = & m_{\rm H} \times \dot{N} _{\rm rec} \\
								& = & m_{\rm H} \times \frac{\mathcal{L}_{\rm H\alpha}}{0.45 \times E}_{\rm H\alpha} \\
								& = & 1263 \times \frac{\mathcal{L}_{\rm H\alpha}}{6.5 \times 10^{33} \ \rm W} \quad \rm M_{\odot}~yr^{-1} \ \ ,
\end{eqnarray}
where $ \dot{N} _{\rm rec}$ is the number of hydrogen atoms that recombine per second, $E_{\rm H \alpha}$ is the energy of an H$\, \alpha$ photon and $\mathcal{L}_{\rm H\alpha}$ is the $H\alpha$ luminosity \citep[see][for details]{Heckman1989}. To match the observed H$_2$ line fluxes with this model would require a mass flow of recombining gas of $\approx 30300$~M$_{\odot}$~yr$^{-1}$, which is not compatible with the mass rate derived from  H$\alpha$ observations ($\approx 1260$~M$_{\odot}$~yr$^{-1}$). This model also fails to reproduce the [O{\sc i}]~6300\AA~line emission by a factor of 7.
Therefore, the cooling of the warm H$\,${\sc i} gas produced by H$\,${\sc ii} recombination
cannot reproduce the observed H$_2$ emission nor the  [O{\sc i}]~6300\AA~line emission. A larger amount of energy needs to be dissipated within the molecular gas to account for the
H$_2$ emission. 

\subsubsection{Shocks}

In \hyperref[paper_SQ_H2]{paper~{\sc i}} we show that, within the \textit{IRS} SH slit aperture centered in the ridge, the dominant energy reservoir of the postshock gas is the bulk kinetic energy of the molecular gas. In sect.~\ref{subsec:mass-energ-budget} we have extended the mass and energy budget of the SQ galaxy collision and shown that the mechanical H$_2$ energy is at least a factor of 2 higher than the thermal energy of the hot plasma for the whole shock. If we assume the equipartition of energy between cosmic rays and magnetic energy, the mean magnetic field strength in the SQ shocked region is $\simeq 10\, \mu$G \citep{Xu2003}.  The
magnetic energy and thereby the cosmic ray energy $\displaystyle \frac{B^2}{8\, \pi} \, V_{\rm MS}$, contained in the volume of the main shock region, $V_{\rm MS} = \mathcal{A}_{\rm MS} \times l_z$, is 
\begin{equation}
\mathcal{E}_{\rm mag.} (\rm SQ) \approx 2 \times 10^{56} \left( \frac{B}{10\,\mu \rm G} \right) ^{2} \frac{\mathcal{A}_{\rm MS}}{446\,\rm kpc^2} \, \frac{l_z}{4 \, \rm kpc} \ ,
\end{equation}
where $l_z$ is the dimension along the line-of-sight and $\mathcal{A}_{\rm MS}$ the area of the main shock region defined in \hyperref[subsec:paper_Cluver]{paper~{\sc ii}} and sect.~\ref{spectral_IRSmap_H2}. Assuming $l_z \approx 4$~kpc \citep{O'Sullivan2009}, we conclude that the equipartition energy  of cosmic-rays is one order of magnitude smaller than the bulk
kinetic energy of the $\rm H_2$ gas.  So the bulk kinetic energy of the warm H$_2$ gas greatly domines the overall energy budget of the entire SQ ridge structure.

This conclusion drives us to the following interpretation of the extreme H$_2$ power in the SQ ridge: we propose that the H$_2$ emission is powered by the dissipation of turbulent kinetic energy into molecular gas. In the turbulent postshock medium, we expect that the relative motion between the backgroung plasma and the H$_2$ gas, but perhaps also the relative motion between the molecular clouds themselves, supply supersonic turbulence in the H$_2$ gas. Since the postshock medium is
magnetized (as indicated by radio synchrotron observations, see sect.~\ref{subsubsec:SQ-radio-observations}), we phenomenologically model this dissipation of mechanic energy by shocks. 
In sect.~\ref{subsec:shocks-molecular-cooling} we have discussed extensively the molecular cooling in  shocks. Non-dissociative MHD shocks are a very efficient way to channel the mechanical energy of the shock to the rotational excitation of H$_2$. Then we favor this process. In the absence of a magnetic field, shocks are rapidly dissociative and much of the cooling of the shocked gas  occurs through atomic, rather than H$_2$, line emission.

\subsection{Low-velocity MHD shocks and H$_{\bf 2}$ excitation}
\label{subsec:low-velShocksH2excitation}

We quantify the required gas densities and shock velocities to account for the $\rm H_2$ rotational line fluxes. In \hyperref[paper_SQ_H2]{paper~{\sc i}}, we present in details the result for the single-pointing  observations by \citet{Appleton2006}. Here we use our spectral mapping data to extend our analysis on the whole region of the SQ ridge. We compare these new results with our previous analysis in the center of the SQ shock.

The $\rm H_2$ emission induced by low velocity shocks is calculated using an updated version of the shock model of \citet{Flower2003}, presented in chapter~\ref{chapter:shocks}, sect.~\ref{sec:MHD-shock-models-molecular-gas}.
We use the same grid of models presented in \hyperref[paper_SQ_H2]{paper~{\sc i}}, i.e.  shock
velocities from $3 \ \rm to \ 40$~km~s$^{-1}$ with steps of $1$~km~s$^{-1}$, two values of the initial $\rm H_2$ ortho to para  ratio ($3$ and $10^{-2}$), and three different preshock densities ($n_{\rm H} = 10^2$, $10^3$, and $10^4$~cm$^{-3}$).  
In this models, we assume a standard value for the cosmic ray ionization rate of $\zeta = 5 \times 10^{-17}$~s$^{-1}$. In its initial state, the gas is molecular and cold ($\sim 10$~K), with molecular abundances resulting from the output of our model for the isobaric cooling (see sect.~\ref{subsubsec:gas-cooling-dust-survival-H2-formation} and \hyperref[paper_SQ_H2]{paper~{\sc i}} for details). We adopt the scaling of the preshock magnetic strength with the gas density, $B(\mu G) = b\, \sqrt{n_{\rm H}~[\rm cm^{-3}]}$ \citep{Crutcher1999, Hennebelle2008}. In our models, $b=1$. Since the molecular gas colder than 50~K does not contribute to the H$_2$ emission, we stop the integration when the gas has cooled to 50~K, and we calculate the line fluxes at this point. 

In \hyperref[paper_SQ_H2]{paper~{\sc i}}  we present the result in terms of H$_2$ excitation diagram (see sect.~\ref{subsec:H2excdiagrams} for the details about how it is calculated). In this manuscript we present the observed and model spectral energy distribution (SED) of the H$_2$ line fluxes for \textit{(i)} the same data as in \hyperref[paper_SQ_H2]{paper~{\sc i}}, and \textit{(ii)} the new data of \hyperref[subsec:paper_Cluver]{paper~{\sc ii}} for the main shock region.

We detail how we fit the H$_2$ line fluxes with shock models. In the models presented here, the preshock gas density is $n_{\rm H} = 10^4$~cm$^{-3}$. The shock velocity is the only shock parameter that we allow to vary.  
From the output of the shock model we derive the H$_2$ line brightnesses $\mathcal{B}^{\rm 50\,K} (\rm H_2)$ at 50~K, in units of [W~m$^{-2}$~sr$^{-1}$]. We perform a least-squares fit of the observed $\rm H_2$ line fluxes to a linear combination of two MHD shocks at different velocities $V_{s1}$ and $V_{s2}$ has been performed. These fits provide a scaling factor for each of the shocks which represents a mass flow $\dot{M} = dM/dt$, the amount of gas traversing the shocks per unit time. 
The $\rm H_2$ gas luminosity is proportional to this mass flow.  
For a given H$_2$  line of upper rotational  level $J$, the model line flux is derived as follows:
\begin{equation}
\mathcal{F}^{\rm mod.}_{J} (\rm H_2) = \frac{1}{4 \pi d^2} \,\frac{\dot{M}}{m_{\rm H}} \, \frac{4 \pi \mathcal{B}^{\rm 50\,K}_{J} (\rm H_2)}{n_{\rm H} V_{\rm s}} \quad \rm [W~m^{-2}],
\end{equation} 
where $d$ is the distance to the source (94~Mpc for SQ), $m_{\rm H }$ is the mass of an hydrogen atom, $\dot{M}$ the hydrogen mass flow [kg~s$^{-1}$],  and $n_{\rm H} V_{\rm s}$ represents the flow of hydrogen atoms per unit surface [H~m$^{-2}$~s$^{-1}$].

\begin{figure}
   \centering
    \includegraphics[angle=90, width=0.495\textwidth]{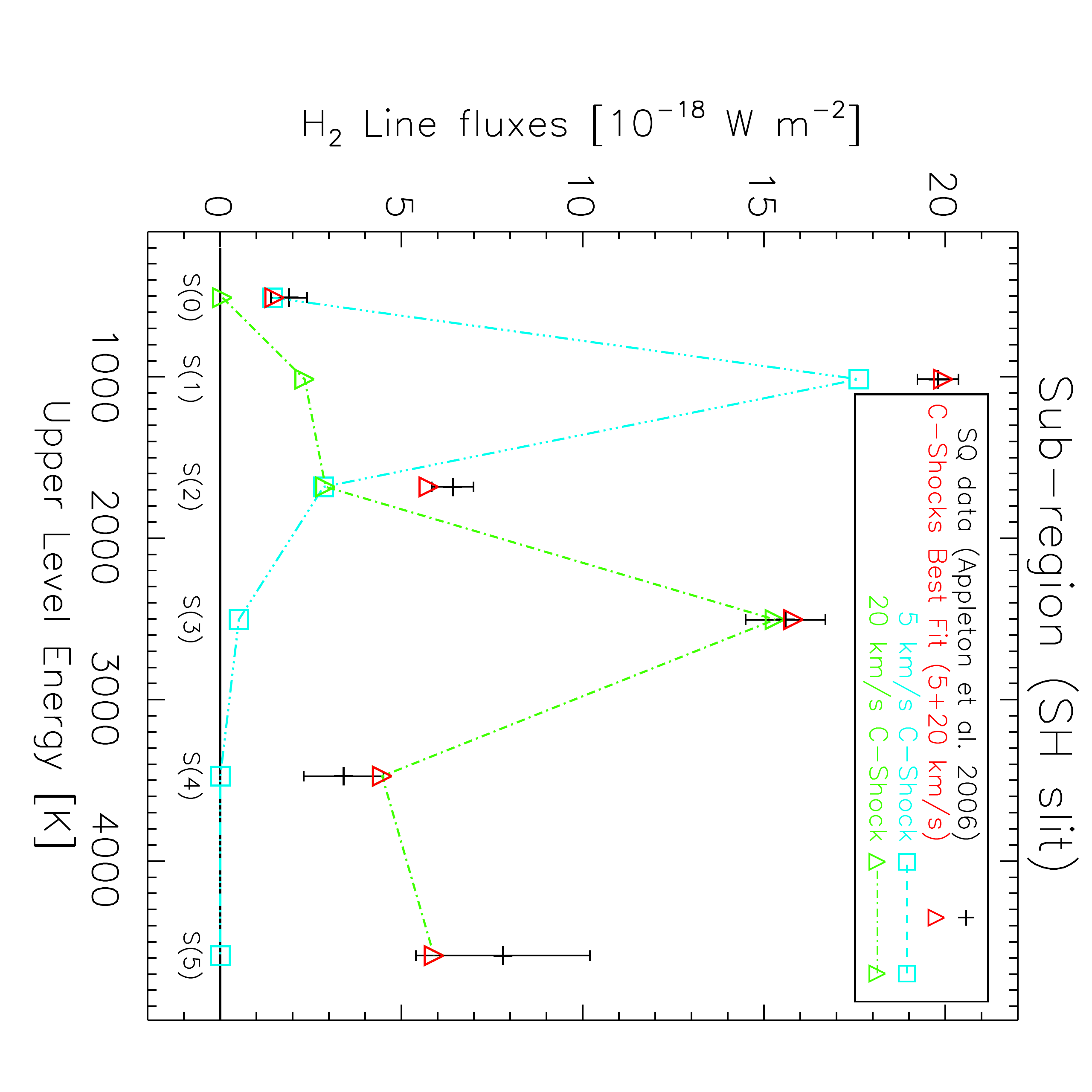}
    \includegraphics[angle=90, width=0.495\textwidth]{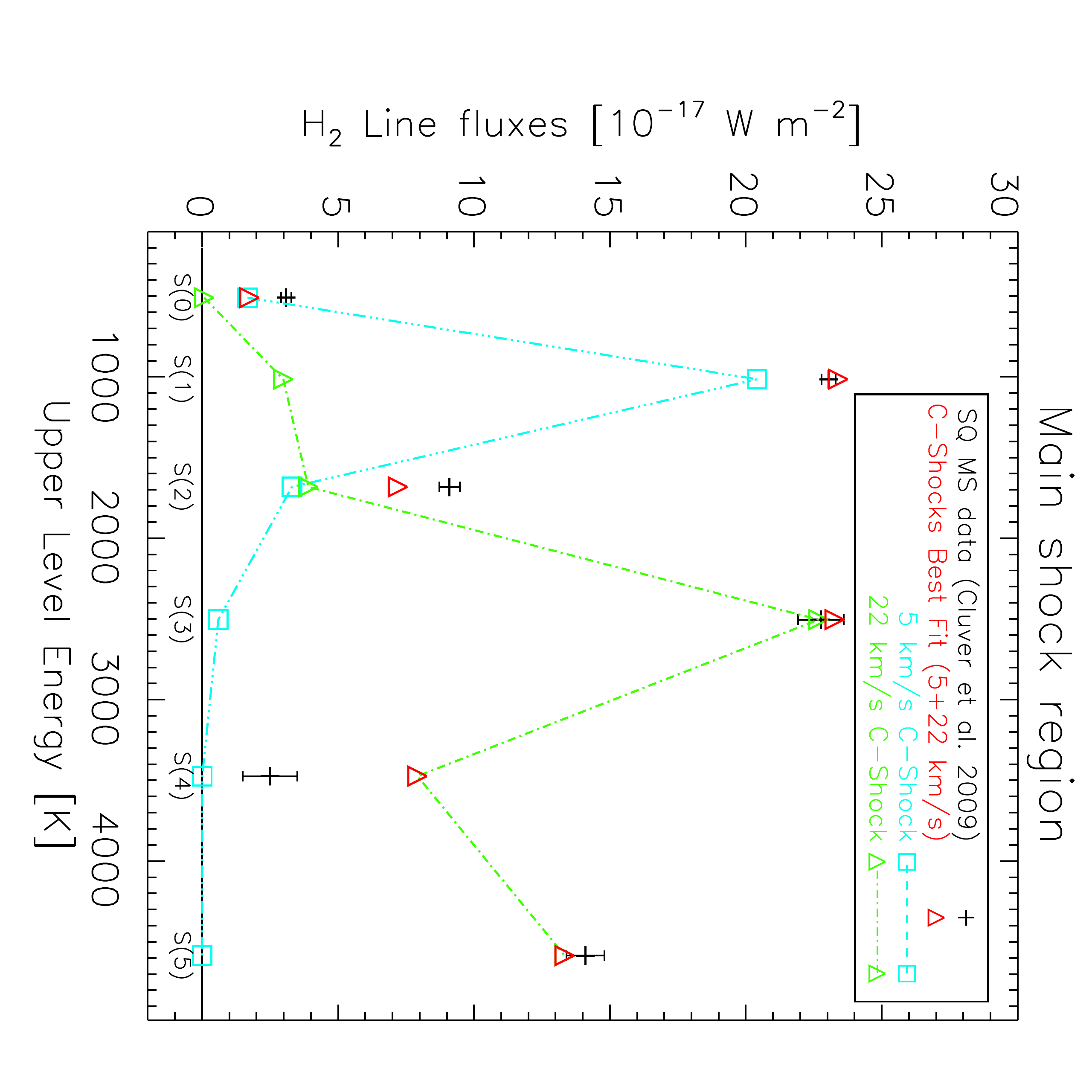}
      \caption[Fit of the Stephan's Quintet H$_2$ SED with 2 shocks]{Fit of the Stephan's Quintet H$_2$ spectral energy distribution with 2 shocks at fixed density $n_{\rm H} = 10^4$~cm$^{-3}$.
 \textit{Left:} data from \citet{Appleton2006}, extracted within the \textit{IRS} SH slit aperture in the center of the ridge.  \textit{Right:} data from \citet{Cluver2009} (\hyperref[subsec:paper_Cluver]{paper~{\sc ii}}) for the main shock region.
 The 2 shock components account for comparable fractions of the total H$_2$ luminosity but the less energetic component accounts for most of the warm H$_2$ mass (see table~\ref{tab_shock_mass_SQ_SH} and \ref{tab_shock_mass_SQ_MS}).}
       \label{fig_H2line_flux_SQ}
\end{figure}
\begin{figure}
   \centering
    \includegraphics[angle=90, width=0.495\textwidth]{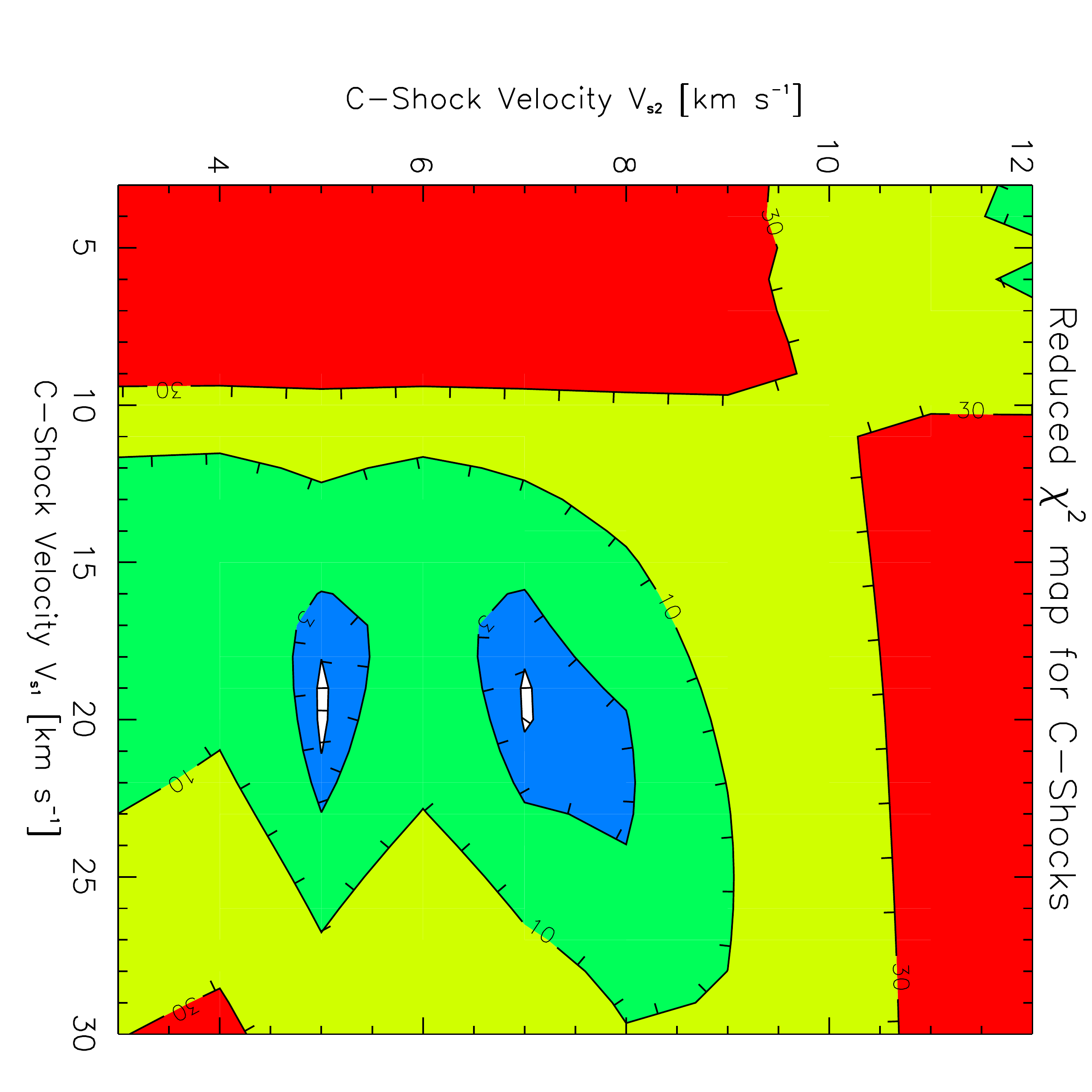}
    \includegraphics[angle=90, width=0.495\textwidth]{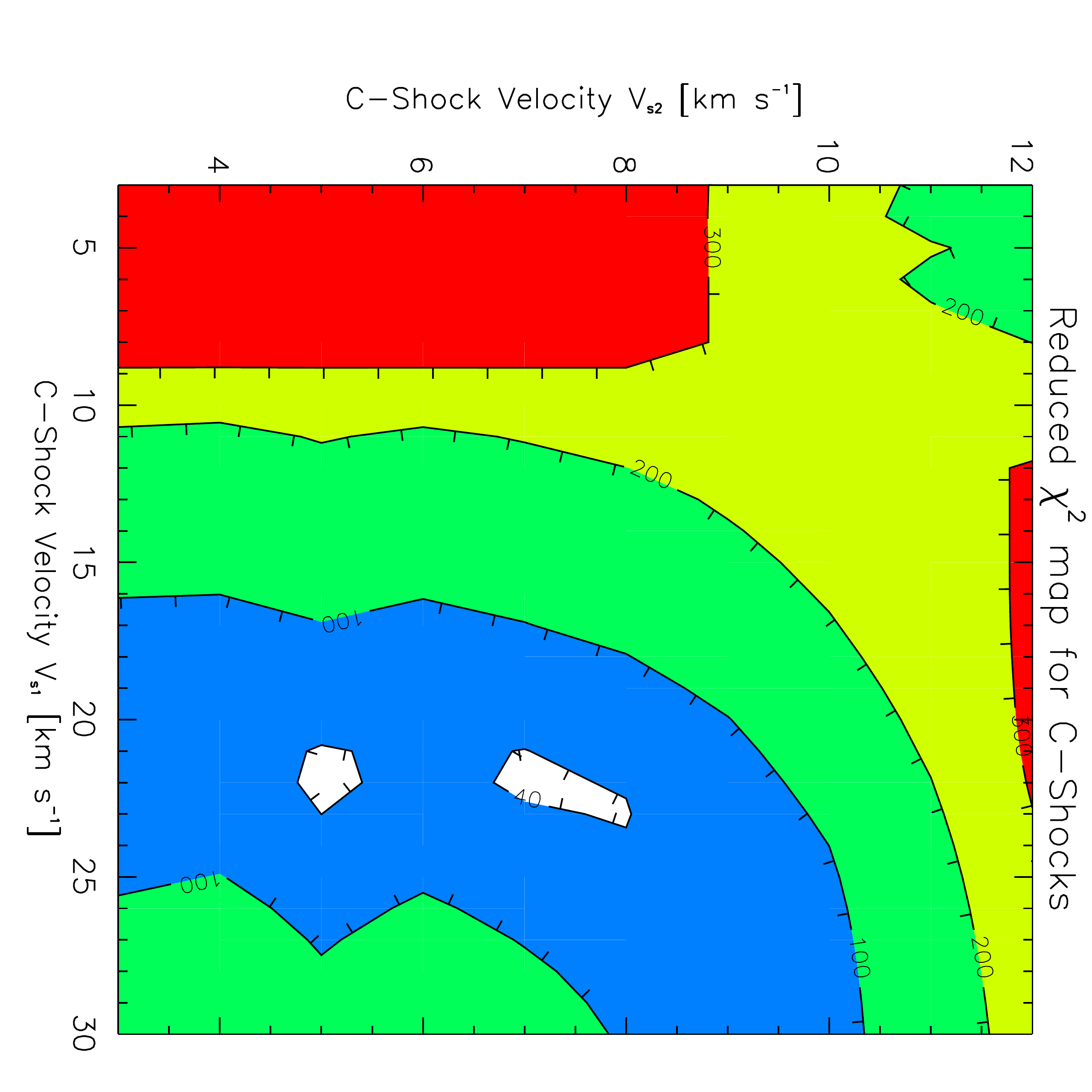}
      \caption[$\chi ^2$ map for C-shocks fit of the SQ H$_2$ SED]{Reduced $\chi ^2$ map for the fit of the Stephan's Quintet H$_2$ spectral energy distribution with 2 shocks at fixed density $n_{\rm H} = 10^4$~cm$^{-3}$. \textit{Left:} data from \citet{Appleton2006}, extracted within the \textit{IRS} SH slit aperture in the center of the ridge. The $\chi ^2$ contours are 1.5, 3, 10, 30. \textit{Right:} data from \citet{Cluver2009} (\hyperref[subsec:paper_Cluver]{paper~{\sc ii}}) for the main shock region. The $\chi ^2$ contours are 40, 100, 200, 300.}
       \label{fig_H2chi2map_twoCshocks_flux_SQ}
\end{figure}

The data cannot be fitted with a single shock. For the six H$_2$ lines detected S(0)$-$S(5), we obtain a good fit for two different shock velocities, $V_{s1}$ and $V_{s2}$.
The Fig.~\ref{fig_H2line_flux_SQ} shows the observed and modeled H$_2$ line fluxes for the \citet{Appleton2006} data (left panel, corresponding to the fig.~5 of \hyperref[paper_SQ_H2]{paper~{\sc i}}) and for the main shock region (data from \hyperref[subsec:paper_Cluver]{paper~{\sc ii}}). 
For a preshock density of $10^4$~cm$^{-3}$, the best fit is obtained for a combination of $5$ and $\approx 20$~km~s$^{-1}$ shocks (22~km~s$^{-1}$ for the main shock region)  with an initial value of the ortho to para ratio of $3$. Note  that no satisfactory fit  could be obtained with the low value of the initial H$_2$ ortho-to-para ratio.
The red line shows the weighted sum of the two shock components (best fit). 
The contributions of the $5$ (blue dashed line) and $\approx 20$~km~s$^{-1}$  (green dashed line) shocks to the total column densities are also plotted.
The $5$~km~s$^{-1}$ component dominates the contribution to the column densities of the upper levels of S(0) and S(1) lines.
The $20$~km~s$^{-1}$ component has a major contribution for the upper levels of S(3) to S(5). For S(2), the 2 contributions are comparable.

The fig.~\ref{fig_H2chi2map_twoCshocks_flux_SQ} shows the reduced chi-square contour maps of the two-velocities shock fit for the two apertures used. These maps allow the visualize which are the best combinations of shock velocities to fit the data. As compared to the small aperture in the center of the shock, there are bigger discrepancies between the best-fit model and the data for the extended ridge aperture. We note that the agreement between the model and the data is very good for the odd $J$ values of upper H$_2$ levels, whereas much larger differences are seen for the even-levels.
The error bars on the H$_2$ line fluxes are also much smaller for the main shock aperture because data is much deeper than the previous observations by \citet{Appleton2006}, so the $\chi ^2$ value are thus much higher. The extended ridge aperture is more contaminated by star-forming regions than the small aperture (see chapter~\ref{chapter:SQ_dust} for a discussion of star formation in the ridge), which may explain why the fit is poorer.

\renewcommand{\arraystretch}{1.1} 
\begin{table}
\begin{center}
\begin{minipage}[t]{\textwidth}
\renewcommand{\footnoterule}{}
\def\thefootnote{\alph{footnote}}
\centering
\small
\caption[MHD shock model parameters and mass flows for the \citet{Appleton2006} data.]{MHD shock model parameters, mass flows and cooling times for the \citet{Appleton2006} data  \protect\footnotemark[1].}
    \begin{tabular}{ c c c c c c c }
	\hline
	\hline
  \multirow{2}*{$V_s$ \footnotemark[2] } &  \multirow{2}*{Mass Flow} &   \multicolumn{2}{c}{{\sc Cooling Times}} & &   \multicolumn{2}{c}{{\sc Gas Masses}} \\
\cline{3-4}
\cline{6-7}
    &  & $t_{\rm cool}(150\rm K)$ \footnotemark[3] & $t_{\rm cool}(50\rm K)$ \footnotemark[3] & &  (150$\,$K) &  (50$\,$K)  \\
\hline
$[\rm km~s^{-1}]$ & [M$_{\odot}$~yr$^{-1}$]  & [yr] &  [yr] & & [M$_{\odot}$] & [M$_{\odot}$] \\
\hline
5   & 5870 & 14460 & 18160 & & $8.5 \times 10^7$ & $1.07 \times 10^8$ \\
20 &  260  &   2375 &   3230 & & $6.2 \times 10^5$ & $8.5 \times 10^5$ \\
\hline
\end{tabular}
\label{tab_shock_mass_SQ_SH}
\footnotetext[1]{This table lists the mass flows, cooling time to 150~K and 50~K, and molecular gas masses, associated with each shock velocity components.  The models are the same as those used in the left panel of fig.~\ref{fig_H2line_flux_SQ}.}
\footnotetext[2]{MHD Shock velocity.}
\footnotetext[3]{Cooling time computed down to 150~K and 50~K.}
\normalsize
\end{minipage}
\end{center}
\end{table}
\renewcommand{\arraystretch}{1.0}
\renewcommand{\arraystretch}{1.1} 
\begin{table}
\begin{center}
\begin{minipage}[t]{\textwidth}
\renewcommand{\footnoterule}{}
\def\thefootnote{\alph{footnote}}
\centering
\small
\caption[MHD shock model parameters and mass flows for the \citet{Cluver2009} data.]{MHD shock model parameters, mass flows and cooling times for the \citet{Cluver2009} main shock region data  \protect\footnotemark[1].}
    \begin{tabular}{ c c c c c c c }
	\hline
	\hline
  \multirow{2}*{$V_s$ \footnotemark[2] } &  \multirow{2}*{Mass Flow} &   \multicolumn{2}{c}{{\sc Cooling Times}} & &   \multicolumn{2}{c}{{\sc Gas Masses}} \\
\cline{3-4}
\cline{6-7}
    &  & $t_{\rm cool}(150\rm K)$ \footnotemark[3] & $t_{\rm cool}(50\rm K)$ \footnotemark[3] & &  (150$\,$K) &  (50$\,$K)  \\
\hline
$[\rm km~s^{-1}]$ & [M$_{\odot}$~yr$^{-1}$]  & [yr] &  [yr] & & [M$_{\odot}$] & [M$_{\odot}$] \\
\hline
5   & $6.8\! \times \! 10^4$ & 14460 & 18160 & & $9.8 \times 10^8$ & $1.23 \times 10^9$ \\
22 &  3545                           &  2100  &   2910 & & $7.5 \times 10^6$ & $1.03 \times 10^7$ \\
\hline
\end{tabular}
\label{tab_shock_mass_SQ_MS}
\footnotetext[1]{This table lists the mass flows, cooling time to 150~K and 50~K, and molecular gas masses, associated with each shock velocity components.  The models are the same as those used in the right panel of fig.~\ref{fig_H2line_flux_SQ}.}
\footnotetext[2]{MHD Shock velocity.}
\footnotetext[3]{Cooling time computed down to 150~K and 50~K.}
\normalsize
\end{minipage}
\end{center}
\end{table}
\renewcommand{\arraystretch}{1.0}

Some of the key-parameters derived from the fit are gathered in tables~\ref{tab_shock_mass_SQ_SH} and \ref{tab_shock_mass_SQ_MS} for the two sets of data. We indicate the best-fit shock velocities,  the mass flows $\dot{M}$ that are required to explain the absolute H$_2$ line fluxes, and the gas cooling times $t_{\rm cool.}$ at 150~K and 50~K. The gas masses are derived from $M = \dot{M} \times t_{\rm cool.}$. 
For the main shock region, we find a mass of gas at temperatures larger than
$T=150$~K of $9.8\times 10^8$~M$_{\odot}$, which is a factor of 2 larger than the mass derived from a three-temperatures fit of the excitation diagram (see sect.~4 of  \hyperref[subsec:paper_Cluver]{paper~{\sc ii}}, fig.~9). This difference essentially comes from the fact that multi-temperatures fits assume that the H$_2$ ortho-to-para ratio is thermalized, which is  not the case in shock models. 
In addition, one of the advantages of shock models in estimating the gas masses is that it includes the mass of gas at lower temperatures which has sufficient time too cool before being reheated. The total mass of gas at temperatures above 50~K is $1.2 \times 10^9$~M$_{\odot}$.

\renewcommand{\arraystretch}{1.1} 
\begin{table}
\begin{center}
\begin{minipage}[t]{\textwidth}
\renewcommand{\footnoterule}{}
\def\thefootnote{\alph{footnote}}
\centering
\footnotesize
\caption[MHD shock model parameters and predicted H$_2$ line fluxes for the center SQ region]{MHD shock model parameters and predicted H$_2$ line fluxes for the \citet{Appleton2006} data in the center of the SQ ridge \protect\footnotemark[1] for SQ. }
    \begin{tabular}{ c c c c c c c c c}
	\hline
	\hline
 $V_s$ \footnotemark[2]  &  \multicolumn{6}{c}{{\sc H$_2$ Line Fluxes} [$10^{-18}$ W~m$^{-2}$]} &  \multirow{2}*{$\mathcal{F}_{\rm H_2}$ \footnotemark[4]} & \multirow{2}*{$\mathcal{F}_{\rm bol}$ \footnotemark[5]} \\
\cline{2-7}
$\rm [km~s^{-1}]$   & S(0) & S(1) & S(2) & S(3) & S(4) & S(5)  & & \\
\hline 
5   & 1.44 &  17.61 & 2.85 & 0.51 & 0.0028 & 0.0012 & 22.42  & 38.05  \\
20 &  0.06 &  2.33 & 2.90 & 15.30 & 4.47 &  5.90  & 30.96  &  33.15 \\
 \hline 
2 shocks & 1.50 & 19.94  & 5.75  & 15.81 & 4.47   & 5.90   & 53.38 &  71.20 \\
\hline
\textit{Spitzer} obs. & $ 1.9 \pm 0.5 $ & $19.8 \pm 0.6$ & $6.4 \pm 0.6$ & $15.6 \pm 1.1$ & $3.4 \pm 1.1$  & $7.8 \pm 2.4$ & $54.9 \pm 3.1$ &  \\
\hline
\end{tabular}
\label{tab_shock_H2_fluxes_SQ_SH}
\footnotetext[1]{This table lists the shock model velocities, the predicted H$_2$ rotational line fluxes, and bolometric luminosities associated with each shock velocity components. The models are the same as those used in the lesft panel of Fig.~\ref{fig_H2line_flux_SQ}.}
\footnotetext[2]{MHD best fit shock velocities.}
\footnotetext[4]{Sum of the $\rm H_2$ S(0) to S(5) rotational lines in $10^{-18}$ W~m$^{-2}$}
\footnotetext[5]{Sum over all the lines (bolometric luminosity of the shock)}
\normalsize
\end{minipage}
\end{center}
\end{table}
\renewcommand{\arraystretch}{1.0}
\renewcommand{\arraystretch}{1.1} 
\begin{table}
\begin{center}
\begin{minipage}[t]{\textwidth}
\renewcommand{\footnoterule}{}
\def\thefootnote{\alph{footnote}}
\centering
\scriptsize
\caption[MHD shock model parameters and predicted H$_2$ line fluxes for  the main shock region]{MHD shock model parameters and predicted H$_2$ line fluxes for the \citet{Cluver2009} main shock region \protect\footnotemark[1] for SQ. }
    \begin{tabular}{ c c c c c c c c c}
	\hline
	\hline
 $V_s$ \footnotemark[2]  &  \multicolumn{6}{c}{{\sc H$_2$ Line Fluxes} [$10^{-17}$ W~m$^{-2}$]} &  \multirow{2}*{$\mathcal{F}_{\rm H_2}$ \footnotemark[4]} & \multirow{2}*{$\mathcal{F}_{\rm bol}$ \footnotemark[5]} \\
\cline{2-7}
$\rm [km~s^{-1}]$   & S(0) & S(1) & S(2) & S(3) & S(4) & S(5)  & & \\
\hline 
5   & 1.67 & 20.41 &  3.30 &  0.59 & 0.003  &   0.001 & 26.0  & 44.1  \\
22 &  0.07 & 2.98 & 3.90 &  22.67 &  7.92 & 13.33 & 50.9 &  55.5 \\
 \hline 
2 shocks & 1.74 &  23.4 &   7.20 &    23.26 & 7.92 & 13.34 & 76.9 & 99.6 \\
\hline
\textit{Spitzer} obs. & $ 3.09 \pm 0.19$ & $23.05 \pm 0.26$ & $9.10 \pm 0.38$ & $22.76 \pm 0.84$ & $2.5 \pm 1.0$  & $14.1 \pm 0.7$ & $74.6 \pm 1.6$ &  \\
\hline
\end{tabular}
\label{tab_shock_H2_fluxes_SQ_MS}
\footnotetext[1]{This table lists the shock model velocities, the predicted H$_2$ rotational line fluxes, and bolometric luminosities, associated with each shock velocity components, for the main shock region (aperture $\mathcal{A}_{\rm MS} \approx 2310$~arcsec$^2$). The models are the same as those used in the right panel of Fig.~\ref{fig_H2line_flux_SQ}.}
\footnotetext[2]{MHD Shock velocity.}
\footnotetext[4]{Sum of the $\rm H_2$ S(0) to S(5) rotational lines in $10^{-17}$ W~m$^{-2}$}
\footnotetext[5]{Sum over all the lines (bolometric luminosity of the shock)}
\normalsize
\end{minipage}
\end{center}
\end{table}
\renewcommand{\arraystretch}{1.0}

The gas luminosities are listed in Table~\ref{tab_shock_H2_fluxes_SQ_SH} and \ref{tab_shock_H2_fluxes_SQ_MS} for the \citet{Appleton2006} small aperture and for the extended shock region. 
We list the H$_2$ S(0) to S(5) line fluxes from each best-fit shock model velocity and the sum of the two shocks. The observed values are indicated at the bottom line.  
The two last rows show the total H$_2$ S(0)$-$S(5) flux and the bolometric flux of the shocks obtained by integrating the emission in all lines. The bolometric luminosities are smaller than the kinetic luminosities 
\begin{equation}
\mathcal{L}_{\rm kin.} = \frac{1}{2} \, \dot{M} \, V_{\rm s} ^2 
\end{equation}
because some of the mechanical energy of the shock is transferred to the magnetic energy. The H$_2$ line  luminosities are close to the bolometric values obtained by integrating the emission in all lines because the H$_2$ rotational lines are the main cooling lines of magnetic shocks (see sect.~\ref{subsec:shocks-molecular-cooling} for a discussion of the energetics of magnetic shocks).
Summing over the two shocks,  we estimate a total luminosity of $8.2 \times 10 ^ {34}$~W for the S(0)$-$S(5) H$_2$ lines, which is excellent agreement with the observed value ( $8.0 \pm 0.2 \times 10 ^ {34}$~W).

Our grid of models shows that this solution is not unique. 
If one decreases the density to  $10^3$~cm$^{-3}$, we find that one can fit the observations with a combination of MHD shocks (at $9$ and $35$~km~s$^{-1}$). 
If one decreases the density to  $10^2$~cm$^{-3}$, the rotational $\rm H_2$ excitation cannot be fitted satisfactorily. At such low densities, MHD shocks fail to reproduce the S(3) and S(5) lines because the critical densities for rotational H$_2$ excitation increase steeply with the $J$ rotational quantum number (see sect.~\ref{subsec:H2-thermometer-critical-densities}).
The warm H$_2$ pressures $\rho \, V^{2}$ in the two shocks are $4.5 \times 10^{7}$~K~cm$^{-3}$ and $5.6 \times 10^{8}$~K~cm$^{-3}$) for the $5$~km~s$^{-1}$ and $20$~km~s$^{-1}$ shocks, respectively. These pressure are much higher than the pressure of the background hot plasma ($\approx 2 \times 10^5$~K~cm$^{-3}$).
Both warm and dense (i.e. high pressure)  gas is needed to account for emission from the higher $J$ levels.

Though the fit is not unique, it gives an estimate of the relevant range of shock velocities required to account for the H$_2$ excitation. The velocities ($\approx 5 - 20$~km~s$^{-1}$) are remarkably low compared to the SQ collision velocity ($\approx 1000$~km~s$^{-1}$) and to the velocity dispersion of the warm molecular gas ($\approx 900$~km~s$^{-1}$).
If our interpretation of the H$_2$ emission in terms of shock-driven dissipation of energy within the warm molecular gas is right, such a contrast in velocities is a beautiful illustration of turbulent cascade, from high-velocities at large scales, to low-velocity at small scales.
The fact that energy dissipation occurs over this low velocity range is an essential key to account for the importance of H$_2$ cooling. High velocity shocks ($> 40-50$~km~s$^{-1}$) in dense ($n_{\rm H} > 10^4$~cm$^{-3}$) molecular gas would dissociate H$_2$ molecules.
However, we are far from understanding in details how the transfer of energy from the background plasma (shocked at high speed, $\approx 700$~km~s$^{-1}$, see sect.~\ref{subsubsec:Xray_obs_SQ}) to the molecular gas. This is discussed qualitatively in sect.~5 and 6 of  \hyperref[paper_SQ_H2]{paper~{\sc i}}, and we will come back to this question in sect.~\ref{sec:SQ-Why-H2-important-coolant}.

\subsection{Future observational tests: near-IR H$_{\bf 2}$ line emission}

A missing piece of the puzzle of the SQ shock is the near-IR spectrum. This would allow to have access to  both ro-vibrational H$_2$ cooling lines and atomic lines, and therefore to gain a more complete picture of the total line cooling in the SQ shock. The near-IR ro-vibrational lines sample much hotter H$_2$ gas than the Spitzer data (see chapter~\ref{chapter:H2Molecule}). 
In particular, near-IR spectroscopy will allow us to extend the H$_2$ excitation diagram into the hotter
regime, where we can start to distinguish between dissociative or non-dissociative type of shocks.  This would have important implications for how the energy cascades down from the large
scale turbulence to the small cloud fragments.

A proposal has been written to perform near-IR spectroscopy with the  \textit{TripleSpec} instrument installed on the  Palomar 200-inch telescope (P.I.: P.N. Appleton).
The spectral ($ \mathcal{R}=2600$ or $\Delta v =  115$~km~s$^{-1}$) and spatial (1'') resolution of TripleSpec is far superior to that of the \textit{Spitzer IRS}, and is crucial to determine if the large velocity
spread seen in the Mid-IR H$_2$ lines ($\sigma \approx 800-900$~km~s$^{-1}$) represents a single broad line  or two separate narrow lines indicating both pre-and post shocked material. 

\begin{figure}
   \centering
    \includegraphics[angle=90, width=0.9\textwidth]{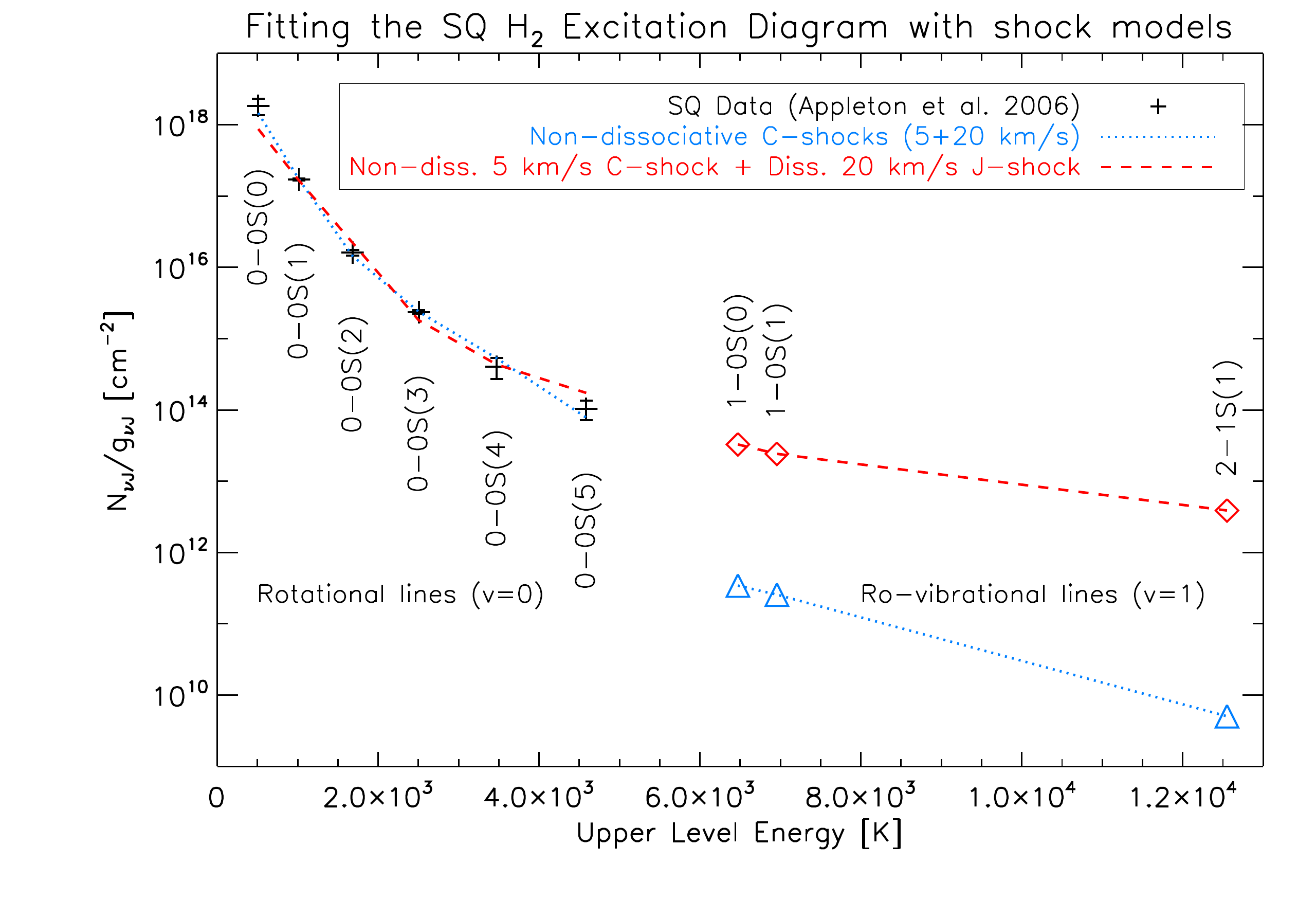}
      \caption[Model prediction of H$_2$ ro-vibrational emission in SQ]{The excitation diagram of H$_2$ emission in Stephan's Quintet from observations \citep[black symbols from][]{Appleton2006} compared with a model of the post-shock gas from \citep{Guillard2009}. The model predicts fluxes for the ro-vibrational lines in the range $0.2 - 14 \times 10^{-18}$~W~m$^{-2}$ over a solid angle of $5 \times 10$ arcsec$^{2}$, depending on the kind of shock. Observations in the near-IR can significantly limit the kinds of shocks capable of explaining the excitation diagram. These will influence our picture of how eneregy is funnelled down to 20~km~s$^{-1}$ shocks from the large-shock ($V_{\rm s} \sim 600$~km~s$^{-1}$) which creates the X-ray emission.}
       \label{H2excitDiag_SQ_Rovib_lines}
\end{figure}

 Fig~\ref{H2excitDiag_SQ_Rovib_lines} shows example
of dissociative and non-dissociative shock predictions from the shock model grid presented in  \hyperref[paper_SQ_H2]{paper~{\sc i}}. The blue dotted curve is our best fit model for C-shocks presented in sect.~\ref{subsec:low-velShocksH2excitation}, Fig.~\ref{fig_H2line_flux_SQ} (left panel). 
The red curve is a model where the higher velocity component (20~km~s$^{-1}$) is dissociative (J-type). In this case, the ro-vibrational line fluxes are much higher than the non-dissociative case because the gas reaches higher temperatures. Near-IR spectroscopy should provide constrains on the contribution from non-ionizing shocks to the H$_2$ emission. This would help in determining if these shocks can contribute to other line emission, like [O$\,${\sc i}] 6300\AA~for instance.

I participated to our first observation run at Palomar observatory to perform near-IR spectroscopy with the newly-commissioned \textit{TripleSpec} instrument in November 2008 with Phil Appleton and Patrick Ogle. Unfortunately the weather was very bad and we have only detected, through the clouds, some H$_2$ emission from the NGC~7319 Seyfert galaxy. The proposal was re-submitted and accepted for a new observation run in August 2009. 

\section{Why is H$_{\bf 2}$ such an important coolant?}
\label{sec:SQ-Why-H2-important-coolant}

Our interpretation of mid-IR line emission, but also optical and X-ray emission, within the context of a multiphase postshock gas, suggests that observations can be understood in terms of dynamical interactions between the ISM phases (sect.~\ref{subsec:mass_energy_transfers}). Exchanges of mass and energy between the gas phases seem a key to elucidate why the H$_2$ power is so high in the ridge. An efficient transfer of the bulk kinetic energy of the gas to turbulent motions of much lower velocities within molecular clouds seems required to make H$_2$ a dominant coolant of the postshock gas (sect.~\ref{subsec:SQ-energy-transfer-H2-power}). Here we briefly summarize and update the discussion of \hyperref[paper_SQ_H2]{paper~{\sc i}} (sect.~5 and 6).

\subsection{Mass and energy cycle between ISM phases}
\label{subsec:mass_energy_transfers}

In the Stephan's Quintet shock region, observations and modeling suggest that the dissipation of the mechanical energy of the collision between a galaxy and a tidail tail is the main source to power the X-ray emission, the optical, and mid-IR line emission from the gas in the ridge. The relative luminosities from the different gas components depend on the dynamical coupling between the gas phases. In this section we propose a scenario for the postshock gas evolution that sketches the dynamical interaction between ISM phases. This interpretation introduces a physical framework that may apply to other astrophysical situations (AGN-driven jets in radio-galaxies, supernova remnants, starburst-driven winds, etc.).

\begin{figure}
 \centering
       \includegraphics[width=0.7\textwidth]{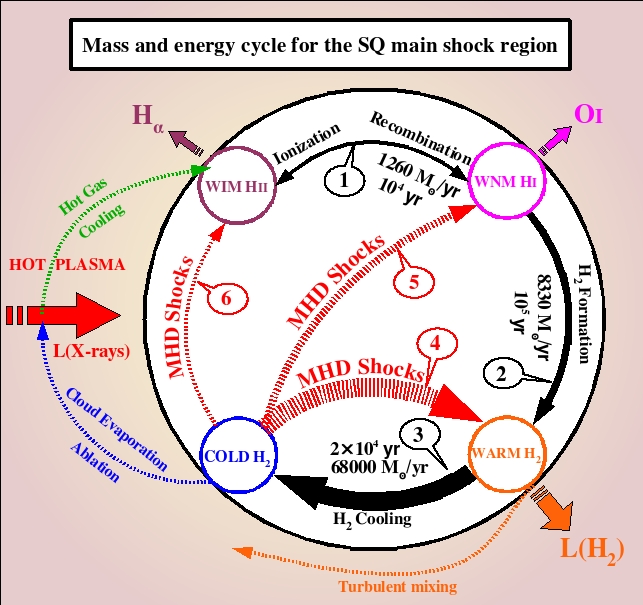}
   \caption[Sketch of the mass and energy cycle between ISM phases]{Schematic view at the gas evolutionary cycle proposed in our interpretation of Stephan's Quintet optical and H$_2$ observations.
Arrows represent the mass flows between the H~II, warm H~I, warm and cold H$_2$ gas components. They are numbered for clarity. The dynamical interaction between gas phases drives the cycle.
The mass flow values and associated timescales are derived from the $\rm H_{\alpha}$, O{\sc i}, and $\rm H_2$ luminosities and model calculations. Heating of the cold H$_2$ gas (red arrows) is necessary to account for the increasing mass flow from the ionized gas to cold H$_2$  phases. }
  \label{fig:mass_flows}
   \end{figure}

 Fig.~\ref{fig:mass_flows} sketches our view at the mass and energy exchanges between the different gas phases that control the evolution of the postshock gas.  This cartoon is the same as Fig.~6 of \hyperref[paper_SQ_H2]{paper~{\sc i}} except that the values of the mass flows are updated for the full SQ ridge.
 Black and red arrows represent the mass flows between the H$\,${\sc ii}, H$\,${\sc i}, warm and cold H$_2$ gas components of the post-shock gas. The large arrow to the left symbolizes the relative motion between the warm and cold gas and the surrounding plasma.
Each of the black arrows is labeled with its main associated process: gas recombination and ionization (arrow number 1), H$_2$ formation (2) and H$_2$ cooling (3). The values of the mass flows and the associated timescales are derived from observations and our model calculations.

The total mass flow associated with the low-velocity component of MHD shocks required to reproduce the H$_2$ line fluxes is much larger ($\approx 68000$~M$_{\odot}$~yr$^{-1}$) than the mass flow of recombining gas derived from H$\alpha$ observations ($\approx 1260$~M$_{\odot}$~yr$^{-1}$).
The mass rates needed to account for the $\rm H_{\alpha}$, O{\sc i}, and $\rm H_2$ luminosities increases. Therefore, one cannot explain the $\rm H_2$ luminosity with only a continuous cycle through the gas components, from H$\,${\sc ii} to cold $\rm H_2$. 
Heating of the cold H$_2$ gas towards warmer gas states (red arrows) needs to occur.

The main energy reservoir that can power these red arrows is the mechanical energy of the galaxy collision. We propose that mass and energy are transfered from cold molecular gas via shocks. 
The post-shock molecular cloud fragments are likely to experience a distribution of shock velocities, depending on their size and density. Arrow number 4 represents the low velocity magnetic shocks excitation of H$_2$ gas that can account for the H$_2$ emission (described in sect.~\ref{subsec:low-velShocksH2excitation}).
More energetic shocks may dissociate the molecular gas (arrows number 5).
They are necessary to account for the low H$ _{\alpha}$ to [O~{\sc i}]~6300~\AA~luminosity ratio.
Even more energetic shocks may ionize the molecular gas (arrow number 6).
This would bring cold $\rm H_2$ directly into the H~{\sc ii} reservoir.


The H$\,${\sc ii} gas can also be produced by the turbulent mixing of the warm and cold gas with the hot background plasma (see sect.~\ref{turbulent-mixing} for a description of turbulent mixing layers). This is represented by the thin blue and orange arrows from the cold and warm molecular gas reservoirs to the hot plasma. This mixing produces gas at intermediate temperature (a few $10^5$~K) that is thermally instable. At these temperatures, the thermal sputtering of the dust is uneffective, so dust is preserved. This gas cools back to produce H$\,${\sc ii} gas that re-enters the cycle (thin green arrow). 

An important question is to know whether there is a significant mass of cold ($< 50$~K) molecular gas  associated with the warm H$_2$. This cold gas does not contribute to the H$_2$ emission.
The mass accumulated in this cold H$_2$ reservoir depends on the ratio between the timescale of mechanical energy dissipation  (on which gas is excited by shocks through one of the red arrows 4, 5 and 6) and the cooling timescale along the black arrow 3. 

Following the calculation made in sect.~6 of \hyperref[paper_SQ_H2]{paper~{\sc i}}, we estimate the timescale of dissipation of the turbulent energy, $t_{\rm turb}$, and compare it to the cooling time of the warm H$_2$ gas. We assume that the  dissipation of turbulent energy  powers the H$_2$ emission, so the energy dissipation rate $P_{\rm turb}$ must be at least equal to the power radiated by H$_2$, $\mathcal{L}_{\rm H_2}$.
Thus we write the turbulent dissipation timescale as the ratio of the total turbulent energy of the warm H$_2$ gas, $E_{\rm turb}$,  to the total H$_2$ power:
\begin{eqnarray}
t_{\rm turb} & = & \frac{E_{\rm turb}}{P_{\rm turb}} \\
                     & = & \dfrac{\frac{3}{2} M_{\rm H_2} \sigma _{\rm turb}^{2}}{\mathcal{L}_{\rm H_2}} \\
					 & = & 1.14 \times 10^4 \, \left( \frac{M_{\rm H_2}}{1.2 \times 10^9 \, \rm M_{\odot}}\right) \left( \frac{ \sigma _{\rm turb}}{10 \, \rm km \, s^{-1}}\right) ^{2} \left( \frac{10^{35} \, \rm W}{\mathcal{L}_{\rm H_2}} \right) \quad \rm [yr]
\label{eq:turbulent-diss-time}
\end{eqnarray}
In Eq.~\ref{eq:turbulent-diss-time}, $M_{\rm H_2}$ is the total warm H$_2$ mass and $\sigma _{\rm turb}$ is the turbulent velocity dispersion within the warm H$_2$ gas.
Using the  H$_2$ luminosity and the warm H$_2$ mass in Table~\ref{table_mass_NRJ_budgets_SQ}, we
estimate a turbulent dissipation timescale of $\approx 10^4$~yr for a typical dispersion velocity of $\approx 10$~km~s$^{-1}$. This estimate of the  turbulent dissipation timescale is a lower limit since the dissipated power $P_{\rm turb}$ may be larger if it powers the dissociative and ionizing shocks represented by the arrows 5 and 6 in Fig.~\ref{fig:mass_flows}.
Therefore, cold H$_2$ gas would be heated by turbulent dissipation on timescales comparable to the cooling time of the warm H$_2$ gas that accounts for the S(0) and S(1) line emission ($1.8 \times 10^4$~yr, see table~\ref{tab_shock_mass_SQ_MS}). 
If this is correct, the cold molecular gas mass should not be much higher than the warm H$_2$ mass inferred from the \textit{Spitzer} observations. 

We compare the turbulent dissipation timescale to the dissipation timescale of the total bulk kinetic energy of the warm H$_2$ gas, $t_{\rm TDK}$, which can be written as the ratio between the H$_2$  gas bulk kinetic energy and H$_2$ luminosity:
\begin{equation}
t_{\rm TDK} = 5.7 \times 10^8 \left( \frac{E_{\rm kin}(\rm H_2)}{1.8 \times 10^{51} \, \rm J} \right) \left(\frac{10^{35} \, \rm W}{\mathcal{L}_{\rm H_2}} \right)  \quad \rm [yr]
\end{equation}
The turbulence dissipation timescale  $t_{\rm turb}$ is more than three orders of magnitude smaller than  $t_{\rm TDK}$. Therefore, the dissipation of the mechanical energy of the galaxy collision must involve a large ($\gtrsim 1\,000$) number of cycles where H$_2$ gas fragments are accelerated and dissipate their kinetic energy in shocks. 

\subsection{Physics of the energy transfer: qualitative discussion}
\label{subsec:SQ-energy-transfer-H2-power}

We have shown that the bulk kinetic energy of the gas accelerated by the shock induced by the galaxy collision has to be transfered efficiently within molecular gas to explain the powerful H$_2$ emission. 
However, how this energy transfer occurs is far from being known. We discuss here two possible processes: \textit{(i)} the momentum transfer induced by the motion of clouds in the hot plasma, and \textit{(ii)} the production of cosmic rays in the turbulent mixing layers. The first process is discussed in more details in sect.~6.1 of \hyperref[paper_SQ_H2]{paper~{\sc i}}.

\subsubsection{Supersonic turbulence sustained by the motion of clouds}

We remind that in our interpretation of the H$_2$ emission in the SQ halo, the formation of H$_2$ gas arises from the compression of shocked H$\,${\sc i} gas. The two pre-collision H$\,${\sc i} complexes, one associated with the intruding galaxy NGC~7318b and the other with NGC~7319's tidal tail, move at a relative velocity of $\approx 900$~km~s$^{-1}$ with respect to each other. 
While they are compressed and becoming H$_2$ clouds, the clouds  move at high velocity with respect to the background plasma because they keep their preshock momentum. 
In the collision, the clouds do not share the same dynamics as the lower density intercloud gas, and the kinetic energy recorded in the  $\approx 900~\rm km\,s^{-1}$ H$_2$ line width  is dominated by the bulk motion of the H$_2$ clouds in the plasma rest frame.

The background flow applies drag forces on the clouds  \citep{Murray2004}. 
Because of the viscosity of the fluid, the dynamical friction (shear forces) between the cloud and the hot flow also lead to a transport of momentum from the fast flow into the cloud.
Analytical and numerical studies of shock-cloud interactions, or studies of the interaction between  an outflow and molecular clouds seem to show that the injection of momentum within cloud can feed supersonic turbulence within clouds \citep[e.g.][]{Kornreich2000, Matzner2007, Carroll2009}. The turbulent energy injected into the clouds is dissipated by low-velocity shocks into the dense molecular material. 
Besides, the relative motion between cloud fragments  can lead to cloud-cloud collisions, which  results in the dissipation of  kinetic energy  preferentially in the molecular gas because it is the coldest component with the lowest sound speed.

In addition to dynamics, the thermodynamics of the multiphase gas is a key to understand the energy transfers between the phases.
First, the cooling of the gas transfers the turbulent energy of the hotter  H$\,${\sc ii} and H$\,${\sc i} phases to the H$_2$ gas. Secondly, the thermal instability is also an efficient mechanism to convert the thermal energy to turbulent kinetic energy. Numerical simulations \citep{Sutherland2003, Audit2005} show that the thermal instability inhibits energy-loss through radiation from the warmer gas and feeds turbulence in the cooler gas. The turbulent velocities generated by the thermal instability are found to be of the order of the sound speed in the warmer gas and are thus supersonic for the colder gas.

\subsubsection{H$_{\bf 2}$ heating by cosmic rays produced in turbulent mixing layers}

An alternative mecanism to transfer the bulk kinetic energy of the gas to H$_2$ heating is the production of cosmic-ray particles in the turbulent mixing layers that would scatter within the molecular clouds and excite H$_2$ molecules (see sect.~\ref{subsec:H2excitation-mechanisms} for a description of H$_2$ excitation by cosmic rays). 
In sect.~\ref{subsec:SQ-possible_H2-excitation-mec} we show that the average cosmic-ray ionization rate in the SQ ridge is likely to be too low to balance the H$_2$ cooling. The advantage of cosmic-ray acceleration in mixing layers is that cosmic-ray particles are produced \textit{locally} and can diffuse into the H$_2$ gas. Cosmic rays are accelerated by the Fermi mechanism and can in turn generate turbulence and MHD waves into the molecular gas if they are accelerated to velocities faster than the local Alfvén speed. These processes are reviewed by \citet{Scalo2004} (see their sect.~4) and we direct the reader to this review for references. 

The two processes discussed above (shocks induced by momentum transfer or cosmic ray excitation produced by turbulent mixing) are two alternative means to cascade the bulk kinetic energy of the gas on large scales to small-scale dissipation withinthe H$_2$ gas. From an astrophysical point of view, the most important point is that the H$_2$ emission traces the dissipation of the bulk kinetic energy of the system.

\section{Optical and mid-IR fine structure line diagnostics}
\label{sec:opt-midIR-line-excit}

In the previous sections we have shown that the dominance of the H$_2$ line cooling in the SQ shock region would require an efficient cascade of energy from the large-scale, bulk kinetic energy of the gas to small-scale turbulent motions within the H$_2$ gas. 
However, the H$_2$ rotational lines observed by \textit{Spitzer} do not provide a complete budget of the kinetic energy dissipation in the SQ ridge. These low-excitation lines trace gas between $\approx 50 - 1000$~K, and therefore they only sample a narrow range of shock velocities. 
Within the framework of our interpretation of the H$_2$ emission (the high-speed collision of two flows of multiphase gas), we expect a broad distribution of shock velocities. 

Fast shocks may lead to the line emission from ionized gas, which, up to now, has not been considered. This section adresses the following questions: is the emission from the ionized gas an important component of the energetics in the SQ ridge? 
Is H$_2$ really a dominant coolant? 
We first describe optical imaging and spectroscopy, as well as mid-IR fine-struture line maps from our new \textit{Spitzer IRS} observations. We use line ratios diagnostics to infer the shock properties from the ionized gas.

\subsection{H${\bf \alpha}$ imaging and optical line spectroscopy}
\label{subsec:opticallineemission_shocks}

\begin{figure}
   \centering
   \includegraphics[width=0.55\textwidth]{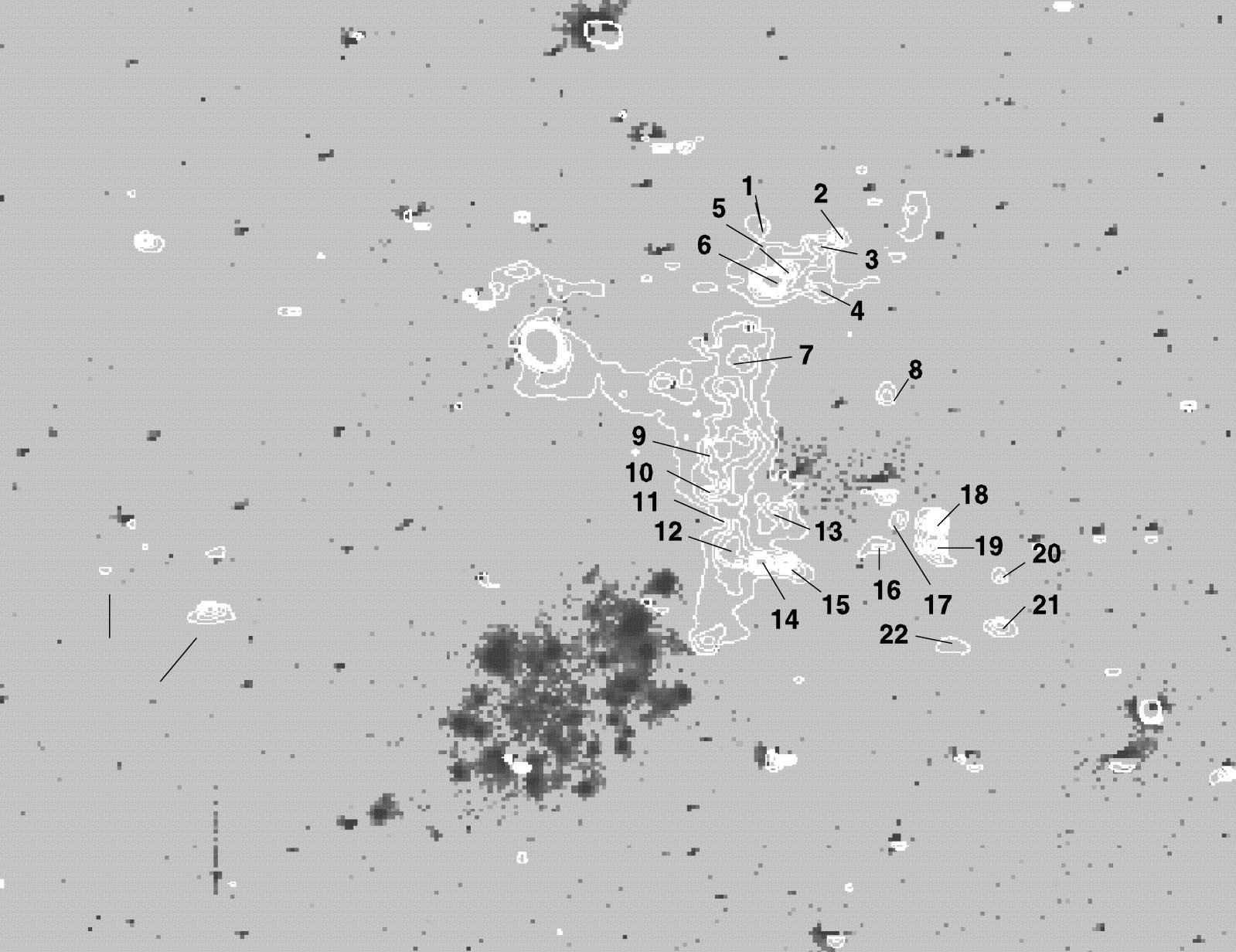}
     \includegraphics[width=0.44\textwidth]{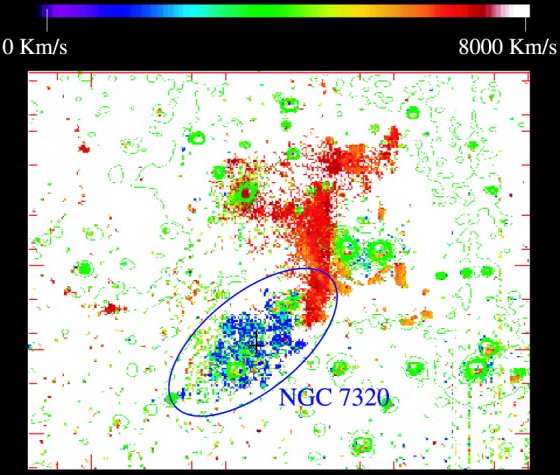}
      \caption[H$\alpha$ contours and velocity maps of SQ]{H$\alpha$ contours and velocity maps of SQ. \textit{Left:} High-velocity ($> 5500$~km~s$^{-1}$, white contours) and low-velocity ($\sim 800$~km~s$^{-1}$, gray scale) H$\alpha$ maps of Stephan's Quintet. \textit{Right:} Velocity map obtained from the low-$v$ and high-$v$ H$\alpha$ images. Figures taken from \citet{Gutierrez2002}.}
       \label{fig_SQ_Ha}
   \end{figure}

\citet{Moles1997, Vilchez1998, Xu1999, Plana1999, Sulentic2001, Gutierrez2002} present narrowband filter and Fabry-Pérot images that show a chaotic morphology of emission-line regions and complex H$\,${\sc ii} gas kinematics. 
The left panel of Fig.~\ref{fig_SQ_Ha} shows H$\alpha$ maps of SQ from \citet{Gutierrez2002}. The white contours result from combination of four narrowband filter images centered at different wavelengths (from 6667\AA~to 6737\AA), covering the full velocity range of the SQ galaxies ($4760-7960$~km~s$^{-1}$). The grey-scale background image is a combination of two images centered at 6569\AA~($280$~km~s$^{-1}$) and 6611\AA~($2200$~km~s$^{-1}$), and thus mainly exhibits the NGC~7320 foreground galaxy. The white contours clearly show an extended and clumpy North-South H$\alpha$ filament that spatially correlates with the radio emission in the ridge, as well as an East-West feature (we called it the ``bridge'') that connects the nucleus of the Seyfert 2 NGC~7319 with the North-South ridge. In addition, emission-line regions associated with the new intruder and with the northern SQ-A region are detected.

The two background images on Fig.~\ref{fig_SQ_HI_contours_Sulentic} show H$\alpha$ Fabry-Pérot images from \citet{Sulentic2001}, centered at $\sim 6600$~km~s$^{-1}$ (left panel) and $\sim 5600$~km~s$^{-1}$ (right panel). 
These images show different morphologies. The $6600$~km~s$^{-1}$ image shows the North-South clumpy, linear shock structure, the core of NGC~7319 and its ``bridge'', and the SQ-A starburst. On the $5600$~km~s$^{-1}$ image, the intra-group emission has an ``arc shape'', with brighter emission in the southern region. This southern arc is comprised of a string of H$\alpha$ knots that connects to the east end of the faint bar in the new intruder NGC~7318b. This southern arc may be material disrupted from NGC~7318b that has been compressed by the shock. This arc almost connects to a fainter string starting from the core of NGC~7318a, then forming a peculiar ``smily face'' (the cores of NGC~7318 a and b would be the eyes of this giant smiley!), also clearly visible on the UV (see the central panel of  Fig.~\ref{fig_SQ_radio_GMRT_OSullivan2009}) and mid-IR (see chapter~\ref{chapter:SQ_dust}) images.
Note that in some regions, the two velocities components have similar morphologies \citep[see also Fig.~2 of][]{Xu1999}. This emission may arise from gas at intermediate velocities ($\sim 6000 $~km~s$^{-1}$) that may be shock-accelerated material from the new intruder.

Up to now, very few high signal-to noise optical spectra in the SQ ridge have been published.  \citet{Xu2003} performed optical spectroscopy with the Palomar 200''  telescope at several positions along the SQ ridge. In the center of the shock the flux ratios are:
\begin{eqnarray}
\mathcal{F}([\rm O\, I]~\lambda 6300) / \mathcal{F}(\rm H\,\alpha + [N\, II]) & = & 0.53 \\
\mathcal{I}(\rm [S\, II]~\lambda 6717, \lambda 6731) / \mathcal{I}(\rm H\,\alpha + [N\, II]) &= &1.09 \\
 \mathcal{I}([\rm O\, III]~\lambda 5007) / \mathcal{I}(\rm H\,\beta) &>& 1 \ .
\end{eqnarray}
These ratios are different from those observed in H$\,${\sc ii} regions, and suggest shock excitation. Based on the shock models by \citet{Dopita1995}, \citet{Xu2003} conclude that these lines are predominantly excited by fast ($V_{\rm s} > 100$~km~s$^{-1}$) shocks. 
In fact, the [O$\,${\sc i}] / H$\alpha$ line ratio is so high that it is difficult to explain this ratio only with fast shock models (see sect.~\ref{subsec:SQ-lines-fast-shocks}). 

Optical emission lines are mainly detected at three velocities (6800, 6100, and 6400~km~s$^{-1}$). The
H$\,\alpha$, [N$\,${\sc ii}],  [S$\,${\sc ii}] and  [O$\,${\sc ii}] lines are detected predominently at 6100~km~s$^{-1}$, and the [O$\,${\sc iii}]~$\lambda 5007$ line is detected at 6800~km~s$^{-1}$.
The [O$\,${\sc i}]~$\lambda 6300\,$\AA~ line is centered at 6400~km~s$^{-1}$, and is remarkably broad, with a FWHM $\Delta v \simeq 1130$~km~s$^{-1}$. The width of the line is commensurate with the relative velocity between the intruder and the tidal tail.

\citet{Dopita1996} give  the following  relation for the total energy flux of a radiative shock: 
\begin{equation}
\mathcal{F}_{\rm tot} = 2.28 \times 10^{-6} \left( \frac{V_{\rm s}}{100~\rm km\,s^{-1}} \right) ^3 \left( \frac{n_{\rm H}}{\rm cm^{-3}}\right) ~\rm W~m^{-2} \ ,
\end{equation}
where $n_{\rm H}$ is the preshock hydrogen number density and $V_{\rm s}$ the shock velocity.
We use the aperture $\mathcal{A}_{\rm MS}$ corresponding to the SQ ``main shock region'' ($30 \times 77$~arcsec$^{2}$, corresponding to $\sim 13.7 \times  35.1$~kpc$^{2}$). 
Therefore, the total mechanical luminosity of the SQ high-speed shock propagating into the tenuous intra-group medium is given by: 
\begin{equation}
\mathcal{L}_{\rm tot} = 5.48 \times 10^{36} \left( \frac{V_{\rm s}}{700~\rm km\,s^{-1}} \right) ^3 \left( \frac{n_{\rm H}}{\rm 3.2 \times 10^{-2}~cm^{-3}}\right) \left( \frac{\mathcal{A}_{\rm MS}}{2310~\rm kpc^2} \right)  ~\rm W
\end{equation}
From X-ray observations, \citet{Trinchieri2003, Trinchieri2005, O'Sullivan2009} estimate the average hydrogen density of the hot plasma in the ridge to be $n_{\rm H} \simeq 1.17 \times 10^{-2}$~cm$^{-3}$. For a fast shock, the postshock density is about 4 times the preshock density (sect.~\ref{sec:MHD_shocks}), so we estimate a preshock density of the tenuous gas of $n_{\rm H} \simeq 3.2 \times 10^{-3}$~cm$^{-3}$. The total mechanical energy flux for the SQ main shock is thus $\mathcal{F}_{\rm tot} \simeq 2.6 \times  10^{-6}$~W~m$^{-2}$.  

If we assume that this luminosity is radiated over a galaxy crossing time (the shock age) of $5 \times 10^6$~yr, the intruder/IGM collision deposits $\sim 8.7 \times 10^{57}$~erg of energy in the SQ shock region. This energy is  equivalent to the kinematic energy of $\sim 10^7$ supernovae.

\subsection{Distribution of the fine-structure line emission}
\label{subsec:IRSmap_fine_struct_lines}

\citet{Appleton2006} reported the detection of the [Ne$\,${\sc ii}] $\lambda 12.8\,\mu$m and [Si$\,${\sc ii}] $\lambda 34.8\,\mu$m lines. Thanks to the deeper spectral maps, we have access to the  
spatial distribution of emission from the [Fe$\,${\sc ii}] $\lambda 25.99\,\mu$m and [O$\,${\sc iv}] $\lambda 25.89\,\mu$m blend, [S$\,${\sc iii}] $\lambda 33.48\,\mu$m, [Si$\,${\sc ii}] $\lambda 34.82\,\mu$m,
[Ne$\,${\sc ii}] $\lambda 12.8\,\mu$m and [Ne$\,${\sc iii}] $\lambda 15.56\,\mu$m lines. 

Fig.~4 and 5 of  \hyperref[subsec:paper_Cluver]{paper~{\sc ii}} show the intensity contours of these fine-structure lines overlaid on H$_2$  S(1) and X-ray maps. The strong  [Si$\,${\sc ii}]$\lambda 34.82\,\mu$m emission spatially correlates with the H$_2$ and X-ray emission, whereas the [Fe$\,${\sc ii}]$\lambda 25.99\,\mu$m line emission is only detected towards the AGN NGC~7319 and the center of the ridge, where there is a bright spot of X-ray emission. We direct the reader to sect.~3.2 of \hyperref[subsec:paper_Cluver]{paper~{\sc ii}} for a detailed description of these fine-structure line emission maps. We now discuss the interpretation of these observations with models that allow to constrain the gas physical conditions and the dominant excitation mechanisms (shocks, X-rays, photoionization).

\subsection{Optical and mid-IR fine structure lines as shock diagnostics}
\label{subsec:SQ-lines-fast-shocks}

The deep mid-IR spectral mapping data presented above allow to complement the analysis of the rotational H$_2$ lines with the fine-structure line emission from the SQ ridge (see sect.~\ref{spectral_IRSmap_H2} for a description of the observational results). 
These metal lines provide complementary diagnostics to probe the physical conditions of the emitting gas and more energetic excitation mechanisms. Since the main shock region (the X-ray ridge) also contains star forming regions, we have selected a sub-region of our extended \textit{IRS} spectral mapping, so-called \textit{``center''} because it is located in the central region of the ridge, that has less contamination by star-forming regions (see sect.~4, Fig.~7 of  \hyperref[subsec:paper_Cluver]{paper~{\sc ii}}). We compare the line ratio diagnostics within these two apertures. 

Optical spectroscopy diagnostics show that the SQ shock can be classified as a LINER (Low Ionization Excitation Region). We will show in this section that the gas must be heated to $\gtrsim 10^5$~K by collisional excitation to explain the optical and mid-IR fine-structure line emission.

We use the \citet{Allen2008}  shock model library described in sect.~\ref{sec:fast-shock-models} to compare models with the fine structure line ratios measured with \textit{Spitzer IRS} data. The model results depend on the preshock magnetic parameter $b=B/\sqrt{n_{\rm H}}$, where $B$ is the magnetic field strength and $n_{\rm H}$ the preshock gas density. In the models presented here, we assume $b=1$, which is slightly lower than the equipartition value of 3.23. We compare the model results with and without the radiative precursor. The radiative precursor of a fast shock moving into low-density
gas adds an emission component which has the spectral characteristics of photoionized gas with a high ionization parameter.
To neglect the precursor may be relevant if the shock is propagating in a clumpy medium because only a fraction of the ionizing photons will interact with the molecular gas (see sect.~\ref{sec:shocksinmultiphasemedia}).

\begin{figure}
   \centering
    \includegraphics[angle=90, width=0.49\textwidth]{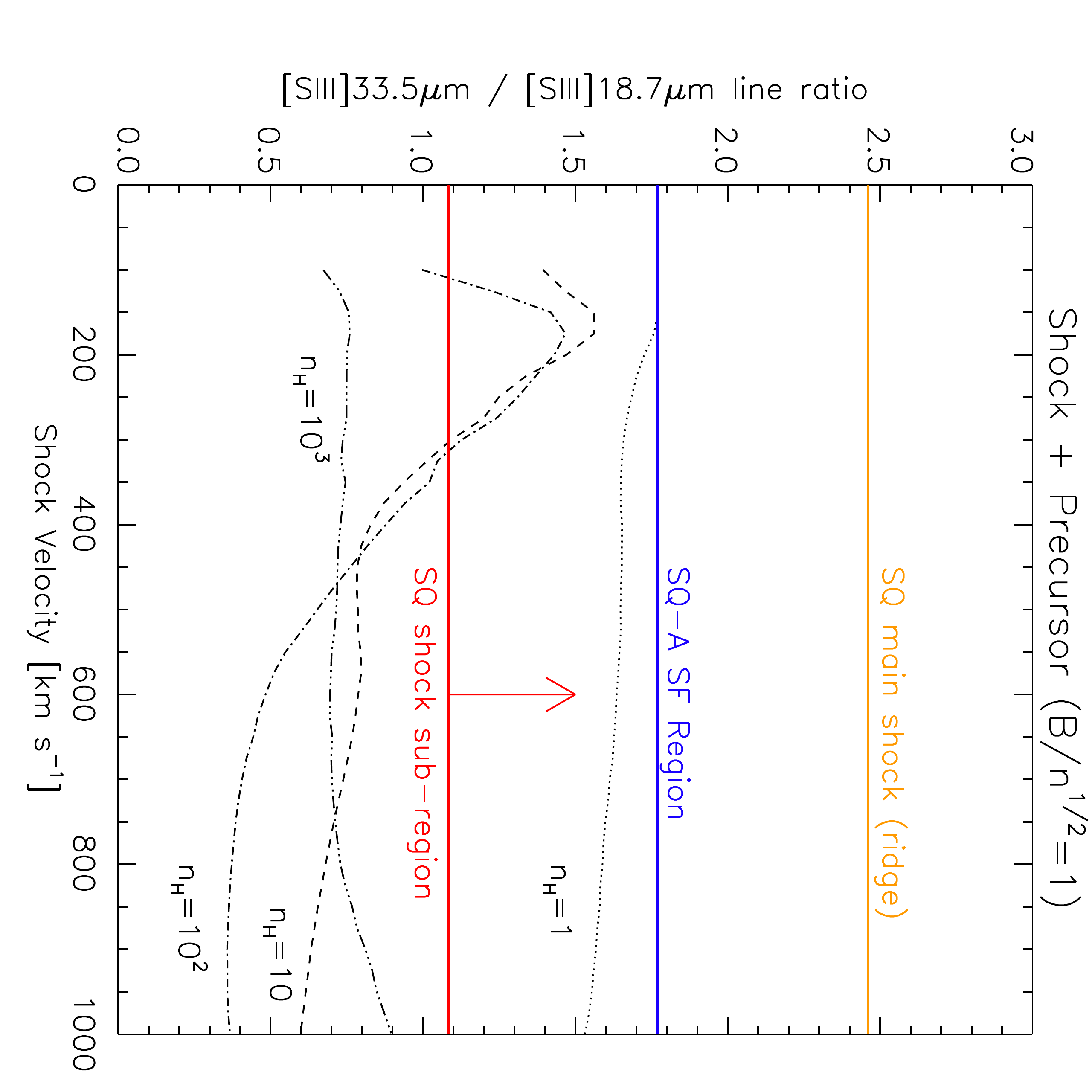}
    \includegraphics[angle=90, width=0.49\textwidth]{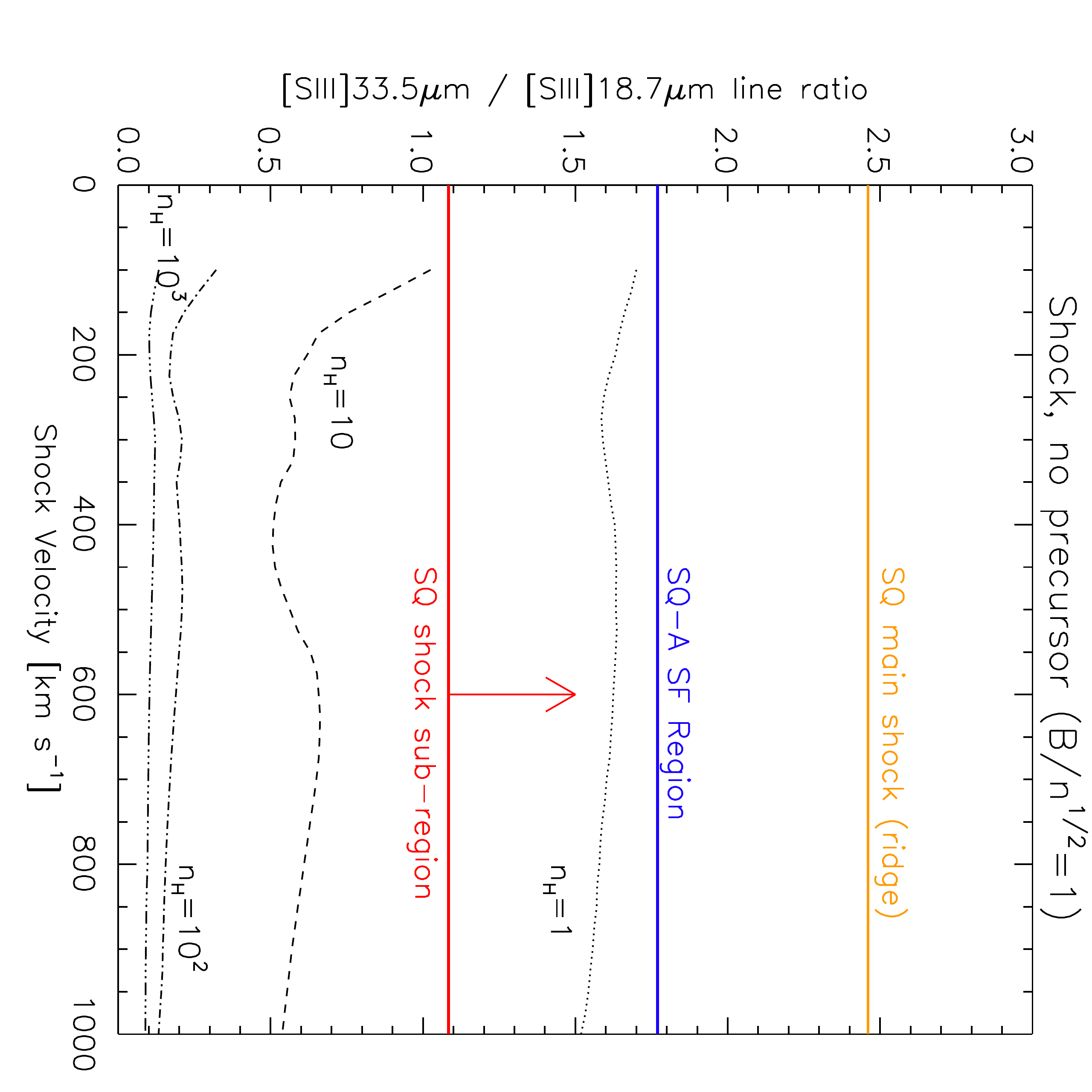}
      \caption[Model and observed fine-structure line ratios for SQ]{The fine-structure line ratios measured in the SQ shock (full ridge and sub-region in the center) are  compared with shock calculations (broken curves)  from \citet{Allen2008}. The model values are shown for shock velocities between 100 and 1000~km~s$^{-1}$ (with steps of 25~km~s$^{-1}$ ), four preshock gas densities ($n_{\rm H} = 1$, 10, 100 and 1000~cm$^{-3}$) for a magnetic parameter $B/\sqrt{n_{\rm H}} = 1$.   Observed values are the horizontal solid lines in colors.}
       \label{fig_SIII33_SIII18_Vsh_SQ}
\end{figure}

\begin{figure}
   \centering
    \includegraphics[width=\textwidth]{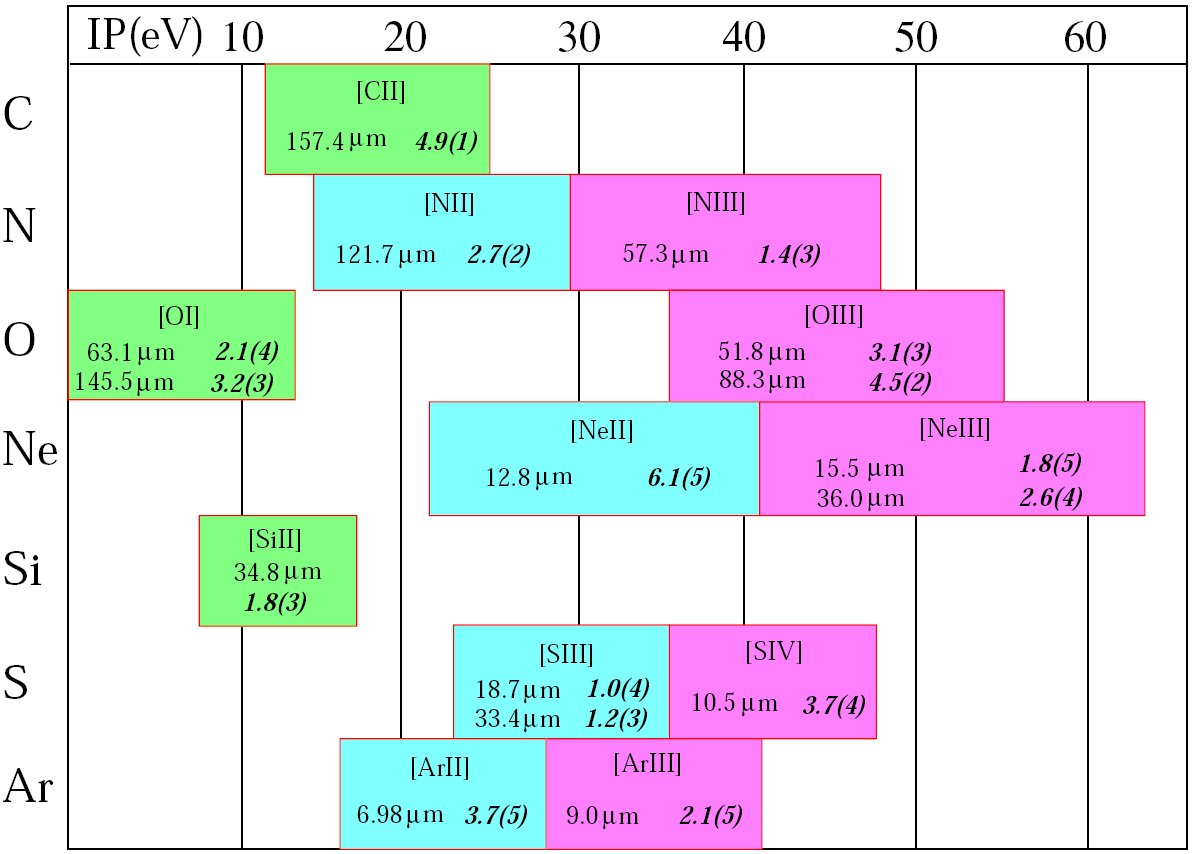}
      \caption[Ionization potentials and critical densities of mid-IR fine structure lines]{Ionization potentials and critical densities of important mid-IR fine structure lines. The electron critical densities, indicated for every line in italic, are given in units of cm$^{-3}$ and are expressed as $a(b) = a \times 10^b$. Figure taken from \citet{Mart'in-Hern'andez2002}.}
       \label{fig_ioniz_pot_fine_str_Mid-IRlines}
\end{figure}

The [S$\,${\sc iii}] lines ratios are shown on Fig.~\ref{fig_SIII33_SIII18_Vsh_SQ} as a function of the shock velocity. The broken curves are the shock models for different preshock gas densities ($1-1000$~cm$^{-3}$). The solid coloured lines are the observed values.
The electron  critical densities of the [S$\,${\sc iii}]$\lambda 33.48\,\mu$m and [S$\,${\sc iii}]$\lambda 18.71\,\mu$ lines are $n_{e} = 1.2 \times 10^{3}$~cm$^{-3}$ and $n_{e} = 10^{4}$~cm$^{-3}$, respectively. The observed ratio for the full ridge (2.46) is close to the low density limit \citep[$n_{e} < 100$~cm$^{-3}$, see Fig.~\ref{fig_ioniz_pot_fine_str_Mid-IRlines} and details in][]{Mart'in-Hern'andez2002}. The comparison with shock models show that the gas preshock density is smaller than $\sim 1$~cm$^{-3}$. The lower limit for the shock sub-region is compatible with preshock gas densities smaller than $100$~cm$^{-3}$ and shock velocities $100 < V_{\rm s} < 300$~km~s$^{-1}$.
The [S$\,${\sc iii}] lines ratio diagnostic clearly shows that the mid-IR line emission from the ionized gas originates from moderate shocks (of the order of 100~km~s$^{-1}$) in low-density ($\approx 1$~km~s$^{-1}$). Faster shocks in denser gas would not be compatible with the available ram pressure in the surrounding hot plasma ($\approx 2 \times 10^5$~K~cm$^{-3}$).

\begin{figure}
   \centering
    \includegraphics[angle=90, width=0.49\textwidth]{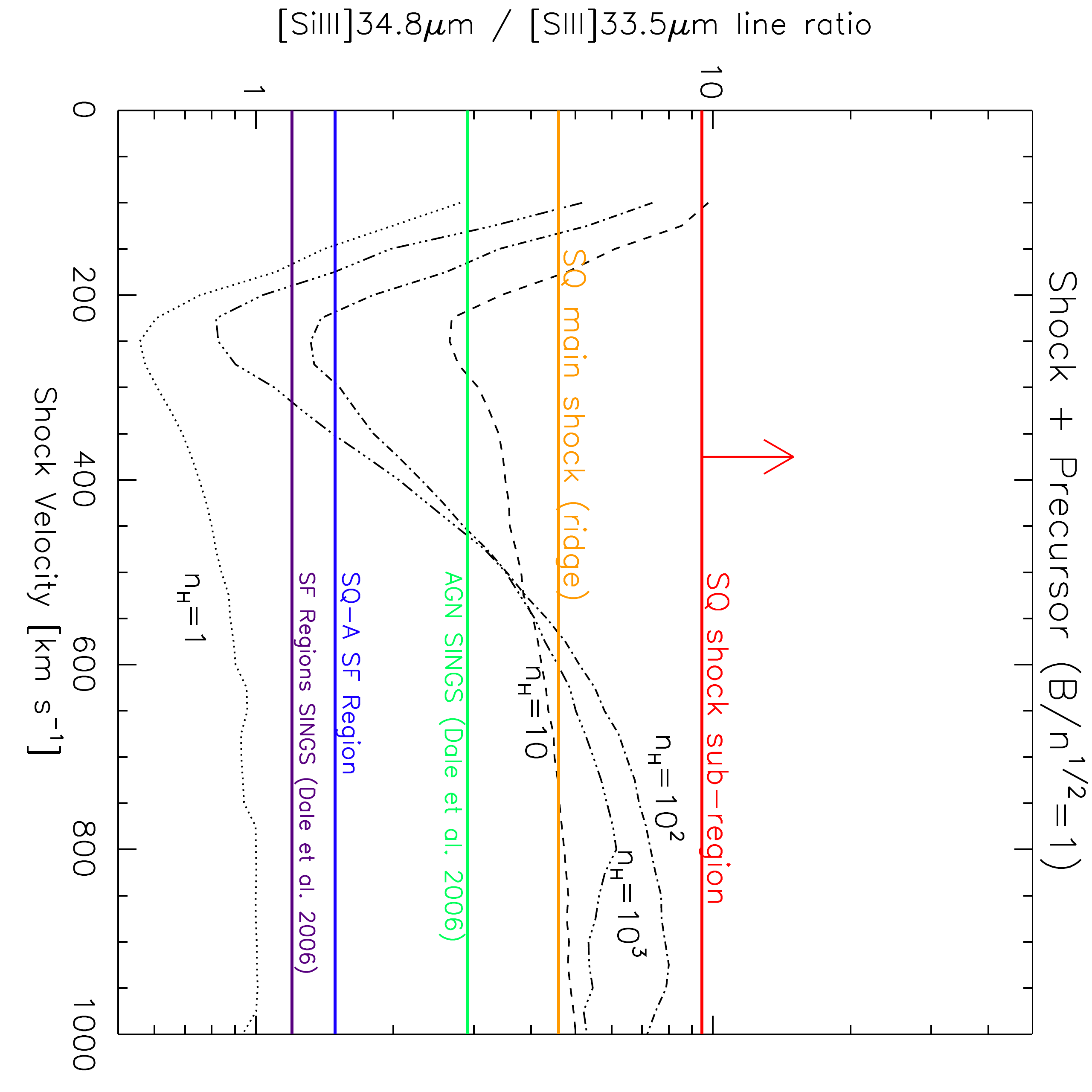}
    \includegraphics[angle=90, width=0.49\textwidth]{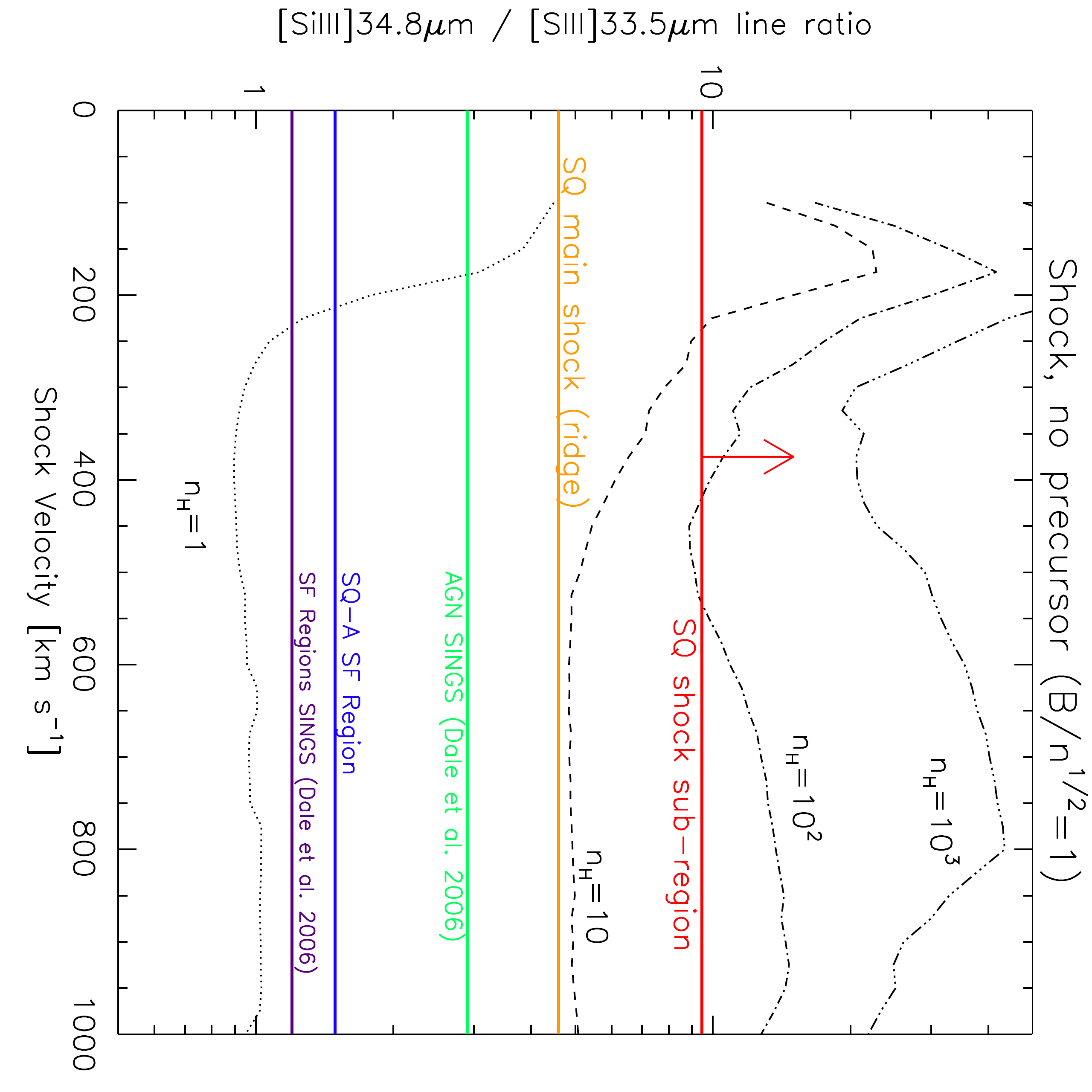}
      \caption[Model and observed fine-structure line ratios for SQ]{The fine-structure line ratios measured in the SQ shock (full ridge and sub-region in the center) are  compared with shock calculations (broken curves)  from \citet{Allen2008}. The model values are shown for shock velocities between 100 and 1000~km~s$^{-1}$ (with steps of 25~km~s$^{-1}$ ), four preshock gas densities ($n_{\rm H} = 1$, 10, 100 and 1000~cm$^{-3}$) for a magnetic parameter $B/\sqrt{n_{\rm H}} = 1$.   Observed values are the horizontal solid lines in colors.}
       \label{fig_SiII_SIII33_Vsh_SQ}
\end{figure}

\begin{figure}
   \centering
    \includegraphics[angle=90, width=0.49\textwidth]{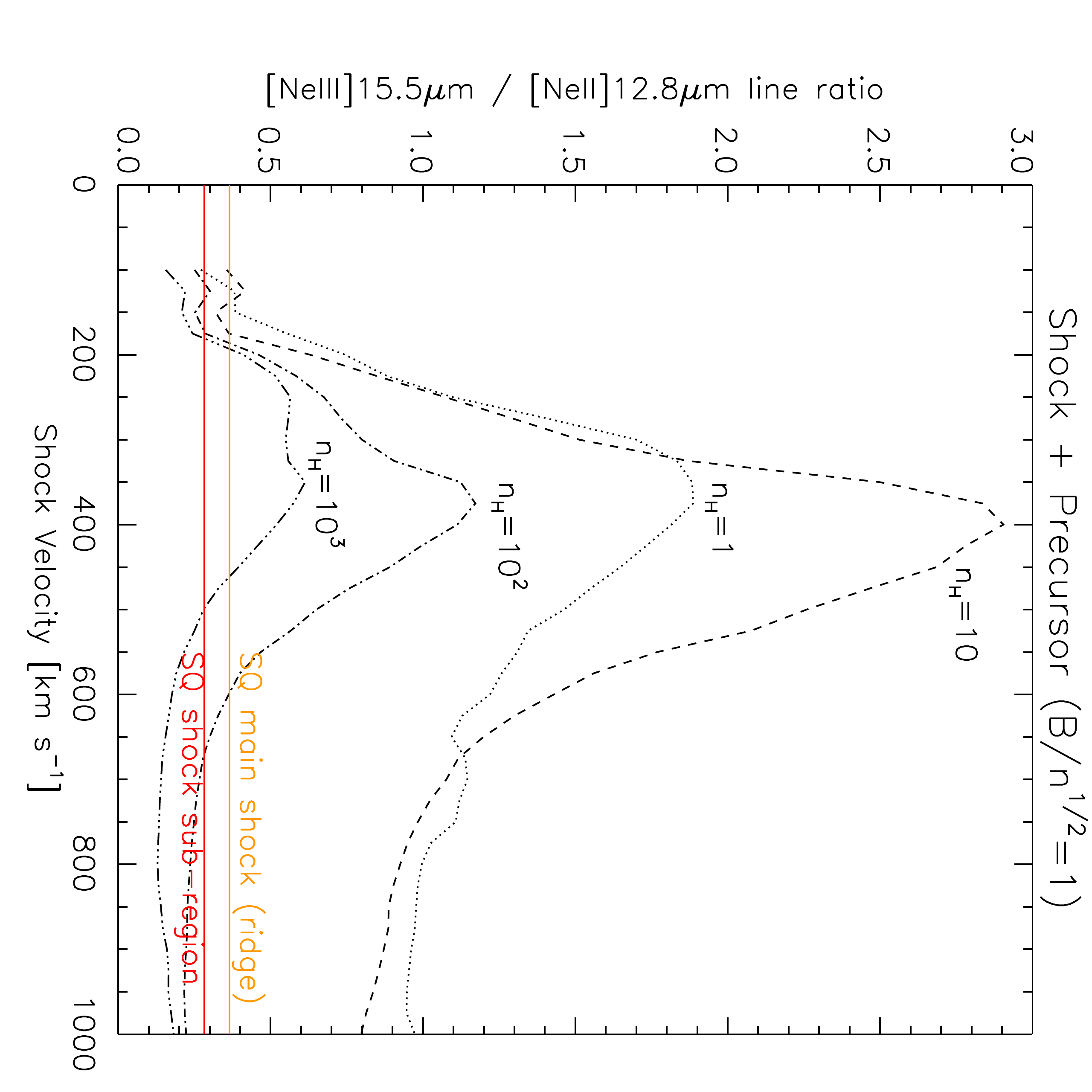}
    \includegraphics[angle=90, width=0.49\textwidth]{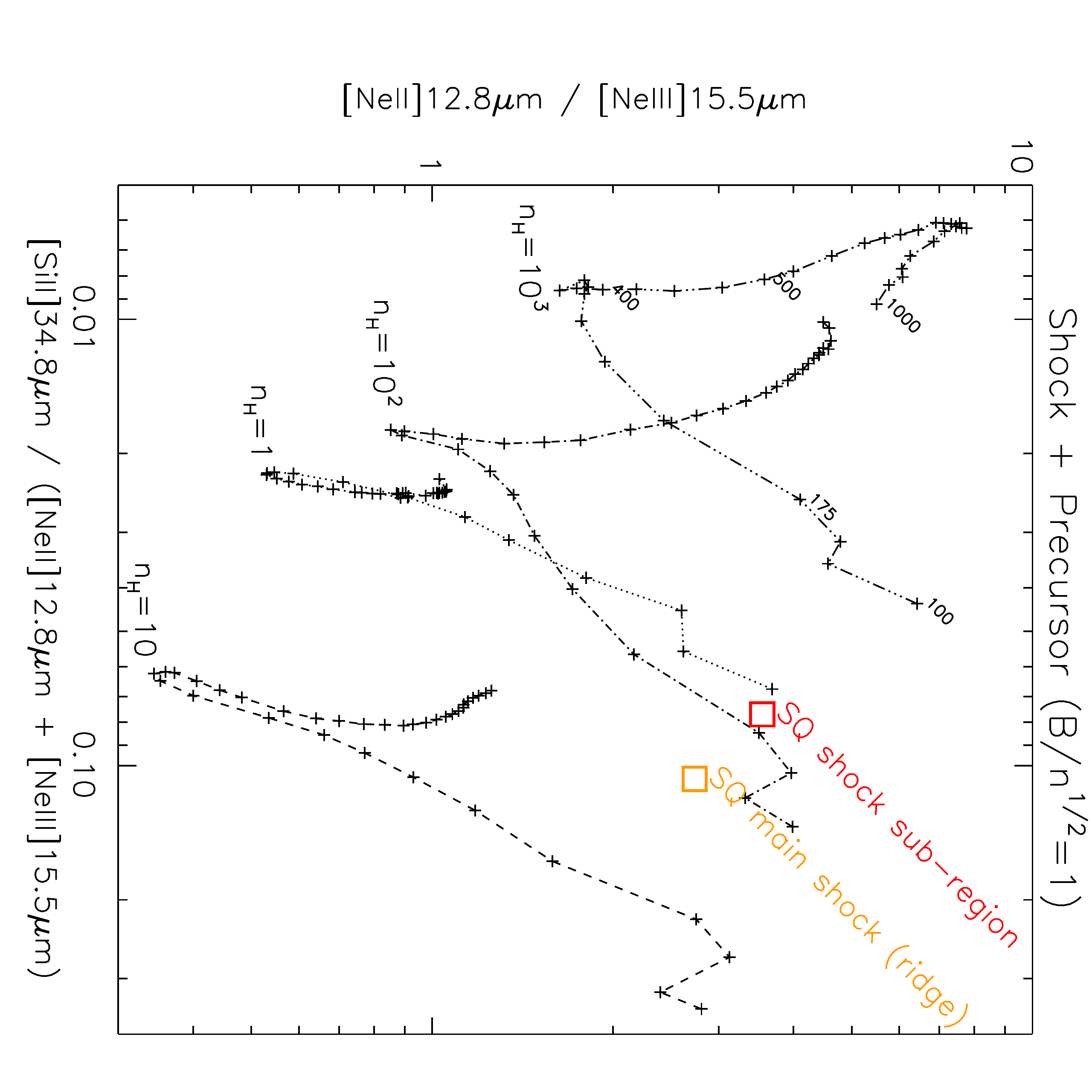}
      \caption[Model and observed fine-structure line ratios for SQ]{Observed mid-IR line ratios compared with shock models (same as Fig.~\ref{fig_SiII_SIII33_Vsh_SQ}.  The model values are shown for shock velocities between 100 and 1000~km~s$^{-1}$ (with steps of 25~km~s$^{-1}$ ), four preshock gas densities ($n_{\rm H} = 1$, 10, 100 and 1000~cm$^{-3}$) for a magnetic parameter $B/\sqrt{n_{\rm H}} = 1$.   Observed values are the horizontal solid lines in colors. The Ne line ratios shown on the \textit{left} panel contrain the shock velocities to be of order of $100-200$~km~s$^{-1}$. This result is confirmed by the Ne and S line diagnostic plot on the \textit{right panel}. Note that the radiative precursor is included in both panels.  }
       \label{fig_NeIII_NeII_Vsh_SQ}
\end{figure}

XDRs\footnote{X-ray Dissociation Regions} models show that the  [Si$\,${\sc ii}] line is one of the major mid-IR cooling line from gas exposed to X-rays \citep[see Fig.~8 of][]{Maloney1996}. 
XDRs models have been used to interpret the line emission from AGN.
Powerfull [Si$\,${\sc ii}] line emission can also be produced by shock excitation, whereas the  [S$\,${\sc iii}] mid-IR lines are mostly tracers of H$\,${\sc ii} regions. Therefore, the [Si$\,${\sc ii}] /  [S$\,${\sc iii}] line ratio measures the ratio between the X-rays or shocks and photoionization excitation powers.

Fig.~\ref{fig_SiII_SIII33_Vsh_SQ} shows the observed (solid, coloured lines) and model (broken curves) [Si$\,${\sc ii}]$\lambda 34.82\,\mu$m /  [S$\,${\sc iii}]$\lambda 33.48\,\mu$m and  [S$\,${\sc iii}]$\lambda 33.48\,\mu$m /  [S$\,${\sc iii}]$\lambda 18.71\,\mu$m line flux ratios as a function of the shock velocity, and for different preshock gas densities ($1-1000$~cm$^{-3}$). The [S$\,${\sc iii}]$\lambda 33.48\,\mu$m line is not detected in the sub-region in the center of the shock, so we indicate an lower limit for the line ratios (red line). For comparison, the average line ratios observed in local AGN and extragalactic star-forming regions (from the \textit{SINGS}\footnote{\textit{Spitzer} Infrared Nearby Galaxies Survey, \url{http://sings.stsci.edu/}} survey) are shown.

The [Si$\,${\sc ii}] /  [S$\,${\sc iii}] SQ line ratios stand out above typical values observed in star-forming regions or AGN, and cannot be accounted for by PDRs or XDRs models. They are rather consistent with the line ratios observed in supernova remnants (see sect.~5 of \hyperref[subsec:paper_Cluver]{paper~{\sc ii}}), suggesting shock-excitation. 
The  lower limit of the  [Si$\,${\sc ii}]$\lambda 34.82\,\mu$m /  [S$\,${\sc iii}]$\lambda 33.48\,\mu$m line ratio observed in the SQ shock sub-region seems to be consistent with $\sim 100$~km~s$^{-1}$ shocks into $n_{\rm H} \simeq 10$~cm$^{-3}$ gas. Unlike the sub-region, the SQ-A starburst line ratios are consistent with the values observed in standard extragalactic star-forming regions. This shows that photoionization must be the dominant process in this region, with a minor contribution from shocks. 

The Fig.~\ref{fig_NeIII_NeII_Vsh_SQ} shows the [Ne$\,${\sc iii}]$\lambda 15.56$ / [Ne$\,${\sc ii}]$\lambda 12.8$ line flux ratio as a function of the shock velocity (left panel) or as a function of the  [Si$\,${\sc ii}] / ([Ne$\,${\sc ii}] + [Ne$\,${\sc iii}]) line ratio. 
For our range of preshock densities ($1-1000$~cm$^{-3}$), the shock models show that the observed line  ratios are consistent with shock velocities of $\sim 100 - 200$~km~s$^{-1}$.

\begin{figure}
   \centering
    \includegraphics[angle=90, width=0.49\textwidth]{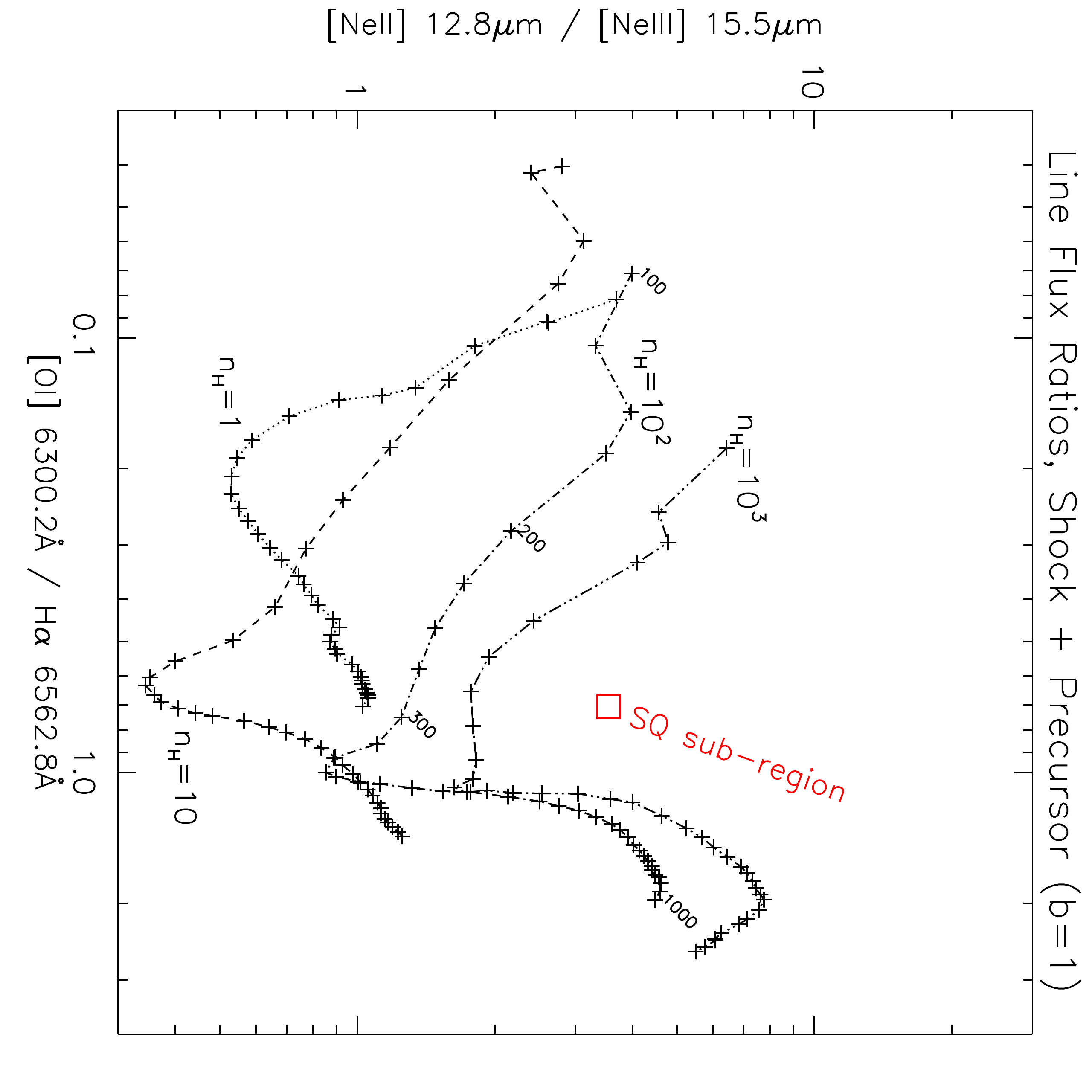}
    \includegraphics[angle=90, width=0.49\textwidth]{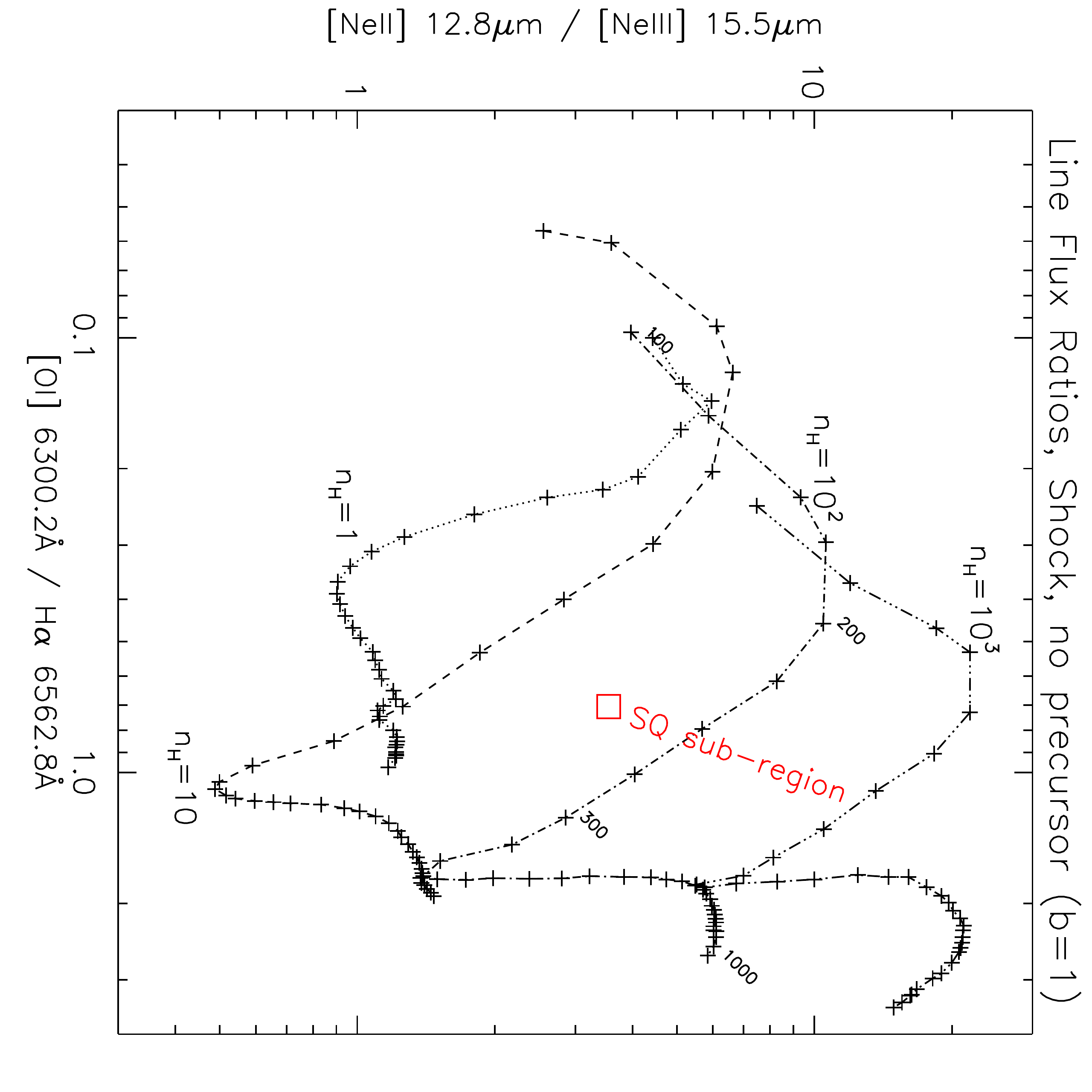}
      \caption[Model and observed optical and fine-structure line ratios for SQ]{Ne fine-structure line ratio vs. optical line ratio measured in the SQ shock (sub-region in the center) as  compared with shock calculations (broken curves)  from \citet{Allen2008}. The model values are shown for shock velocities between 100 and 1000~km~s$^{-1}$ (with steps of 25~km~s$^{-1}$ ), four preshock gas densities ($n_{\rm H} = 1$, 10, 100 and 1000~cm$^{-3}$) for a magnetic parameter $B/\sqrt{n_{\rm H}} = 1$. In the \textit{left} panel, the radiative precursor is included, whereas in the \textit{right} panel, only the shock contribution is plotted.}
       \label{fig_LineRatios_OIHa_NeII_NeIII_SQ}
\end{figure}

\citet{Xu2003} suggest that the optical lines ratios observed in the center of the ridge (see sect.~\ref{subsec:opticallineemission_shocks}) are evidence for a dominant shock excitation.
The mid-IR fine structure lines detected by \textit{Spitzer} confirm this conclusion. 
However, the large value of the [O$\,${\sc i}] / H$\alpha$ line ratio observed in the ridge, 
$\mathcal{F}([\rm O\, I]~\lambda 6300) / \mathcal{F}(\rm H\,\alpha + [N\, II])  =  0.53$, is difficult to explain with fast shocks models. Fig.~\ref{fig_LineRatios_OIHa_NeII_NeIII_SQ} shows the [Ne$\,${\sc ii}] to  [Ne$\,${\sc iii}] ratio as a  function of the [O$\,${\sc i}] / H$\alpha$ optical line ratio. 
In the left panel, where the precursor is included, the observed value do not match with the models.
The mid-IR [Ne$\,${\sc ii}] to  [Ne$\,${\sc iii}] line ratio in SQ suggest a low ionization parameter, whereas the radiative precursor produces a highly-ionized medium ahead of the shock. This apparent contradiction can be solved if the SQ intragroup medium is clumpy. In this case, a large fraction of the ionizing photons can escape from the medium without interacting with the dense phase.
 If we neglect the radiative precursor (right panel), we find a much better agreement with the models. The   line ratios are reproduced for a preshock density of $n_{\rm H} = 10-10^2$~cm$^{-3}$ and a shock velocity of $V_{\rm s} = 200 - 250$~km~s$^{-1}$. These conditions imply high values of the postshock pressure as compared with the ambient value. 
Note that a significant contribution from non-ionizing $J$-shocks contribute to the optical lines, in particular to the [O$\,${sc i}] 6300\AA~line (see sect.~\ref{subsec:shocks-molecular-cooling}). 
This will be further discussed in sect.~\ref{subsec:mass_energy_transfers}.

In sect.~6 of \hyperref[subsec:paper_Cluver]{paper~{\sc ii}} we discuss the [Fe$\,${\sc ii}] $\lambda 25.99\,\mu$m and  [Si$\,${\sc ii}]$\lambda 34.82\,\mu$m line emission. The fact that the observed [Fe$\,${\sc ii}] / [Ne$\,${\sc ii}]  and  [Si$\,${\sc ii}] /  [Ne$\,${\sc ii}] line ratios are both smaller than the
shock model values can be interpreted as evidence for Si and Fe depletion onto dust grains (see Fig.~11 of \hyperref[subsec:paper_Cluver]{paper~{\sc ii}}). This supports our idea that dust survives in the dense gas phases that are the sites for H$_2$ formation. This will be further discussed in chapter~\ref{chapter:SQ_dust} where we model the dust emission from the molecular gas in SQ.

\subsection{Conclusion: a distribution of shock velocities}

The X-ray, optical and mid-IR line emissions from the SQ shock suggest that the gas experiences a wide distribution of shock velocities. We relate this distribution of shock velocities with a distribution of preshock gas densities, due to the multiphasic nature of the preshock medium (\hyperref[paper_SQ_H2]{paper~{\sc i}}).  The denser gas phases are associated with the smaller shock velocities. 

Very fast shocks  ($V_{\rm s} \sim 700$~km~s$^{-1}$) driven into preshock gas of a few $10^{-3}$~cm$^{-3}$ are responsible for the bright X-ray emission in the ridge. Optical lines and mid-IR fine-structure lines are likely to be produced by intermediate ($100 < V_{\rm s} < 300$~km~s$^{-1}$) ionizing $J$-shocks. A significant contribution from non-ionizing $J$-shocks can also contribute to these lines, in particular to the [O$\,${\sc i}] 6300\AA ~(which may help in explaining the very high value of  [O$\,${\sc i}]/ H$\,\alpha$), [Fe$\,${\sc ii}] $\lambda 25.99\,\mu$m and  [Si$\,${\sc ii}]$\lambda 34.82\,\mu$m lines. And last but not least, low-velocity ($5-20$~km~s$^{-1}$), non-dissociative $C$-shocks driven into dense gas ($>10^3$~cm$^{-3}$) are associated to the rotational H$_2$ emission seen by \textit{Spitzer}.

\section{Remarks about the NGC~7319 Seyfert galaxy}
\label{SQ-NGC7319-comments}
\index{NGC 7319!outflow}
\index{NGC 7319!bridge}

\subsection{Observational context: and AGN in the SQ group}

In this short paragraph I gather some observational results about the NGC~7319 galaxy. Although not related to the SQ main shock itself, these results are helpful to interpret our recent, extended IR and CO observations of SQ.

\citet{VanDerHulst1981} first detected the NGC~7319 AGN with radio continuum VLA observations of
SQ at 20cm. They detected a bright compact source with a jet-like extension, and suggested this might be related to those found in Seyfert galaxies. \citet{Huchra1982} obtained the optical spectrum which shows clearly Sy2-type emission-line features. Based on the stellar velocity dispersion, \citet{Woo2002} estimated a black hole mass of $2.4 \times 10^7$~M$_{\odot}$.
\citet{Aoki1999} made VLA A-array radio continuum observations, complimented with an
archival HST optical image. They found a chain of 3 radio sources, interpreted as the nucleus
and its two jets on opposite sides. Optical features are found in the HST image closely related
to the radio jets, interpreted as gas compressed and excited by bow shocks driven into the
ambient medium by the jets. This is different from the so-called extended emission line regions
(EELR) which are supposedly excited by the AGN radiation \citep{Aoki1996}. 

Even higher
resolution (0.16'') radio continuum observations at 1658 MHz, using MERLIN, were reported
by \citet{Xanthopoulos2004}. They compared the data with an HST/ACS U-band (F330W)
image, and found extended UV emission around the nucleus and the northern jet. They argued
that this indicates star formation triggered by the jet/ISM interaction. By assuming the diffuse
radio emission outside the compact sources is due to the star formation, they estimated that
the SFR in the circum-nuclear region is 8.4~M$_{\odot}$~yr$^{-1}$. However, all of the optical spectra of emission line regions in the circum-nuclear region show Seyfert or LINER line ratios and none is H$\,${\sc ii} region-like \citep{Aoki1996}, indicating that most of the radio and UV radiation is related to the AGN and/or the shocks. Comparing the UV and MIR observations, \citet{Xu2005} found the AGN
and the surrounding region is highly obscured, with an FUV extinction of $A_{\rm FUV} = 5.4$~mag, consistent with the Sy2 classification.

\subsection{The H$_{\bf 2}$ bridge: AGN-driven outflow or tidal interactions?}

\index{NGC 7319!outflow}
\index{NGC 7319!bridge}

Our new \textit{Spitzer} spectral maps presented in sect.~\ref{spectral_IRSmap_H2} show that the H$_2$ emission extends from the ridge to the core of the AGN NGC~7319. 
This ``H$_2$  bridge'' feature is also seen on the 1.4~GHz radio image \citep[see Fig.~\ref{fig_SQ_Xrays_radio_contours} and][]{Williams2002}, on H$\,\alpha$ maps \citep[see Fig.~\ref{fig_SQ_Ha} and][]{Xu1999, Plana1999, Sulentic2001, Gutierrez2002}, and faintly detected in X-rays \citep[see Fig.~\ref{fig_SQ_CFHT_Xray} and][]{Trinchieri2003, Trinchieri2005}. The distribution of the H$_2$ emission seems to correlate with that of the radio emission. 

The origin of the ``H$_2$  bridge''  is unclear. Is it related to the AGN activity of NGC~7319, or is it related to past or recent galaxy interactions? Optical spectroscopy \citep{Aoki1996} and VLA radio observations \citep{Aoki1999} seem to support the first hypothesis. Indeed, \citet{Aoki1996} interpret the large velocity gradients between the AGN core and the South-West bridge as an high-speed ($\sim 500$~km~s$^{-1}$) jet-induced outflow of gas. The very high electron density in the outflow ($\sim 600$~cm$^{-3}$, twice that of the nucleus) and the multiple radio structure in the jet suggest that the outflow is interacting with interstellar matter. The H$_2$ emission in this bridge may then be powered by jet-driven shocks within dense clouds or/and AGN radiation. \citet{Cluver2009} propose an alternative explanation: the bridge is a remnant of previous tidal interactions. This scenario is suggested by the similarity between the mid-IR spectrum and X-ray properties of the \textit{ridge} and \textit{bridge} regions. This picture fits in the context of multiple shock heating events proposed by \citet{Moles1997, Sulentic2001, O'Sullivan2009}.
The two scenarios (AGN-driven outflow and galaxy-wide shock) are difficult to disentangle because they lead to similar observational signatures. Our new CO observations in the ridge and in the bridge set a new light on this question, and we will discuss this further in sect.~\ref{sec:CO_EMIR}.

\section{Summary and conclusions}
\label{sec:SQ-conclusions}

One of the most interesting features of the Stephan's Quintet compact group is the presence of a galaxy-wide shock in the halo of the group, created by an intruding galaxy colliding with a tidal tail at a relative velocity of $\sim 1\,000$~km~s$^{-1}$.  Evidence for a group-wide shock comes from observations of X-rays from the hot postshock gas in the ridge strong radio synchrotron emission from the radio emitting plasma and shocked-gas excitation diagnostics from optical emission lines. \textit{Spitzer} observations show that this gas also contains molecular hydrogen with an extreme velocity dispersion. The $\rm H_2$ luminosity  is larger than the X-ray emission from the same region, thus the $\rm H_2$ line emission dominates over X-ray cooling in the shock. As such, it plays a major role in the energy dissipation and evolution of the postshock gas. 

The interpretation of these observations is the core of my PhD work. I present a scenario where H$_2$ gas forms out of the postshock gas that results from the supersonic collision between two multiphase gas streams. 
This collision leads to a multiphase medium with a distribution of shock velocities, related to the variations of the density in the preshock medium. The H$_2$ formation in the SQ ridge environment is quantified by calculating the time-dependent isobaric cooling of the gas, including dust destruction mechanisms. I show that H$_2$ forms out of the dense regions where dust survives the shocks. 
The bulk kinetic energy of the molecular gas is the main energy reservoir of the postshock gas, and it has to be dissipated within the H$_2$ gas to explain the poserful H$_2$ excitation. I propose that supersonic turbulence is sustained within the molecular gas by a transfer of momentum induced by cloud-cloud collisions and the relative shear motions between the molecular gas and the hot plasma. A phenomenological model of low-velocity MHD shocks driven into the dense H$_2$ gas is capable of explaining the observed excitation characteristics.

Within this framework, dust is a key ingredient. It is required for H$_2$ formation. Our model has implications on the expected dust emission in the SQ halo and on dust survival in a multiphase environment. These points are adressed in chapter~\ref{chapter:SQ_dust}, as well as a discussion of star formation in the SQ group.

\section{Publication: paper II}
\label{sec:paper_Cluver}

In the following, we reproduce the \citet{Cluver2009} paper, entitled \textit{Powerful H$_2$ line-cooling in Stephan's Quintet : I- Mapping the dominant cooling pathways in group-wide shocks}, accepted for publication in ApJ. This paper contain some figures I have discussed in this chapter, and a lot of details that I  have intentionally not repeated in this manuscrit.

 \includepdf[noautoscale=true, scale=0.9, frame=false, pages=-, fitpaper=true, link=true]{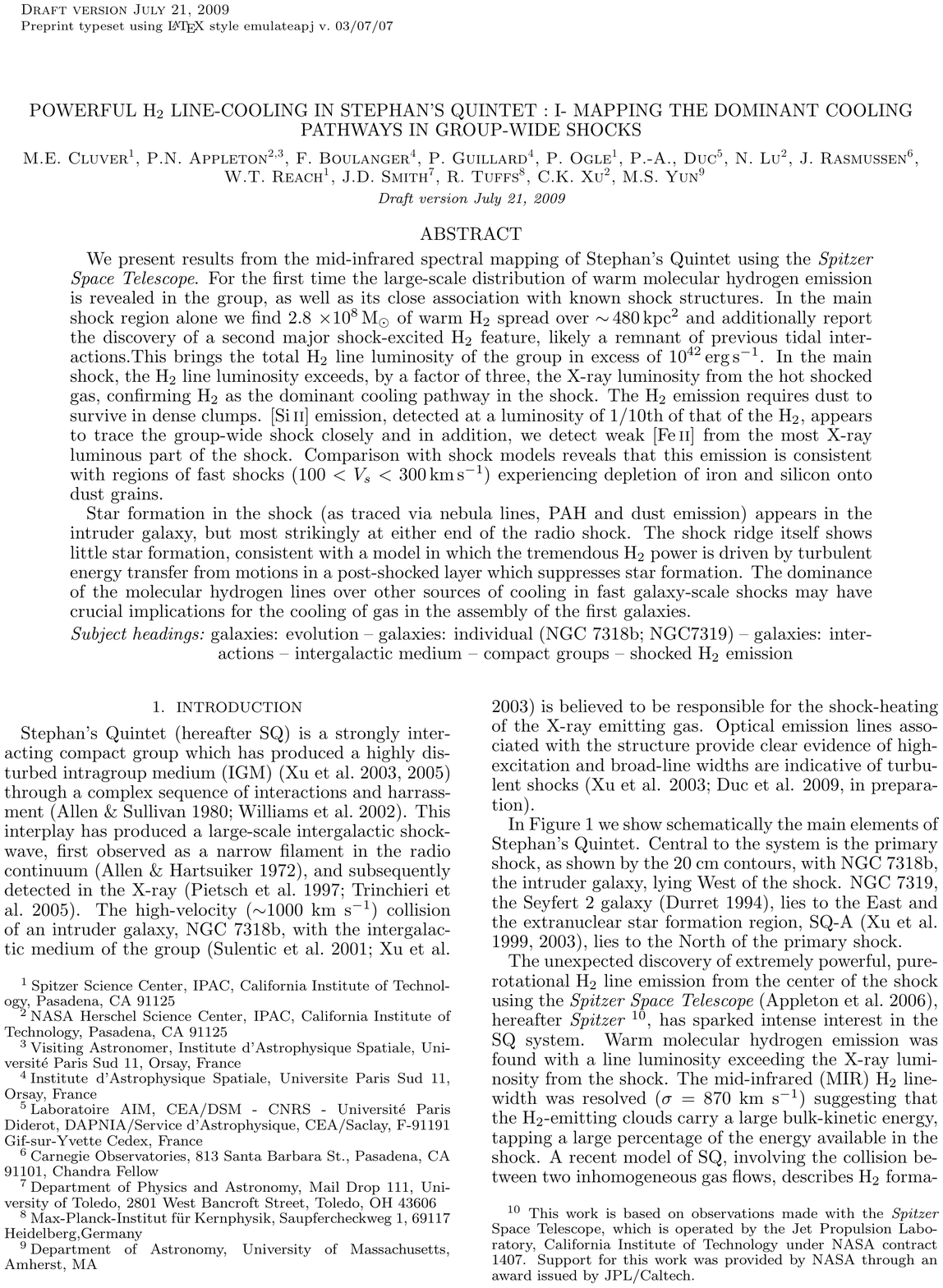}


\chapter{Cold molecular gas in Stephan's Quintet}
\label{chapter:SQ_CO}

\epigraph{Continuous eloquence wearies. Grandeur must be abandoned to be appreciated. Continuity in everything is unpleasant. Cold is agreeable, that we may get warm.}{Blaise Pascal}



\begin{Abstract}

The warm ($> 50$~K) H$_2$ gas detected by Spitzer in the halo of the Stephan's Quintet (SQ) group does not provide a complete inventory of the molecular gas. Gas colder than $\approx 50$~K does not contribute to the mid-IR H$_2$ emission. In order to determine the physical state, the mass and kinematics of the molecular gas in SQ, it is crucial to look for cold molecular gas. This chapter presents the previous CO obervations of the group, and the results of two observation campaigns at the IRAM 30m telescope that I have conducted from the writing of the proposal to the data analysis.

\end{Abstract}

\minitoc


\section{Introduction}
\label{sec:SQ-CO-intro}

\PARstart{W}ithin the evolutionary picture of the Stephan's Quintet (SQ) postshock gas presented in sect.~\ref{subsec:mass_energy_transfers}, it is fundamental to know whether cold molecular gas ($< 50$~K) is associated with the warm H$_2$ gas. In star forming galaxies, the warm molecular gas accounts for about a few percent of the total cold H$_2$ gas inferred from the CO luminosity \citep{Roussel2007}. \textit{What is the cold to warm H$_2$ gas mass ratio in SQ?}
This is a major open question about the physical state and mass budget of the molecular gas. 

According to our model of formation of molecules in the SQ shock, CO should form on the same timescale as H$_2$. However, because the cold gas may be continuously energized by turbulent energy dissipation, the physical state of the shocked gas is likely to be different from that of GMCs in galactic disks.
These observations are crucial to test our scenario of H$_2$ formation in the postshock gas and to obtain 
the kinematics of the molecular gas, which cannot be obtained from the low-resolution Spitzer observations of the  H$_2$ alone.

Past interferometric observations do not show any CO signal associated with the H$_2$-emitting ridge. Previous single-dish observations are mostly concentrated on the southern tidal features or on the northern starburst region (SQ-A). Therefore we decided to embark on an observation campaign with the IRAM 30-meter telescope to search for CO emission in the SQ ridge. I have carried on this campain from the beginning (writing of the proposal) to the end (data reduction and analysis), through the observations at the telescope. This was one of the great experiences I had the chance to live during my PhD!

Before presenting our new observations with the IRAM 30-meter telescope (sect.~\ref{sec:CO_EMIR}) and some preliminary results (sect.~\ref{sec:SQ-CO-EMIR-results}), I first review the context of the previous CO observation in the SQ group. 

\section[Previous CO observations of Stephan's Quintet]{Previous CO observations of Stephan's Quintet: molecular gas outside galaxies}
\index{Stephan's Quintet!molecular gas content}
\label{sec:SQ-previous-CO-obs}

Before the discovery of warm ($T > 150$~K) molecular hydrogen in SQ (sect.~\ref{sec:H2obsSQ}), 
CO emitting gas was detected in the group. I summarize here the history and main results of these CO detections.

The first millimeter CO observations only allowed to detect cold molecular gas in NGC 7319 \citep{Yun1997, Verdes-Montenegro1998, Leon1998}, which contains $4.8 \times 10^9$ M$_{\odot}$ \citet{Smith2001a}. Given its blue luminosity of $\mathcal{L}_ {\rm B} = 5.5 \times 10^{10}$~L$_{\odot}$ \citep{Verdes-Montenegro1998}, this yields a normal ratio of $\mathcal{L}_ {\rm B} / M_{\rm H_2} = 11$ \citep{Lisenfeld2002}. 
Thanks to the BIMA\footnote{Berkeley Illinois Maryland Association, \url{http://bima.astro.umd.edu/bima.html}} interferometer, \citet{Gao2000, Petitpas2005} showed that $3/4$ of the CO gas in NGC~7319 is lying outside its nuclei, mostly in an extended northern region.

\begin{figure}
   \centering
    \includegraphics[width=\textwidth]{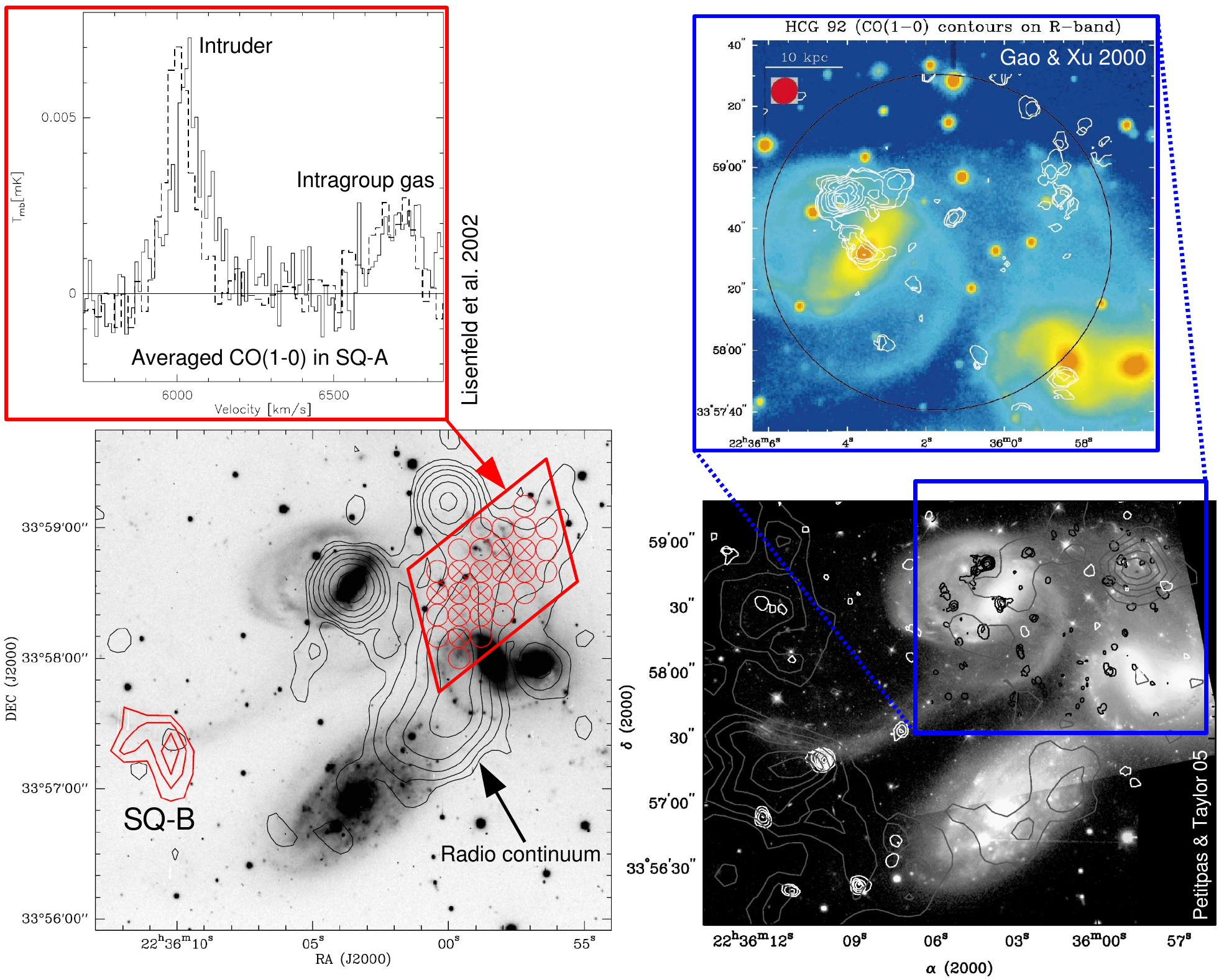}
      \caption[Single-dish and interferometric CO observations of Stephan's Quintet]{Single-dish and interferometric CO observations of Stephan's Quintet. \textit{Left:} The bottom is an optical image overlaid with contours of the 20 cm radio continuum emission \citep{Williams2002}. The thick red contours on SQ B and the circles on SQ A show the area observed by \citet{Lisenfeld2002} with the IRAM 30m telescope. The crosses indicate $>3\sigma$ CO(1-0) detections. The top inset shows the CO(1-0) spectrum (full line) in SQ A, with a velocity resolution of 10.6~km~s$^{-1}$, averaged over the total observed area. The dashed line represents the H$\,${\sc i} emission with a velocity resolution of 21.4~km~s$^{-1}$ in arbitrary units from \citet{Williams2002}, averaged
over the same area. \textit{Right:} The top image \citep{Gao2000} show CO(1-0) BIMA contours overlaid on an R-optical image. The first contour is 2.4$\sigma$, the next one is 3$\sigma$, and then the contours increase by 1$\sigma$. The black circle indicate the FWHM BIMA primary beam of 110'', and the filled red circle the FWHM synthesized beam of 6''.The white contours on the bottom image are from \citet{Petitpas2005} BIMA observations, overlaid on an \textit{HST} image. Levels are 0.5, 1.0, 1.5, etc [Jy~beam$^{-1}$~km~s$^{-1}$.] The black contours are the \citep{Gao2000} observations. Gray contours show the H$\,${\sc i} gas. }
       \label{fig_CO_Lisenfeld30m_GaoBIMA}
\end{figure}

\begin{table}
\begin{center}
\begin{minipage}{\textwidth}
 \renewcommand{\footnoterule}{}
\def\thefootnote{\alph{footnote}}
 \caption[CO properties in Stephan's Quintet]{Observed and derived CO properties in Stephan's Quintet from the litterature}
\centering
\begin{tabular}{l c c c c c}
\hline
\hline
 \multirow{2}*{Target} & Size & $v_{\rm CO}$  & $\Delta v_{\rm CO}$  &	$\mathcal{I}_{\rm CO}$ & $M_{\rm H_2}$  \footnotemark[1] \\	
										&  [arcsec$^{2}$] & [km s$^{-1}$] &   [km s$^{-1}$] &	  [K km s$^{-1}$] &  [$\times 10^8$~M$_{\odot}$]  \\	
\hline
NGC 7318a	 \footnotemark[2]  & $55 \times 55$  &  &  &$<2.13$	 &   	\\
NGC 7318b \footnotemark[3]	 	& $55 \times 55$  & $6250$ \footnotemark[3]  & $50$   & $0.3 \pm 0.1$ & 2.8	\\
NGC 7319  \footnotemark[2]	 	& $55 \times 55$   & 6730  & $\sim 470$ &	$1.90 \pm 0.15$ &  48	\\
SQ-A (6000) \footnotemark[4]	  & $60 \times 80$  & 6030 & 70 &	$0.84 \pm 0.04$ & 22 \\
SQ-A (6700) \footnotemark[4]	  & $60 \times 80$  & 6700  & 70 &$0.33 \pm 0.05$	 & 8.6  \\
SQ-B \footnotemark[4]			  & $60 \times 40$ & 6625 & 35 &	$0.54 \pm 0.02$	 & 7.0 \\
NGC 7320 \footnotemark[5] & $55 \times 55$  & 780 & 22 &	0.3 & 0.1  \\
\hline
\end{tabular}
\footnotetext[1]{All the H$_2$ gas masses are converted using the Galactic conversion factor of $N(\rm H_2)/\mathcal{I}_{\rm CO} = 2 \times 10^{20}$~cm$^{-2}$ [K km s$^{-1}$]$^{-1}$. }
\footnotetext[2]{from observations by \citet{Smith2001a} with the NRAO 12m telescope}
\footnotetext[3]{from interferometric observations by \citet{Gao2000} with the BIMA array. The velocity of the CO clump do not match with the optical redshift of the galaxy (5760~km~s$^{-1}$). The emission is also offset from the nucleus of NGC~7318b.}
\footnotetext[4]{from observations by \citet{Lisenfeld2002} with the IRAM 30m telescope}
\footnotetext[5]{foreground galaxy, from  \citet{Verdes-Montenegro1998}}
\label{table_CO_SQ}
\end{minipage}
\end{center}
\end{table}

Molecular gas outside galactic disks, associated with the IGM starburst SQ-A, was first detected by \citet{Gao2000} using BIMA. Then, single dish observations by \citet{Smith2001a} and \citet{Lisenfeld2002} confirmed this spectacular result. 
Fig.~\ref{fig_CO_Lisenfeld30m_GaoBIMA} presents an overview of these observations and table~\ref{table_CO_SQ} gathers observational results from the litterature. 
All the cold H$_2$ masses in this table are derived assuming a Galactic CO-to-H$_2$ conversion factor of $N(\rm H_2)/\mathcal{I}_{\rm CO} = 2 \times 10^{20}$~cm$^{-2}$ [K km s$^{-1}$]$^{-1}$, which explains why the values quoted here may be different from those of the litterature.
Two velocity components were detected, one centered at
$\sim 6000$~km~s$^{-1}$, and the other at $\sim 6700$~km~s$^{-1}$. These velocities match the redshifts of
the two H$\,${\sc i} gas systems found in the same region (see sect.~\ref{subsubsec:SQ_HI_kinematics}). 
\citet{Lisenfeld2002} found that there is more molecular gas ($3.1 \times 10^9$~M$_{\odot}$)
in SQ-A than in the H$\,${\sc i} gas ($1.6 \times 10^9$~M$_{\odot}$). 

Interestingly, past observations with interferometers have not detected any CO gas  in the main shock region. The single-dish observations by \citet{Lisenfeld2002} with the IRAM 30m telescope partially overlap the main shock region and show $> 3\sigma$ detection of CO in the northern region of the ridge. We note that they include this emission in the SQ-A budget. Their observations do not cover entirely the shock region. 
 \citet{Gao2000, Smith2001a, Petitpas2005} reported CO emission close to NGC~7318b nucleus (see table~\ref{table_CO_SQ}) and from several other regions in the group. 

An other interesting feature of SQ is the presence of a CO-rich \citep[$7 \times 10^8$~M$_{\odot}$]{Braine2001, Lisenfeld2002, Lisenfeld2004} region, SQ-B, located within the young tidail tail, to the south of NGC~7319. This region is also bright in mid-IR  \citep{Xu1999}, H$\alpha$ \citep{Arp1973a} and UV \citep{Xu2005}. The H$_2$ to H$\,${\sc i} gas mass ratio of 0.5 indicates that SQ-B is a tidal dwarf galaxy (TDG) candidate, i.e. a small galaxy which is in the process of formation  from the material of the NGC~7319's tidail tail. TDGs will not be further discussed, so we direct the reader to the concise review by \citet{Xu2006} and references therein.

The gas metallicity in SQ-A and SQ-B is slightly higher than solar \citep{Xu2003, Lisenfeld2004}
This is a strong indication that the gas in these regions has been pulled out from the inner part of a galaxy disk (or maybe several disks) by tidal interactions.

\section{New CO observations in the SQ ridge}
\label{sec:CO_EMIR}
\index{Stephan's Quintet!CO observations with EMIR}

Since none of the past CO observations in SQ yielded detections from the main area of interaction between the new intruder (NGC~7318b) and the intragroup tidal tail, we decided to search for the CO counterpart of the warm H$_2$ gas seen by \textit{Spitzer}. 
Our recent results are the result of an iterative process that has involved three observing runs! Before showing that these results significantly change our view at the molecular gas in SQ (sect.~\ref{sec:SQ-CO-EMIR-results}), I first quickly review the history of our observing runs.

\subsection{History of our CO observations}

In September 2007, we submitted a proposal (P.I.: P. Guillard) to observe the (2-1)CO line from the SQ shock with the multi-beam ($3 \times 3$ pixels) \textit{HERA} receiver\footnote{HEterodyne Receiver Array, \url{http://www.iram.es/IRAMES/mainWiki/HeraforAstronomers}} installed on the IRAM\footnote{Institut de Radio Astronomie Millimétrique, \url{http://www.iram-institute.org}} 30-meter single-dish telescope in Pico Veleta, Spain. The previous observations by \citet{Lisenfeld2002} did not cover entirely the SQ ridge. Moreover, based on the fact that the H$_2$ emission is extended, we argued that much of the CO emission associated with the shock has been missed by previous  interferometric observations. We thought that \textit{HERA} would provide us both the spatial distribution and kinematics of the molecular gas across the shock front.

\begin{figure}
   \centering
    \includegraphics[width=\textwidth]{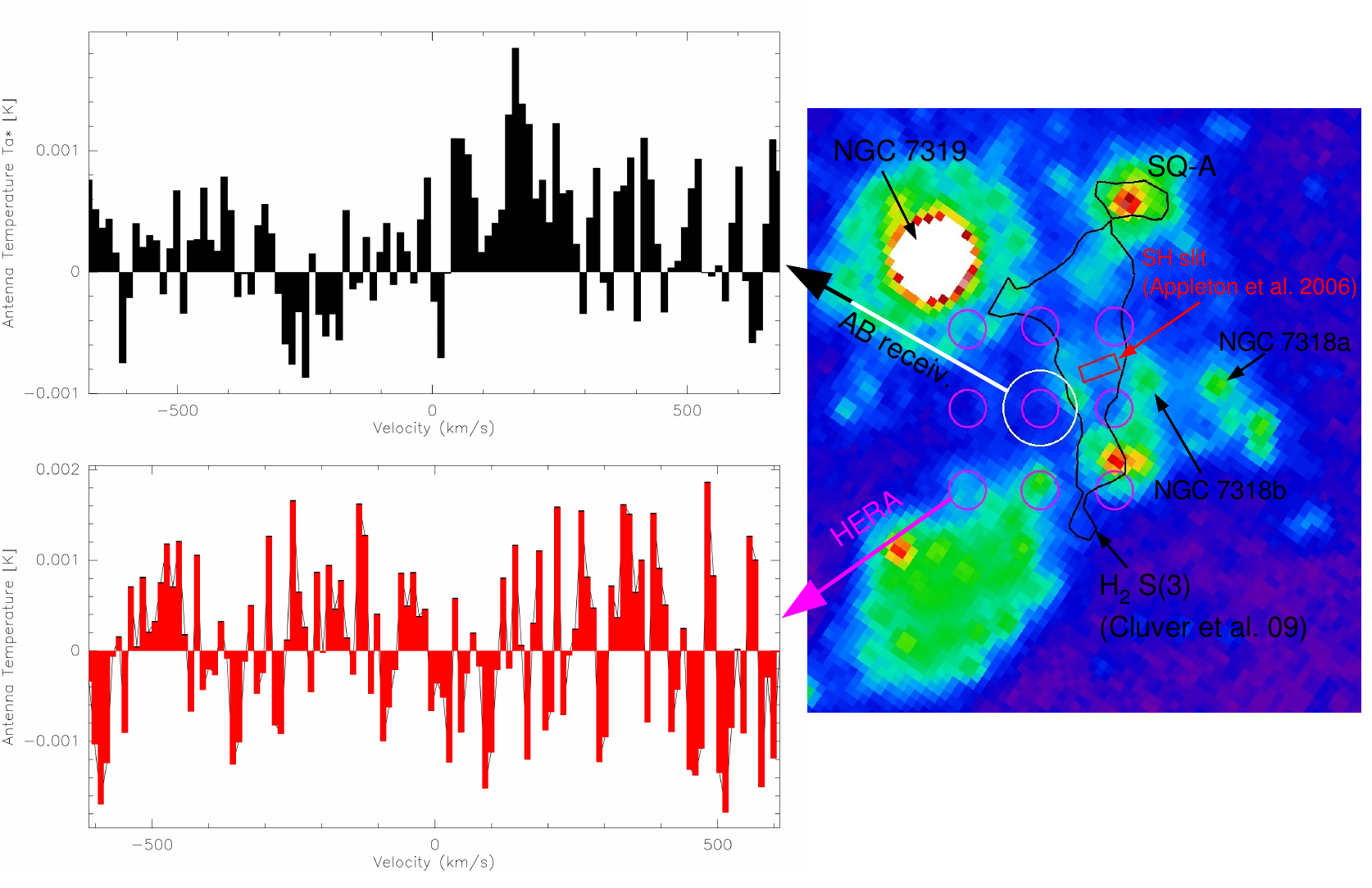}
      \caption[First CO observations with the AB or HERA receivers]{Pointing positions and averaged spectra of our two first runs of CO observations, overlaid on a \textit{Spitzer MIPS} 24$\,\mu$m image. The small magenta circles show the CO(2-1) 11" half-power beams of the HERA $3 \times 3$ matrix. The white circle show the CO(1-0) 22'' HP beam of the AB receivers for our DDT program. 
The black contours is the $4\sigma$ (0.3~MJy~sr$^{-1}$) H$_2$ S(3) line emission from our \hyperref[subsec:paper_Cluver]{paper~{\sc ii}}, which was not available at the time of these first CO observations. 
The central velocity is 6400~km~s$^{-1}$ on the spectra. Weak lines at 6050 and 6130~km~s$^{-1}$ are perhaps detected ($< 3\sigma$) on the CO(1-0) spectrum. On the HERA CO(2-1)  spectrum no clear detection of CO gas is seen.}
       \label{fig:CO_AB_HERA_results_beams}
   \end{figure}

François and I performed the observations in January 2008  but we did not detect any  obvious signal. No line  was clearly detected. 
The Fig.~\ref{fig:CO_AB_HERA_results_beams} shows the CO(2-1) HERA spectra (in red) averaged on the 9 pixels. Only a constant  baseline was subtracted  in the data reduction.
In the eastern and western (group/intruder)  regions there may be weak ($T_A^* \sim 1$~mK) and broad (full width $\sim 500$~km~s$^{-1}$) lines at the right and left ends of the bandpass, which was $\sim 1200 $~km~s$^{-1}$.
In the shock region, these observations set an upper limit on the cold component of the molecular gas of a few $10^{20}$~H$_2$~cm$^{-2}$ for a standard $\rm H_{2} / CO$ conversion factor. This upper limit is roughly the surface density of the warm ($\sim 200$~K) molecular gas detected by Spitzer. 

To confirm or not this puzzling result, we wanted to make sure that the CO emission had not been subtracted in the baseline in case it had a large linewidth ($\gtrsim 1000$~km~s$^{-1}$) comparable to that 
of the ionized gas as measured by optical lines (e.g. the 6300\AA~ O$\,${\sc i} line). Therefore, we submitted a DDT proposal (P.I.: P. Guillard) to observed the (1-0)CO line with the single-pixel \textit{A-B} receivers, arguing that we would double the bandwidth we had with \textit{HERA} (2-1)CO observations. The proposal was accepted and observed in March 2008 by an astronomer on duty. However, when writing up the proposal, we did not realize that the bandwidth of the \textit{A-B} receivers is half (512~MHz) the bandwidth of \textit{HERA} (1~GHz). So we faced the same problem of baseline fitting!  The \textit{top} spectrum in Fig.~\ref{fig:CO_AB_HERA_results_beams}
shows the averaged CO(1-0) spectrum obtained with the AB receivers. The spectral resolution is $\approx 13$~km~s$^{-1}$. Two very weak lines at 6050 and 6130~km~s$^{-1}$ are perhaps detected, but this hypothetic detection is unreliable ($< 3\sigma$). 

\subsection{Observations with the new \textit{EMIR} reveiver at IRAM 30m telescope}
\label{subsec:SQ-CO-EMIR-obs}

In March-April 2009, a new heterodyne receiver, \textit{EMIR} \footnote{Eight MIxer Receiver, see \url{ http://www.iram.es/IRAMES/mainWiki/EmirforAstronomers} for details about the instrument}, was installed and commissioned at the 30m telescope. It replaces the former \textit{ABCD} single-pixel receivers. \textit{EMIR} provides a bandwidth of 4~GHz in each of the two orthogonal linear polarizations for the 3, 2, 1.3 and 0.9~mm atmospheric windows. This increase in bandwith is crucial for our study since we expect that the CO line width to be broad ($\approx 1\,000$~km~s$^{-1}$). The four \textit{EMIR} bands are designated as E090, E150, E230, and E330 according to their approximate center frequencies in GHz.

A proposal was submitted in March 2009 (P.I.: P. Guillard) to observe the CO(1-0) transition. We required 32 hours of observations. Our goal was to reach a 0.3 mK rms noise on the $T_{A}^{*}$ scale at 113~GHz, for a spectral resolution of 100~km/s. The proposal was accepted but only 16h were allocated. I carried on the observations in June 28-30, 2009. I briefly give some observational details here and some preliminary results are given in the next paragraph,  sect.~\ref{sec:SQ-CO-EMIR-results}.

Our observations were concentrated on the CO(1-0) line, because of summer weather conditions. In addition,   we were able to connect the third and fourth parts of the WILMA backend to the horizontal and vertical polarizations of the E230 band. So we observed the CO(1-0) and CO(2-1) lines at 115.27~GHz and 230.54~GHz. To increase the redundance for the CO(1-0) data, we used in parallel the 4~MHz filterbank with both polarizations connected to the E090 band. The receivers were tuned to 112.86~GHz and 225.72~GHz, which corresponds to a ``redshift'' of 6400~km~s$^{-1}$. The observations were done in wobbler switching mode, with a wobbler throw of 120'' in azimuthal direction.
This observation mode provides accurate background subtraction and good quality baselines, which is crucial for our study.
 Pointing and focus were monitored every 2h in stable conditions and every 1h during sunrise. System temperatures were $\approx 150-200$~K at 115 GHz on the $T_{A}^{*}$ scale. The  forward effciency of the telescope was 0.94 and 0.91 at 115 and 230 GHz and the beam effciency was 0.78 and 0.58, respectively.
Our half-power beam size was 22'' at 115 GHz and 11'' at 230 GHz.

The data was reduced with the IRAM GILDAS \footnote{Grenoble Image and Line Data Analysis System, \url{http://www.iram.fr/IRAMFR/GILDAS/}} software package (version of July 2009). The lowest quality spectra (typically with $\sigma \gtrsim 30$~mK for a 3~min on-source integration)  were rejected. We only subtract a constant  when doing baseline fitting. Unfortunately, we experienced problems with the vertical polarization. These spectra show weird features and have been simply  removed for this preliminary analysis. 

\begin{table}
\begin{center}
\begin{minipage}{\textwidth}
 \renewcommand{\footnoterule}{}
\def\thefootnote{\alph{footnote}}
 \caption[Coordinates of the CO pointings with EMIR]{Pointing positions and integration times of our recent CO observations with the new EMIR receiver at the IRAM 30m telescope.}
\centering
\begin{tabular}{c c c c}
\hline
\hline
position & RA (J2000) & DEC (J2000) & integration time [min] \\
\hline
SQ-A \footnotemark[1]	  & 22:35:58.88 & +33:58:50.69   & 150 \\
Bridge \footnotemark[2]	  & 22:36:01.56 &  +33:58:23.30  & 250 \\
Ridge 1   						     & 22:35:59.85  & +33:58:16.55   &  400 \\
Ridge 2  						    & 22:35:59.85   &  +33:58:04.00  & 400 \\
Ridge 3   							& 22:35:59.73   & +33:57:37.16   & 120 \\
\hline
\end{tabular}
\footnotetext[1]{Starburst region at the northern tip of the SQ ridge}
\footnotetext[2]{Bridge feature that connects the ridge to the AGN NGC~7319. }
\label{table_CO_EMIR_positions}
\end{minipage}
\end{center}
\end{table}

The results are beyond our expectations! The quality of the data was such that we were able to observed five different positions in the SQ shock region. The image in the center of Fig.~\ref{fig_CO_EMIR_results_beams} shows the positions of our pointing (blue circles of diameter 22 arcsec, corresponding to the half-power CO(1-0) beam) overlaid on a \textit{Spitzer} 24$\,\mu$m \textit{MIPS} image of SQ. 
The three positions in the ridge are designated as ``Ridge 1, 2 and 3'', from North to South. The two other positions are centered on SQ-A and the middle of the bridge feature in between NGC~7319 and the ridge.
The black line shows the 4$\sigma$ contours of the H$_2$ 0-0S(3) line emission detected by \textit{Spitzer} (see sect.~\ref{spectral_IRSmap_H2}). 
Note that the half power beam width at 115~GHz matches very well the width of the H$_2$ emission in the ridge.
The red rectangle shows the position of the \textit{IRS} SH slit of the \citet{Appleton2006} observations. The table~\ref{table_CO_EMIR_positions} list the coordinates of these positions and the total ON+OFF time spent at each position. A total of $\approx 20$ hours has been spent for ON+OFF observations towards SQ.

\subsubsection{Coming back to the previous observation run\dots}

The reason why we did not detect any clear signal during our previous observations with HERA or the ABCD receivers is not completely clear for us. The possible explanations are the following.

First, for the CO(1-0) ABCD position (see Fig.~\ref{fig:CO_AB_HERA_results_beams})  we may have missed the molecular gas rich ridge, because the position was not exactly centered on the H$_2$ ridge. The central beam position was offset to the East of the \citet{Appleton2006} position because we wanted to avoid the contamination from star-forming regions. However,  we did not have the full \textit{IRS} mapping and the mid-IR imaging at that time, so did not have any idea about the extent of the CO emission. So we may have offset our central position too much, thus diluting the signal in the beam. This may explain the non detection if the CO gas is confined to a $\approx 25''$-wide ridge. However, since the H$_2$ emission seems a bit more extended, the beam dilution does not seem to be a very strong argument. 

Second, during our first run, we had lots of problems with the WILMA backend. Several pixels of HERA were not usable, especially pixels 9 and 7, which were, unfortunately, located in the ridge. For the second DDT run with the ABCD receiver, a lot of spectra also show weird features, and I had to throw away $\approx 30\%$ of spurious spectra. These technical problems may also have entrained calibration issues in the data.

The non-detection in our previous HERA and ABCD observations led me to double-check my data reduction for the new EMIR observations. I have checked that the signal is reproductible from one day of observation to another, and search for any spurious features in the spectra. These checks, and the fact that the line velocities are in very good agreement with H$\,${\sc i} velocities, allow to confirm our new results presented in the following.

\section{Preliminary results}
\label{sec:SQ-CO-EMIR-results}

\begin{figure}
   \centering
    \includegraphics[angle=90, width=0.9\textwidth]{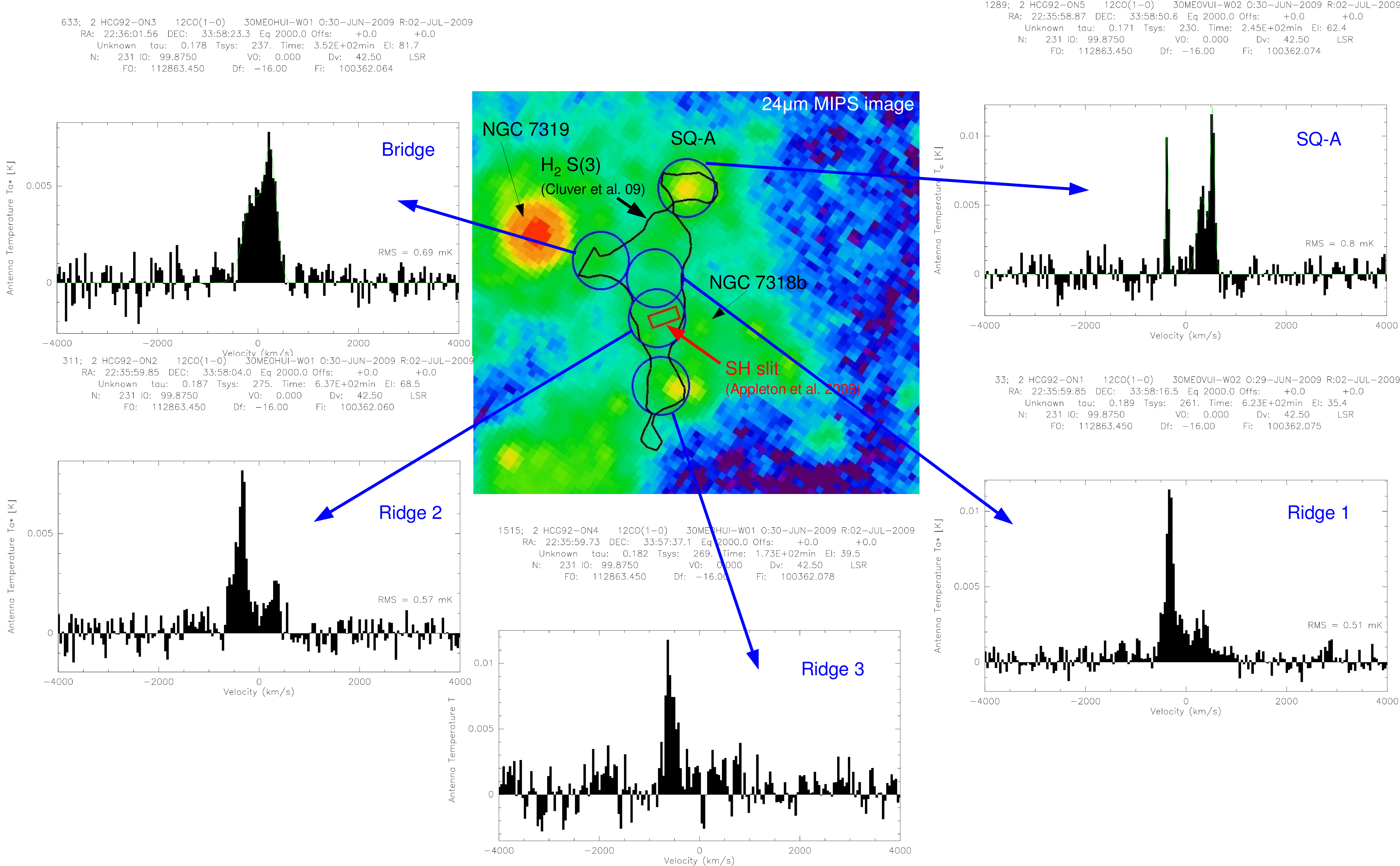}
      \caption[Complex kinematics of the CO gas in the Stephan's Quintet H$_2$-emitting ridge]{New results of CO observations in the SQ group. Extended CO emission is detected in the \textit{ridge} and in the \textit{bridge} feature towards NGC~7319. The spectra show complex kinematics of the CO gas in the SQ halo.}
       \label{fig_CO_EMIR_results_beams}
   \end{figure}

Since the data has been taken very recently, we present here a preliminary analysis that points out the main observational results about the mass of molecular gas in the system and its kinematics.

\subsection{Distribution and mass of molecular gas}

In Fig.~\ref{fig_CO_EMIR_results_beams} we show the averaged CO(1-0) spectra for each of the five observed positions. The spectra are plotted over the full available bandwidth ($\approx 8000$~km~s$^{-1}$ at 115~GHz) at a spectral resolution of 42~km~s$^{-1}$. CO is detected with a high signal-to-noise ratio for all the pointings. This result is in sharp contrast with the previous interferometric observations, where almost no CO was detected in the ridge. It shows that the CO emission is extended and associated with the extended H$_2$ emission. The noise in the spectra are $\approx 0.5 - 0.7$~mK at a resolution of 42~km~s$^{-1}$. The CO(2-1) is also detected but the quality of the spectra is much poorer than the CO(1-0) because of a shorter integration time and moderate weather conditions. This is why we only show the CO(1-0) spectra here.

\begin{figure}
   \centering
    \includegraphics[width=0.9\textwidth]{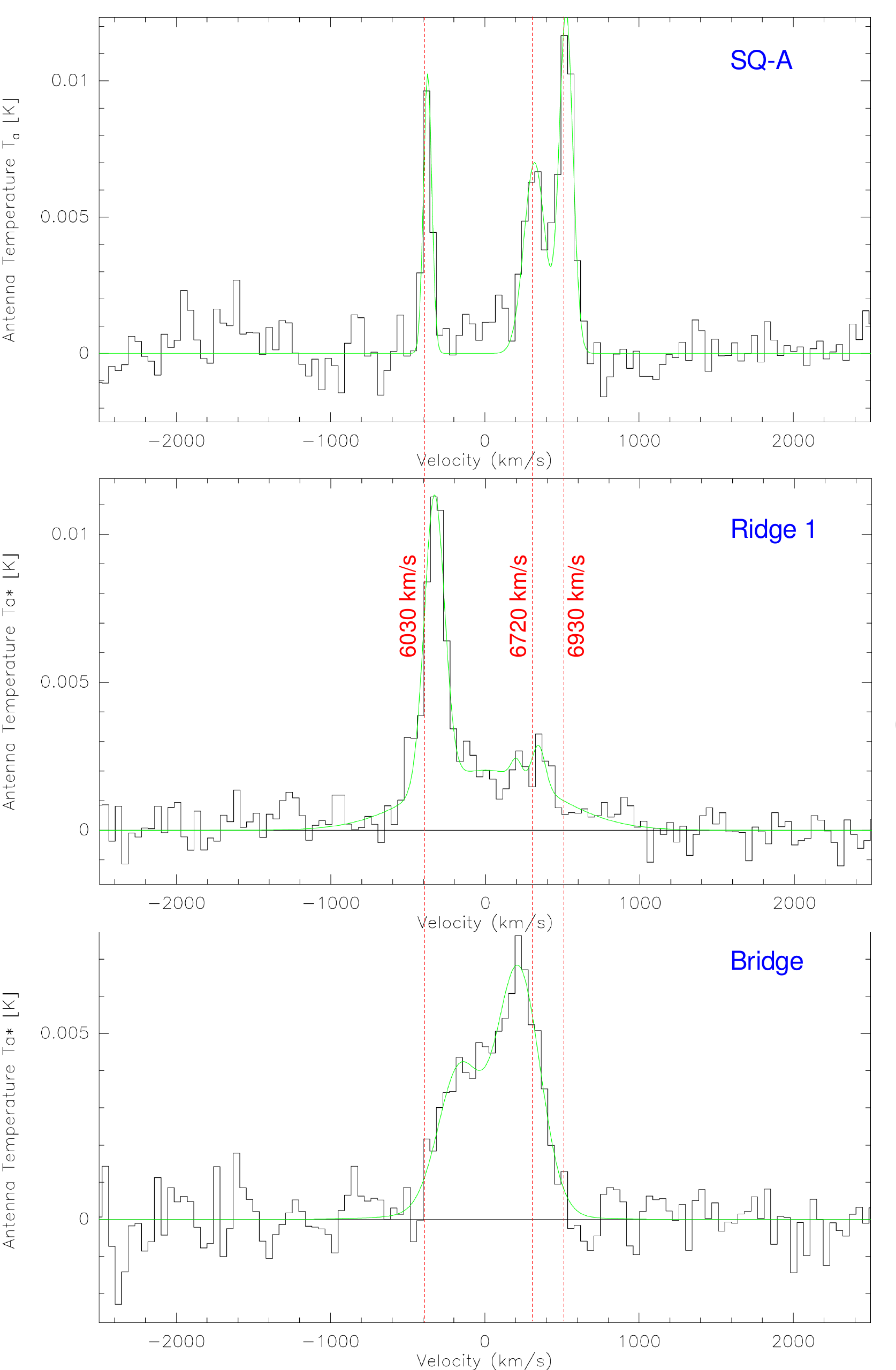}
      \caption[CO(1-0) spectra at three positions in the SQ group]{CO(1-0) spectra  at three positions in the SQ group: SQ-A, Ridge 1, and the in the bridge, from top to bottom. The red dashed lines indicate reference velocities on the SQ-A spectrum, at 6030, 6720 and 6930~km~s$^{-1}$. The green line shows the result of the fitting with three gaussians.}
       \label{fig_SQ-CO-EMIR-spectra-fit-lines}
   \end{figure}

\begin{figure}
   \centering
    \includegraphics[width=\textwidth]{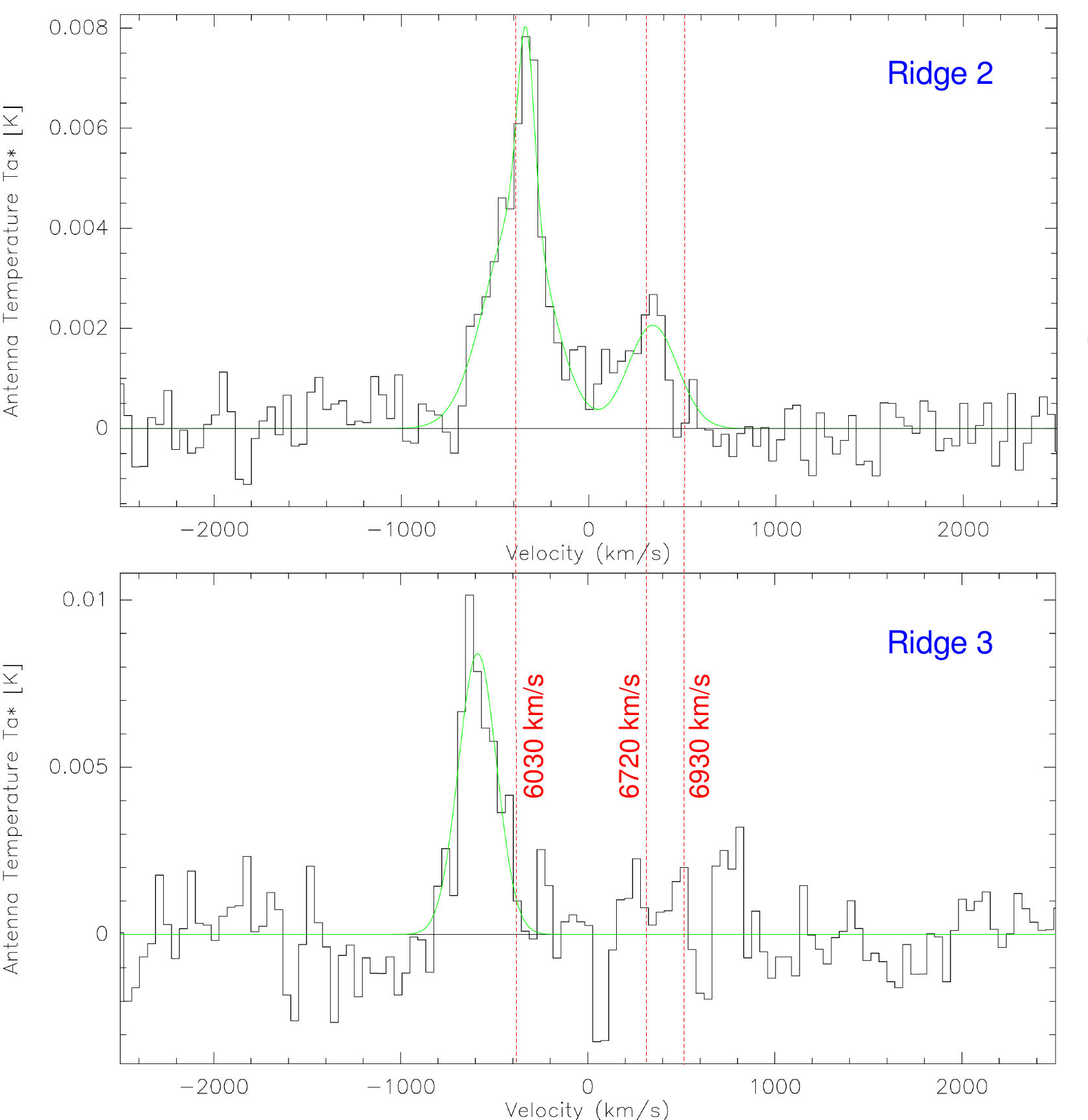}
      \caption[CO(1-0) spectra at two positions in the SQ group]{Idem Fig.~\ref{fig_SQ-CO-EMIR-spectra-fit-lines} for the CO(1-0) spectra  at the Ridge 2 and 3 positions.}
       \label{fig_SQ-CO-EMIR-spectra-fit-lines2}
   \end{figure}

\begin{table}
\begin{center}
\begin{minipage}{\textwidth}
 \renewcommand{\footnoterule}{}
\def\thefootnote{\alph{footnote}}
 \caption[New CO observations of Stephan's Quintet]{New CO observations of the center of the SQ group\footnotemark[1]: observational results}
\centering
\begin{tabular}{l c c c c c}
\hline
\hline
 \multirow{2}*{Target} & Area & $v_{\rm CO}$  & $\Delta v_{\rm CO}$  &	$\mathcal{I}_{\rm CO}$ & $M_{\rm H_2}$  \footnotemark[2] \\	
										&  [arcsec$^{2}$] & [km s$^{-1}$] &   [km s$^{-1}$] &	  [K km s$^{-1}$] &  [$\times 10^8$~M$_{\odot}$]  \\	
\hline
SQ-A (6000) \footnotemark[3]	  & \multirow{9}*{\begin{sideways}$1.13 \times 22\,''\,^2 = 547 \,''\,^2$\end{sideways}}  & $6028 \pm 5$ & $63 \pm 8$ &	$0.60 \pm 0.07$ & $2.0 \pm 0.2$ \\
SQ-A (6700) \footnotemark[3]	  &   & $6723 \pm 12$ & $161 \pm 31$ &	$0.96 \pm 0.15$ & $ 3.2 \pm 0.5 $ \\
SQ-A (6900) \footnotemark[3]	  &   & $6927 \pm 5$ & $104 \pm 12$ &	$1.23 \pm 0.12$ & $ 4.1 \pm 0.4 $ \\
\cline{3-6}
\cline{1-1}
SQ-A (tot) \footnotemark[4]	  							  &    & &  													&	$2.8 \pm 0.2$ & $9.2 \pm 0.7$ \\
\cline{3-6}
\cline{1-1}
Ridge 1 (6000)                           &   & $6069 \pm 6$ & $154 \pm 10$  & $1.61 \pm 0.10$  & $5.3 \pm 0.3$ \\
\cline{3-6}
\cline{1-1}
Ridge 1 (tot) \footnotemark[4]   							  &    & &  													&	$1.8 \pm 0.1$ & $5.9 \pm 0.3$ \\
\cline{3-6}
\cline{1-1}
Ridge 2 (6000)                           &   & $6052 \pm 24$ & $405 \pm 15$  & $2.3 \pm 0.1$  & $7.7 \pm 0.4$ \\
\cline{3-6}
\cline{1-1}
Ridge 2 (tot) \footnotemark[4]   							  &    & &  													&	$3.0 \pm 0.1$ & $9.9 \pm 0.4$ \\
\cline{3-6}
\cline{1-1}
Ridge 3 (5700)   & & $5767 \pm 35$   & $241 \pm 29$ & 	$2.1 \pm 0.2$ & $6.8 \pm 0.7 $ \\
\cline{3-6}
\cline{1-1}
Bridge (tot) \footnotemark[4]  &   &  &  &$3.9 \pm 0.4$	 & $12.7 \pm 1.4$  \\
\hline
\end{tabular}
\footnotetext[1]{Recent observations performed with the new EMIR receiver at the IRAM 30m telescope by \citet{Guillard2009b}}
\footnotetext[2]{H$_2$ gas masses are calculated using the Galactic conversion factor of $N(\rm H_2)/\mathcal{I}_{\rm CO} = 2 \times 10^{20}$~cm$^{-2}$ [K km s$^{-1}$]$^{-1}$}
\footnotetext[3]{Results are given for each velocity components, centered at $\sim 6000$, 6700 and 6900 km s$^{-1}$. The last component at 6900 km s$^{-1}$ is a newly discovered feature (see text for details).}
\footnotetext[4]{All velocity components}
\label{table_CO_SQ_EMIR_masses}
\end{minipage}
\end{center}
\end{table}

The central velocity (0~km~s$^{-1}$ on the plots) corresponds to the recession velocity of 6400~km~s$^{-1}$  at which the receivers were tuned. The CO(1-0) signal extends over the velocity range $\approx 5700 - 7000$~km~s$^{-1}$ (and perhaps until $7400$~km~s$^{-1}$ for the ridge 1 position, see comment below). Multiple velocity components are detected, pointing out the complexity of the kinematics of the CO gas in the system (see next paragraph below for a discussion of the kinematics).
The figures~\ref{fig_SQ-CO-EMIR-spectra-fit-lines} and \ref{fig_SQ-CO-EMIR-spectra-fit-lines2} shows a zoom (bandwidth of $-2500 - 2500$~km s$^{-1}$) on the spectra with reference velocities of SQ-A overlaid (red dashed lines). The green line shows the result of the multiple gaussian fitting. This fitting is used to derive the central velocities of the CO components,  the line widths and integrated intensities. We check that the residuals are below the 2$\sigma$ level.

In Table~\ref{table_CO_SQ_EMIR_masses} we gather  the CO(1-0) central line velocities, the integrated line intensities, and the derived H$_2$ gas masses for the different observed regions shown on Fig.~\ref{fig_CO_EMIR_results_beams} (see Table~\ref{table_CO_EMIR_positions} for the coordinates of these positions and the integration times). The molecular gas masses in this table are calculated using the Galactic conversion factor of
$N(\rm H_2) / \mathcal{I}_{\rm CO} = 2 \times  10^{20}~\rm cm^{-2}~(K~km~s^{-1})^{-1}$:
\begin{eqnarray}
M_{\rm H_2} & = &  \mathcal{I}_{\rm CO} ^{\rm obs} \times \frac{N(\rm H_2) }{\mathcal{I}_{\rm CO}} \times d^2 \times \Omega \times 2 \, m_{\rm H} \\
						& = & 3.3 \times 10^8 \left( \frac{ \mathcal{I}_{\rm CO} ^{\rm obs}}{\rm K \, km \, s^{-1}} \right) \left(\frac{N(\rm H_2) / \mathcal{I}_{\rm CO}}{2 \times  10^{20}} \right) \left(\frac{d}{94 \, \rm Mpc} \right)^2 \left(\frac{\theta _{\rm {\tiny FWHM}}}{21''} \right) ^2 \ \rm M_{\odot}
\end{eqnarray}
where $ \mathcal{I}_{\rm CO} ^{\rm obs}$ is the observed velocity integrated CO line intensity expressed
in K~km~s$^{-1}$, $d$ is the distance to the source in Mpc and $\Omega$ is the area covered
by the observations in arcsec $^2$, with $\Omega = 1.13 \times \theta _{\rm \tiny HP}$ for a single
pointing with a Gaussian beam of Half Power beam width of $\theta _{\rm \tiny HP}$. Note that to derive the total mass of molecular gas, one has to multiply $M_{\rm H_2}$ by a factor 1.36 to take into account Helium.
In the following we briefly discuss the masses of molecular gas detected in the three main area, the ridge, SQ-A, and the bridge.

\begin{description}
\item[CO in the SQ ridge]

Summing over all the velocity components, we derive $\approx 6 \times 10^8$ and $10^9$~M$_{\odot}$ of cold molecular gas in the positions ``ridge 1'' and 2. For the ``ridge 3'' position, the spectrum is more noisy because of limited integration time, and we detect $\approx 7$~M$_{\odot}$ of cold H$_2$ at 5700~km~s$^{-1}$. Our pointings do not cover entirely the ridge, but we can safely say that the total mass of molecular gas in the ridge is at least $2.5 \times 10^9 $~M$_{\odot}$.  It is only a factor of 2 larger than the total  warm H$_2$ mass seen by \textit{Spitzer} in the ridge ($1.2 \times 10^9 $~M$_{\odot}$, see table~\ref{table_mass_NRJ_budgets_SQ}).
The mass of molecular gas in the SQ shock is larger than in the Milky Way!

\item[CO in SQ-A]

Within our aperture  $\Omega = 547$~arcsec$^2$ centered on SQ-A, we find a total cold H$_2$ mass of $9.2 \times 10^8$~M$_{\odot}$. \citet{Lisenfeld2002} find $\approx 3 \times 10^9$~M$_{\odot}$ over a much larger area ($60'' \times 80''$) that partially overlaps the north of the ridge.
However, because of a limited bandwith, they only detect the 6000 and 6700~km~s$^{-1}$ velocity components, and they miss the higher velocity component at 6900~km~s$^{-1}$ that  seem to contain almost half of the mass in the core region of SQ-A (see table~\ref{table_CO_SQ_EMIR_masses}). 
Based on our new IR maps, we estimate the area of SQ-A to be $\approx 40'' \times 40''$. If we scale the mass we find in our aperture $\Omega$ to this larger aperture, we find a total mass of cold H$_2$ of $2.6 \times 10^9$~M$_{\odot}$. This is comparable to the mass derived by \citet{Lisenfeld2002}, which is the result of a lucky compensation between two effects, the over-estimate of the mass in SQ-A because of a large aperture, and the under-estimate because of the non-detection of the 6900~km~s$^{-1}$  component.

\item[CO in NGC~7319's bridge]

Summing over all velocities, we find $M_{\rm H_2} = 1.2 \times 10^9$~M$_{\odot}$ within our aperture $\Omega$. Note that our aperture  matches quite well the width of the bridge, but does not cover its whole extension towards the AGN. Therefore, the total mass of molecular gas in the bridge may be a factor of $\approx 2$ larger. \citet{Smith2001a} find $\approx 5 \times 10^9$~M$_{\odot}$ within a half-power beam of 55'' centered on the nucleus of NGC~7319. This beam overlaps our bridge region.

\end{description}

These observations represent a substantial revision of the mass and energy budgets of the collision. 
The first striking result is that we detect in the ridge and in the bridge features a comparable amount of molecular gas to that of SQ-A. The CO emission in the intragroup medium is then very extended, and in total (SQ-A + ridge + bridge), $\gtrsim 6.5 \times 10^9$~M$_{\odot}$ of cold H$_2$ gas is lying outside the galactic disks of the group. 
Although it represents a huge mass of molecular gas, most of this gas appears to be inefficient at forming stars, except in SQ-A (see chapter~\ref{chapter:SQ_dust}).

Interestingly, assuming a Galactic CO-to-H$_2$ conversion factor, we find that the ratio of the cold to warm H$_2$ masses is $\approx 2$ in the ridge.  This ratio is much smaller than in galaxies, where the mass of cold molecular gas is $1-2$ orders of magnitude greater than the warm H$_2$ mass: $M_{\rm H_2}(\rm cold) =10 - 100 \times M_{\rm H_2}(\rm warm)$ \citep{Roussel2007}. This preliminary result seems in agreement with our interpretation of the H$_2$ emission in terms of a continuous heating by dissipation of turbulent energy (see sect.~\ref{subsec:mass_energy_transfers}).


\subsection{Complex kinematics of the CO gas}
\index{Stephan's Quintet!CO kinematics}

Fig.~\ref{fig_CO_EMIR_results_beams} clearly show the complexity of the  kinematics of the CO gas in the SQ group. 
The shape of the spectra varies a lot from one position to another.
Multiple velocity components are detected (see Fig.~\ref{fig_SQ-CO-EMIR-spectra-fit-lines} and \ref{fig_SQ-CO-EMIR-spectra-fit-lines2} for detailed view at the spectra with some reference velocities overlaid). The four main components are at 5700, 6000, 6700 and 6900~km~s$^{-1}$ (see table~\ref{table_CO_SQ_EMIR_masses} for the precise central line velocities derived from line fitting). Note that the H$\,${\sc i} velocities detected in SQ are 5700, 6000 (associated with the intruder gas), and 6700~km~s$^{-1}$  (associated with the tidal tail to the south of NGC~7319). The velocity component at 6900~km~s$^{-1}$ is a new feature that we detect only in the SQ-A region. It was not seen before in CO or H$\,${\sc i} because of limited velocity coverage. 
The main general results are the following:

\begin{itemize}
\item the gas associated with the intruder NGC~7318b ($5700-6000$~km~s$^{-1}$) and with the intragroup medium (the tidal tail, $6700$~km~s$^{-1}$) are both detected along the full North-South extension of the ridge, and in the bridge eastern feature towards NGC~7319 as well.
\item gas at intermediate velocities, in between that of the intruder and that of the intragroup, is detected. There is a broad component centered at $\approx 6400$~km~s$^{-1}$ in the ridge 1, 2 and in the bridge positions. Interestingly, this intermediate component seems to be absent (or very weak, $\approx 2\sigma$) in SQ-A and vanishes at the southern end of the main shock region (ridge 3). The $6400$~km~s$^{-1}$ CO velocity component is consistent with the central velocity of the broad, resolved H$_2$ S(1) line at  $6360 \pm 100$~km~s$^{-1}$ (see Appendix~B of \hyperref[subsec:paper_Cluver]{paper~{\sc ii}} for a re-analysis of the \citet{Appleton2006} high-resolution \textit{IRS} spectrum). This is the first time CO gas is detected at these intermediate velocities, and this gas is associated with the warm H$_2$ seen by \textit{Spitzer}. This detection support the idea that molecular gas is formed out of gas accelerated in the shock.

\item 
The CO line velocity components are very broad, suggesting that the kinematics of the CO gas is highly disturbed.
The FWHM of the main CO(1-0) lines are of the order of $100-400$~km~s$^{-1}$, with an even broader signal ($\approx 1000$~km~s$^{-1}$) at intermediate velocities (see coments below). This suggests that huge velocity gradients are present in this region. These broad line widths are in agreement with the \citet{Appleton2006} interpretation that the mid-IR H$_2$ lines are intrinsically very broad and resolved by the high resolution module of the \textit{IRS.} 

\end{itemize}

Some specific comments about each regions are given in the following:

\begin{description}
\item[SQ ridge]

In the ridge, the line widths are of the order of $100-400$~km~s$^{-1}$, and lines seem to be broader in the center (ridge 2) than at the edges (ridge 1 and 3). In the ridge 1 position (and perhaps ridge 2), the underlying intermediate component at $6400$~km~s$^{-1}$ is extremely broad ($\approx 1000$~km~s$^{-1}$, comparable to the relative velocity between the intruder and the intragroup medium), with a hint of emission between 6900 and 7400~km~s$^{-1}$ (see ridge 1 spectrum on Fig.~\ref{fig_SQ-CO-EMIR-spectra-fit-lines}). 

\item[SQ-A]

In SQ-A, three velocity components are detected at 6000, 6700 and 6900~km~s$^{-1}$. The line widths are of the order of 100~km~s$^{-1}$, with the smallest dispersion for the 6000~km~s$^{-1}$ line. The broadest component at 6700~km~s$^{-1}$ seems to have a left wing extending towards 6400~km~s$^{-1}$. 
Interestingly, the CO(1-0) line widths are significantly smaller in SQ-A than in the (b)ridge, and this is perhaps a clue to understand why there is much more star formation in this region than in the ridge. 
This observational picture fits within our interpretation of two colliding gas flows (presented in sect.~\ref{subsec:SQ-H2formation-multiphase-gas}), where the shear velocities between the flows are maximum in the central region of the contact discontinuity (ridge), and smaller at the edges of the main shock structure \citep[see chapter~\ref{chapter:shocks}, sect.~\ref{3D-simulations-collision-two-gas-streams} and Fig~5 of][]{Lee1996}. 

\item[NGC~7319's bridge]
\label{subsubsec:bridge_NGC7319_CO}
\index{NGC 7319!outflow}
\index{NGC 7319!bridge}

The spectrum in the bridge is perhaps the most striking. It shows a very broad signal in between the extreme velocities detected in SQ-A (6000 and 6900~km~s$^{-1}$), see Fig.~\ref{fig_SQ-CO-EMIR-spectra-fit-lines}.  This profile is well fitted by a sum of 3 gaussians. As discussed in sect.~\ref{SQ-NGC7319-comments}, the origin of the molecular gas in the bridge structure is an open question. 
If an outflow from the AGN is present \citep[as suggested by][]{Aoki1996}, molecular may be entrained from the disk or formed directly in the outflow (see also sect.~\ref{SQ-NGC7319-comments} for observational details about NGC~7319). 
The molecular gas in the bridge may also be the result of a previous tidal interaction (with NGC~7320c for instance) that would have stripped some material from the galactic disk. Then the interaction between the new intruder and this ``tidal bridge'' would trigger further molecular gas formation in this region. We do not see any clear-cut argument to favor one of the two possibilities.

\end{description}

\section{Astrophysical questions raised by CO observations}

These CO observations essentially raise two open questions, one related to the dynamical coupling of the molecular gas to the lower density phases, and the other related to star formation. Here are the tracks we are following to tackle these two issues.

\subsection{How is the molecular gas accelerated?}

Observations shows that some of the CO gas lies at intermediate velocities between that of the intruder (NGC~7318b) and that of the target, i.e. NGC~7319's tidal tail. 
The CO line widths match that of the [O$\,${\sc i}]$\,\lambda 6300\,$\AA~line from warm atomic gas. 
 This suggests that the cold molecular gas is dynamically coupled to the lower density components.
If the picture of two colliding multiphase gas flows is correct, these observations mean that, in the reference frame of the surface discontinuity  between the two gas streams (which may be materialized by the X-ray bright ridge), the dense clouds has been significantly decelerated. 

This coupling seems difficult to explain if most of the molecular gas  was preshock gas. The high density contrast between cold clouds and the hot, tenuous flow would lead to deceleration timescales longer than the shock age (see sect.~\ref{evolution-shocked-molecular-cloud}), and therefore a weak coupling.
In our model of H$_2$ formation in the shock, the molecular gas is formed out of moderate density medium that is compressed and decelerated while becoming molecular. 
Therefore, in this context, the coupling is much more efficient because the density contrast is much lower.
This \textsl{in-situ} formation is perhaps a solution to explain the high velocity dispersion seen in the spectra, although we do not exclude the presence of preshock dense clouds.

\subsection{Driving Schmidt-Kennicut into a corner}
\label{subsec:kennicutrelation}
\index{Schmidt-Kennicut relation}

\begin{figure}
   \centering
    \includegraphics[width=0.8\textwidth]{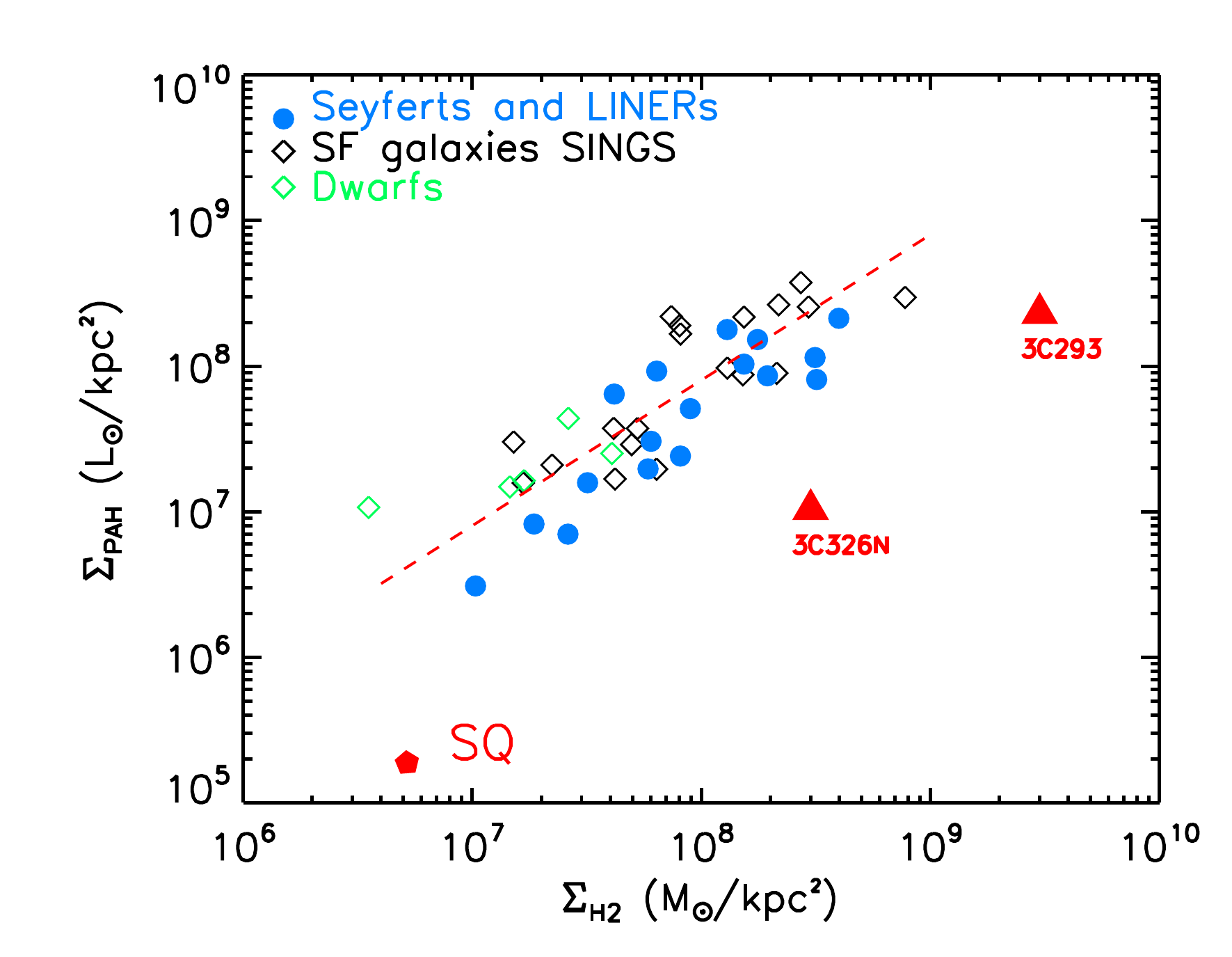}
      \caption[PAH emission vs. H$_2$ mass plot]{Star-formation efficiency (assumed to be traced by the surface luminosity of the PAH emission) as a function of mass of cold H$_2$, derived from the CO(1-0) line intensity. The hatched line shows the relationship obtained for the SINGS sample, where empty diamonds and filled dots mark dwarf galaxies, star-forming galaxies, and AGN. The SINGS data is from \citet{Roussel2007}. The filled pentagon shows the center of the SQ shock. For comparison, the filled triangles mark the H$_2$-luminous radio galaxies 3C326~N and 3C293.}
       \label{fig:KSplot_PAH}
\end{figure}

Many galaxy collisions are observed to trigger IR-luminous bursts of star formation. This is not the case in SQ, although a large reservoir of cold molecular gas is present in the area of interaction between the two galaxies. 
Fig.~\ref{fig:KSplot_PAH}  shows a diagram similar to the classical Schmidt-Kennicutt relationship.
Based on the empirical correlation between PAH emission and star formation in starbursts \citep{Calzetti2007, Pope2008}, the $7.7\,\mu$m PAH emission is used as a tracer of star formation. The cold H$_2$ mass is derived from the standard CO(1-0) intensity to H$_2$ mass conversion factor (see chapter~\ref{chapter:H2Molecule},  Eq.~\ref{eq:CO10-to-H2mass}).
The SQ point corresponds to the ridge 1 position.
The PAH luminosity in the SQ shock is integrated over a $18''  \times 15''$ region centered on the CO(1-0) beam of the ridge 1 position (see chapter~\ref{chapter:SQ_dust} for details about how the PAH luminosity is calculated).

We find that SQ is offset from the  correlation obtained with the SINGs sources \citep[data taken from][]{Roussel2007}. Interestingly, SQ has a similar offset as the H$_2$-luminous radio galaxies 3C326N and 3C293 (see chapter~\ref{chapter:perspectives} for a discussion of 3C326N). This plot clearly shows that star formation is suppressed in the center of the shock as compared to star forming galaxies.   

The molecular gas is observed to be more turbulent in the ridge than in SQ-A, where star formation is occuring. This suggests that the dissipation of the mechanical energy may be an important process in heating the molecular gas and perhaps quenching star formation (see  \hyperref[paper_SQ_H2]{paper~{\sc i}} and chapter~\ref{chapter:SQ_dust} for a more detailed discussion of star formation in SQ).

\section{Concluding remarks}

These new CO observations provide a more complete inventory of the molecular gas in the SQ intragroup medium. Large amounts of CO emitting gas is detected, both in the main shock region and in the eastern bridge feature. This gas is coexisting  with warm H$_2$, H$\,${\sc ii} gas, and X-ray emitting hot plasma. The CO(1-0) lines are intrinsically broader outside the SQ-A and southern star-forming regions, suggesting that the galaxy collision has disturbed its kinematics within the group halo.

A more detailed analysis and a paper is in preparation to estimate as precisely as possible the mass, excitation characteristics (in particular the CO(2-1) to CO(1-0) line ratio), and the full kinetic energy of the molecular gas in the SQ shock. This will lead to re-evaluate our prior estimate of the dissipation timescale through H$_2$ line emission. The velocity structure will be used to distinguish between preshock molecular gas and molecular gas formed out of gas that has been shocked and decelerated in the collision. In addition, we will use MHD shock models (similar to those used to interpret the H$_2$ line emission) to compute the CO emission in shocks, and compare the results with the observations presented here.

\chapter{Dust emission in Stephan's Quintet}
\label{chapter:SQ_dust}

\epigraph{I close my eyes \\ Only for a moment, and the moment's gone\\ All my dreams\\ Pass before my eyes with curiosity\\ Dust in the wind \\All they are is dust in the wind}{Kerry Livgren, Kansas}


\begin{Abstract}

In chapters~\ref{chapter:H2_SQ} and \ref{chapter:H2_SQ_mapping}, I propose an interpretation to the surprisingly  bright H$_2$ rotational line emission from the Stephan's Quintet (SQ) shock, in which H$_2$ gas forms out of the multiphase postshock gas.  In this scenario, dust is a key-element because its presence is required for H$_2$ to form. Therefore we expect that some of the dust emission comes from molecular gas.
This chapter is a brief introduction to the work I did on the analysis and modeling of new \textit{Spitzer} imaging and spectroscopy of dust emission in the SQ ridge. The associated paper is reproduced in the manuscript. I have analysed the characteristics of the PAH emission in the shock,  built a complete IR spectral energy distribution of the shock, and compare it to that of the diffuse Galactic ISM. In order to try to constrain the physical structure of the molecular gas, a detailed modeling of the dust emission in the SQ ridge has been performed, and the model results compared to observations.
\end{Abstract}

\minitoc


\index{Stephan's Quintet!dust emission}

\section{Introduction}

\PARstart{D}ust shapes the spectra of galaxies: dust absorbs\footnote{The existence of  the extinction of stellar light by interstellar dust was found by \citet{Trumpler1930} by measuring the angular diameter of star clusters. They introduce a corrective term, $A_{\lambda}$, in their estimate of the distance to the source, in order to take into account its overestimation due to the presence of dust. \citet{Stebbins1934, Stebbins1935} first showed that the extinction follows a $\lambda ^{-1}$ law in the U, B, V bands.  This law in the optical shows that the grain size is of the order of the wavelength, i.e. 0.1 to 0.5 $\mu$m.}, scatters stellar light, and re-emits this energy in the infrared domain. It is estimated that 30\% or
more of the energy emitted as starlight in the Universe is re-radiated by dust in the
infrared \citep[see][for a review about the cosmic IR background]{Lagache2005}.

Molecular hydrogen  forms on dust grains. Therefore, dust is a key-element for our interpretation of H$_2$ emission from the SQ X-ray bright shock structure. In \hyperref[paper_SQ_H2]{paper~{\sc i}}, we show that H$_2$ gas can form  out of the dense gas that is shocked to velocities sufficiently low ($V_{\rm s} < 300$~km~s$^{-1}$) for dust to survive. 
One of the consequences of this interpretation is that some of the dust emission comes from molecular gas. Therefore, in order to test this prediction, I have embarked on a study and modeling of the dust emission from the SQ shock, which has been written in \hyperref[paper_SQ_dust]{paper~{\sc iii}} \citep{Guillard2009a}, which will be soon submitted to A\&A. This chapter will not review the content of the paper, but rather introduce the context of the past observations of dust emission from SQ, as well as their interpretation (sect~\ref{past-obs-dust-SQ}).  Then, I briefly present the context of the new \textit{Spitzer} observations  (sect~\ref{new-spitzer-dust-obs}, as well as the \hyperref[paper_SQ_dust]{paper~{\sc iii}}. The section~\ref{revisiting-dust-survival} concludes this chapter by opening the discussion about dust survival in the violent SQ environment, and more generally in the multiphase ISM.

\section{Observations of dust emission from Stephan's Quintet}

\subsection{Past observations (\textit{ISO})}
\label{past-obs-dust-SQ}

\begin{figure}
   \centering
    \includegraphics[width=\textwidth]{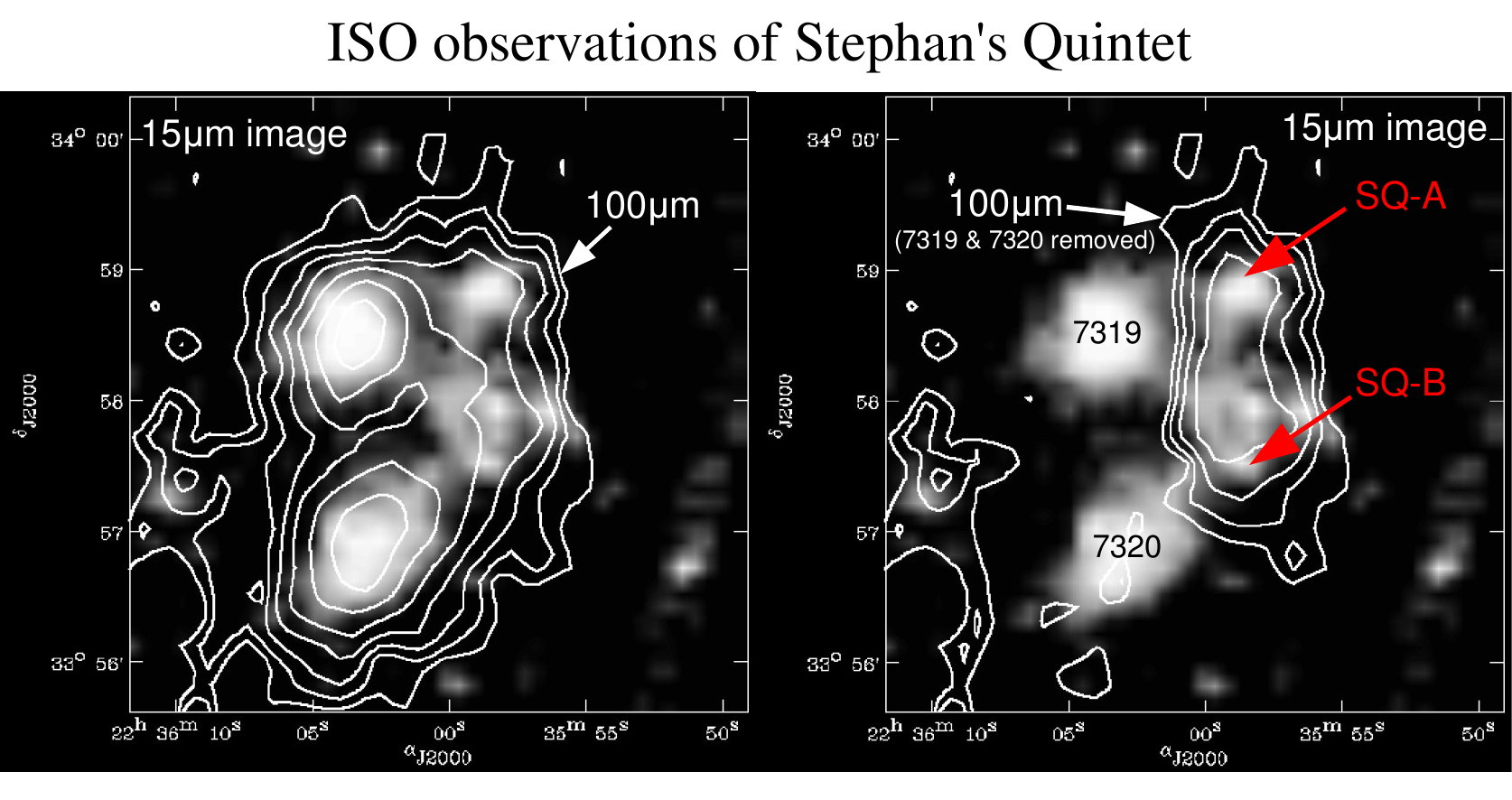}
      \caption[\textit{ISO} observations of SQ (15 \& 100$\,\mu$m) by \citet{Xu2003}]{\textit{ISO} observations of SQ (15 \& 100$\,\mu$m) by \citet{Xu2003}. Contour maps of the IR emission (background subtracted) at 100$\,\mu$m are overlaid on the \textit{ISO} 15$\,\mu$m image. \textit{Left:} The contour levels for the 100$\,\mu$m map are 0.8, 1.12, 1.6,  2.24, 3.2, 4.48,  6.4 MJy~sr$^{-1}$. \textit{Right:} 100$\,\mu$m emission after the subtraction of NGC 7319 and NGC 7320. The
contour levels  are 0.8, 1.12, 1.6,  2.24 MJy~sr$^{-1}$. }
       \label{fig:SQ_ISO_obs_Xu03}
\end{figure}

\textit{IRAS} first detected SQ in the infrared \citep{Allam1996, Yun1997}, but it was barely resolved. 
The first quantitative study of dust emission in the SQ group was made possible by the \textit{ISO}\footnote{Infrared Space Observatory, \url{http://iso.esac.esa.int/}.} space observatory. \citet{Xu1999, Xu2003} reported \textit{ISO} observations 
that show that most of the mid-IR (11.3 and 15$\,\mu$m) and Far-IR (60 and 100$\,\mu$m) emission comes from the disks of NGC~7319 and of the foreground galaxy NGC~7320 (see Fig.~\ref{fig:SQ_ISO_obs_Xu03}). Two intra-group medium (IGM) starbursts SQ-A and SQ-B, both including several previously detected H$\,${\sc ii} regions, stand out in the 15$\,\mu$m image due to their strong mid-IR emission. There is evidence for dust emission in the shock front. \citet{Xu2003} have modeled the  \textit{ISOPHOT}\footnote{Infrared Space Observatory Photopolarimeter \url{http://www.mpia-hd.mpg.de/ISO/welcome.html}.} maps at 60 and 100$\,\mu$m to isolate emission from the shock (Fig.~\ref{fig:SQ_ISO_obs_Xu03}). However, the angular resolution is marginally sufficient to separate the shock itself from the SQ-A and SQ-B star forming regions.

 Based on \textit{ISO} observations, the total dust luminosity of the shock front is $\mathcal{L}_{\rm dust} = 1.9 \times 10^{35}$~W \citep{Xu2003}. This is about an order of magnitude higher than the X-ray luminosity. \citet{Xu2003} concluded that dust cooling may be a significant cooling mechanism for the shock. 
They propose that the  hot gas is predominantly cooled by electronic collisions with dust grains. This conclusion is based on the comparison between the sputtering time scale of dust grains and the gas cooling time. Assuming a constant, Galactic mass-to-dust ratio,  the sputtering time of an $a = 0.1\,\mu$m grain is $3.7 \times 10^6$~yr, compared to the gas cooling time of $2.1 \times 10^6$~yr. Hence the big grains can indeed survive the shock long enough. However, as shown in \citet{Smith1996} and \citet{Guillard2009}, the sputtering of dust grains in the hot ($ T > 10^6$~K) phase should have reduced significantly the dust to gas mass ratio on a timescale shorter than the gas cooling time (see chapter~\ref{chapter:dust_gas_galaxies} and discussion later in \hyperref[paper_SQ_dust]{paper~{\sc iii}}).

\subsection{New  Spitzer observations}
\label{new-spitzer-dust-obs}

\textit{ISO} observations have been followed up by \textit{Spitzer} observations.
\citet{Appleton2006} release a spectacular composite image with IRAC (see the cover front of this manuscript!) at 3.6, 4.5, 5.8 and 8$\,\mu$m.
Images at 16 and 24$\,\mu$m were obtained by our team and presented in \citet{Cluver2009} and \citet{Guillard2009a}. 
Longer wavelength images at 70 and 160$\,\mu$m were obtained by \citet{Xu2008}. They claim to detect a significant diffuse dust emission outside individual galaxies, amounting to more than 50\% of the total $\mathcal{L}_{160\,\mu\rm m}$ of SQ (see also Natale, in preparation). 
Because of the better correspondence between the morphology of the 160$\,\mu$m map and that of the X-ray map, compared to that between 160$\,\mu$m and H$\,${\sc i}, the authors favor the possibility that the grains are collisionally heated in the hot intra-group medium.  

The presence of H$_2$ gas in the intragroup set a new perspective on the origin of the dust emission in the ridge. As introduced above, we expect some dust emission from the H$_2$ gas. My work presented in  \hyperref[paper_SQ_dust]{paper~{\sc iii}} explore this possibility.

\section{Publication: Paper III}
\label{paper_SQ_dust}

I reproduce here the paper \textit{``Observations and modeling of the dust emission from the H$_2$-bright galaxy-wide shock in Stephan's Quintet''}, P. Guillard, F. Boulanger, M. Cluver, P.N. Appleton, G. Pineau des Forêts, submitted to A\&A. I have intentionally not repeated the content of the paper in this manuscript.

 \includepdf[noautoscale=true, scale=0.9,link=true,frame=false,pages=-]{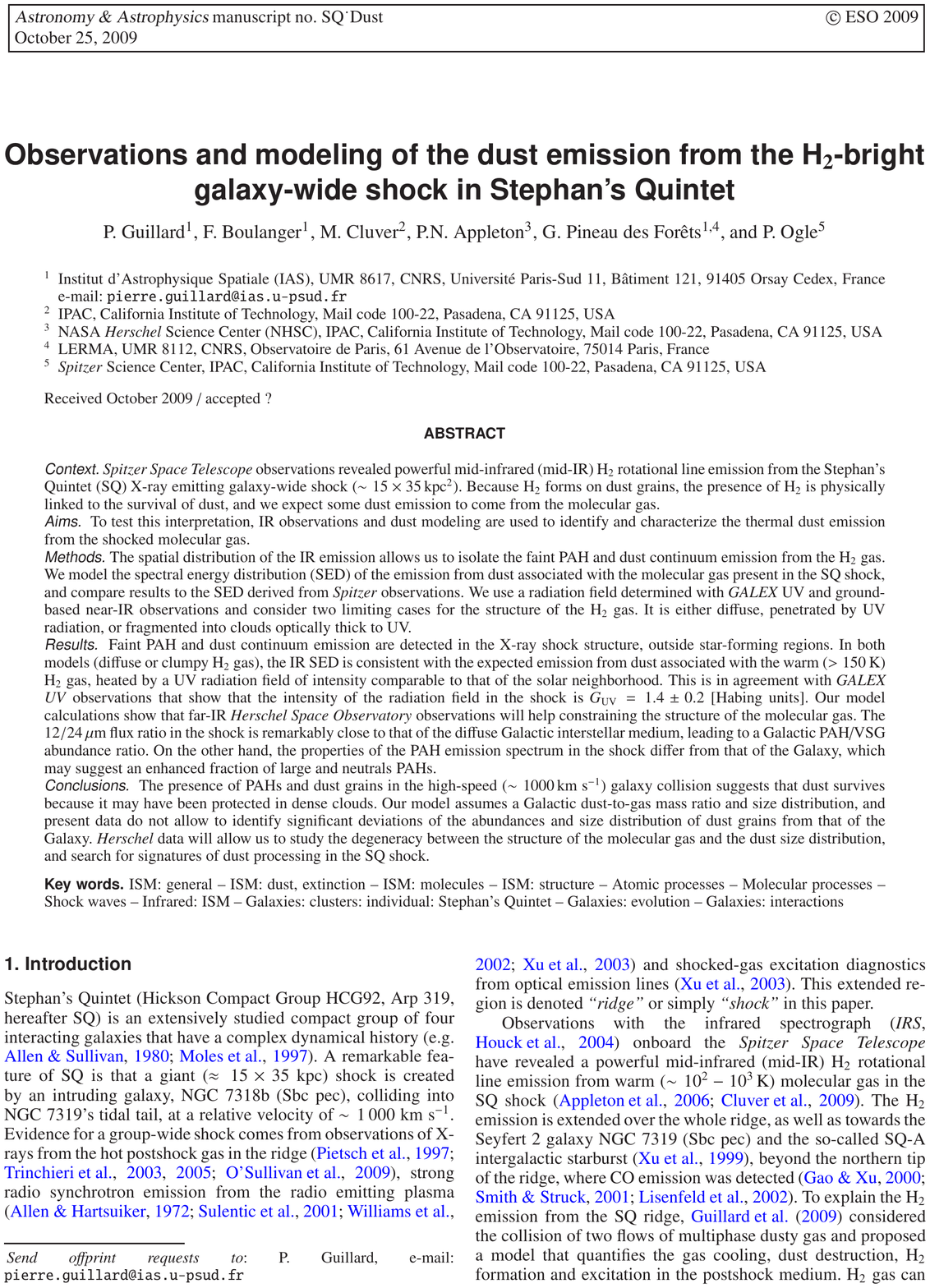}

\section[Perspectives on revisiting dust survival timescales in a multiphase ISM]{Perspectives on revisiting dust survival timescales: impact of the multiphase ISM structure}
\label{revisiting-dust-survival}

Observations of dust, from PAHs to BGs, in the violent SQ galaxy collision is certainly a challenge for dust evolution models, and has potentially important consequences for the estimate of dust survival in the ISM.  

So far, dust survival timescales are estimated by assuming that the shocked medium is homogeneous. Within the context of the SQ shock, the temperature of the hot plasma in constrains the shock velocity in the tenuous gas to be $\approx 700$~km~s$^{-1}$. 
If we assume that the shocked medium is homogeneous, all the dusty gas is shocked at these very high velocities, and thus the dust destruction timescale would be much shorter than the dust re-formation time. 

This issue is also true in the galactic ISM. Let us give a rough estimate of the dust survival timescale in an homogeneous medium, following \citet{Jones2004}.
Once a dust grain finds itself in the ISM, it is subject to
destruction by supernova shock waves. We calculate the the timescale for SN shock 
waves to destroy all the dust in the ISM, $t_{SNR}$. If $\epsilon (v_s)$ is
the efficiency of grain destruction by a shock of velocity $v_s$, 
$M_s (v_s)$ is the mass of gas shocked to a velocity of at least
$v_s$, $\tau _{SN}$ the interval between supernovae, and $M_{ISM}$ the
total mass of the galactic ISM, then the timescale for grain destruction is given by:
\begin{equation}\label{eq_timescale_dust_destruction}
  \frac{M_{ISM}}{t_{SNR}} = \frac{1}{\tau _{SN}} \int \epsilon (v_s)
  \textrm{d} M_s (v_s) \approx  \frac{\bar{\epsilon} M_s(\small{100\,
  \mbox{km}\,\mbox{s}^{-1}})}{\tau _{SN}} \; ,
\end{equation}
where $\bar{\epsilon}$ is the mean grain destruction parameter, and 
$ M_s(\small{100\, \mbox{km}\,\mbox{s}^{-1}})$ corresponds to a typical SN shock
that affects a sufficient volume of the ISM. Now, to determine the
mass shocked to a velocity of at least $v_s$, we exploit the
conservation of the energy $E$ released by the explosion of the SN and
write (McKee 1989, \cite{McKee1989}):
\begin{equation}
  M_s (v_s) = \frac{E}{\sigma \, v_s^2} \; , \quad \mbox{and then} \quad 
  t_{SNR} = \sigma \frac{M_{ISM} \, \tau _{SN} }{\bar{\epsilon} \,
  E}\, v_s^2  \approx 4 \times 10^8 \, \mbox{yr}\; ,
\end{equation}
where $\sigma = 0.736$ (Ostriker \& McKee 1988), and $E =
10^{44}\,\mbox{J}$. 
I use $\bar{\epsilon} = 0.5$, , $M_{ISM} = 5\,10^9$~M$_{\odot}$ and
$\tau _{SN} = 30$~yr. 
On the other hand, \citet{Jones2004} estimate the dust formation timescale to be $t_f \approx 3 \times 10^9 \, \mbox{yr}$.
Clearly, by comparing $t_f$ and $t_{SNR}$, dust destruction is faster
than stardust formation. This timescale discrepancy is clearly at odds
with the observations of dust in the ISM! 
The conclusion that is generally given  is that dust must be  (re)formed \emph{in situ} in the ISM. This may be true, but the exact way dust is re-accreted and re-coagulated is still
poorly known .

Stephan's Quintet observations may suggest an alternative solution to the issue of dust survival in the ISM. 
As discussed in \hyperref[paper_SQ_H2]{paper~{\sc i}}, the density inhomogeneities in the preshock medium allow the dust to survive, provided that the density contrast is high enough to reduce the transmitted shock velocity. The multiphase nature of the ISM may increase significantly the overall dust survival timescale.


\chapter{H$_{\bf 2}$ in galaxy evolution}
\label{chapter:perspectives}



\epigraph{There are $10^{11}$ stars in the galaxy. That used to be a huge number. But it's only a hundred billion. It's less than the national deficit! We used to call them astronomical numbers. Now we should call them economical numbers.}{Richard Feynman}




\begin{Abstract}
Molecular gas is the reservoir for star formation, and thus plays a key role in galaxy evolution. 
However, so far, the molecular gas has been largely ignored in the description of galaxy evolution in a cosmological context.  In particular, the response of the molecular gas to gas accretion onto galactic disks, galaxy merging,  feedback, and its impact on the energetics of these major evolutionary phases have so far not been addressed in detail. H$_2$-luminous galaxies open a new perspective for the study of molecular gas in active phases of galaxy evolution. Our interpretation of the H$_2$ emission from the Stephan's Quintet galaxy collision  set a theoretical framework that may apply to other astrophysical situations. 
The H$_2$-luminous radio-galaxy 3C326 is one example, where H$_2$ observations allow for the first time  to peer at the impact of the AGN-driven jet onto the molecular gas of the host galaxy. Does the AGN feedback regulate or even suppress star formation (negative feedback), or does it trigger star formation (positive)? This is a key question in galaxy evolution.
\end{Abstract}

\minitoc


\section{Introduction}
\label{sec:intr_perspectives}

\PARstart{U}nderstanding how galaxies formed and evolved is one of the major goals of
modern extragalactic astronomy. This dissertation is a first step to address the role of  molecular gas within this context.
The emerging population of H$_2$-luminous galaxies (with enhanced mid-IR H$_2$ line emission and relatively weak star formation),  includes galaxies in several key phases of their evolution, dominated by gas accretion,
galaxy interactions, or galactic winds driven by star formation and AGN. The analysis presented in this manuscript suggests that the warm H$_2$ contributes significantly to the overall energy budget of these galaxies, and
may potentially play a central role for galaxy evolution.

This chapter  introduces the role of the molecular gas for the energetics of star formation and galaxy mass build-up (sect.~\ref{sec:molecular-gas-in-galaxy-evolution}). I point out some major astrophysical questions in this context, with an emphasis on AGN feedback (sect.~\ref{H2-AGN-feedback}). Then I present the case of 3C326, a spectacular example of an H$_2$-luminous radio-galaxy for which I have applied the same analytical and numerical approach as for Stephan's Quintet to model the H$_2$ emission and derive gas masses in the system. These results, together with a detailed analysis of the energetics of the system, are gathered in \citet{Nesvadba2009} (hereafter \hyperref[paper_3C326]{paper~{\sc iv}}), submitted to A\&A. The observational context of 3C326 is given in sect.~\ref{observational-context-3C326} and the paper is reproduced in sect.~\ref{paper_3C326}.
This chapter ends with a presentation of some observational (sect.~\ref{observational-perspectives}) and theoretical perspectives (sect.~\ref{theoretical-perspectives}).

\section{H$_{\bf 2}$ and feedback in galaxy evolution}
\label{sec:molecular-gas-in-galaxy-evolution}

\index{Galaxy formation}
\index{Molecular gas!galaxy formation}
\index{Galaxy formation!molecular gas}
\index{Feedback!and galaxy evolution}

Thanks to ever-improving cosmological
simulations, we now have a sound comprehension of structure formation on the
largest scales. Spectacular progress is made in describing the structure of dark matter, and the history of
star formation through cosmic time.  However, describing galaxy evolution within this picture
is still a challenge.

\begin{figure}
   \centering
    \includegraphics[width=\textwidth]{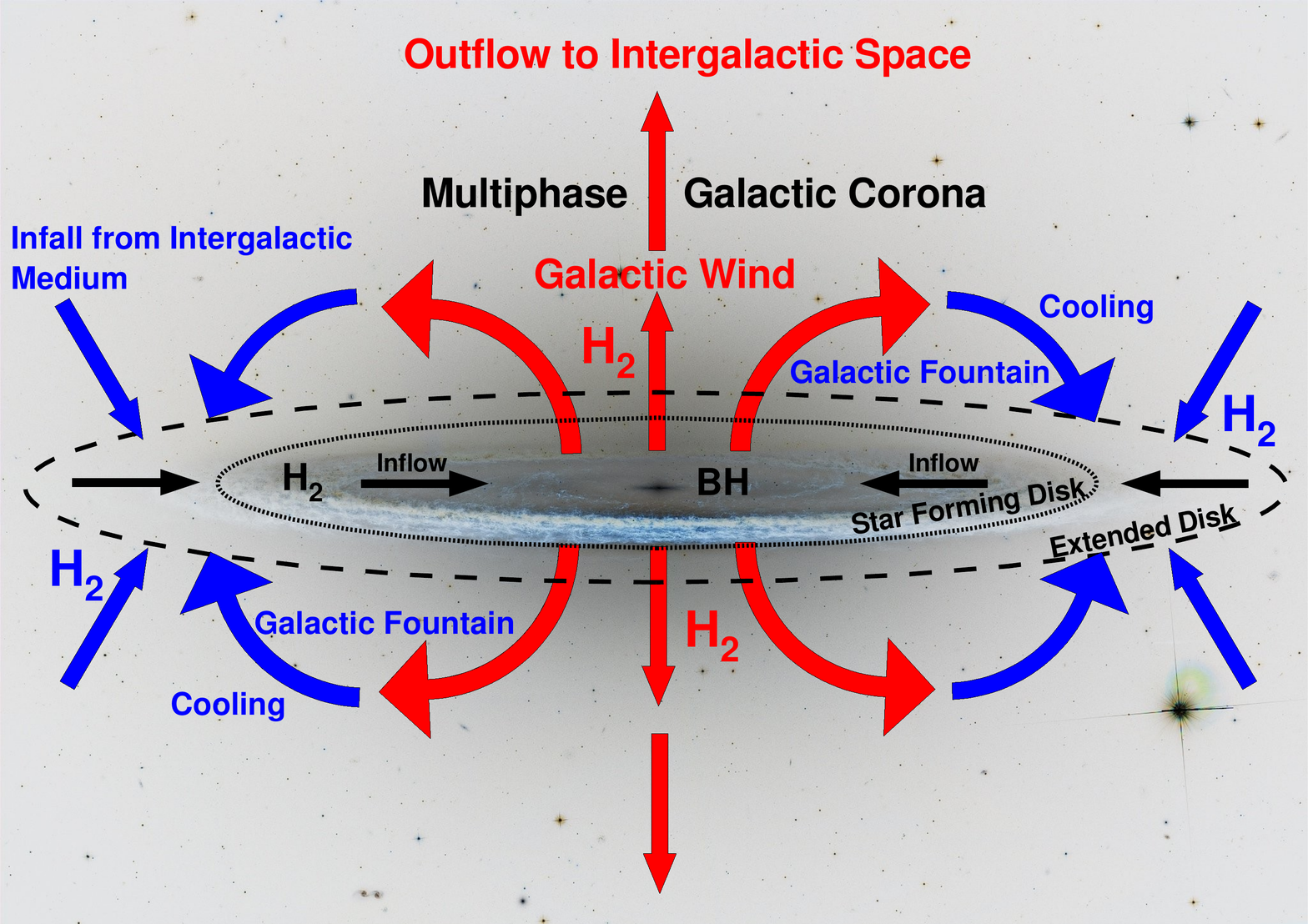}
      \caption[Circulation of the multiphase interstellar matter during galaxy mass build-up]{Sketch of the complex circulation of the multiphase interstellar matter during galaxy mass build-up. The gas disk of the galaxy is fed by infall from intergalactic space and by gas raining down from the halo. This gas flows inwards and fuel star formation as well as the activity
of the central black hole.  The blue and red arrows schematically outline
these inflows and outflows, and the gas cooling and heating.  Observations
show that H$_2$ is a key actor in all stages. Adapted from \citet{Boulanger2009}. }
       \label{fig_galaxy_formation}
   \end{figure}
  
\index{Superwinds}
\index{Galactic winds}
The reason is that the build-up of baryonic mass in
galaxies is regulated by a complex interplay between gravitational collapse and
feedback related to AGN and star formation. 
The  dynamical interactions between the ISM phases and their thermal properties play a key role in this regulation. 
Fig.~\ref{fig_galaxy_formation} illustrates schematically the mass and energy exchange between the ISM of a galaxy and its halo's environment. In the context of galaxy baryonic mass build-up, two processes compete.
 On one hand, the activity of the black hole, and the bursts of star formation, create ``galactic winds'', or ``superwinds'' on galactic scales \citep[e.g.][]{Heckman1990} that push the hot/warm gas, and possibly also molecular gas, into the galactic corona, or even further out to the intergalactic space  (\textit{outflow}). The relative importance between the impact of black holes or stars depends on the mass of the galaxy, more massive ones being dominated by the AGN activity \citep[e.g.][]{Dekel1986}. 
On the other hand, part of this matter does not escape the gravitational potential well of the galaxy, and falls back onto the galactic disk. In addition, matter from the intergalactic medium, or from a companion galaxy, is accreted onto the disk. These processes are particularly important during the early phases of galaxy evolution at high redshift, when most of the stars in the universe were
formed.

In this picture of galaxy formation and evolution, a lot of astrophysical questions remains. We still do not know how gas that falls within dark matter haloes is accreting onto the galactic disk, and how it comes to form stars.  
On the other hand, it is not clear what is the impact and efficiency of  feedback processes from stars or black holes on the molecular gas, and thus on star formation. 
These questions are related to the regulation of mass build-up in galaxies described above. I will only focus on the second aspect, which is closely related to the theoretical framework developed in my PhD work: \textit{What is the impact of feedback on galaxy evolution?}


Since H$_2$ formation is a natural outcome of gas cooling, and since molecular gas represents the reservoir for star formation, H$_2$ gas must play an important role in all the processes listed above.
Therefore, understanding  the role that cold interstellar matter plays in the evolution of galaxies is a key to elucidate the above questions.
 However, very little is known about its \textit{impact} on the energetics of feedback, and its \textit{response} to it. 
In a sense, the \textit{Spitzer Space Telescope} has opened a new observational window into the ``molecular Universe'' through mid-infrared observations of warm H$_2$.
In the following I will emphasize what can we learn from H$_2$ observations  that may help in answering these questions. I distinguish two types of feedback, depending on the powering sources: either star formation or black hole.



\subsection{Feedback from star formation: H$_{\bf 2}$ in ``superwinds''}
\index{Feedback!star formation}


Star formation is known from optical to X-ray observations of warm and hot gas to 
drive galactic winds \citep{Heckman1990, Heckman2000}. 
Superwinds are generated when the kinetic energy from stellar winds and supernovae is thermalized, generating a region of very hot ($T \approx 10^7 - 10^8$~K) high-pressure plasma in the ISM of a starburst galaxy \citep{Chevalier1985}. As the bubble breaks out of the disk of the galaxy, the plasma expands producing
a weakly-collimated bipolar outflow into the galaxy halo. The outflowing plasma sweeps up and shocks ambient material, creating a superwind.

To understand the impact of these winds on star formation, and thereby on galaxy evolution in general, it is crucial  to constrain their impact on the multiphase interstellar medium of 
galaxies. 
\textit{Spitzer} mid-IR imaging and spectroscopy have revealed  that
winds are loaded with molecular gas and dust (see the illustration for the wind of M82 in Fig.~\ref{fig_H2_M82_NGC6240}). 
However, the origin of this molecular gas is still unclear. Is it material lifted up from the galactic disk, or is the molecular gas formed \textit{in situ}, by compression of dusty atomic gas that is advected in the flow? We do not favor the first interpretation because the hot wind is too tenuous to lift up dense molecular clouds from the disk. If one considers a molecular cloud of radius 10~pc and density $10^4$~cm$^{-3}$, entrained in a 500~km~s$^{-1}$ superwind at density $10^{-2}$~cm$^{-3}$, the acceleration timescale of the cloud would be (Eq.~\ref{eq:cloud-acceleration-timescale}) $\sim 3 \times 10^{10}$~yr, which is much larger than the dynamical timescale of the outflow ($\sim 10^7$~yr).
Following our scenario of H$_2$ formation in the Stephan's Quintet shock (chapter~\ref{chapter:H2_SQ}), we favor  the second interpretation, in which the H$_2$ gas is formed outside galactic disks as
the result of the dynamical interaction of the wind with gas in the halo.  

The molecular component has previously been  ignored in
galactic winds studies. Is the molecular gas  a
dominant mass component of galactic winds?  Does it share the outward
flow motion of the lighter and warmer gas? Does it fall back on the
disk? Is the emission from molecular gas the main dissipative channel that sets
the energetic efficiency of galactic winds? The answers to these
questions about molecular gas are essential to determine the impact of feedback from
star formation on the evolution of galaxies. 

\subsection{H$_{\bf 2}$ and AGN feedback}
\label{H2-AGN-feedback}
\index{Feedback!AGN}


An even more powerful source for feedback is the central engine of AGN, the supermassive black hole in the center of the host galaxy. A small fraction of the  energy released by the growth of the black hole, if 
absorbed by interstellar matter in the host galaxy, could 
regulate or even quench star formation by heating and ejecting ambient gas from the disk of the host galaxy under the action of an AGN-driven outflow. This is the so-called \textit{negative AGN feedback}.  On the other hand, the impact of the AGN could in principle be also \textit{positive}, by triggering star formation through the interaction of the AGN-driven jet with the ISM of the host galaxy.

\subsubsection{Monsters needed to break the hierarchy!}
This negative AGN feedback has been often invoked to explain the decline of star formation in massive, early-type galaxies \citep[e.g.][]{Silk1998, Springel2005, Croton2006, Hopkins2006}. 
The large majority of these galaxies appears to have formed
most of their stars at high redshift\footnote{This is indicated by the redshift evolution of luminosity and mass functions of galaxies, and by the characterization of 
stellar populations, chemical abundances, and structural  properties of giant ellipticals in the local universe.  Even a small fraction of the gas returned by dying-stars 
gas should result in star formation rates much larger than observational estimates.}, so that they are now ``old, red, and dead''. 
These observations are in apparent contradiction with a gradual, ``bottom-up'', mass assembly. 
Under the assumption of  hierarchical galaxy evolution, one would naïvely think that the gas  cools and is accreted gradually onto massive halos. So we should find many more massive,  blue, star-forming galaxies, particularly at the centers of groups and clusters. 
This is called the \textit{hierarchical problem}. AGN ``monsters'' are then called to break the hierarchical scenario!


It is worth mentioning  that an other alternative mechanism, called \textit{morphological quenching}, has been recently proposed by  \citet{Martig2009} to account for ``red'' early-type galaxies. Based on cosmological simulations, they propose that star formation is quenched when the galaxy experiences a morphological transition from a rotating stellar disk to a pressure-dominated stellar spheroid, which induces a steeper potential well and reduces the disk selfgravity. In their simulations, this transition results  from the growth of a stellar spheroid, for instance by galaxy merging. The gas in the disk of the galaxy becomes stable against fragmentation, thus reducing star formation efficiency. 
Though interesting, this process may not be directly related to H$_2$-luminous objects since it appears that a powerful source of mechanical energy is needed to explain the H$_2$ emission.

\subsubsection{H$_2$-luminous galaxies as a tool to study AGN feedback}
Despite its potential importance, \textsl{dixit} Matthew D. Lehnert, ``AGN feedback has remained in the realm of theoretical \textsl{deus ex machina}''.
We still do not understand how  the transfer of energy from the AGN to the 
surrounding gas occurs. In particular, a major question is to determine the impact of the AGN-driven jet to the molecular gas settled in the disk of the host galaxy. Is the ISM of the host galaxy blown up in the outflow, thus supressing star formation (\textit{negative feedback})? Or does the jet trigger shock-compression of clouds, thus enhancing star formation (\textit{positive feedback})? Or is it both?




Evidences for AGN feedback are now growing. Observations suggest that AGN feedback may be associated mostly with radio-loud AGN. Perhaps the most striking examples are nearby clusters of galaxies, in which the central black  hole(s) inject a huge amount of energy (typically $\approx 10^{60}$~erg for the most massive ones) over a rather short timescale ($\approx 10^7$~yr), which seems enough to prevent the cooling of the gas through X-ray emission. This may be an explanation to solve for the ``cooling flow problem'', in which the amount of hot gas that cools radiatively (inferred from soft X-ray observations) is much less than expected from the harder X-ray luminosity \citep[see][and references therein for a review]{McNamara2007}. A number of observations show that cool-core clusters contain large amounts of molecular gas at their center. H$_2$ was first detected
through its near-IR ro-vibrational lines. Warm H$_2$ has since also been observed
through its mid-IR rotational lines with Spitzer \citep[see chapter~\ref{chapter:H2_galaxies},  sect.~\ref{H2-in-cooling-flows-galaxy-clusters}, Fig.~\ref{fig_H2_Perseus_Z3146} and e.g.][]{Egami2006}. Numerous CO detections have also been reported but the spatial distribution and
kinematics of the cold molecular gas has only been imaged
in a few clusters. For instance 
CO observations reveal  a large amount ($\approx 10^{11}$~M$_{\odot}$) of cold molecular gas extending in the halo to several tens of kpc from the central elliptical galaxy NGC ~1275 \citep{Salom'e2006} in the Perseus A galaxy cluster.
This molecular gas is observed to be surprisingly inefficient at forming stars.

\textit{Radio galaxies} are  interesting targets to elucidate the physical mechanisms whereby the AGN can regulate the star formation in its host galaxy. Indeed, radio-galaxies have an AGN and an obvious source of mechanical energy which reaches large scales. They also show signs of outflows \citep{Nesvadba2006, Nesvadba2008a, Nesvadba2009, Holt2008}, strong turbulence, and dissipation of mechanical energy \citep{Ogle2007, Ogle2009}.
For the first time,   observations of warm molecular gas in radio-galaxies allow to peer at the impact of the radio jet on the energetics of the dense molecular gas, and thus on star formation. 
\citet{Ogle2009} find that about 30\% of nearby powerful radio galaxies may fall into the class of H$_2$-luminous galaxies presented in chapter~\ref{chapter:H2_galaxies}, which have significant amounts of relatively dense, warm, and surprisingly turbulent molecular gas, but very little star formation.

In the following I shall focus on the example of the 3C326 radio-galaxy, which has been studied in detail by applying the methods developed during my PhD work.

\section{Are we witnessing negative AGN feedback in the radio galaxy 3C326?}

The radio galaxy 3C326 is an outstanding source to study the impact of the radio jets on its molecular gas, and  \hyperref[paper_3C326]{paper~{\sc iv}} presents such a study. 
We report new CO observations of that source that complement the mass and energy budget of the molecular  gas in 3C326. We interpret line emission from the ionized and molecular gas within the framework I have developed for the Stephan's Quintet.

The  \hyperref[paper_3C326]{paper~{\sc iv}} is reproduced at the end of this chapter. I only give here a brief summary of the main points, as well as additional figures and results from my modeling of H$_2$ excitation in that source. 
We find that the ISM of this galaxy has very unusual properties: the molecular gas budget is dominated by warm gas at temperatures $T>100$~K, while diagnostic line ratios suggest that the ionized and molecular gas is mostly excited through shocks. 

We also identify a significant outflow of neutral gas which cannot be explained by star formation. We explain these observations through a common physical framework where a fraction of the mechanical energy of the radio jet is being dissipated on small scales within the molecular gas, which powers the observed H$_2$ emission. This is an extension of the classical ``cocoon'' model, in which the AGN-driven jet inflates a ``cocoon'' of hot gas that accelerates the warm medium. We now explicitly take into account the multiphase character of the ISM and find that only a small fraction ($\sim 10$\%) of the jet mechanical energy is need to power the H$_2$ emission. 
The timescale of dissipation of the jet mechanical energy ($10^7 - 10^8$~yr) suggests that the jet could heat the molecular on timescales comparable, or even longer, than the jet lifetime, thus limiting star formation over a long-lasting period of time.
Studies of repeated jet activity in radio galaxies suggest that the phse of ``quiessence'' between two activity cycles could be of similar duration.  Thus, the physical gas conditions in AGN host galaxies may play an important role in regulating this cyclical jet activity. 

To introduce the \hyperref[paper_3C326]{paper~{\sc iv}} that presents a detailed analysis of the energetics of the molecular gas in 3C326, I briefly present the observational status of 3C326.

\subsection{Observational context}
\label{observational-context-3C326}

\begin{figure}
   \centering
    \includegraphics[width=\textwidth]{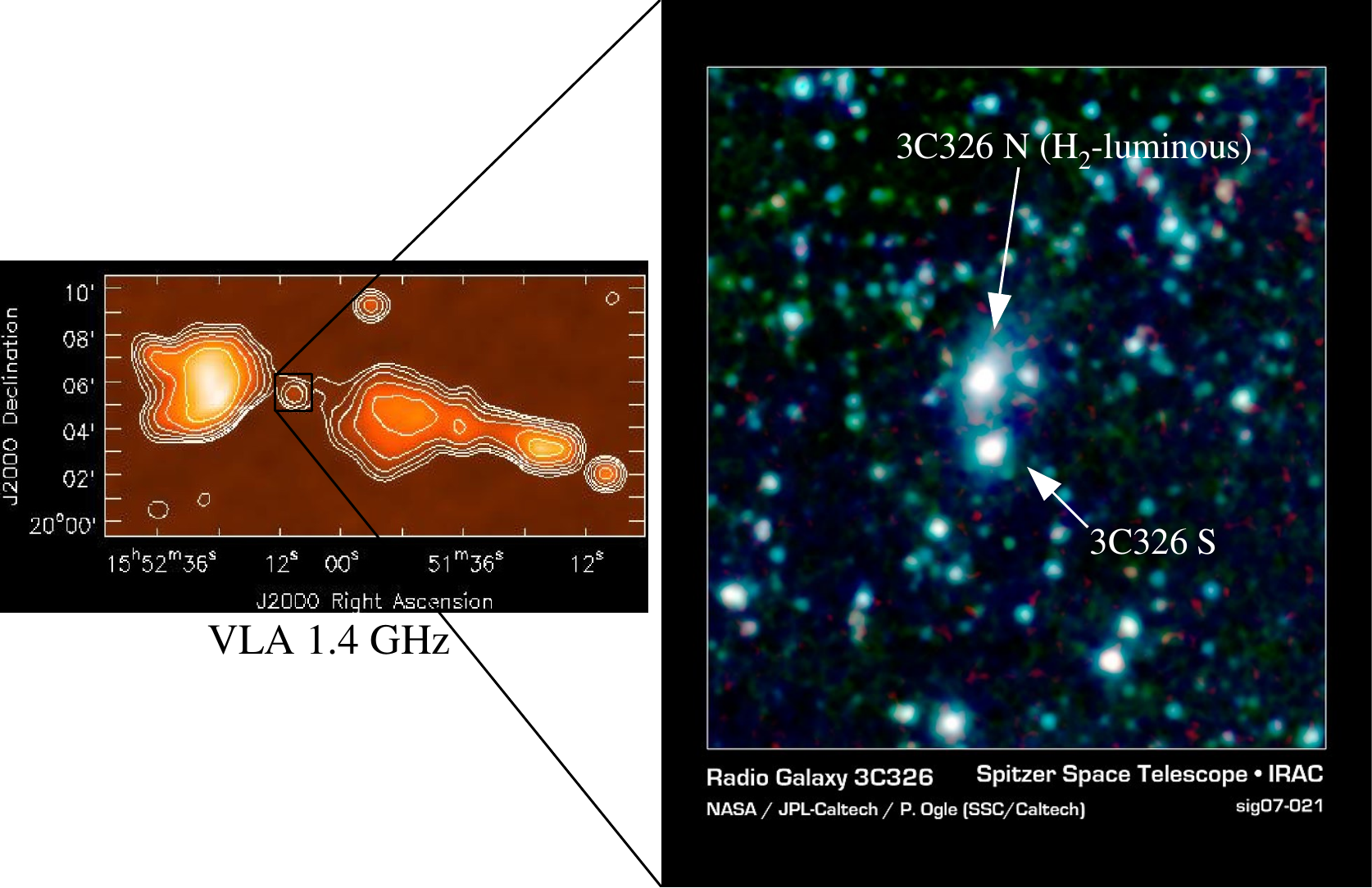}
      \caption[Radio VLA and Spitzer images of the 3C326 radio-galaxy]{Radio VLA (1.4 GHz)   \citep{Klein1994a, Mack1997} and Spitzer images \citep{Ogle2007} of the 3C326 radio-galaxy. The system comprises of two galaxies (3C326 North and South).  The projected separation between the two is 24.9'' (41~kpc). It is not clear which one creates the jet, materialized by the Mpc-scale radio lobes seen on the VLA image. }
       \label{fig_3C326/3C326_VLA_Spitzer}
\end{figure}

\subsubsection{Two companions\dots}
\index{3C326 radio-galaxy!description}
3C326 is one of the largest  radio sources known \citep{Willis1978}, with an angular size of 19.5'' (1.9~Mpc). It is a system of two companions, 3C326~N (N being for North) and 3C326~S (South), at a redshift of $z = 0.089 \pm 0.001$.
Fig.~\ref{fig_3C326/3C326_VLA_Spitzer} shows a radio \textit{(left)} and a mid-IR  \textit{(right)}  image of the pair. 
It is not clear which of the two galaxies  hosts the AGN that produced the Mpc-scale radio lobes \citep[see][]{Rawlings1990}.


The galaxy 3C326~N has been classified as a LINER from optical spectroscopy \citep{Rawlings1990}. It has strong [O$\,${\sc ii}], H$\alpha$, [S$\,${\sc ii}], and weak [O$\,${\sc iii}] lines, whereas 3C~326 S has no detected emission lines \citep{Rawlings1990, Simpson1996}. \citet{Rawlings1990} suggest that 3C~326N is more typical of FR II radio galaxy hosts than its South companion. 


\subsubsection{\dots but one H$_{\bf 2}$-luminous source}
\index{3C326 radio-galaxy!H$_2$ masses}

Based on this observational context, \citet{Ogle2007} targeted the northern companion, 3C326~N, with the \textit{Spitzer IRS}, and found extraordinary powerful H$_2$ rotational line emission. The total emission-line luminosity of 3C326~N is $\mathcal{L(\rm H_2)} = 8.0 \pm 0.4 \times 10^{34}$~W (integrated over the lines S(1) to S(7)). 
The ratio of the H$_2$ luminosity of the $8-70\,\mu$m integrated infrared luminosity is $0.17 \pm 0.02$, which is $1-2$ orders of magnitude higher than normal star-forming galaxies (see discussion in chapter ~\ref{chapter:H2_galaxies}).  

The optical and mid-IR spectra of 3C326~N \citep[see Fig.~\ref{fig_H2_galaxies_H2_7_7PAH_L24uml}, Fig.~\ref{fig_Ogle09_spectra_RGs} and][]{Ogle2007, Ogle2009} suggest a very weak star forming rate ($\sim 0.07$~M$_{\odot}$~yr$^{-1}$) although the galaxy holds a large mass of H$_2$ gas ($\approx 10^9$~M$_{\odot}$).

\subsection{H$_2$ excitation in 3C326}
\index{3C326 radio-galaxy!H$_2$ excitation}

Like in Stephan's Quintet, the H$_2$ emission in 3C326 is associated with the dissipation of mechanical energy. I have used the grid of MHD shocks presented in chapter~\ref{chapter:shocks} to quantify the range of preshock density and shock velocities that fits the oberved H$_2$ line fluxes. 

\begin{figure}
   \centering
    \includegraphics[angle=90, height=0.6\textwidth]{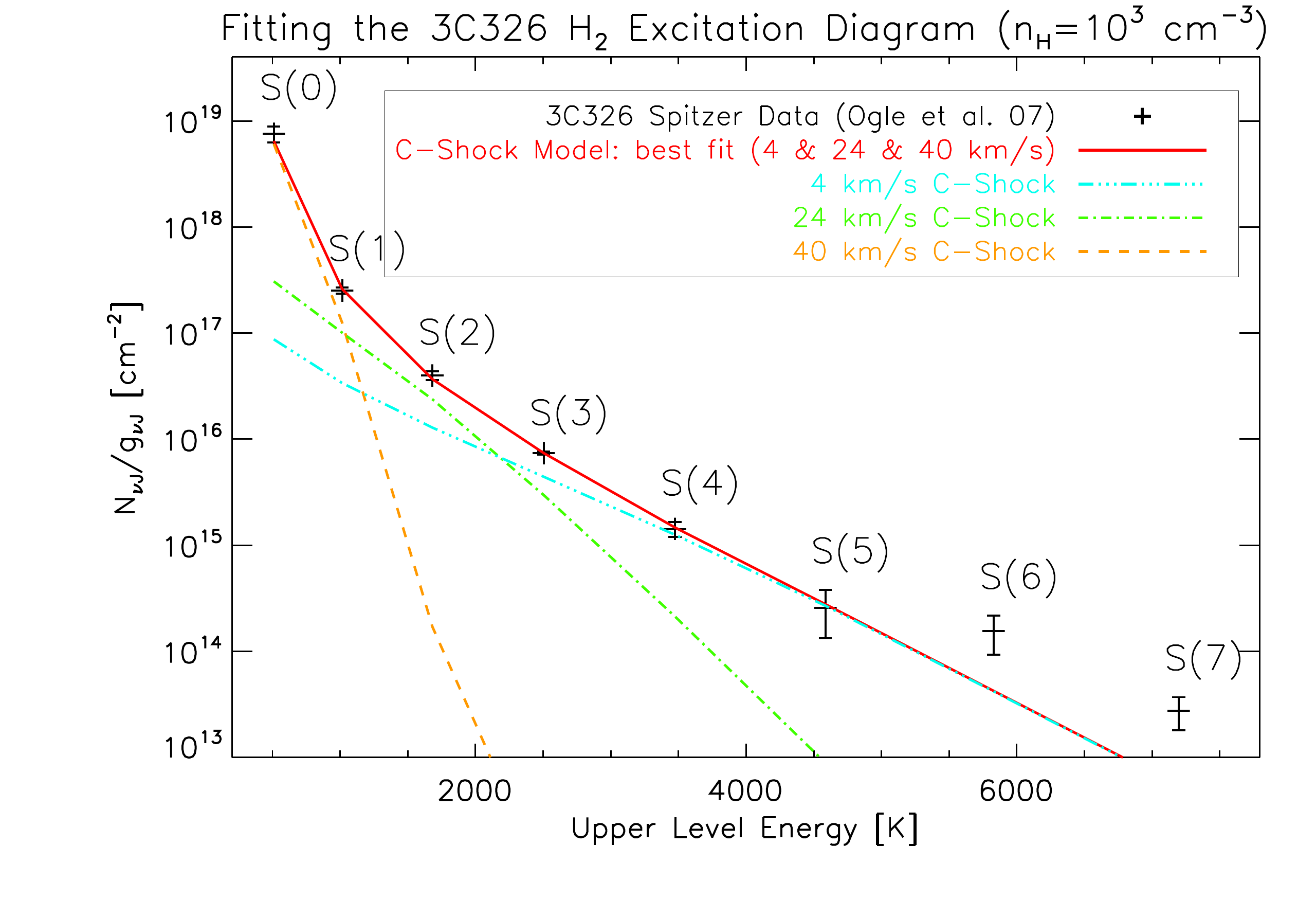}
  \includegraphics[angle=90, height=0.6\textwidth]{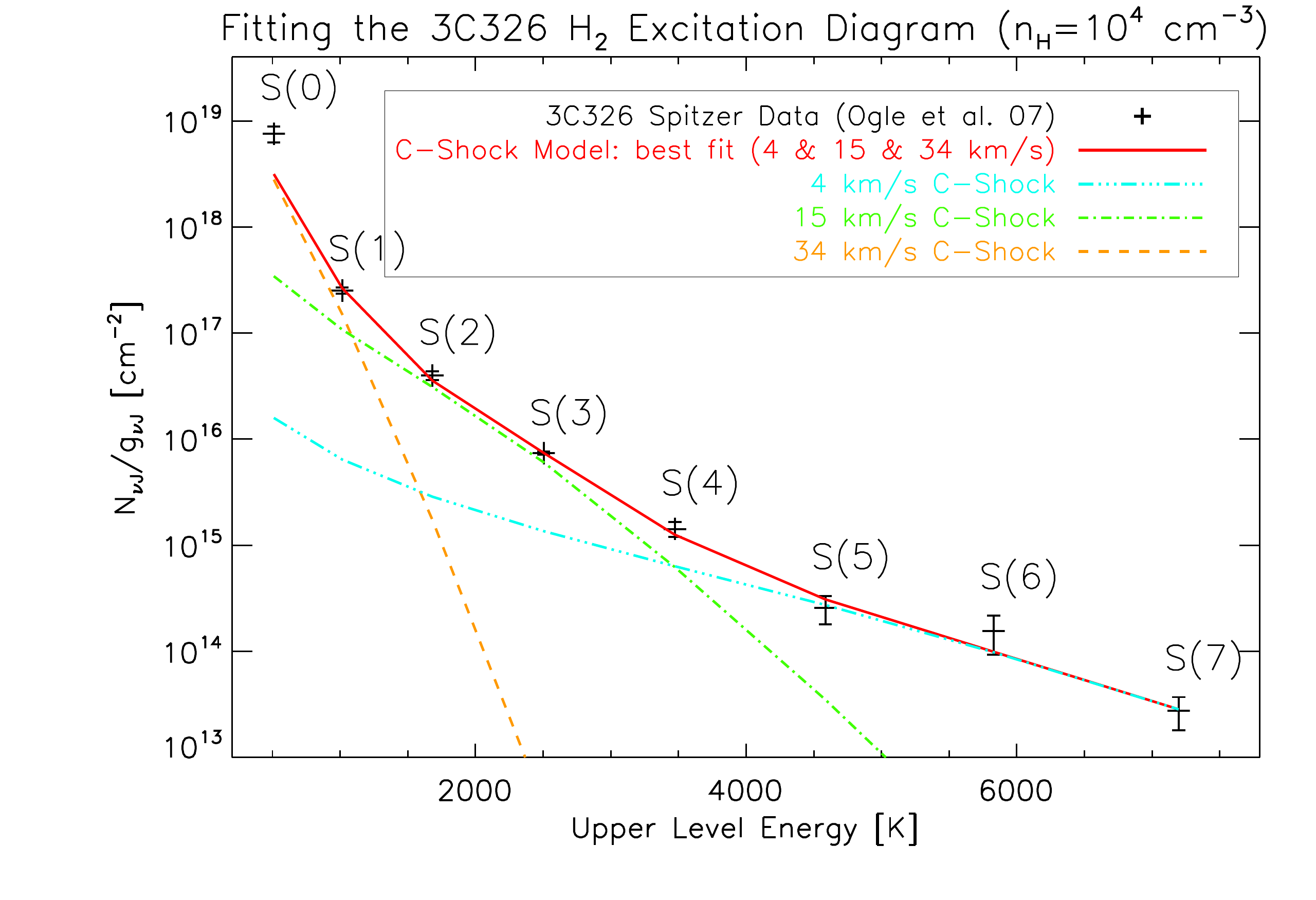}
      \caption[Fit of the 3C326 H$_2$ excitation diagram with 3 MHD shocks]{Fit of the 3C326 H$_2$ excitation diagram with 3 shocks MHD for a preshock density $n_{\rm H} = 10^3$~cm$^{-3}$ \textit{(above)} and $n_{\rm H} = 10^4$~cm$^{-3}$ \textit{(bottom)}. The shock model used is described in chapter~\ref{chapter:shocks}. The 3 shock velocities components account for comparable fractions of the total H$_2$ luminosity but the less energetic component accounts for most of the warm H$_2$ mass (Table~7 of \hyperref[paper_3C326]{paper~{\sc iv}}). }
       \label{fig_H2excitDiag_Cshock3C326}
\end{figure}

The results of the modeling of the H$_2$ excitation are gathered in Tables~6 and 7 of \hyperref[paper_3C326]{paper~{\sc iv}}. To complement these tables,  Fig.~\ref{fig_H2excitDiag_Cshock3C326} shows the observed and modeled H$_2$ excitation diagram for 3C326. As discussed in chapter~\ref{chapter:H2_SQ_mapping}, these  fits are not unique. I show here the results for two preshock densities, $n_{\rm H} = 10^3$ and $10^4$~cm$^{-3}$. The lowest preshock density ($10^3$~cm$^{-3}$) is need to fit the 0-0S(0) line, where higher excitation lines S(6) and S(7) require higher densities or higher velocity shocks. Please see  \hyperref[paper_3C326]{paper~{\sc iv}} for a discussion of the results, in particular the mass and energy budgets of the molecular gas in this galaxy.


\subsection{Publication: paper IV}
\label{paper_3C326}
 \includepdf[noautoscale=true, scale=0.9,link=true,frame=false,pages=-]{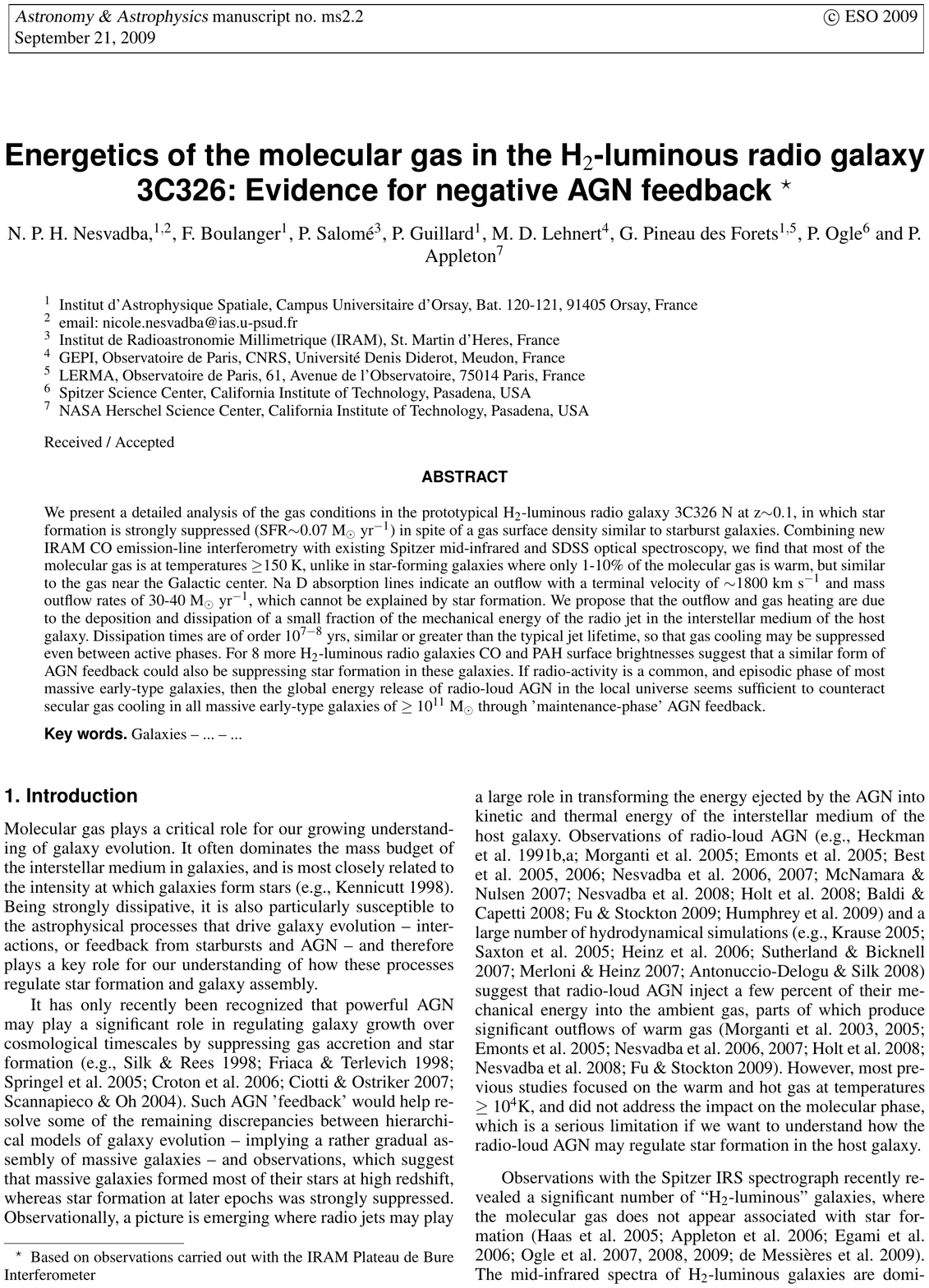}

\part[The next-generation tool to study H$_2$ in space: the JWST]{The next-generation tool to study (among a few other things\dots) H$_{\bf 2}$ in space: the JWST}
 
\chapter{The JWST observatory and its Mid-Infrared Instrument (MIRI)}
\label{chapter:JWST}

\epigraph{The thing's hollow - it goes on forever - and - oh my God! - it's full of stars!}{Arthur C. Clarke (2001: A Space Odyssey)}

\index{JWST}


\begin{Abstract}

Shortly after the beginning of my PhD, I got involved in the european consortium of the Mid-InfraRed Instrument (MIRI), part of the scientific payload of the  \textit{James Webb Telescope} (JWST). 
I participated to the cryogenic tests of the instrument and to the preparation of a science proposal related to the physics of H$_2$ emission in distant galaxies. 
This chapter present the necessary background needed for the next chapters that discuss my contribution to the testing of MIRI (chap.~\ref{chapter:miri_test}) and to its science applications (chap.~\ref{chapter:science_JWST}).
The JWST is a large ($6.5\,$m in diameter) near- and mid-infrared space telescope that will succeed to the \textit{Hubble} and \textit{Spitzer} space telescopes. We present an overview of the JWST and its science instruments, with emphasis on MIRI. MIRI provides imaging, coronagraphy and spectroscopy over the $5 - 28 \, \mu$m band. Its characteristics (data acquisition, observing modes, etc.) and performance (sensitivity) are described.

\end{Abstract}

\minitoc



\section{Introduction}

\PARstart{T}he \textit{James Webb Telescope} (hereafter JWST), giant successor to the \textit{Hubble} and \textit{Spitzer} space telescopes, is in preparation since 20 years. We are now in the phase where the flight model instruments have been built and are being tested. The french community, led by the CEA Saclay, IAS Orsay, Meudon and Marseille institutes, took the responsibility of the mid-infrared ($5-29\,\mu$m) camera, a sub-component of the Mid-InfraRed Instrument (MIRI), which is one of the four instruments onboard the JWST.

Six months after the beginning of my PhD, Alain Abergel, who had worked in collaboration with CEA to the definition of the optical performance tests of the instrument, offered me to be part of the MIRI test team, in order to participate in the optical test of the MIRI camera taking place at the CEA, Saclay. Needless to say that I accepted without hesitating! I was interested in opening my skills to instrumentation, but also in gaining knowledge of the MIRI instrument, that will be a milestone of future mid-IR astronomy, and an ideal tool to study the  H$_2$-luminous objects we have been speaking about in the previous chapters. In a sense, this project integrates very well in the science topic of this thesis manuscript.

I soon realized that I was a a tiny grain of sand in an enormous project! In order to have my own and little contribution to the project, I decided to focus on the tests of the Point Spread Function of the instrument, one of the tests designed to check the optical quality of the instrument. I participated in  the tests and took the lead of the data analysis of the PSF measurements. This work is described in the next chapter (chap.~\ref{chapter:miri_test}).

In this chapter we provide the necessary background to the next two chapters that present my contributions to the MIRI testing and to the preparation of scientific projects related to the physics of the H$_2$ emission in active phases of galaxy evolution (chap.~\ref{chapter:science_JWST}). 
I first give a brief overview of the JWST mission, with emphasis on its science instruments. Then I describe in more detail the MIRI instrument I have been working on (sect.~\ref{sec:MIRI_description}). Imaging and spectroscopic observation modes and sensitivities are discussed.  This chapter can also be viewed as an up-to-date working document useful to prepare observations.

\section{JWST mission overview}
\label{sec: JWST-mission-overview}
\index{JWST!mission overview}

\subsection{Impressive numbers!}

\begin{figure}
  \begin{center}
 \includegraphics[width= 0.505\textwidth]{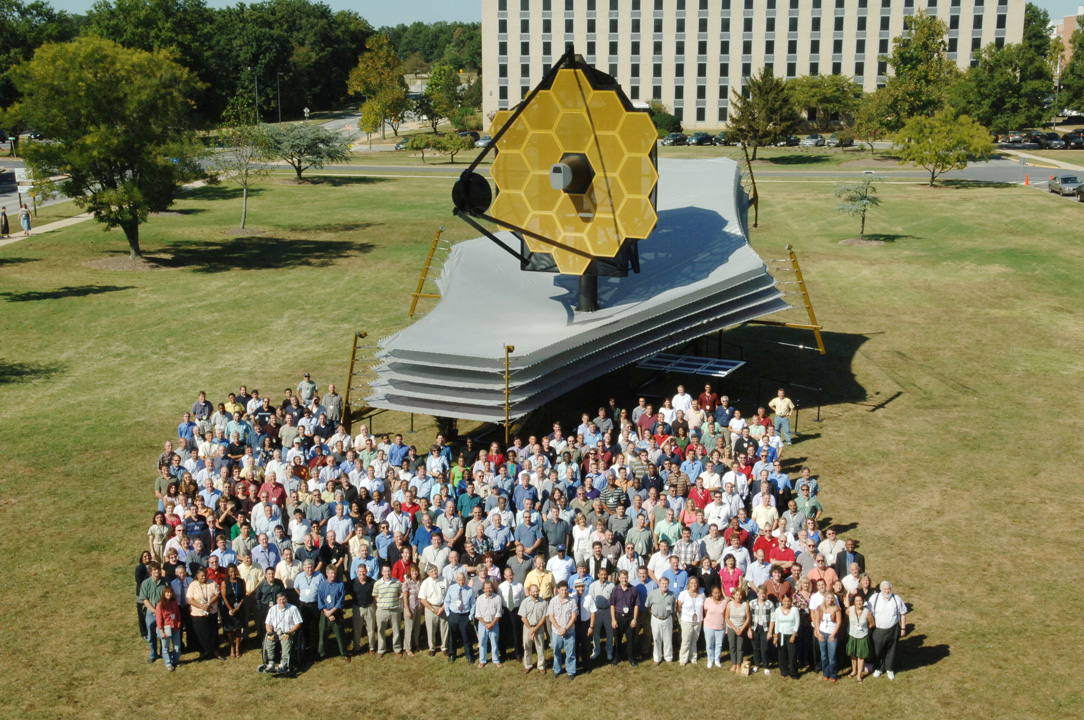}
     \includegraphics[width= 0.485\textwidth]{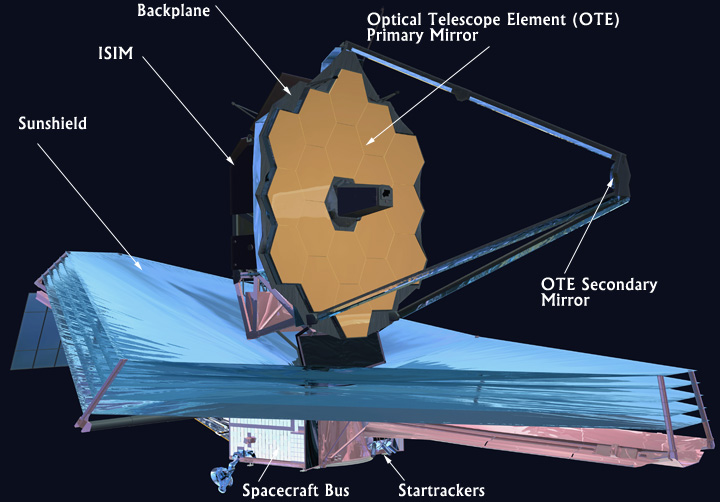}
     \includegraphics[width= \textwidth]{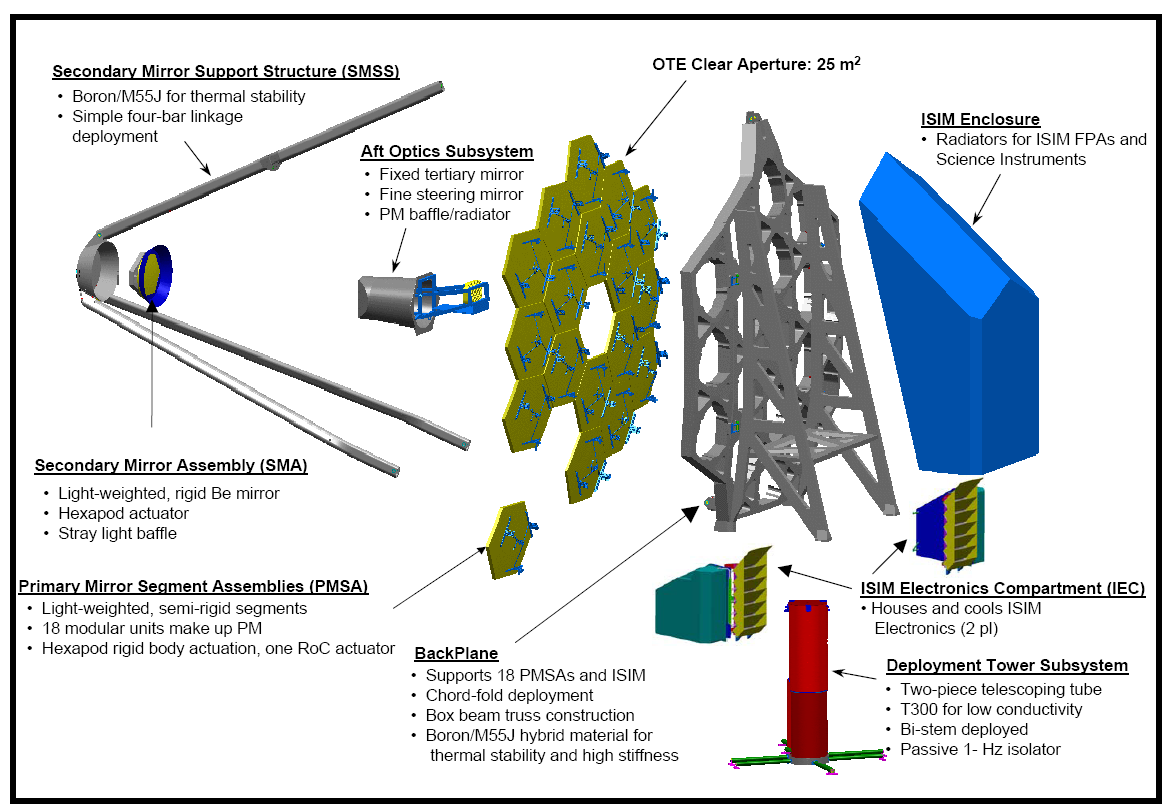}
    \caption{The JWST observatory and its optical telescope}
    \label{fig_JWST_observatory_separated}
  \end{center}
\end{figure}

The \textit{James Webb Space Telescope} (\textit{JWST}) is a large IR ($0.6 - 28\,\mu$m) telescope with a $6.5$~m primary mirror \citep[see][for a review of the JWST implementation]{Gardner2006}. The Fig.~\ref{fig_JWST_observatory_separated} shows the main components of the JWST observatory. The launch is planned in 2014, on an Ariane~5 rocket.  The JWST will be put in a Lissajous orbit around the Earth-Sun L2 point. 
Here are a few more figures just to say that it's an enormous project! The total mass of the observatory is $\sim 6\,200\,$kg. JWST is an international collaboration between NASA, the European Space Agency (ESA), and the Canadian Space Agency (CSA). The NASA Goddard Space Flight Center is managing the development effort and the prime contractor is Northrop Grumman. The Space Telescope Science Institute (STScI) will operate JWST after launch. The life-cycle cost of the project is estimated at about US\$4.5 billion, for a lifetime of 5 years minimum.

The JWST project started in 1989 (already 20 years ago!), when STScI decided to think about a follow-up to the \textit{Hubble Space Telescope} (HST).
Prior to September 10, 2002, the JWST was known as the Next Generation Space Telescope (NGST).
JWST was renamed to honor the NASA's second administrator, James Webb (1906-1992), who supervised the Gemini program, pathfinder to the preparation of the Apollo moon program. 

The JWST observatory is comprised of the following elements: the spacecraft, the optical telescope element (OTE), and the integrated science instrument module (ISIM). 

\subsection{The spacecraft}
\index{JWST!spacecraft}

The JWST spacecraft is comprised of two elements, the spacecraft bus and the sunshield (see the top right panel of Fig.~\ref{fig_JWST_observatory_separated}):
\begin{enumerate}
\item 
The sunshield subsystem separates the observatory into a warm sun-facing side (spacecraft bus) and a cold anti-sun side (OTE and ISIM). It enables passive cooling of telescope and instruments (the operating temperature has to be kept under 50 K). The sunshield consists of five layers of thin membranes made from a polymer-based film (Kapton with aluminum and doped-silicon coatings) and supporting equipment such as spreader bars, booms, cabling, and containment shells. When fully deployed, the sunshield that will be about the size of a tennis court!
\item
The spacecraft bus provides the support functions for the operation of the JWST Observatory. The bus houses the six major subsystems needed to operate the spacecraft: the Electrical Power Subsystem, the Attitude Control Subsystem, the Communication Subsystem, the Command and Data Handling Subsystem, the Propulsion Subsystem, and the Thermal Control Subsystem. 
\end{enumerate}

The JWST architecture, its expected performance and the plans for integration and testing are described  in more detail in \citet{Nella2004}. Even though many details in the design have changed since the article was written, many of the design choices and procedures are still valid.

\subsection{The eye of the JWST}
\index{JWST!mirror}
\index{JWST!fine steering mirror}
\begin{figure}
  \begin{center}
    \includegraphics[width=0.49\textwidth]{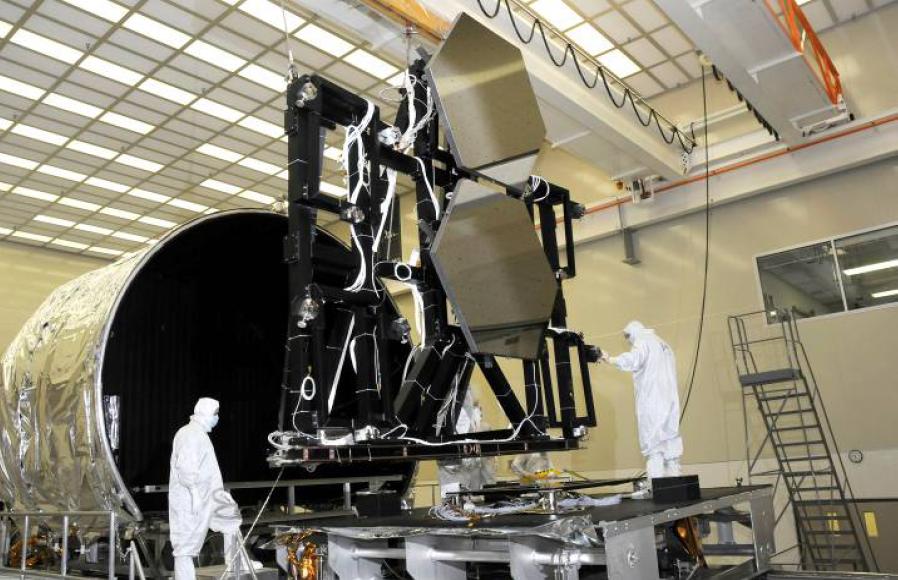}
    \includegraphics[width=0.49\textwidth]{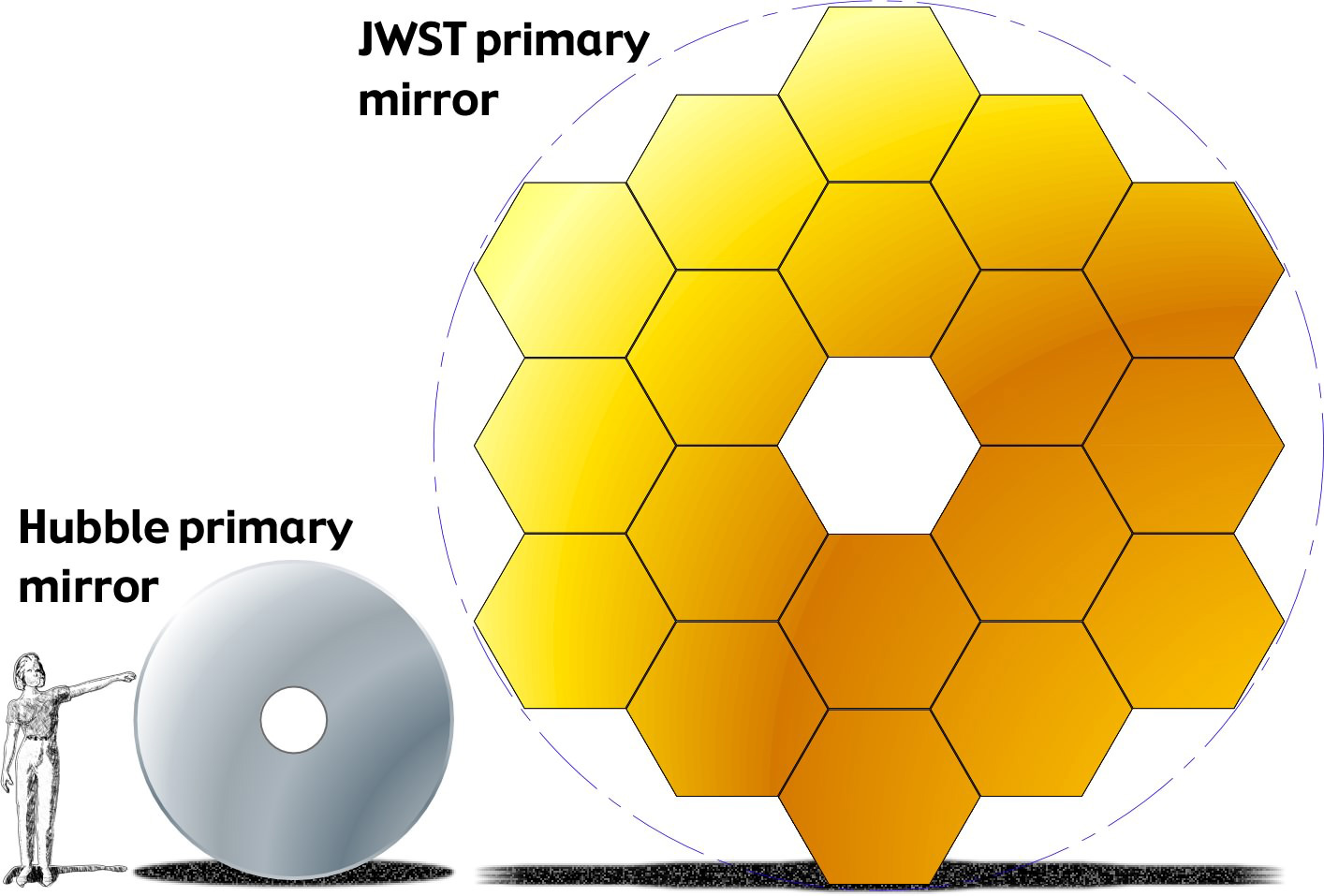}
    \includegraphics[width=0.5\textwidth]{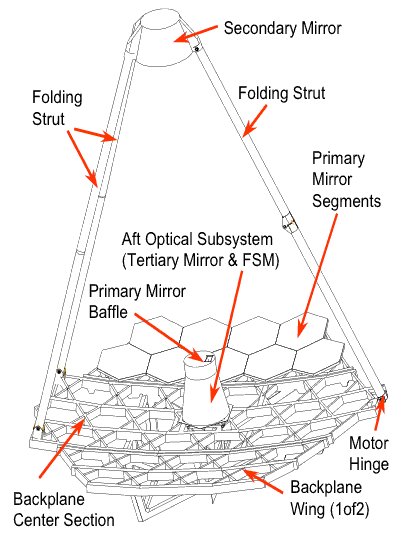}
    \caption[The ``eye'' of the JWST]{The ``eye'' of the JWST: a 6.5~m segmented mirror. \textit{top left:} cryogenic tests of the Flight Model mirrors (April 2009). \textit{top right:} comparison between \textit{HST} and \textit{JWST} mirrors ($2.4$ vs. $6.5$~m).  \textit{bottom:} scheme of the JWST optical telescope element. 
    Two primary mirror ''wings'' and a tripod structure supporting the secondary mirror are deployed during orbit insertion. A Fine Steering Mirror (FSM) provides accurate pointing, image stabilization ($< 7.3\,$milliarcsec) and dithering. }
    \label{fig_JWST_mirror}
  \end{center}
\end{figure}

The Fig.~\ref{fig_JWST_mirror} shows some views at the JWST mirror.
The JWST telescope has a three-mirror anastigmatic design, with a $25\,$m$^{2}$ collecting area (equivalent to a $\sim 6\,$m circular primary), which is $\sim 7$ times the collecting area of the \textit{HST}. The effective focal ratio is $f/16.67$, and the effective focal length is $131.4\,$m. 
The primary mirror consists of 18 hexagonal segments ($\sim 1.3\,$m flat-to-flat side), in two rings around the center, resulting in a 6.5~m flat-to-flat diameter. Each segment is made of Beryllium, and has the correct off-axis surface at the nominal cryogenic 40~K temperature of the primary. The mirrors are gold coated, providing a broad spectral bandpass, from 0.6 to 29$\,\mu$m.

\index{JWST!wavefront sensing}
The wavefront sensing and control subsystem is used to align the segments so that their wavefronts match properly, creating a diffraction-limited image. Each segment will have six actuators to change its focal length. The secondary mirror has 6-degrees of adjustment for collimation and overall focus. Once aligned (about $2-3$ months after launch), the telescope is designed to be diffraction limited at $2\,\mu$m (Strehl ratio of 0.80) and have an encircled energy of 75\% within a 0.15 arcsecond radius at $1\,\mu$m (dominated by sub-segment errors). It is estimated that the backplane will be sufficiently stable during slews that the wavefront control adjustments will be needed less than once every two weeks.

\subsection{Deployment and commissioning}
\index{JWST!deployment}

\begin{figure}
  \begin{center}
    \includegraphics[width=\textwidth]{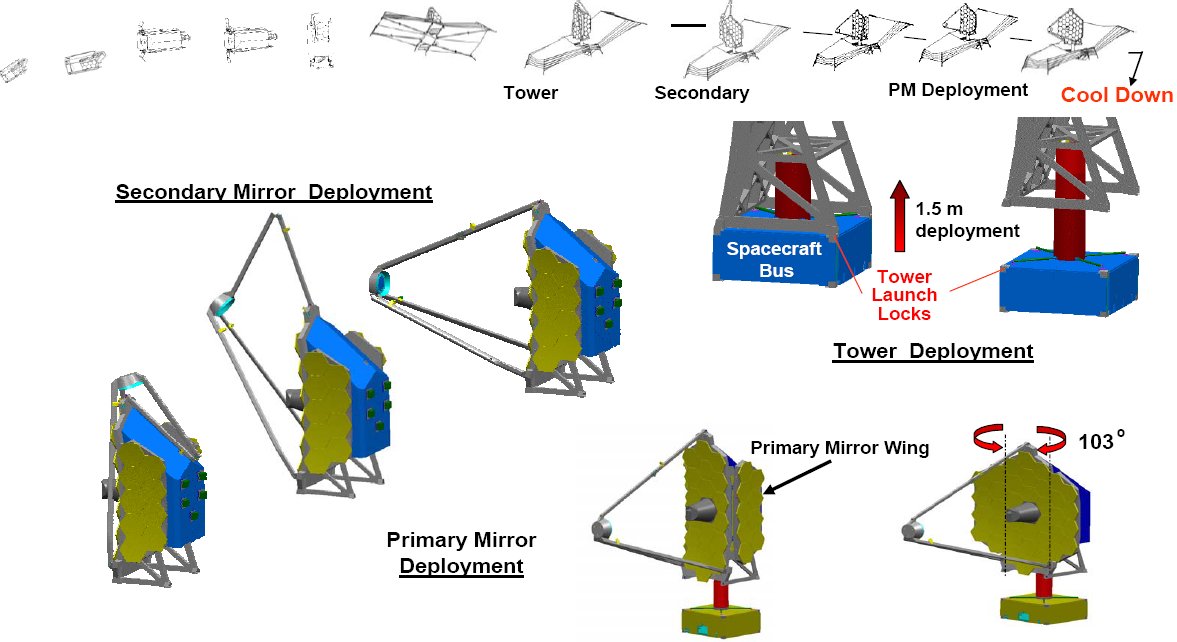}
    \caption[The ``joints'' of the JWST]{The ``joints'' of the JWST: deploying the sunshield and the optical telescope.}
    \label{fig_JWST_deploy}
  \end{center}
\end{figure}

A 6~m telescope does not fit into any launch rocket! The JWST is thus a deployable satellite (Fig.~\ref{fig_JWST_deploy}).
During transfer to L2 the different elements of JWST will be deployed and commissioning will start. The observatory has five deployment stages\footnote{\url{http://www.jwst.nasa.gov/videos_deploy.html}} involving the following elements: 1) spacecraft appendages (solar arrays, high gain antenna), 2) sunshield, 3) extend tower, 4) secondary mirror, and 5) primary mirror.  All critical deployment points have heaters and can be unlatched and re-latched to relieve any residual long term stress in the structure, even in the cold of outer space. 

Deployment of the observatory will start fairly soon after launch, with the sunshade unfolding after 2 days and the secondary and primary mirror deploying after approximately 4 days. First light will occur at about 28 days after launch, initiating wavefront sensing and control activities to align the mirror segments. Instrument checkout will start 37 days after launch, well before the final L2 orbit insertion is obtained after 106 days. Hereafter the full commissioning starts and the observatory will be ready for normal science operations approximately 6 months after launch.

\subsection{Science instruments on-board JWST}
\label{subsec:JWST-science-instruments}
\index{JWST!instruments}

\begin{figure}
  \begin{center}
   \includegraphics[width=\textwidth]{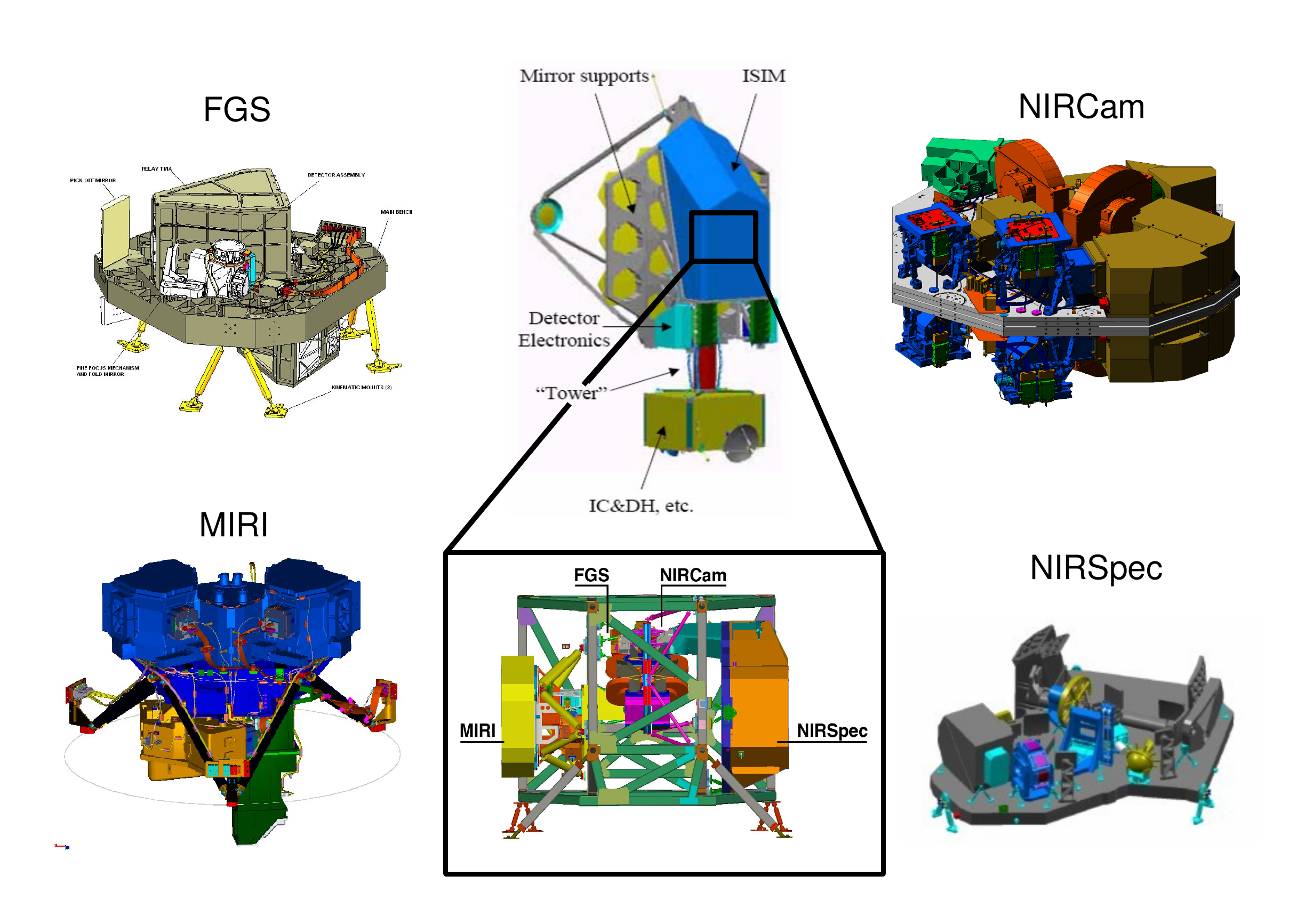}
    \caption[The ``heart'' of the JWST]{The ``heart'' of the JWST: its four science instruments. \textit{From top left, clockwise:} The Fine Guidance Sensor and Tunable Filter Imager (\textit{FGS-TFI}), the Near-Infrared Camera (\textit{NIRCam}), The Near-Infrared Spectrograph (\textit{NIRSpec}) and the Mid-Infrared Instrument (\textit{MIRI}).  The four instruments are integrated into the \textit{ISIM} (Integrated Science Instrument Module), the ``chassis'' that is supporting the instruments (blue cover behind the telescope, and zoom inset). }
    \label{fig_JWST_4instruments}
  \end{center}
\end{figure}

\begin{figure}
  \begin{center}
   \includegraphics[width=\textwidth]{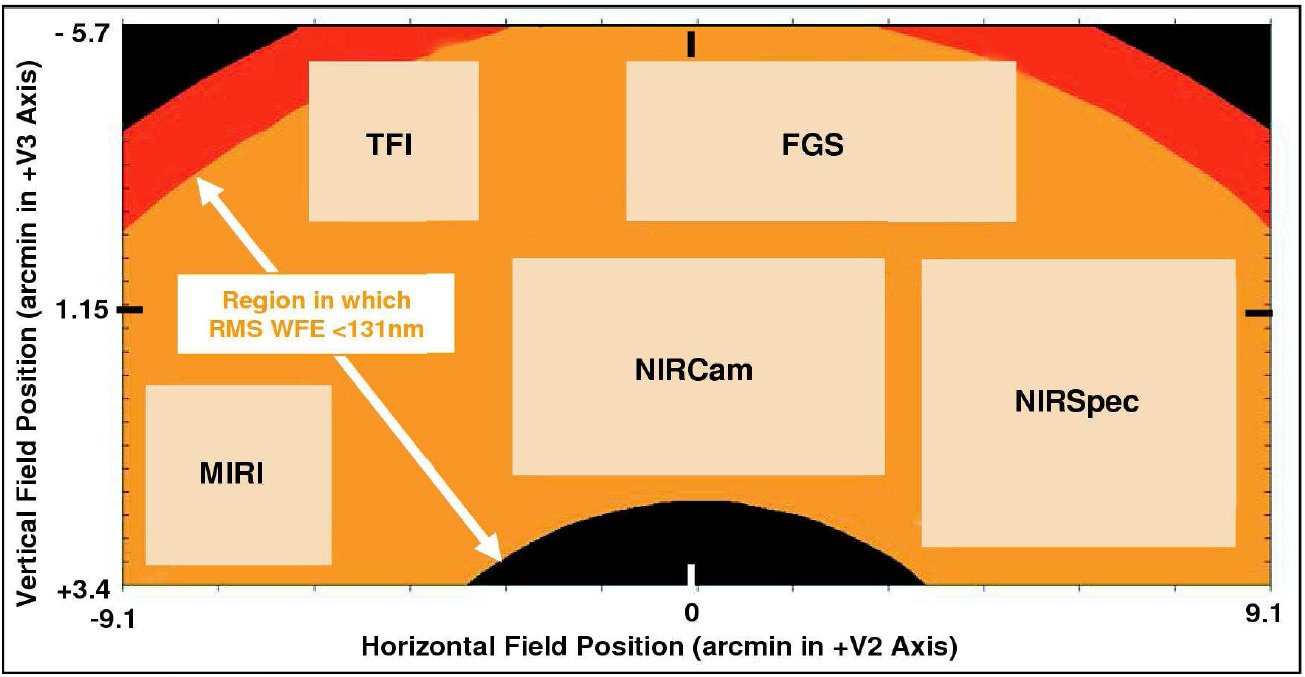}
    \caption[The JWST focal plane]{The JWST focal plane. The placement of the science instruments in the telescope field of view (FOV) is shown. Instruments share a $\sim 166\,$arcmin$^{2}$ FOV. The orange area shows where the wavefront error (WFE), that results from residuals in the optical design, is smaller than 131 nm rms. This is required for the NIRCam field, to be diffraction-limited  at 2$\,\mu$m. The outer black area in the figure represents portions of the field of view that have some amount of vignetting.}
    \label{fig_JWST_FOV}
  \end{center}
\end{figure}

Fig.~\ref{fig_JWST_4instruments} shows the four science instruments gathered into the integrated science instrument module (\textit{ISIM}):

\begin{description}
\item[\textbf{NIRCam: }] A Near-Infrared Camera ($0.6 - 5\,\mu$m), provided by the University of Arizona. It consists of two broad- and intermediate-band imaging modules, each with a $2.16 \times 2.16$ arcmin field of view. The modules have a short and a long wavelength channel, taking images simultaneously with light split by a dichroic at about $2.35 \,\mu$m. The short wavelength channel ($0.6 - 2.3\,\mu$m) is  sampled at $4096 \times 4096$ pixels (0.0317 arcsec/pixel), the long wavelength channels by $2048 \times 2048$ pixels (0.0648 arcsec/pixel). The short and long wavelength arms are Nyquist sampled at 2 and 4$\,\mu$m respectively. \textit{NIRCam} has ten mercury-cadmium-telluride (HgCdTe) detector arrays, analogous to CCDs found in ordinary digital cameras.
Coronagraphs are available in both modules. At the same time it serves as a wavefront sensor for the observatory, measuring the alignment of the 18 mirror pieces of the 6.5m primary mirror.

\item[\textbf{NIRSpec: }] A Near-Infrared Spectrograph ($0.6 - 5\,\mu$m), mainly provided by the European Space Agency,  under an industrial contract with Astrium.  \textit{NIRSpec} provides 3 observing modes: a low resolution $\mathcal{R} \sim 100$ prism mode, an $\mathcal{R} \sim 1000$ multi-object mode and an $\mathcal{R} \sim 3000$ integral field unit or long-slit spectroscopy mode. 

In the $\mathcal{R} \sim 100$ and $\mathcal{R} \sim 1000$ modes, \textit{NIRSpec} uses a micro-electromechanical system (MEMS) to provide dynamic aperture shutter masks ("microshutter array"). These masks allow to obtain simultaneous spectra of more than 100 objects in a $>9$ arcmin$^{2}$ field of view ($\sim 3.4 \times 3.4$~arcmin$^{2}$ at 1~$\mu$m). At $\mathcal{R} \sim 100$ one prism spectrum covers the full $0.6 - 5 \,\mu$m wavelength range. At $\mathcal{R} \sim 1000$, three gratings cover $1 - 5 \,\mu$m. \textit{NIRSpec} detectors are two HgCdTe arrays. 

\item[\textbf{MIRI: }]  A Mid-InfraRed Instrument, developed by a European consortium of more than 20 scientific institutes, nationally funded, with IR detectors and the active cooling system provided by NASA/JPL.
MIRI  provides imaging and spectroscopy over the $5-27\,\mu$m wavelength range. Its design consists of two main modules, an imager and an Integral Field Unit (IFU) medium resolution spectrograph (MRS).
Section~\ref{sec:MIRI_description} describes the instrument in more details.

\item[\textbf{FGS-TFI: }] 
The Fine Guidance Sensor, provided by the Canadian Space Agency, contains a dedicated Guider and a Tunable Filter Camera. The guider is a sensitive camera that provides support for the observatory's attitude control system (ACS). It is used for both "guide star" acquisition and fine pointing. The sensor operates over a wavelength range of 1 to 5$\,\mu$m and has two HgCdTe detector arrays. 
The camera can image two adjacent fields of view, each approximately $2.4 \times 2.4$ arcmin$^{2}$ in size.

A science instrument known as the FGS Tunable Filter Imager (FGS-TFI) is packaged with the guide camera, but is functionally independent. The FGS-TFI will be used solely for science observations. It has a selectable band of resolution $\mathcal{R} \sim 100$ across its $2.2 \times 2.2$ arcmin$^{2}$ field between $1.5-5\,\mu$m, with a gap between 2.6 and 3.1 micrometers. Tunable Fabry-Perot etalons illuminate the detector array with a single order of interference at a user-selected wavelength. The camera has a single HgCdTe detector array.
\end{description}

\begin{figure}
\centering
\includegraphics[height=\textwidth, angle=90]{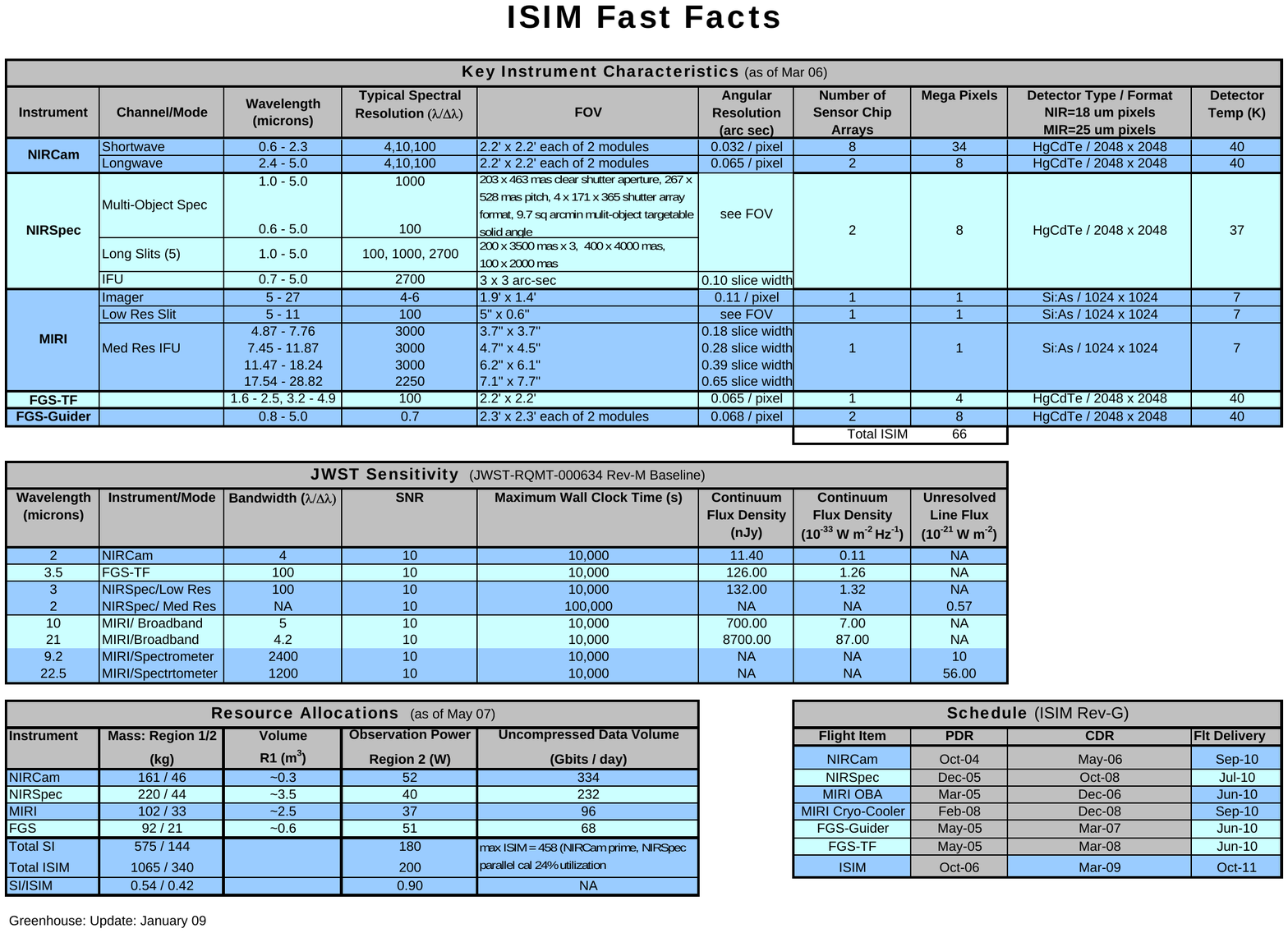}
\caption[Summary of the JWST instruments facts]{Summary of the main characteristics and sensitivities of the JWST instruments.}
 \label{fig_JWST_all_instruments_facts}
\end{figure}

\section{The Mid-Infrared Instrument (MIRI)}
\label{sec:MIRI_description}
\index{JWST!MIRI}

\subsection{Overview of MIRI}

\begin{figure}
  \begin{center}
    \includegraphics[width=\textwidth]{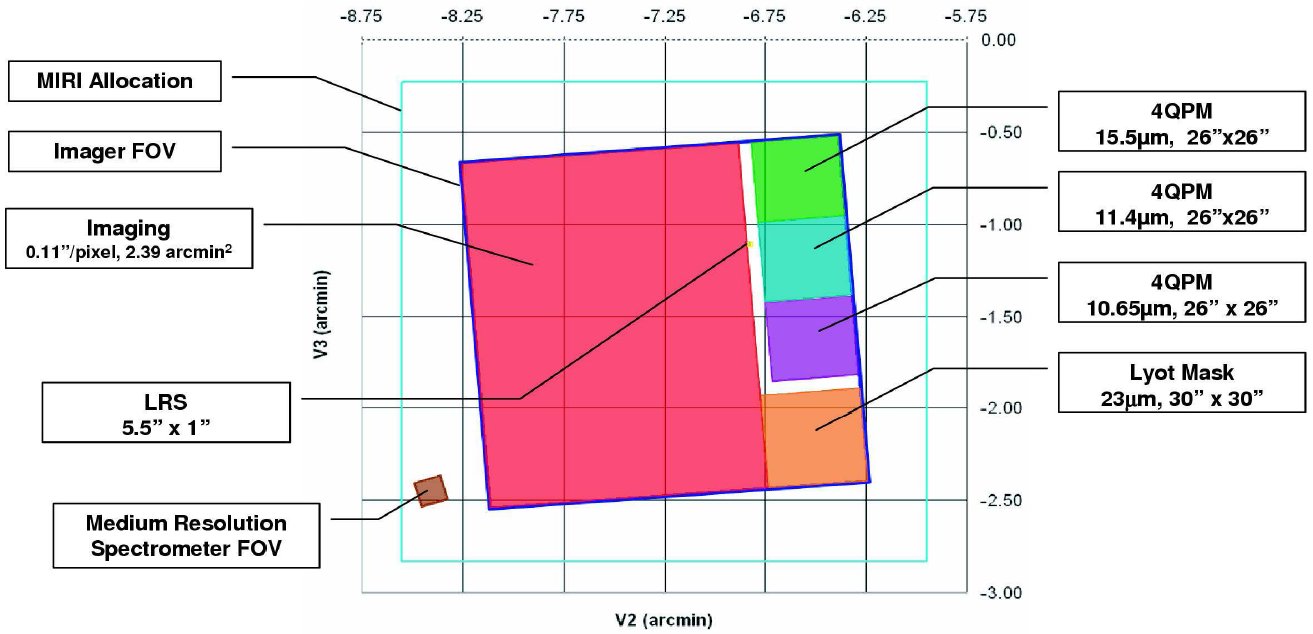}
    \caption[Schematic view of the MIRI's focal plane]{Schematic view of the MIRI's focal plane. 
The biggest $1.9\times  1.9$ arcmin$^{2}$ square is the imager field of view. 
One quarter of the field of view is devoted to coronagraphy (with three 4-Quadrant Phase Mask and one Lyot Mask coronographs), a 5 arcsec wide strip is used for low-resolution spectroscopy (\textit{LRS}), and the remaining $1.4\times  1.9$ arcmin$^{2}$ is available for broad-band imaging. The smallest square on the bottom left is the field of view of the medium resolution spectrometer.  \citep[From][]{Gardner2006}.}
    \label{fig_MIRI_FOV}
  \end{center}
\end{figure}

Fig.~\ref{fig_MIRI_FOV} shows the whole field of view (FOV) of the Mid-Infrared Instrument (MIRI) focal plane. 
MIRI is comprised of an imager (henceforth MIRIM) and an IFU spectrograph (MRS, \S~\ref{subsec:MIRI_MRS}) that cover the wavelength range $\sim 5-28\,\mu$m. 
MIRIM features three observing modes plus an alignment check mode:
\begin{description}
\item
[Imaging mode:] the  camera is equiped with a filter wheel, where broad-band filters are mounted (\S~\ref{subsec:MIRIcamera}).
\item
[Coronagraphic mode:] one Lyot mask ($23\,\mu$m) and three 4-Quadrant Phase-Masks (4QPM at 10.65, 11.4, 15.5$\,\mu$m) are fixed in a sub-area of the focal plane. The 4 corresponding pupil diaphragms are located in the filter wheel. 
The Lyot mask is a classical coronograph with an occulting disk that blocks the direct starlight.
The 4QPM coronagraph is based the peculiar design of binary phase mask (0, $\pi$) dividing the full field of view at the focal plane in four quadrants. The mutual destructive interferences of the coherent light of a source, perfectly centered on the mask, produce a total nulling within the pupil image \citep[see e.g.][for details]{Rouan2000}. The coronographic mode of MIRI will not be further discussed here.
\item
[Low-Resolution Spectrometer:] The slit is fixed in the focal plane and the dispersive element (a double prism) in the filter wheel  (\textit{LRS}, \S~\ref{subsec:MIRI_LRS}).
\item
[Alignment check mode:] a tool to check the alignment of MIRIM with respect to the telescope. It includes a fused silica lens in the filter wheel to defocus the image of an on-axis point source on the detector.
\end{description}

\begin{figure}
  \begin{center}
    \includegraphics[width=\textwidth]{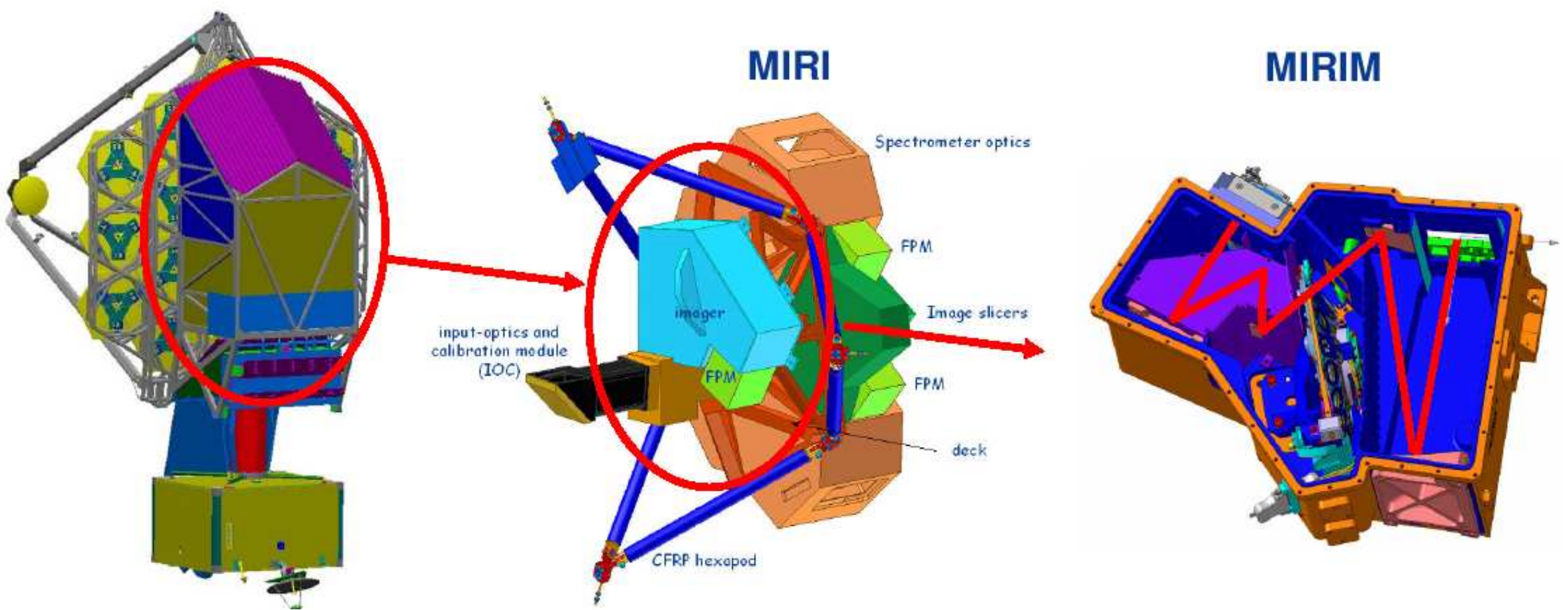}
 \includegraphics[width=\textwidth]{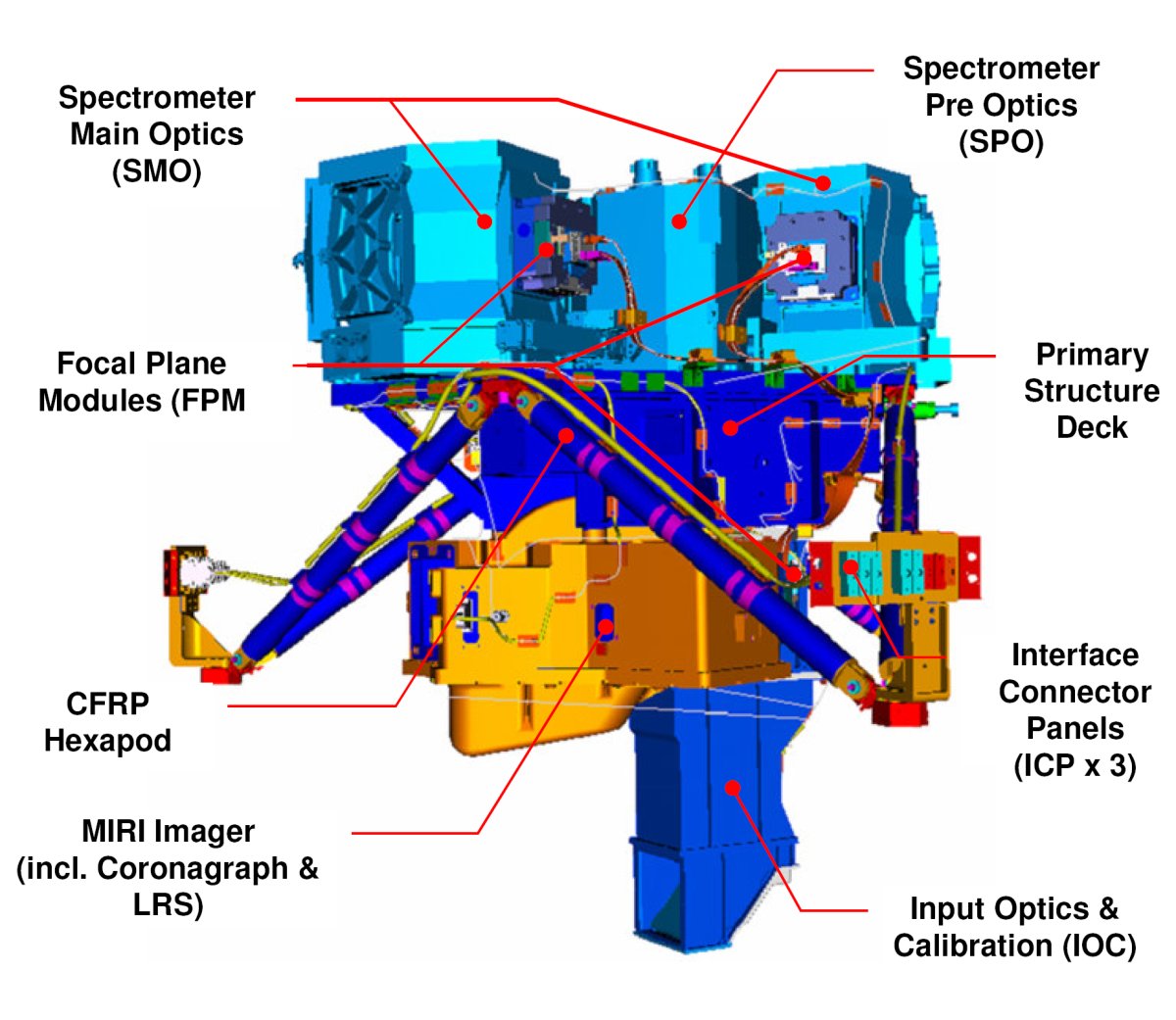}
    \caption[MIRI as part of the ISIM and MIRI mechanical layout]{\textit{Top panel, from left to right:} The Mid Infra Red Instrument (MIRI) as part of the scientific payload of the JWST, the MIRI Optical Bench Assembly (MIRI-OBA), with the imager in blue, and the MIRIM (Mid Infra Red IMager) mechanical layout, with the filter wheel assembly (FWA). \textit{Bottom:} MIRI's structural concept. The total weigth of MIRI is $\approx 95$~kg and its dimensions $\approx 0.9 \times 1 \times 0.8$~m.}
    \label{fig_MIRI_payload_design}
  \end{center}
\end{figure}

Fig.~\ref{fig_MIRI_payload_design} shows the whole Mid-Infrared Instrument (MIRI) and its structural design. 
The MIRI structure design uses a six-strut (hexapod) mount that provide the degrees of freedom for kinematic mounting to the ISIM. With the exception of the struts and some filters and prisms, the instrument is made entirely from 6061 aluminium alloy.
The nominal operating temperature for the MIRI is 7K. This level of cooling cannot be attained using the passive cooling. Instead, a cryocooler is used. First, a pulse tube precooler gets the instrument down to 18K, and then a Joule-Thomson loop heat exchanger cools it down to 7K. The cryocooler compressor assembly and control electronics are housed in the Spacecraft Bus (see Fig.~\ref{fig_JWST_observatory_separated}).

\subsection{MIRIM: broad-band imaging and coronography}
 \label{subsec:MIRIcamera}
\index{MIRI!imager}

\begin{figure}
  \begin{center}
     \includegraphics[width=0.33\textwidth]{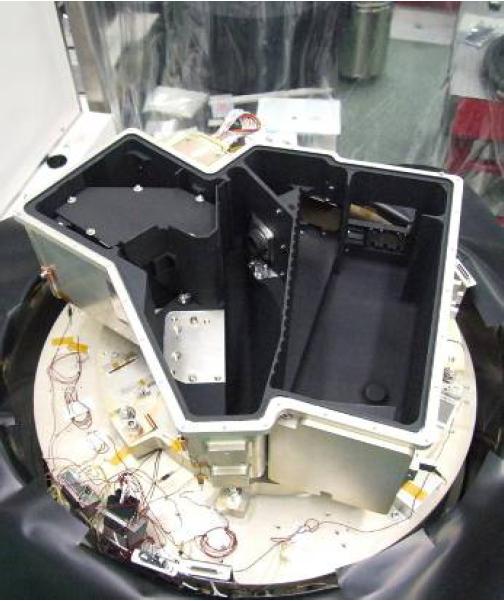}
     \includegraphics[width=0.3\textwidth]{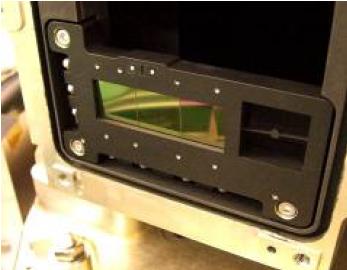}
    \includegraphics[width=0.27\textwidth]{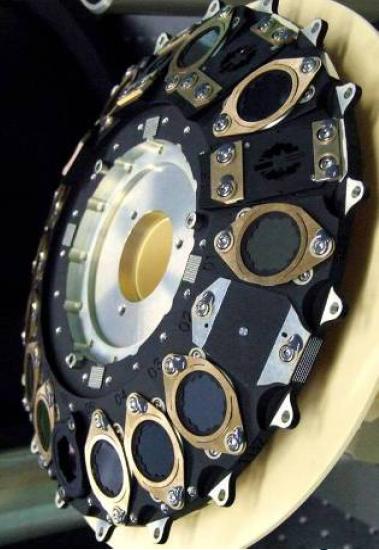}
    \caption[MIRIM open box and its the filter wheel]{\textit{Left:} the MIRIM open box. \textit{Middle:} the focal plane, with the four coronograph masks (one Lyot and three 4QPMs). \textit{Right:} the filter wheel, providing the broad-band filters for imaging, the coronograph filters, the double prisms for the LRS, and the alignment lens.}
    \label{fig_MIRI_photos}
  \end{center}
\end{figure}

\begin{figure}
  \begin{center}
    \includegraphics[width=\textwidth]{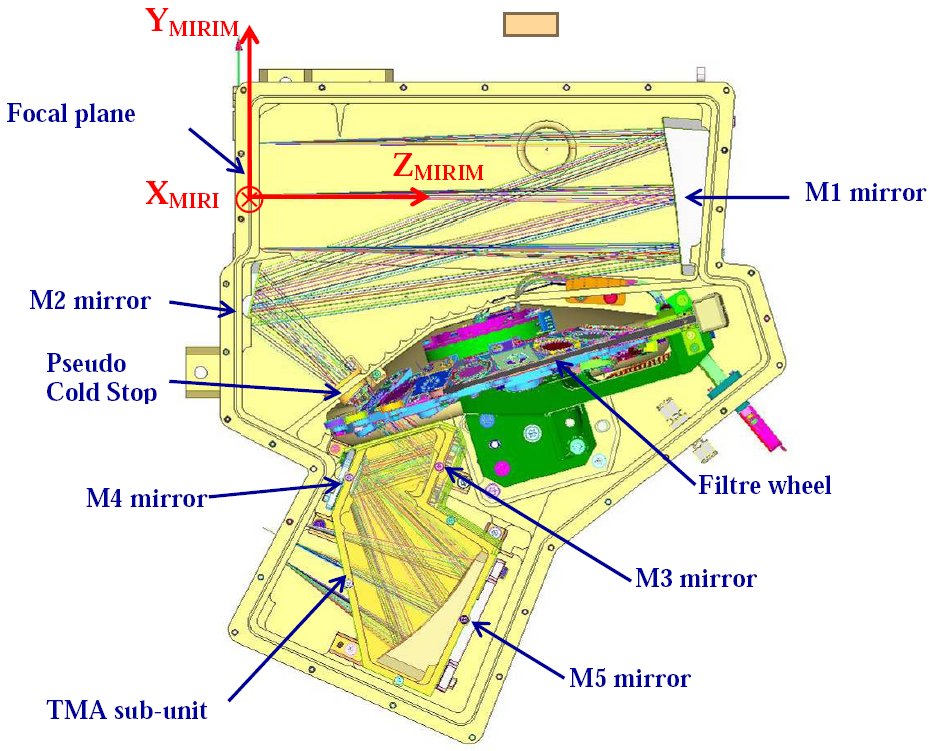}
    \includegraphics[width=0.5\textwidth]{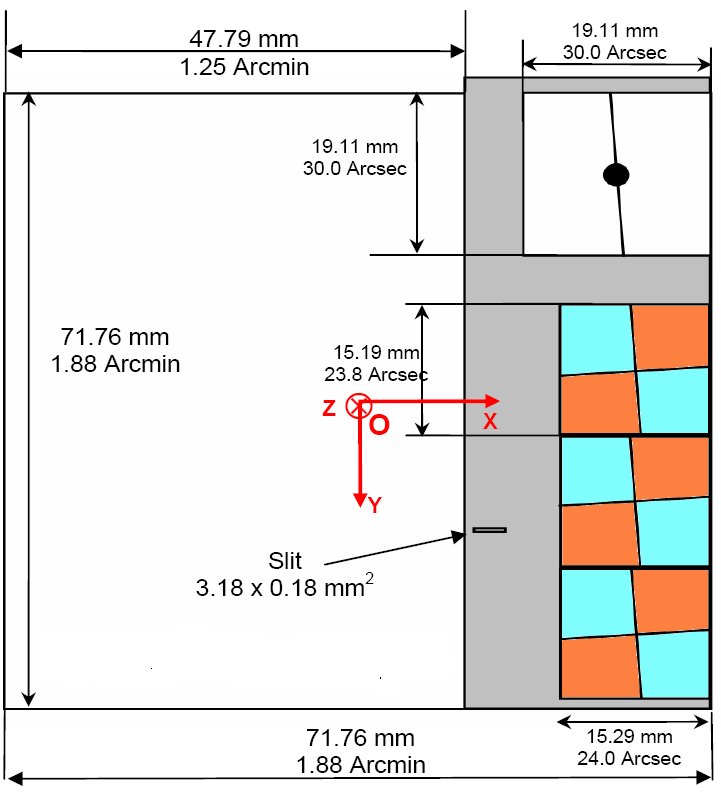}
    \caption[MIRI's mechanical and optical layout with its reference frame]{\textit{top:} MIRI's mechanical and optical layout with its reference frame. \textit{Bottom:} schematic view of the MIRIM (imager) focal plane, showing the position of the imaging field, the LRS slit, and the coronograph masks (Lyot and the three 4QPM masks).}
    \label{fig_MIRI_layout}
  \end{center}
\end{figure}

\begin{table}
\begin{center}
\begin{minipage}{\textwidth}
 \renewcommand{\footnoterule}{}
\def\thefootnote{\alph{footnote}}
 \caption[MIRIM filters and components of the filter wheel]{MIRIM filters\footnotemark[1] and components of the filter wheel}
\centering
\begin{tabular}{c c c c c c}
\hline
\hline
Filter & $\lambda _0$ [$\mu$m] & $\Delta \lambda$ [$\mu$m]  & FoV [arcsec$^{2}$] & Material & Comment \\
\hline
F560W   & 5.6 & 1.2 & \multirow{10}*{\begin{sideways}$84 \times 113$\end{sideways}} & CaF$_2$ & Broad band \\
F770W   & 7.7 & 2.2 &  & CaF$_2$ & PAH, broad band \\
F1000W & 10.0 & 2.0 &  & ZnSe & Silicate, broad band \\
F1130W & 11.3 & 0.7 &  & ZnSe & PAH, broad band \\
F1280W & 12.8 & 2.4 &  & ZnSe & Broad band \\
F1500W & 15.0 & 3.0 &  & ZnSe & Broad band \\
F1800W & 18.0 & 3.0 &  & CdTe & Silicate, broad band \\
F2100W & 21.0 & 5.0 &  & CdTe & Broad band \\
F2550W & 25.5 & 4.0 &  & CdTe & Broad band  \\
F2550WR & 25.5 & 4.0 &  &CdTe  & Redundant filter \footnotemark[2] \\
\hline
F1065C & 10.65 & 0.53 & $24 \times 24$ & ZnSe & 4QPM Coronagraph  \\
F1140C & 11.4 & 0.57 & $24 \times 24$ & ZnSe  & 4QPM Coronagraph   \\
F1550C & 15.5 & 0.78 &$24 \times 24$ & ZnSe  & 4QPM Coronagraph   \\
F2300C & 23.0 & 4.6 &$30 \times 30$  & CdTe  & Lyot Coronagraph \\
\hline
DPA & $5-11$ & & $5 \times 0.28$ & Ge, ZnS & Double Prism (LRS) \\
\hline
FLENS & & & & Fused silica & Alignment lens \\
OPAQUE & & & & & Closed position (dark) \\
FND  & & & & & Neutral density \\
\hline
\end{tabular}
\footnotetext[1]{The filter name, central wavelength, bandwidth, Field of View are presented.}
\footnotetext[2]{Additional $25.5\,\mu$m spare filter to mitigate risks associated with the technological challenge of manufacturing this element.}
\label{table_MIRIM_filters}
\end{minipage}
\end{center}
\end{table}

\index{MIRI!filters}
\begin{figure}
  \begin{center}
    \includegraphics[width=\textwidth]{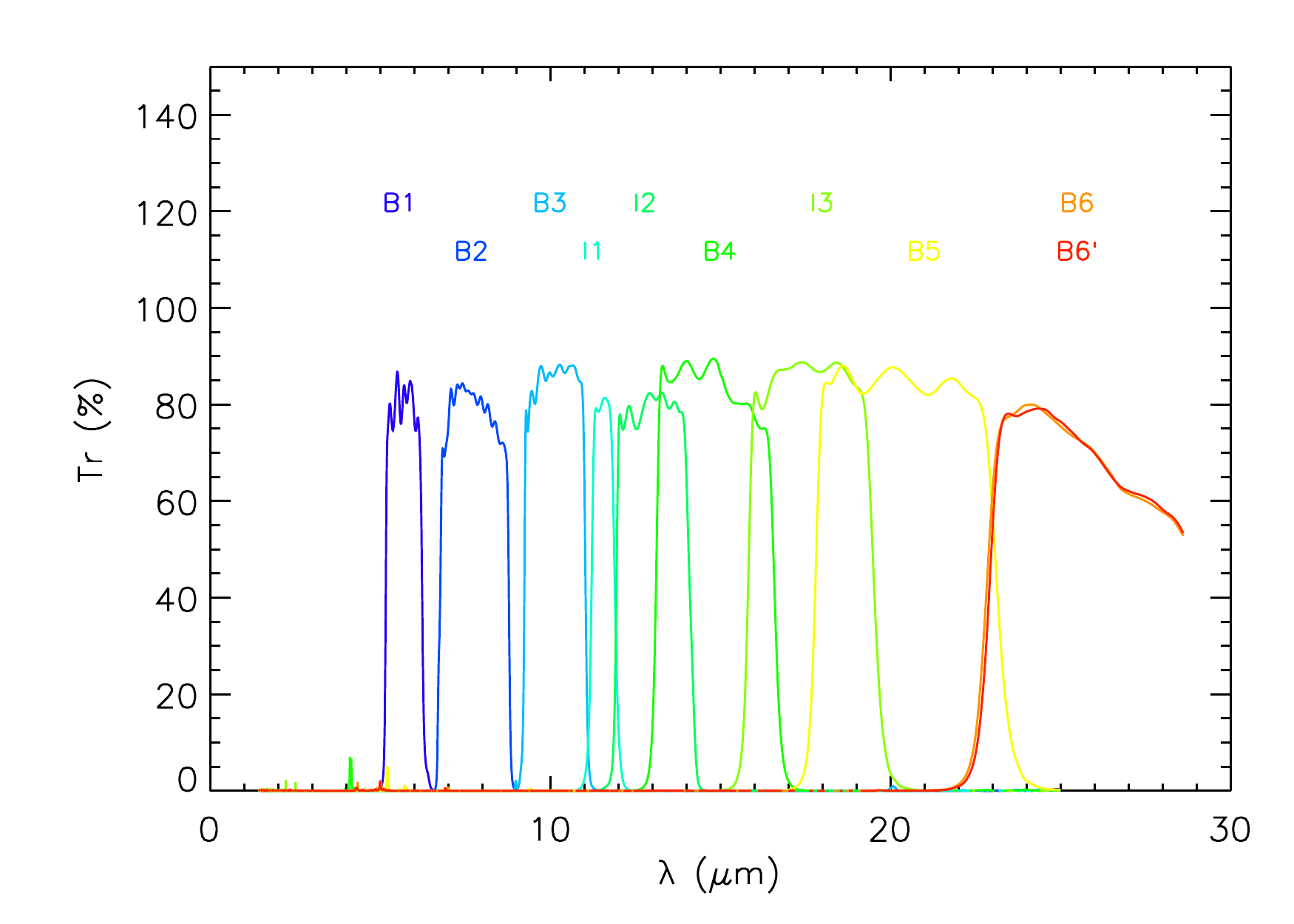}
    \caption[Spectral response of the MIRI broad-band filters]{Spectral response of the MIRI broad-band filters, measured by K. Justtanont, Onsala Observatory.}
    \label{fig_MIRI_filters}
  \end{center}
\end{figure}

\paragraph{Optical layout}
We briefly describe the optical layout of the camera, shown in Fig.~\ref{fig_MIRI_layout}. The  footprints of the beams for 3 different off-axis fields are overlaid on the mechanical layout. 
The JWST input beam enters MIRIM's focal plane via the Input Optics and Calibration unit (IOC, see Fig.~\ref{fig_MIRI_payload_design}). The cold MIRIM FOV is $72.76\,\rm mm \times 72.76 mm$ at focal plane, where the coronograph masks are fixed. M1 is a collimating mirror (focal length = 352 mm), and M2 is a flat folding mirror that directs the beam to the cold stop and the filter wheel. The cold stop is a diaphragm required at the OTE pupil image location to reduce the amount of stray light. Because the shape of the diaphragms used for imaging or coronography are different, the use of a single cold stop for all the modes is not possible. Thus, a ``pseudo cold stop'' is located 10 mm along the optical axis before the OTE pupil image. The diaphragms (mean diameter value around 19 mm), adapted for each mode and attached to the filter wheel, are located exactly at the OTE pupil image position.
The M3, M4 and M5 mirrors form a Three-Mirror Anastigmatic (TMA) objective that reflects the light on the detector array.

\paragraph{Imager and coronographic filters}
The useful area for the imaging mode is $47.79 \times 71.76$~mm$^{2}$, corresponding to $1.25 \times 1.88 = 2.35$~arcmin$^{2}$ on the sky (see Fig.~\ref{fig_MIRI_layout}, bottom). The imager will have a pixel scale of 0.11 arcsec/pixel. As the coronagraphic masks are transparent, the area used for the coronagraphic mode could be marginally used in the imaging mode. 
The only moving part of the imager is an 18-position filter wheel. Filter positions are allocated as follows: 12 filters for imaging, 4 filter and diaphragm combinations for coronagraphy, 1 ZnS-Ge double prism for the low-resolution spectroscopic mode, and 1 dark position. The list and properties of the MIRIM filters are given in table~\ref{table_MIRIM_filters}. The transmission curves of the imager filters are plotted in Fig.~\ref{fig_MIRI_filters}. The coronographic mode will not be detailed here, and we direct the reader to \citet{Boccaletti2005} for details.

\paragraph{Calibration}
A calibration facility is dispatched over the IOC and MIRIM. The calibration is achieved through a dedicated optical path, by imaging a point source at the internal edge of the OTE pupil image. The light is reflected through the TMA to illuminate uniformly the detector.

\subsection{MIRIM: the Low-Resolution Spectrometer (LRS)}
\index{MIRI!low-resolution spectrometer}
 \label{subsec:MIRI_LRS}

The low resolution spectroscopic mode is achieved by a  couple of Ge and ZnS prisms,
attached to the filter wheel. The first Ge prism cancels the deviation due to the ZnS
dispersive prism.
The slit is located at the edge of the coronagraphic mask mounting (see Fig.~\ref{fig_MIRI_layout}, bottom).
The spectrum ($5-10\,\mu$m) is spread over $\sim 3.5$~mm over the detector array, so $\sim 140$~pixels. The nominal resolving power at $7.5\,\mu$m is $\mathcal{R} = 98.5 \pm 6.0$.

\subsection{The Medium-Resolution Spectrometer (\textit{MRS})}
\index{MIRI!medium-resolution spectrometer}
 \label{subsec:MIRI_MRS}

\begin{figure}
  \begin{center}
   \includegraphics[width=\textwidth]{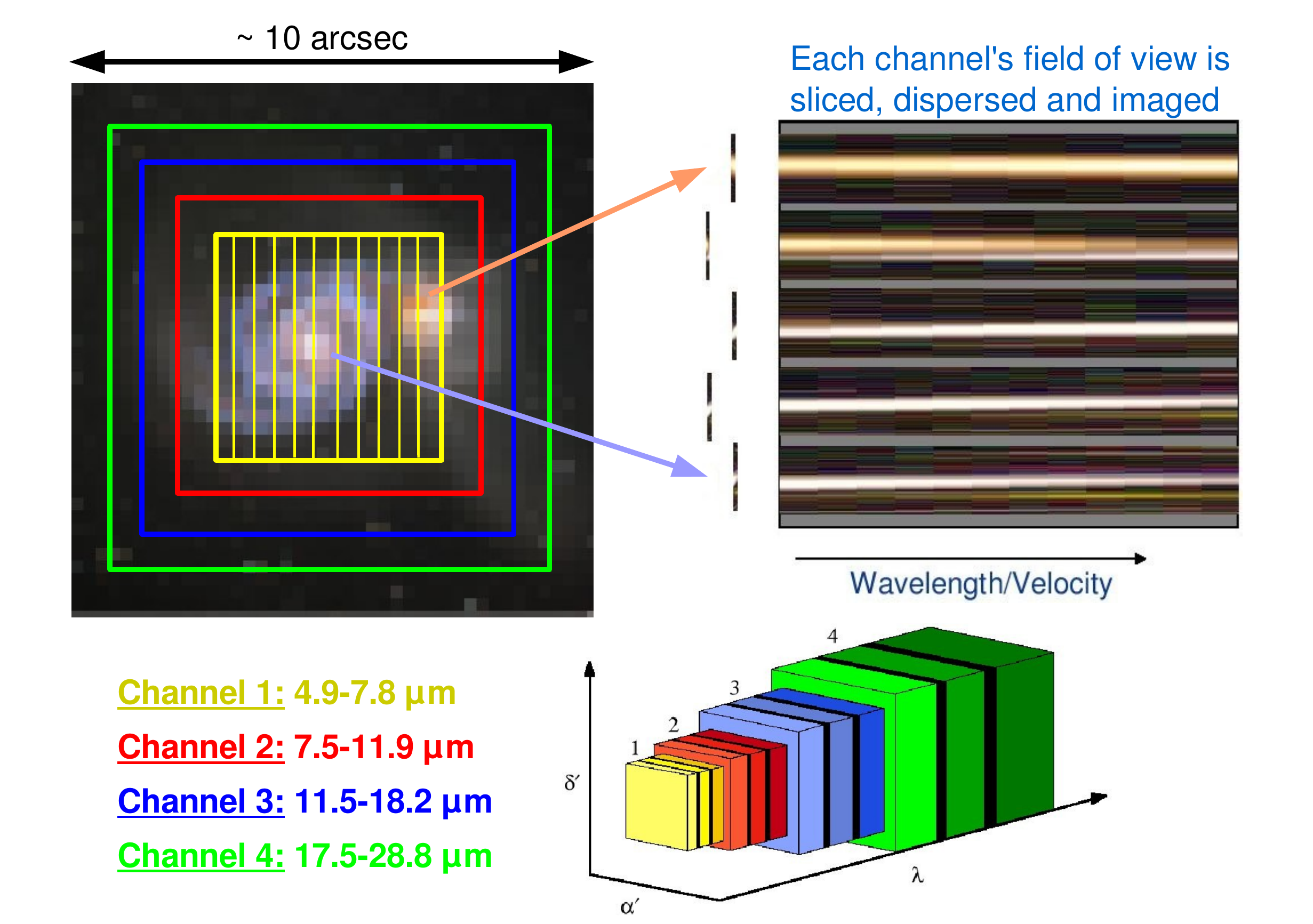}
    \caption[Sketch of the working principle of the MRS]{\textit{Top:} Sketch of the working principle of the MRS. \textit{Bottom right inset:} The field-of-view and bandwidth coverage (to scale) of MIRI-MRS channels 1 to 4 \citep[from][]{Lorente2006}. The gradation of colour from light to dark in each channel denotes the three exposures A, B and C. The black bands between channels are those
spectral regions where the wavelength coverage of adjacent exposures overlap.}
    \label{fig_MRS_howitwork}
  \end{center}
\end{figure}

\begin{figure}
  \begin{center}
   \includegraphics[width=\textwidth]{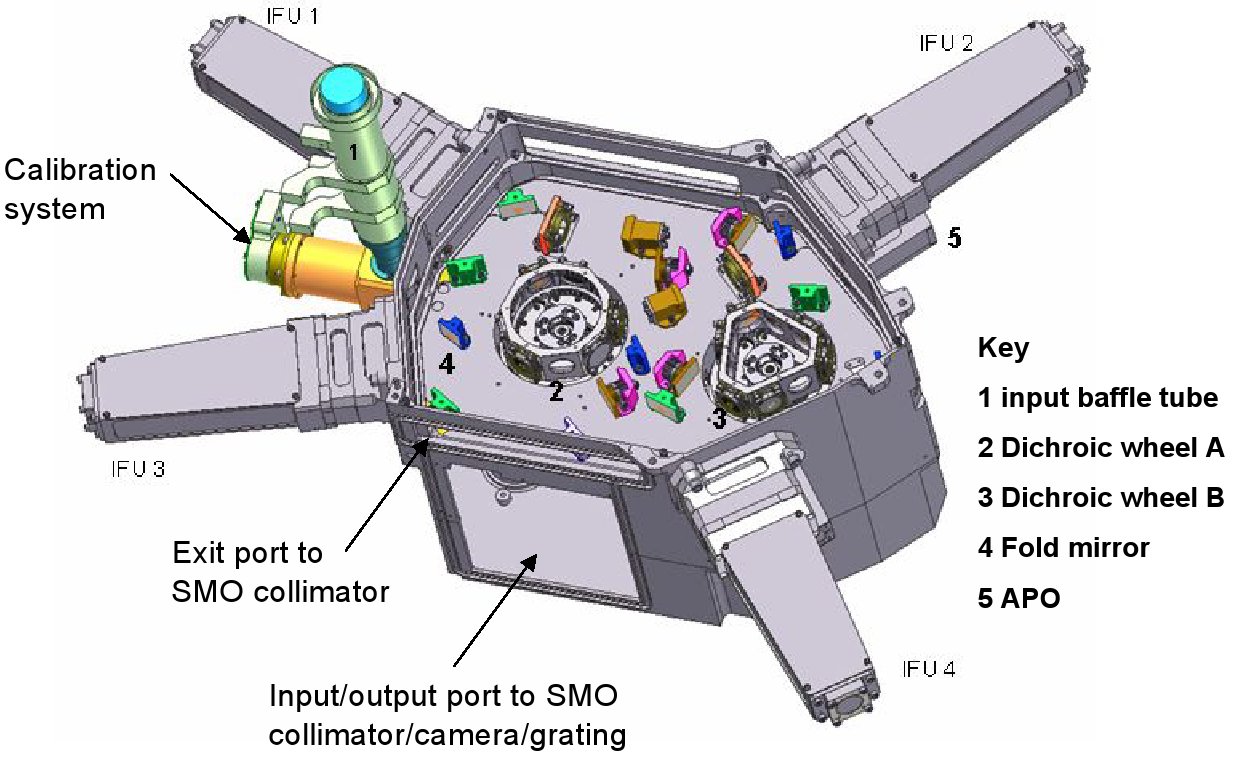}
\includegraphics[width=0.9\textwidth]{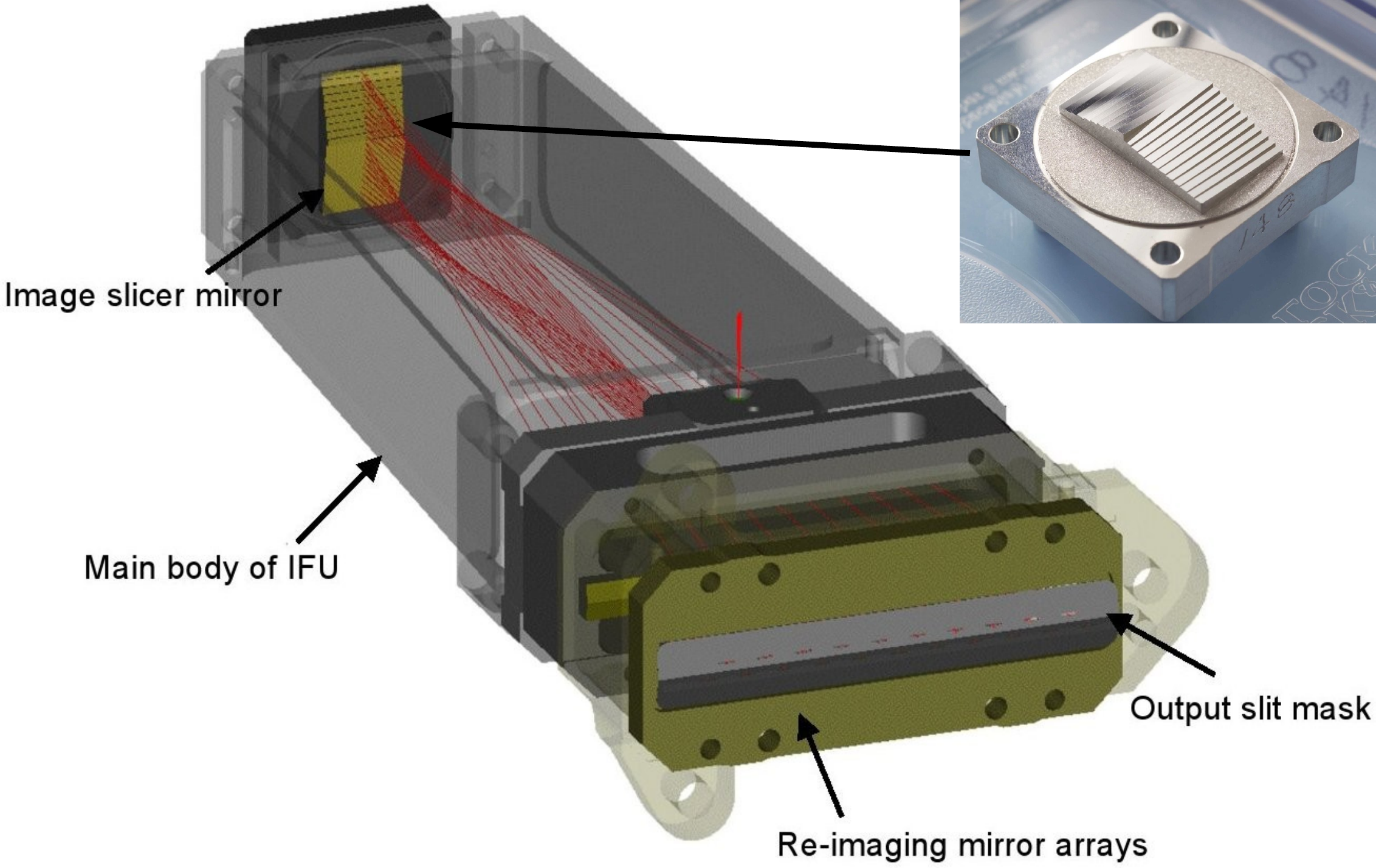}
    \caption[Opto-mechanical layout of the MRS and its IFUs]{\textit{Top:} the opto-mechanical layout of the Medium Resolution Spectrometer (MRS) Pre-Optics. The main beam (1) is separated into the 4 IFU channels. The dichroics sub-divide each channel into 3 sub-bands. \textit{Bottom:} the Integral Field Unit (IFU) and its image slicer mirror.}
    \label{fig_MRS_IFU_slicer}
  \end{center}
\end{figure}

\index{MIRI!IFU}
The Medium Resolution Spectrometer (MRS) is an Integral Field Unit (IFU) that allows spectroscopy to be carried out on a 2-dimensional area of sky in
a single observation. Fig.~\ref{fig_MRS_howitwork} illustrates its working principle and Fig.~\ref{fig_MRS_IFU_slicer} shows the optical and mechanical layout of the MRS. The rectangular field of view is divided into narrow slices by mean of a image-slicer mirror. The image slices are arranged along the entrance slit mask (bottom panel of Fig.~\ref{fig_MRS_IFU_slicer}) of a first-order diffraction grating, which carries out the dispersion, and imaged onto two $1024 \times 1024$ pixels Si:As chips.

The band ($5 - 29\,\mu$m) is divided into 4 IFU channels (Fig.~\ref{fig_MRS_IFU_slicer}), which are observed simultaneously. Each channel is equipped with an IFU image slicer designed to match the width of each slice to the diffraction-limited point-spread function of the telescope, at
the wavelength of each channel. The spectral band of each IFU channel is  subdivided into 3 sub-bands (A, B and C) by means of dichroic filters. Thus, an observation over the entire spectral band is carried out in a set of three
exposures.

\begin{table}
\begin{center}
\begin{minipage}[t]{\textwidth}
 \renewcommand{\footnoterule}{}
\def\thefootnote{\alph{footnote}}
\centering
 \caption{Summary of the MIRI-MRS parameters}
    \begin{tabular}{c c c c c c c }
	\hline
	\hline
  \multirow{5}*{\begin{sideways}Channel\end{sideways}} & FoV in a single & Spatial sample & \multirow{5}*{\begin{sideways}Slices\end{sideways}} &  &  \multirow{5}*{$\mathcal{R}_{\rm spectral}$} &  \multirow{5}*{\begin{sideways}Exposure\end{sideways}} \\
 & integration & dimensions &  &  & & \\
\cline{2-3}
 & \multirow{2}*{across$\ \times$\ along} & across$\ \times\ $along &  & $\lambda$ & & \\
 & & slice width$\ / \ $pixel  &  & & & \\
 & [arcsec$^{2}$] &  [arcsec$^{2}$] &  & $[\mu$m] & & \\
	\hline
\multirow{3}*{1} & \multirow{3}*{$3.70 \times 3.70$} & \multirow{3}*{$0.18 \times 0.20$} &  \multirow{3}*{21} & $4.87 - 5.82$ & $2450 - 3710$ & A \\
 & & & & $5.62 - 6.73$ &  $2450 - 3710$ &  B \\
 & & & & $6.49 - 7.76$ &  $2450 - 3710$ &  C \\
\hline
\multirow{3}*{2}  & \multirow{3}*{$ 4.51 \times 4.70$} & \multirow{3}*{$0.28 \times 0.20$} &  \multirow{3}*{ 17} & $7.45 - 8.90$ & $2480 - 3690$ & A \\
 & & & & $8.61 - 10.28$ &  $2480 - 3690$ & B \\
 & & & & $9.94 - 11.87$ & $2480 - 3690$ & C \\
\hline
\multirow{3}*{3} & \multirow{3}*{$6.13 \times 6.20$} & \multirow{3}*{$0.39 \times 0.25$} &  \multirow{3}*{16} & $11.47 - 13.67$ & $2510 - 3730$ & A \\
 & & & & $13.25 - 15.80$ & $2510 - 3730$ & B \\
 & & & & $15.30 - 18.24$ & $2510 - 3730$ & C \\
\hline
\multirow{3}*{4} & \multirow{3}*{$7.93\times 7.74$} & \multirow{3}*{$0.64 \times 0.27$} & \multirow{3}*{12} & $17.54 - 21.10$ & $2070 - 2490$ &  A \\
 & & & & $20.44 - 24.72$ & $2070 - 2490$ &  B \\
& & & & $23.84 - 28.82$ & $2070 -  2490$ &  C \\
\hline
\end{tabular}
    \label{table_MRS_param}
\end{minipage}
\end{center}
\end{table}

The table~\ref{table_MRS_param} gathers the main functional parameters of the MRS. The spectral resolution is $\mathcal{R} \sim 3000$ over the $5-29\,\mu$m wavelength range. The field of view ranges from $3.7 \times 3.7$~arcsec to $7.7 \times 7.7$~arcsec$^{2}$ with increasing wavelength, with pixel sizes  from 0.2 to 0.65 arcsec. Within this overall field of view, the region of sky corresponding to a spatial sample is set by the slice width in the perpendicular to the slice direction (across), and by the pixel field of view in the  slice direction (along). The slice widths have been chosen to have odd integer multiples of their half widths matched. The result is that a single telescope offset (of 0.97 arcseconds in the across slice direction and a multiple of 0.10 arcseconds in the along-slice direction) will provide full spatial sampling in all channels simultaneously.

\subsection{MIRI's detectors, chip readout and sub-arrays}
\index{MIRI!detectors}
 \label{subsec:MIRI_detectors}

\begin{figure}
  \begin{center}
   \includegraphics[width=\textwidth]{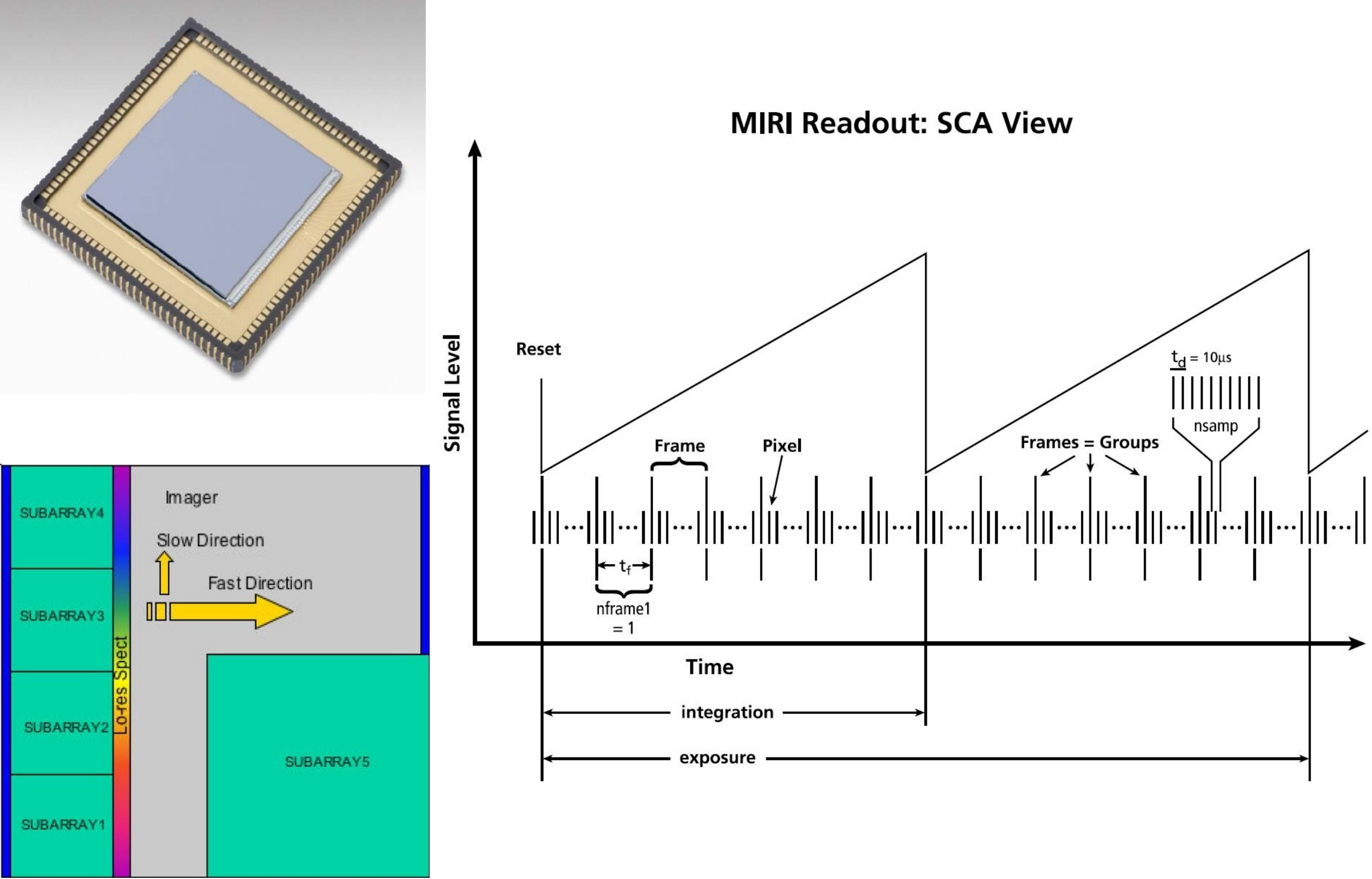}
    \caption[MIRI's detector, its sub-arrays and readout scheme]{\textit{Top left:} MIRI's detector: a SB305 $1024 \times 1024$ pixels Si:As Sensor Chip Array (SCA), delivered by JPL / Raytheon Vision Systems. \textit{Bottom left:} Pre-defined sub-arrays locations. \textit{Right:} the readout scheme of the MIRI SCA. }
    \label{fig_MIRI_detector}
  \end{center}
\end{figure}

MIRI has three $1024 \times 1024$ pixels Arsenic-doped Silicon (Si:As) sensor chip arrays (SCA), two for the MRS, one for the imager (Fig.~\ref{fig_MIRI_detector}, \textit{top left}). With $25\,\mu \rm m \times 25\mu m$ pixels, the detecting area in approximately 1 inch $\times$ 1 inch. The previous generation's similar IR detectors at these wavelengths ($5 - 28\,\mu$m) were the $256 \times 256$ pixels, $30\,\mu$m-pitch, Si:As devices built for the IRAC instrument on-board \textit{Spitzer}. The main characteristics of the MIRI detector are gathered in table~\ref{table_MIRI_detector}. We direct the reader to \citet{Love2005, Love2006} for more details.

\begin{table}
\begin{center}
\begin{minipage}[t]{\textwidth}
 \renewcommand{\footnoterule}{}
\def\thefootnote{\alph{footnote}}
\centering
 \caption[MIRI detector parameters]{MIRI detector parameters \footnotemark[1]}
    \begin{tabular}{c c c}
\hline
\hline
 \multicolumn{2}{c}{Parameter} & Measured value \\
\hline
\multicolumn{2}{c}{Format} & $1024 \times 1024$ pixels \\
\multicolumn{2}{c}{Material} & Si:As Inpurity Band Conduction \\
\multicolumn{2}{c}{Pixel size} & 25$\,\mu$m \\
\multicolumn{2}{c}{Well capacity} & $1-2\times 10^5$ e$^{-}$ \\
\multicolumn{2}{c}{Read-out noise} & $10-20$ e$^{-}$ @ 7.1 K \\
\multicolumn{2}{c}{Dark current} & $< 0.1$ e$^{-}$ s$^{-1}$ @ 7.1 K \\
\hline
\multirow{5}*{\begin{sideways}RQE \footnotemark[2]\end{sideways}}  & $5-6\,\mu$m & $\sim 50$\% \footnotemark[3] \\
& $\ 6-12\ \,\mu$m & $\sim 60$\% \footnotemark[3] \\
& $12-24\ \,\mu$m & $\sim 70$\% \footnotemark[3] \\
& $24-26\ \,\mu$m & $\sim 30$\% \footnotemark[3] \\
& $26-28.2\,\mu$m & $\sim 5$\% \footnotemark[3] \\
\hline
\end{tabular}
\label{table_MIRI_detector}
\footnotetext[1]{These figures are from \citet{Love2006} and should be close to the characteristics of the flight model detectors that were selected in 2008. }
\footnotetext[2]{Response Quantum Efficiency}
\footnotetext[3]{The efficiency of the detector is smaller than the intrisic quantum efficiency of the material because of the reflection of light on the Si layer. An anti-reflection coating is used to limit this loss.}
\end{minipage}
\end{center}
\end{table}

\index{MIRI!detector readout}
The right panel of Fig.~\ref{fig_MIRI_detector} shows the MIRI detector readout pattern, i.e. the signal on a MIRI pixel evolving with time. The fundamental unit of time is the pixel rate clock; all other times can be calculated from that time base. This time
between pixel samples, $t_{\rm d} = 10\,\mu$s, is set by the Focal Plane Electronics master 100 kHz clock. 
A ``frame'', the next unit of time, is the number of clock cycles needed to scan through the array, which represents
\begin{equation}
t_{\rm frame} = \rm (256\  pixels + 8\ for\  references + 5 \  for\  reset) \times (1024 \  rows) \times 10\,\mu s = 2.755\ s
\end{equation}
A ``group'' is a number of frames that are read and coadded together. MIRI does not generally use this
feature, so that there is only one frame per group.
An ``integration'' is the length of time the detector is allowed to collect light before resetting to zero, and is
defined to be an integer number of frame times in length. Single integrations may range from 3 seconds (2.755 to be precise) up to $4\,000$ seconds.
An ``exposure'' is the length of time MIRI is allowed to collect light before some intervention is needed, 
e.g. dithering the telescope, changing filters, etc. An exposure is the level at which MIRI is commanded: a command is sent with the number of frames, groups, and integrations that sets up MIRI, and a ``GO'' command is received to begin the actual exposure.

\index{MIRI!readout modes}
Two readout modes are implemented, the ``Fast'' and ``Slow'' modes. They apply to different kinds of astrophysical observations:
\begin{description}
\item[Fast Mode] 
For MIRI observations of bright objects and long wavelength imaging, short source- or background-limited
exposures will be used. Observations of the following types will utilize the Fast Mode:
\begin{itemize}
\item Broadband imaging at longer wavelengths ($\lambda >12\,\mu$m)
\item Bright object imaging at all wavelengths
\item Bright objects with MRS spectroscopy
\end{itemize}
Some examples of bright objects include nearby star forming regions, the Galactic center, and more
generally targets that lie in the Galactic plane.
The Fast Mode is simply reading out the array at the fastest possible speed. Each pixel is addressed during
the $10\,\mu$s window, sampled once after an appropriate amount of settling time, and the digitised data is
sent out to the ISIM. Frame times in fast mode are $\sim 3$ seconds.
\item[Slow Mode] 
For many MIRI observations involving faint objects, deep imaging and MRS spectroscopy, long
background-limited exposures will be required to achieve the needed sensitivity. Observations using
SLOWMode will include the following:
\begin{itemize}
\item Broadband imaging at short wavelengths ($\lambda < 12\,\mu$m)
\item $\mathcal{R} \sim 100$ LRS spectroscopy
\item MRS spectroscopy
\end{itemize}
The Slow Mode increases the overall integration time with the SCAs in the following way. The FPE dwells on
a pixel for 10 sample periods (or $100\,\mu$s) resulting in a frame time that is $\sim 30$ seconds long. The last eight
samples are coadded and bit-shifted to a 16-bit value within the FPE before being sent along to the ISIM;
the first two samples are discarded. This mode allows the potential of reducing the power dissipation of
the detectors quite a bit if it is found to be necessary for stability concerns.
\end{description}

\index{MIRI!sub-arrays}
MIRI's arrays have the ability to read out partial frames, or subarrays, through the manipulation of the
clocking patterns. This facilitates target acquisition imaging and selective use of the four coronagraph
masks, and allows a reduction in the frame time and integration times of $< 3$ seconds. Subarray modes
are only useful for the MIRI Imager. 
The subarrays are shown in the bottom left panel of Fig.~\ref{fig_MIRI_detector}. Four $256 \times 256$ subarrays are defined 
The four subarrays that match the coronagraphic locations are defined by the MIRI Imager optics and are approximately $256 \times 256$ in size. The $512 \times 512$ quadrant subarray's position is determined by the user. Subarrays can be defined in
any part of the imaging area, with a minimum size of $64 \times 64$ pixels. 
In order to ensure consistent calibration of subarray modes, only a few subarrays will be allowed and
the user will select from these predefined sets. 

\subsection{Observation modes: operating MIRI}
\index{MIRI!operating}
 \label{subsec:MIRI_operating}

The basic principles underlying the operation of the MIRI focal plane system (FPS)  are as follows:
\begin{itemize}
\item
 the FPS enables four principal types of scientific observation: imaging, coronagraphic imaging, low-resolution slit spectroscopy, and medium resolution integral field spectroscopy.
\item 
all MIRI data are sent to the ISIM Focal Plane Array Processor cards for further processing. There is no processing within the FPS, and on-orbit processing of the data is limited to frame coaddition in some high data rate circumstances.
\item
 For bright object imaging and for imaging at the longest wavelengths (i.e. largest background emission from telescope), the array is read out at the fastest rate possible and subarrays may be used.
\item
the MIRI imager and spectrograph may be run in parallel for calibration purposes or target acquisitions.
\item
Data are taken in a  redundant fashion. Sources will be dithered on the focal plane with the Fine Steering Mirror so that each point on the sky is sampled by many different areas of the sensor chip assembly (SCA). For the medium resolution Spectrometer (MRS), this accomplishes spectral as well as spatial dithering.
\item
Calibration of MIRI will be done periodically to monitor its stability and performance, as well as diagnostic tests and maintenance.
\end{itemize}

\subsection{MIRI's sensitivity}
\index{MIRI!sensitivity}
 \label{subsec:MIRI_sensitivity}

\begin{figure}
  \begin{center}
   \includegraphics[width=\textwidth]{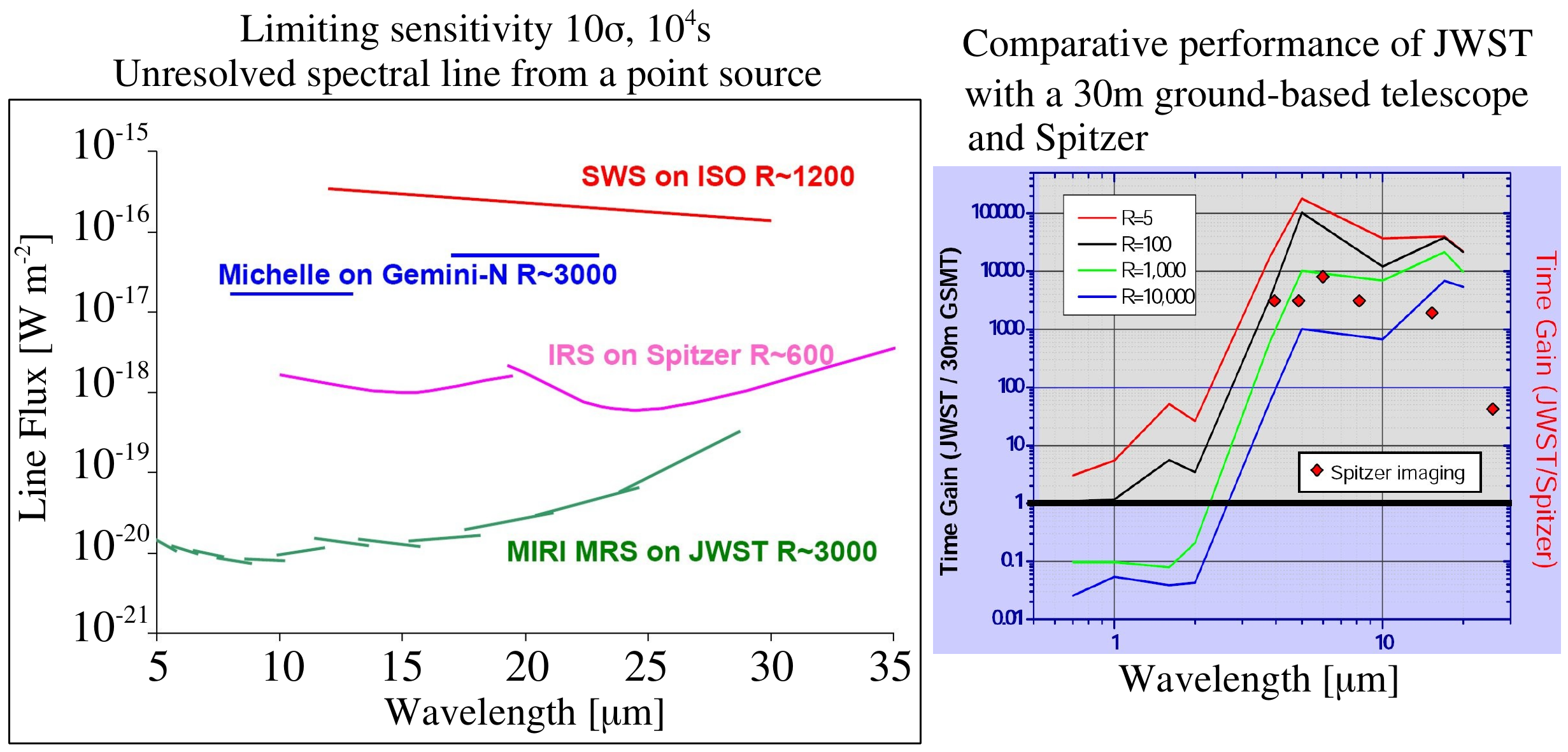}
    \caption[The JWST sensitivity]{\textit{Left:} Comparison of the spectral sensitivities between MIRI/MRS and other facilities. \textit{Right:} Relative time gain of JWST compared to a $30\,$m Giant Segmented Mirror Telescope (GSMT) and \textit{Spitzer}. The vertical axis is in relative units, where 1.0 means an observation with both JWST and GSMT (and \textit{Spitzer}) will take the same time to reach the same S/N on a point source; a larger number means JWST is faster. Both $Y$-axis are the same.}
    \label{fig_JWST_sensitivity2}
  \end{center}
\end{figure}

\renewcommand{\arraystretch}{1.1} 
\begin{table}
\begin{center}
\begin{minipage}[t]{\textwidth}
 \renewcommand{\footnoterule}{}
\def\thefootnote{\alph{footnote}}
 \caption[Sensitivity of the broad-band MIRI imager (MIRIM)]{Sensitivity of the broad-band MIRI imager (MIRIM) \footnotemark[1]}
\begin{tabular}{c c c c c c}
\hline
\hline
\multirow{4}*{Filter} & \multirow{3}*{Wavelength} & \multirow{3}*{Bandwidth} & \multicolumn{2}{c}{Sensitivity S/N = 10 in $10\,000$ s} & \multirow{2}*{Saturation limit} \\
& & & \multicolumn{2}{c}{on-chip integration} & \\
\cline{4-6}
& & & point source & extended source & point source \\
& [$\mu$m] & [$\mu$m] & [$\mu$Jy] &  [$\mu$Jy arcsec$^{-2}$] & mJy \\ 
\hline
F560W   & 5.6 & 1.2 & 0.15 & 0.80 & 14 \\
F770W   & 7.7 & 2.2 & 0.23 & 0.66 & 8 \\
F1000W & 10.0 & 2.0 & 0.50 & 0.85 & 11 \\
F1130W & 11.3 & 0.7 & 1.24 & 1.63 & 37 \\
F1280W & 12.8 & 2.4 & 0.87 & 0.85 & 12 \\
F1500W & 15.0 & 3.0 & 1.21 & 0.91 & 11 \\
F1800W & 18.0 & 3.0 & 3.0 & 1.6 & 14 \\
F2100W & 21.0 & 5.0 & 7.0 & 2.7 & 16 \\
F2550W & 25.5 & 4.0 & 20.0 & 5.2 & 22 \\
\hline
\end{tabular}
\footnotetext[1]{These figures do not include overheads due to target acquisition and calibration. The sensitivities quoted in this table are updated values of \citet{Swinyard2004}.}
\footnotetext[2]{Saturation based on 10\% of flux falling within the brightest pixel at 8 microns and a full frame exposure of 3 seconds integration time.}
\label{table_MIRIM_sensitivity}
\end{minipage}
\end{center}
\end{table}
\renewcommand{\arraystretch}{1}

\begin{table}
\begin{center}
\begin{minipage}[t]{\textwidth}
 \renewcommand{\footnoterule}{}
\def\thefootnote{\alph{footnote}}
\centering
 \caption{Sensitivity of the Low-Resolution Spectrometer (LRS)}
    \begin{tabular}{c c c c c}
\hline
\hline
\multirow{2}*{Wavelength}  & \multirow{2}*{Image FWHM} & Spectral Resolving  & \multicolumn{2}{c}{10$\,\sigma \ 10\,000$ sec sensitivity at EOL} \\
  &  & Power $\mathcal{R} = \lambda / \delta \lambda $ & \multicolumn{2}{c}{(Point source)} \\
\cline{4-5}
$[\mu$m] &   [arcsec] &  & Predicted [$\mu$Jy] & Requirement [$\mu$Jy] \\
\hline
7.5 & 0.24 & 100 & 2.8  & 2.9 \\
\hline
\end{tabular}
    \label{table_LRS_sensitivity}
\end{minipage}
\end{center}
\end{table}

\renewcommand{\arraystretch}{1.1} 
\begin{sidewaystable}
\begin{center}
\begin{minipage}{\textwidth}
 \renewcommand{\footnoterule}{}
\def\thefootnote{\alph{footnote}}
 \caption[Sensitivity of the Medium-Resolution Spectrometer (MRS)]{Sensitivity of the Medium-Resolution Spectrometer (MRS)  \footnotemark[1]}
\begin{tabular}{c c c c c c c c c}
\hline
\hline
\multirow{6}*{\begin{sideways}Sub-band\end{sideways}} & Wavelength & Spectral &  \multicolumn{2}{c}{Pixels per}  &  \multicolumn{3}{c}{\multirow{2}*{Sensitivity S/N = 10 in $10\,000$ second onchip integration}} & Saturation \\
& Coverage & Resolving & \multicolumn{2}{c}{Resolution} &  & & & Limit \\
\cline{6-9}
& $[\mu$m] & Power & \multicolumn{2}{c}{element} & Point source + & Extended source + & Point source + & Point s \\
& & &  \multirow{4}*{\begin{sideways}Spectral\end{sideways}} & \multirow{4}*{\begin{sideways}Spatial\end{sideways}}  & narrow spectral line & narrow spectral line & continuum & source \\
\cline{6-9}
& &  \multirow{2}*{$\mathcal{R} = \lambda / \delta \lambda$} & &  & \multirow{2}*{$[\times 10^{-20}\,$W$\,$m$^{-2}]$} &  $[\times 10^{-20}\,$W$\,$m$^{-2}$ & \multirow{2}*{[mJy]} & \multirow{2}*{[Jy]} \\
& & & & & & arcsec$^{-2}]$ & & \\
\hline
1A & $4.9 - 5.8$ &  \multirow{3}*{$2\,400 - 3\,700$}  & $0.9 - 1.1$ & $1.1 - 1.7$  & $2.8 - 2.2$ &  $9.7 - 5.6$ & $0.11 - 0.15$ & \multirow{3}*{5} \\
1B & $5.6 - 6.7$ & & $0.9 - 1.2$ & $1.2 - 1.6$ & $2.2 - 2.0$ &  $6.2 - 3.9$ & $0.10 - 0.16$ & \\
1C & $6.5 - 7.7$ & & $0.9 - 1.3$ & $1.2 - 1.6$ & $1.8 - 1.8$ &  $4.1 - 2.7$ & $0.10 - 0.18$ & \\
\hline
2A & $7.5 - 8.8$ &  \multirow{3}*{$2\,400 - 3\,600$}  & $1.1 - 3.1$ & $1.2 - 1.7$  & $1.1 - 0.95$ &  $1.7 - 1.1$ & $0.07 - 0.10$ & \multirow{3}*{3} \\
2B & $8.6 - 10.2$ & & $1.1 - 3.7$ & $1.3 - 1.9$ & $0.93 - 1.0$ &  $1.1 - 0.86$ & $0.07 - 0.162$ & \\
2C & $10.0 - 11.8$ & & $1.2 - 4.1$ & $1.6 - 2.2$ & $1.0 - 1.1$ &  $0.92 - 0.70$ & $0.08 - 0.16$ & \\
\hline
3A & $11.5 - 13.6$ &  \multirow{3}*{$2\,400 - 3\,600$}  & $1.0 - 2.1$ & $1.6 - 2.0$  & $1.1 - 1.2$ &  $0.73 - 0.58$ & $0.10 - 0.19$ & \multirow{3}*{2} \\
3B & $13.3 - 15.7$ & & $1.1 - 2.2$ & $1.9 - 2.3$ & $1.1 -1.0$ &  $0.54 - 0.36$ & $0.12 - 0.19$ & \\
3C & $15.3 - 18.1$ & & $1.2 - 2.5$ & $2.2 - 2.6$ & $0.92 - 1.3$ &  $0.36 - 0.36$ & $0.12 - 0.29$ & \\
\hline
4A & $17.6 - 21.0$ &  \multirow{3}*{$2\,000 - 2\,400$}  & $1.7 - 2.1$ & $2.2 - 2.7$  & $1.6 - 3.3$ &  $0.44 - 0.66$ & $0.18 - 0.56$ & \multirow{3}*{$2 - 10$ \footnotemark[2] } \\
4B & $20.5 - 24.5$ & & $1.9 - 2.4$ & $2.6 - 4.0$ & $2.8 - 7.4$ &  $0.59 - 1.1$ & $0.4 - 1.4$ & \\
4C & $23.9 - 28.6$ & & $2.2 - 2.7$ & $3.1 - 3.7$ & $5.7 - 30.0$ \footnotemark[2] &  $0.88 - 3.1$ \footnotemark[2]  & $0.9 - 6.5$ \footnotemark[2]  & \\
\hline
\end{tabular}
\footnotetext[1]{The sensitivities quoted in this table are updated values of \citet{Swinyard2004}.}
\footnotetext[2]{The quoted sensitivity is for $\lambda = 28.3\,\mu$m.}
\label{table_MRS_sensitivity}
\end{minipage}

\end{center}
\end{sidewaystable}
\renewcommand{\arraystretch}{1}

All the sensitivities of the broad-band imager, the low-resolution spectrometer, and the medium-resolution IFU spectrometer are gathered in tables~\ref{table_MIRIM_sensitivity}, \ref{table_LRS_sensitivity}, and \ref{table_MRS_sensitivity}, respectively. 
These figures are particularly helpful to prepare observation proposals (see chapter~\ref{chapter:science_JWST}).
The results are from modeling \citep{Swinyard2004} that takes into account the different factors affecting the sensitivity: the  background flux from the telescope and surrounding structure, from the instrument optics and structure, and from the zodiacal light, the cosmic ray hitting rate, the detector characteristics, the optical efficiency of the instrument, etc. 
The imaging channel of the instrument is limited by the zodiacal background at wavelengths below $\sim 12\,\mu$m and by the emission from the telescope and sunshields at longer wavelengths. In the case of the MRS spectrometer, the instrument is almost entirely detector noise limited, except in the longest wavelength channel where the background is highest.

In the left panel of Fig.~\ref{fig_JWST_sensitivity2} we compare the sensitivity of the MIRI Medium Resolution Spectrometer to other facilities. In particular, up to $\sim 20\,\mu$m, the MRS will be about two orders of magnitude more sensitive than was the \textit{IRS} spectrometer on-board \textit{Spitzer}, with a spectral resolution $\sim 5$ times higher. The right panel of Fig.~\ref{fig_JWST_sensitivity2} summarizes the relative time gain between JWST instruments and \textit{Spitzer}, or an hypothetical $30\,$m ground-based telescope. Such a gain in sensitivity will certainly lead to unexpexted discoveries\dots


\chapter{First results from JWST/MIRI testing}
\label{chapter:miri_test}

\epigraph{Pour explorer le champ des possibles, le bricolage est la méthode la plus efficace. ``It was like magic. The fact that a few calculations and a bit of do-it-yourself could make the invisible visible\dots''}
{Hubert Reeves}

\index{MIRI!Tests}


\begin{Abstract}
This chapter presents my contribution to the testing of the Mid Infra Red Instrument (MIRI), part of the scientific payload that will be integrated onto the James Webb Space Telescope (JWST).
The first tests at cryogenic temperatures and infrared wavelengths were performed on the Flight Model (FM) of the Mid-InfraRed IMager (MIRIM), from December 2008 to April 2009 at CEA, France. 
My work is focused on the ``microscanning'' test, which allows us to characterize the MIRIM optical quality with an unprecedent accuracy. 
I describe the instrumental test setup and discuss the results of the microscanning analysis that allows to obtain a ``high-resolution'' Point Spreag Function of the instrument. This analysis is of particular importance, not only for the optical quality check and the MIRI calibration, but also for potential scientific applications. 

\end{Abstract}

\minitoc
\section{Introduction}
\label{sec:intro_MIRIMtests}

\PARstart{G}round tests are essential to ensure the functionality and the optical performance of the instrument. They often lead to design changes from the first models built to the final flight model (FM). The characterization of the instrument (stray light, point spread function, flat field, etc.) has to be as accurate as possible during ground test campaigns since these tests are extremely limited in space.

In this chapter I present my contribution to the extensive test campaign of the MIRI imager (MIRIM, see chap.~\ref{chapter:JWST} for a description of the instrument). 
The first MIRIM optical performance tests in the infrared, at cryogenic temperatures, were performed at CEA, Saclay, during the December 2008 $-$ April 2009 test campaign. I had the chance to participate in these tests, and I took the responsability of the analysis of one of these tests, so-called  \textit{microscanning}, which allows  to characterize the Point Spread Function\footnote{The Point Spread Function is the image of a point source through the instrument. The PSF characterizes the imaging system's spatial response, i.e. its spatial resolution. In technical words, this is the impulse-response of the optical system.} (hereafter PSF) of the instrument very accurately. 
This was a great experience, and I had the opportunity to interact with engineers, and to attend meetings of the European Consortium, in which I had several opportunities to present my work.

It would be impossible to review all the tests and results that were performed on MIRIM. I will rather focus on the PSF analysis. After introducing some generalities about the PSF of the JWST and the context of MIRI testing, I describe in sect.~\ref{sec:instru_setup}  the instrumental setup used to simulate the JWST beam and to perform the cold tests. Section~\ref{sec:overview_MIRIMtests}  presents a very brief overview of the tests performed on MIRIM at CEA, Saclay.  In sect~\ref{sec:MIRIMdataReduction}, we discuss the response curve of the detector and how we correct the images for the non-linearity of the detector. Then, Sect.~\ref{sec:microscan} describes the microscanning test and the method used to reconstruct the high-resolution PSF of MIRIM. Sect.~\ref{sec:results} discusses the PSF characteristics determined by the microscan analysis. 
Sect.~\ref{sec:microscan-conclusion} summarizes our results and in sect.~\ref{sec:MIRItestsNext} I briefly give the next steps of MIRI/JWST testing.

\subsubsection{Introduction to the PSF of the JWST}
\index{JWST!Point Spread Function}

\begin{figure}
\begin{minipage}{\textwidth}
    \def\thefootnote{\alph{footnote}}
      \setlength{\footnotesep}{0pt}
  \centering
    \includegraphics[width=\textwidth]{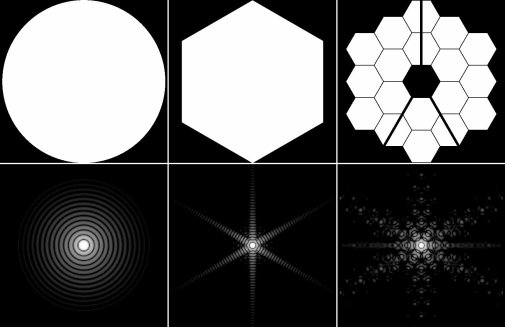}
    \caption[PSF shapes for different apertures]{The PSF corresponding to each aperture is displayed on a logarithmic grayscale from $10^{-7}$ to $10^{-3}$ counts. The total number of counts in each of these simulated PSFs is 1. These PSF have been modeled with the JWPSF software\footnotemark[1].}
    \label{fig:JWST_Aperture-PSFComparison}
\footnotetext[1]{\url{http://www.stsci.edu/jwst/software/jwpsf}}
\end{minipage}
\end{figure}

From a basic point of of view, the  PSF of an imaging system is determined by two things: the pupil shape and the WaveFront Errors (henceforth WFE). Often, the shape of the pupil is well known and relatively simple. For instance, in the case of a circular or annular aperture, the PSF is an Airy disk (the square of a Bessel function). However, for more sophisticated and larger systems, like the JWST or the Keck telescope, pupils are intrinsically a lot more complex. 
In Fig.~\ref{fig:JWST_Aperture-PSFComparison}, we illustrate how the pupil shape affects the PSF, building from an open circular pupil of diameter 6.5~m \textit{(left)}, a hexagonal pupil of diameter 6.5~m \textit{(center)}, and a realistic representation of the JWST pupil \textit{(right)}. These PSF have been computed with the JWPSF software described in \citet{Cox2006}.

\subsubsection{Why is it important to characterize the PSF with precision?}

The PSF translates the optical quality of the system. A careful measurement of the PSF characteristics is first needed to check the optical performance of the instrument, in order to verify that the instrument is nominally built. This is the main goal of the tests described in this chapter.
A precise knowledge of the PSF is also crucial for the photometric calibration of the instrument and for scientific applications (photometry, source extraction, etc.). 

\begin{figure}
  \begin{center}
    \includegraphics[width=\textwidth]{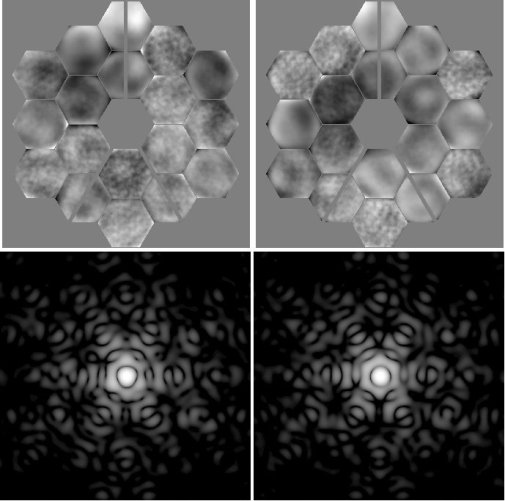}
    \caption[Variations of the JWST PSFs with the WFE]{Comparison of two JWST error budget Optical Path Difference (OPD) files \textit{(top)} and their associated PSFs \textit{(bottom)} for a wavelength of $2\,\mu$m. OPDs are displayed on a linear grayscale from $-400$ to $400$~nm, while PSFs are displayed on a logarithmic grayscale from $10^{-6}$ to $10^{-3}$. The OPDs include wavefront error (WFE) contributions from the JWST Optical Telescope Element (OTE), the Integrated Science Instrument Module (ISIM), and the Near-Infrared Camera (NIRCam). The RMS WFE of both OPDs is $\approx 110$~nm. While the differences in total RMS WFE are small, the impact of these differences on the resultant PSFs is measurable. These PSF have been modeled with the JWPSF software.}
    \label{fig:JWST-OPDs-PSFs-2micron}
  \end{center}
\end{figure}

Before determining if the PSF of the instrument is nominal or not, an important step is to study the impact of aberrations on the PSF. Errors in the wavefront can arise from a variety of sources, including imperfections in the system's optics or atmospheric variations  in the case for ground-based observations. These WFE can be extremely difficult to determine.
The Fig.~\ref{fig:JWST-OPDs-PSFs-2micron} illustrates the impact of variations of the WFE on the PSF at $2\,\mu$m with the NirCam instrument (see sect.~\ref{subsec:JWST-science-instruments} for a description of the instrument). Two realizations of the WFE and corresponding JWST PSFs are shown. These PSFs differ in the specific distribution of the wavefront error across the pupil. As long as the error budget is satisfied, comparing these PSFs provides an estimate of the PSF differences that could be encountered in the normal course of observations. 

\subsubsection{A long path to the FM, and a European adventure!}

To design and build an instrument like MIRI is a hard and long route!
 The development and qualification of the optical bench of the MIRI imager (MIRIM) started in 2004, and the first tests of the MIRIM flight model (henceforth FM) started in December 2008. Some hardware parts of the MIRI FM are still being, and will be, manufactured and tested in 2009-2010. MIRI is expected to be delivered to NASA at the end of 2010. Then MIRI will be integrated into the Instrument Science Module, at the back of the JWST primary mirror (see chapter~\ref{chapter:JWST}), and basic testing (alignment, focus, etc.) will be performed in the giant cryogenic chamber at Johnson Space Center, Houston, USA (see \ref{sec:MIRItestsNext}). 

Before the FM, several models of MIRI were built. A first mechanical model (the Structure and Thermal Model, STM) was to test thermal balance, integration and vibration at MIRI level. The first optical model was the Verification model (VM), and it was used to test the integration procedures and the optical performance. Then, the Engineering and Test Model (ETM) was an optically fully representative model, used to verify the optical performance and thermal behaviour of the instrument at cryogenic temperatures at CEA. 
The different steps leading to the FM integration of the Mid-Infra Red IMager
Optical Bench (MIRIM-OB), and the principal results associated with the two test models, the STM and the ETM, were presented in \citet{Amiaux2008}. In this chapter, I will focus on the FM.

The different parts of the MIRIM-OB system are built within the different institutes of the European Consortium, then
gathered together and integrated at CEA-Saclay. The filters are under the responsibility of University of Sweden. The Double Prism Assembly for the Low-Resolution Spectrometer (LRS, see \S~\ref{subsec:MIRI_LRS}) is a Belgium (CSL) and a Germany (University of Köln) collaboration. The coronographic masks are manufactured at Observatoire de Meudon (LESIA) in France. Optical elements are then integrated onto the Filter Wheel Disk under CEA responsibility (see \S~\ref{subsec:MIRIcamera} for a description of the filter wheel and its filters). The Filter Wheel Disk is delivered to MPIA in Germany for integration on the Filter Wheel cryo-mechanism. Filter Wheel Assembly is then returned to CEA-Saclay for further integration into MIRIM Optical Bench.

\section{Instrumental setup: the MIRIM test bench at CEA}
\label{sec:instru_setup}
\index{Microscanning test!test bench}

\begin{figure}
  \begin{center}
    \includegraphics[width=\textwidth]{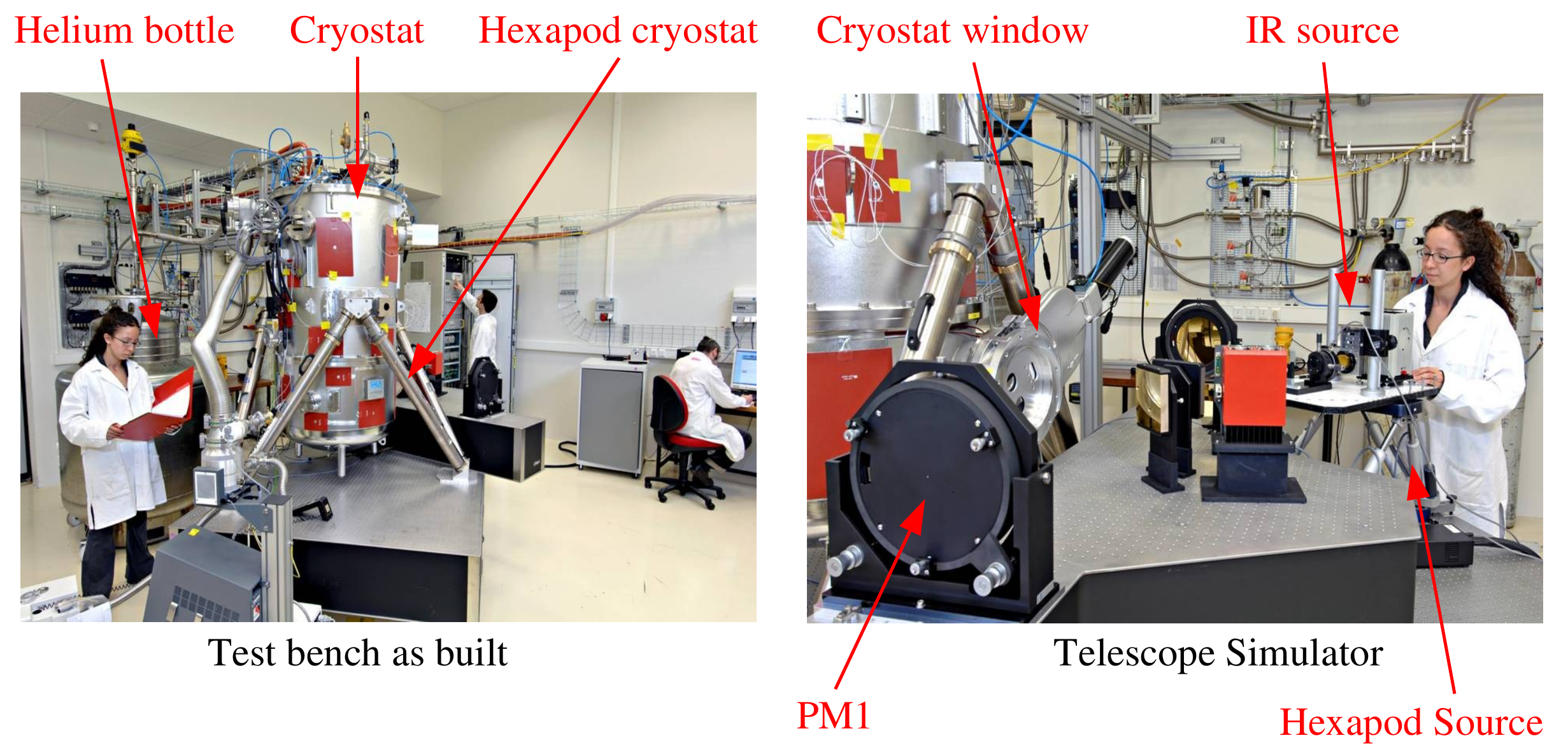}
    \caption[Photos of the test bench for testing of MIRIM at CEA, Saclay]{Photos of the test bench for testing of MIRIM at CEA, Saclay. The telescope simulator (TS) is lying next to the cryostat where MIRIM is installed. The IR blackbody source is mounted on an hexapod that allows fine motion scanning in the object plane of the TS.}
    \label{fig_MIRIM_testbench_photos}
  \end{center}
\end{figure}

\begin{figure}
  \begin{center}
    \includegraphics[width=0.7\textwidth]{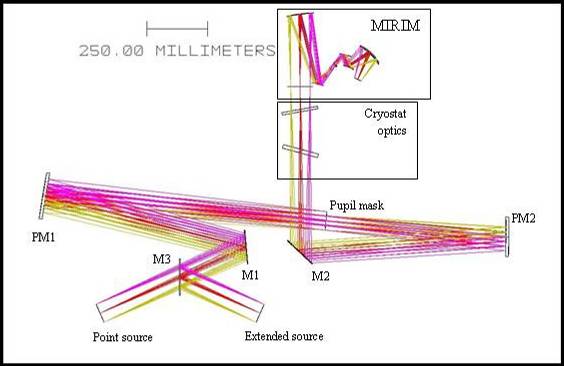}
    \includegraphics[width=0.29\textwidth]{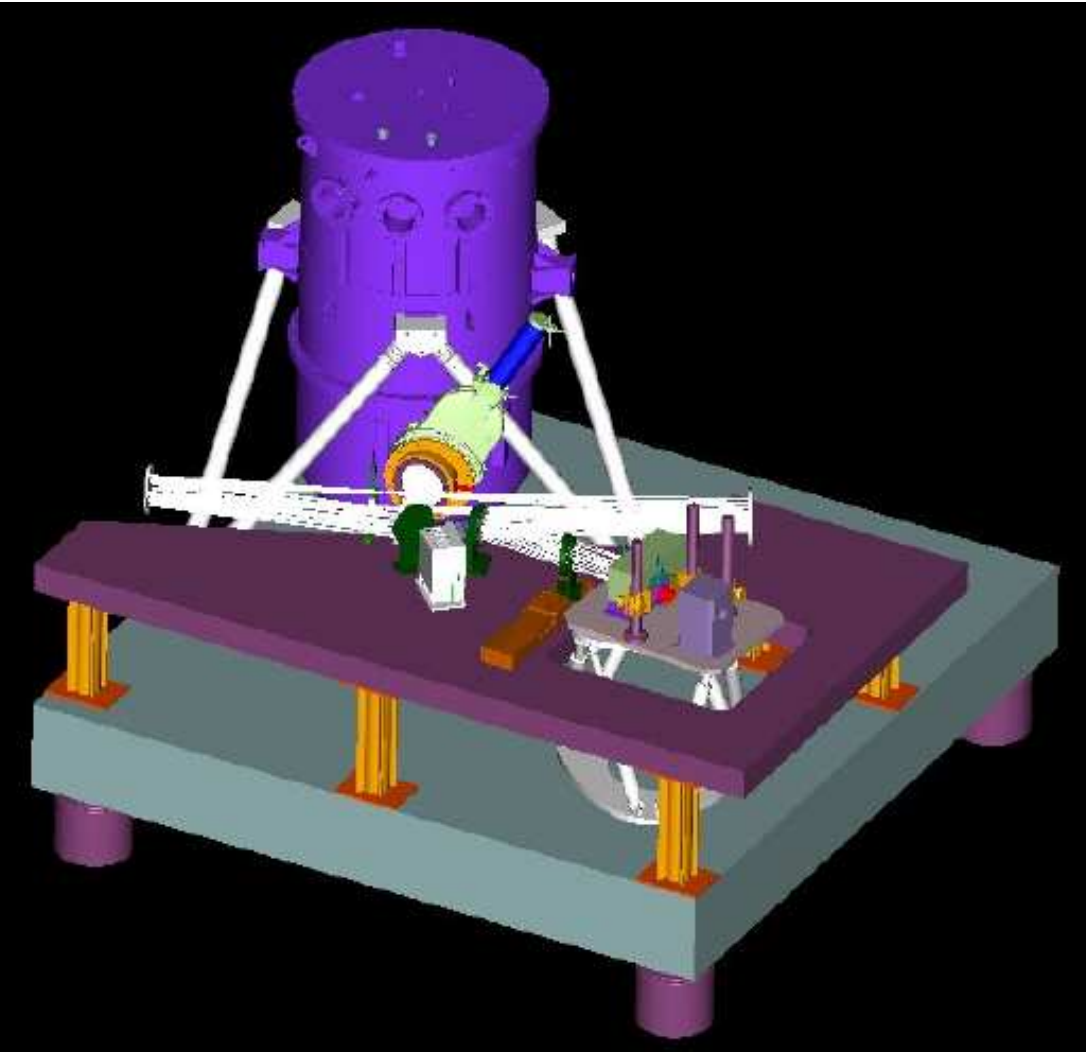}
    \caption[The MIRIM test bench at CEA, Saclay]{The MIRIM test bench. The left panel shows the optical scheme of the Telescope Simulator (TS). The warm TS is installed outside the cryostat (in blue on the right panel) that contains MIRIM. The TS simulates the JWST beam and pupil. It allows the use of a point or extended source, in order to characterize MIRIM in image or coronographic mode.}
    \label{fig_testbed_simu_telescope}
  \end{center}
\end{figure}

The tests of the MIRIM optical quality have been performed with an ambient temperature Telescope Simulator (hereafter TS), installed outside an helium-cooled cryostat that contains MIRIM-OB (Fig.~\ref{fig_MIRIM_testbench_photos}). The TS simulates the optical beam delivered by the JWST. The optical conception of the TS is based on two off-axis parabolic mirrors (PM1 and PM2) and a pupil mask in between (Fig.~\ref{fig_testbed_simu_telescope}). The pupil mask is the pupil of the telescope simulator (so-called the ``STOP'') and it reproduces the pupil of JWST. It is mounted on motorised translation and rotation stages (4 axis: 3 translations and one rotation). This allows fine adjustment to co-align the telescope simulator and MIRIM pupils. These degrees of freedom are also used to simulate pupil shear, which occurs when the pupil stop is mis-aligned with the telescope pupil image. 

The telescope simulator was mainly designed by Yuying Longval at IAS, and installed at CEA, Saclay. 
The field of view of the TS is $80\, \rm mm \times 80\, mm$ for a MIRIM field of view of $72\, \rm mm \times 72\, mm$. An analysis of the  image quality\footnote{The design and optical quality analysis are reported in a document by Y. Longval \& A. Abergel (not reproduced here).}  shows that the distortion amounts to 0.33\% for the useful field.
The aperture is $F/20$ in the image plane. The exit pupil location is 3017.5~mm from the image plane for a diameter of 151.6~mm. The  pupil is 75.4 mm in diameter. To implement the MIRIM cryostat in a minimum space, folding mirrors are used.

An IR point source with a shutter is mounted on a remotely controlled hexapod, that can be moved in 3 directions ($X$, $Y$ and $Z$). The minimum diameter of the point source is 30$\, \mu$m, and the temperature of the source is 1150 K for the imaging mode, or 2000 K for the coronographic mode.
An extended black body source (400 K) can also be used.
The 10K screen in the cryostat is fitted with a neutral density (ND) that reduces the flux due to the 295K background.
All the tests were performed when the instrument is cold. During each of the tests,
the temperature of the instrument must be between 5 and 7K and stable at $\pm 250$~mK. The detector temperature is 6.7K and is stable at $\pm 10$~mK.
 For a more detailed description of the test bench, see \citet{Amiaux2008}.

\begin{table}
\begin{center}
\begin{minipage}{\textwidth}
 \renewcommand{\footnoterule}{}
\def\thefootnote{\alph{footnote}}
 \caption[MIRIM images type]{MIRIM images type\footnotemark[1]}
\centering
\begin{tabular}{l l}
\hline
\hline
Name & Description \\
\hline
DIT & single image from one detector integration time  \\
HCYC1 &  addition of all the DIT images taken during the first half-cycle of the shutter (open)  \\
HCYC2 & addition of all the DIT images taken during the 2nd half-cycle of the shutter (closed)  \\
INTERM & intermediate integrated image from one full cycle of the chopper (HCYC1$-$HCYC2)  \\
INT & integrated image (addition of all the INTERM images)  \\
\hline
\end{tabular}
\footnotetext[1]{We list the different types of images generated with the chopper. The type of images to produce depends also on the test being performed.}
\label{table_MIRIM_imagetype}
\end{minipage}
\end{center}
\end{table}

\index{MIRI!images type}

At the time of the FM test campaign, the filter wheel assembly was not available because of vibration testing, and only one filter ($5.6\,\mu$m) was available. The filter wheel was replaced by a ``mono-filter wheel'' (often called a ``cyclop tool''). Since there was no dark position available, a chopper has been used. 
The different types of images produced are listed in Table~\ref{table_MIRIM_imagetype}.
A measurement cycle consists in two sub-cycles (shutter closed, shutter open). The exposition time, or the total duration of a measurement, $T_{\rm exp}$, (i.e. total time to get one INT image) is calculated as follows:
\begin{equation}
T_{\rm exp} = N_{\rm cycles} \times T_{\rm cycle} \ ,
\end{equation}
where $T_{\rm cycle}$ is the duration of a cycle. We have $ N_{\rm cycles} = N / N_{\rm DIT}$, where $N$ is the total number of DIT images (see Table~\ref{table_MIRIM_imagetype} for definition) required to achieve a given signal ratio and $N_{\rm DIT}$ is the number of DIT images per half cycle. The duration of a cycle is given by
\begin{equation}
\label{eq_tcycle}
T_{\rm cycle} = 2 \times (N_{\rm DITSKIP} + N_{\rm DIT}) \times DIT = 1/ f_{\rm CHOP} \ , 
\end{equation}
where $N_{\rm DITSKIP}$ is the number of images to skip at the
beginning of the cycle (delay for the shutter to close), $DIT$ the detector integration time, and $f_{\rm CHOP}$ the chopping frequency. The factor of 2 is because there are two half-cycles per cycle: shutter close and shutter open.

\section{Overview of the MIRIM optical performance tests}
\label{sec:overview_MIRIMtests}
\index{MIRI!test overview}

\begin{figure}
  \begin{center}
    \includegraphics[width=0.49\textwidth]{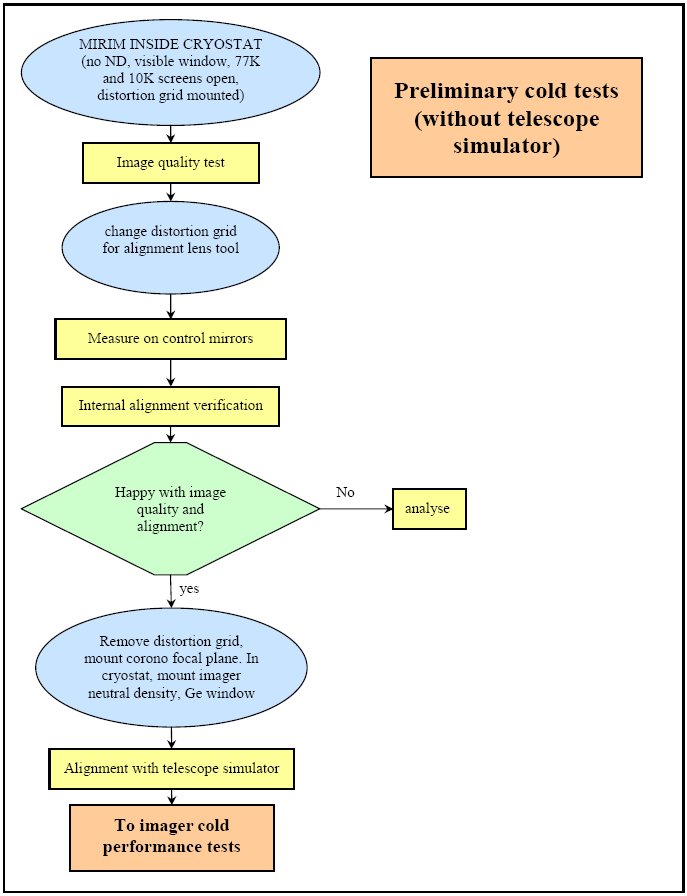}
    \includegraphics[width=0.49\textwidth]{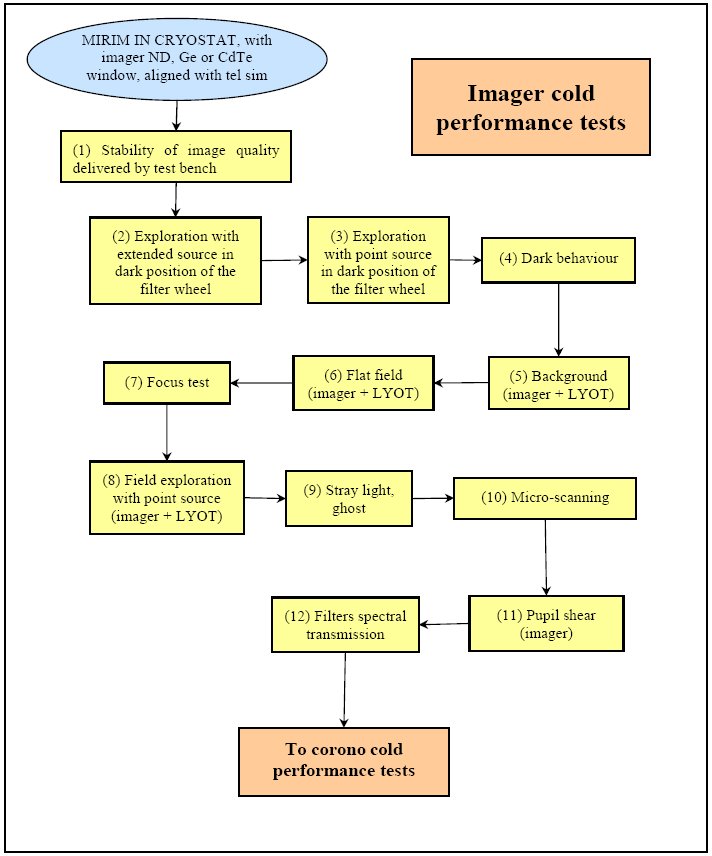}
    \caption[Flowchart of the tests performed on the MIRIM ETM]{Flowchart of the tests performed on the MIRIM Engineering Test Model (ETM) and the Flight Model (FM).}
    \label{fig_TestFlow_cold}
  \end{center}
\end{figure}

The first tests in the infrared, at cryogenic temperatures, were performed on the ETM at CEA, between February and April 2008. An overview of the optical performance tests is presented in the flowchart of Fig.~\ref{fig_TestFlow_cold}. The 
ETM model is fully representative in terms of optical path, but not in terms of alignment of the optical path. Indeed, the ETM FWA is a proto-type without requirements on alignment. Below is a brief description of these tests. Similar tests have been performed on the Flight Model (Dec. 2008$-$April 2009).

\paragraph*{Distortion grid at ambient and cold temperatures:}
a grid of $87 \times 87$ pinholes ($5\,\mu$m in diameter), with a regular pitch, is mounted at the entrance focal plane, illuminated by an optical LED, and imaged with a $2048 \times 2048$ CCD chip. This setup allows to measure the magnification and the distorsion over the whole field of view in one image, at ambient temperature ($\sim 300$~K). Shims\footnote{a shim is a thin and often tapered or wedged piece of material, used to fill spaces between the mirror and its supporting structure. Shims are  used in order to adjust the  positions of the mirrors.} are used to simulate a possible telescope defocus. The CCD camera tool is equipped with a micrometric translation stage along the optical axis, which allows to explore and to search the best focus position. This test is also performed at cryogenic temperatures and with infrared light, which allows to assess image quality and perform flat field measurements.
\paragraph*{Alignment check at ambient and cold:}
 The alignment consists in pointing the coronagraphic diaphragms with an alignment telescope that is coaligned with MIRIM optical axis. A correction is applied by shimming the M1 in order to minimize the shift between the optical axis and the centre of the diaphragms. This test has been performed at ambient and croygenic temperatures.
\paragraph*{Image quality with telescope simulator:}
\begin{itemize*}
\item Field of View exploration with a point source: a point source ($30\,\mu$m in diameter) is scanned in the FoV to measure its extent, its vignetting, straylight, and search for ghost images.
\item Stray light: a very bright point source ($150\,\mu$m pinhole) is scanned out of the field of view (on the mechanical aperture of the focal plane and across its edges).  
\item Pixel micro-scanning: for accurate measurement of the PSF (see sect.~\ref{sec:microscan}).
\end{itemize*}
\paragraph*{Coronagraphy performance:}
\begin{itemize*}
\item Peak-up: validation of the peak-up mode (accurate positioning of a source at the centre of a
coronographic mask)
\item Rejection factor: measurement of the attenuation of the coronagraphs.
\end{itemize*}

\section{Data reduction}
\label{sec:MIRIMdataReduction}
\index{MIRI!data reduction}

\subsection{Basic reduction steps}

For the microscanning, the shutter is used in chopping mode (see sect.~\ref{sec:instru_setup}). A total of 8 $DIT$ frames is taken for each cycle (open$-$close shutter). A $72 \times 64$ sub-array is read out, with a detector integration time $DIT = 0.827$~s. Thus, according to Eq.~\ref{eq_tcycle}, the duration of a cycle is $T_{\rm cycle} = 13.23$~s. For each cycle that is beginning, the first frame is skipped because during this frame the shutter is closing. Then, the following 3 $DIT$ frames are skipped because of latency effects on the detector since the point source is very bright. Therefore, we use the last 4 frames of the cycle. Each exposure  comprises  10 cycles, so the exposure time is $T_{\rm exp} = 132.3$~s. A median image is created from these 10 cycles. 

\subsection{Linearity correction for the response of the detector}
\label{subsec:linearity}
\index{MIRI!linearity}

JPL and CEA performed tests of the response of the detector and it appears that the detector is non linear over a wide range of ADU values. To quantify the impact of the non-linear response curve on the PSF measurements, S. Ronayette and myself  provide two sets of test data: one that is not corrected for the detector response, and the other which is corrected for the non-linearity. I have performed the PSF analysis on the two data sets to  compare the results. 

The MIRIM detector is a $1024 \times 1024$ pixels SiAs array. JPL has delivered a Raytheon SB305 detector to CEA. The detector configuration will be a Sensor Chip Assembly (SCA), i.e. an ultra-light mechanical \& electrical housing. CEA has developed an appropriate housing (similar to the MIRI focal plane) to mount the IR detector onto the imager for the cold tests. Thus, qualitative analysis of the optical properties, including stray light performances, are possible.

\begin{figure}
      \includegraphics[height=\textwidth, angle = 90]{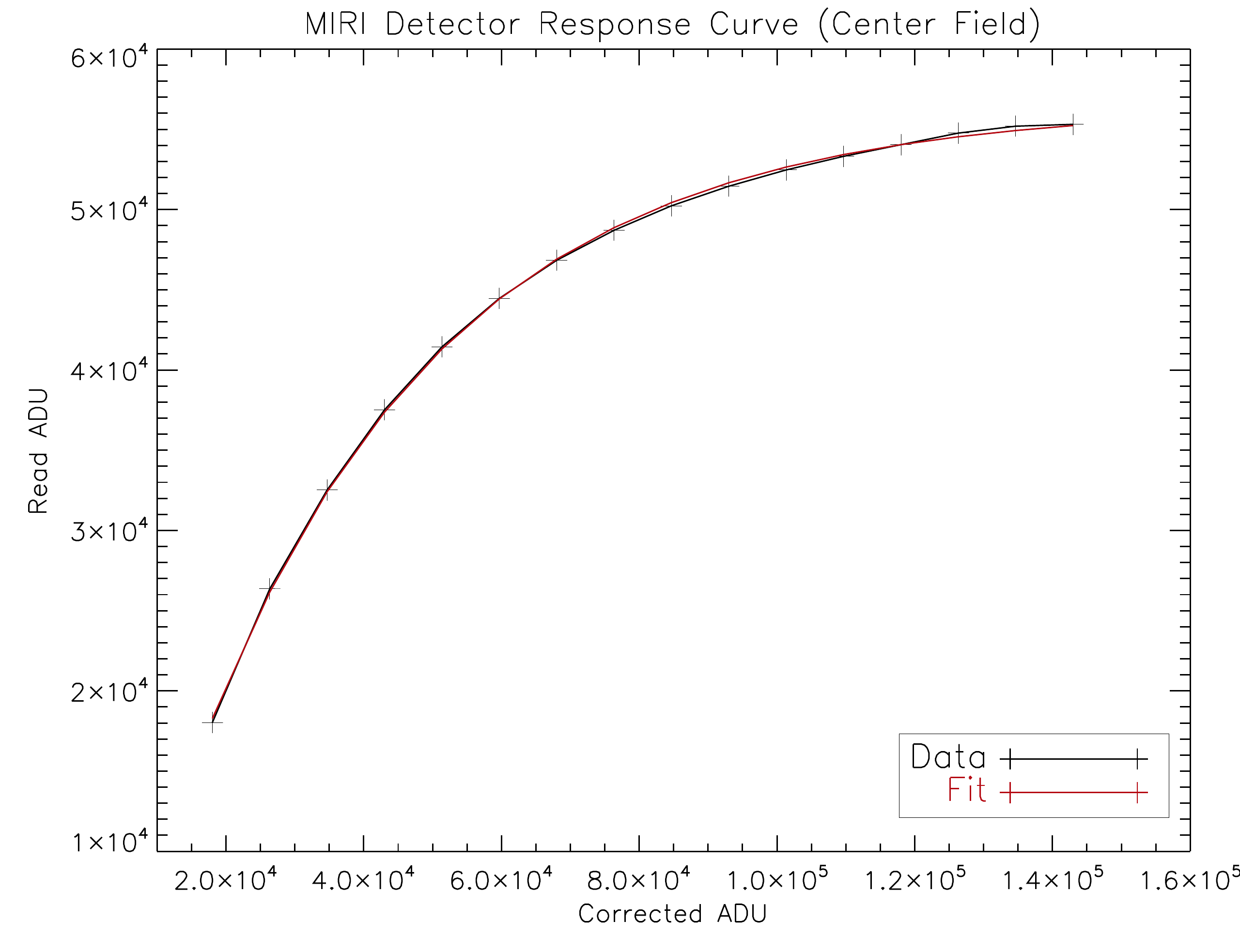}
    \caption[The response curve of the MIRI SB305 detector obtained at CEA]{The response curve of the MIRI SB305 detector obtained at CEA. Measurements have been done at 5~K. The back curve show the average read ADU values of the pixels in the center of the field as a function of the corrected values (see text for details). The red curve indicates an exponential fit (see \S~\ref{subsec:linearity}, Eq.~\ref{eq:response}).}
    \label{fig_MIRI_Response_Curve}
\end{figure}

 \begin{figure}
      \includegraphics[height=\textwidth, angle = 90]{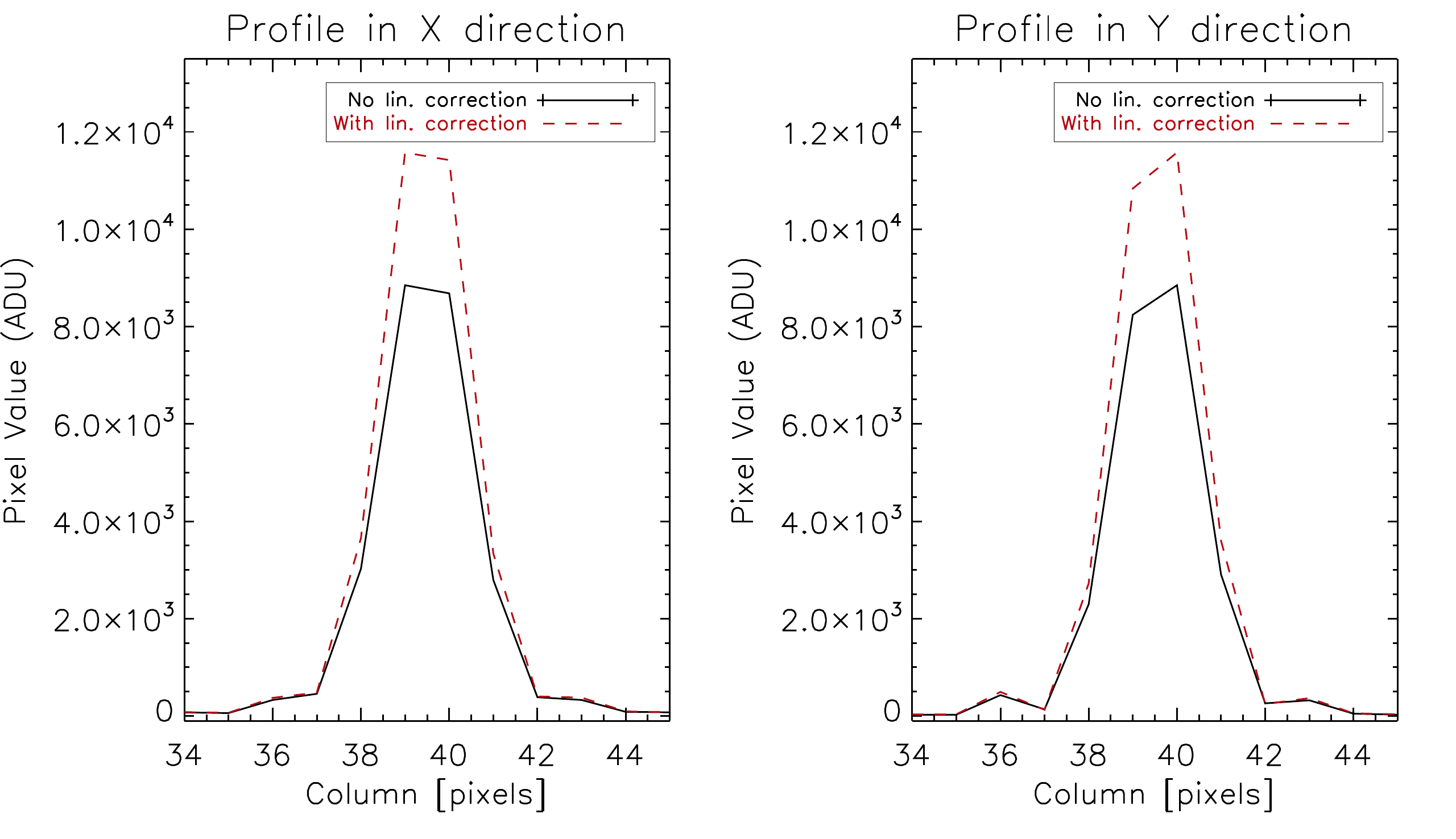}
      \caption[The impact of the linearity correction on the low-resolution PSFs]{The impact of the linearity correction on the low-resolution PSFs. The plots show MIRIM PSFs profiles at $5.6\, \mu$m in two perpendicular directions. The pixel size is $25\, \mu$m. The black line represents data that have not been corrected for the detector response. The red dashed line shows the same PSF profile when the linearity correction is applied.}
    \label{fig_comp_lin_correction}
\end{figure}

Fig.~\ref{fig_MIRI_Response_Curve} presents the detector response. 
Measurements have been done by V. Moreau at CEA on the SB305 detector at 5~K, with an extended blackbody source in front of the detector.
The source is at constant temperature (30~K) and the integration time is the only parameter that changes. The longer the integration time, the higher the signal in units of  ADU. The 30~K blackbody source does not allow us to explore the linear part of the response curve, at low ADU values, since with the smallest integration time (2.7~s, see chapter~\ref{chapter:JWST}, sect.~\ref{subsec:MIRI_detectors}), we reach a signal of $26\,360$~ADU (see Fig.~\ref{fig_MIRI_Response_Curve}).

Fine measurements a low integrated fluxes were performed 
with colder black body sources (not shown here). An accurate sampling of the response curve at low ADU values  shows that the response is linear for ADU values lower than $\approx 16\,000$ (V. Moreau, private communication) and follows the law:
\begin{eqnarray}
\label{eq:signal-linear-MIRI}
S_{\rm lin} = \alpha \times t_{\rm int} + \beta \ \ ,
\end{eqnarray}
where $S_{\rm lin}$ is the signal in ADU in the linear part of the response, $\alpha = 3067$~ADU~s$^{-1}$ is the slope of the linear response and $\beta = 18030$~ADU the offset.

The low background signal ($18\,030$~ADU on Fig.~\ref{fig_MIRI_Response_Curve}) is measured more accurately with colder black body sources. It allows to do a fine sampling of the beginning of the response curve, at low levels, where the response is approximately linear.
Fig.~\ref{fig_MIRI_Response_Curve} shows the read value as a function of the corrected value.
To estimate the corrected value,  the difference between the linear response (linear fit done at low signal given by Eq.~\ref{eq:signal-linear-MIRI}) and the read value is calculated.
To perform the linearity correction, I have fitted the response curve by an exponential function:
\begin{equation}\label{eq:response}
S_{\rm read} = A - B \, e^{-S_{\rm cor} / C} \ ,
\end{equation}
where $S_{\rm read }$ is the read signal in ADU and $S_{\rm cor}$ the corrected one. I found $A = 56\,418.5$, $B = 62\,939.1$ and $C = 35\, 990.5$~ADU. The red line on Fig.~\ref{fig_MIRI_Response_Curve} shows the result of the fit. 

To apply the linearity correction, we derive $S_{\rm cor}$ from $S_{\rm read }$ and correct all the individual images.
Fig.~\ref{fig_comp_lin_correction} compares the PSF profiles when the linearity correction is applied or not. The correction enhances the highest values of the signal, so it improves the sharpness of the PSFs.

\section{The micro-scanning test}
\label{sec:microscan}

\index{MIRI!microscanning}

In this section we first describe the goals, principle and experimental method to perform the microscanning. Then, results are discussed. This test, together with field of view measurements, led to the discovery of a defect in the instrument after the first FM1 test campaign. This defect was a tilt of the M4 mirror (see Fig.~\ref{fig_MIRI_layout} for the position of this mirror in the optical layout of the instrument). This anomaly was corrected and a second test campaign was performed (FM2). We show the differences in the optical quality of the instrument, before and after correction for the two FM test campaigns (FM1 and FM2).

\subsection{Aims and experimental method}

Since the physical size\footnote{The angular size for one pixel is 0.11''} of the pixels is $25\, \mu$m, the Point Spread Function (PSF) of the instrument is not Nyquist-sampled at a wavelength of 5.6$\, \mu$m (the theoretical FWHM of the PSF at 5.6$\, \mu$m is 0.185'', i.e. is less than 2 pixels), which is the shortest wavelength for MIRI, and the most critical for the assessment of the optical quality. Therefore, it is difficult to check the optical quality of the instrument at this wavelength with a single point source image.

The goal of the microscanning test is to obtain a \textit{``high-resolution''} (HR) image of the PSF from multiple  ``low-resolution'' (LR) raw images. 
In other words, the microscanning aims at providing a highly-sampled image of the PSF.
Such a resolution enhancement approach has been an active research area, in particular in image or video processing, and it is sometimes called \textit{over-resolution} image reconstruction, or simply resolution enhancement. Here, the HR PSFs are used to perform an accurate measurement of the MIRIM image quality. 

The basis of the inversion method used to analyse the microscanning data has been developed to perform high spectral and spatial resolution spectroscopy with \textit{Spitzer} telescope InfraRed Spectrometer (IRS) data \citep{Rodet2009}. This work relies on a fruitful collaboration between the LSS\footnote{The LSS is the ``Laboratoire des Signaux et Systèmes'', an institute specialized in signal processing, see \url{http://www.lss.supelec.fr/}.} and the IAS institutes in Orsay.
With the help of T. Rodet (LSS), I have applied Matlab algorithms of data inversion to process  the microscanning data, and analyse the results with standard tools (IRAF, IDL).

\begin{figure}
  \begin{center}
    \includegraphics[width=9cm]{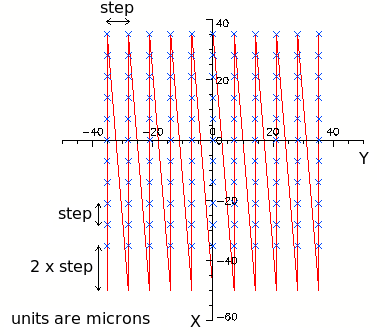}
    \caption[Pattern of the microscan]{{\it Pattern of the microscan. A $11 \times 11$ points scanning grid has been performed, corresponding to a total scanning area of 1 pixel on the detector plane. The source has been moved by steps of $7$~$\mu$m on the hexapod mount of the telescope simulator. This corresponds to individual displacements of  $2.5$~$\mu$m on the detector plane.}}
    \label{fig_schema_scan}
  \end{center}
\end{figure}

The microscanning test is performed at cryogenic temperatures and at a wavelength of 5.6$\, \mu$m. No other filter was available at that time because the filter wheel was not delivered to CEA.
The method consists in scanning a point source in a fine spatial resolution.  The microscan pattern is given in Fig.~\ref{fig_schema_scan}. A $11 \times 11$ points grid has been done, corresponding to a total scanning area of 1 MIRIM pixel on the detector plane, i.e. an area of  $10^2$~pixels$^2$. The source has been moved by steps of $7$~$\mu$m on the hexapod mount. This corresponds to individual displacements of  $2.5$~$\mu$m on the detector plane. In the following sub-sections the method we apply to reconstruct the high-resolution images is detailed.

\subsection{Direct and inverse problem}
\label{subsec:directinverse_pb}
\index{Microscanning test!direct problem}
\begin{figure}
  \begin{center}
    \includegraphics[width=0.8\textwidth]{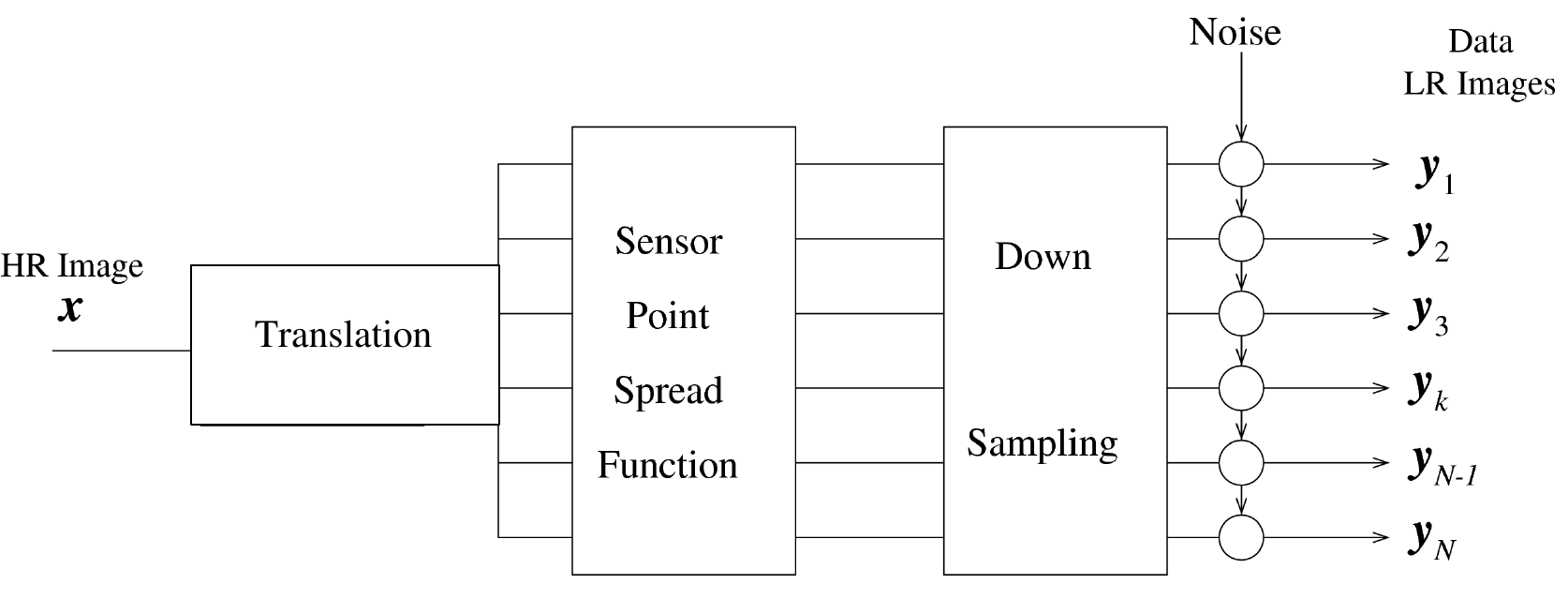}
    \caption[Sketch of the direct problem of generation of images]{Sketch of the direct problem of generation of images. This figure links the high-resolution optical PSF $\xb$  to the low-resolution ones $\yb_k$ that we observe. }
    \label{fig_directModel}
  \end{center}
\end{figure}

\index{Point Spread Function!definition}

The total point spread function $PSF_{\rm tot}$ of an imaging system
consists of several components. Firstly, the light is diffracted  within the optics. This part will be denoted $PSF_{\rm opt}$. For instance, $PSF_{\rm opt}$ is an Airy disk in case of a circular aperture.
Once the light has passed through the optical elements (mirrors, filters, gratings, etc.), the signal is integrated on the sensor (array) composed of squared detectors (pixels). Because the area of a pixel is not infinitely small, this causes blurring ($PSF_{\rm det}$) which can be modeled by a uniform square function if the spatial response of a pixel is flat over its entire area.
The total PSF can be computed by convolution of its different parts\footnote{There are additional parts that contribute to the final PSF that we do not consider here. For instance, motion of the sensor platform causes blurring. In addition the electronic components
cause smoothing  by applying a low-pass filter to reduce noise.}:
\begin{equation}
PSF_{\rm tot} = PSF_{\rm opt} \star PSF_{\rm det}
\end{equation}
The process of generation of the images is schematically shown on Fig.~\ref{fig_directModel}. 
We denote $\xb$ the high-resolution (HR) optical PSF. 
The aim of the deconvolution method is to deduce $\xb$ from multiple low-resolution images 
$\yb_k$ (the observed data). 

Let us describe the \textit{direct} problem by assuming that we start from the high-resolution image $\xb$. 
The high-resolution optical PSF $\xb$ that we want to estimate is first translated because of the motions of the source. Then, the observed and translated images $\yb_k$ are the result
of the convolution by the PSF of the detector, $PSF_{\rm det}$. In the schematic Fig.~\ref{fig_directModel}, we separate the integration of the detector in two steps:  \textit{(1)} a convolution by the PSF of an over-sampled (fine grid) detector (a detector that would have smaller pixels than the real detector of the instrument), and \textit{(2)} a down-sampling operation needed to degrade the resolution of the fine detector to the real,  low-resolution of the obervations. The ratio between the pixel size on the fine grid and the physical size on the real detector is named \textit{over-sampling factor}, $f$, which represents the gain in resolution.

We formalize these operations by matrix multiplications. We define three matrices,  $\Tb_k$, $\Rb$ et $\Sb$, which are associated with the operations of translation, convolution by the impulse-response of a pixel, and down-sampling, respectively. Thus, the direct problem can be summarized as
\begin{equation}
\label{eq:directproblem}
  \yb_k =  \Sb \Rb \Tb_k\xb + \nb_k
\end{equation}
with $k\in \{1,\ldots,N\}$. $N$ is the total number of images taken at different positions on the detector. $\nb$ is the noise, whose level is assumed to be commesurate for all the images. Note that in Eq.~\ref{eq:directproblem}, the only operation that is different from an image to another is the translation $\Tb_k$.

To obtain high values of the over-sampling factor (gain in resolution), one needs to know very precisely the translations between images. Unfortunately, this is not the case for our experimental setup. Although the hexapod (where the point source is mounted) is designed to allow fine motions of the source, we note significant mismatches between the expected position and the effective observed location of the source.
Therefore we decided to estimate the relative positions between images by a technique which is independant of the mechanical control of the hexapod. 
We perform cross-correlation calculations between a reference images and the other translated images in order to estimate the shifts between images. This is the first step of our inversion problem.

\begin{figure}
  \begin{center}
    \includegraphics[width=0.8\textwidth]{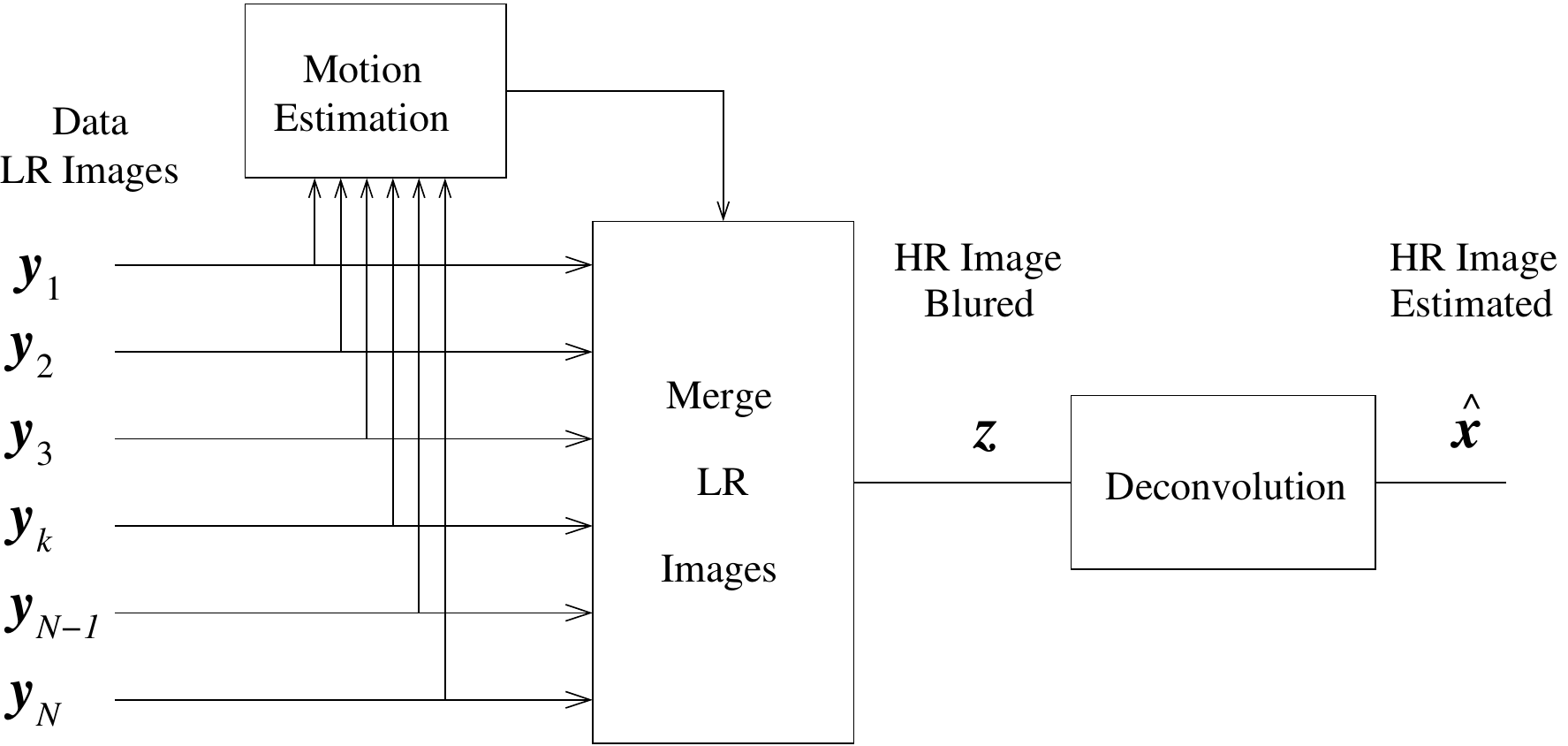}
    \caption[Sketch of the different steps to reconstruct the high-resolution image]{Sketch of the different steps to reconstruct the high-resolution image (inverse problem).}
    \label{fig_inverseModel}
  \end{center}
\end{figure}

\index{Microscanning test!inverse problem}
Fig.~\ref{fig_inverseModel} shows a sketch of the inverse problem. To reconstruct a high-resolution image from $N$ low-resolution ones, we proceed in three steps:
\begin{enumerate*}
\item we first estimate the translation between the low-resolution (LR) images by cross correlation (see sect.~\ref{subsec:translations} for details). 
\item then, we co-add the LR images on a fine grid, assuming that there is no convolution by the PSF of the detector.
\item finally, we solve the deconvolution problem. 
\end{enumerate*}
This algorithm is a simplified version of a more complex algorithm developped by \citet{Rodet2009}. 

\subsection{Estimate of the translations between images and co-addition}
\label{subsec:translations}
\index{Microscanning test!cross-correlation}
\index{Microscanning test!motions estimate}
Although the translation between the point source images is mechanically controlled, we note that the effective position of the source is slightly different from the expected position. Therefore, we compute the shifts between a reference image and the current translated image, and we compare the result to the expected position that the hexapod should have given.

We estimate the relative motions between the images with a cross-correlation technique. 
To determine sub-pixel translations between images, we perform a bilinear interpolation on an HR grid at a resolution equal to three times the over-sampling factor ($3 \times f$).
We define the first image as the the reference image $\Ib_{\rm ref}$ and we search, in an exhaustive manner, the displacement that maximizes the cross-correlation factor, $\rm Cor$, between the reference image and the current translated one, $\Ib$.  
The cross-correlation is computed as:
\begin{equation}
\label{eq:crosscorelation}
\rm Cor(\Ib,  \Ib_{\rm ref})= \frac{1}{\sigma _{\Ib_{\rm ref}} \, \sigma _{\Ib}} \sum _{i_{\rm r}, \, j_{\rm r}} \sum _{i, \, j} \left( \Ib _{i, \, j} - \langle \Ib \rangle \right) \left( \Ib _{\rm ref, \, i, \, j} - \langle \Ib_{\rm ref} \rangle \right) 
\end{equation}
In Eq.~\ref{eq:crosscorelation}, $\sigma _{\Ib_{\rm ref}}$ and $ \sigma _{\Ib}$ are the standard deviations of the matrices $\Ib_{\rm ref}$ and $\Ib$, respectively. $\langle \Ib_{\rm ref} \rangle$ and $\langle \Ib \rangle$ are the averages of $\Ib_{\rm ref}$ and $\Ib$, respectively.

\begin{figure}
  \begin{center}
    \includegraphics[width=0.49\textwidth]{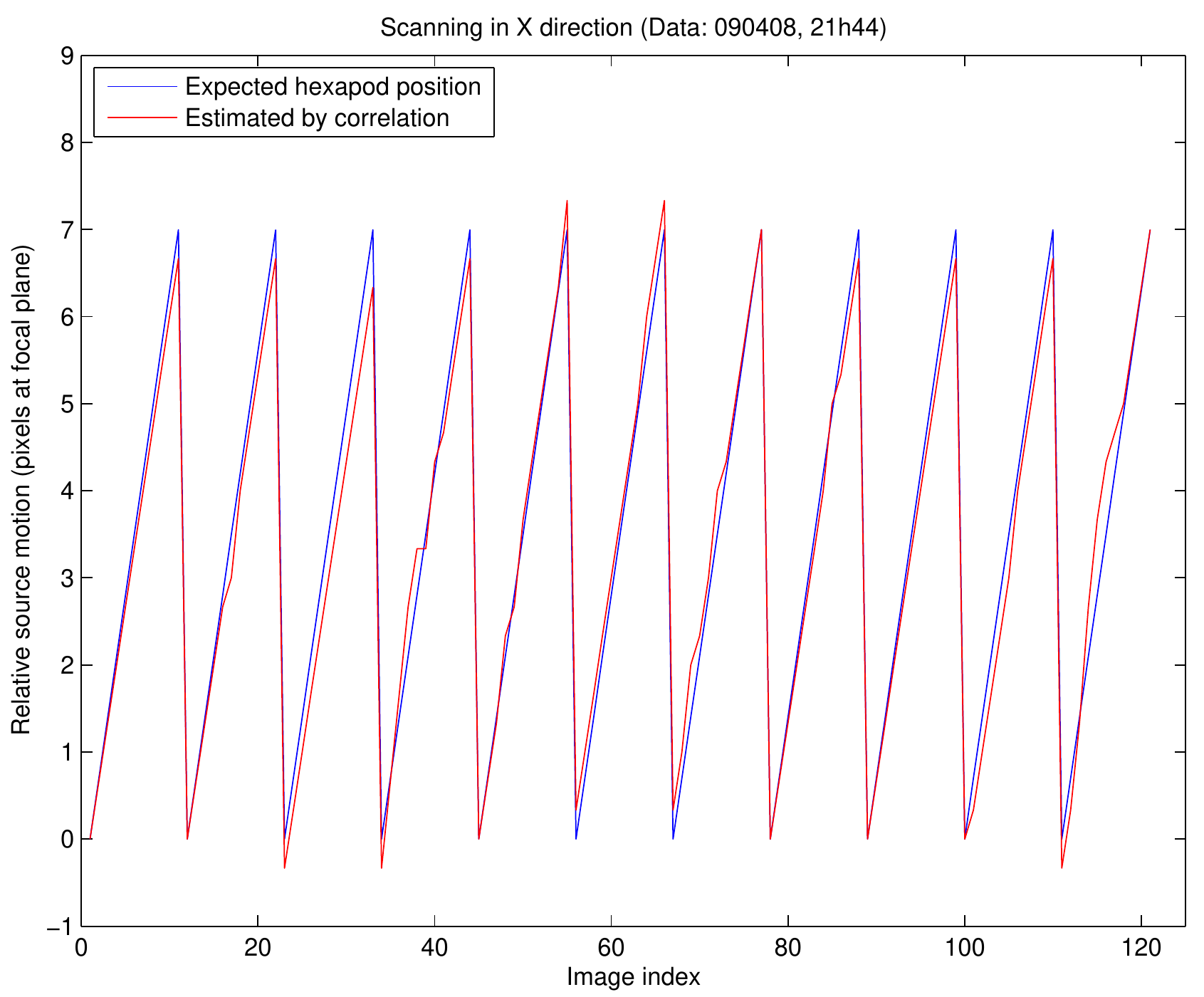}
	\includegraphics[width=0.49\textwidth]{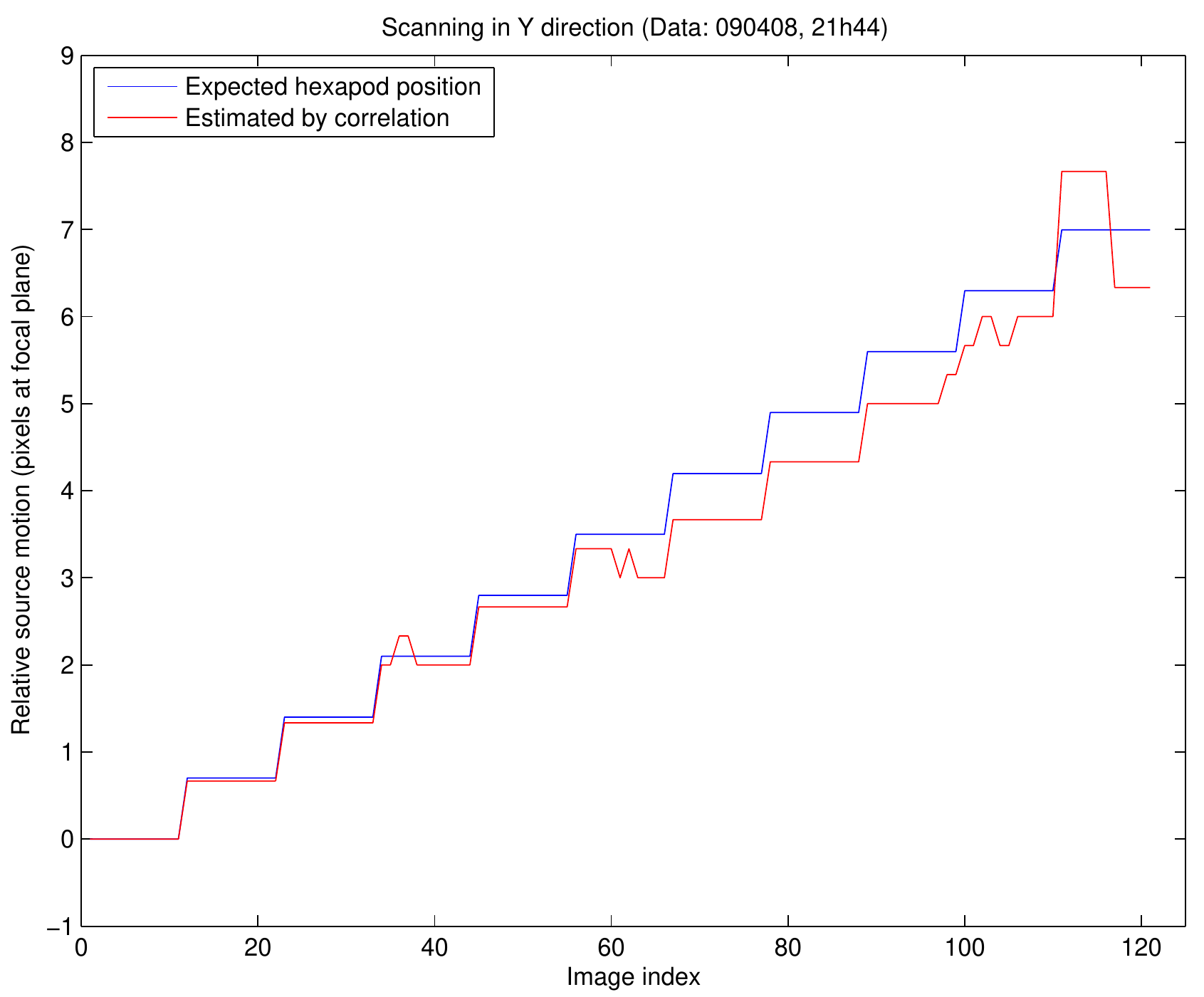}
    \caption[Expected vs. Estimated displacements of the source]{Expected vs. Estimated displacements of the source in the X and Y directions during the microscanning.  A slight discrepancy is seen between the estimated moves on the image and the expected position of the source for the $Y$-direction because, in practice, to control the absolute position of the hexapod is difficult. For the $X$-direction, the agreement is excellent. }
    \label{fig_Deplacement_source}
  \end{center}
\end{figure}

Then, these motions are compared with the expected values measured with the hexapod. A sub-pixel precision is obtained. 
Fig.~\ref{fig_Deplacement_source} shows an example of the comparison between the read position of the hexapod (expected position in blue) and the position estimated by cross-correlation (in red).
The agreement between expected and estimated positions for the displacement in the $X$-direction is very good. However, a slight discrepancy is seen for the $Y$-direction. The hexapod did not move at the expected positions in this direction. This issue is likely to be due to the control system of the hexapod support, which is not optimally designed to perform translations along the $X$ and $Y$ axis.  

The mismatch between estimated and expected relative motions between images introduces errors in the reconstruction method if one uses the expected positions. Therefore, the estimated motions are used to reconstruct the high resolution image. We also compare the results if one uses the expected motions.

Assuming that the translations $\Tb_k$ are known, we co-add low-resolution images $\yb_k$ onto an over-resolved image (fine grid). We choose the over-sampling factor $f$ such that $7 \leqslant f \leqslant 10$. Thus, a HR image $\zb$ is created but it is still blurred because of the response of the detector. The slight discrepancy between expected and estimated positions $\Tb_k$ represents a lack of information, in particular in the $Y$-direction. Some pixels of the HR image $\zb$ are not present in the data $\yb_k$.

\subsection{Deconvolution}
\index{Point Spread Function!deconvolution}
\index{Microscanning test!deconvolution}


\begin{figure}
  \begin{center}
    \includegraphics[width=0.49\textwidth]{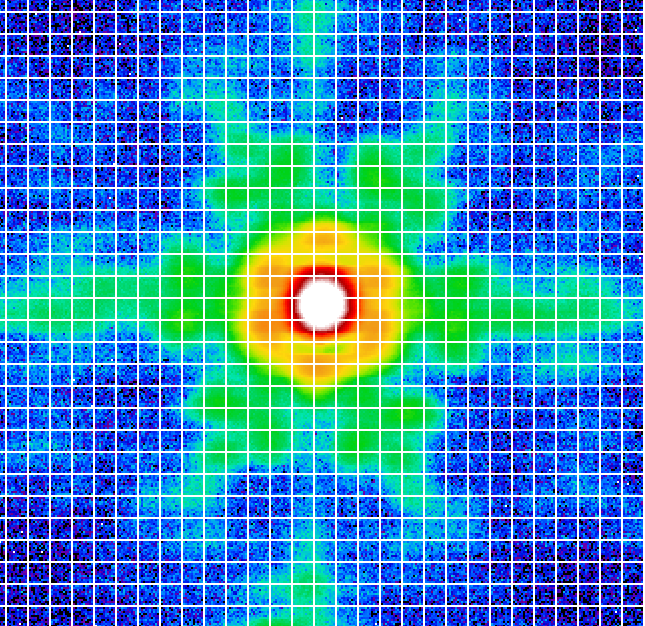}
	\includegraphics[width=0.49\textwidth]{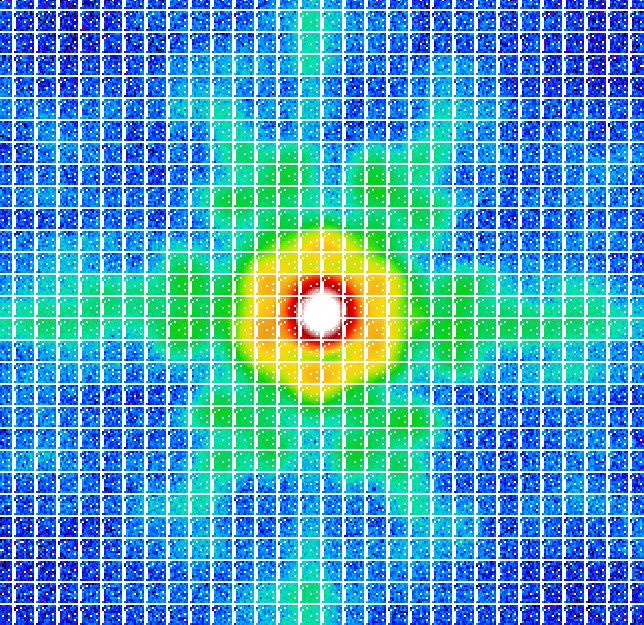}
    \caption[PSF data co-added on a fine grid]{PSF data co-added on a fine grid for an over-sampling factor of 11. The white pixels show the truncation of the data (lack of information).  The \textit{left} panel shows the co-added image when the expected shifts between images are used. 
Since the microscan comprises 10 shifts, the truncation of the data is a cross-hatching with 11 pixels between each white line. On the \textit{right} panel, the effective shifts estimated by cross-correlation are used. Additional truncation can be seen, due to the mismatch of the expected and effective translations between images (see text for details). }
    \label{fig_Data081206_00h15DecalLuFac11}
  \end{center}
\end{figure}

We denote $\zb$ the LR image that have been over-sampled on a fine grid. This image is still ``blurred'', but its number of pixels is a factor $f^2$ higher than the observed LR image.   
$\zb$ is the convolution of the HR image $\xb$ that we want to calculate by the impulse-response (PSF) of the detector:
\begin{equation}
\label{eq:HRimage}
\zb = \Rb \xb + \nb
\end{equation}
One cannot apply the classical methods of deconvolution because part of the data is not observed (due to uncertainties on the translations, see sect.~\ref{subsec:translations}). 
Indeed, there are missing parts, ``holes'', in the data because the source is not exactly at its expected position on the fine grid image.
The Fig.~\ref{fig_Data081206_00h15DecalLuFac11} shows the PSF data co-added onto a fine grid  with an over-sampling factor of 11. We illustrate the truncation of the data by comparing the results of the co-addition when using the expected positions of the source, to the  results when using the shifts estimated by cross-correlation. The additional truncation for the effective translations comes from the mismatch between expected and real position of the source.

We take into account the truncation of the data by introducing a matrix $\Tb$ in Eq.~\ref{eq:HRimage}:
\begin{equation}
\label{eq:HRimage_T}
\zb = \Tb \Rb \xb + \nb = \Hb \xb +  \nb
\end{equation}
$ \Tb$ is a mask with pixels value of $1$ where the data exists, and 0 otherwise. The operator $\Hb =  \Tb \Rb $ returns the measurements $\zb$ from the HR image $\xb$.

To solve the deconvolution problem, a Bayesian formalism is used. In practice, the deconvolution consists of minimizing a least-square criterion. The formalism and the method used to solve the deconvolution problem and estimate $\xb$ is detailed in Appendix~\ref{appendix_microscan}.

\section{Results}
\label{sec:results}
\index{MIRI!PSF}

We first describe the results of the first test campaign of the Flight Model (FM1), and show that the PSF characteristics of the intrument were out of specifications. After correction of the defect exhibited with these tests, we compare the results between the FM1 and FM2 test campaigns.

\subsection{Analysis of the MIRIM PSF}
\index{Point Spread Function (PSF)!analysis of MIRI's PSF}

The analysis of the 5.6~$\mu$m images taken during the microscanning  test was done with IDL routines. The results have been compared and complemented with the use of the IRAF software. Different measurements of the FWHM of the PSFs have been performed (Gauss, Bessel and moffat fits, radial profiles, etc). 
I have written a technical document for the MIRI European Consortium (EC) about the focus and FOV exploration with a point source tests was written (document MIRI-TN-00743-IAS). In this document (not reproduced here), the techniques used to measure the FWHM of the PSF are detailed. Here we rather focus on the main results of the microscanning test. These results are gathered in an other EC document (see Appendix~\ref{appendix:publications}, sect.~\ref{publications-technical-reports} for a list of these documents) and I have written a SPIE paper that will be submitted for the next SPIE conference ``Observational Frontiers of Astronomy for the New Decade'' in June 2010.

\begin{figure}
  \begin{center}
	\includegraphics[width=\textwidth]{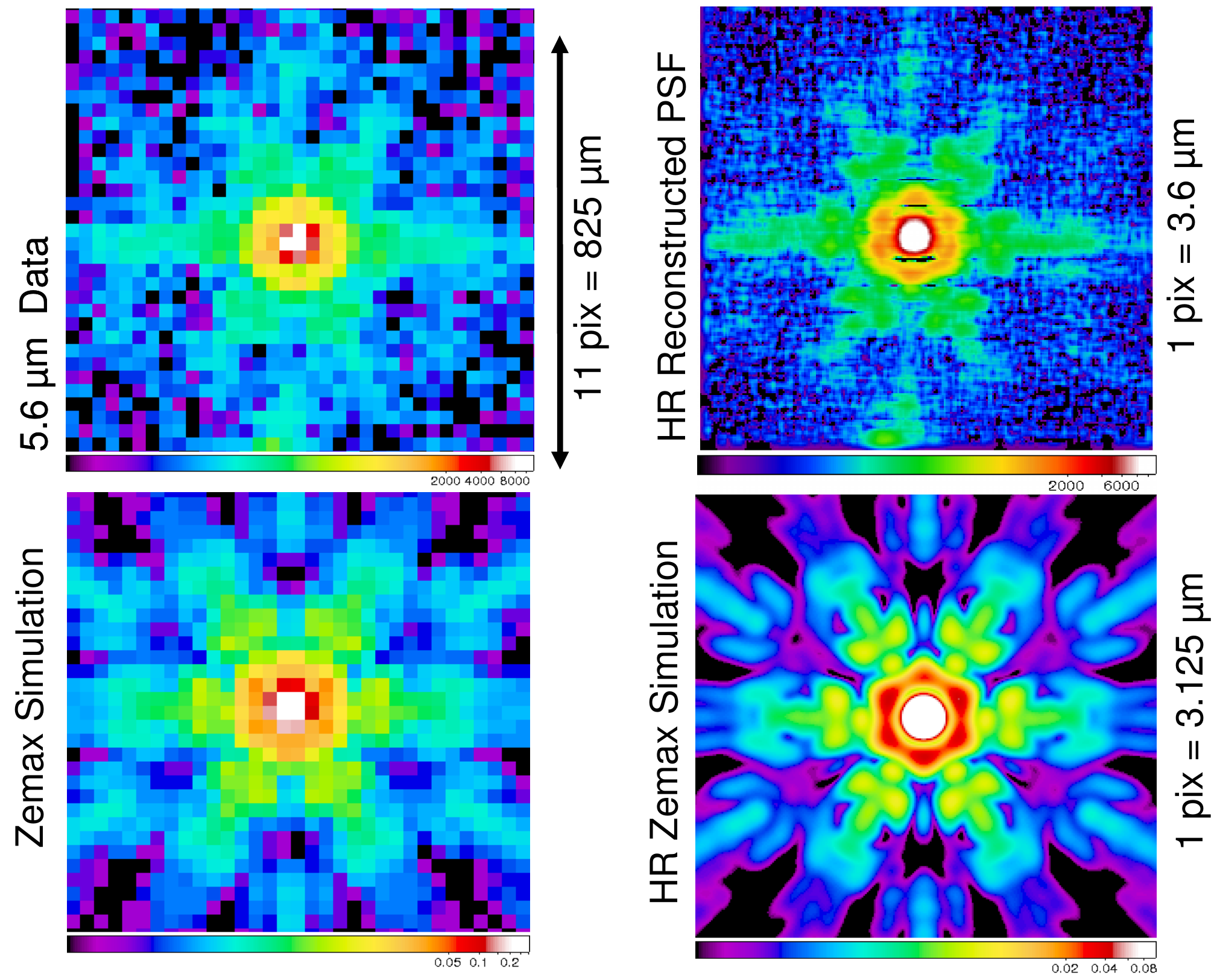}
    \caption[Low- and High-Resolution images]{Comparison between Low- and High-Resolution images (data in the upper panel, simulations at the bottom). The upper right image is the high-resolution image reconstructed from the data, with an over-sampling factor of 7. The bottom right image show a HR PSF image simulated with Zemax. Note that the shape of the secondary lobes agrees very well between the data and the simulation.}
    \label{fig_HR_vs_LR_images}
  \end{center}
\end{figure}

\begin{figure}
  \begin{center}
	\includegraphics[width=\textwidth]{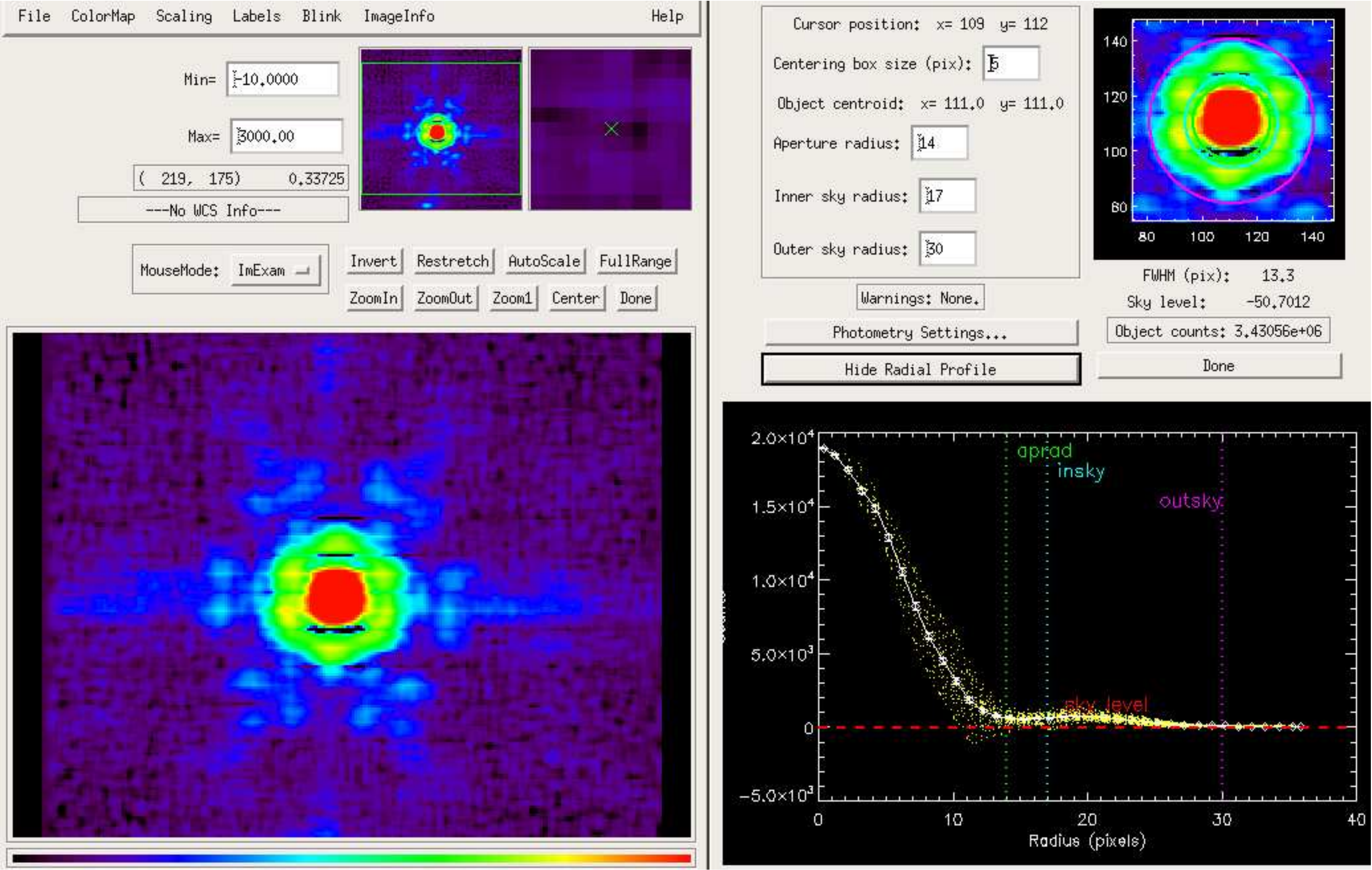}
    \caption[Image of the HR PSF and its radial 1D profile]{Image of the HR PSF and its radial 1D profile. The oversampling factor is 7 and the linearity correction has been applied. The pixel size equals to 3.6~$\mu$m. The radial profile has been fitted with an Airy disk  function. It gives a FWHM of $47.16 \, \mu$m.}
    \label{fig_PSF_microscan_radial_profile}
  \end{center}
\end{figure}

Fig.~\ref{fig_PSF_microscan_radial_profile} shows an example of reconstructed HR PSF. The linearity correction is applied and the over-sampling factor is set to 7, i.e. the size of a pixel is 3.6~$\mu$m instead of 25~$\mu$m for a standard MIRI image. The secondary Airy diffraction ring, which is barely visible on low-resolution images, is exhibited on the HR images, as well as the shape of the lobes. The radial profile of the HR PSF is shown on the right of Fig.~\ref{fig_PSF_microscan_radial_profile}. 

Different high-resolution PSF images and all the quantitative results for the FM1 tests are gathered in Table~\ref{tab_results_superPSF_FM1}. The results are shown  for images that have  been corrected for the detector response (linearity correction). 
These images correspond to over-sampling factors equal to 7, 10 and 11. The first value corresponds to a reasonable compromise between the gain in resolution and the introduction of artefacts in the reconstructed images. The last one (11) is chosen to show the limits of the method. In this case, the pixel size ($2.3$~$\mu$m) is smaller than the microscanning step ($2.5$~$\mu$m). For an over-sampling factor of $7$ (resp. $11$), one original MIRIM pixel equals to $7^2$ (resp. $11^2$) pixels in the super-resolution image. The pixel sizes in the HR images are  $\sim 3.6$~$\mu$m and $\sim 2.3$~$\mu$m, respectively. 

Table~\ref{tab_results_superPSF_FM1} also shows the results of the simulations of MIRIM PSFs with Zemax. These simulations are done at two resolutions, the standard one ($25$~$\mu$m / pixel) and a fine, re-sampled one ($3.125$~$\mu$m / pixel) which allows the comparison with the HR reconstructed PSFs.


The correction for the response of the detector decreases the width of the computed MIRIM PSFs. The reason is that the deviation to the linearity is greater at high fluxes than that at low values (see Fig.~\ref{fig_MIRI_Response_Curve} and \ref{fig_comp_lin_correction}). The FWHM of the corrected PSFs are $\sim 5\,$\% lower than the raw ones.
The $X$, $Y$, or radial profiles of the SR PSFs are well fitted by an Airy disk   (Fig.~\ref{fig_PSF_microscan_radial_profile}).
The measured FWHM with Airy fits on the HR PSF are of the order of $46-50 \ \mu$m for both $X$- and $Y$- directions. Gaussian fits to the core of the PSF give slightly lower values ($\sim 44-47$~$\mu$m).  

\begin{figure}
  \begin{center}
	\includegraphics[height=8.5cm, angle = 90]{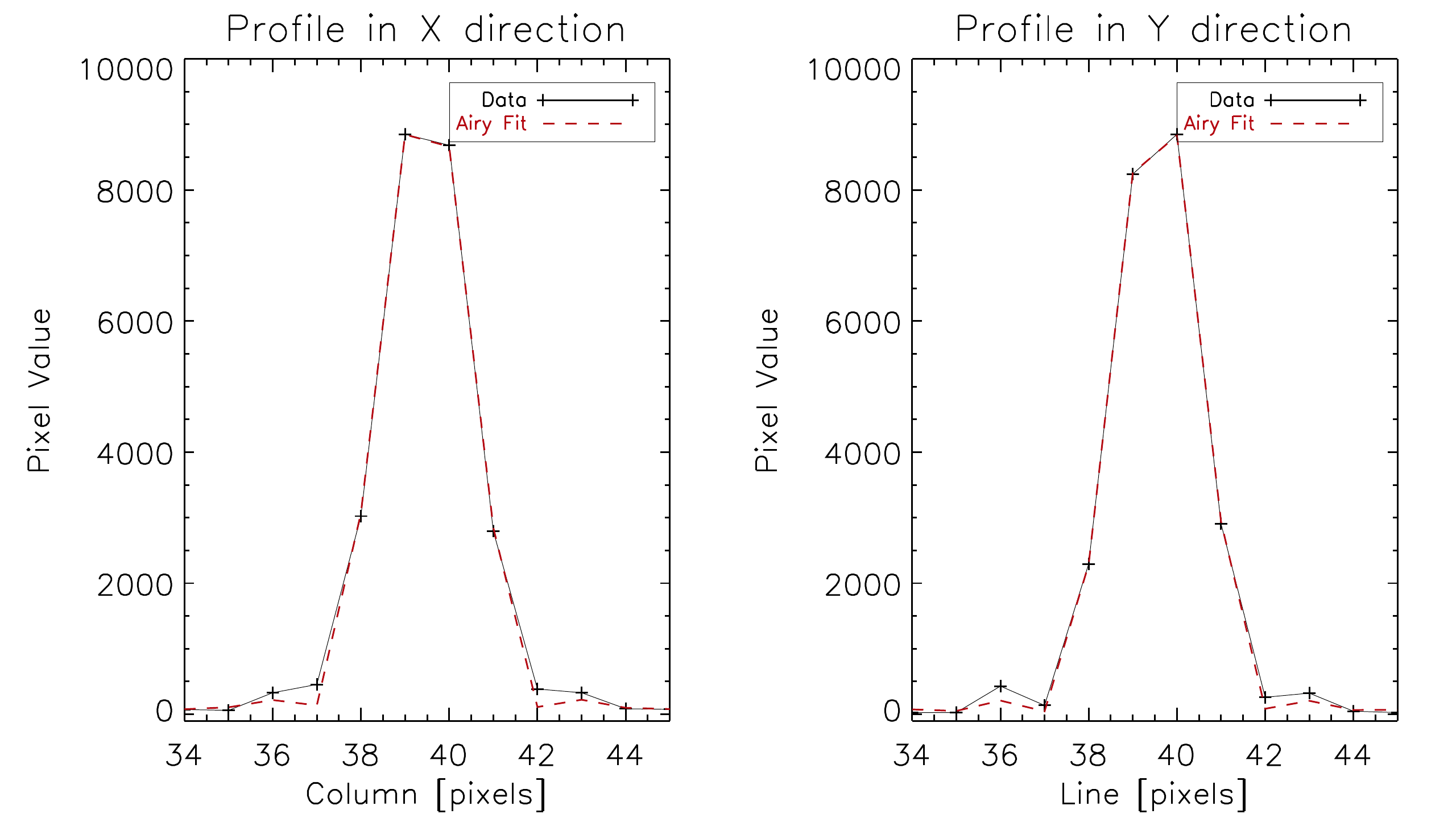}
    \includegraphics[height=8.5cm, angle = 90]{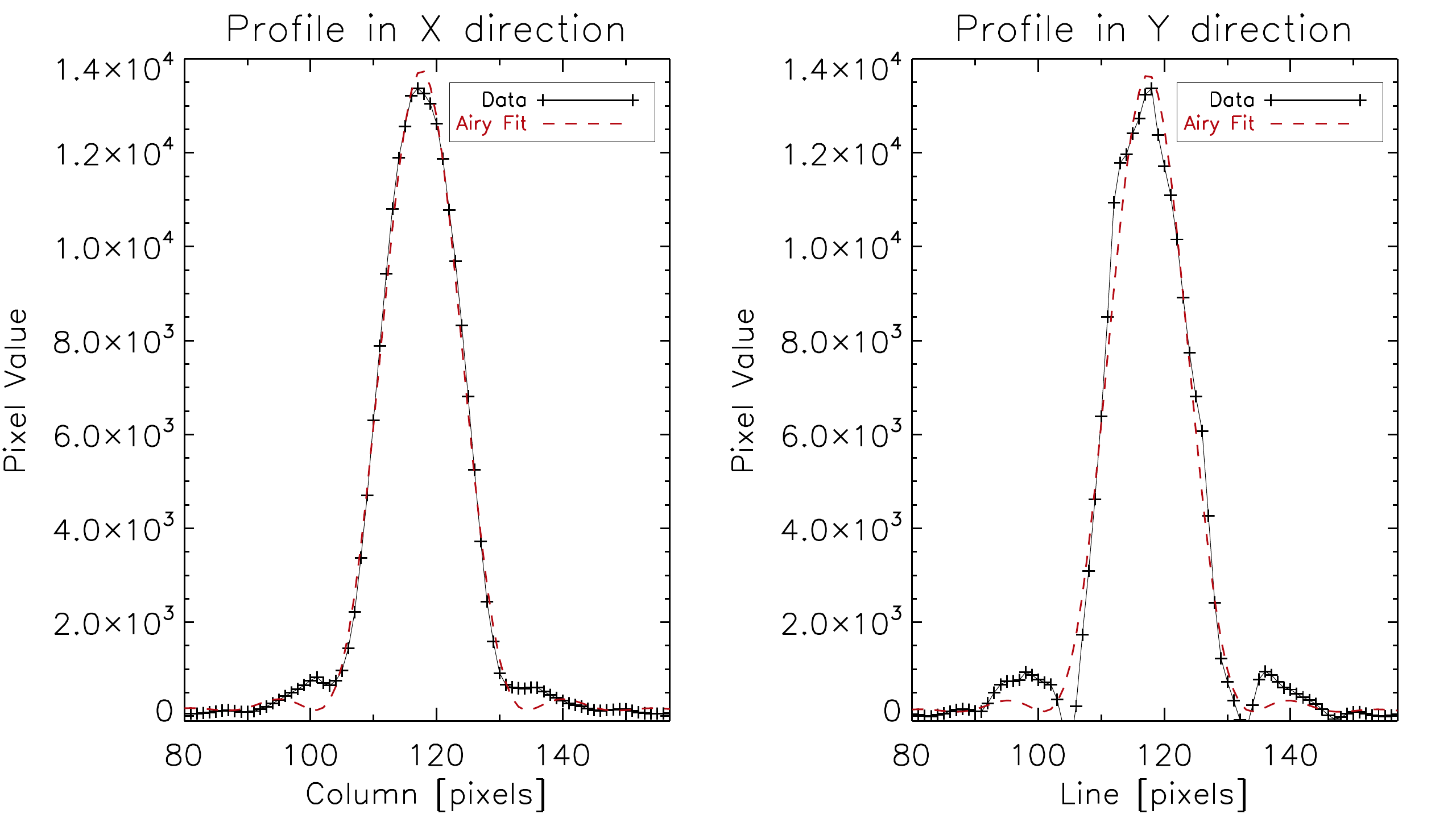}
    \caption[Low- and high-resolution MIRIM PSF profiles]{Low (2 upper plots ) and super-resolution (2 bottom plots) PSF profiles in the X and Y directions. The low-resolution PSF correspond to one single position MIRIM image. For the super-resolution PSF, the sampling is increased by a factor of 7. Red dashed lines are the results of the Airy fits.}
    \label{fig_PSF_profiles}
  \end{center}
\end{figure}

\begin{figure}
  \begin{center}
    \includegraphics[height=8.5cm, angle = 90]{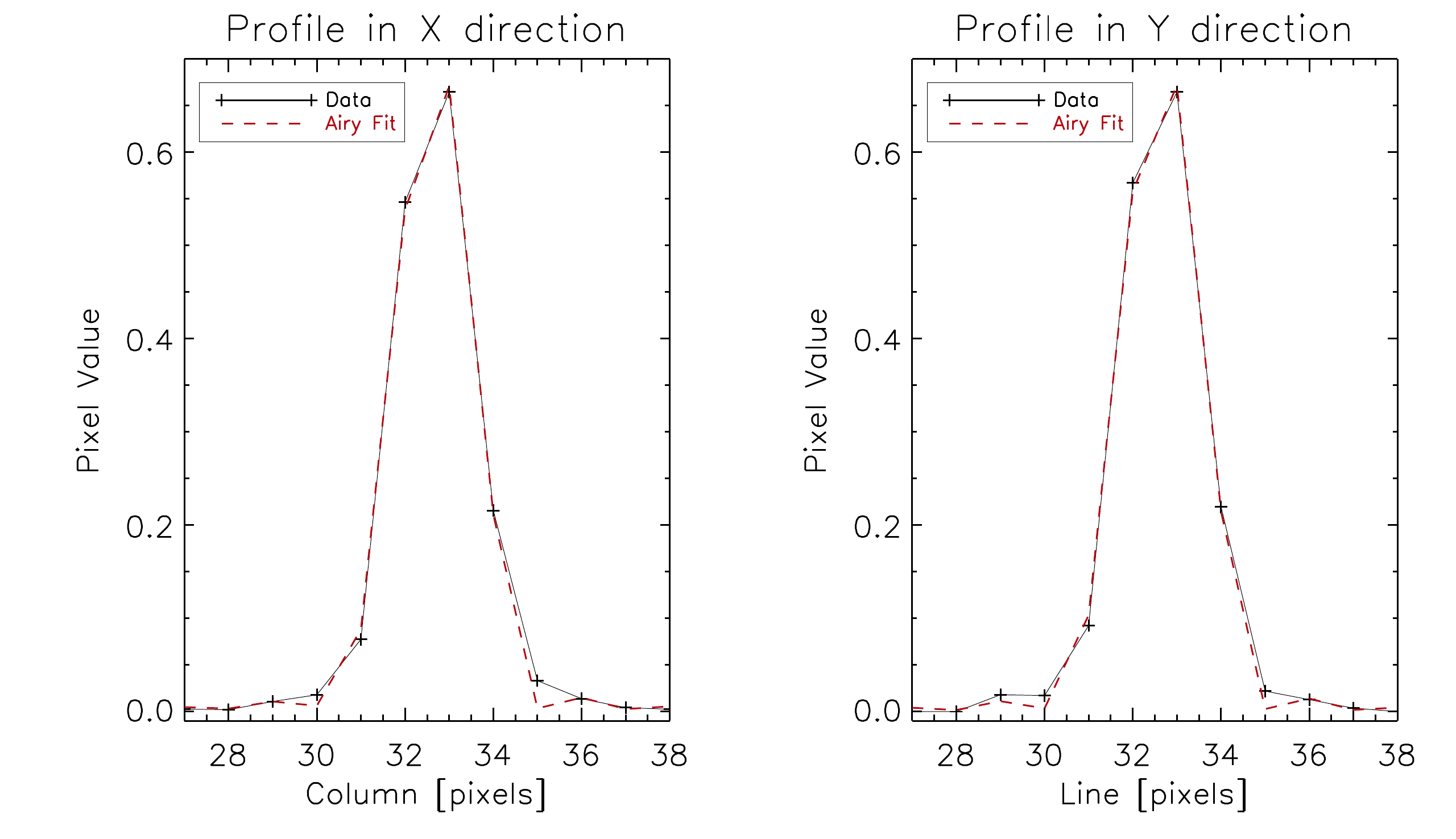}
    \includegraphics[height=8.5cm, angle = 90]{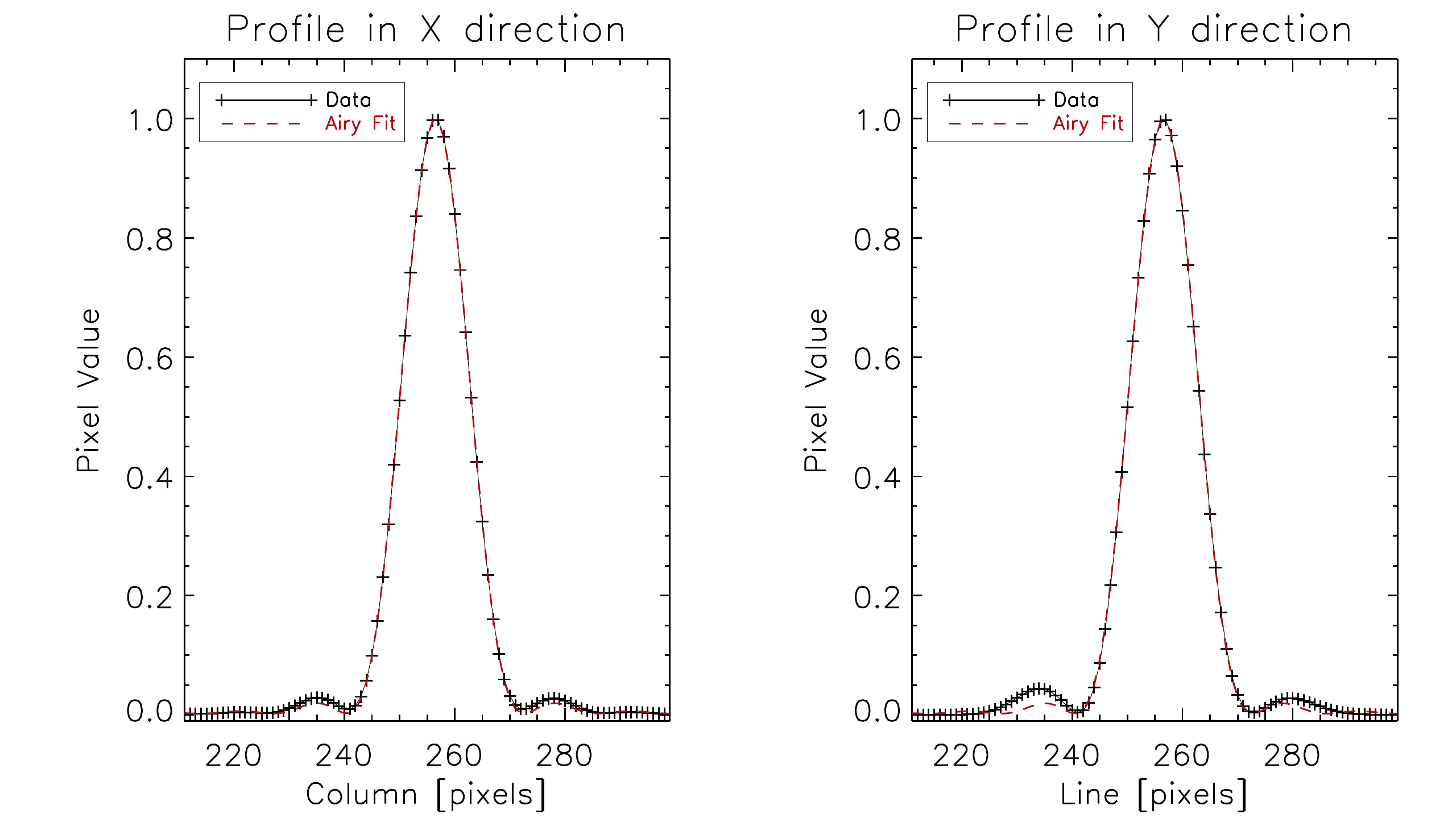}
    \caption[Zemax simulations of the MIRIM PSF]{Simulated (Zemax, with tolerances) PSF profiles in the X and Y directions at a low-resolution of 25~$\mu$m / pixel (2 first plots) and a high-resolution of $3.125$~$\mu$m / pixel (2 last plots).  Red dashed lines are the results of the Airy fits.}
    \label{fig_ZemaxPSF_profiles}
  \end{center}
\end{figure}

\begin{figure}
  \begin{center}
    \includegraphics[height=8.5cm, angle = 90]{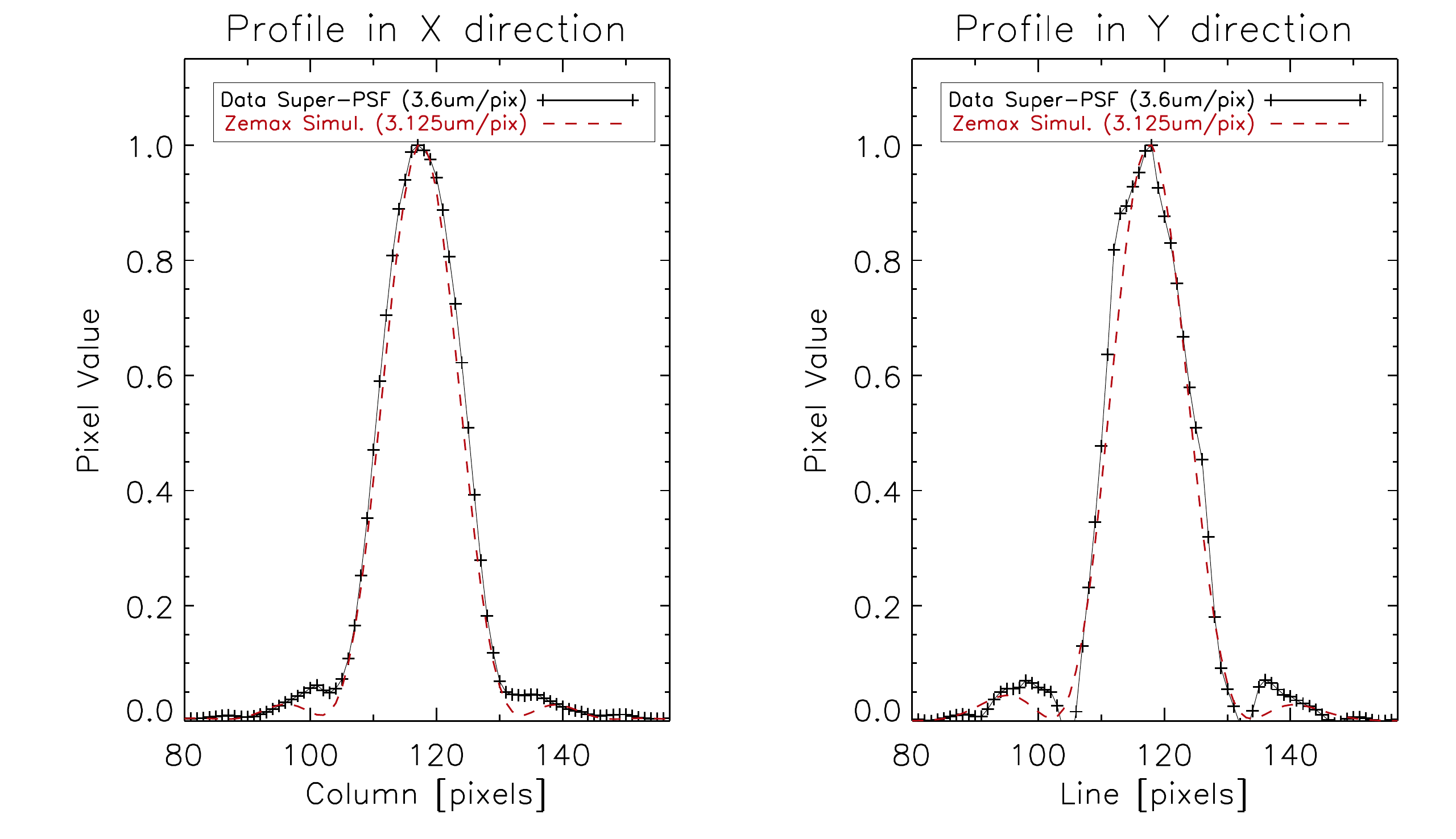}
    \includegraphics[height=8.5cm, angle = 90]{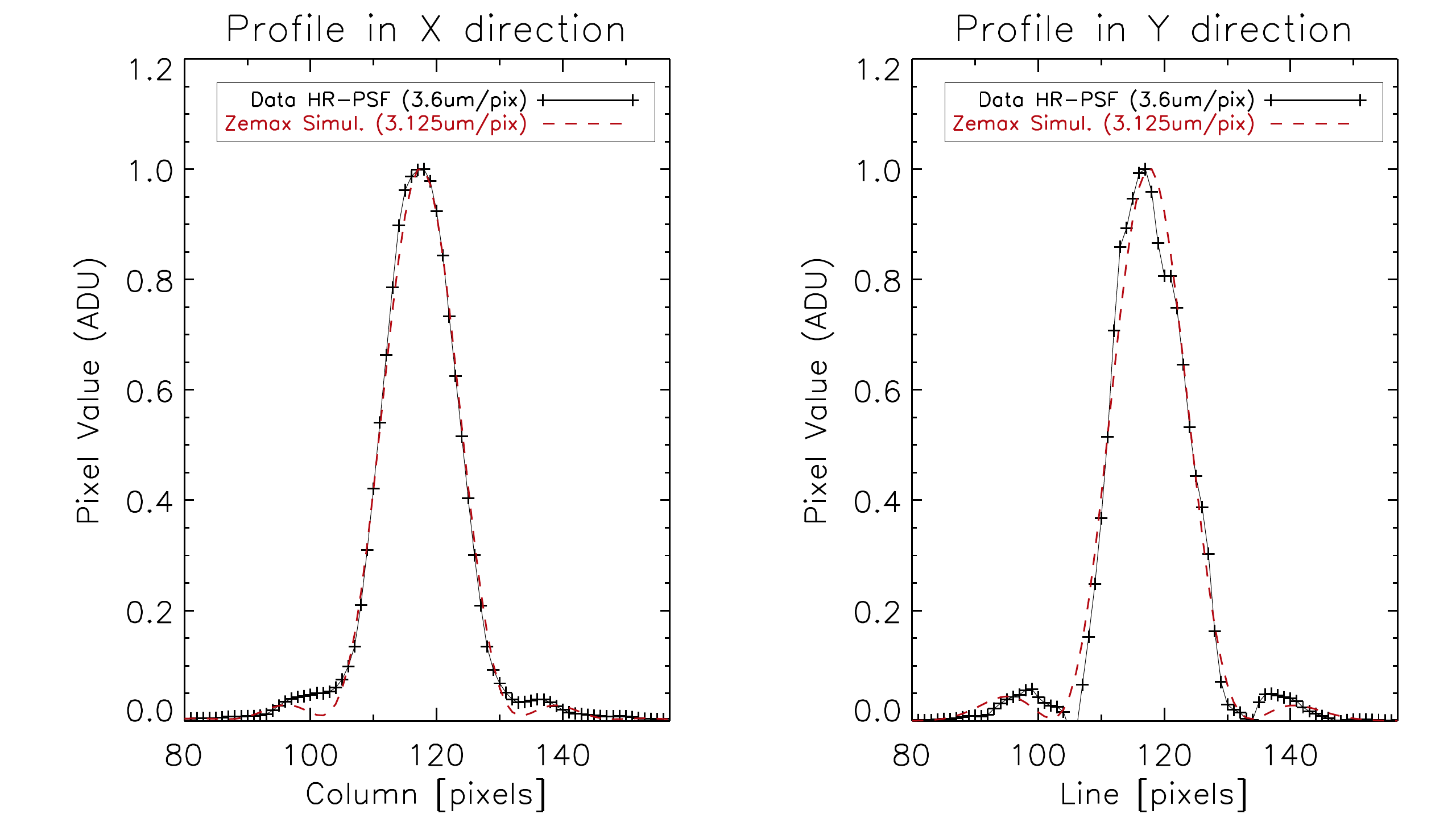}
    \caption[High-resolution data and simulations of MIRIM PSF]{ Comparison between the high-resolution reconstructed data ($3.6$~$\mu$m / pixel, black line) and the simulated (Zemax, with tolerances, 3.125~$\mu$m / pixel) PSF profiles (red dashed lines) in the X and Y directions. 
On the two first plots on the left, we do not correct for the response of the detector. On the two last plots on the right, linearity correction is applied. }
    \label{fig_ZemaxPSF_profiles2}
  \end{center}
\end{figure}

HR PSF profiles in the $X$- and $Y$- directions are shown on Fig.~\ref{fig_PSF_profiles}, \ref{fig_ZemaxPSF_profiles} and \ref{fig_ZemaxPSF_profiles2}. For comparison, the PSF profiles of one low-resolution image (one single position MIRIM image) is plotted for the corresponding number of pixels. Airy and Gaussian fits have been performed in the $X$- and $Y$- directions. We show on Fig.~\ref{fig_PSF_profiles} and \ref{fig_ZemaxPSF_profiles} the results of the Airy fits on the data and the Zemax simulations. The better sampling of the PSF helps in reconstructing more accurately the main peak of the PSF. The secondary diffraction lobes are also resolved. We note that the positions and amplitudes of the secondary lobes do not match exactly with the Airy fits. 

For an over-sampling factor of 11, we reach the limit of the method, and one can see artefacts on the reconstructed image (in both $X$- and $Y$- directions, see Table~\ref{tab_results_superPSF_FM1}). They may be due to errors in the motions estimations, signal-to-noise limitations, and simplifications made in the model of the detector and of the pixels (uniform response, etc.). For an over-sampling factor of 7, most of the artefacts are removed in the $X$- directions. The remaining artefacts in the $Y$-direction are due to the errors in the motion estimation  (see Fig.~\ref{fig_Deplacement_source}). The $Y$- direction profile exhibit a slight distorsion of the shape of the main peak (Fig.~\ref{fig_ZemaxPSF_profiles2}).


\renewcommand{\arraystretch}{2} 
\begin{sidewaystable}
\begin{center}
\begin{tabular}{|c|c c|c c|c c|c c|c c|}
\hline
\hline
Method &  \multicolumn{6}{c |}{Data} & \multicolumn{4}{c |}{Simulations} \\
\hline
&  \multicolumn{2}{c |}{Low-Resolution} & \multicolumn{2}{c |}{Oversampling 7} & \multicolumn{2}{c|}{Oversampling 11} & \multicolumn{2}{c|}{Zemax Low-R}& \multicolumn{2}{c|}{Zemax High-R}\\ 
\hline
 FWHM &  X &  Y &  X &  Y &  X &  Y &  X &  Y &  X & Y \\
\hline
Gauss & 51.54 & 51.27 & 47.47 & 48.51 & 47.40 & 48.42 & 46.33 & 46.20 & 40.15 & 40.07 \\
\hline
Airy    & 56.84 & 55.61 & 50.57 & 49.91 & 50.69 & 49.78 & 51.08 & 52.10 & 42.37 & 42.38 \\
\hline
Radial  &  \multicolumn{2}{c|}{53.75} &  \multicolumn{2}{c|}{51.43} & \multicolumn{2}{c|}{47.64} &  \multicolumn{2}{c|}{49.75} &  \multicolumn{2}{c|}{43.60}\\
\hline
Direct  &  \multicolumn{2}{c|}{50.50} &  \multicolumn{2}{c|}{50.03} & \multicolumn{2}{c|}{49.71} &  \multicolumn{2}{c|}{47.75}&  \multicolumn{2}{c|}{41.88}\\
\hline
 &  \multicolumn{2}{c}{\includegraphics[width=3cm]{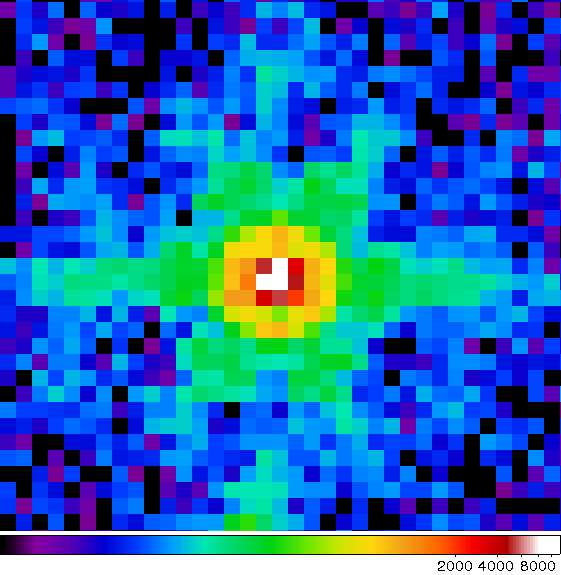}} &  \multicolumn{2}{c}{\includegraphics[width=3cm]{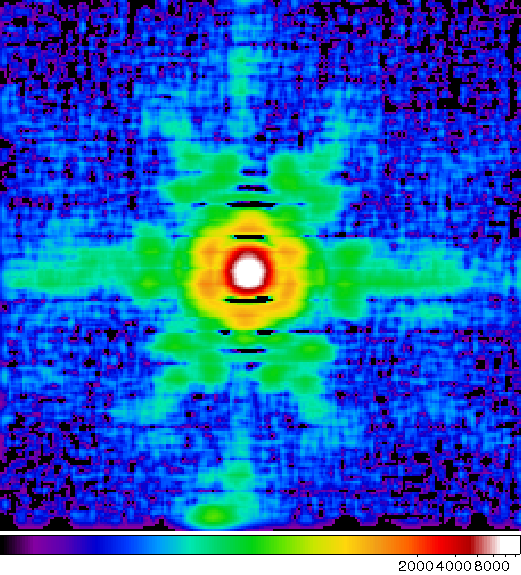}} &  \multicolumn{2}{c}{\includegraphics[width=3cm]{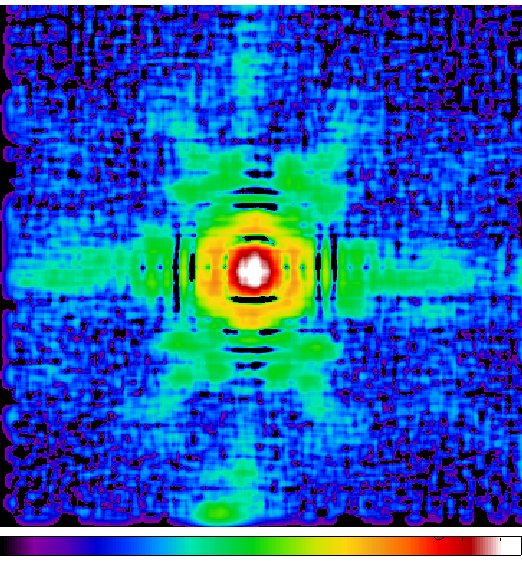}} &  \multicolumn{2}{c}{\includegraphics[width=3cm]{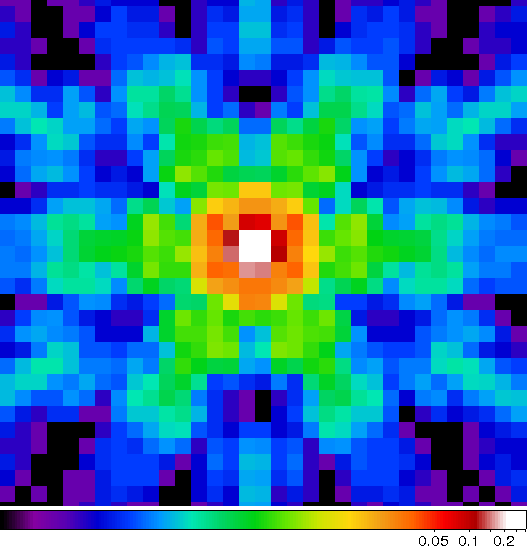}} &  \multicolumn{2}{c|}{\includegraphics[width=3cm]{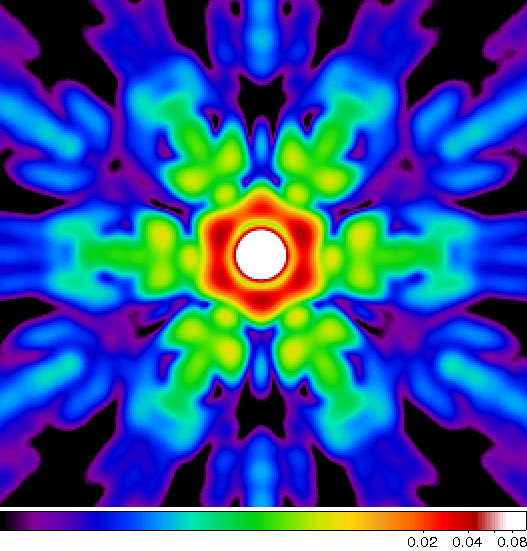}}\\
\hline
\end{tabular}
 \caption[MIRIM PSF measurements for FM1 test campaign]{{\it Summary of the results for the FM1 campaign. The linearity correction is applied. FWHM measurements (in $\mu$m)  on the low-, super-resolution (resulting image from the microscanning test), and the simulated PSF. For the super-resolved image, the sampling is increased by a factor of 7 or 11 with respect to a single MIRIM image, i.e. one original MIRIM pixel equals to $7^2$ or $11^2$ pixels in the super-resolution image. The images shows a 825$\, \mu$m wide area. Pixels sizes are respectively 25$\, \mu$m, 3.6$\, \mu$m, 2.3$\, \mu$m, 25$\, \mu$m, and 3.125$\, \mu$m. The display scale is logarithmic. The results of the PSF FWHM are indicated for 4 different measurements: a 2D gaussian elliptical fit (longest FWHM and perpendicular),  a 2D Airy disk fit ($X$ and $Y$ directions), a gaussian fit of the radial profile of the source, and a direct measurement of the FWHM, without assuming a profile shape. The $1\, \sigma$ uncertainty on the FWHM values are of the order of $0.03\, \mu$m. The Zemax simulations were performed on a case where mechanical tolerances were applied in the Monte-Carlos.}}
\label{tab_results_superPSF_FM1}
\end{center}
\end{sidewaystable}
\renewcommand{\arraystretch}{1} 

\subsubsection{Conclusion of the FM1 tests: a defect in the instrument}

\begin{figure}
  \begin{center}
    \includegraphics[width=\textwidth]{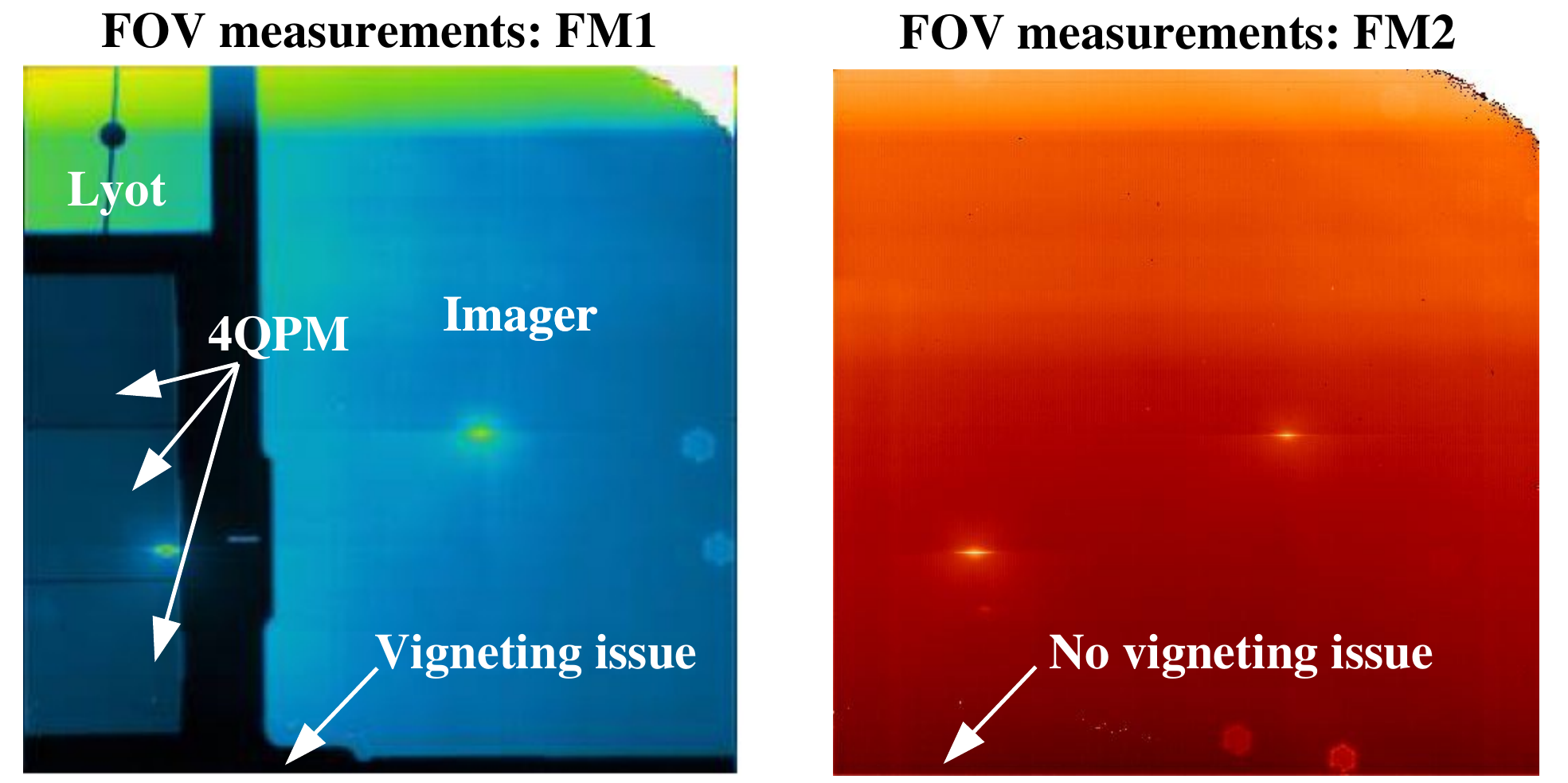}
    \caption[Field of View measurements for the two MIRIM FM test campaigns]{Comparison of the Field of View measurement between the two Flight model test campaigns, FM1 and FM2. 
On the left plot (FM1), the complete MIRIM focal plane is mounted, with the coronograph masks and the LRS slit (see Fig.~\ref{fig_MIRI_layout} for the pattern of the focal plane). The vignetting issue is clearly seen at the bottom of the image. The black stripe at the bottom shows that a significant fraction of the field of view is lost. On the right plot (FM2), the vignetting issue has been corrected (no black stripe anymore). Note that for the FM2 tests, the coronograph masks are not mounted. }
    \label{fig_FOV_measurements_FM1_2}
  \end{center}
\end{figure}

From the results gathered in Table~\ref{tab_results_superPSF_FM1}, we note that the FWHM of the HR PSFs are systematically larger than the measured FWHM on the HR simulated Zemax PSF. The discrepancy is about $+8$~$\mu$m after correction for the non-linearity of the detector. This mismatch is out of the specifications of the instrument. 
This out-of-specification issue was also revealed independently by an another test. The measurement of the MIRIM field of view shows a vignetting issue, which seems to confirm that a defect is present in the instrument. The left panel of Fig.~\ref{fig_FOV_measurements_FM1_2} shows the field of view of the imager during the FM1 test campaign. Note that the bottom of the FoV is lost. 

Motivated by these tests, a new metrology analysis of the instrument was done by engineers from the CEA, Saclay and IAS, Orsay. The engineers from the IAS provided a ``Faro arm'', a special tool to perfom the metrology.  
It was first thought that the root cause of the out-of-specification optical quality of the instrument was due to a misalignment of the detector. It turned out that this was not the case. In-depth investigations concluded to a slight interference between the structure of the Three-Mirror-Anastigmatic (TMA) objective and the M4 mirror (see Fig.~\ref{fig_MIRI_layout} and sect.~\ref{subsec:MIRIcamera} for a description of the optical layout of MIRIM). This design problem caused a tilt of the M4 mirror with respect to its nominal positioning.
The M4 mirror has been found damaged, with some bending of its mounting flexible pad. 
 This tilt results in a degradation of the nominal PSF properties and the FoV area.

The cause of the interference between the TMA structure and the M4 mirror has been removed manually. The damaged M4 mirror was replaced by the Flight Spare model. 
A second test campaign (FM2) was launched, and we performed the same microscanning analysis to check the optical quality of the PSF.
This story illustrates well the need to perform accurate tests on the ground! 
Indeed, the discrepancy on the PSF characteristics could not have been exhibited without the deconvolution of high-resolution images from the microscanning data. 

\subsection{Results of the FM2 tests after correction of the M4 tilt}

\subsubsection{Comparison between FM1 and FM2 tests}

\renewcommand{\arraystretch}{2} 
\begin{sidewaystable}
\begin{center}
\begin{tabular}{|c|c c|c c|c c|c c|c c|}
\hline
\hline
Method &  \multicolumn{6}{c |}{Data} & \multicolumn{4}{c |}{Simulations} \\
\hline
&  \multicolumn{2}{c |}{Low-Resolution} & \multicolumn{2}{c |}{Oversampling 7} & \multicolumn{2}{c|}{Oversampling 10} & \multicolumn{2}{c|}{Zemax Low-R}& \multicolumn{2}{c|}{Zemax High-R}\\ 
\hline
FWHM &  X &  Y &  X &  Y &  X &  Y &  X &  Y &  X &  \\
\hline
Gauss & 51.54 & 51.27 & 44.88 & 45.70 & 44.87 & 45.70 & 46.33 & 46.20 & 40.15 & 40.07 \\
\hline
Airy    & 56.84 & 55.61 & 46.93 & 47.29 & 46.63 & 47.14 & 51.08 & 52.10 & 42.37 & 42.38 \\
\hline
Radial  &  \multicolumn{2}{c|}{53.75} &  \multicolumn{2}{c|}{47.10} & \multicolumn{2}{c|}{46.15} &  \multicolumn{2}{c|}{49.75} &  \multicolumn{2}{c|}{43.60}\\
\hline
Direct  &  \multicolumn{2}{c|}{50.50} &  \multicolumn{2}{c|}{48.67} & \multicolumn{2}{c|}{48.77} &  \multicolumn{2}{c|}{47.75}&  \multicolumn{2}{c|}{41.88}\\
\hline
 &  \multicolumn{2}{c}{\includegraphics[width=3cm]{figures/microscan/MICROSCAN_081206_00h15_0013.png}} &  \multicolumn{2}{c}{\includegraphics[width=3cm]{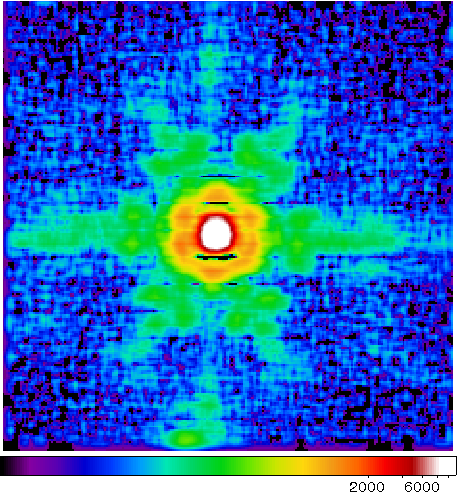}} &  \multicolumn{2}{c}{\includegraphics[width=3cm]{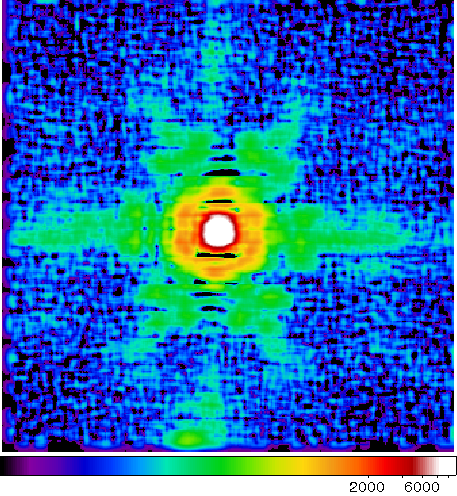}} &  \multicolumn{2}{c}{\includegraphics[width=3cm]{figures/microscan/PSF_poly_5um6_pixsize25um_zone5_TOL_FM_sqrt.png}} &  \multicolumn{2}{c|}{\includegraphics[width=3cm]{figures/microscan/PSF_poly_5um6_pixsize3um125_zone5_TOL_FM_sqrt.png}}\\
\hline
\end{tabular}
 \caption[MIRIM PSF measurements for FM2 test campaign]{{\it Summary of the results for the FM2 test campaign. The linearity correction is applied. FWHM measurements (in $\mu$m)  on the low-, high-resolution (resulting image from the microscanning test), and the simulated PSF. For the HR image, the sampling is increased by a factor of 7 or 11 with respect to a single MIRIM image, i.e. one original MIRIM pixel equals to $7^2$ or $11^2$ pixels in the super-resolution image. The images shows a 825$\, \mu$m wide area. Pixels sizes are respectively 25$\, \mu$m, 3.6$\, \mu$m, 2.3$\, \mu$m, 25$\, \mu$m, and 3.125$\, \mu$m. The display scale is logarithmic. The results of the PSF FWHM are indicated for 4 different measurements: a 2D gaussian elliptical fit (longest FWHM and perpendicular),  a 2D Airy disk fit ($X$ and $Y$ directions), a gaussian fit of the radial profile of the source, and a direct measurement of the FWHM, without assuming a profile shape. The $1\, \sigma$ uncertainty on the FWHM values are of the order of $0.03\, \mu$m. The Zemax simulations were performed on a case where mechanical tolerances were applied in the Monte-Carlos.}}
\label{tab_results_superPSF_FM2}
\end{center}
\end{sidewaystable}
\renewcommand{\arraystretch}{1} 

The results of the microscanning analysis and PSF measurements for the FM2 test campaign are gathered in Table~\ref{tab_results_superPSF_FM2}.
Gaussian fits to the PSFs show that the FWHM are of the order of 44$\,\mu$m (0.19~arcsec) at 5.6$\,\mu$m, which brings this value inside the specification range\footnote{The specification on the PSF is not expressed in terms of FWHM, but in terms of wavefront errors (WFE).  A zemax calculation, performed by S. Ronayette (CEA), was used to translate the WFE into FWHL tolerances.}. The effect of the correction of the tilf of the M4 mirror is clearly visible on the HR reconstructed PSFs.  Note that this effect is not measurable of the   raw (LR) images, which emphasizes the importance of this microscanning test and analysis.

\subsubsection{Encircled energy of a point source image}
\index{Point Spread Function (PSF)!encircled energy}

\begin{figure}
  \begin{center}
    \includegraphics[angle=90,width=0.8 \textwidth]{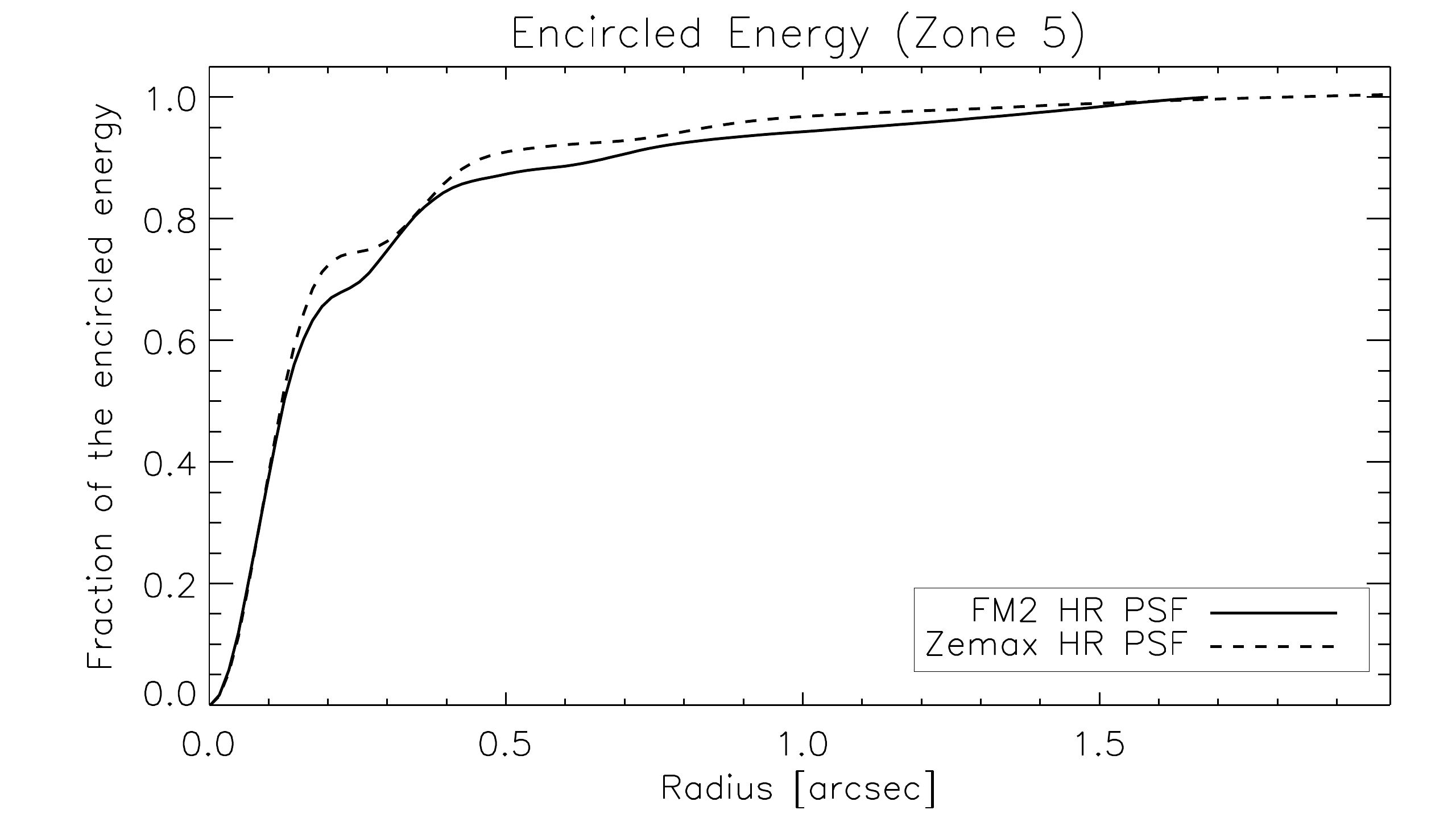}
    \caption[Encircled  energy of the MIRI high-resolution PSF]{Encircled energy of the MIRI high-resolution (HR) PSF  as a function of radius in arcsec (1 pixel = 0.11''). The solid line is the result of our deconvolution algorithm to reconstruct a HR image from the data. The dashed line shows a PSF simulated with the Zemax software. The results are shown for a position close to the center of the field fo view of the imager. The inflection point corresponds to the first dark ring of the Airy disk of the PSF.}
    \label{fig:MIRI_PSF_EE_Z5}
  \end{center}
\end{figure}

The encircled energy is another way of estimating the optical quality of the PSF. The requirements for MIRI are the following. The MIRI imager data shall include more than  $56\,$\% of the encircled energy of the image of a point source at wavelengths longward of $5.6\,\mu$m within the angular diameter of the first dark Airy ring. Note that in the MIRI requirement document, the encircled energy is defined relative to the energy within a radius of 5~arcsec equivalent projected onto the sky.

The encircled energy $E(r)$ of a radially symmetric point spread function $\phi(r)$ can be defined as the integral
\begin{equation}
E(r) = 2 \pi \int_{0}^{r} \phi (r') d r'
\end{equation}
For the purpose of comparison to the requirement document, we wish to normalise $E(r)$ to 1 at $r =5~$arcseconds. However, our microscanning data acquisition was done for a sub-array of $\approx 2'' \times 2''$, so we have normalized the encircled energy to a 2~arcsec radius. Therefore it is difficult to relate our results to the specification requirements at 5~arcsec.

Fig.~\ref{fig:MIRI_PSF_EE_Z5} shows the encircled energy of a high-resolution PSF in the center of the field of view of the imager, as a function of radius in arcsec (1 pixel = 0.11''). The dashed line shows the encircled energy of a high-resolution MIRIM PSF simulated with the Zemax software. The agreement between the two curse is very good, suggesting that the instrument is close to its nominal behaviour. The diameter of the first dark ring can be  interpreted as the first point of inflection in $E(r)$, i.e the value of $r$ for which $d^2E/dr^2 = 0$. We find an encircled energy of 73$\,$\% inside the central lobe. 

Note that the measured variation of encircled energy with radial distance from a point image is very sensitive to the flux levels at large radial distances. One must ensure that $dE/dr \geqslant 0$ for all values of $r$ for the measurement to make sense. Here, this condition is satisfied because we work at high background values. The telescope simulator is at ambient temperatures, thus producing a high thermal background.


\section{Summary and conclusions}
\label{sec:microscan-conclusion}


This chapter reports the first detailed analysis of the characteristics of the Point Spread Function (PSF) of the Flight Model of the JWST Mid-InfraRed IMager (MIRIM). The data are taken at  5.6$\, \mu$m, the only filter available at that time. 
We discuss the results of the \textit{``microscanning''} test that has been performed at cryogenic temperatures in December 2008 - March 2009 at CEA, Saclay. This \textit{high-resolution} (HR)  technique is used  to characterize the MIRIM optical quality.  The microscanning consists in a fine, sub-pixel scanning of a point source on the focal plane. A deconvolution algorithm is used to reconstruct HR PSFs. The main results are the following.

\begin{enumerate}
\item The microscanning test provides a significant improvement for the analysis of the PSF characteristics. This test allows to resolve the diffraction pattern of MIRIM and to measure accurately the width of the PSF for the FM. We confirm the global shape of the PSF secondary lobes that is seen on HR Zemax simulations.

\item 
 The FWHM of the high-resolution, reconstructed images are $\sim 48-50\,\mu$m, i.e. $0.21 - 0.22\ $" for the first test campaign. We note a discrepancy of $\sim + 8$~$\mu$m between the data and the high-resolution Zemax simulations. The microscanning test, together with field of view measurements, showed that a defect was present in the instrument. The main cause for the mismatch between simulations and MIRIM data has been identified. Metrology measurements of the MIRIM mechanical and optical layout have shown a out-of-specification tilt of the M4 mirror. This tilt was due to an interference between the MIRIM structure and the mirror. This manufacturing fault has been corrected, and a second test campaign began in February 2009. 
The results show that the FWHM of the HR PSFs are $44-46\,\mu$m, i.e. $0.19'' - 0.20''$, which is within the specifications. The FWHM are only $\sim + 4$~$\mu$m compared to high-resolution simulations. 

\item Slight differences between the simulated PSF patterns and the MIRIM HR reconstructed data are exhibited. We discuss the limits of the method due to errors in the relative positions of the images and the lack of information (signal-to-noise limitations, assumptions made on the characteristics of the detector, etc.). 

\item The MIRI detector shows a non-linear response curve. This result is confirmed by independant measurements by the Jet Propulsion Laboratory (JPL), responsible for the detector array. We show that the correction for the response of the detector improves the quality of the MIRIM PSFs. The FWHM of the corrected PSFs are $\sim 5\,$\% lower than the raw ones.

\end{enumerate}

\section{What's next? }
\label{sec:MIRItestsNext}

\begin{figure}
  \begin{center}
    \includegraphics[width=\textwidth]{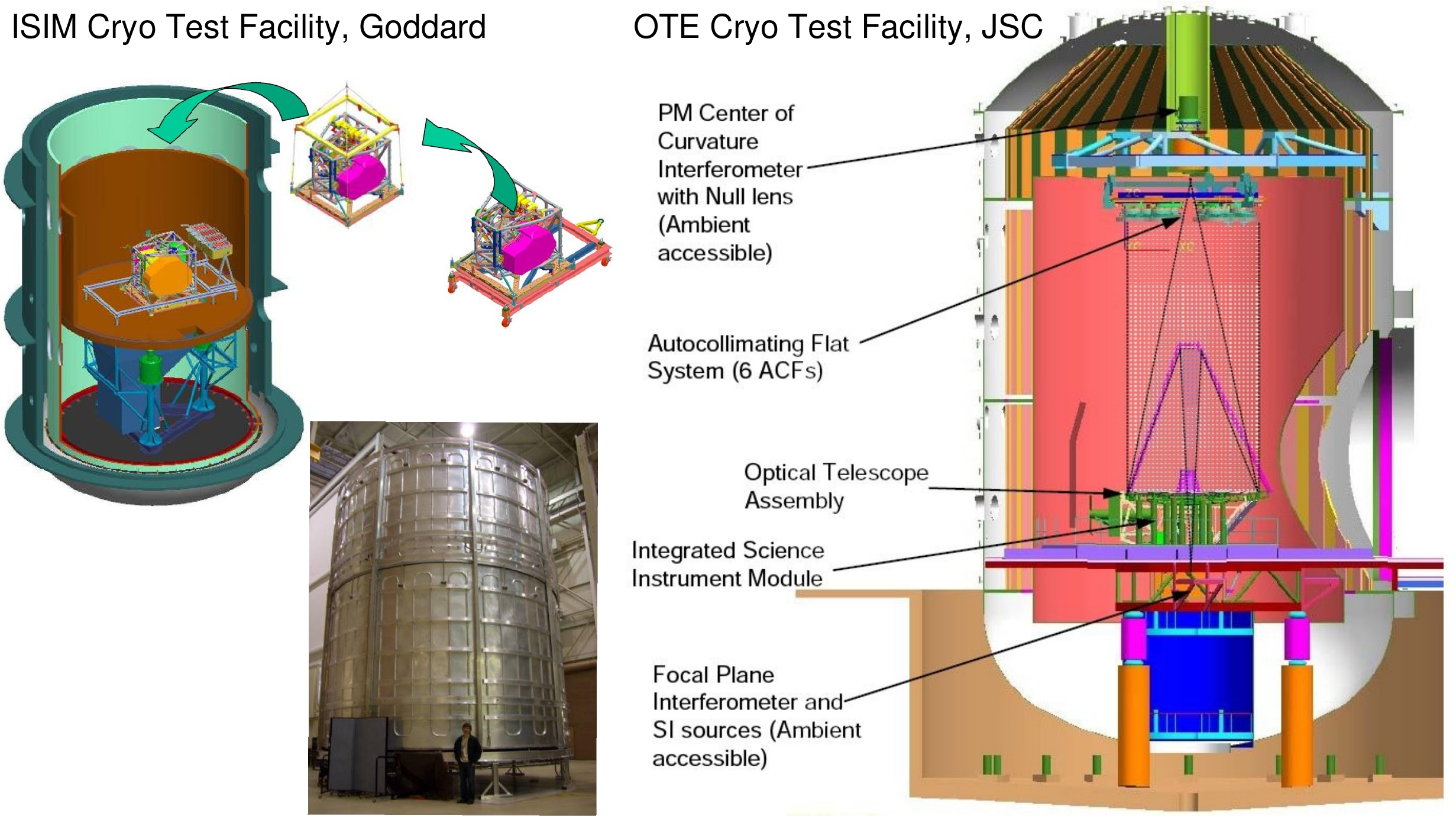}
    \caption[Giant cryo-facilities for testing MIRI and JWST]{\textit{Left:} The cryogenic test chamber at NASA Goddard is being built to perform cryo tests of the Integrated Science Instrument Module (ISIM) containing the 4 JWST instruments. \textit{Right:} The giant cryo test facility at the Johnson Space Center (JSC). The whole JWST Optical Telescope Element (OTE) and the ISIM will be tested at cryogenic temperatures. Optical alignment of the instrument, as well as functional tests and basic optical performance checks will be performed.}
    \label{fig_JWST_tests}
  \end{center}
\end{figure}

MIRI testing is not finished! Today (end of august), the instrument is still at CEA and will be tested with the filter wheel. MIRIM will be delivered to RAL\footnote{Rutherford Appleton Laboratory, \url{http://www.sstd.rl.ac.uk/}} for additional testing (functional and optical performance tests with a cold telescope simulator, as opposed to the warm simulator used at CEA). 

After testing of the Optical System at RAL, the Optical Bench Assembly (OBA, see sect.~\ref{sec:MIRI_description}) will be delivered to NASA Goddard, where the MIRI OBA will be integrated into the Integrated Science Instrument Module (see sect.~\ref{sec:MIRI_description}) for testing at cryogenic temperatures (see Fig.~\ref{fig_JWST_tests}).
Then, the ISIM is mounted onto the whole telescope (Optical Telescope Element, OTE), and the 6m telescope is tested in the giant cryo test facility at the Johnson Space Center (JSC), in Houston (see Fig.~\ref{fig_JWST_tests}). Finally, the OTE will be delivered to Kourou for final integration and launch. Let's hope I'll be there!


\chapter{Science with the JWST and MIRI}
\label{chapter:science_JWST}

\epigraph{They spend their time mostly looking forward to the past.}
{John Osborne, Look Back in Anger}




\begin{Abstract}

In addition to my participation to the tests of the Mid-InfraRed instrument (MIRI) onboard the JWST, I am involved in the scientific preparation of MIRI, as a member of the European Science Team. 
The MIRI instrument will be next future of mid-IR infrared observations, and a ideal tool to look at H$_2$ in active phases of galaxy evolution. 
This chapter presents a specific proposal (``white paper'') I have submitted to the working group ``Galaxy evolution between $z=0.6 - 6$'', in preparation of the guaranteed time observations. This proposal is based on our on-going program addressing the multiphase gas physics during highly dissipative phases of galaxy evolution at low and high redshift, which was introduced in chapter~\ref{chapter:perspectives}.

\end{Abstract}

\minitoc


\section{Overview of the JWST scientific goals}

The science of the JWST is definitively oriented towards high-redshift astrophysics. If one looks at NASA's website, the JWST science drivers  are divided into four main topics:
\begin{enumerate}
\item First Light (after the big bang);
To find and study the first luminous objects; proto-galaxies, supernova, black holes
\item Assembly of Galaxies;
To study the merging of proto-galaxies, effects of black holes, history of star formations
\item  Birth of Stars and Planetary Systems;
JWST will study how stars form in dust clouds, and how chemical elements are produced and re-circulated.
\item Planetary Systems and Origin of Life; To study the formation of planets and obtain direct observations of other planetary systems, as well as study of outer solar system.
\end{enumerate}
Based on these themes, several science working groups were created in January 2009, to start working on precise proposals of observations with MIRI, in preparation of the guaranteed time projects. I ``signed up'' into two of them: the working group ``Galaxy evolution between $z=0.6 - 6$'', led by P. Van Der Werf and the working group ``Interstellar medium'' led by Alain Abergel. 

With F. Boulanger and N. Nesvadba we started to work on a proposal to observe H$_2$ luminous objects with MIRI. The next section reproduces the proposal I have written and submitted to the ``Galaxy Evolution'' group. I also participated to a proposal written by Alain Abergel on the observations of nearby PDRs to probe the gas and dust physics at the cloud/inter-cloud interface.  

\section[H$_2$-luminous sources with MIRI]{Tracing the kinetic energy dissipation associated with galaxy merging and AGN feedback at $\bf z \sim 1.5 - 3.5$ with MIRI}

Here is the proposal I have written with F. Boulanger and N. Nesvadba and submitted to the ``Galaxy Evolution'' working group.

\subsection{Background}


The build-up of baryonic mass in galaxies is regulated by a complex interplay between gravitational collapse, galaxy merging and feedback related to AGN and star formation (Fig.~\ref{fig_galaxy_formation}). These
processes may inject sufficient mechanical energy into the interstellar medium (ISM) to
potentially have a major impact on star formation and galaxy assembly.
This will be particularly important during the early phases of galaxy
evolution at high redshift, when galaxies were gas-rich and most of the stars in the universe were
formed.



The molecular gas plays a critical role in this context since it represents an important, if not dominant, cooling agent in the energetics of galaxy merging and feedback.
Recent observations by the InfraRed Spectrograph (IRS) onboard the  {\it Spitzer Space Telescope} have discovered a significant and diverse population of low-$z$ objects where the
mid-infrared rotational line emission of H$_2$ is strongly enhanced ($L_{\rm H_2} \sim 10^{40} - 10^{43}$~erg~s$^{-1}$), while star formation is suppressed (Fig.~\ref{fig_H2_galaxies_H2_7_7PAH_L24uml}). This suggest that shocks are the primary cause of the H$_2$ emission \citep{Ogle2007, Guillard2009}. This sample of {\it H$_2$-luminous sources} includes galaxies in several key phases of their evolution, dominated by gas accretion \citep{Egami2006}, galaxy interactions \citep{Appleton2006}, or galactic winds driven by star formation \citep[e.g. M82][]{Engelbracht2006} and AGN \citep{Ogle2009}. 

Constraining the impact of merging and AGN feedback on the formation and evolution of massive galaxies
can only be addressed through direct observations at $z \sim 2$, near the cosmologically most active period of star formation, galaxy interactions and AGN activity. 
By analogy to our models on local H$_2$-luminous objects \citep{Guillard2009}, we expect the mid-IR lines to be the dominant cooling lines for warm, $10^{2-3}$~K, gas in the strongly shocked, highly turbulent, colliding flows in galaxy interactions (e.g. the galaxy-wide shock in Stephan's Quintet), but also, e.g.,  in AGN-driven outflows.
High gas velocity dispersions measured in $z \sim 2$ actively star-forming galaxies show that the gas kinematics in these systems was strongly disturbed compared to galaxies today \citep[e.g.][]{Lehnert2009}. We expect the molecular gas to be highly turbulent and therefore the warm H$_2$ emission to be more frequent and more  powerful than at low-$z$. 



\subsection{The proposal}


The MIRI Medium Resolution Spectrometer (MRS) will be the first IFU instrument to provide the sensitivity and resolving power to spatially and spectrally resolve  H$_2$ and forbidden ionized gas lines at rest-frame near-IR and mid-IR wavelengths, out to $z = 1.5 - 3.5$. For the first time, this will allow us to directly investigate  the physical state and the kinematics of the ionized gas and the warm ($> 150$~K) molecular gas that is dynamically heated by the dissipation of mechanical energy associated with galaxy merging and AGN feedback. 
We propose MIRI/MRS observations of selected sources to address the following questions:

\begin{itemize}
\addtolength{\itemsep}{-0.6\baselineskip} 
\item Are AGN driving massive multiphase outflows carrying away the bulk of the ISM of the galaxy  and therefore quenching star formation in the host galaxy? If this is the case, we should observe broad, luminous H$_2$ emission lines, with widths of 1000 km/s or more, by analogy to the large velocity shears and dispersions that have been revealed by near-infrared integral field spectroscopy of the rest-frame optical emission lines in HzRGs \citep{Nesvadba2006, Nesvadba2008}. The propose MIRI observations will allow to quantify the energy radiated and carried away by molecular outflows.
\item Does the H$_2$ emission trace ``pre-starburst'' phases of interacting galaxies? 
 \textit{Spitzer} observations of the Stephan's Quintet revealed that H$_2$ line emission represents a major  fraction of the  bolometric luminosity of the galaxy-wide shock \citep{Appleton2006}. This is a nearby example of an H$_2$-bright, early phase of a galaxy interaction, and we expect these systems to be more frequent and more active at high-$z$. The proposed MIRI spectroscopy will explore galaxy evolution  beyond the active starburst and obscured/unobscured AGN phases, highlighted by luminous dust emission, and test whether H$_2$ emission is scaled-up at high-$z$ or not.
\end{itemize}

Which lines will we look at? To establish the energy budget of the warm molecular gas and shock diagnostics, we will use, in most cases, the full MIRI spectral range. In particular:
\begin{itemize}
\addtolength{\itemsep}{-0.6\baselineskip} 
\item we will concentrate on the near-IR ro-vibrational lines, e.g. the H$_2$ 1-0 S(1) $2.12 \, \mu$m line, and the mid-IR pure rotational H$_2$ 0-0 S(3) $9.7 \, \mu$m and S(5) $6.9 \, \mu$m lines. These lines will be redshifted in the MRS spectral range ($4.9 - 28.6\, \mu$m) for $z = 1.5 - 3.5$.
\item the forbidden ionized gas lines (e.g. [Ne II], [Ne III]) will be used to compare the kinematics of the molecular gas with that of the ionized gas.
\item the synergy with NIRSPEC will be helpful to observe the CO bandheads and Ca$\,${\sc II} triplet to estimate the stellar kinematics. This will allow to estimate the ratio between the bulk galaxy rotation and the gas velocity dispersion in these high-$z$ objects, and provide an absolute rest-frame in which to interpret the gas motions as blueshift or redshift.
\end{itemize}

\subsection{The sample}

We propose to observe a representative sample of order 20-30 strongly star forming and AGN host galaxies at $z \sim 1.5-3.5$, using the MIRI medium-resolution ($R \sim 2200$) spectrometer (MRS, IFU mode). Our sample will include powerful radio galaxies, as well as radio-loud and radio-quiet obscured and unobscured AGN. We will also include powerful, dusty starburst galaxies like submillimeter galaxies and infrared-selected galaxies, many of which will be mergers. Suitable parent samples already exist, e.g. Nesvadba et al. 2009, in preparation for HzRGs with deep rest-optical IFU observations; \citet{Greve2005} for SMGs with CO observations; \citet{Yan2005} for powerful infrared galaxies at $z \sim 2$; \citet{Farrah2008} for $24\,\mu$m selected starbursts with PAH detections, etc), and will be extended with the upcoming \textit{Herschel} mission. The development of our models will also help in selecting high-$z$, non IR-luminous, interacting systems that may be ``pre-starbursts''.


\subsection{Observing mode and integration times}

Galaxies at high-$z$ are gas-rich and large gas velocity dispersions show that the gas is not fully settled in a rotating disk. This surplus in mechanical energy must be dissipated. Therefore, we expect H$_2$ emission to be more powerful at high-$z$ than for local sources.

We base the required sensitivity on the H$_2$ luminosity of the most luminous low-redshift sources at low redshift (a few $10^{43}$~erg~s$^{-1}$). With the resolving power of the MRS, ($R  \sim 2200$), corresponding to a resolution of FWHM~$\sim 140$~km~s$^{-1}$, and typical observed line widths of FWHM~$\sim 800$~km~s$^{-1}$, the lines will be resolved by a factor $\sim 6$. Typical sizes of the outflows are $\sim 5"$, for an IFU field of view of $\sim 8" \times 8"$ for $\lambda > 17 \, \mu$m. 
Given that the spatial sampling is $0.64"$ across the slice and $0.27"$ along the slice, we will be able to resolve  the molecular gas emission spatially and spectrally and perform the first spectral mappings of molecular  AGN outflows at high-redshift. Using the MIRI performance estimation, and assuming a narrow line plus extended source, we can reach an RMS of $10^{-20}$~W~m$^{-2}$ in 1 hour of on-source observing time for a $\sim 3" \times 3"$ field. At $z \sim 2.5$ this will correspond to a $10 \,\sigma$ detection limit for an H$_2$ luminosity of $3 \times 10^{43}$~erg~s$^{-1}$, for the integrated spectrum.  The typical observing time will be $\sim 1$~hr on-source per exposure, i.e. 3~hrs to cover the full wavelength range (A, B and C exposures). The total observing time for this project, excluding overheads, would be of the order of 60~hrs for $\sim 20$ sources.

\chapter{Perspectives}
\label{chapter:perspectives2}



\epigraph{That is the essence of science: Ask an impertinent question, and you are on the way to a pertinent answer.}{Jacob Bronowski (Ascent of man)}




\begin{Abstract}
Observations of H$_2$ luminous galaxies 
set molecular gas in a new context where one has to account for the presence of H$_2$
outside galactic disks, characterize its physical state and
describe its relation to star formation. These observations and the theoretical framework presented in this manuscript motivate new observations and theoretical studies that are discussed here. 
In particular, as for Stephan's Quintet, in many of the H$_2$-luminous galaxies the molecular gas has been detected through the mid-IR H$_{2}$ rotational lines prior to any CO observation. Thus we propose to complement our knowledge of the mass and kinematics of the molecular gas in these sources by CO observations. 
Then I present the theoretical perspectives, with emphasis on new numerical studies that are needed to understand how the bulk kinetic energy released in the multiphase ISM during violent phases of galaxy interactions is transferred into the warm molecular gas.

\end{Abstract}

\minitoc



\section{Observational perspectives}
\label{observational-perspectives}

\PARstart{T}his thesis work has rich observational perspectives. The flow of the discovery of H$_2$-luminous galaxies, and the observational and theoretical work presented in this manuscript, has inspired a large number of follow-up observations. Here are a few examples that will complement our view of molecular gas in galaxy evolution. 

\subsection{The search for molecular gas in active phases of galaxy evolution}

A series of upcoming  observations are proposed to search for warm H$_2$ or associated CO gas in\dots

\begin{description}
\item[compact groups of galaxies:]  a \textit{Spitzer} proposal (P.I.: P.N. Appleton) was prepared during my PhD to search for H$_2$ in HCGs\footnote{Hickson Compact Groups}. This project has been motivated by the discovery of powerful H$_2$ emission from the Stephan's Quintet group. It has been observed and the data is being reduced. I will participate to the interpretation of this data during my postdoc. 

\item[radio-galaxies:] we propose to look for CO emission in a sample of radio-galaxies with the IRAM 30-m telescope (P.I.: M. Lehnert). The proposal has been submitted in September 2009. These observations will  determine the amount of kinetic energy carried by the cold molecular gas in these radio-galaxies. 

Observations at high spatial resolution of two specific radio galaxies, 3C293 and 3C433, with the IRAM Plateau de Bure Interferometer (PdBI) have been proposed (P.I.: N.~Nesvadba). These two sources, where powerful H$_2$ is detected, are more complex than 3C326 because they exhibit star formation.  

In addition, a proposal has been submitted (P.I.: N.~Nesvadba) to look at the ``Minkowski's object'', a nearby and young starburst that might be triggered by the jet of a nearby radio galaxy. This is our best candidate for ``positive'' feedback. The proposal aims at determining the mass and kinematics of the cold molecular gas, in order to check whether these objects verifies the Schmidt law. We plan to complement these observations with VIMOS\footnote{VIMOS, the VIsible MultiObject Spectrograph, is an optical Integral Field Unit spectrometer mounted on the Very Large Telescope, see \url{http://www.eso.org/sci/facilities/paranal/instruments/vimos/} for details.} observations. 

\item[the M82 starburst driven wind:] We have submitted (P.I.: F. Boulanger) a proposal in September 2009 to look for CO emission in the M82 superwind. We obtained deep \textit{Spitzer IRS} observations of the warm molecular gas in the wind (P.I: L. Armus) but the spectral resolution is not high enough to measure outflow velocities or line widths. The CO data will allow to derive the kinematics of the cold gas in the wind, which will help in elucidating some of the questions raised above. 

\end{description}

\subsection{Observing H$_{\bf 2}$-luminous galaxies with the JWST}
\index{JWST!observing H$_2$-luminous galaxies}
\label{subsec:H2-luminous-galaxies-MIRI}

During my PhD work, I have been working on the tests and the scientific preparation of the Mid-Infrared Instrument (MIRI) that will be part of the scientific payload of the James Webb Space Telescope (see chapter~\ref{chapter:miri_test} and \ref{chapter:science_JWST}).

\subsubsection{H$_{\bf 2}$ in active phases of galaxy evolution at high-${\bf z}$}

\begin{figure}
   \centering
    \includegraphics[angle=90, width=\textwidth]{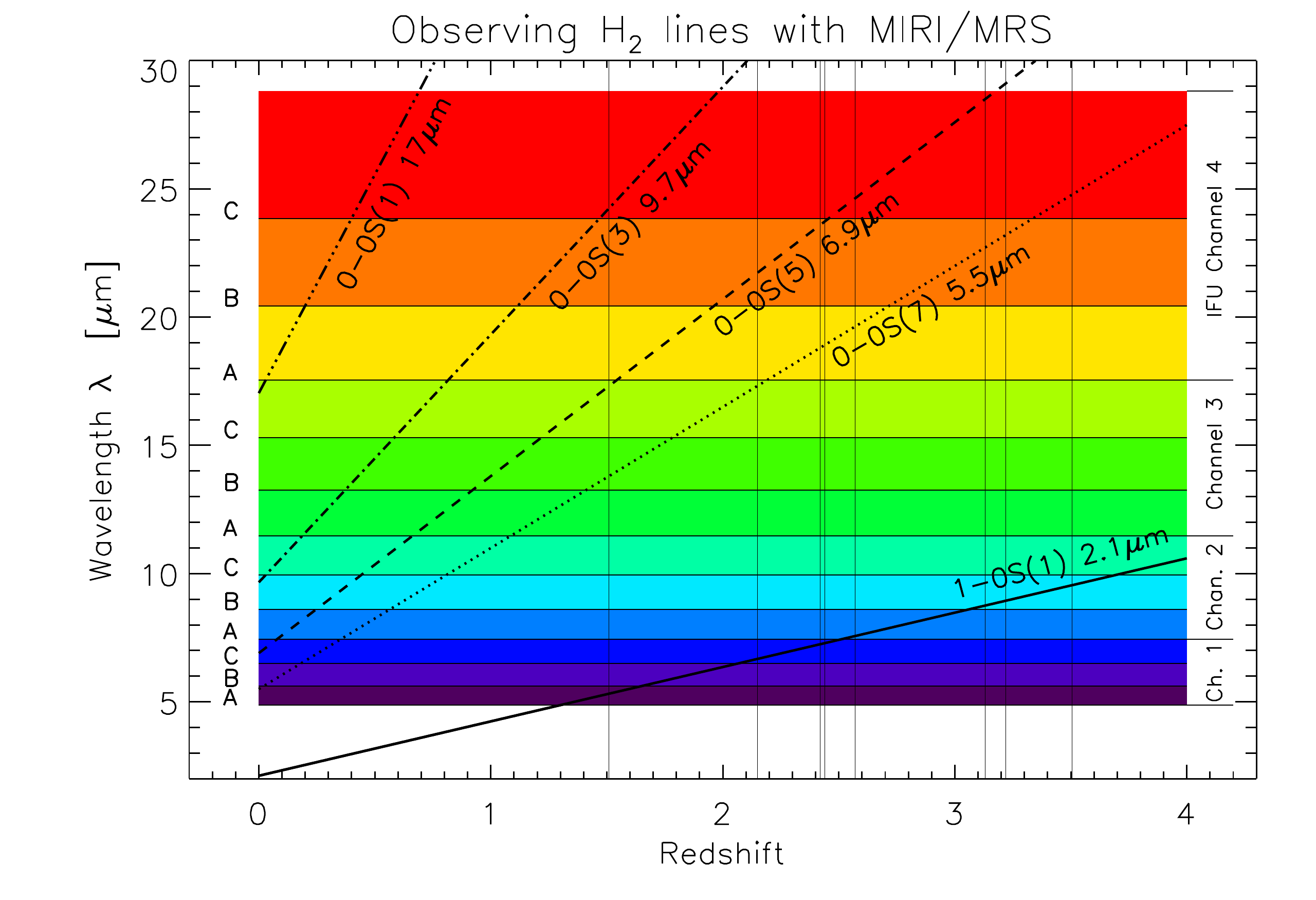}
      \caption[Observing H$_2$ lines at high-redshift with JWST/MIRI]{Observing H$_2$ lines at high-redshift with JWST/MIRI. The observed wavelengths of some H$_2$ lines are shown as a function of the redshift. The colored bars indicate the channels and bands of the Medium Resolution Spectrometer (MRS) of the MIRI instrument. One observation corresponds to four sub-bands, like 1A, 2A, 3A, 4A for instance (see sect.~\ref{subsec:MIRI_MRS} for technical details about how the MRS is working). The vertical lines indicate the redshifts of some high-$z$ radio-galaxies that might be interesting to look at with MIRI.}
       \label{fig:MIRI_lambda_H2}
\end{figure}

In particular, I have proposed   a specific project to look at high-$z$ H$_2$-luminous objects with MIRI to the MIRI European Consortium. The proposal is reproduced in  chapter~\ref{chapter:science_JWST}. It is specific to the MIRI instrument and relies on the background presented in chapter~\ref{chapter:JWST}.
For the first time, MIRI will allow us the study the kinematics and excitation of the warm H$_2$ gas in high-$z$ objects. To collect a sample of high-$z$ objects, we will use the analogy between the characteristics of VLT/SINFONI\footnote{Spectrograph for INtegral Field Observations in the Near Infrared, \url{http://www.eso.org/sci/facilities/paranal/instruments/sinfoni/}.} and JWST/MIRI. I will use high-$z$ radio-galaxies SINFONI and \textit{Spitzer} data (PI: N. Nesvadba, co-I: P. Guillard) as a starting point to define the observing strategy. MIRI will allow integral field spectroscopy of redshifted optical or $\rm H_2$ lines (Fig.~\ref{fig:MIRI_lambda_H2}), at high spatial (1-2 kpc/pix) and spectral ($\lambda/\Delta \lambda \sim 3\,000$) resolution. This spectral resolution is crucial to measure the kinematics of the

The Fig.~\ref{fig:MIRI_lambda_H2} shows the visibility of the some H$_2$ lines as a function of redshift. I also indicate, as a function of redshift,  in which sub-band of the Medium Resolution Spectrometer (MRS) these lines are falling. This allows to configure the instrument for observations and estimate the integration time. Fig.~\ref{fig:MIRI_lambda_H2} is thus useful to prepare observation proposals.


\subsubsection{An infrared high-resolution look at Stephan's Quintet with JWST}
\label{JWST-MIRI-SQ-observations}
\index{JWST!observing SQ}

In addition, in the near future, I will propose an observational program to investigate Stephan's Quintet with NirCam/NirSpec and MIRI, the near-IR and Mid-IR instruments onboard JWST (see chapter~\ref{chapter:JWST} for an introduction to these instruments). 
The NirCam and MIRI imagers will provide an unprecedented spatial resolution (comparable to the HST at 7$\,\mu$m) to study small scales structures in the shock. MIRI will allow us to clearly separate the dust emission associated with star-forming regions from that associated with the shock itself. 
Multi-band imaging with NirCam and MIRI will allow to build a full map of spectral energy distributions from 0.6 to 28$\,\mu$m. This may be used to search for variations of the dust properties across the shock. These variations may be interpreted in terms of dust processing in shocks, and possibly related to velocity gradients in the warm molecular gas. 

The spectral resolution of NirSpec (see sect.~\ref{subsec:JWST-science-instruments} for technical details) and  MIRI's Medium Resolution Spectrometer (sect.~\ref{subsec:MIRI_MRS}) will provide for the first time the kinematics of the warm H$_2$ gas across the  shock.  This will allow us to map the relative velocities and  dispersion velocities over the SQ ridge, which we will compare with the kinematics of the CO gas presented in chapter~\ref{chapter:SQ_CO}. This comparison will be particularly useful to estimate the efficiency of the dynamical coupling between the warm and cold molecular gas. 


\section{Theoretical perspectives} 
\label{theoretical-perspectives}

The detection of powerful $\rm H_2$ emission in a diverse set of extragalactic
sources with little or no star formation highlights the need to understand the
role of $\rm H_2$ formation and cooling for the dissipation of mechanical
energy in a multiphase ISM.  
The previous sections have shown that these observations provide insights in the physics of galaxy interactions or of the regulation of star forming activity in radio galaxies.
An efficient transfer of the bulk
kinetic energy to molecular gas is required to make H$_2$ an important coolant
and thus explain these observations.  The dissipation of mechanical energy 
within cold molecular gas, $\rm H_2$ formation and its collisional excitation,
play a key-role in the physics of turbulent mixing of gas phases.

\subsection{A novel numerical study of the dynamical interactions between gas phases} \label{subsec_simulations}

In my thesis work, I reached a qualitative understanding on how the energy
transfer may occur. The fragmented, multiphase structure of the ISM is likely
to be a main aspect. The dynamical interactions between gas phases drive a
mass cycle which contributes to the energy transfer in two ways: (1) Gas cooling transfers the turbulent energy of the warm
H~{\sc ii} and H~{\sc i} to the H$_2$ gas. (2) The thermal instability induced
by the gas cooling transfers a significant fraction of the thermal energy of
the gas to turbulent energy. The turbulent velocities of the motions generated
by the thermal instability are expected to be commensurate with the sound
speed in the warmer gas and thus are supersonic for the colder gas.  These
non-linear aspects of the dynamical interaction between gas phases can only be
investigated quantatively with numerical simulations.

Numerical simulations \citep[e.g.][]{Audit2005} show that the thermal
instability prevents energy from being radiated by the warmer gas to feed
turbulence in the cooler gas.  I have started a collaboration with P. Lesaffre
and E. Audit which I will pursue and extend during my postdoc. We will use 
the RAMSES magnetohydrodynamical (MHD) adaptive Mesh Refinement (AMR) 3-D code.
It is essential to resolve thermal instability to quantify energy and mass
exchanges.  This can only be achieved by including chemical evolution and its
impact on the cooling efficiencies, and we are currently writing a routine to
introduce time-dependent chemistry and thermal diffusion in the code. The main
coolants are C$^+$, O, $\rm H_2$, and CO.  This code will be used to model mass cycling and mechanical energy dissipation in a
mutiphase ISM. This will be the first simulation of the physical and chemical
evolution of a thermally unstable multiphase gas including a self-consistent
treatment of the $\rm H_2$, H{\sc i}, H{\sc ii} and hot phases of the ISM.
Including all phases requires a large dynamic
range and  is beyond the capabilities of state-of-the-art computing.  We will hence describe the processes sequentially, starting
at the large scales. As we proceed to smaller scales we will include a more
detailed description of the thermal budget and chemistry of the gas.

 We will
first study the effects of a large scale shock on a set of H{\sc i} clouds
embedded in a hot plasma.  Then we will study the fate of one of these clouds
in greater details as it is being fragmented. We will pay special attention to
the energy transfer between these phases. Thanks to our AMR
approach and with a careful treatment of the thermal and chemical diffusion accross
the boundaries between these various phases, this will be possible.

We will also explore the momentum and energy transfer from a hot stream of gas into an initially static molecular cloud. In both stellar and AGN feedback, it is of crucial importance to determine whether the dynamical coupling between the tenuous outflow and the cloud is efficient, and whether it can power the amount of H$_2$ emission seen in these H$_2$-luminous sources.  

Due to their proximity, starburst-driven winds or galactic supernova remnants provide a unique opportunity to confront the model with data in great detail. Our interpretation
will be based on the physical conditions inferred from these numerical
models. They will serve as input for detailed calculations of gas cooling and
chemistry in shock models. This method relates the observational approach
(sect.~\ref{observational-perspectives}) with our numerical and analytical models, and will
also be applied to extragalactic sources.

\subsection[Towards a phenomenological prescription of $\rm H_2$ in galaxy evolution]{Towards a phenomenological prescription of the impact of H$_{\bf 2}$ on the energetics of galaxy evolution}

Though $\rm H_2$ emission is related to small scale processes, its formation
and cooling power play a major role in the overall energetics of violent
phases of galaxy evolution.  Numerical studies of galaxy formation cannot
properly account for these processes. Small scale simulations with sufficient
resolution to include the detailed \textit{multiphase} gas physics do not
include the overall context of feedback processes. Inversely, large-scale
simulations can only include gas physics through phenomenological recipes.

From the results obtained from the observational (sect.~\ref{subsec:H2-luminous-galaxies-MIRI})
and numerical (sect~\ref{subsec_simulations}) approach of feedback, an ultimate goal would be to  
distill reasonably simple, but robust phenomenological equations, which would 
approximate the physics of energy dissipation, and allow to predict $\rm H_2$
emission as a function of the physical parameters of a galaxy. These equations
will represent a crucial input to cosmological simulations of galaxy formation
and the growth of structure.  Ultimately, this will allow to determine the
role of the gas for the build-up of galaxies, and the importance of starburst
and AGN driven winds on the growth and properties of the ensemble of galaxies
across cosmic time.

Since most of the galaxy mass build-up occurred at redshifts between 1 and 2, this prescription will need to be tested with direct observations of sources in this redshift range. This will become possible with JWST/MIRI (see sect.~\ref{subsec:H2-luminous-galaxies-MIRI}). 

\addcontentsline{toc}{chapter}{The last word\dots}


\chapter*{The last word\dots}
\label{chapter:conclusion}

\vspace{-2cm}

\epigraph{I hate quotations. Don't use quotations. Tell me what you know.}{Ralph Waldo Emerson}





\section*{A thesis! What a shock!}

\begin{wrapfigure}{r}{90mm}
   \includegraphics[width=90mm]{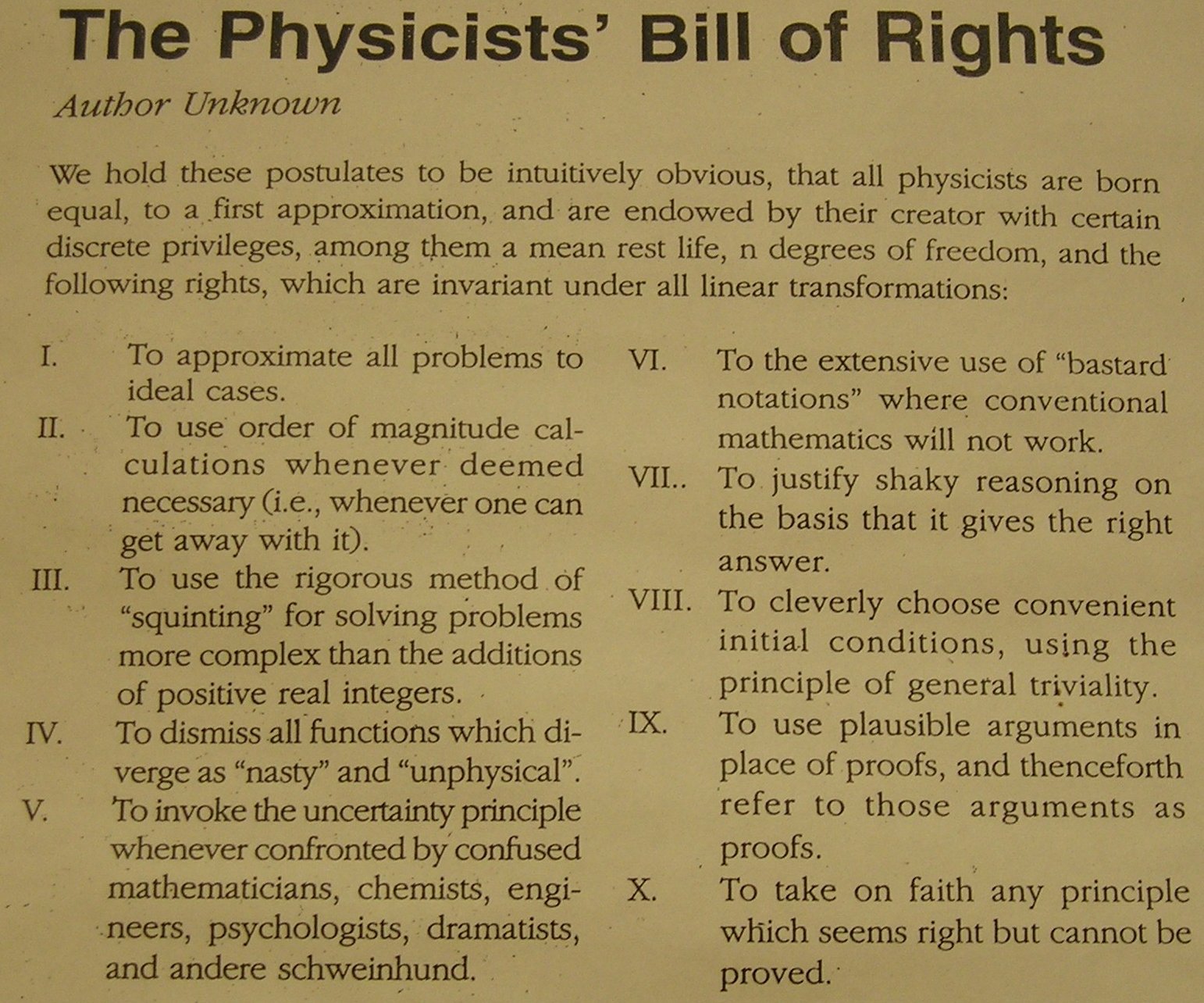}
\end{wrapfigure}

In fact, if you think about it, a thesis \textit{is} a shock. To be a PhD student is to be shocked. You are in the shock front, between what you believe to know, the postshock medium, and the unknown, the preshock medium. 
In this shock front, you are heated to high temperatures, highly excited by your research. Sometimes you feel like going in the postshock, stay there to cool down a little bit, have a fresh shower, to come back to the known. Of course, what you did in the shock front greatly affected the postshock, but also, what you're doing in the shock front affects the future science, but you don't have access to this information\footnote{\dots unless you can travel at the speed of light\dots Obviously, you've understood that I'm talking about the radiative precursor here!}.

Obviously, \textit{la vie n'est pas un long fleuve tranquille}, the preshock medium is inhomogeneous, multiphased, full of surprises. Sometimes you propagate fast, smoothly, in a teneous medium, whereas sometimes you hit a denser problem, and your speed is reduced, you need to think harder!
Your physical state in the shock depends on your progression speed. Be careful of not going too fast, be careful of the $J$-type discontinuity, be careful not to be ionized\dots otherwise you'll lose something\dots

\appendix
\part{Appendices}
\chapter{Modeling dust emission: the DUSTEM code}
\label{appendix_DUSTEM_code}
\dominitoc
\index{Dust!DUSTEM code}

\begin{Abstract}

This appendix  briefly presents the code that I have used to model the dust emission from the Stephan's Quintet galaxy collision. This code, so-called DUSTEM, is an update of the \citet{Desert1990} model, which computes, from an input radiation field, the emission, temperature distribution,  absorption and scattering cross-sections, as a function of the wavelength, for a size distribution of grains including stochastically heated PAHs and very small grains. In the case of optically thick (to UV radiation) molecular clouds, the DUSTEM code is coupled to a radiation transfer code to take into account in a self-consistent way both the extinction and the heating of dust grains by their own IR emission. 
\end{Abstract}

\section{Introduction}

Since the 1970's, a lot of models were built to reproduce the observations of interstellar dust. The \citet{Desert1990} model is one example among others. Since this first publications, a lot of modifications have been done. The most recent ones are reported in \citet{Compiegne2008}. In the following, I briefly give the main features of the \textit{DUSTEM} model, and its key parameters. The interested reader may consult the PhD thesis by \citet{Compiegne2007a}, \citet{Flagey2007} and \citet{Gonzalez2009} for a description of the updates included in the code. 

\section{Dust models: populations and size distributions of grains}

The \textit{DUSTEM} model includes a mixture of three populations of dust grains with increasing sizes; which is the minimum number of components needed to account for interstellar extinction and IR dust emission. These components are the following. 
\begin{itemize}
\item Polycyclic Aromatic Hydrocarbons (PAHs) of radius $a = 0.4 - 1.2$~nm, responsible for the Aromatic Infrared Bands (AIBs) and the FUV non-linear rise in the extinction curve.
Much of the PAH emission is concentrated in features at 3.3, 6.2, 7.7, 8.6, 11.3 and
17 $\mu$m. \citet{Duley1981}, followed by \citet{Leger1984},   identified these features as the optically active vibrational modes of PAH molecules. 

\item Very Small Grains (VSGs, $a = 1 - 4$~nm), which are carbonaceous nanoparticles producing the mid-IR continuum
emission and the extinction bump at 2175~\AA.
\item  Big Grains (BGs, $a = 4 - 110$~nm) of silicates with carbonaceous
mantles or inclusions, which account for the far IR emission and the $1/\lambda$ rise at visible and near-IR wavelengths.
\end{itemize}
The version of the  \textit{DUSTEM} code I have used assumes  the size distributions of all the dust populations to be a power-law,  $n(a) \propto a^{-3.5}$ \citep{Mathis1977}, with minimum and maximum radii listed in the items above. The parameters of the grains used in the version of the code I use are gathered in Table.~\ref{table:DUSTEM_parameters}.

\renewcommand{\arraystretch}{1.1} 
\begin{table}
\begin{center}
\begin{minipage}[t]{\textwidth}
\renewcommand{\footnoterule}{}
\def\thefootnote{\alph{footnote}}
\centering
\caption[Parameters of the DUSTEM code]{Parameters of the DUSTEM code\footnotemark[1].}
    \begin{tabular}{c c c c c c c }
	\hline
	\hline
Component\footnotemark[2] & Abundance & $\alpha$ & $a_{\rm min}$ [nm] & $a_{\rm max}$ [nm] & $\rho$& $\beta$ \\
\hline
PAHs & $4.3 \times 10^{-4}$ & 3.5 & 0.4 & 1.2 & $2.4 \times 10^{-7}$  [g~cm$^{-2}$] & 0 \\
VSGs & $4.7 \times 10^{-4}$ & 3.5 & 1.2 & 4.0 g~cm$^{-3}$& 2.3 & 0 \\
BGs & $6.4 \times 10^{-3}$ & 3.5 & 4.0 & 110 g~cm$^{-3}$& 3.0 & 0.61 \\
\hline
\end{tabular}
\label{table:DUSTEM_parameters}
\footnotetext[1]{We list the mass abundance relative  to Hydrogen, the parameters of the size distribution $n(a)$ for the three components of the DUSTEM model (it is assumed that $n(a) \propto a^{-\alpha}$ with $a$ between $a_{\rm min}$ and $a_{\rm max}$), the mass density $\rho$ of the grains, and their maximum albedo $\beta$.}
\footnotetext[2]{The grain populations are: Polycyclic Aromatic Hydrocarbons (PAHs), Very Small Grains (VSGs), and Big Grains (BGs).}
\normalsize
\end{minipage}
\end{center}
\end{table}
\renewcommand{\arraystretch}{1.0}

\section{Updates and calculations}

The model is designed to fit both the interstellar extinction curve (see chapter.~\ref{chapter:dust_gas_galaxies}, sect.~\ref{dust-composition-size-distribution}, Fig.~\ref{fig:dust_extinction_size_dist}) and the infrared dust emission, in particular that of the diffuse interstellar medium.  
The dust properties, the dust-to-gas mass ratio, and the exciting radiation field are provided by the user. The code then calculates the dust SED $\nu S_{\nu}$ in units of erg~s$^{-1}$~H$^{-1}$, the distribution of dust temperatures,   as well as the UV to IR absorption and scattering cross-sections in units of cm$^{2}$~g$^{-1}$, for each dust grain species, as a function of  the wavelength.

Since the original version of  \citet{Desert1990}, the absorption cross sections of the PAHs (with addition of new Aromatic Infrared Bands, AIBs) and those of the VSGs and BGs, as well as the heat capacities (graphite, PAH C-H, silicate and amorphous carbon) have been updated \citep{Compiegne2008}. 

In the diffuse interstellar medium, small grains (PAHs and VSGs) undergo temperature
fluctuations triggered by the discrete absorption of stellar photons \citep[see e.g.][for a review]{Draine2003}. They cool by
emission in the near and mid-IR range. The code computes the probability distribution of temperatures resulting from the stochastic heating of these grains. Conversely, BGs, which have a longer cooling time and a shorter timescale between absorption of two photons because of their size, stay at constant
temperature. The code computes their emission as a grey body.

\section{Inputs of the DUSTEM code}

The inputs of the code are the following:

\begin{description}
\item[Grain properties:] a file, named \texttt{grain.dat} list the types of grains and their density in [g~cm$^{-3}$] for all types except the PAHs which are in g~cm$^{-2}$. The types of grains may be chosen from the following list: 1 = graphite
grain \citep[Mie theory for spherical grains in the $1/3-2/3$ approximation,
see][]{Draine1984}. 2 = PAH, 3 = BG \citep{Desert1990},
4 = VSGs  \citep{Desert1990}, 5 = astro-silicates (Mie theory for spherical grains)

The file \texttt{propmass.dat} sets the mass abundance of the grains relative to H, \texttt{calor.dat} lists the heat capacities for the different available grain species, and  \texttt{taille.dat} defines the size distribution for each grain population.

\item[Incident radiation field:] the  file  \texttt{isrf.dat} contains the flux of the input radiation field in units of erg~s$^{-1}$~cm$^{-2}$~Hz$^{-1}$, and  \texttt{lambda.dat} is the associated wavelength grid in $\mu$m.

\item[Absorption and scattering coefficients] are listed in  the \texttt{QABS\_*.dat}, one file for each type of grains.

\end{description}

\section{Outputs of the DUSTEM code}

The code provides the following outputs:

\begin{description}
\item[Dust emissivity:] the file  \texttt{dustem.res} gives the dust emissivity in units of erg~s$^{-1}$ per Hydrogen atom, as a function of the wavelength, for each population of grains. 
 \item[Absorption and scattering cross-sections]  are listed, in units of [cm$^{2}$~g$^{-1}$], as a function of the wavelength in \texttt{secteffir.res} for the infrared domain, and in \texttt{secteffuv.res} for the UV domain.
\end{description}

%

\section{When DUSTEM cooperates with the Meudon PDR\protect \footnote{Photon-Dominated Region}  code\dots}

The spectral energy distribution of the IR dust emission does not only depend on the radiation field and the grain properties. It also depends on the physical structure of the molecular gas, and in particular  on its optical thickness to UV radiation. If a molecular cloud illuminated by a UV radiation field is  optically thick to UV photons, the radiation field is attenuated as it penetrates into the cloud, thus reducing the dust heating. In addition, dust emission has a \textit{positive feedback} on its own IR emission. The local IR dust emission contributes to the IR radiation field, and becomes a significant heat source within dark clouds where UV light is attenuated. I briefly describe how these two effects are taken into account in our modeling of dust emission from the Stephan's Quintet galaxy collision presented in chapter~\ref{chapter:SQ_dust}.

\subsubsection{Modeling dust emission from optically thick (to UV) molecular clouds}

To model the dust emission from a  molecular cloud that is optically thick to UV radiation (see chapter~\ref{chapter:SQ_dust} and \hyperref[paper_SQ_dust]{paper~{\sc iii}}), I have used the \textit{Meudon PDR} (Photon Dominated Region) code\footnote{This code,  fully described in \citet{LePetit2006}, is available on the web on \url{http://aristote.obspm.fr/MIS} and will be incorporated in the Virtual Observatory.}  to compute the radiative transfer through the cloud. This 1-dimensional, steady-state model considers a stationary plane-parallel slab of gas and dust, illuminated by UV radiation. 
The radiation field output $I_{\nu}$ of the Meudon PDR code is used as input to
the \textit{DUSTEM} program to compute the spectral energy distribution (SED) of the
dust as a function of the optical depth into the cloud. 

This 2-step process is iterated to take into account dust heating by the dust IR emission. The interface between  the \textit{DUSTEM} code and the \textit{Meudon PDR} code represents a significant update, described in the PhD thesis by \citet{Gonzalez2009}. In the present version, so-called \textit{PDR-DUSTEM}, the thermal balance of the grains resulting from the IR continuum emission of the dust is now solved.
 A first run is done with the   \textit{Meudon PDR} code to solve for the radiative transfer and compute the radiation field $I_{\nu}$ through the slab of gas. Then, the \textit{DUSTEM} program is called to compute the dust temperature distribution and the dust contribution to $I_{\nu}$. This 2-step process is iterated until the calculation of the dust emission has converged (usually $4-5$ iterations).

\chapter{Deconvolution of high-resolution PSF: bayesian formalism}
\label{appendix_microscan}
\dominitoc
\index{Deconvolution!Bayesian formalism}

\begin{Abstract}

In this appendix  I give a brief description of the formalism used to devonvolve a high-resolution image from several low-resolution ones. I have applied this algorithm to reconstruct a high resolution PSF image from the data provided by the microscanning test (see chapter~\ref{chapter:miri_test}). This algorithm was developed by T. Rodet at LSS, Orsay.

\end{Abstract}

\section{The inverse problem}

The aim of the deconvolution is to reconstruct a high-resolution image, denoted by $\xb$, from multiple low-resolution images. These low-resolution images are co-added on a fine grid. We denote $\zb$ the LR image that have been over-sampled on a fine grid. This image is still ``blurred'', but its number of pixels is a factor $f^2$ higher than the observed LR image.   
$\zb$ is the convolution of the HR image $\xb$ that we want to calculate by the impulse-response (PSF) of the detector:
\begin{equation}
\label{eq:HRimage}
\zb = \Rb \xb + \nb
\end{equation}
One cannot apply the classical methods of deconvolution because part of the data is not observed (due to uncertainties on the translations, see sect.~\ref{subsec:translations}). 
Indeed, there are missing parts, ``holes'', in the data because the source is not exactly at its expected position on the fine grid image.
The Fig.~\ref{fig_Data081206_00h15DecalLuFac11} shows the PSF data co-added onto a fine grid  with an over-sampling factor of 11. We illustrate the truncation of the data by comparing the results of the co-addition when using the expected positions of the source, to the  results when using the shifts estimated by cross-correlation. The additional truncation for the effective translations comes from the mismatch between expected and real position of the source.

We recall that, because of truncation of the data (we miss some part of the data on the highly-sampled grid), a truncation matrix $\Tb$ is introduced in Eq.~\ref{eq:HRimage}:
\begin{equation}
\label{eq:HRimage_T}
\zb = \Tb \Rb \xb + \nb = \Hb \xb +  \nb
\end{equation}
$ \Tb$ is a mask with pixels value of $1$ where the data exists, and 0 otherwise. The operator $\Hb =  \Tb \Rb $ returns the measurements $\zb$ from the HR image $\xb$.
To solve the deconvolution problem, a bayesian formalism is used. 
In the following, I briefly give the basis of the formalism we use to estimate $\xb$, the HR image.

\section{Bayesian inference}

Solving inverse problems involves statistical methods that have been used since many decades to perform signal processing. In practice, these methods rely on the minimization of a quadratic criterion, that we will express below. The estimator used to minimize this criterion can be interpreted in a Bayesian framework. The Bayesian inference process, in which observations are used to infer the probability that an hypothesis is true, is based on the Bayes' theorem. We denote 
 $\zb$ the measured data and $\Cb$ a variable that gathers the model of data acquisition and the parameters \textit{a priori}. The Bayes' law is expressed as:
\begin{equation}
\label{eq:bayes}
f(x \vert z , \, C) = \frac{f(z \vert x , \, C) \, f(x \vert C)}{f(z \vert C)} \ .
\end{equation}
The terms of the Bayes' theorem are the following. 
$f(z \vert x , \, C)$ is the \textit{likelihood}, i.e. the probability distribution of the data if we know the physical quantity $x$. This probability gathers all the information about our knowledge on the data acquisition system, and can be determined by solving the \textit{direct} problem (sect.~\ref{subsec:directinverse_pb}). In Eq.~\ref{eq:bayes}, $f(x \vert C)$ is the \textit{prior} probability of $x$, i.e. the knowledge about $x$ before doing the measurements. It is the \textit{prior} in the sense that it does not take into account any infomation about $z$. The Bayes' formula gives $f(x \vert z , \, C) $,  the conditional probability of $x$, given $z$. it represents all the information we have on the problem. It is also called the \textit{posterior probability} because it is derived from the specified value of $z$.
$f(z \vert C)$ is the prior probability of $z$, and acts as a normalizing constant. This constant is chosen as to make the integral of the function $f(x \vert z , \, C)$ equal to 1, so that it is indeed a probability density function. 
Intuitively, the Bayes' theorem in this form describes the way in which one's knowledge about observing the HR data ($\xb$) is updated by having observed the LR data ($\zb$).

\section{Least-square criterion and minimization technique}

The estimate of the value $\hat{\xb}$ is given by the argument of the maximum of the \textit{posterior} probability  function, often called  \textit{posterior} maximum likelihood:
\begin{equation}
\hat{x}_{\rm ML} = {\rm arg \  max} (f(x \vert y, \, C))
\end{equation}
The noise $\nb$ is assumed to be Gaussian, independent and uniform over the whole image. 
Therefore we use a gaussian likelihood:
\begin{equation}
f(z \vert x, \, C) = K \, e ^{-\frac{(z - \Hb x)^{2}}{2\,\sigma ^{2}}}
\end{equation}
Therefore we obtain the least-squares estimator:
\begin{eqnarray}
\hat{x}_{\rm ML} &= & {\rm arg \  max} (f(z \vert x, \, C)) \\
							& = & {\rm arg \  max} \left(  e ^{-\frac{(z - \Hb x)^{2}}{2\,\sigma ^{2}}} \right) \\
							& = & {\rm arg \  max} \left( -\frac{(z - \Hb x)^{2}}{2\,\sigma ^{2}} \right)  \\
							& = & {\rm arg \  min} ( (z - \Hb x)^{2}) \\
							& = &  {\rm arg \  min} ( \Vert z - \Tb \Rb \xb  \Vert ^{2} + \mu \Vert \Db \xb \Vert ^{2} )
\label{eq_reg_LS_crit}
\end{eqnarray}
$\mu$ is the regularization parameter. 
The last equation defines the regularized least-square estimator, where $\Db$ is the finite differences operator (Laplacian). $\Db$ has a size $(N-1)  \times N$ and can be written as:
\begin{equation}
\Db = 
 \begin{bmatrix}
1 & -1 & 0 & 0 & \ldots &  \ldots & 0 \\
0 & 1 & -1 & 0 & \ldots &  \ldots & 0 \\
\vdots & & \ddots & \ddots & & & \vdots \\
\vdots & & & \ddots & \ddots &  & \vdots \\
0 & \cdots & \cdots & 0 & 1 & -1 & 0 \\
0 & \cdots & \cdots & 0 & 0 & 1 & -1 \\
\end{bmatrix}
\end{equation}
The least-square criterion, $Q_{\rm LS}$, to minimize in Eq.~\ref{eq_reg_LS_crit} is quadratic and can be re-written as:
\begin{equation}
Q_{\rm LS}(\xb) = (\yb - \Hb \xb) ^{\rm t} (\yb - \Hb \xb) + \mu \xb ^{\rm t} \Db ^{\rm t} \Db \xb
\end{equation}
The solution is explicit:
\begin{equation}
\hat{x}_{\rm ML} = (\Hb ^{\rm t} \Hb + \mu \Db ^{\rm t} \Db)^{-1} \Hb ^{\rm t} \zb
\end{equation}
The main issue with this explicit solution is that the size of the matrix $\Hb ^{\rm t} \Hb$ can be large and its inversion can be very time-consuming. Since the criterion to minimize is quadratic, it is convex and has a unique minimizor. We use a method called ``gradient descent'' or ``steepest descent'', which consists in starting from whatever value for $\xb$, and then converge to the solution in the direction of the steepest slope (opposite of the gradient's direction). At each step, this algorithm ensures that, after a sufficient number of iterations, we will approach the required minimum value of the criterion. Thus we calculate the gradient of the criterion to minimize:
\begin{equation}
g(\xb) =  \frac{\partial Q_{\rm LS}(x) }{\partial x}  = 2 \, \Hb ^{\rm t} (\Hb x - z) + 2 \mu \Db ^{\rm t} \Db x
\end{equation}
and we update the solution with the following equation:
\begin{equation}
x ^{k+1} = x ^{k} - \alpha _{\rm opt} (\Hb ^{\rm t} (\Hb x ^{k} - y) + \mu \Db ^{\rm t} \Db  x ^{k} ) \ ,
\end{equation}
where $\alpha _{\rm opt}$ is the optimal step to ensure the most efficient descent is the opposite direction of the gradient. This step is calculated by cancelling the gradient with respect to $\alpha$. In our case, we use a variant of the gradient descent, so-called ``conjugate gradient descent'', which consists in having orthogonal directions at each descent steps. 

\chapter{Publications}
\label{appendix:publications}
\section{Refereed Articles}

\begin{enumerate}
\item 
 \textbf{Guillard, P.}, Boulanger, F., Pineau des For\^ets, G., Appleton, P.N., \textit{$\rm H_2$ formation and excitation in the Stephan's Quintet galaxy-wide collision}, 2009, A\&A, 502, 515.
\item 	
Boulanger, F., Maillard, J. P., Appleton, P., Falgarone, E., Lagache, G., Schulz, B., Wakker, B. P., Bressan, A., Cernicharo, J., Charmandaris, V., Drissen, L., Helou, G., Henning, T., Lim, T. L., Valentjin, E. A., Abergel, A., Bourlot, J. Le, Bouzit, M., Cabrit, S., Combes, F., Deharveng, J. M., Desmet, P., Dole, H., Dumesnil, C., Dutrey, A., Fourmond, J. J., Gavila, E., Grangé, R., Gry, C., \textbf{Guillard, P.} et al. 2008, \textit{The Molecular Hydrogen Explorer}, Experimental Astronomy special issue on Cosmic Vision Proposals.
\item
 \textbf{Guillard, P.}, Boulanger, F., Cluver, M., Appleton, P.N., Pineau des For\^ets, G., \textit{Observations and modeling of the dust emission from the H$_2$-bright galaxy-wide shock in Stephan's Quintet}, 2009, A\&A, submitted.

\item
\textbf{Guillard, P.}; Boulanger, F., Lisenfeld, U., Duc, P.A., Appleton, P.N. \& Pineau des Forêts, G., \textit{Complex kinematics of the CO gas in the Stephan's Quintet galaxy collision}, 2009, A\&A, in preparation.

\item
Nesvadba, N.P.H., Boulanger, F., Salomé, P.,  \textbf{Guillard, P.}, Lehnert, M.D.,  Pineau des For\^ets, G., Ogle, P., Appleton, P.N.,\textit{Energetics of the molecular gas in the H$_2$-luminous radio galaxy 3C326: Evidence for negative AGN feedback}, 2009, A\&A, submitted.

\item
M.E. Cluver, P.N. Appleton, \textbf{P. Guillard}, F. Boulanger, P. Ogle, P.-A., Duc, N. Lu, J. Rasmussen,
W.T. Reach, J.D. Smith, R. Tuffs, K. Xu, M. Yun, \textit{Powerful H$_2$ line-cooling in Stephan's Quintet: I- Mapping the dominant cooling pathways in group-wide shocks}, 2009, ApJ, accepted for publication. 

\item
\textbf{Guillard, P.}, Jones, A. P., Tielens, A.G.G.M, \textit{Noble gas implantation into SiC stardust: Constrains on dust evolution processes}, 2009, A\&A, submitted.

\end{enumerate}

\section{Proceedings}

\begin{enumerate}
\item 
\textbf{Guillard, P.} \& Boulanger, F. \textit{$H_{2}$ Energetics in Galaxy-wide Shocks Insights in Starburst Triggering and Galaxy Formation}, 2008, SF2A-2008: Proceedings of the Annual meeting of the French Society of Astronomy and Astrophysics, p371.
\item
 J. Amiaux, F. Alouadi, J.L. Augueres, P. Bouchet, M. Bouzat, C. Cavarroc, C. Cloue, P. De
Antoni, D. Desforges, A. Donati, D. Dubreuil, D. Eppelle, F. Gougnaud, B. Hervieu, P.O.
Lagage, D. Leboeuf, I. Le Mer , Y. Lussignol, P. Mattei, F. Meigner, V. Moreau, E.
Pantin, P. Perrin, S. Ronayette, G. Tauzin, S. Poupar, D. Wright, A. Glasse, G. Wright, E.
Mazy, J.Y. Plesseria, E. Renotte, T. Ray, A. Abergel, \textbf{P. Guillard} et al. \textit{Development approach and first infrared test results of JWST/Mid Infra Red Imager Optical Bench}, 2008, Proc. SPIE, Vol. 7010
\item
\textbf{Guillard, P.}, Boulanger, F., Pineau des For\^ets, G., Appleton, P.N., \textit{$\rm H_2$ formation and excitation in the Stephan's Quintet galaxy-wide shock}, 2007, 4th Spitzer Conference Proceedings, ``The Evolving ISM in the Milky Way \& Nearby Galaxies''.
\item
\textbf{ Guillard, P.}, Jones, A. P., Tielens, A.G.G.M, \textit{The lifecycle of interstellar dust as constrained by noble gas implantation into SiC grains}, 2007, SF2A-2007: Proceedings of the Annual meeting of the French Society of Astronomy and Astrophysics held in Grenoble, France, July 2-6, 2007, Eds.: J. Bouvier, A. Chalabaev, and C. Charbonnel, p.246
\item
Theureau, G., \textbf{Guillard, P.}, Jegouzo, I., Martin, J.-M., Prugniel, Ph., Serres, B. \textit{HI galaxies and the Nançay Archive}, 2004, SF2A-2004: Semaine de l'Astrophysique Francaise, meeting held in Paris, France, June 14-18, 2004. Edited by F. Combes, D. Barret, T. Contini, F. Meynadier and L. Pagani. Published by EdP-Sciences, Conference Series, 2004, p. 577.

\end{enumerate}

\section{Technical Reports}
\label{publications-technical-reports}
\begin{enumerate}
\item 
 Ronayette, S., \textbf{Guillard, P.}, Pantin, E., Cavarroc, C., Amiaux, J.  \textit{JWST/MIRI ETM Optical Results}, 2008,  Reference: MIRI-RP-00676-CEA
\item 
 \textbf{Guillard, P.} \& Augueres, J.L., \textit{Image Quality Analysis of the JWST MIRI Imager: First Cold Tests done at CEA Saclay, Dec. 2007}, 2008, Reference: MIRI-TN-00743-IAS
\item 
 \textbf{Guillard, P.}, Rodet, T. \& Augueres, J.L., \textit{Image Quality Analysis of the JWST MIRI Imager: High-Resolution PSF Analysis: Microscanning Test}, 2008, Reference: MIRI-TN-00853-IAS
\item 
 \textbf{Guillard, P.} \& Augueres, J.L., \textit{Image Quality Analysis of the JWST MIRI Imager: Focus and FOV Exploration with a Point Source Tests}, 2008, Reference: MIRI-TN-00867-IAS

\end{enumerate}

\chapter{Teaching and public outreach}

\begin{Abstract}
During my PhD thesis, I have teached at the University of Paris Sud 11. This teaching validates my ``Agrégation'' diploma. Here is a short summary of my teaching activities. Besides this, I have participated to public outreach activities at the University of Orsay, such as ``Forum faites de la science'', ``La fête de la science'', or ''Année Mondiale de l'Astronomie 2009''.
\end{Abstract}

\begin{description}
\item[2006-2007] $\,$\\

\begin{itemize}
\item TP de Physique fondamentale (fundamental physics experiments), Master 1.   Supervisor: Laurent Verstraete. 45h.
\item Coupole d'Astrophysique (CCD astrophysical observations with a 14'' telescope), Master 1. Supervisor: Hervé Dole. 15h. 
\item TD de Mécanique (classical mechanics course), L1 (BSc). Supervisor: Arne Keller. 
\end{itemize}

\item[2007-2008]  $\,$\\
\begin{itemize}
\item TP de Physique fondamentale (fundamental physics experiments), Master 1.   Supervisor: Laurent Verstraete. 24h.
\item Coupole d'Astrophysique (CCD astrophysical observations with a 14'' telescope), Master 1. Supervisor: Hervé Dole. 15h. 
\item TD de Mécanique (classical mechanics course), L1 (BSc). Supervisor: Arne Keller. 
\item Astrophysical course in data reduction, Mster 1.  Supervisor: Hervé Dole. 16h. 
\item Astrophysical observation with Master 2 students at the Observatoire de Haute Provence. 3 nights (80 cm, 1.2 m, 2 m telescopes), 30h.
\end{itemize}

\item[2008-2009]  $\,$\\

\begin{itemize}
\item Coupole d'Astrophysique (CCD astrophysical observations with a 14'' telescope), Master 1. Supervisor: Hervé Dole. 20h. 
\item Astrophysical course in data reduction, Mster 1.  Supervisor: Hervé Dole. 16h. 
\item Astrophysical observation with Master 2 students at the Observatoire de Haute Provence. 6 nights (80 cm, 1.2 m, 2 m telescopes), 60h.
\end{itemize}

\end{description}

\chapter{The show must go on!}


\begin{Abstract}

This appendix lists the conferences and schools attended during this PhD thesis. The talks,  posters and invited seminars, as well as some associated abstracts, are presented. These contributions are ordered from the most recent to the oldest.

\end{Abstract}

\section{Conferences}

\subsubsection{``Hunting for the dark: The Hidden Side of Galaxy Formation'', 2009, October 19-23, Qwara, Malta}

\begin{itemize}
\item

\textbf{Poster: ``Molecular gas and dust in galaxy halos: impact on the energetics of galaxy formation and evolution''. P. Guillard, F. Boulanger, N. Nesvadba, P. Ogle, P.N. Appleton, G. Pineau des Forêts.}

One of the most surprising results obtained with the Spitzer space telescope was the discovery of powerful mid-IR H$_2$ line emission from a significant number of sources that are in active phases of galaxy evolution. The underlying physics may be a key ingredient to describe the build-up of baryonic mass in galaxies and their evolution. We focus here on the interpretation of observations of interacting/merging systems and AGN host galaxies. They provide insights on the  exchange of mass between galaxies and the intergalactic medium and the energetics of galaxy formation. The Stephan's Quintet (SQ), a compact group of interacting galaxies, is a striking example showing that dust survives in a high-speed galaxy collision and that H$_2$ forms in the halo of the group. This is also true for galactic winds driven by AGN feedback or starbursts. In SQ and in H$_2$-luminous galaxies in general, we propose that the H$_2$ emission is powered by the dissipation of the mechanical energy provided by galaxy merging or feedback. Shocks driven into multiphase gas lead to efficient dissipation of this energy at small scales, within molecular gas. 
\end{itemize}

\subsubsection{Semaine de l'astrophysique française (SF2A), 2008, June 30 - July 4, Paris, France}

\begin{itemize}
\item

\textbf{Talk, PNG Session: ``H$_2$ energetics in Galaxy-wide Shocks: Insights on starburst triggering in galaxy collisions'', P. Guillard, F. Boulanger, G. Pineau des Forêts, P.N. Appleton, P. Ogle.}

How the gas cools and fuels galaxy formation and what processes regulate star formation in galaxies are some major questions of extragalactic astronomy. We focus on galaxy collisions and mergers which are observed to trigger IR-luminous bursts of star formation. Recently, Spitzer observations show that some interacting systems stand out for having a high H$_2$ to IR luminosity ratio. We propose that these systems represent an intermediate phase in the evolution of mergers, prior to the starburst. Observations and modeling contribute to define a physical framework of the mechanical energy dissipation in this H$_2$-luminous stage. The H$_2$ excitation show that highly-turbulent ($10-50$ km/s) molecular gas is formed within hot gas. The cloud turbulence is powered by a slow energy and momentum transfer from the gas bulk motion. We propose that the timescale to dissipate the collision kinetic energy represents the time necessary for the molecular gas to settle in gravitationally bound clouds and therefore to form stars.

\item

\textbf{Poster, PCMI Session: ``H$_2$ Formation and Excitation in the Stephan's Quintet Galaxy-Wide Shock'', P. Guillard, F. Boulanger, G. Pineau des Forêts, P.N. Appleton}

Spitzer has detected an extremely powerful $(L > 10^{41}$~erg~s$^{-1}$), shock-powered, mid-IR H$_2$ emission towards the Stephan's Quintet (SQ) galaxy group (Appleton et al. 06). This is the first time an almost pure H$_2$ line spectrum has been seen in an extragalactic object. The luminosity in the H$_2$ lines exceeds by a factor of $\approx 2$ the surface brightness in X-rays. How can we explain such a huge amount of molecular gas (M$_{\rm H_2} \approx 4\times 10^7$~M$_{\odot}$) coexisting with a X-ray emitting plasma? How come there is no star formation associated with H$_2$? We summarize in this poster the main results of a scenario (P. Guillard et al. 08) where the molecular gas is formed in the shock. The pre-shock gas is assumed to be inhomogeneous because the shock has induced formation of both hot ($5 \times 10^6$~K) plasma and warm ($10^{2-3}$~K) H$_2$. Our model (P. Guillard et al. 2008) computes the physical and chemical evolution of the shock heated gas, including density inhomogeneities in the pre-shock medium and dust destruction. Schematically, low density preshock gas ($n_{\rm H} < 0.01-0.001$~cm$^{-3}$) is shocked at high speed ($\approx 1000$~km/s for SQ) and becomes a dust-free X-ray emitting plasma. Denser gas ($n_{\rm H} > 0.1$~cm$^{-3}$) is heated at lower temperatures and dust survives. In the context of the SQ shock, this gas had time to cool and become molecular. Only a fraction of the collision energy is used to heat the hot plasma. Therefore, a large fraction of the gas energy is stored in bulk gas motions. We propose that, in such conditions, the H$_2$ emission must be powered by a slow transfer of energy from fast bulk gas motions into low velocity ($< 40$ km/s) supersonic turbulent motions within the molecular gas. We phenomenologically model this turbulent motions by J-type and magnetic C-type shocks into the newly-formed molecular gas. We show that the observed H$_2$ line fluxes are very well reproduced for densities $n_{\rm H}=10^3 - 10^4$~cm$^{-3}$ and shock velocities within the $5-35$~km/s range. SQ is a unique object for studying the physics of galaxy-wide shocks and understanding their key-role in the formation and excitation of H$_2$. This detailed study of the SQ shock defines a physical framework to interpret observations of H$_2$-bright galaxies in general. 

\end{itemize}

\subsubsection{4th Spitzer Conference, The Evolving ISM in the Milky Way and Nearby Galaxies, 2007, December 2-5, Pasadena, USA}

\begin{itemize}
\item \textbf{The Stephan's Quintet: a beautiful case of $\rm \bf H_2$ formation in shocks}

Recently, Spitzer IRS (Infra-Red Spectrometer) observations led to the unexpected detection of extremely bright mid-IR ($L > 10^{41}$~erg~s$^{-1}$) $\rm H_2$ rotational line emission (Appleton et al. 2006) from warm gas towards the group-wide shock in Stephan's Quintet (hereafter SQ). This first result was quickly followed by the detection of bright $\rm H_2$ line emission from more distant galaxies (Egami et al. 2006, Ogle et al. 2007) and from the NGC 1275 and NGC 4696 cooling flows (Johnstone et al. 2007).

Because of the absence or relative weakness of star forming signatures (dust features, ionized gas lines) in the mid-infrared Spitzer spectra, the $\rm  H_2$ emission must be powered by large-scale shocks, associated with galaxy interactions but also possibly with gas infall and AGN feedback. But how can we explain such a huge amount of molecular gas coexisting with a hot plasma at $5 \times 10^{6}$~K~? How come there is no (or very few) star formation associated with $\rm  H_2$?

I will focus this talk on SQ, a group of four interacting galaxies. The shock is created by the galaxy NGC~7318b which is colliding with a tidal stream at a relative velocity of $\sim 1\,000$~km/s. SQ is then a unique object where one can study the physics of galactic wide shocks. I will present the results of a model capable of interpreting the observations. Two key-points of our calculations of the physical and chemical evolution of the shock-heated dusty plasma will be emphasized: dust destruction and the inhomogeneity of the galaxy and intragroup gas. Schematically, low density ($n_{\rm H} < 0.1\,$cm$^{-3}$) gas is shocked at high speed. It becomes the X-ray emitting gas where dust is destroyed. Higher density gas ($n_{\rm H} > 1\,$cm$^{-3}$) is heated to lower temperatures and the dust survives. In the context of the SQ shock, this gas had time to cool and become molecular. 

In the near future, this model will be extended to more distant objects belonging to  this newly discovered and potentially important population of $\rm  H_2$-bright galaxies. In particular, these $\rm H_2$ mid-IR lines appear to be a powerful diagnostic of unexplored steps in galaxy evolution. 

\end{itemize}

\subsubsection{Semaine de l'astrophysique française (SF2A), 2007, July 5-6, Grenoble, France}

\begin{itemize}
\item \textbf{Talk, PCMI Session: ``Noble Gas Implantation into SiC Stardust: Constraints on Astrophysical Processes''. P. Guillard, A. Jones, A.G.G.M. Tielens}

Primitive meteorites contain presolar grains that originated in stellar outflows and supernovae ejecta prior to the formation of the solar system. We show that the measured elemental abundances of noble gas atoms trapped into meteoritic SiC grains can be understood in terms of ion irradiation and erosion in circumstellar and interstellar media. A simple atom implantation and grain erosion model has been developed that allows us to provide new insights on dust processing and on the physical conditions that reign in these environments. 

\end{itemize}

\section{Seminars}

\subsubsection{STScI, John Hopkins University, 2008, October, 14, Baltimore, USA}

\begin{itemize}
\item \textbf{H$_2$ Formation and Excitation in Galaxy-wide Shocks}
\end{itemize}

\subsubsection{Princeton University, 2008, October, 10, Princeton NJ, USA}

\begin{itemize}
\item \textbf{H$_2$ Formation and Excitation in Galaxy-wide Shocks}
\end{itemize}

\subsubsection{KIPAC, Stanford University, 2008, October, 8, Stanford, USA}

\begin{itemize}
\item \textbf{H$_2$ Formation and Excitation in Galaxy-wide Shocks}
\end{itemize}

\subsubsection{IPAC, Caltech, 2008; October, 1, Pasadena, USA}

\begin{itemize}
\item \textbf{H$_2$ Formation and Excitation in Galaxy-wide Shocks}

Spitzer space telescope observations led to the surprising detection
of a diverse set of extragalactic sources whose infrared spectra are
dominated by line emission of molecular hydrogen. The absence or relative weakness
of typical signs of star formation (like dust features or lines of
ionized gas) suggest the presence of large quantities of molecular gas with no (or
very little) associated star formation. These observations set a new light on the
contribution of H$_2$ to the cooling of the interstellar medium, on the relation
between molecular gas and star formation, and on the energetics of galaxy formation.

I will focus on the striking example of the galaxy-wide shock in
Stephan's Quintet, where a galaxy is colliding with a tidal tail at a relative velocity
of 1000 km/s. Observations show that exceptionally turbulent H$_2$ gas is coexisting
with a hot, X-ray emitting plasma. The extreme mid-IR H$_2$ emission from the shock
exceeds that of the X-rays. I will present the observations and show how they can
be interpreted by considering that the shock is moving through an inhomogeneous
medium. Observations suggest that most of the shock energy is transferred to bulk kinetic
energy of the H$_2$ gas. The turbulent energy of the post-shock gas drives a mass
cycle across the different gas phases where H$_2$ is forming out of the hot/warm gas.
This interpretation puts the H$_2$ emission into a broader astrophysical context including
optical and X-ray observations.

We propose that the turbulence in the clouds is powered by a slow
energy and momentum transfer from the bulk motion of the gas, and that the dissipation of
this turbulent energy in turn is powering the H$_2$ emission. The timescale to
dissipate the kinetic energy from the collision may set the relaxation time for the molecular
gas to settle in gravitationally bound clouds and ultimately to form
stars.

Our detailed study of Stephan's Quintet represents an important step
in defining a physical framework to describe the dissipation of mechanical energy
in H$_2$-luminous galaxies. I will discuss the relevance of H$_2$ formation for the
energetics of key processes that participate in galaxy formation: galaxy interactions,
gas accretion, and AGN feedback.

\end{itemize}

\subsubsection{GEPI, Observatoire de Paris-Meudon, 2008, September, 8}

\begin{itemize}
\item \textbf{H$_2$ emission in galaxy-wide shocks: insights in starburst triggering}
\end{itemize}

\section{Schools}

\subsubsection{``Supercomputing \& Numerical Techniques in Astrophysics Fluid Flow Modeling'', Evora, Portugal, 2009, February 2-15}

A training school devoted to an integrated overview of the numerical and computational techniques that are most adequate to model different astrophysical fluid flow phenomena by means of supercomputing simulations. I have attended theoretical and laboratory courses and made use of the supercomputer facility at the University of Coimbra and the University of Evora.

\subsubsection{``Indo-French training school in optical astronomical observations'', Pune, India, 2007, February 12-27}

During the two weeks of the school, one week was devoted for observations with the IUCAA Girawali Observatory 2-m telescope, followed by data reduction and presentation of scientific results. Rest of the time was used to introduce the basic fundamentals of optical observations. With Amit Patak, we presented two projects: ``The IUCAA deep field'',  and ``Spectroscopy of AGN''.

\subsubsection{``International Max Planck Research School: Physics of the Interstellar Medium'', 2006, September 5-13, Heidelberg, Germany}

A one-week school with courses and practical exercises on ISM physics. We also visited the MPI. 

\subsubsection{``\'Ecole des Houches: interstellar dust'', 2006, May 1-5, Les Houches, France} 

A one week school devoted to the physics and chemistry of interstellar dust. 

\section{JWST/MIRI European Consortium Meetings}

\subsubsection{EC meeting,  2009, April, 28-30, Leicester, UK}

I have presented my work on the PSF microscanning analysis of the MIRI instrument. 
I also participated in the scientific discussion in the Galaxy Evolution  and the ISM working groups. I have presented a project about observations of H$_2$ galaxies with MIRI within the Galaxy Evolution working group. 

 \subsubsection{EC meeting,  2009, January, 19-23, Leuven, Belgium}

I have presented my work on the PSF microscanning analysis of the MIRI instrument.
The science working groups were formed during this meeting. 
I participated to the first scientific discussions in the Galaxy Evolution and  ISM working groups.

\subsubsection{EC Science team meeting,  2008, May 26-30, Onsala, Sweden}

I have presented a talk entitled ``Observing H$_2$-luminous galaxies with MIRI: what can we learn on galaxy evolution?'' in the framework on scientific discussions for the preparation of the science of the JWST/MIRI instrument.

\subsubsection{EC Test team meeting,  2007, September 4-8, Leiden, Netherlands}

It was my first European Consortium meeting about the JWST Instrument. I remember having been completely lost by the technical talks and overwhelmed by an incredibly long list of acronyms!

\cleardoublepage
\listoftables
\listoffigures
\renewcommand{\publ}{}
\cleardoublepage
\small
\addcontentsline{toc}{chapter}{Bibliography}
\bibliography{/Users/guillard/bibliography/biblio_bibtex/biblio}


\bibliographystyle{/Users/guillard/Library/texmf/tex/latex/bib_styles/nsf.bst}
\cleardoublepage
\addcontentsline{toc}{chapter}{Index}

\printindex

\end{document}